\documentclass[a4paper,11pt,twoside]{article}
\pdfoutput=1
\usepackage[british]{babel}
\usepackage{cmbright}
\usepackage[T1]{fontenc}
\usepackage[utf8]{inputenc}  
\usepackage{amsfonts}
\usepackage[centertags]{amsmath}
\usepackage{amssymb,amscd,amsbsy}
\usepackage{graphicx}
\usepackage{eurosym}
\usepackage{color}
\usepackage{sidecap}
\usepackage{cite}
\usepackage[pscoord=true]{eso-pic}
\usepackage{fancyhdr}

\usepackage{multirow} 

\usepackage{float} 
\usepackage{tikz}
\usetikzlibrary{shapes,arrows}
\usetikzlibrary{positioning}
\usepackage{placeins} 
\usepackage{subfigure}
\usepackage{tabularx}
\usepackage{common/aas_macros} 

\usepackage[percent]{overpic}

\usepackage[font=small]{caption}

\usepackage{hyperref}

\catcode`\@=11 
\def\parenbar{\mathpalette\p@renb@r}
\def\p@renb@r#1#2{\vbox{%
  \ifx#1\scriptscriptstyle \dimen@.7em\dimen@ii.2em\else
  \ifx#1\scriptstyle \dimen@.8em\dimen@ii.25em\else
  \dimen@1em\dimen@ii.4em\fi\fi \offinterlineskip
  \ialign{\hfill##\hfill\cr
    \vbox{\hrule width\dimen@ii}\cr
    \noalign{\vskip-.3ex}%
    \hbox to\dimen@{$\mathchar300\hfil\mathchar301$}\cr
    \noalign{\vskip-.3ex}%
    $#1#2$\cr}}}
\catcode`\@=12 
\def\nuan{\parenbar{\nu}\kern-0.4ex}

\newcommand{\dml}{\Delta m^{2}_{\rm large} }

\newcommand\mysref[1]{Sec.\;\ref{#1}}
\newcommand\myfref[1]{Fig.\;\ref{#1}}
\newcommand\myeref[1]{Eq.\;\ref{#1}}
\newcommand\mytref[1]{Tab.\;\ref{#1}}

\newcommand{\LoIversion}{19th July 2016}
\begin{document}
\selectlanguage{british}
\thispagestyle{empty}
\AddToShipoutPicture*{%
\setlength{\unitlength}{1.mm}%
\put(20,-13){\includegraphics[height=297mm]{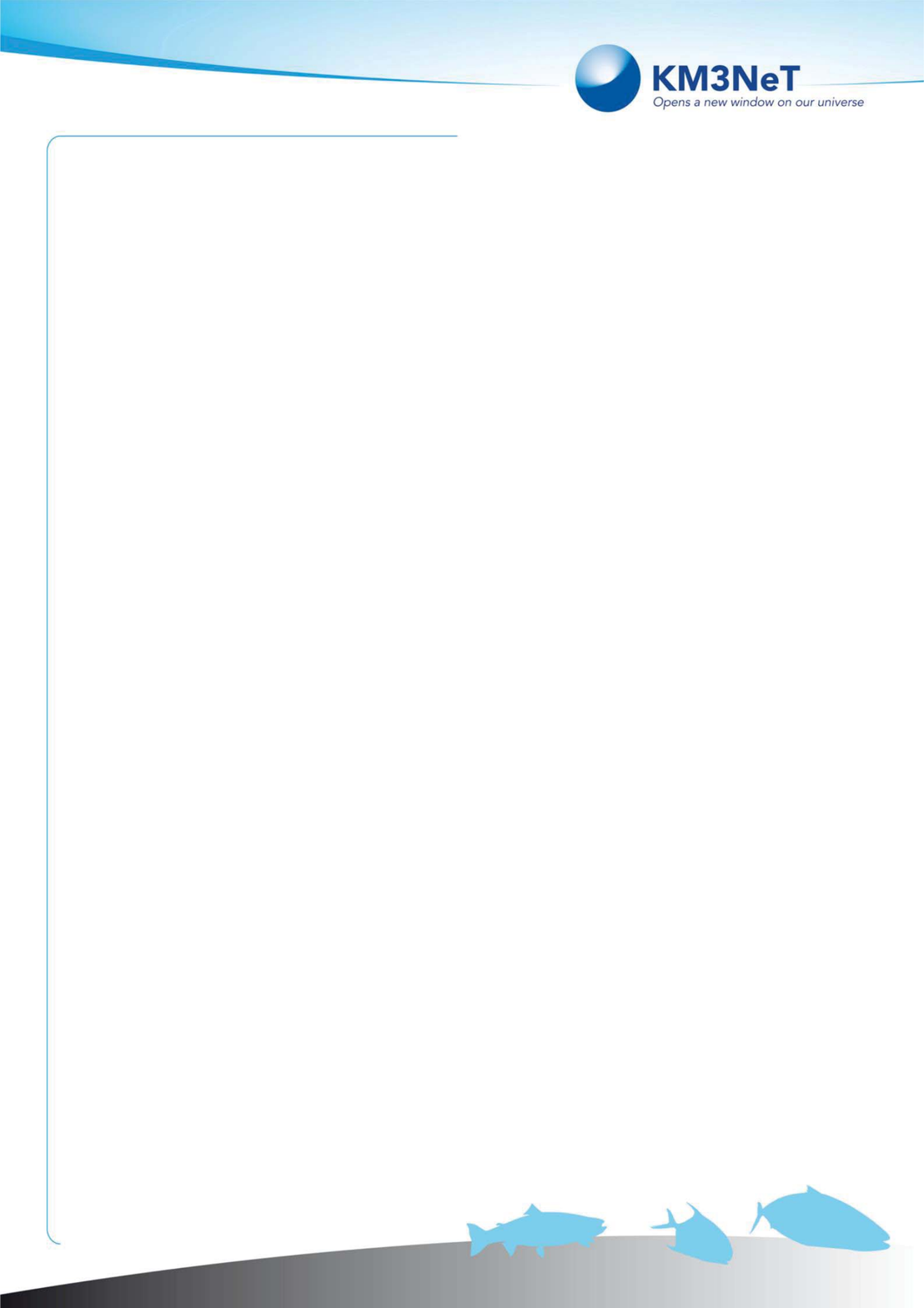}}}
\vspace*{5.cm}

\begin{center}
{\Huge
KM3NeT 2.0\\[2.cm]
Letter of Intent\\
for\\
ARCA and ORCA\\[6.mm]}
{\Large 
-- Astroparticle \& Oscillation Research with Cosmics in the Abyss --\\
[1.cm]
\LoIversion\\
}
\vspace*{8.cm}
Contact: spokesperson@km3net.de
\end{center}
\clearpage
\thispagestyle{empty}

\clearpage
\thispagestyle{empty}

\vspace*{2cm}
\begin{center}
\begin{minipage}{10.05cm}
\begin{center}
\includegraphics*[width=10cm]{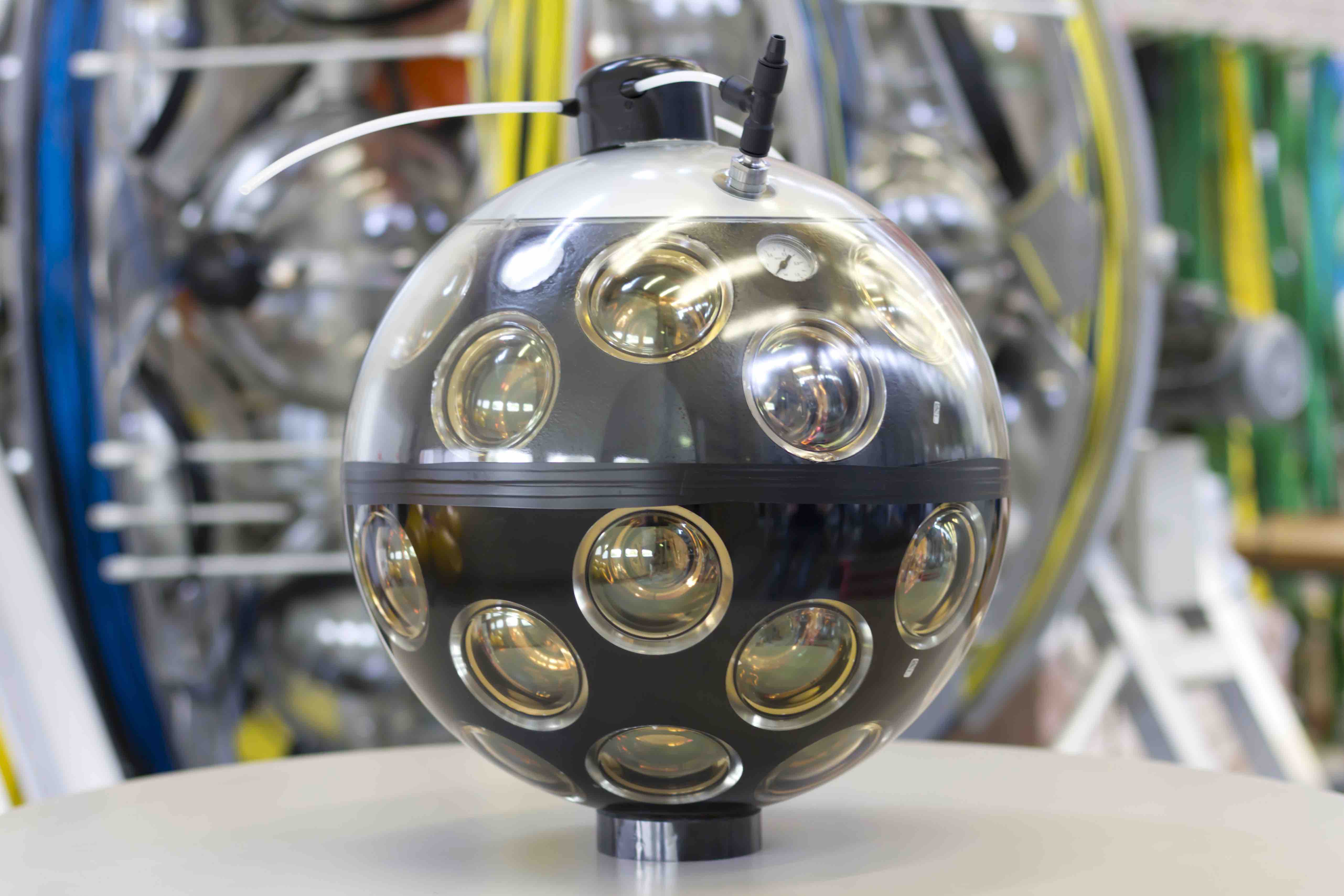}
\end{center}
\noindent
The KM3NeT optical module.
The KM3NeT infrastructure comprises several thousand identical optical modules
arranged in three-dimensional spatial arrays located in the deep waters of the
Mediterranean Sea. The spacing between the optical modules is different for the
ARCA and ORCA detectors to optimally detect neutrinos with the targeted energies. 
Each optical module consists of a glass sphere with a diameter of 42\,cm,
housing 31 photo-sensors (yellowish disks). The glass sphere can withstand
the pressure of the water and is transparent to the faint light that must be
detected to see neutrinos.
\end{minipage}
\end{center}

\clearpage
\thispagestyle{empty}

\cleardoublepage

\pagenumbering{roman}
\pagestyle{fancy}
\fancyhead{}
\fancyfoot{}
\fancyhead[C]{KM3NeT 2.0: Letter of Intent for ARCA and ORCA}
\fancyfoot[L]{\LoIversion}
\fancyfoot[R]{Page \thepage}




\renewcommand{\author}[1]{#1\vspace*{3.mm}}
\providecommand{\address}[1]{\small #1\\}
{\noindent \Large The KM3NeT Collaboration:}\\
{\raggedright

\author{
S.~Adri{\'a}n-Mart{\'\i}nez$^{39}$, 
M.~Ageron$^{2}$, 
F.~Aharonian$^{33}$, 
S.~Aiello$^{18}$, 
A.~Albert$^{40}$, 
F.~Ameli$^{22}$, 
E.~Anassontzis$^{31}$, 
M.~Andre$^{38}$, 
G.~Androulakis$^{29}$, 
M.~Anghinolfi$^{19}$, 
G.~Anton$^{7}$, 
M.~Ardid$^{39}$, 
T.~Avgitas$^{3}$, 
G.~Barbarino$^{20,48}$, 
E.~Barbarito$^{16}$, 
B.~Baret$^{3}$, 
J.~Barrios-Mart\'{i}$^{11}$, 
B.~Belhorma$^{27}$, 
A.~Belias$^{29}$, 
E.~Berbee$^{6}$, 
A.~van~den~Berg$^{25}$, 
V.~Bertin$^{2}$, 
S.~Beurthey$^{2}$, 
V.~van~Beveren$^{6}$, 
N.~Beverini$^{36,21}$, 
S.~Biagi$^{14}$, 
A.~Biagioni$^{22}$, 
M.~Billault$^{2}$, 
M.~Bond{\`\i}$^{18}$, 
R.~Bormuth$^{6,26}$, 
B.~Bouhadef$^{21}$, 
G.~Bourlis$^{10}$, 
S.~Bourret$^{3}$, 
C.~Boutonnet$^{3}$, 
M.~Bouwhuis$^{6}$, 
C.~Bozza$^{49}$, 
R.~Bruijn$^{43}$, 
J.~Brunner$^{2}$, 
E.~Buis$^{34}$, 
J.~Busto$^{2}$, 
G.~Cacopardo$^{14}$, 
L.~Caillat$^{2}$, 
M.~Calamai$^{21}$, 
D.~Calvo$^{11}$, 
A.~Capone$^{37,22}$, 
L.~Caramete$^{24}$, 
S.~Cecchini$^{17}$, 
S.~Celli$^{37,22,9}$, 
C.~Champion$^{3}$, 
R.~Cherkaoui~El~Moursli$^{42}$, 
S.~Cherubini$^{14,45}$, 
T.~Chiarusi$^{17}$, 
M.~Circella$^{16}$, 
L.~Classen$^{7}$, 
R.~Cocimano$^{14}$, 
J.\,A.\,B.~Coelho$^{3}$, 
A.~Coleiro$^{3}$, 
S.~Colonges$^{3}$, 
R.~Coniglione$^{14}$, 
M.~Cordelli$^{15}$, 
A.~Cosquer$^{2}$, 
P.~Coyle$^{2}$, 
A.~Creusot$^{3}$, 
G.~Cuttone$^{14}$, 
A.~D'Amico$^{6}$, 
G.~De~Bonis$^{22}$, 
G.~De~Rosa$^{20,48}$, 
C.~De~Sio$^{49}$, 
F.~Di~Capua$^{20}$, 
I.~Di~Palma$^{37,22}$, 
A.\,F.~D\'\i{}az~Garc\'\i{}a$^{35}$, 
C.~Distefano$^{14}$, 
C.~Donzaud$^{3}$, 
D.~Dornic$^{2}$, 
Q.~Dorosti-Hasankiadeh$^{25}$, 
E.~Drakopoulou$^{29}$, 
D.~Drouhin$^{40}$, 
L.~Drury$^{33}$, 
M.~Durocher$^{14,9}$, 
T.~Eberl$^{7}$, 
S.~Eichie$^{7,8}$, 
D.~van~Eijk$^{6}$, 
I.~El~Bojaddaini$^{41}$, 
N.~El~Khayati$^{42}$, 
D.~Elsaesser$^{51}$, 
A.~Enzenh\"ofer$^{2}$, 
F.~Fassi$^{42}$, 
P.~Favali$^{23}$, 
P.~Fermani$^{22}$, 
G.~Ferrara$^{14,45}$, 
C.~Filippidis$^{29}$, 
G.~Frascadore$^{14}$, 
L.\,A.~Fusco$^{17,44}$, 
T.~Gal$^{7}$, 
S.~Galat\`a$^{3}$, 
F.~Garufi$^{20,48}$, 
P.~Gay$^{3,13}$, 
M.~Gebyehu$^{6}$, 
V.~Giordano$^{18}$, 
N.~Gizani$^{10}$, 
R.~Gracia$^{3}$, 
K.~Graf$^{7}$, 
T.~Gr{\'e}goire$^{3}$, 
G.~Grella$^{49}$, 
R.~Habel$^{15}$, 
S.~Hallmann$^{7}$, 
H.~van~Haren$^{30}$, 
S.~Harissopulos$^{29}$, 
T.~Heid$^{7}$, 
A.~Heijboer$^{6}$, 
E.~Heine$^{6}$, 
S.~Henry$^{2}$, 
J.\,J.~Hern{\'a}ndez-Rey$^{11}$, 
M.~Hevinga$^{25}$, 
J.~Hofest\"adt$^{7}$, 
C.\,M.\,F.~Hugon$^{19}$, 
G.~Illuminati$^{11}$, 
C.\,W.~James$^{7}$, 
P.~Jansweijer$^{6}$, 
M.~Jongen$^{6}$, 
M.~de~Jong$^{6}$, 
M.~Kadler$^{51}$, 
O.~Kalekin$^{7}$, 
A.~Kappes$^{7}$, 
U.\,F.~Katz$^{7}$, 
P.~Keller$^{2}$, 
G.~Kieft$^{6}$, 
D.~Kie{\ss}ling$^{7}$, 
E.\,N.~Koffeman$^{6}$, 
P.~Kooijman$^{43,52}$, 
A.~Kouchner$^{3}$, 
V.~Kulikovskiy$^{14}$, 
R.~Lahmann$^{7}$, 
P.~Lamare$^{2}$, 
A.~Leisos$^{10}$, 
E.~Leonora$^{18}$, 
M.~Lindsey~Clark$^{3}$, 
A.~Liolios$^{4}$, 
C.\,D.~Llorens~Alvarez$^{39}$, 
D.~Lo~Presti$^{18}$, 
H.~L{\"o}hner$^{25}$, 
A.~Lonardo$^{22}$, 
M.~Lotze$^{11}$, 
S.~Loucatos$^{3}$, 
E.~Maccioni$^{36,21}$, 
K.~Mannheim$^{51}$, 
A.~Margiotta$^{17,44}$, 
A.~Marinelli$^{36,21}$, 
O.~Mari\c{s}$^{24}$, 
C.~Markou$^{29}$, 
J.\,A.~Mart{\'\i}nez-Mora$^{39}$, 
A.~Martini$^{15}$, 
R.~Mele$^{20,48}$, 
K.\,W.~Melis$^{6}$, 
T.~Michael$^{6}$, 
P.~Migliozzi$^{20}$, 
E.~Migneco$^{14}$, 
P.~Mijakowski$^{28}$, 
A.~Miraglia$^{14}$, 
C.\,M.~Mollo$^{20}$, 
M.~Mongelli$^{16}$, 
M.~Morganti$^{21,1}$, 
A.~Moussa$^{41}$, 
P.~Musico$^{19}$, 
M.~Musumeci$^{14}$, 
S.~Navas$^{35}$, 
C.\,A.~Nicolau$^{22}$, 
I.~Olcina$^{11}$, 
C.~Olivetto$^{3}$, 
A.~Orlando$^{14}$, 
A.~Papaikonomou$^{10}$, 
R.~Papaleo$^{14}$, 
G.\,E.~P\u{a}v\u{a}la\c{s}$^{24}$, 
H.~Peek$^{6}$, 
C.~Pellegrino$^{17,44}$, 
C.~Perrina$^{37,22}$, 
M.~Pfutzner$^{6}$, 
P.~Piattelli$^{14}$, 
K.~Pikounis$^{29}$, 
G.\,E.~Poma$^{14,45}$, 
V.~Popa$^{24}$, 
T.~Pradier$^{12}$, 
F.~Pratolongo$^{19}$, 
G.~P{\"u}hlhofer$^{5}$, 
S.~Pulvirenti$^{14}$, 
L.~Quinn$^{2}$, 
C.~Racca$^{40}$, 
F.~Raffaelli$^{21}$, 
N.~Randazzo$^{18}$, 
P.~Rapidis$^{29}$, 
P.~Razis$^{46}$, 
D.~Real$^{11}$, 
L.~Resvanis$^{31}$, 
J.~Reubelt$^{7}$, 
G.~Riccobene$^{14}$, 
C.~Rossi$^{19}$, 
A.~Rovelli$^{14}$, 
M.~Salda{\~n}a$^{39}$, 
I.~Salvadori$^{2}$, 
D.\,F.\,E.~Samtleben$^{6,26}$, 
A.~S{\'a}nchez~Garc{\'\i}a$^{11}$, 
A.~S{\'a}nchez~Losa$^{16}$, 
M.~Sanguineti$^{19}$, 
A.~Santangelo$^{5}$, 
D.~Santonocito$^{14}$, 
P.~Sapienza$^{14}$, 
F.~Schimmel$^{6}$, 
J.~Schmelling$^{6}$, 
V.~Sciacca$^{14}$, 
M.~Sedita$^{14}$, 
T.~Seitz$^{7}$, 
I.~Sgura$^{16}$, 
F.~Simeone$^{22}$, 
I.~Siotis$^{29}$, 
V.~Sipala$^{18}$, 
B.~Spisso$^{20}$, 
M.~Spurio$^{17,44}$, 
G.~Stavropoulos$^{29}$, 
J.~Steijger$^{6}$, 
S.\,M.~Stellacci$^{49}$, 
D.~Stransky$^{7}$, 
M.~Taiuti$^{19,47}$, 
Y.~Tayalati$^{41,42}$, 
D.~T{\'e}zier$^{2}$, 
S.~Theraube$^{2}$, 
L.~Thompson$^{50}$, 
P.~Timmer$^{6}$, 
C.~T\"onnis$^{11}$, 
L.~Trasatti$^{15}$, 
A.~Trovato$^{14}$, 
A.~Tsirigotis$^{10}$, 
S.~Tzamarias$^{10}$, 
E.~Tzamariudaki$^{29}$, 
B.~Vallage$^{3}$, 
V.~Van~Elewyck$^{3}$, 
J.~Vermeulen$^{6}$, 
P.~Vicini$^{22}$, 
S.~Viola$^{14}$, 
D.~Vivolo$^{20,48}$, 
M.~Volkert$^{7}$, 
G.~Voulgaris$^{31}$, 
L.~Wiggers$^{6}$, 
J.~Wilms$^{8}$, 
E.~de~Wolf$^{6,43}$, 
K.~Zachariadou$^{32}$, 
J.\,D.~Zornoza$^{11}$, 
J.~Z{\'u}{\~n}iga$^{11}$

}
}

\vspace*{0.5cm}

{\noindent \Large Additional Supporters:}\\
{\raggedright

\author{D.~Allard$^{\text a}$,
A.~Esmaili$^{\text{b,i}}$,
G.~Fogli$^{\text{f,m}}$,
O.~Mena$^{\text e}$,
K.~Murase$^{\text c}$,
S.~Palomares Ruiz$^{\text e}$,
E.~Parizot$^{\text a}$,
S.~T.~Petcov$^{\text{h,k}}$,
S.~Razzaque$^{\text n}$,
D.~Semikoz$^{\text a}$,
A.~Yu.~Smirnov$^{\text{g,j}}$,
F.L.~Villante$^{\text{i,o}}$
F.~Vissani$^{\text{d,i}}$,
R.~Zukanovich Funchal$^{\text l}$
}
}
\newpage
{\noindent \Large Affiliations of KM3NeT members:}\\
{\raggedright

\address{$^{1}$~Accademia Navale di Livorno, Viale Italia 72, Livorno, 57100 Italy}
\address{$^{2}$~Aix-Marseille Universit{\'e},~CNRS/IN2P3,~CPPM~UMR~7346,~13288,~Marseille,~France}
\address{$^{3}$~APC, Universit{\'e} Paris Diderot, CNRS/IN2P3, CEA/IRFU, Observatoire de Paris, Sorbonne Paris Cit\'e, 75205 Paris, France}
\address{$^{4}$~Aristotle University Thessaloniki, University Campus, Thessaloniki, 54124 Greece}
\address{$^{5}$~Eberhard Karls Universit{\"a}t T{\"u}bingen, Institut f{\"u}r Astronomie und Astrophysik, Sand 1, 72076 T{\"u}bingen, Germany}
\address{$^{6}$~FOM, Nikhef, PO Box 41882, Amsterdam, 1098 DB Netherlands}
\address{$^{7}$~Friedrich-Alexander-Universit{\"a}t Erlangen-N{\"u}rnberg, Erlangen Centre for Astroparticle Physics, Erwin-Rommel-Stra{\ss}e 1, 91058 Erlangen, Germany}
\address{$^{8}$~Friedrich-Alexander-Universit{\"a}t Erlangen-N{\"u}rnberg, Remeis Sternwarte, Sternwartstra{\ss}e 7, 96049 Bamberg, Germany}
\address{$^{9}$~Gran Sasso Science Institute, GSSI, Viale Francesco Crispi 7, L'Aquila, 67100  Italy}
\address{$^{10}$~Hellenic Open University, School of Science / Technology, Natural Sciences, Sahtouri St. / Ag. Andreou St. 16, Patra, 26222 Greece}
\address{$^{11}$~IFIC - Instituto de F{\'\i}sica Corpuscular (CSIC - Universitat de Val{\`e}ncia), c/Catedr{\'a}tico Jos{\'e} Beltr{\'a}n, 2, 46980 Paterna, Valencia, Spain}
\address{$^{12}$~IN2P3, IPHC, 23 rue du Loess, Strasbourg, 67037 France}
\address{$^{13}$~IN2P3, LPC, Campus des C{\'e}zeaux 24, avenue des Landais BP 80026, Aubi{\`e}re Cedex, 63171 France}
\address{$^{14}$~INFN, Laboratori Nazionali del Sud, Via S. Sofia 62, Catania, 95123 Italy}
\address{$^{15}$~INFN, LNF, Via Enrico Fermi , 40, Frascati, 00044 Italy}
\address{$^{16}$~INFN, Sezione di Bari, Via Amendola 173, Bari, 70126 Italy}
\address{$^{17}$~INFN, Sezione di Bologna, v.le C. Berti-Pichat, 6/2, Bologna, 40127 Italy}
\address{$^{18}$~INFN, Sezione di Catania, Via Santa Sofia 64, Catania, 95123 Italy}
\address{$^{19}$~INFN, Sezione di Genova, Via Dodecaneso 33, Genova, 16146 Italy}
\address{$^{20}$~INFN, Sezione di Napoli, Complesso Universitario di Monte S. Angelo, Via Cintia ed. G, Napoli, 80126 Italy}
\address{$^{21}$~INFN, Sezione di Pisa, Largo Bruno Pontecorvo 3, Pisa, 56127 Italy}
\address{$^{22}$~INFN, Sezione di Roma, Piazzale Aldo Moro 2, Roma, 00185 Italy}
\address{$^{23}$~INGV, Via di Vigna Murata, 605, Rome, 00143 Italy}
\address{$^{24}$~ISS, 242, Vacaresti, Bucharest, 40061 Romania}
\address{$^{25}$~KVI-CART~University~of~Groningen,~Groningen,~The~Netherlands}
\address{$^{26}$~Leiden University, Leiden Institute of Physics, PO Box 9504, Leiden, 2300 RA Netherlands}
\address{$^{27}$~National Center for Energy Sciences and Nuclear Techniques, B.P.~1382, R.P.~10001 Rabat, Morocco}
\address{$^{28}$~National~Centre~for~Nuclear~Research,~00-681~Warsaw,~Poland}
\address{$^{29}$~NCSR Demokritos, Institute of Nuclear and Particle Physics, Ag. Paraskevi Attikis, Athens, 15310 Greece}
\address{$^{30}$~NIOZ, PO Box 59, Den Burg, Texel, 1790 AB Netherlands}
\address{$^{31}$~Physics~Department,~N.~and~K.~University~of~Athens,~Athens,~Greece}
\address{$^{32}$~Technological Education Institute of Pireaus, Thivon and P. Ralli Str. 250, Egaleo - Athens, 12244 Greece}
\address{$^{33}$~The Dublin Institute for Advanced Studies, 10 Burlington Road, Dublin, 4 Ireland}
\address{$^{34}$~TNO, Technical Sciences, PO Box 155, Delft, 2600 AD Netherlands}
\address{$^{35}$~Universidad~de~Granada~\&~C.A.F.P.E, Av.~del Hospicio s/n, 18071~Granada, Spain}
\address{$^{36}$~Universit{\`a} di Pisa, Dipartimento di Fisica, Largo Bruno Pontecorvo 3, Pisa, 56127 Italy}
\address{$^{37}$~Universit{\`a} La Sapienza, Dipartimento di Fisica, Piazzale Aldo Moro 2, Roma, 00185 Italy}
\address{$^{38}$~Universitat Polit{\`e}cnica de Catalunya, Laboratori d'Aplicacions Bioac{\'u}stiques, Centre Tecnol{\`o}gic de Vilanova i la Geltr{\'u}, Avda. Rambla Exposici{\'o}, s/n, Vilanova i la Geltr{\'u}, 08800 Spain}
\address{$^{39}$~Universitat Polit{\`e}cnica de Val{\`e}ncia, Instituto de Investigaci{\'o}n para la Gesti{\'o}n Integrada de las Zonas Costeras, C/ Paranimf, 1, Gandia, 46730 Spain}
\address{$^{40}$~Universit{\'e} de Strasbourg, Universit{\'e} de Haute Alsace, GRPHE, 34, Rue du Grillenbreit, Colmar, 68008 France}
\address{$^{41}$~University Mohammed I, Faculty of Sciences, BV Mohammed VI, B.P.~717, R.P.~60000 Oujda, Morocco}
\address{$^{42}$~University Mohammed V in Rabat, Faculty of Sciences, 4 av.~Ibn Battouta, B.P.~1014, R.P.~10000 Rabat, Morocco}
\address{$^{43}$~University of Amsterdam, Institute of Physics/IHEF, PO Box 94216, Amsterdam, 1090 GE Netherlands}
\address{$^{44}$~University of Bologna, Dipartimento di Fisica e Astronomia, v.le C. Berti-Pichat, 6/2, Bologna, 40127 Italy}
\address{$^{45}$~University of Catania, Dipartimento di Fisica ed Astronomia di Catania, Via Santa Sofia 64, Catania, 95123 Italy}
\address{$^{46}$~University of Cyprus, Physics, Kallipoleos 75, Nicosia, 1678 Cyprus}
\address{$^{47}$~University of Genova, Via Dodecaneso 33, Genova, 16146 Italy}
\address{$^{48}$~University of Napoli, Dip. Scienze fisiche, Complesso Universitario di Monte S. Angelo, Via Cintia ed. G, Napoli, 80126 Italy}
\address{$^{49}$~University of Salerno, Department of Physics, Via Giovanni Paolo II 132, Fisciano, 84084 Italy}
\address{$^{50}$~University of Sheffield, Department of Physics and Astronomy, Hounsfield Road, Sheffield, S3 7RH United Kingdom}
\address{$^{51}$~University W{\"u}rzburg, Emil-Fischer-Stra{\ss}e 31, 97074 W{\"u}rzburg, Germany}
\address{$^{52}$~Utrecht University, Department of Physics and Astronomy, PO Box 80000, Utrecht, 3508 TA Netherlands}
}
\vspace*{0.5cm}
{\noindent \Large Affiliations of additional supporters:}\\
{\raggedright

\address{$^{\text a}$~APC, Universit{\'e} Paris Diderot, CNRS/IN2P3,
CEA/IRFU, Observatoire de Paris, Sorbonne Paris Cit\'e, 75205 Paris,
France} 
\address{$^{\text b}$~Departamento de F\'isica, Pontif\'icia Universidade Cat\'olica do Rio de
Janeiro, C.P. 38071, 22452-970, Rio de Janeiro, Brazil} 
\address{$^{\text c}$~Department of Physics; Department of Astronomy \&
Astrophysics, The Pennsylvania State University, University Park,
Pennsylvania 16802, USA} 
\address{$^{\text d}$~Gran Sasso Science Institute, GSSI, Viale Francesco
Crispi 7, L'Aquila, 67100 Italy} 
\address{$^{\text e}$~IFIC - Instituto de F{\'\i}sica Corpuscular (CSIC -
Universitat de Val{\`e}ncia), c/Catedr{\'a}tico Jos{\'e} Beltr{\'a}n,
2, 46980 Paterna, Valencia, Spain} 
\address{$^{\text f}$~INFN, Sezione di Bari, Via Amendola 173, 70126 Bari, Italy}
\address{$^{\text g}$~International Centre for Theoretical Physics, Strada
Costiera 11, I-34100 Trieste, Italy} 
\address{$^{\text h}$~Kavli IPMU (WPI), University of Tokyo, 5-1-5
Kashiwanoha, 277-8583 Kashiwa, Japan} 
\address{$^{\text i}$~Laboratori Nazionali del Gran Sasso, INFN, Assergi
(AQ), Italy}
\address{$^{\text j}$~Max-Planck-Institut für Kernphysik, Saupfercheckweg 1,
69117 Heidelberg, Germany} 
\address{$^{\text k}$~SISSA/INFN, Via Bonomea 265, 34136 Trieste, Italy} 
\address{$^{\text l}$~Universidade de S\~{a}o Paulo, Instituto de
F{\'\i}sica, Departamento de F{\'\i}sica Matem{\'a}tica, C.P.66318,
05315-970 S\~{a}o Paulo, Brazil} 
\address{$^{\text m}$~University of Bari, Dipartimento di Fisica, Via
Amendola 173, 70126 Bari, Italy}
\address{$^{\text n}$~University of Johannesburg, Department of Physics, PO
Box 524, Auckland Park 2006, South Africa} 
\address{$^{\text o}$~University of L'Aquila, Dip. Scienze Fisiche e Chimiche, via Vetoio, 67100
L'Aquila, Italy} 
}
\small

\cleardoublepage

\tableofcontents
\cleardoublepage
\pagenumbering{arabic}
\fancyhead{}
\fancyfoot{}
\fancyhead[CE]{KM3NeT 2.0: Letter of Intent for ARCA and ORCA}
\renewcommand\sectionmark[1]{\markboth{\thesection\ #1}{}}
\fancyhead[CO]{Executive Summary}
\fancyfoot[L]{\LoIversion}
\fancyfoot[R]{Page \thepage~of~\pageref{lastpage}}
\section*{Executive Summary}
\addcontentsline{toc}{section}{Executive Summary}

The main objectives of the KM3NeT\footnote{http://www.km3net.org} Collaboration
are {\em i)} the discovery and subsequent observation of high-energy neutrino
sources in the Universe and {\em ii)} the determination of the mass hierarchy of
neutrinos. These objectives are strongly motivated by two recent important
discoveries, namely: {\em 1)} The high-energy astrophysical neutrino signal
reported by IceCube and {\em 2)} the sizeable contribution of electron neutrinos
to the third neutrino mass eigenstate as reported by Daya Bay, Reno and others. 
To meet these objectives, the
KM3NeT Collaboration plans to build a new Research Infrastructure consisting of
a network of deep-sea neutrino telescopes in the Mediterranean Sea. A phased and
distributed implementation is pursued which maximises the access to regional funds, 
the availability of human resources and the synergistic opportunities 
for the earth and sea sciences community. 
Three deep-sea sites  are selected for the optical properties of the water, 
distance to shore and local infrastructure, 
namely off-shore Toulon (France), Capo Passero (Sicily, Italy) and Pylos (Peloponnese, Greece).

The infrastructure will consist of three so-called building blocks. A building
block comprises 115~strings, each string comprises 18 optical modules and each
optical module comprises 31~photo-multiplier tubes (PMTs). Each building block
thus constitutes a 3-dimensional array of photo sensors that can be used to
detect the Cherenkov light produced by relativistic particles emerging from
neutrino interactions. Two building blocks will be sparsely configured to fully explore
the IceCube signal with comparable instrumented volume, different methodology, 
improved resolution and complementary field of view, including the Galactic plane.
Collectively, these
building blocks are referred to as ARCA: Astroparticle Research with
Cosmics in the Abyss. One building block will be densely configured to precisely
measure atmospheric neutrino oscillations. This building block is referred to as
ORCA: Oscillation Research with Cosmics in the Abyss. ARCA will be
realised at the Capo Passero site and ORCA at the Toulon site. Due to KM3NeT's
flexible design, the technical implementation of ARCA and ORCA is almost
identical. The deep-sea sites are linked to shore with a network of cables for
electrical power and high-bandwidth data communication. On site, shore stations
are equipped to provide power, computing and a high-bandwidth internet
connection to the data repositories. The readout of the detectors is based on
the ``All-data-to-shore'' concept, pioneered in ANTARES. \mbox{~}
The overall design allows for 
a flexible and cost-effective implementation of the Research Infrastructure and 
its low-cost operation. 
The costs remaining to realise ARCA and ORCA amount to 95\,M\euro. 
The operational costs are estimated at about 2\,M\euro~per year, 
equivalent to less than 2\% of the total investment costs.

The whole project is organised in a single Collaboration with a central
management and common data analysis and repository centres. A Memorandum of
Understanding for the first phase (Phase-1), covering the currently available
budget of about 31\,M\euro, has been signed by the representatives of the
corresponding funding agencies. 
During Phase-1, 
the technical design has been validated through in situ prototypes; 
data analysis tools have been developed; 
assembly sites for the production of optical modules and strings have been setup;
and deployment and connection of strings in the deep 
sea are being optimised for speed and reliability.
During the next phase
(Phase-2.0), the Collaboration will complete the construction of ARCA and ORCA
by 2020. The ultimate goal is to fully develop the KM3NeT Research
Infrastructure to comprise a distributed installation at the three foreseen
sites (Phase-3)  and operate it for ten years or more. The phased implementation of the KM3NeT Research Infrastructure
is summarised in Table \ref{tab:phased_implementation}. 
The Collaboration aspires to establish a European Research Infrastructure Consortium (ERIC)
hosted in the Netherlands. 

\begin{figure}[H]
\centering
\includegraphics[width=0.55\textwidth, trim=0 4mm 0 8mm]{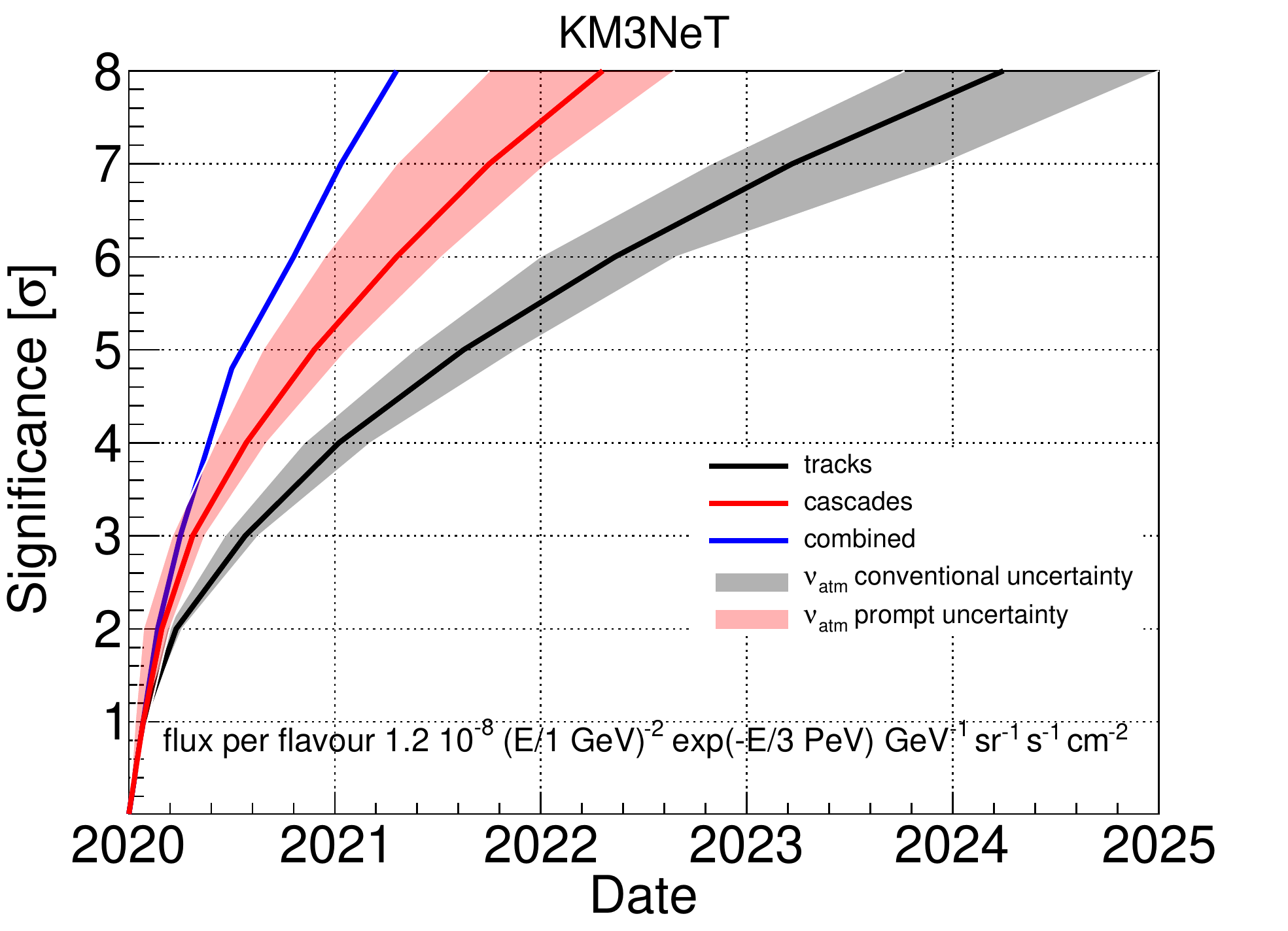}
\caption{%
Significance as a function of time for the detection of a
diffuse flux of neutrinos corresponding to the signal reported by IceCube, for
cascade-like events (red line) and track-like events (black line). The black and
red bands represent the uncertainties due to the conventional and prompt
component of the atmospheric neutrino flux, respectively. The blue line
indicates the result of the combined analysis.}
\label{fig.SignificanceExecSum_arca}
\end{figure}

\begin{figure}[H]
\centering
\includegraphics[width=0.55\textwidth, trim=0 4mm 0 8mm]{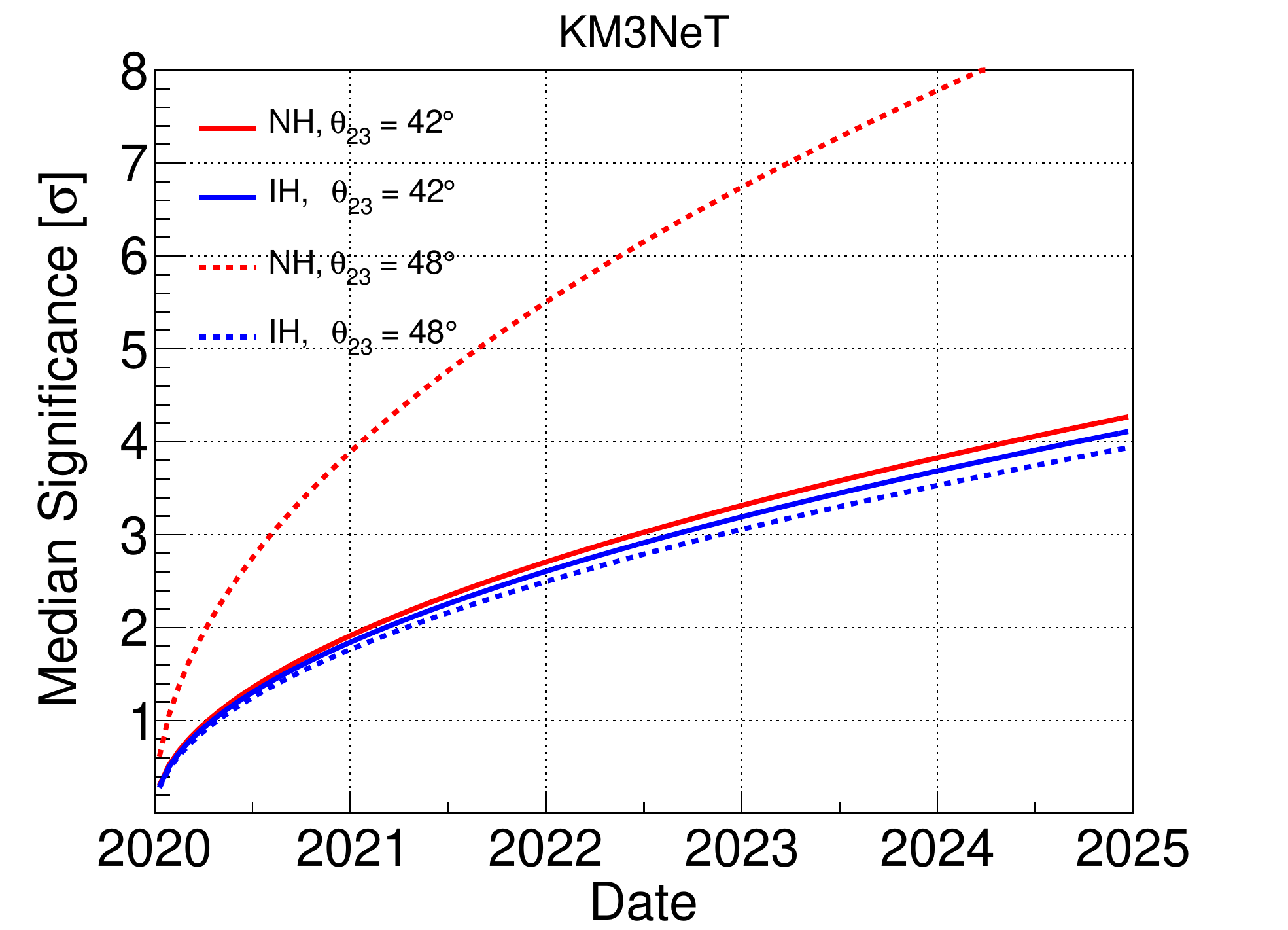}
\caption[]{Median significance as a function of time for the 
determination of the neutrino mass hierarchy. The different lines
denote expectations for different combinations of hierarchy and
atmospheric mixing angle $\theta_{23}$. Note that the CP-violating
phase $\delta_{CP}$ has been assumed to be zero.}
\label{fig.SignificanceExecSum_orca}
\end{figure}

The first part of this document focuses on the technical design of the infrastructure.
As a preview to the science objectives presented later in this document, 
\myfref{fig.SignificanceExecSum_arca} shows the significance as a function of
time for the detection of a diffuse, flavour-symmetric neutrino flux
corresponding to the result reported by IceCube. Thanks to the purity of the
event sample, a high-significance detection of this neutrino flux will be
possible for both track-like and cascade-like events within one year of
operation. The excellent angular and energy resolutions, combined with the large
effective mass, provide for a significant discovery potential to find neutrino
sources in the Universe.
\myfref{fig.SignificanceExecSum_orca} shows the significance as a function of observation time
for the determination of the neutrino mass hierarchy.
A determination of the neutrino mass hierarchy with at least 3-sigma significance can be made after three years of operation,
i.e.\ as early as 2023. This precedes results of other experiments and provides timely input
for experiments aiming at a measurement of the CP-violation phase with high sensitivity.
In addition, ORCA will provide improved measurements of some of the neutrino oscillation parameters.

\begin{table*}[h!]
\small
\begin{center}
\begin{tabular}{|c|c|c|c|l|}
\hline
{\bf Phase}      & {\bf Total costs} & {\bf Building blocks} & {\bf Start} & {\bf Primary deliverables}                         \\
\mbox{}          &    (cumulative)   &                       &             &                                                    \\
\mbox{}          &    M\euro         &                       &             &                                                    \\
\hline \hline
\mbox{}  1       &       31          &          0.2          &    2013     &  Proof of feasibility and first science results.   \\
\hline
\mbox{} 2.0      &      125          &
\begin{tabular}{@{}c@{}}
    2 \\
    1 \\
\end{tabular}                                                &    2017     &
\begin{tabular}{@{}l}
  Study of the neutrino signal reported by IceCube;  \\
  Determination of the neutrino mass hierarchy.      \\
  \end{tabular}                                                                                                                 \\
\hline
\mbox{}  3       &    220--250       &            6          &    2025     &  Neutrino astronomy including Galactic sources.    \\
\hline
\end{tabular}
\end{center}
\caption{%
Phased implementation of the KM3NeT Research Infrastructure (see text). 
The quoted costs for each phase include the costs of the previous phase(s). 
The funds for Phase-1 are secured.}
\label{tab:phased_implementation}
\end{table*}

\cleardoublepage
\fancyhead[CO]{\leftmark}


\section{Detector Design and Technology}
\label{sec-tec}

The successful deployment and operation of the ANTARES neutrino
telescope~\cite{ANT-Detector-2011} has demonstrated the
feasibility of performing neutrino studies with large volume detectors in the
deep sea. The detection of neutrinos is based on the detection of Cherenkov
light produced by relativistic particles emerging from a neutrino interaction.
The same technology can be used for studying neutrinos from GeV (for KM3NeT/ORCA)
to PeV energies and above (for KM3NeT/ARCA).

The goal of the KM3NeT technology is to instrument, at minimal cost and maximal
reliability, the largest possible volume of seawater with a three dimensional
spatial grid of ultra-sensitive photo-sensors, while remaining sensitive to
neutrino interactions in the target energy range.
The KM3NeT design builds upon the ANTARES experience and 
improves the cost effectiveness of its design by about a factor four.
All components are designed for at least ten years
of operation with negligible loss of efficiency.
The system should provide
nanosecond precision on the arrival time of single photons, while the position
and orientation of the photo-sensors must be known to a few centimetres and few
degrees, respectively. The photo-sensors and the readout electronics are hosted
within pressure-resistant glass spheres, so called digital optical modules
(DOMs). The DOMs are distributed in space along flexible strings, one end of
which is fixed to the sea floor and the other end is held close to vertical by a
submerged buoy. The concept of strings is modular by design. The construction and
operation of the research infrastructure thus allows for a phased and
distributed implementation.

A collection of 115 strings forms a single KM3NeT building block. 
The modular design allows building blocks with different 
spacings between lines/DOMs to be constructed, in order to target
different neutrino energies. The full KM3NeT telescope comprises seven building blocks
distributed on three sites. For Phase-2.0, three building blocks are planned:
two KM3NeT/ARCA blocks, with a large spacing to target astrophysical neutrinos at
TeV energies and above; and one KM3NeT/ORCA block, to target atmospheric neutrinos
in the few-GeV range.

\begin{figure}[!hbt]
\centering
   \includegraphics[width=0.65\linewidth] {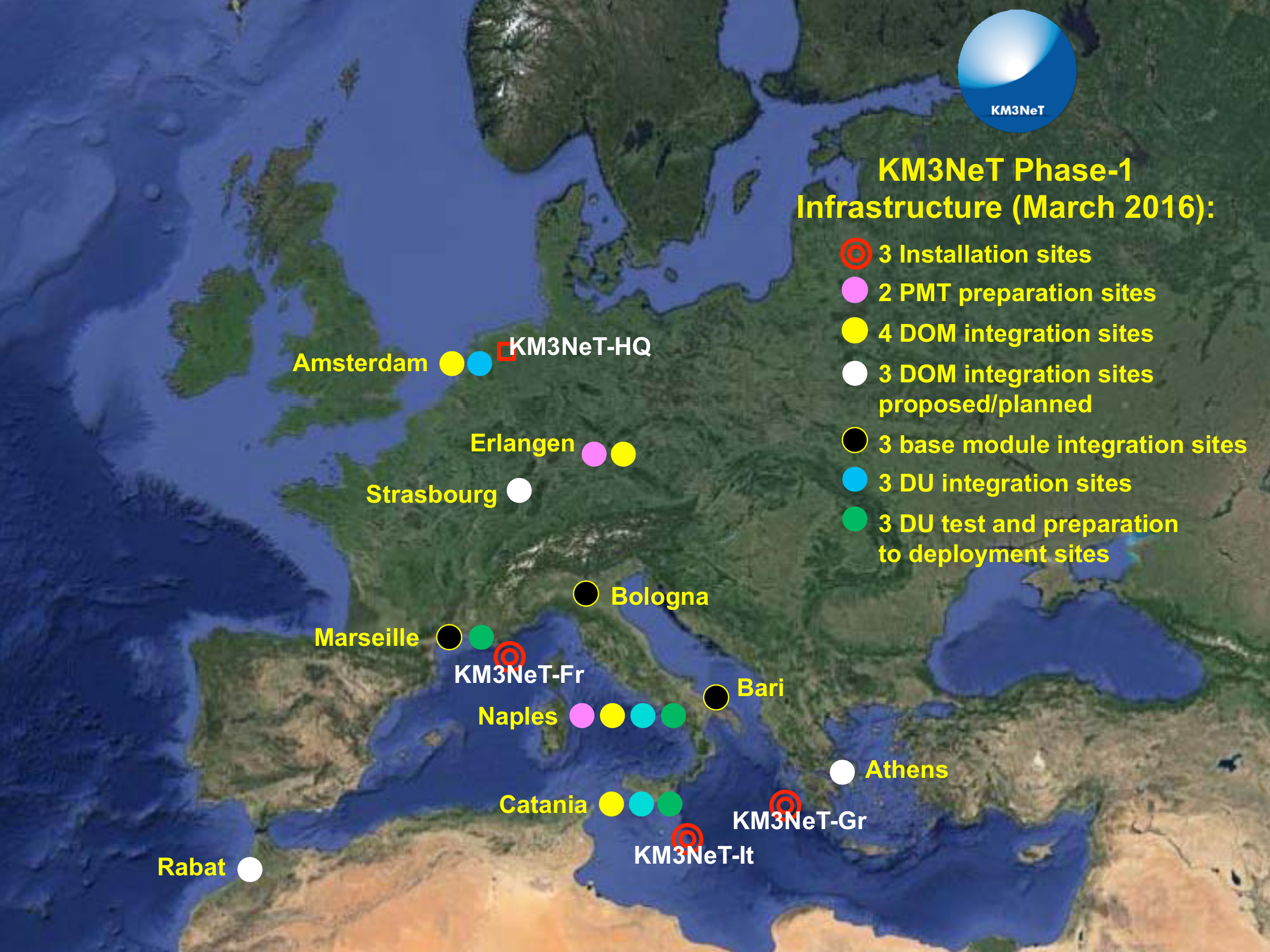}
\caption{%
Map of the various preparation, integration and installation sites at the time of this writing.
}
\label{fig:production_sites}
\end{figure}

\myfref{fig:production_sites} indicates the location of the KM3NeT deep sea sites and 
the location of the various institutes which are currently involved in the PMT testing, the DOM integration, 
the string integration and the deployment of strings for KM3NeT Phase-1.

\subsection{KM3NeT/ARCA: deep sea and onshore infrastructures}
\label{sec:deepsea_arca}

\begin{figure}[!hbt]
  \raisebox{-0.5\height}{\includegraphics[height=2.in]{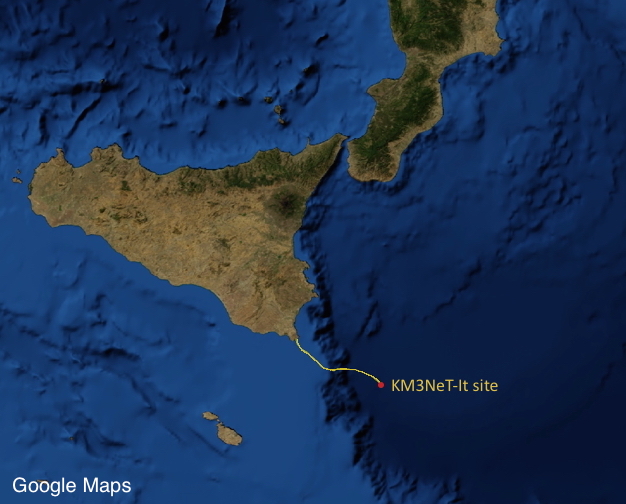}}
  \hspace*{.8in}
  \raisebox{-0.5\height}{\includegraphics[height=3.5in]{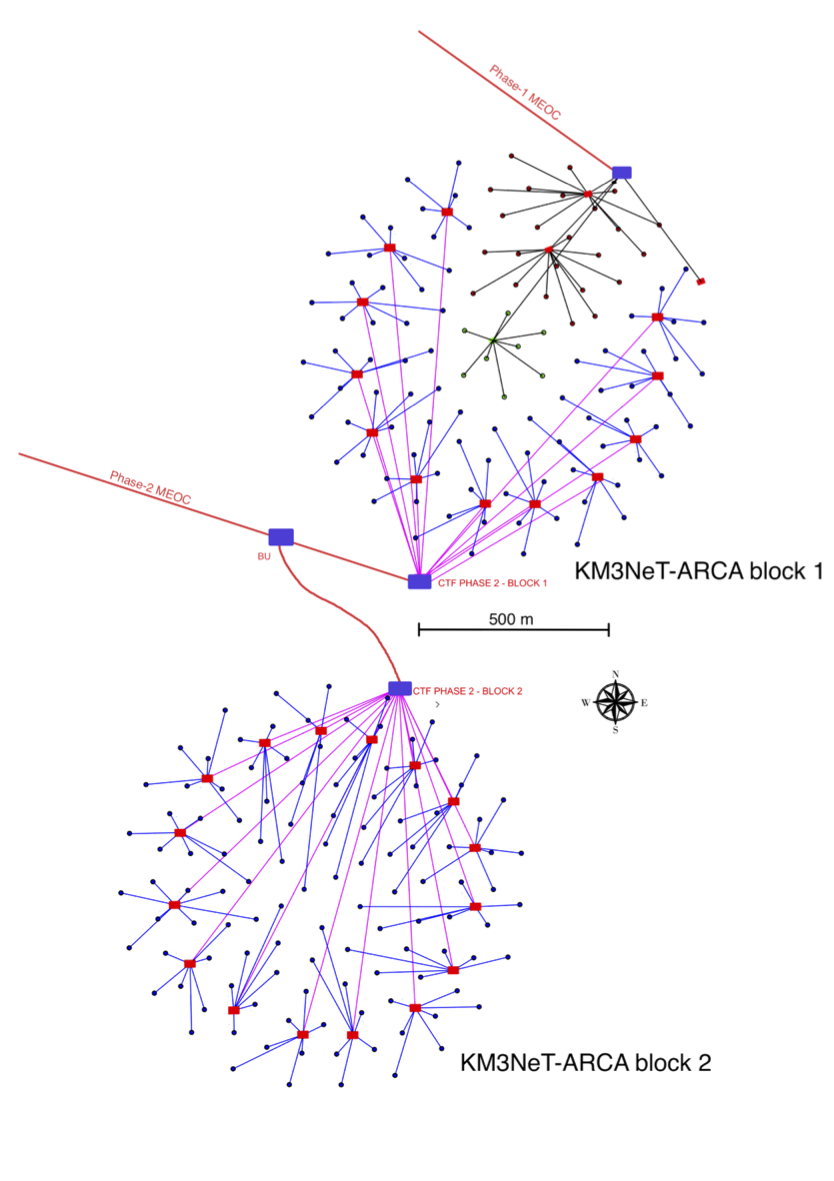}}
\caption{%
Map of the Mediterranean Sea close to Sicily, Italy. The cable and the
location of the KM3NeT-Italy installation are indicated (left).
Layout of the two ARCA building blocks (right).} 
\label{fig:arca_layout}
\end{figure}

\begin{figure}[!hbt]
  \includegraphics[width=0.45\linewidth] {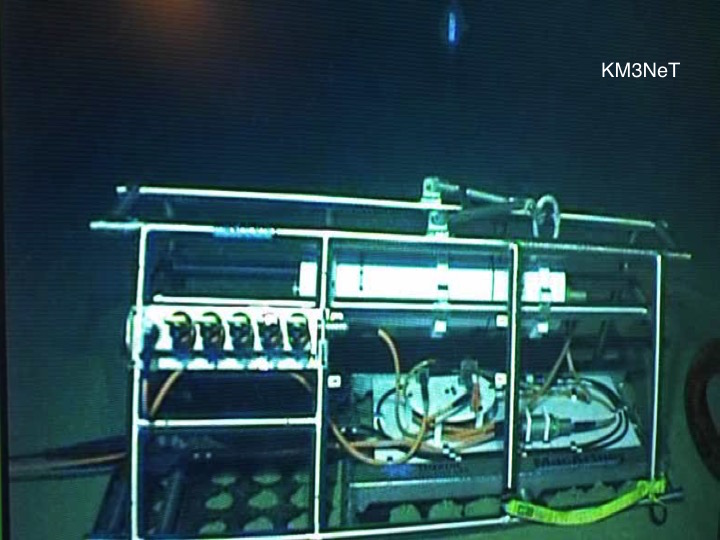}
\hfil
   \includegraphics[width=0.45\linewidth] {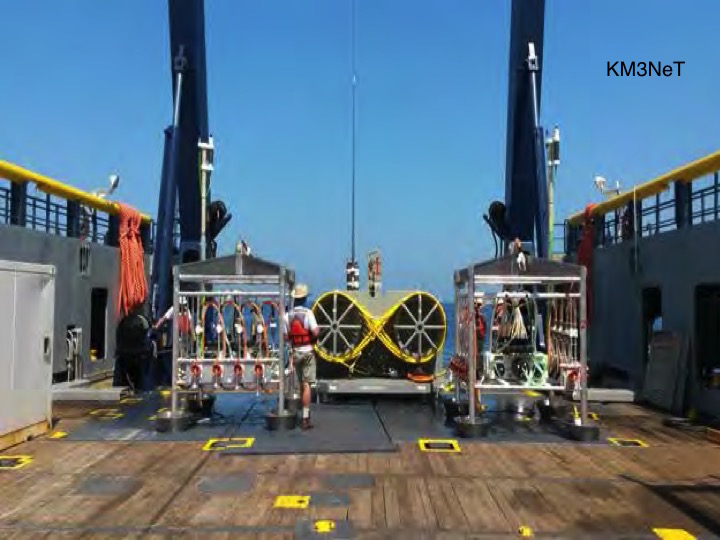}
\caption{%
Photograph of the CTF after deployment on the seabed (left). 
Photograph of two secondary junction boxes on the boat prior to
deployment (right).} 
\label{fig:arca_node}
\end{figure}

The KM3NeT-Italy infrastructure is located at 36$^\circ$ 16' N 16$^\circ$ 06' E
at a depth of 3500\,m, about 100\,km offshore from Porto Palo di Capo Passero,
Sicily, Italy (\myfref{fig:arca_layout}, left). The site is the former NEMO site and is
shared with the EMSO facility for Earth and Sea science research.

The ARCA installation comprises two KM3NeT building blocks. \myfref{fig:arca_layout} right illustrates the layout.
The power/data are transferred to/from the infrastructure via two main electro-optic cables.
In addition to the already operating cable serving the Phase-1 detector a new cable will be installed.
This Phase-2 cable will comprise 48 optical fibres.
Close to the underwater installation the cable is split by means of a
Branching Unit (BU) in two branches,
each one terminated with a Cable
Termination Frame (CTF) (\myfref{fig:arca_node}, left).
Each CTF is connected to secondary junction boxes, 12 for the ARCA block 1 and 16 for the ARCA block 2.
Each secondary junction box allows the connection of up to 7 KM3NeT detection strings. The underwater connection of the strings to
the junction boxes is via interlink cables running along the seabed. For the ARCA configuration, the average
horizontal spacing between detection strings is about 95\,m.
On-shore each main electro-optic cable is connected to a power feeding equipment located in the shore station at
Porto Palo di Capo Passero. Power is transferred at 10 kVDC and is converted to 375 VDC at the CTF
for transmission, via the secondary junction boxes, along the interlink cables to the strings.
The shore station also hosts the data acquisition electronics and a
commodity PC farm used for data filtering.

In December, 2008, the first main electro-optic cable was deployed. 
A CTF and two secondary junction boxes were successfully connected in summer 2015.

\subsection{KM3NeT/ORCA: deep sea and onshore infrastructures}
\label{sec:deepsea_orca}

\begin{figure}[!hbt]
  \raisebox{-0.5\height}{\includegraphics[height=2.in]{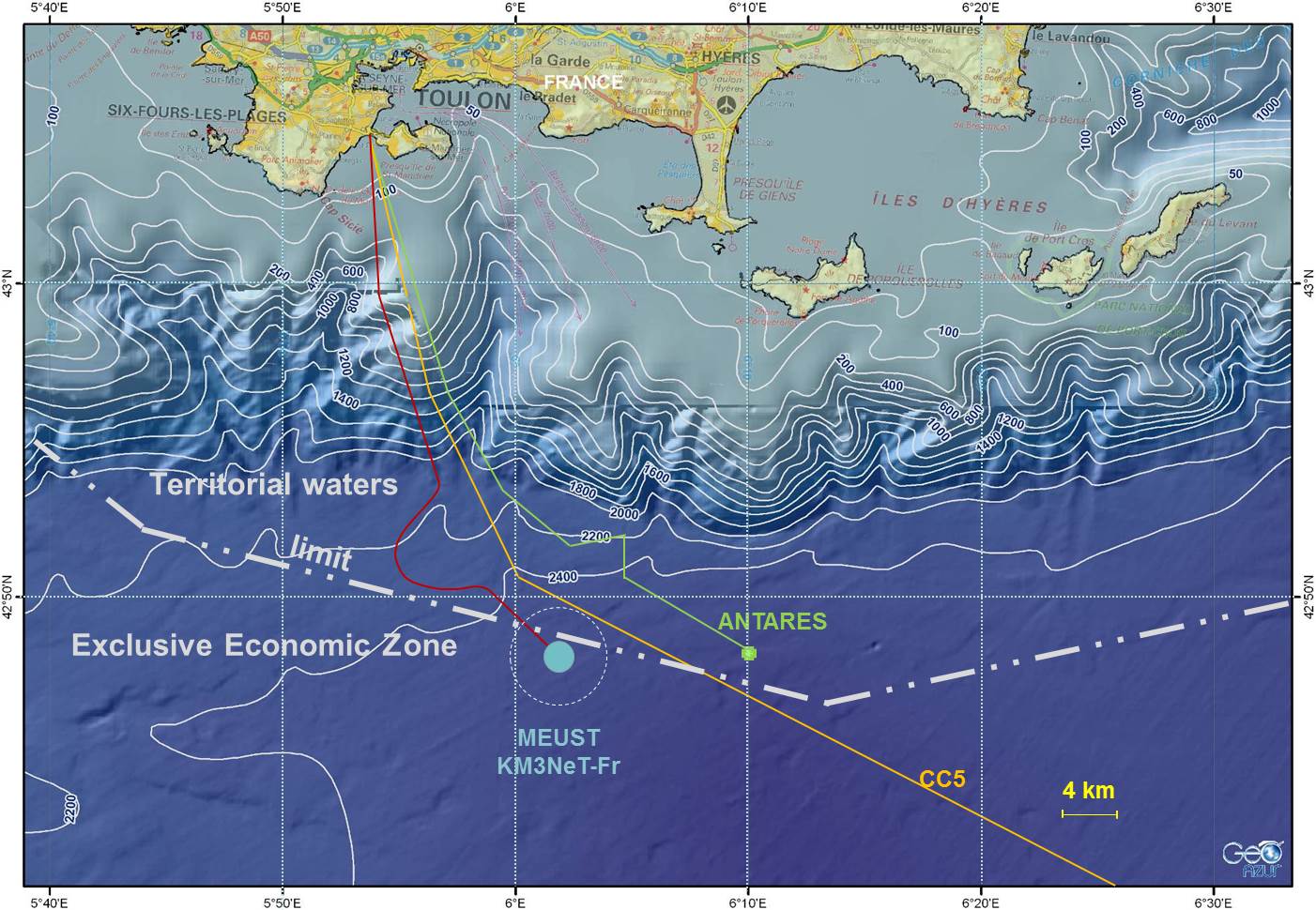}}
  \hspace*{.2in}
  \raisebox{-0.5\height}{\includegraphics[height=2.8in]{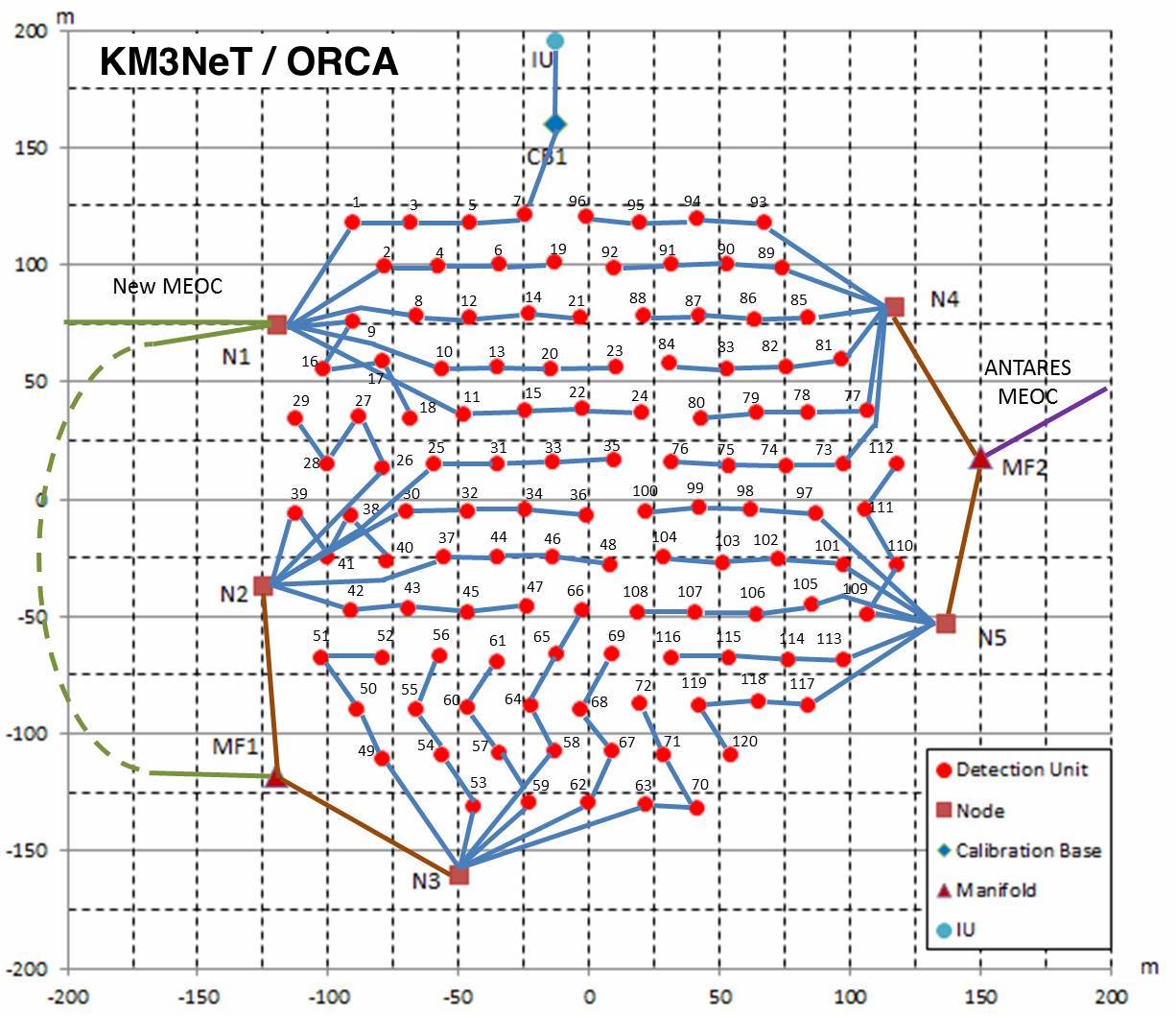}}
\caption{
 Map of the Mediterranean Sea south of Toulon, France. 
 The location of the KM3NeT-France and ANTARES installations are indicated (left).
Layout of the ORCA array (right), depicting the
115 (+5 contingency) Detection Units, cables and connection devices of the full array.}
 \label{fig:orca_layout} 
\end{figure}

\begin{figure}[!hbt]
\centering
	  \raisebox{-0.5\height}{\includegraphics[width=0.45\linewidth] {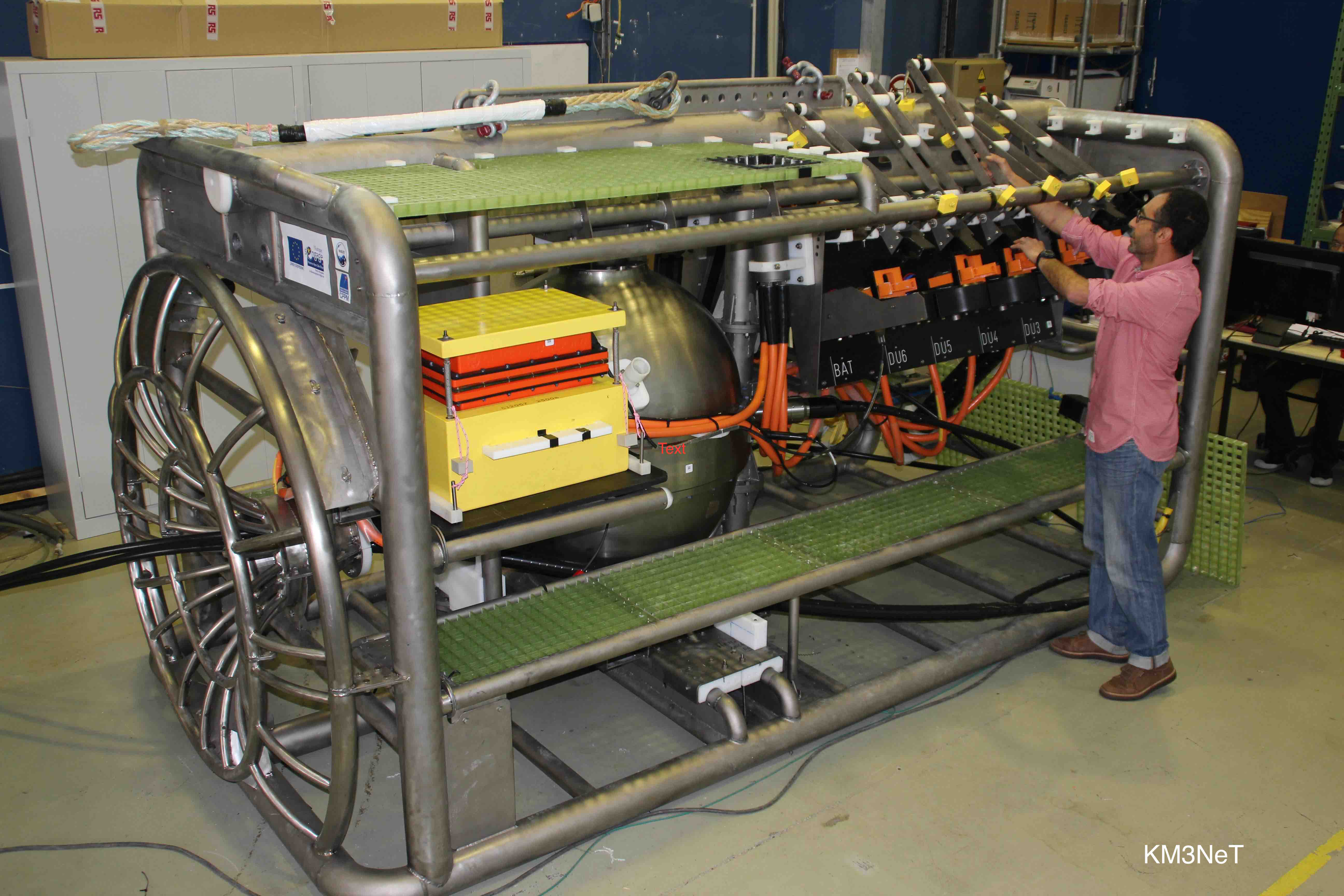}}
\hspace*{.2in}
	\raisebox{-0.5\height}{\includegraphics[width=0.50\linewidth] {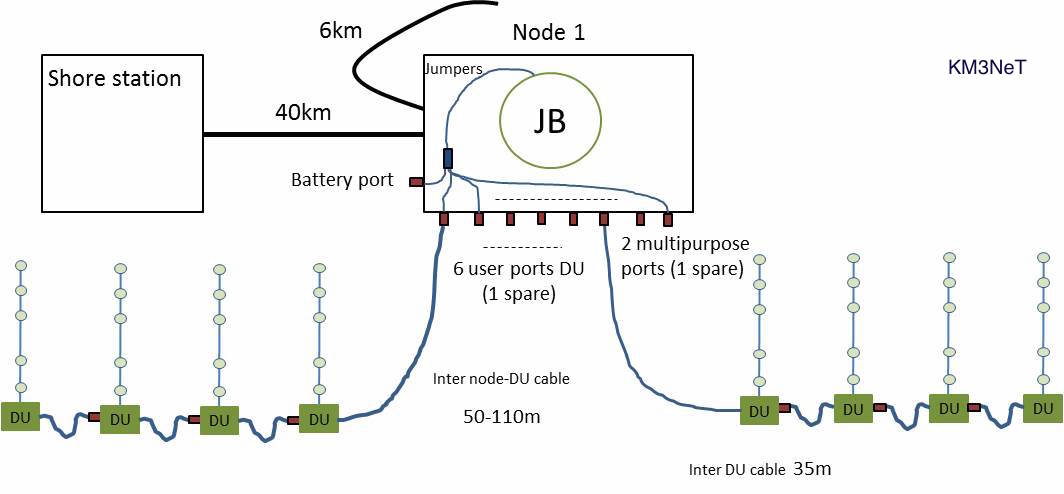}}
\caption{Photograph of a KM3NeT-France junction box (left). Schematic of the connections to the junction box (right).}
 \label{fig:orca_node} 
\end{figure}

The KM3NeT-France infrastructure is located at 42$^\circ$\,48'\,N
06$^\circ$\,02'\,E at a depth of 2450\,m, about 40\,km offshore from
Toulon, France (see \myfref{fig:orca_layout}, left). The site is outside of the French territorial waters and about 10\,km west of the site of the existing ANTARES telescope. 

\myfref{fig:orca_layout} right illustrates the layout of the full ORCA array; a single KM3NeT building block of 115 strings. The power/data are transferred to/from the infrastructure via two main electro-optic cables comprising 36/48 optical fibres and a single power conductor (the return is via the sea). 

The strings are connected to five junction boxes (\myfref{fig:orca_node}, left), located on the periphery of the array. Each junction box has eight connectors, each of which can power four strings daisy chained in series (\myfref{fig:orca_node}, right). Some daisy chains include calibration units, which incorporate laser beacons and/or hydrophone acoustic emitters. In the baseline design, five connectors on the junction box are dedicated for the neutrino array and one is dedicated for Earth and Sea science sensors and two are spares. The underwater connection of the strings to the junction box is via interlink cables running along the seabed. For the ORCA configuration, the average horizontal spacing between detection strings is about 20 m. 

Due to the shorter transmission distance involved in the ORCA
  configuration power is transferred in Alternating Current. The power station, dimensioned for a single building block (92 KVA) is located at the shore end of the main cable near the 'Les Sablettes' beach. Power is transferred at 3500 VAC. The offshore junction boxes use a AC transformer to convert this to 400 VAC for transmission along the interlink cables to the strings. The control room is located at the Institute Michel Pacha, La Seyne-sur-Mer, and hosts the data acquisition electronics and a commodity PC farm used for data filtering. 

In December, 2014, the first main electro-optic cable was successfully deployed by Orange Marine. Once ANTARES is decommissioned, its main electro-optic cable will be reused for ORCA. The first junction box was connected in spring 2015.  

\subsection{Detection string}

\begin{figure}[!hbt]
\hfil
  \includegraphics[width=0.2\linewidth, trim=0 0 0 1.5cm, clip=true] {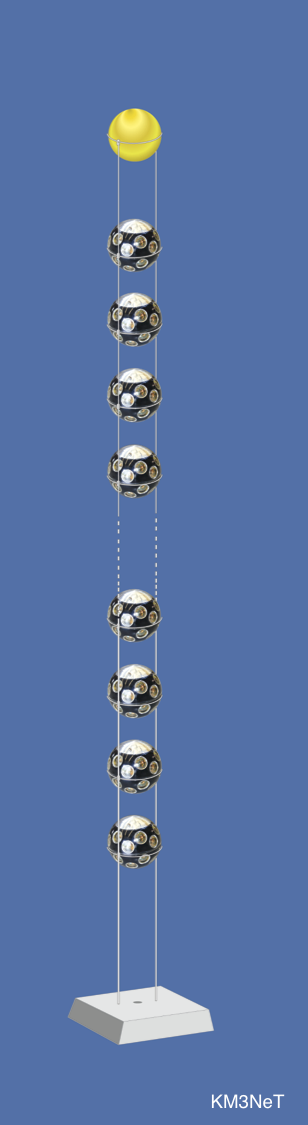}         
\hfil
 \includegraphics[width=0.5\linewidth] {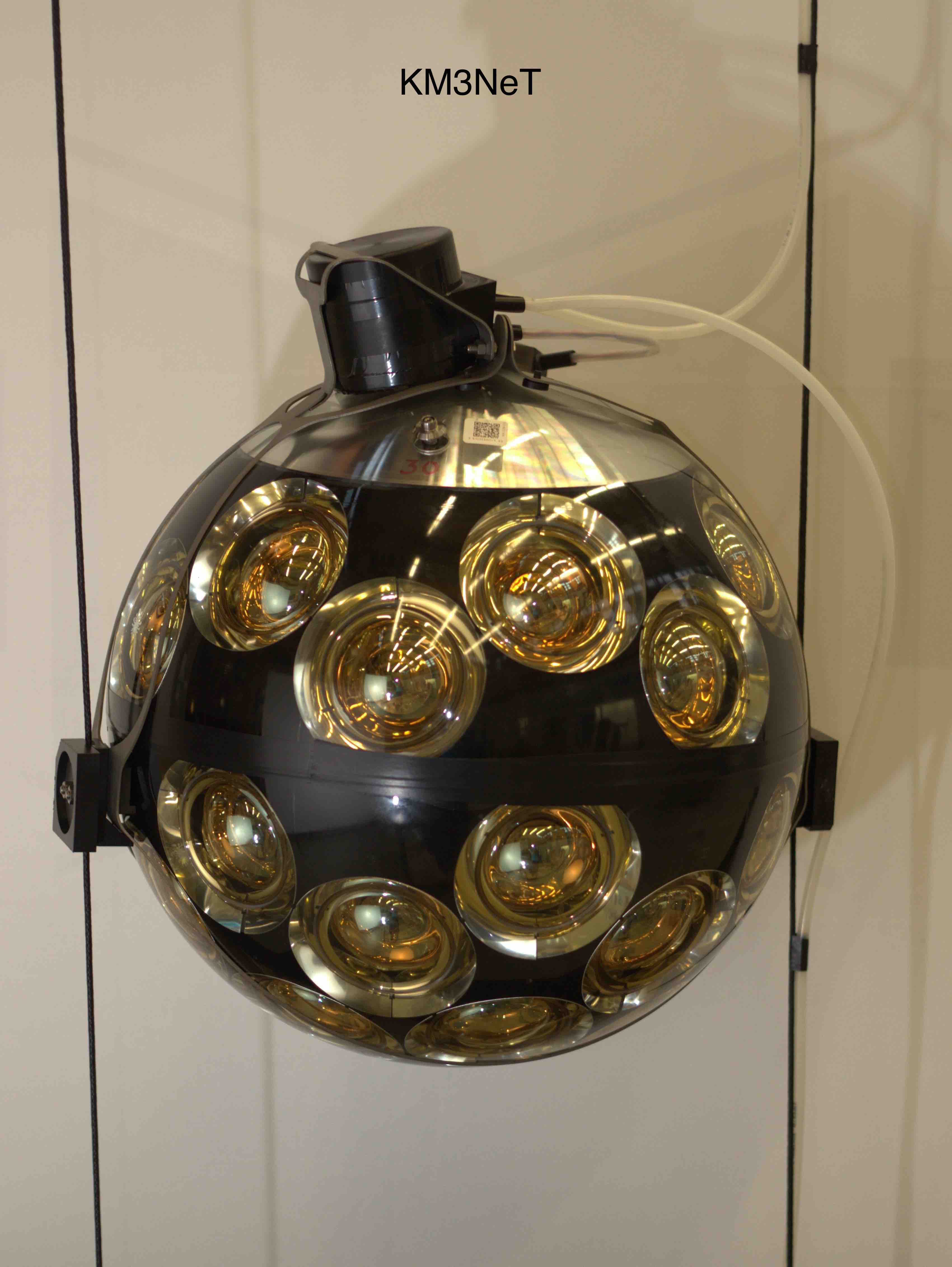}
\hfil
\caption{%
The detection string (left) and the breakout box and the fixation of the DOM
on the two parallel Dyneema\textsuperscript{\textregistered} ropes (right).}
\label{fig:string}
\end{figure}

The detection strings \cite{STRING} (\myfref{fig:string}) each host 18 DOMs.
For KM3NeT/ARCA, each is about 700\,m in height, with DOMs spaced 36\,m apart in
the vertical direction, starting about 80\,m from the sea floor. For KM3NeT/ORCA,
each string is 200 m in height with DOMs spaced 9\,m apart in the vertical direction,
starting about 40 m from the sea floor. Each string comprises two thin (4\,mm diameter) parallel
Dyneema\textsuperscript{\textregistered} ropes to which the DOMs are attached via a titanium collar. Additional
spacers are added in between the DOMs to maintain the ropes parallel. Attached
to the ropes is the vertical electro-optical cable, a pressure balanced,
oil-filled, plastic tube that contains two copper wires for the power
transmission (400 VDC) and 18 optical fibres for the data transmission. At each
storey two power conductors and a single fibre are branched out via the breakout
box. The breakout box also contains a DC/DC converter (400\,V to 12\,V). The power
conductors and optical fibre enter the glass sphere via a penetrator.

Even though the string design minimises drag and itself is buoyant, additional
buoyancy is introduced at the top of the string to reduce the horizontal
displacement of the top relative to the base for the case of large sea currents.

\begin{figure}[!hbt]
 \includegraphics[trim=0 5cm 0 5cm,clip=true,height=10cm]{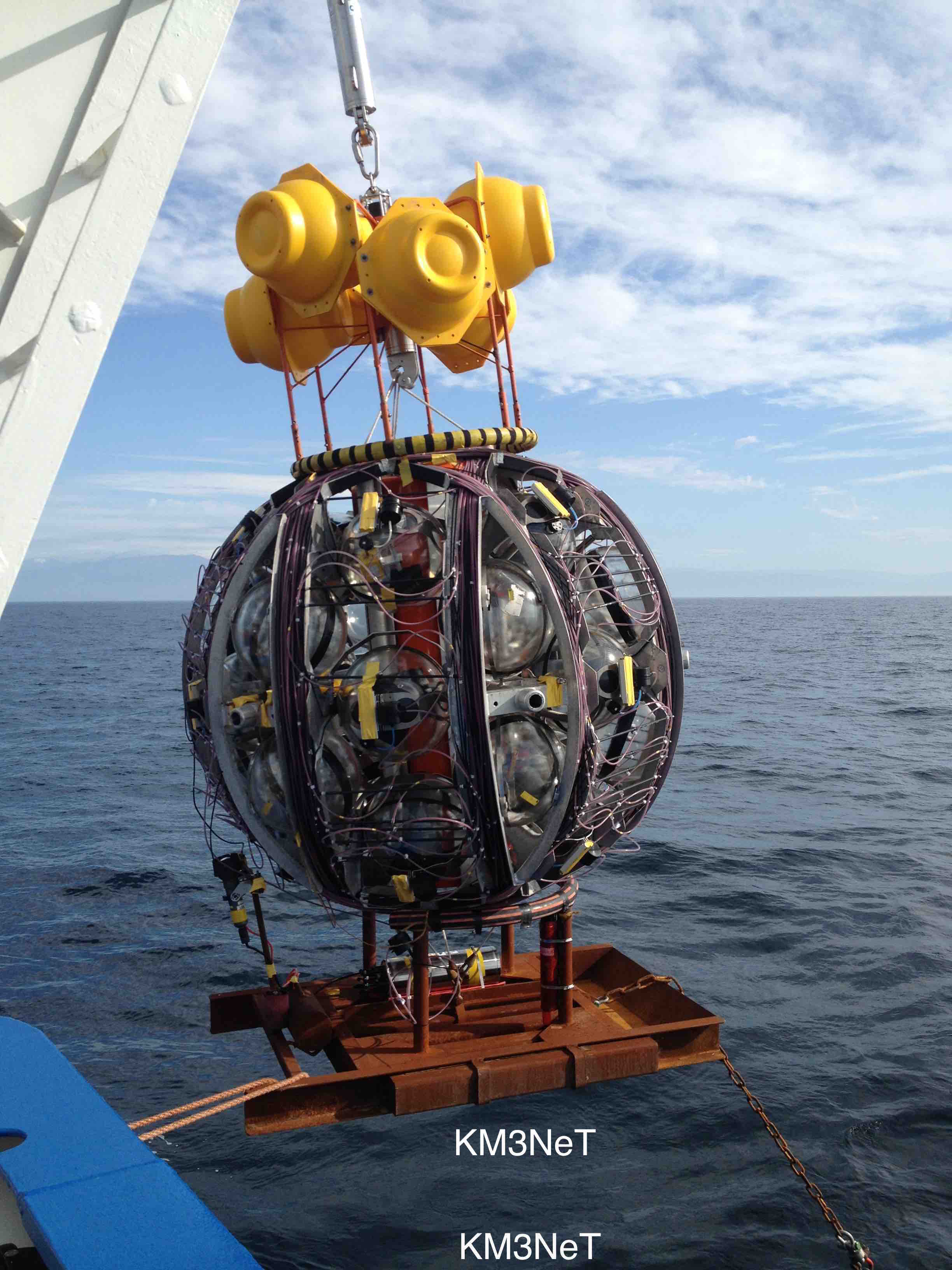}
\hfil
 \includegraphics[trim=0 3cm 0 3cm,clip=true,height=10cm]{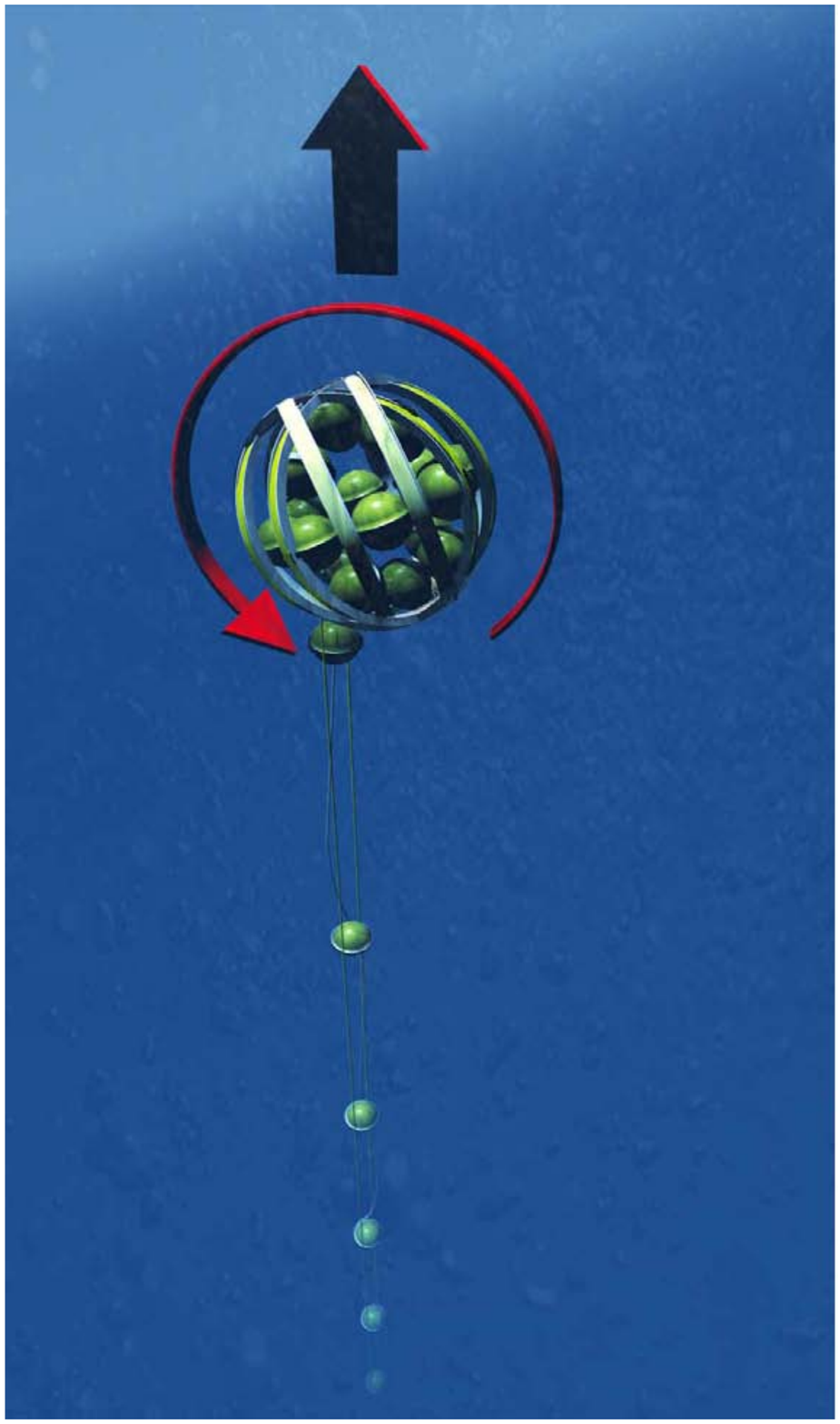}
\caption {%
Photo of a launch vehicle deployment (left). Principle of the launch
vehicle unfurling (right, picture courtesy Marijn van der Meer/Quest).}
\label{fig:LOM} 
\end{figure}

For deployment and storage, the string is coiled around a large spherical frame,
the so-called launcher vehicle, in which the DOMs slot into dedicated cavities
(see \myfref{fig:LOM}). The anchor at the bottom of the string is the
interface with the seabed infrastructure. It is external to the launcher vehicle
and is sufficiently heavy to keep the string fixed on the seabed. The anchor
houses an interlink cable, equipped with a wet-mateable connectors, and the base
container. The base container incorporates dedicated optical components and an
acoustic receiver used for positioning of the detector elements.

\begin{figure}[!hbt]
   \raisebox{-0.5\height}{\includegraphics[height=5cm,width=0.5\linewidth, trim=0 1.5cm 0 0,clip=true]{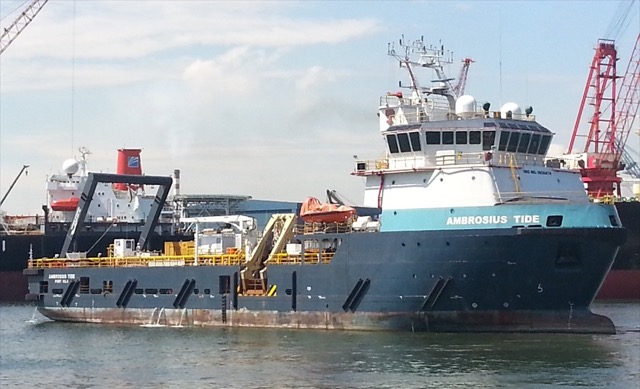}}
\hspace*{.6in}
  \raisebox{-0.5\height}{\includegraphics[width=0.35\linewidth, trim= 0 1.5cm 0 0,clip=true]{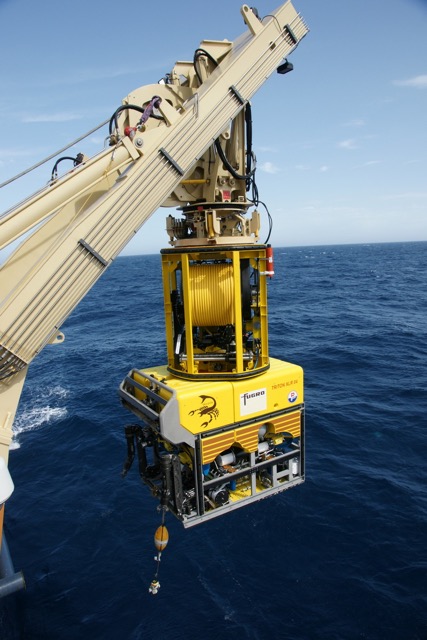}}
\caption {%
Photograph of the Ambrosius Tide boat, used for the KM3NeT/ARCA string deployment (left). 
Photograph of the remote operated vehicle, used for the string
connection (right).}
\label{fig:boats1} 
\end{figure}

\begin{figure}[!hbt]
	  \raisebox{-0.5\height}{\includegraphics[height=4cm,width=0.55\linewidth]{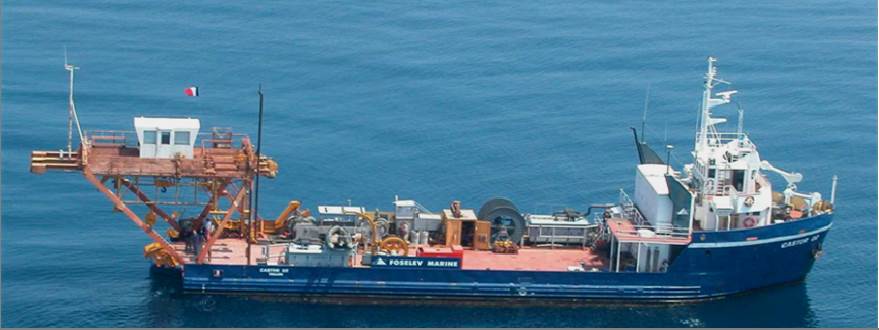}}
\hspace*{.2in}
	  \raisebox{-0.5\height}{\includegraphics[height=4cm,width=0.4\linewidth, trim=-1cm 0 0 1cm, clip]{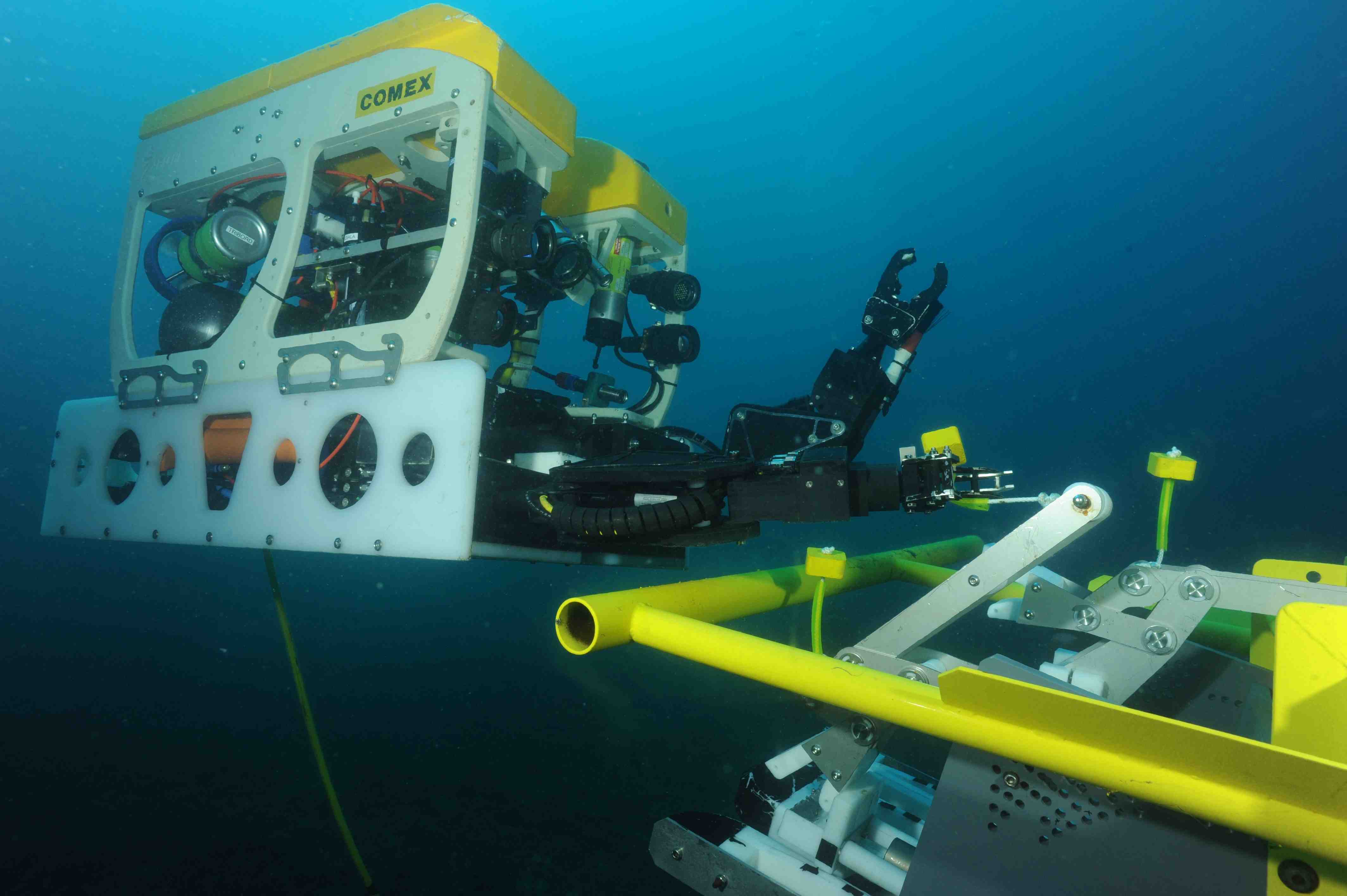}}
\caption {Photograph of the Castor boat, used for the KM3NeT/ORCA
  string deployment (left). Photograph of the Comex Apache ROV used for the
  KM3NeT/ORCA string connection (right).} 
\label{fig:boats2} 
\end{figure}

A surface vessel (\myfref{fig:boats1} (left), \myfref{fig:boats2} (left)), with dynamic positioning capability,
is used at each site to deploy the launcher vehicle at its designated position on
the seabed with an accuracy of 1\,m. A remotely operated vehicle
(\myfref{fig:boats1}, \myfref{fig:boats2}, right) is used to deploy and connect the interlink
cables from the base of a string to the junction box. Once the connection to the
string has been verified onshore, an acoustic signal from the boat triggers the
unfurling of the string. During this process, the launcher vehicle starts to
rise to the surface while slowly rotating and releasing the DOMs. The empty
launcher vehicle floats to the surface and is recovered by the surface vessel. 
The use of compact strings allows for transportation of many units on board and
thus multiple deployments during a single cruise. This method reduces costs and
also has advantages in terms of risk reduction for ship personnel and material
during the deployment. It also improves tolerance to rough sea conditions.

In May 2014, a prototype string comprising three active DOMs was successfully
deployed and connected to the KM3NeT-Italy site and operated for more than one
year \cite{km3net-ppmdu-2015}. This test deployment validated many aspects of
the deployment scheme. The first ORCA-style string will be connected to
KM3NeT-France infrastructure spring 2016.

\subsection{Digital optical module}
\label{sec-tec-dom}

\begin{figure}[!hbt]
 \includegraphics[height=7cm, width=0.45\linewidth] {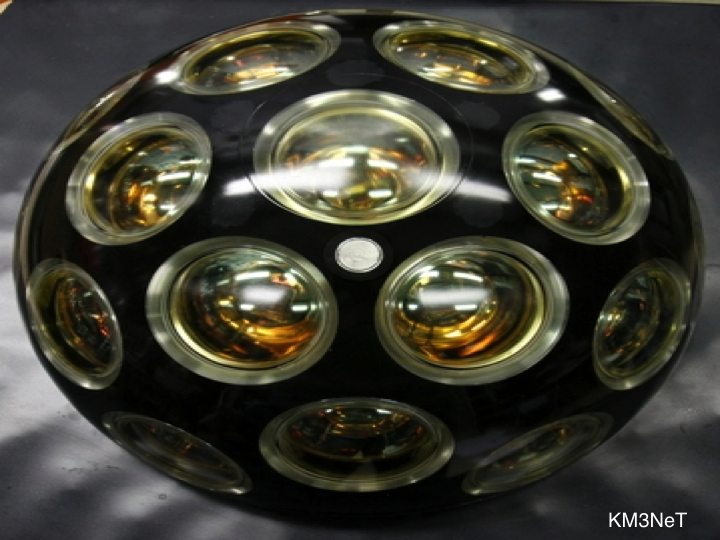}
\hfil
  \includegraphics[height=7cm, width=0.45\linewidth] {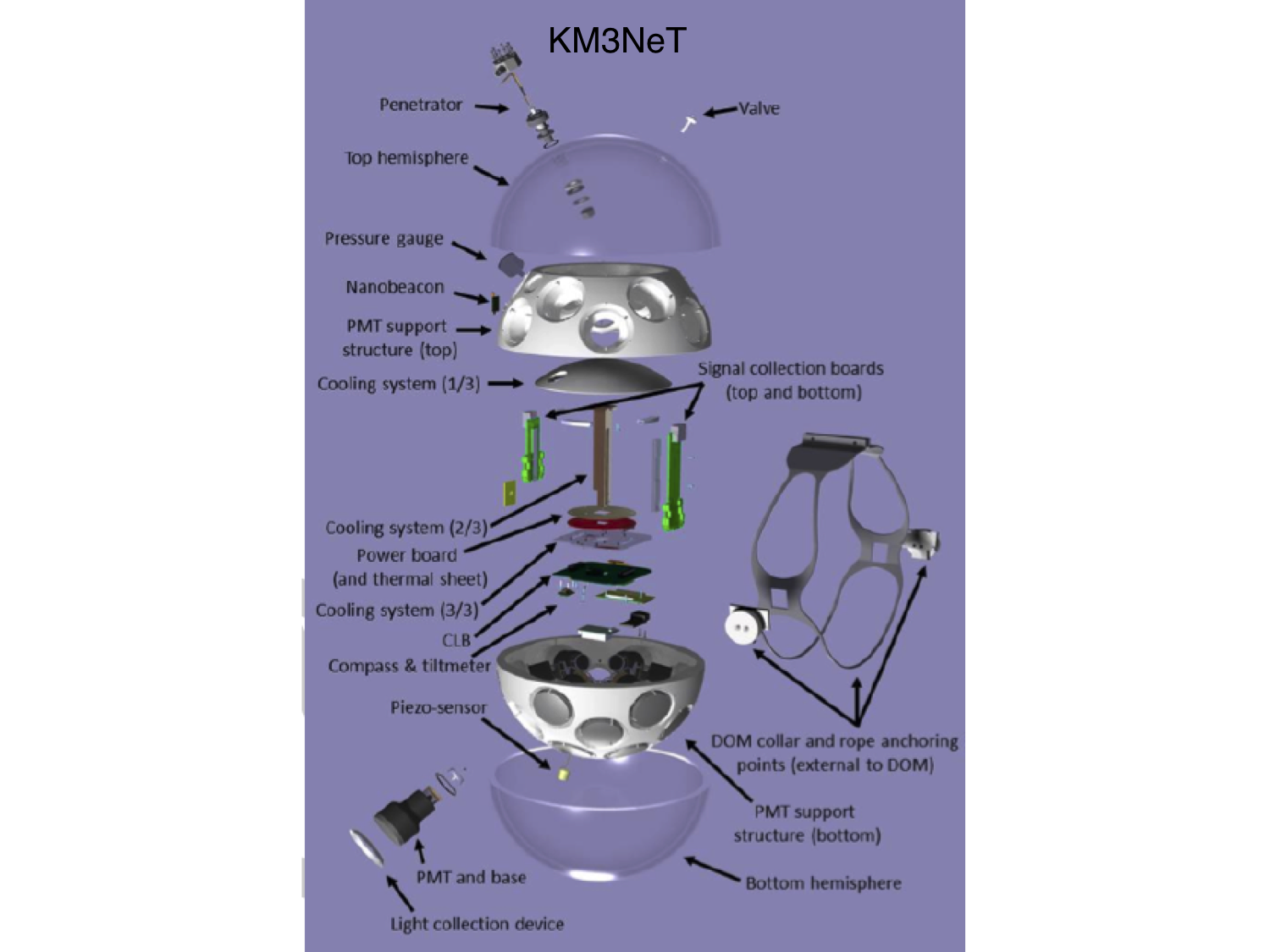}
\caption{%
A photograph of a completed Digital Optical Module. The central white spot is the
acoustic piezo sensor (left). Exploded view of the inside of a DOM (right).}
\label{fig:DOM}
\end{figure}

The Digital Optical Module \cite{DOM} (\myfref{fig:DOM} left) is a transparent
17\,inch diameter glass sphere comprising two separate hemispheres, housing 31
photo-multiplier tubes (PMT) and their associated readout electronics. The
design of the DOM has several advantages over traditional optical modules using
single large PMTs, as it houses three to four times the photo-cathode area in a
single sphere and has an almost uniform angular coverage. As the photo-cathode
is segmented, the identification of more than one photon arriving at the DOM can
be done with high efficiency and purity. In addition, the directional
information provides improved rejection of optical background.
    
The PMTs are arranged in 5 rings of 6 PMTs plus a single PMT at the bottom
pointing vertically downwards. The PMTs are spaced at $60^\circ$ in azimuth and
successive rings are staggered by $30^\circ$. There are 19~PMTs in the lower
hemisphere and 12 PMTs in the upper hemisphere. The PMTs are held in place by a
3D printed support. The photon collection efficiency is increased by 20--40\%
by a reflector ring around the face of each PMT. In order to assure optical
contact, an optical gel fills the cavity between the support and the glass. The
support and the gel are sufficiently flexible to allow for the deformation of
the glass sphere under the hydrostatic pressure.
     
Each PMT has an individual low-power high-voltage base with integrated
amplification and tuneable discrimination. The arrival time and the
time-over-threshold (ToT) of each PMT, are recorded by an individual
time-to-digital converter implemented in a FPGA. The threshold is set at the
level of 0.3 of the mean single photon pulse height and the high voltage is set
to provide an amplification of $3\times10^6$. The FPGA is mounted on the central
logic board, which transfers the data to shore via an Ethernet network of
optical fibres. Each DOM in a string has a dedicated wavelength to be later
multiplexed with other DOM wavelengths for transfer via a single optical fibre
to the shore. The broadcast of the onshore clock signal, needed for time
stamping in each DOM, is embedded in the Gb Ethernet protocol. The white rabbit
protocol has been modified to implement the broadcast of the clock signal. The
power consumption of a single DOM is about 7\,W.
                                       
\begin{figure}[!hbt]
\hfil
  \includegraphics[height=4cm, width=0.3\linewidth] {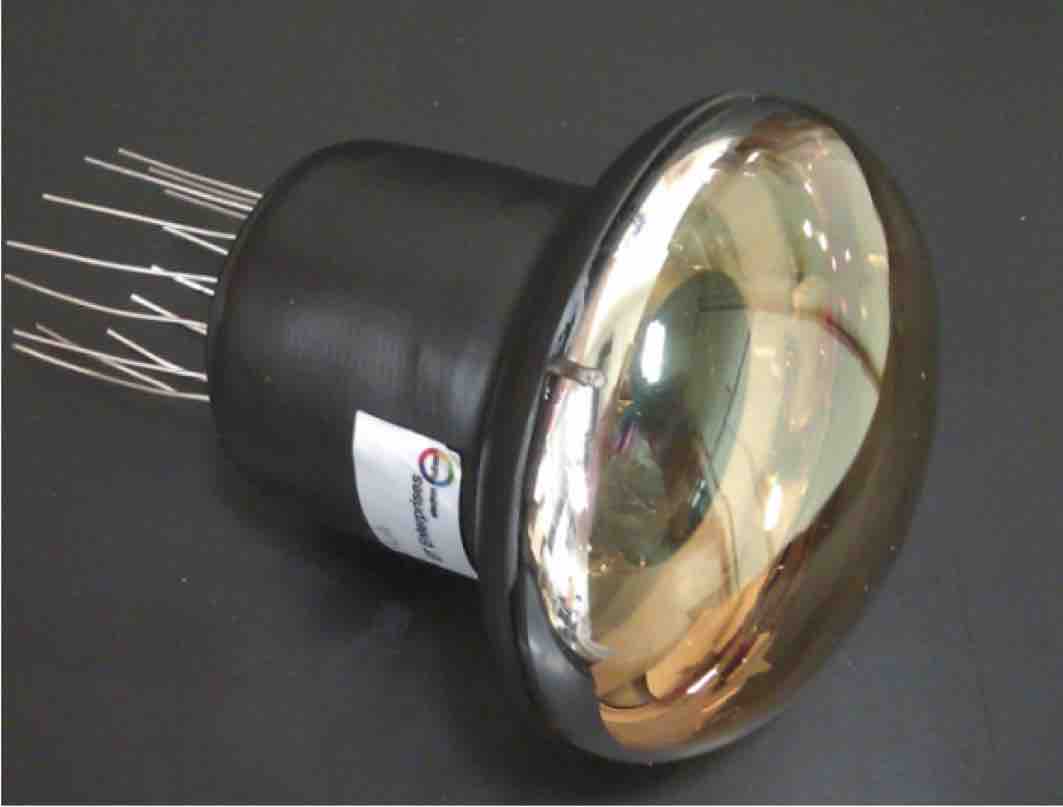}
\hfil
  \includegraphics[height=4cm, width=0.3\linewidth] {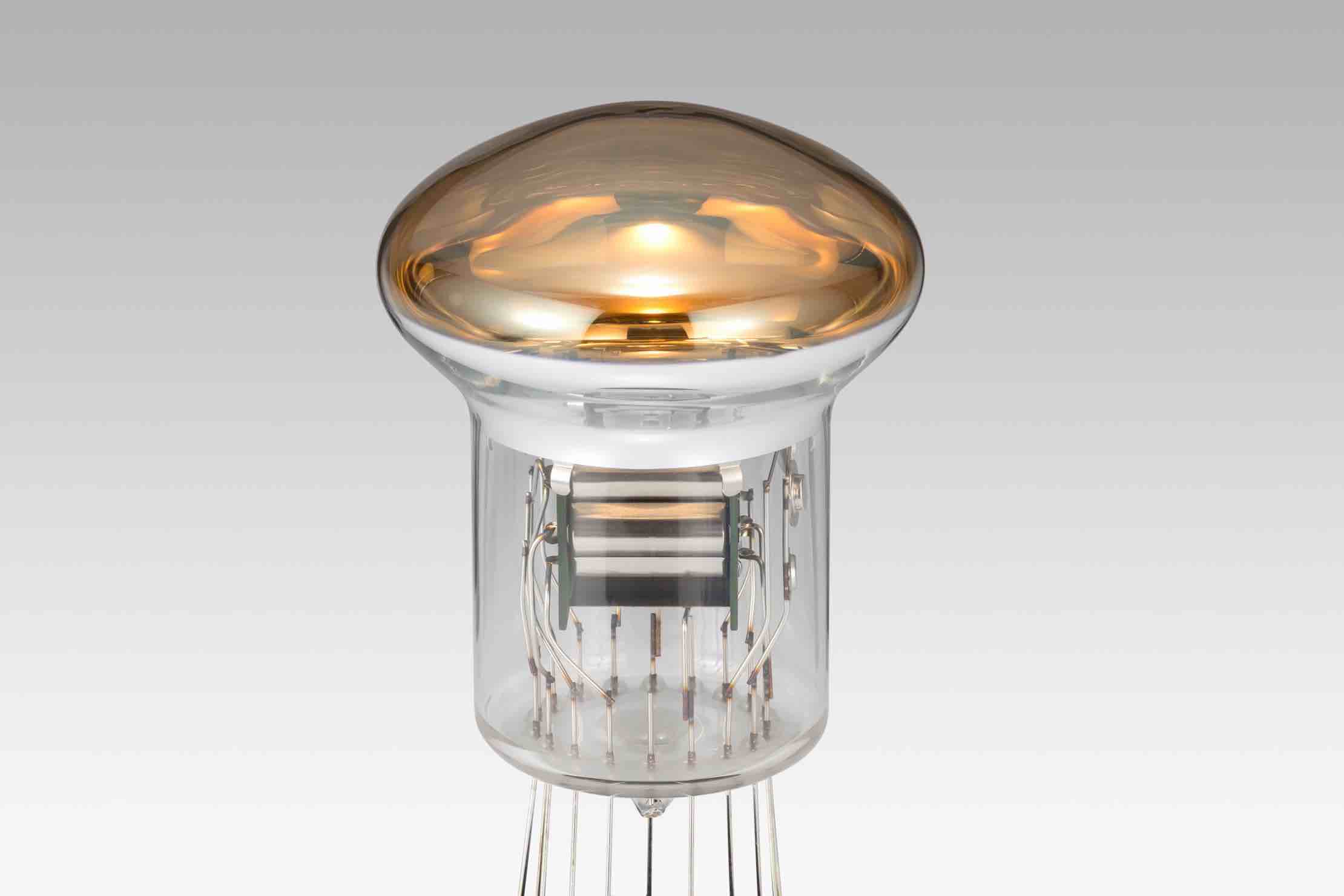}
\hfil
\caption{%
Photographs of the PMTs. ETEL D792KFL (left) and Hamamatsu R12199-02 (right).} 
\label{fig:PMT}
\end{figure}

The specification for the PMTs are summarised in \mytref{table:PMT}. 
Prototype PMTs from Hamamatsu and ETEL have been developed and satisfy the requirements 
(see \myfref{fig:PMT}).
The PMTs have a photo-cathode diameter of at least 72\,mm and a length of less than
122\,mm. The reflector effectively increases the diameter to about 85\,mm. The
photo-multiplier tube has a ten stage dynode structure with a minimum gain of
$10^6$. The front face of the photo-multiplier tube is convex with a radius
smaller than the inner radius of the glass sphere. Due to the small size of the
PMT, the influence of the Earth's magnetic field is negligible and a mu-metal
shield is not required.

\begin{table}
\small
\centering
\begin{tabular} {lr}
\hline
Radiant blue sensitivity at 404\,nm                      & 130 \,mA/W      \\
Quantum efficiency (QE)                                  & 20\% $@$ 470\,nm and 28\% $@$ 404\,nm    \\
Inhomogeneity of cathode response                        & 10\%      \\
Supply voltage for a gain of $3\times10^6$               & 900--1300\,V    \\
Dark count at 15$^{\circ}$C and 0.3\,photo-electron threshold    & 1.5\,kHz     \\
Transit time spread (TTS)                                & 4.5\,ns (FWHM)     \\
Peak to valley ratio                                     & 2.5      \\
\hline 
\end{tabular}    
\caption{Specification of the PMTs.}                                
\label{table:PMT}  
\end{table}
                             
The optical module also contains three calibration sensors: 1) The LED
nano-beacon, which illuminates the optical module(s) vertically above; 2) A
compass and tilt-meter for orientation calibration; 3) An acoustic piezo sensor
glued to the inner surface of the glass sphere for position calibration.

In May 2013, a prototype DOM was successfully installed on an ANTARES detection
line and operated in-situ for over a year \cite{km3net-ppmdom-2014} . Starting in spring
2014, three prototype DOMs were operated for over a year at the KM3NeT-Italy site \cite{km3net-ppmdu-2015}. 
In December 2015, a first production string of 18 DOMs was successfully operated at the KM3NeT-Italy site.

\subsection{Fibre-optic data transmission system}

The KM3NeT fibre-optic data transmission system performs the following
functions:
\begin{itemize}
\item 
Transfers all the data to shore: The bandwidth per DOM is 1\,Gb/s. The observed
singles rate, dominated by $^{40}$K, is typically 6--8\,kHz per PMT \cite{km3net-ppmdom-2014,km3net-ppmdu-2015} or
190--250\,kHz per DOM, which amounts to 9--12\,Mb/s per DOM. Additional contributions
from bioluminescence can be accommodated up to levels of a factor of 10 compared
to $^{40}$K.
\item 
Provide timing synchronisation: Relative time offsets between any pair of DOMs
are stable within 1\,ns;
\item 
Provide individual control for each DOM: Setting the HV of a PMT, turn off/on a
single PMT, turn on/off nano-beacon, update soft- and firmware;
\item 
Provide individual control for each base of a string: Turn string power on/off,
control optical amplifiers, monitor AC/DC converter;
\item 
Provide slow control for the junction boxes. \end{itemize}

The slow-control system is implemented via a broadcast mechanism (same as that
of the clock), in which control information for all strings is sent on a single
common wavelength. If it is a message for just a single string or DOM it is ignored by all the others. 
The communication from offshore exploit a Dense Wavelength Division Multiplexing (DWDM) technique. 
The return signals for the slow control are
transmitted on 34 wavelengths via the slow-control fibre(s). The data return
path is based on a 50\,GHz spacing system with a 72 wavelengths uplink. Each DOM
of 4 strings produces a unique wavelength that is combined on one fibre. EDFA
optical amplifiers are introduced onshore and at the base of a string to
maintain the optical margins above 10 dB. 

\subsection{Data acquisition}

The readout \cite{DAQ} of the KM3NeT detector is based on the
``All-data-to-shore'' concept in which all analogue signals from the PMTs that
pass a preset threshold (typically 0.3\,photo-electrons) are digitised and all digital data
are sent to shore where they are processed in real time. The physics events are
filtered from the background using designated software. To maintain all
available information for the offline analyses, each event will contain a
snapshot of all the data in the detector during the event. Different filters can
be applied to the data simultaneously.

The optical data contain the time of the leading edge and the time over
threshold of every analogue pulse, commonly referred to as a hit. Each hit
corresponds to 6\,Bytes of data (1\,B for PMT address, 4\,B for time and 1\,B for
time over threshold). The least significant bit of the time information
corresponds to 1 ns. The total data rate for a single building block amounts to
about 25\,Gb/s. A reduction of the data rate by a factor of about $10^5$ is thus
required to store the filtered data on disk. In addition to physics data,
summary data containing the singles rates of all PMTs in the detector are stored
with a sampling frequency of 10 Hz. This information is used in the simulations
and the reconstruction to take into account the actual status and optical
background conditions of the detector.

In parallel to the optical data, the data from the acoustics positioning system
are processed and represents a data volume of about one third of that of the
optical data.

\subsubsection{Event trigger}
\label{sec-tec-com-tri}

For the detection of muons and showers, the time-position correlations that are
used to filter the data follow from causality. In the following, the level-zero
filter (L0) refers to the threshold for the analogue pulses which is applied off
shore. All other filtering is applied on shore. The level-one filter (L1) refers
to a coincidence of two (or more) L0 hits from different PMTs in the same
optical module within a fixed time window. The scattering of light in deep-sea
water is such that the time window can be very small. A typical value is $\Delta
T = 10$\,ns. The estimated L1 rate per optical module is then about 1000\,Hz, of
which about 600\,Hz is due to genuine coincidences from $^{40}$K decays. The
remaining part arises from random coincidences which can be reduced by a factor
of two by making use of the known orientations of the PMTs. This is referred to
as the level-two filter (L2). Separate trigger algorithms operate in 
parallel on this data, each optimised for a different event topology.

A general solution to trigger on a muon track event consists of a scan of the sky combined
with a directional filter \cite{km3net-trigger-note}. In the directional filter, 
the direction of the muon is assumed. For each direction, an intersection of a
cylinder with the 3D array of optical modules can be considered. The diameter of
this cylinder (i.e.\ road width) corresponds to the maximal distance traveled by
the light. It can safely be set to a few times the absorption length without a
significant loss of the signal. The number of PMTs to be considered is then
reduced by a factor of 100 or more, depending on the assumed direction. 
Furthermore, the time window that follows from causality is reduced by a similar
factor\footnote{Only the transverse distance between the PMTs should be taken into
account because the propagation time of the muon can be corrected for.}. This
improves the signal-to-noise ratio (S/N) of an L1 hit by a factor of (at least)
$10^4$ compared to the general causality relation. With a requirement of five
(or more) L1 hits, this filter shows a very small contribution of random
coincidences.

The field of view of the directional filter is about 10 degrees. So, a set of
200 directions is sufficient to cover the full sky. By design, this trigger can
be applied to any detector configuration. Furthermore, the minimum number of L1
hits to trigger an event can be lowered for a limited number of directions. A
set of astrophysical sources can thus be tracked continuously with higher
detection efficiency for each source.

For shower events, triggering is simpler, since the maximal 3D-distance
between PMTs can be applied without consideration of the direction of the
shower. 

A maximum distance traveled by the light can be assumed, limiting the maximum distance $D$ between hit PMTs. 
This reduces the number of PMTs to be considered and the time window that follows from causality. 
Hence, an improvement of the S/N ratio compared to the general causality relation can be obtained.

Alternative signals with different time-position correlations, such as slow magnetic
monopoles, can be searched for in parallel. It is obvious but worth noting that
the number of computers and the speed of the algorithms determine the
performance of the system and hence the physics output of KM3NeT.

\subsubsection{Performance}
\label{sec-tec-com-per}

The performance of the online data filter can be summarised by the effective
volume, the event purity and the time needed to process a time slice of raw
data. The effective volume is the volume in which a neutrino interaction would
trigger the event to be written to disk and the event purity is the fraction of
events that contain a neutrino interaction or atmospheric muon bundle.
The effective volume of the ARCA and ORCA detectors are presented in  
\mysref{sec-sci-too-sim} and \mysref{simu:event}, respectively.

To process the data, the concept of time slicing is applied. In this, the data
from each optical module are stored in a frame corresponding to a preset time
period. All data frames corresponding to the same time period are sent to a
single CPU core based on IP level 2 switching. A complete set of data frames is
referred to as a time slice. Data corresponding to subsequent time periods are
sent to different CPU cores until the number of available CPU cores is
exhausted. The first CPU core should then be ready to receive and process the
data from the next time period.

\begin{figure}
\begin{minipage}[tb]{1\textwidth}
\centering
\begin{overpic}[width=0.49\textwidth]{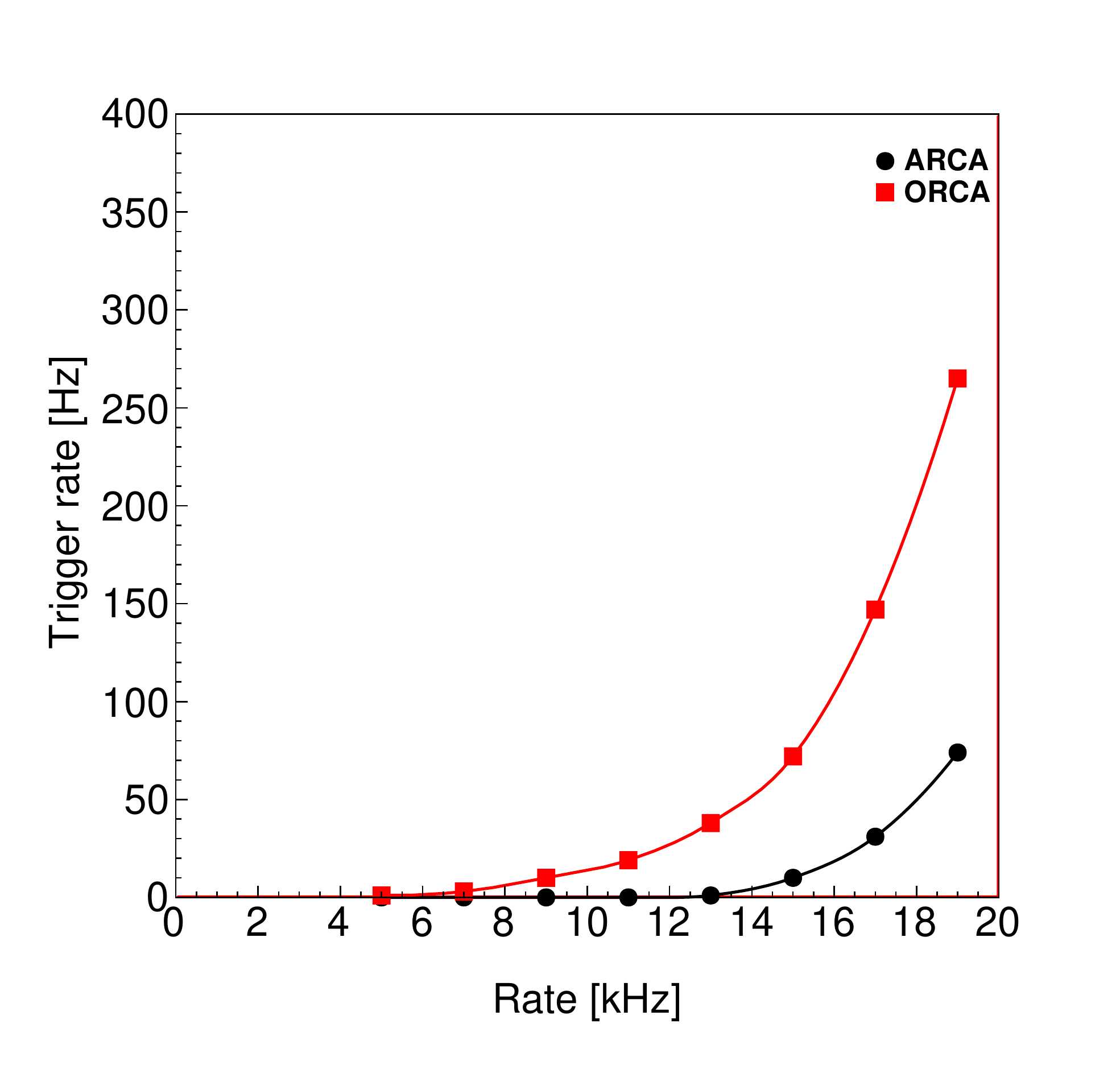}
\put (20,78) {\bf KM3NeT}
\end{overpic}
\begin{overpic}[width=0.49\textwidth]{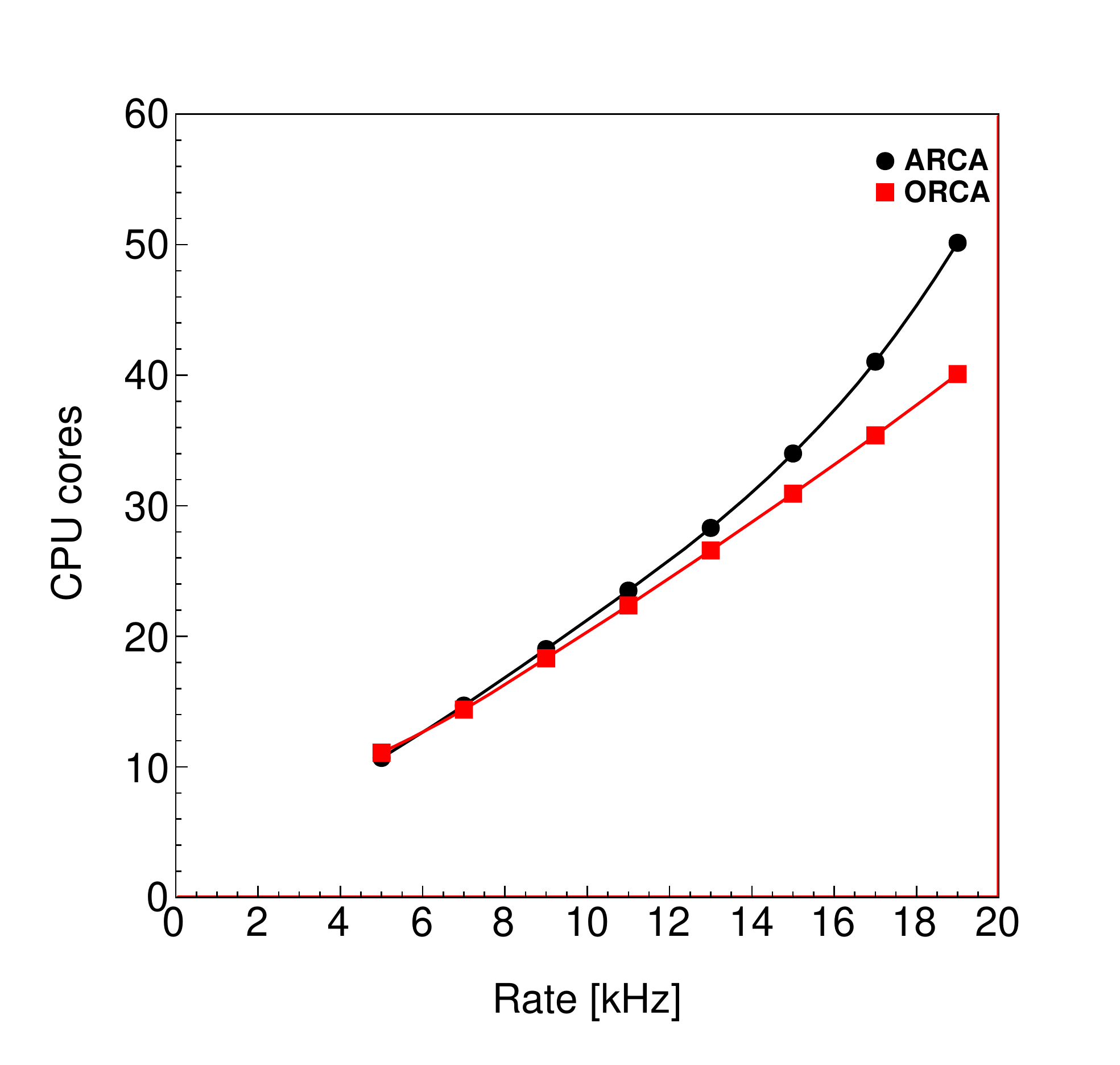}
\put (20,78) {\bf KM3NeT}
\end{overpic}
\end{minipage}
\caption{Trigger rate due to random coincidences (left) and required
  number of CPU cores (right) as a function of the singles rate for
  one building block of ARCA (black circles) and ORCA (red squares).}
\label{fig:trigger_rates}
\end{figure}

In the following, the performance of the online data filter is presented 
for one ARCA and one ORCA building block. 
In this, different triggers are operated in parallel.
The typical trigger settings correspond to  
a L1 time window of $\Delta T = 10$~ns, 
a maximum space angle between the PMT axes of 90 degrees (L2), 
and a minimum number of L1 hits of 4 or 5.
The detection threshold thus corresponds to 8 or 10 photons.
The trigger rate due to random coincidences and the number of CPU cores are 
shown in \myfref{fig:trigger_rates} as a function of the singles rate.

The typical singles rate due to radioactive decays in the sea water is about 6--8\,kHz per PMT \cite{km3net-ppmdom-2014,km3net-ppmdu-2015},
including the dark count rate.
In addition, there are occasional bursts of bioluminescence.
To limit the effect of excursions of the singles rate, short bursts of bioluminescence can be filtered in real-time.
The probability of the occurrence of bioluminescent bursts depends on the site and is found to be correlated with 
the velocity of the sea current  \cite{NEMO_CPsite2016,ANT-Biolum-2013} presumably due to the influence of bioluminescent organisms induced by turbulence or 
impacts on the infrastructure.
An enhanced level of bioluminescence has been observed in the ANTARES detector during the spring period of some years \cite{ANT-Biolum-2013}.
Averaged over the live time of the ANTARES detector, the overall inefficiency due to bioluminescence is about 10\%.
Due to the slender design of KM3NeT, it is expected that the turbulence and impacts on the infrastructure 
are significantly less and so is this inefficiency.

As can be seen from \myfref{fig:trigger_rates}, the number of CPU cores needed to process the data in real time 
is less than 50 up to singles rates of 20\,kHz (three times the nominal rate).
It should be noted that the number of CPU cores may be larger than one for a modern PC.
So, this result provides for a cost-effective implementation of the ``All-data-to-shore'' concept.  
Moreover, the trigger software is the same for the ARCA and ORCA detectors; 
only the settings of the trigger parameters are adjusted to optimally detect 
neutrinos with the targeted energies.

\cleardoublepage

\section{Astroparticle Research with Cosmics in the Abyss (ARCA)}

\subsection{Introduction}
\label{sec-sci}

The main science objective of KM3NeT/ARCA is the detection of high-energy
neutrinos of cosmic origin. Since neutrinos propagate directly from their sources
to the Earth, even modest numbers of detected neutrinos can be of
utmost scientific relevance, by
indicating the astrophysical objects in which cosmic rays are accelerated, or
pointing to places where dark matter particles annihilate or decay. The prospect of such
fundamental physics discoveries have led the astroparticle and astrophysics
communities to include KM3NeT as a high priority in their respective European
road maps (APPEC/ASPERA, AstroNet) and the European Strategy Forum on Research
Infrastructures (ESFRI) to include it in their list of priority projects. The
KM3NeT Research Infrastructure will also provide user ports for real-time,
long-term Earth and Sea science measurements in the deep-sea environment.

One priority goal of KM3NeT/ARCA is indisputably to find neutrinos from the cosmic
ray accelerators in our Galaxy. 
In a neutrino telescope the two simplest event topologies that can be identified are: a ``shower" topology that includes the NC interaction of all three neutrino flavours, the CC interaction of $\nu_e$,  and a subset of $\nu_\tau$ interactions; and a ``track" topology that indicates the presence of muons produced in $\nu_\mu$ and $\nu_\tau$ CC interactions
(see \mysref{sec-sci-too-rec} for a detailed explanation).

The preferred search strategy is to identify upward-moving
muons, which unambiguously indicates neutrino reactions since only neutrinos can
traverse the Earth without being absorbed. A neutrino telescope in the
Mediterranean Sea is ideal for this purpose, since most of the potential Galactic
sources are in the Southern sky; in contrast, the IceCube detector at the South
Pole is much less sensitive to these individual sources, at least in the energy
range where the signal is expected (a few TeV to a few tens of 10 TeV -- see
\mysref{sec-sci-too-poi}). The KM3NeT/ARCA design has been carefully
optimised to maximise the sensitivity to these Galactic sources. One of the
findings in this process is that the overall sensitivity is not reduced if the
neutrino telescope is split into separate building blocks, provided they are
large enough, at least 0.5 cubic kilometres each \cite{km3net-icrc-2013}. It has
thus been decided to make a distributed infrastructure, thereby maximising the
influx of regional funding and human resources. Furthermore, the concept of
independent building blocks complies with the technical specifications for the
construction and operation of the Research Infrastructure.

Currently, the KM3NeT Collaboration is proceeding with the first construction
phase (Phase-1). Until 2017, 31~strings equipped with 558 optical modules (see
\mysref{sec-tec}) will be assembled and deployed. Of these, 24~strings will
be configured for ARCA and deployed at the Italian site. The resulting array
will provide the equivalent of 10--20\% of the size of the IceCube detector. The
recent experience from a combined analysis of ANTARES and IceCube data
\cite{antares-icecube-combined-2015}, increasing the sensitivity to point-like
neutrino sources by up to a factor of two with respect to the individual
analyses, indicates that Phase-1 will already have a decent discovery potential
and provide significant new data.

\subsubsection{Cosmic neutrinos}

A new situation has emerged since IceCube has presented evidence for a neutrino
signal of cosmic origin. This signal includes upward- and downward-going events
with neutrino energies from a few 10\,TeV to above 1\,PeV. Even though the
signal is statistically very significant, its astrophysical implications are
not yet clear. This signal is the first
high-energy extra-terrestrial neutrino signal ever observed and thus marks a
major turning point in the history of neutrino astronomy. Detailed studies have
been and are being conducted to estimate the sensitivity of KM3NeT/ARCA to a neutrino
flux with the reported properties, to investigate the consequences of a
re-optimisation of the detector for such a signal (in particular in terms of
geometry parameters and building block size) and to evaluate the prospects of
Phase-2.0. Results of these studies are presented in the following.

IceCube's high-energy starting event (HESE) analysis \cite{icecube-evidence-2013}
has now observed 54~events with a reconstructed energy above 30\,TeV, 39 of
which are identified as cascades and 14 as track events
\cite{icecube-hese-icrc-fouryears}\footnote{One of them has been identified as a
coincident air-shower event.}. Most of the observed events originate from the
Southern hemisphere, corresponding to down-going neutrinos in the IceCube
detector. Due to the different topologies of the events, the angular resolution
is roughly $10$--$15^\circ$ for cascades and $1^\circ$ for muons. The expected
background due to atmospheric muons and neutrinos is about $12$ and $9$ events
respectively, resulting in a significance of over $5\sigma$ for the observation
to be incompatible with the background. This significance has been obtained by
applying designated event selection cuts using the outer layers of the detector
as veto against incoming charged particles. The best constraints on an (assumed
diffuse) astrophysical spectrum come from a maximum likelihood analysis using
both HESE and other event samples \cite{icecube-max-likelihood}, finding a
neutrino flux proportional to $E^{-2.5}$, disfavouring at $2.1\sigma$ an
$E^{-2}$ spectrum with a cutoff at a few PeV. The distribution of the neutrino
directions combined with the angular resolution does not (yet) allow for the
identification of one or more point sources. Deviations from flavour-uniformity
are only weakly constrained \cite{icecube-flavour-2015}, and tau neutrino events
have not yet been identified \cite{icecube-tau-uhe,icecube-tau-recent}.

The prime physics case for KM3NeT Phase-2.0 ARCA is to measure and investigate
the signal of neutrinos observed by IceCube with different methodology, improved
resolution and a complementary field of view.

\subsubsection{Assumptions}
\label{sec-sci-ass}

\begin{figure}
\begin{center}
\includegraphics[width=10cm]{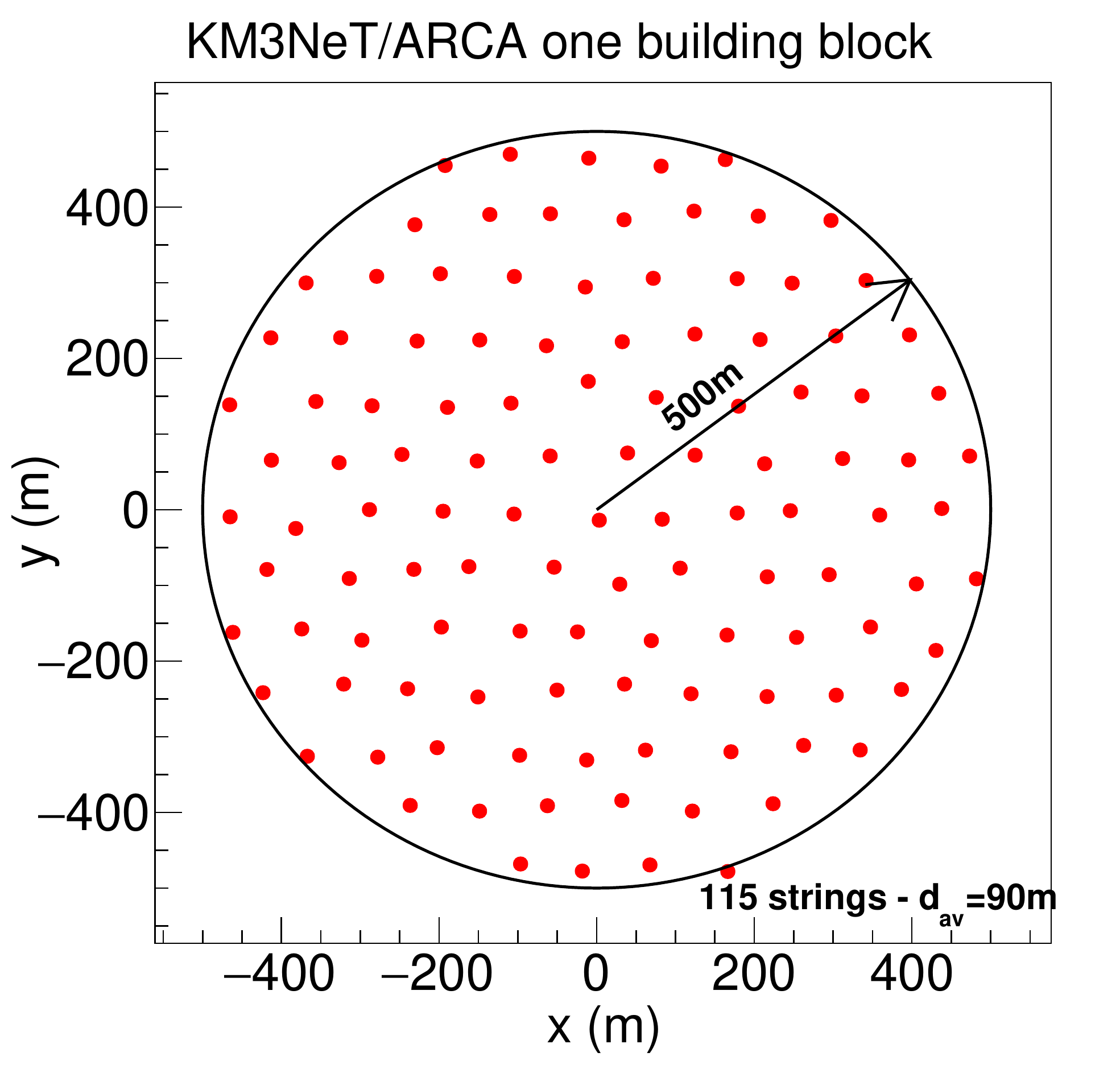}
\end{center}
\caption{Footprint of one building block of the ARCA benchmark detector (top view), with 115 strings (90\,m average spacing), with 18 OMs each (36\,m spacing). The instrumented volume is 0.48 $\mathrm{km}^{3}$ (R=500\,m, z=612\,m).}
\label{fig.footprint}
\end{figure}

The basic assumption in the following studies is that the ARCA detector will
comprise two KM3NeT building blocks, providing an instrumented volume of about
one cubic kilometre, i.e.\ of similar size as the IceCube detector. All analyses
reported in this document are performed for a horizontal distance between
strings of 90\,m and vertical distance between adjacent optical modules of
36\,m. The footprint of one block is shown in \myfref{fig.footprint}. To
estimate the dependence of the sensitivity on the geometrical detector
configuration, an alternative layout with 120\,m distance between strings but
unchanged vertical distances is being investigated; this configuration
corresponds to an increase of the instrumented volume to 1.7\,km$^3$. In both
cases a water depth of 3.5\,km and a latitude of $36^\circ\;16'$\,N were
assumed, corresponding to the Italian KM3NeT site (KM3NeT-It, see
\mysref{sec:deepsea_arca}).

\begin{figure}
\begin{center}
\includegraphics[width=0.7\textwidth]{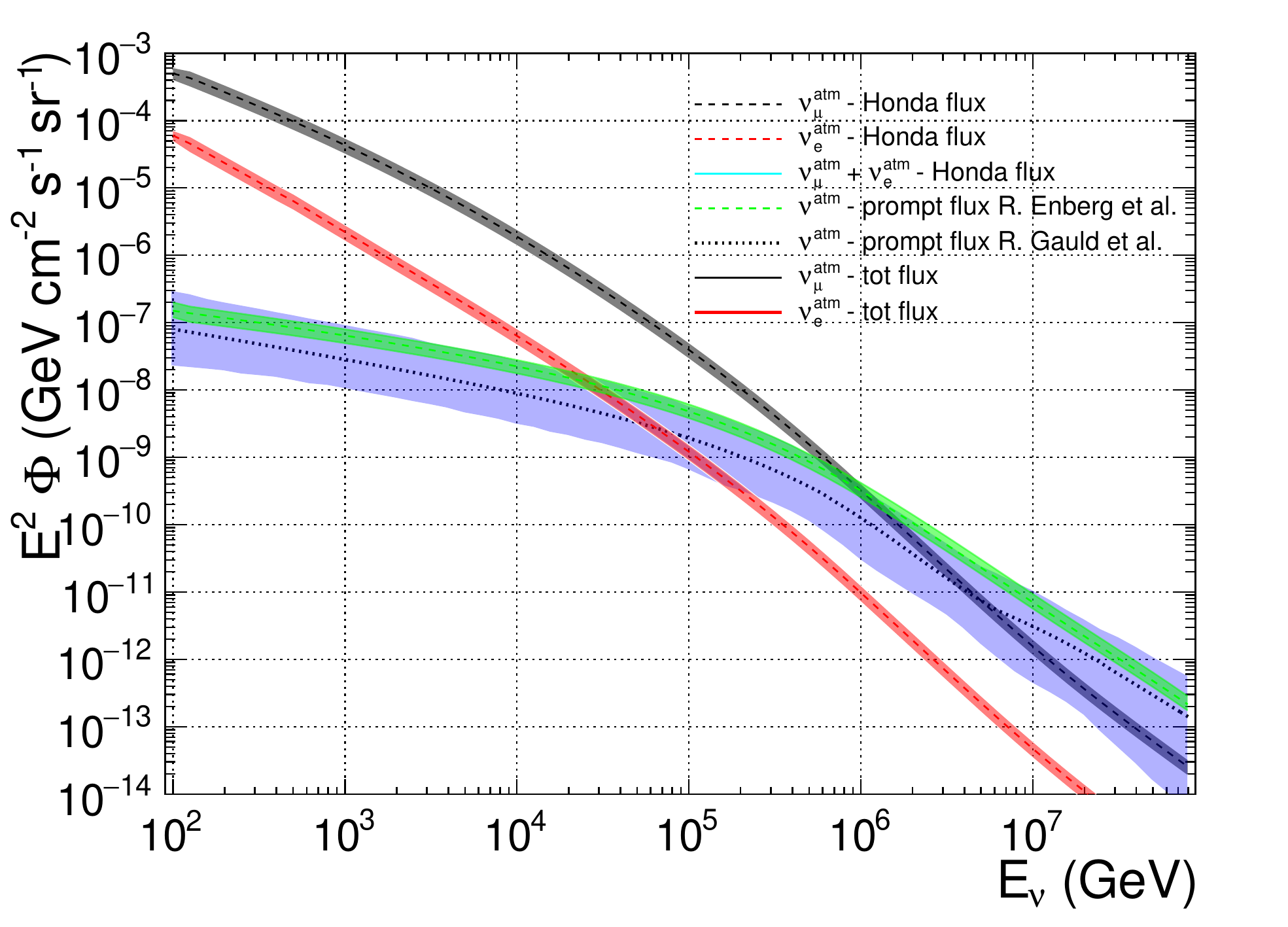}
\end{center}
\caption{%
Atmospheric neutrino fluxes as a function of the neutrino energy. The bands
represent the uncertainties in the conventional (red and black bands) and in the
prompt (green and blue bands) components assumed in this work (see text).}
\label{fig.NuAtmSpectra}
\end{figure}

The following sensitivity studies are discussed in the following:
\begin{itemize}
\item
{\bf Cascade events from a diffuse flux, including high-energy starting muon tracks} 

This analysis includes all neutrino flavours. Owing to an efficient suppression
of the atmospheric muon and neutrino backgrounds (see below), a $4\pi$ angular
coverage has been achieved.
\item 
{\bf Up-going, diffuse flux of muon (anti-)neutrinos} 

This analysis is usually referred to as the ``conventional'' diffuse flux
analysis. Traditionally, it does not include the upper hemisphere, with the
exception of a small zenith region above the horizon.
\item 
{\bf Muon (anti-)neutrinos from a diffuse Galactic plane flux}

Up-going muon track events are used for an analysis covering an extended region
of the Galactic plane near the Galactic centre in the Southern sky.
\item 
{\bf Up-going flux of muon (anti-)neutrinos from point sources}

In order to quantify the sensitivity of KM3NeT Phase-2.0 to extragalactic and
Galactic point sources of neutrinos, both a generic $E^{-2}$ spectrum from point
sources and spectra with energy cut-off for specific Galactic sources with
non-zero radial extension have been considered.
\item 
{\bf Cascade events from point sources}

KM3NeT/ARCA's resolution in the cascade channel will allow us to use these
events in point-source searches. The sensitivity of such an analysis is
evaluated against generic $E^{-2}$ point-sources.
\end{itemize}

The background of atmospheric neutrinos assumed in these analyses corresponds to
the so-called Honda flux \cite{honda-2007} with a prompt component as calculated
by Enberg \cite{enberg-2008}. A correction taking into account the ``knee'' of
the cosmic ray spectrum has been applied to both conventional and prompt
atmospheric neutrino fluxes according to the prescription in
\cite{icecube-diffusemuon-2014} and references therein. The Honda
parameterisation includes an anisotropy caused by the Earth's magnetic field,
while the prompt component is assumed to be isotropic in the full solid angle. 
Moreover, in the sensitivity studies the effect of the uncertainties on the
atmospheric neutrino flux has been estimated. An uncertainty of $\pm25\%$ was
assumed for the intensity of the conventional Honda flux. For the prompt
component, the uncertainty band estimated in \cite{enberg-2008} has been used. 

Recently, new calculations of the prompt neutrino component have been reported in \cite{Bhattacharya2015, Garzelli2015, Gauld2016}. The calculation of \cite{Gauld2016} followed that in \cite{Garzelli2015}, from which it differs mainly in the use of different input, in particular the parton distributions functions (PDFs). The PDFs in \cite{Gauld2016} were further constrained by taking into account LHCb measurements at 7 TeV.)

In \myfref{fig.NuAtmSpectra} the different components of the atmospheric
neutrino flux are reported for $\nu_e$ and $\nu_\mu$; see
\mysref{sec-sci-too-sim} for details on the background from atmospheric muons.
 
It should be noted that the results reported in the following are preliminary
and some analysis details are not yet fully completed and optimised. Also, the
analyses reported above do not reflect the full physics potential of
ARCA; the event resolutions shown in \mysref{sec-sci-too-rec} can be
used to characterise ARCA's ability to probe any assumed extraterrestrial
neutrino fluxes.

\subsection{Simulations}
\label{sec-sci-too-sim}

Monte Carlo simulations have been used to simulate the detector response to
particles incident on the detector, their interaction with the medium
surrounding the detector and subsequent Cherenkov light production, and the
detector response in terms of the PMT data sent to shore. The software packages
described in this section have mostly been developed in the ANTARES
Collaboration and then adapted to KM3NeT. The simulation is based on the nominal
detector geometry described in \mysref{sec-sci-ass} and \myfref{fig.footprint}
-- see \mysref{sec:deepsea_arca} for further details. Each of the two ARCA blocks are 
treated identically and independently -- simulations are performed for a single
block, and the effective lifetime (event rate) is multiplied by two. The effects of
position and orientation calibration uncertainties are estimated using dedicated
simulations, as described in \mysref{sec-sci-sys}.

\subsubsection{Event generation}

The relevant volume for Cherenkov light production is defined as a cylinder with
height and radius of about 3 absorption lengths larger than the instrumented
volume (the ``can''), limited by the seabed below. The first step in the
simulation chain is the generation of particle fluxes incident on the can --
neutrinos from astrophysical sources, and the atmospheric muon and neutrino
backgrounds -- within which a detailed description of particle behaviour and
Cherenkov light production is required.

Astrophysical and atmospheric fluxes of (anti-)neutrinos of all three flavours
($\nu_e$, $\nu_\mu$, and $\nu_\tau$) are simulated with a code propagating
neutrinos through the Earth (density profile from \cite{earthprofile-1996}) and
generating their interactions in rock and sea water. For reactions outside the
can, long-range interaction products (muons and taus) are subsequently
propagated to the can. Both neutral-current (NC) and charged-current (CC)
reactions are simulated. The deep inelastic scattering (DIS) cross-sections,
which are dominant in the energy range relevant to this study, are implemented
using the LEPTO code \cite{lepto-1997}. The CTEQ6D table of parton distribution functions
is used, and the resulting behaviour -- especially in the small-$x$ region --
validated up to $10$~PeV. Quasi-elastic
scattering and resonance production are also taken into account, by using
RSQ \cite{barr-1987} below $300$~GeV. Reactions of
$\overline\nu_e$ with electrons in the atmosphere are relevant in the energy
regime of resonant $W$ production (``Glashow resonance'') around 6.3\,PeV and
are simulated according to the leading-order electroweak cross sections. The
propagation of muons in rock and water is performed with MUSIC
\cite{music-1997}. Tau leptons, which have a life time of $2.9\times10^{-13}\,$s
and thus typically travel only very short distances before decaying, are
propagated by assuming them to be minimally ionising particles, and decayed
using TAUOLA \cite{tauola-1991}.

\begin{figure*}
\begin{center}
\includegraphics[width=0.6\textwidth]{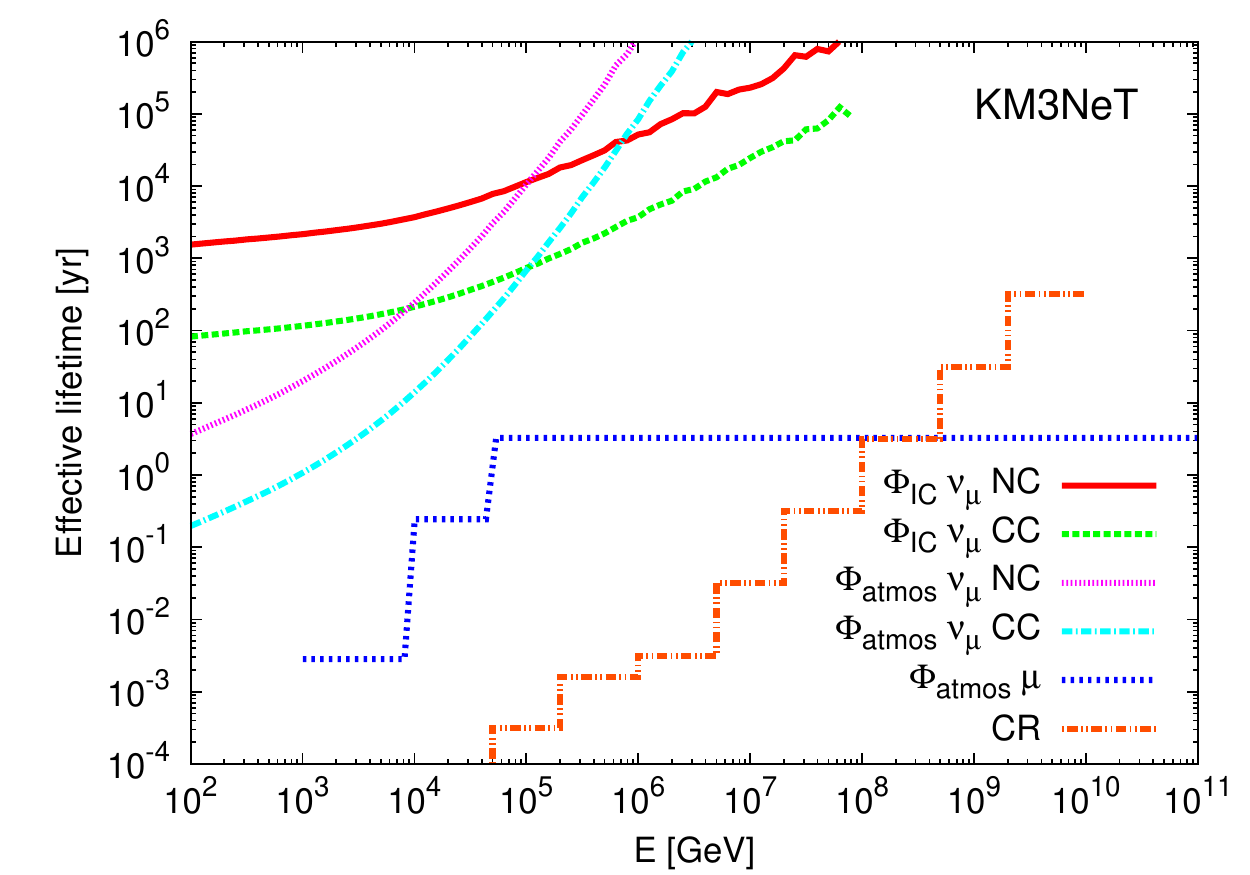}
\end{center}
\caption{%
Effective simulated lifetimes for neutrinos: $\nu_{\mu}$~NC (cascade-like),
$\nu_{\mu}$~CC (track-like), for both the IceCube diffuse flux
\cite{icecube-max-likelihood} and atmospheric spectra; atmospheric $\mu$ events;
and cosmic ray (CR) events from CORSIKA, as a function of neutrino / muon-bundle
/ cosmic-ray energy E. The lifetimes for other neutrino channels ($\nu_e$ and
$\nu_\tau$) are similar to that of $\nu_\mu$~NC, except for the atmospheric
$\nu_\tau$ events, which have effectively infinite lifetime (since the estimated
flux is very small).}
\label{fig:sim-livetimes}
\end{figure*}

Atmospheric muons, produced in cosmic-ray interactions in the atmosphere, can
penetrate to the detector volume if their energy at the sea surface is in the
TeV range or above. This is frequently the case, both for single muons, and muon
`bundles' up to several hundred muons from a primary cosmic ray event. 
Atmospheric muons therefore establish an important, high-rate background that is
simulated using MUPAGE \cite{mupage-2008,mupage_becher}. Single and multiple atmospheric muon
events are generated using a parameterisation of the flux of muon bundles at
different depths and zenith angles. In the present analysis, three simulated
atmospheric muon event samples are used, with muon bundle energies exceeding
1\,TeV, 10\,TeV, and 50\,TeV, respectively, in order to provide sufficient
coverage in the high-energy regime. The corresponding lifetimes of these and the
neutrino productions are shown in \myfref{fig:sim-livetimes}.

The correlated flux of atmospheric neutrinos and muons from the same primary
cosmic ray interaction is simulated with CORSIKA~v7.4001
\cite{corsika-web}, in order to investigate the `self-veto' effect
\cite{self-veto-2009} for high-energy studies. GHEISHA \cite{gheisha} and
QGSJET01 \cite{qgsjet} were respectively used to model low- and high-energy
hadronic interactions, and the curvature of the Earth was accounted for. Both
muons and neutrinos are recorded at sea-level; muons are propagated to and
through the can with MUSIC, while one neutrino from each event is forced to
interact. The intention was to estimate the effect of accompanying muons on
high-energy atmospheric neutrino events (\mysref{sec-self-veto}) -- thus, only
events with at least one muon at can level, and one neutrino above 10\,TeV, are
kept, which excludes all up-coming neutrino events\footnote{Events with only a
neutrino or atmospheric muon bundle are already simulated using standard
methods}. The resulting event sample forms only a small fraction of all
atmospheric muon bundles, but a significant fraction of all down-going
atmospheric neutrino background events above 10\,TeV. Therefore, analyses using
CORSIKA events down-weight the standard atmospheric neutrino events to avoid
double-counting. Additionally, CORSIKA underestimates the expected atmospheric
neutrino flux at high energies \cite{honda-2007,enberg-2008}, and this is
corrected for as per \cite{icrc_selfveto}.

\subsubsection{Detector response}

A quantity often used to characterise the detector response for neutrino telescopes is the neutrino effective area, $A_{\rm eff}$, defined here such that the rate, $R_{\rm trig}$, of particles being detected at trigger level is equal to the flux of particles through $A_{\rm eff}$. Here, $A_{\rm eff}$ is calculated as a function of neutrino flavour, $\ell$, and energy, $E_{\nu_\ell}$, relative to the flux $\Phi$ incident upon the Earth, i.e.:
\begin{eqnarray}
A_{\rm eff}(E_{\nu_{\ell}}) & \equiv & \frac{R_{\rm trig}(E_{\nu_{\ell}})}{\Phi(E_{\nu_{\ell}})}. \label{eq.Event_from_source}
\end{eqnarray}
For a point-like source, $A_{\rm eff}$ is calculated relative to the rate $R_{\rm trig}$ [s$^{-1}$ GeV$^{-1}$] and flux $\Phi$ [m$^{-2}$ s$^{-1}$ GeV$^{-1}$] from that source, while for a diffuse flux, the solid-angle-integrated values of $R_{\rm trig}$ and $\Phi$ are used. Along with the detector efficiency, $A_{\rm eff}$ also includes the neutrino cross-section, and the probability for neutrinos to be absorbed in the Earth, resulting in a smaller value of $A_{\rm eff}$ than the physical cross-sectional area of the instrument.

The generated particles propagated to the can level, or generated inside the can volume,
are then tracked in the sea water using tabulated results from full GEANT\,3.21
simulations of relativistic muons and electromagnetic cascades to
generate the number of Cherenkov photons detected by the KM3NeT photomultiplier
tubes (PMTs). The light production from hadronic or mixed
hadronic/electromagnetic cascades is scaled to that from purely electromagnetic
cascades according to the energy and type of constituent particles. The program
takes into account the full wavelength dependence of Cherenkov light production,
propagation, scattering and absorption; and the response of the PMTs as
described in \mysref{sec-tec} and modelled in \cite{icrc_geant}, including
absorption in the glass and optical gel, the PMT quantum efficiency, the reduced
effective area for photons arriving off-axis, and the effect of the reflecting
expansion cones \cite{km3net-econe-2013}.

\begin{figure*}
\begin{center}
\includegraphics[width=0.7\textwidth]{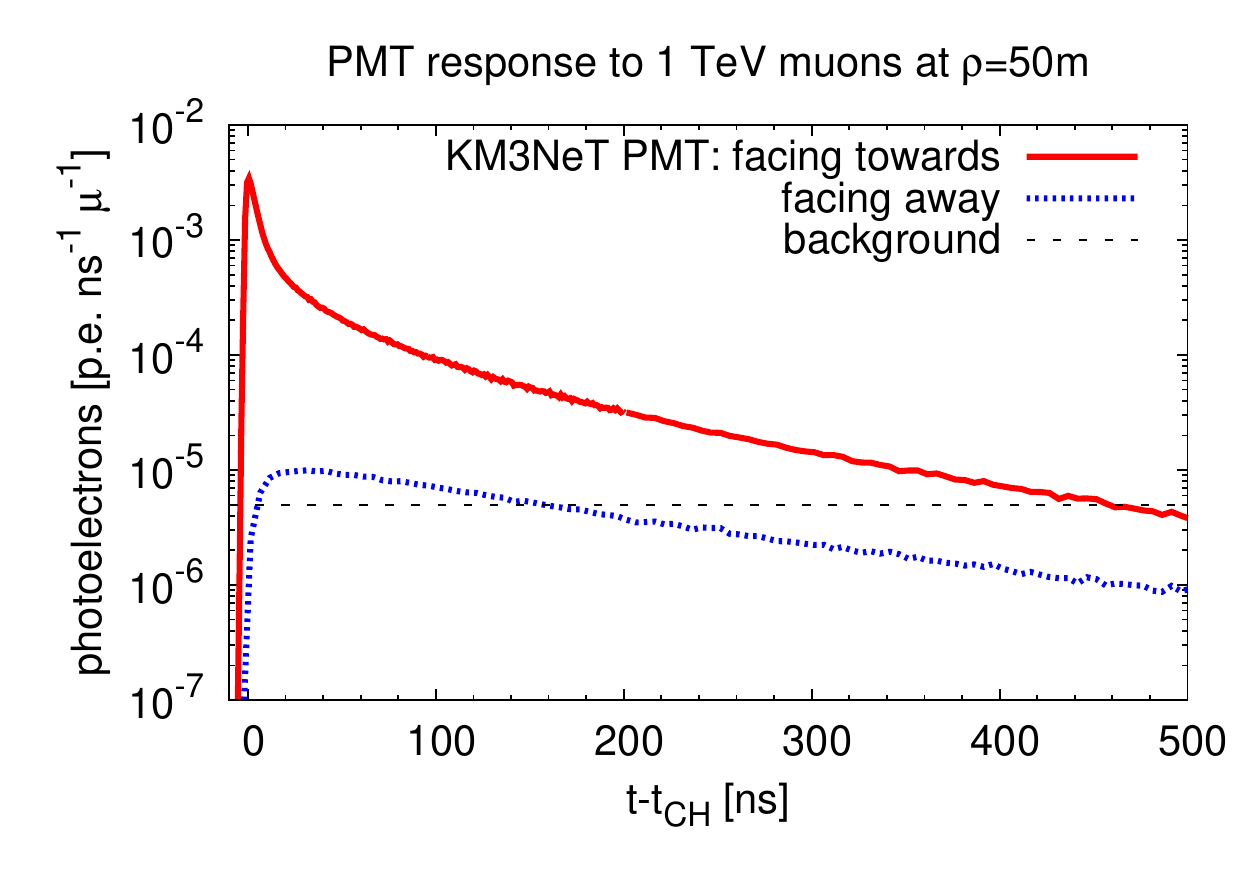}
\end{center}
\caption{
Simulated time distribution (relative to the nominal time of the Cherenkov shock
front) of photoelectrons detected by a KM3NeT PMT from a 1\,TeV muon track with
closest approach distance $\rho=50$\,m . The PMT is simulated facing both
towards (red) and away (blue) from the track. The optical background rate is
also shown for comparison. The PMT response (time-smearing of $\sim 2$\,ns) is
not included.}
\label{fig:muon_dom}
\end{figure*}

Hits from background photons (mostly due to $^{40}$K decays in the sea water) in
an event are simulated by adding random noise hits with a rate of 5\,kHz per
PMT. Correlated hits over multiple PMTs on the same optical module from single
$^{40}$K decays are also included, with $\{2,3,4\}$-fold coincidences at rates
of $\{500,50,5\}$\,Hz per DOM. The singles and coincidence rates as well as the
angular dependence are in reasonable agreement with the results from the
prototype detection unit deployed at the KM3NeT-It site
\cite{km3net-ppmdu-2015}. An example of the simulated time-distribution of
photons detected by a KM3NeT PMT from a 1\,TeV muon 50\,m from the track is
given in \myfref{fig:muon_dom}.

KM3NeT PMT hits are recorded via the start time and the duration of the signal
above a predefined threshold (time-over-threshold, or ToT). This scheme is
implemented in the detector simulation, with the simulated response of
individual PMTs to photon hits being based on laboratory measurements. The full
transit-time distribution is implemented on a per-photon basis, corresponding
approximately to a 2\,ns Gaussian smearing for the majority of photons. Hit
amplitudes are smeared, and the start time and ToT are calculated by accounting
for sequences of photo-electrons on PMTs that cannot be resolved in time,
saturation effects at around 40 simultaneous photoelectrons, and a maximum ToT
readout of 255\,ns. After this step, each event contains a complete and unbiased
snapshot of all hits recorded during a time window around the event,
representing a part of the stream of data sent to shore.

\begin{figure*}
\begin{center}
\includegraphics[width=0.6\textwidth]{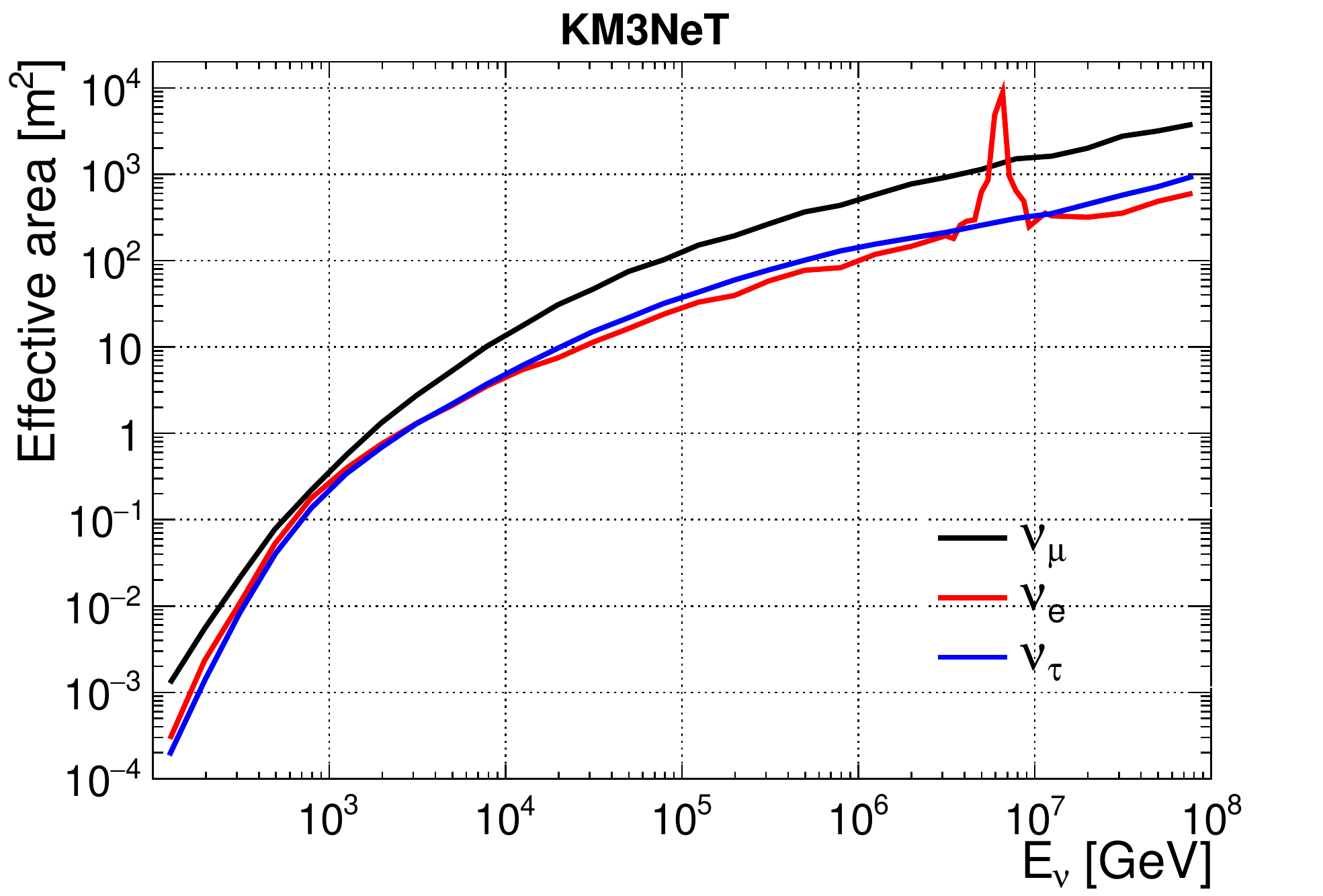}
\end{center}
\caption{%
Effective areas of ARCA (two blocks) at trigger level for $\nu_\mu$, $\nu_e$,
and $\nu_\tau$, as a function of neutrino energy $E_\nu$. The effective area is
defined relative to an isotropic neutrino flux incident on the Earth, is
averaged over both $\nu$ and $\overline{\nu}$, and includes both NC and CC
interactions. The peak at $6.3$~PeV is due to the Glashow resonance of $\overline{\nu}_e$.} 
\label{fig:sim-effareas}
\end{figure*}

The final stage is to simulate on-shore triggering, as described in
\mysref{sec-tec-com-tri}. This process takes filtered L1 hits
(photon hits on multiple PMTs within a short time window on the same
OM) and generates a trigger if multiple nearby OMs record such events at causally
connected times within a spherical (cascade) or cylindrical (track) geometry.
Trigger parameters have been tuned so as to minimise false triggers on optical
backgrounds, while registering all reconstructible physics events. In the case of
ARCA, the real-time trigger rate is dominated by down-going atmospheric muons,
and trigger settings were set to keep the corresponding data rate manageable.

The trigger settings correspond to a coincidence (L1) time window of $\Delta T = 10\,\rm{ns}$, 
and a minimum number of 5 L1 hits for both the shower trigger and the muon trigger.
Only Monte Carlo events which pass either trigger condition are available for further
analysis, as is the case for the on-shore trigger. The resulting effective areas
are given in \myfref{fig:sim-effareas}.

Following  \myeref{eq.Event_from_source}  to evaluate the number of detectable events from a specific neutrino source that maximises the significance (see \mysref{sec-sci-sen}) these effective areas have to be corrected for the number of events that survive the cuts of the analysis.

\begin{figure*}
\begin{center}
\includegraphics[width=0.49\textwidth]{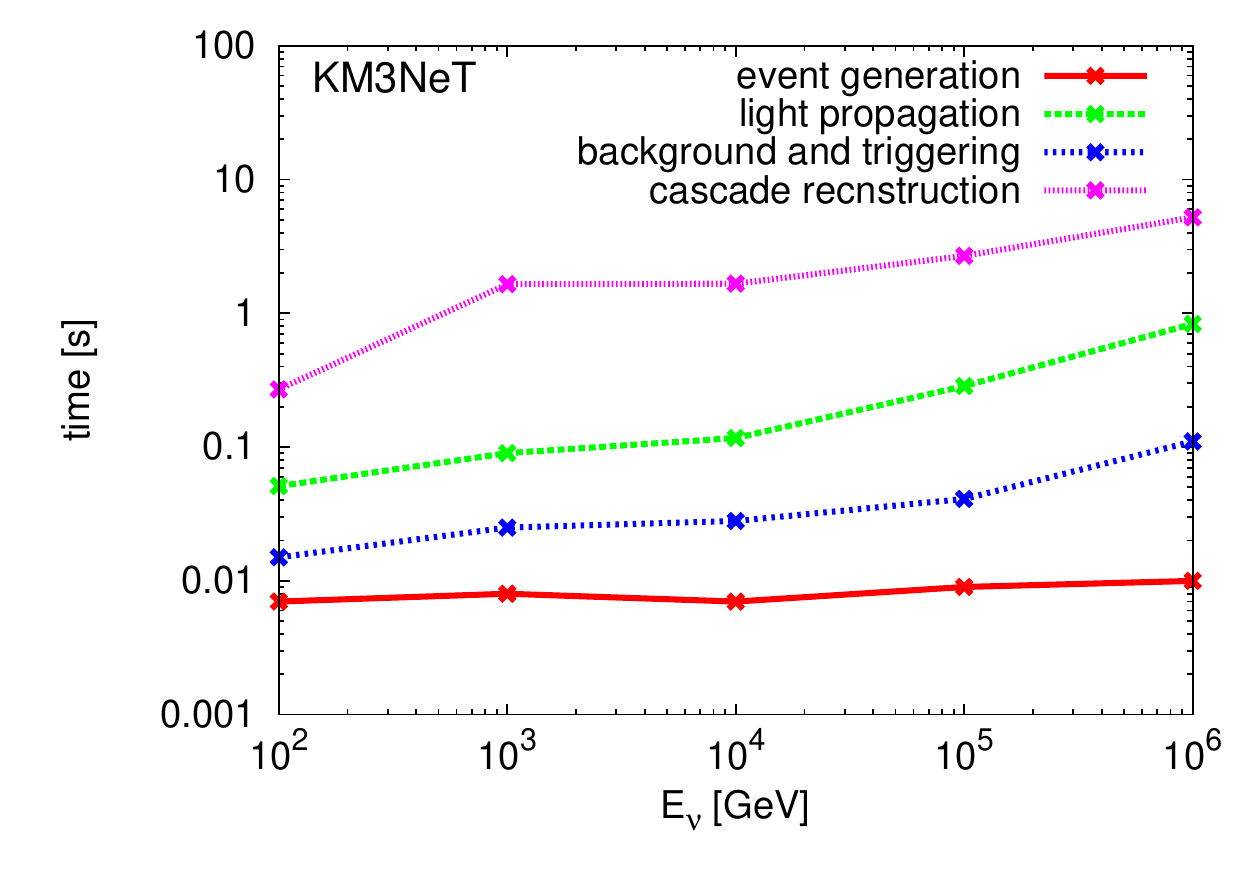} 
\hfill
\includegraphics[width=0.49\textwidth]{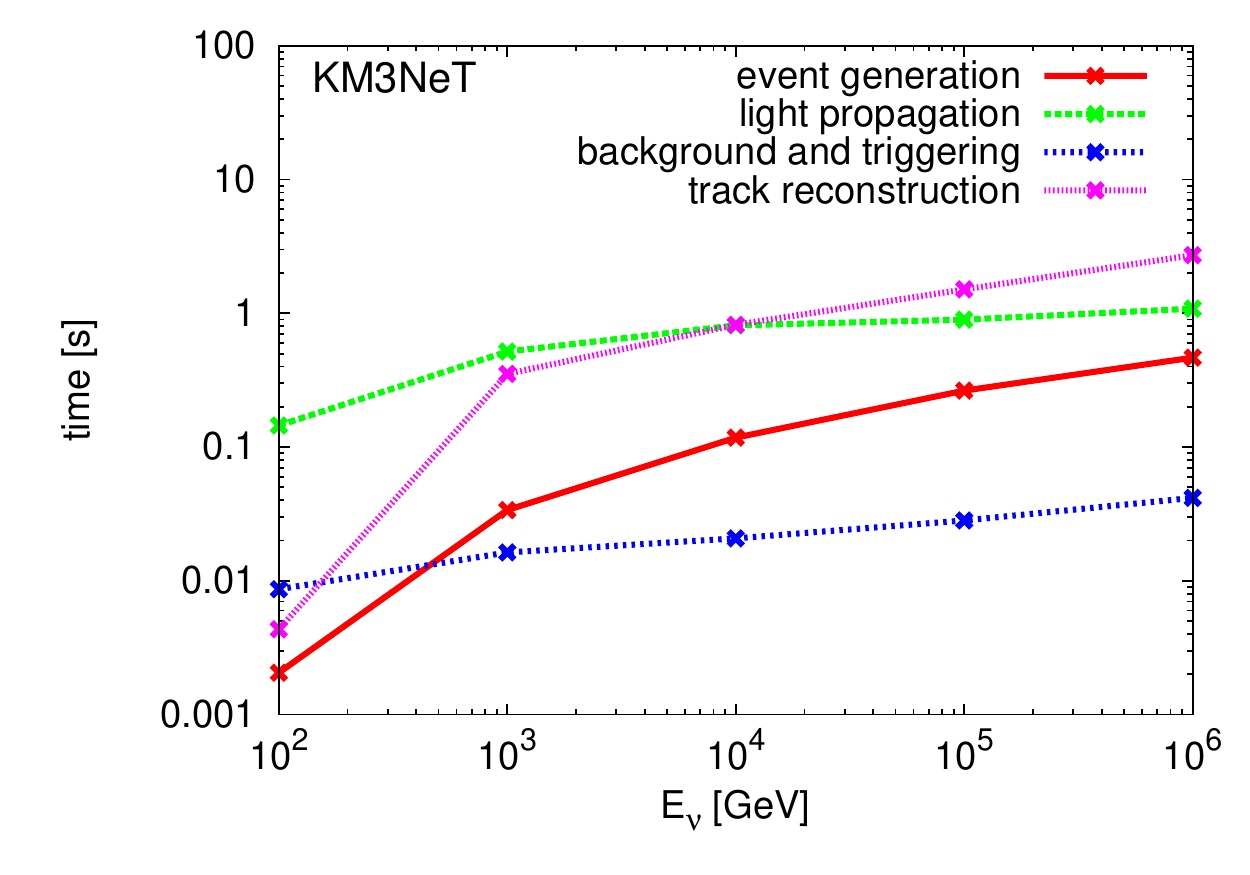}
\end{center}
\caption{
Simulation times per triggered event for different simulation stages, for
$\nu_e$ CC (cascade) events (left), and $\nu_{\mu}$ CC (track) events (right). 
The times are calculated for simulations run on a single dedicated CPU at the
Centre de Calcul de l'IN2P3/CNRS.}
\label{fig:sim-times}
\end{figure*}

The simulation times per event for different stages are shown in
\myfref{fig:sim-times}. The simulation time is dominated by event
reconstruction and light propagation, which can reach up to a few seconds per
event at high energies. The cascade reconstruction time does not reduce quickly
at low energies, since it includes in the likelihood fit PMTs which have no
detections.

The MC events simulated with the described codes have been compared with the
data from a prototype of the string that was deployed at the Italian site and
that took data for about one year \cite{km3net-ppmdu-2015}. The very good
agreement between the data and the MC simulation have demonstrated the high
reliability of the MC simulation chain.

\subsubsection{Further improvements}

The simulation chain for ARCA is mature, but not complete, and several additions
will be required for future data analysis. These are:

\begin{itemize}

\item 
The simulation of tau (anti-)neutrinos is performed using some simplifications. 
Charged-current tau interactions within the Earth are treated as absorbing the
neutrino, i.e.\ the tau `regeneration' effect is not included. Additionally,
only two- and three-body tau decay modes (approximately three quarters of all
decays) are currently implemented -- the branching ratio of $\sim 17.4$\% for
the decay to a muon is kept constant, while other modes are re-normalised to the
remaining $83.6\%$, and result in almost identical event topologies at high
energies.

\item 
The MUPAGE package for generation of atmospheric muons does not contain a prompt
component originating from charm decays in cosmic-ray-induced air showers. The
flux of atmospheric muons with energies above roughly 10\,TeV is therefore
underestimated, although likely only by a small amount. A refined simulation has
recently been provided in the CORSIKA \cite{corsika-web} framework, where the
correlations between conventional and prompt muon and neutrino fluxes are
adequately included at the event-by-event level. While a production with the new
CORSIKA\,v7.4005 has begun, the high CPU demand has so far prevented this
simulation from being fully processed through the Monte Carlo chain and used for
analysis.

\item 
Atmospheric muon events which coincidentally arrive simultaneously with
neutrino events have not been simulated, since it is anticipated that resolving
multiple components will prove feasible for ARCA. An explicit production of
coincident muon events will need to be produced in order to verify this.

\end{itemize}

\subsubsection{Event reconstruction}
\label{sec-sci-too-rec}

Two broad event classes can be identified in a high energy neutrino telescope:
track-like events and cascade-like events:
\begin{itemize}
\item 
The track-like events are generated by muons that are produced in the matter
inside or surrounding the detector through CC interactions of
$\nu_\mu$\,($\overline\nu_\mu$) and $\nu_\tau$\,($\overline\nu_\tau$). CC
reactions of $\nu_\tau$\,($\overline\nu_\tau$) produce a muon with a branching
ratio of 17\%, when the emerging $\tau$ decays into a $\mu$.
\item 
The cascade-like events are produced in the matter near or inside the detector
volume through CC interactions of $\nu_e$\,($\overline\nu_e$) and
$\nu_\tau$\,($\overline\nu_\tau$) and in NC interactions of neutrinos of all
flavours. CC $\nu_\tau$\,($\overline\nu_\tau$) interactions produce cascade
events with a branching ratio of $83\%$.
\end{itemize}

\begin{figure}
\begin{center}
\includegraphics[width=0.49\textwidth]{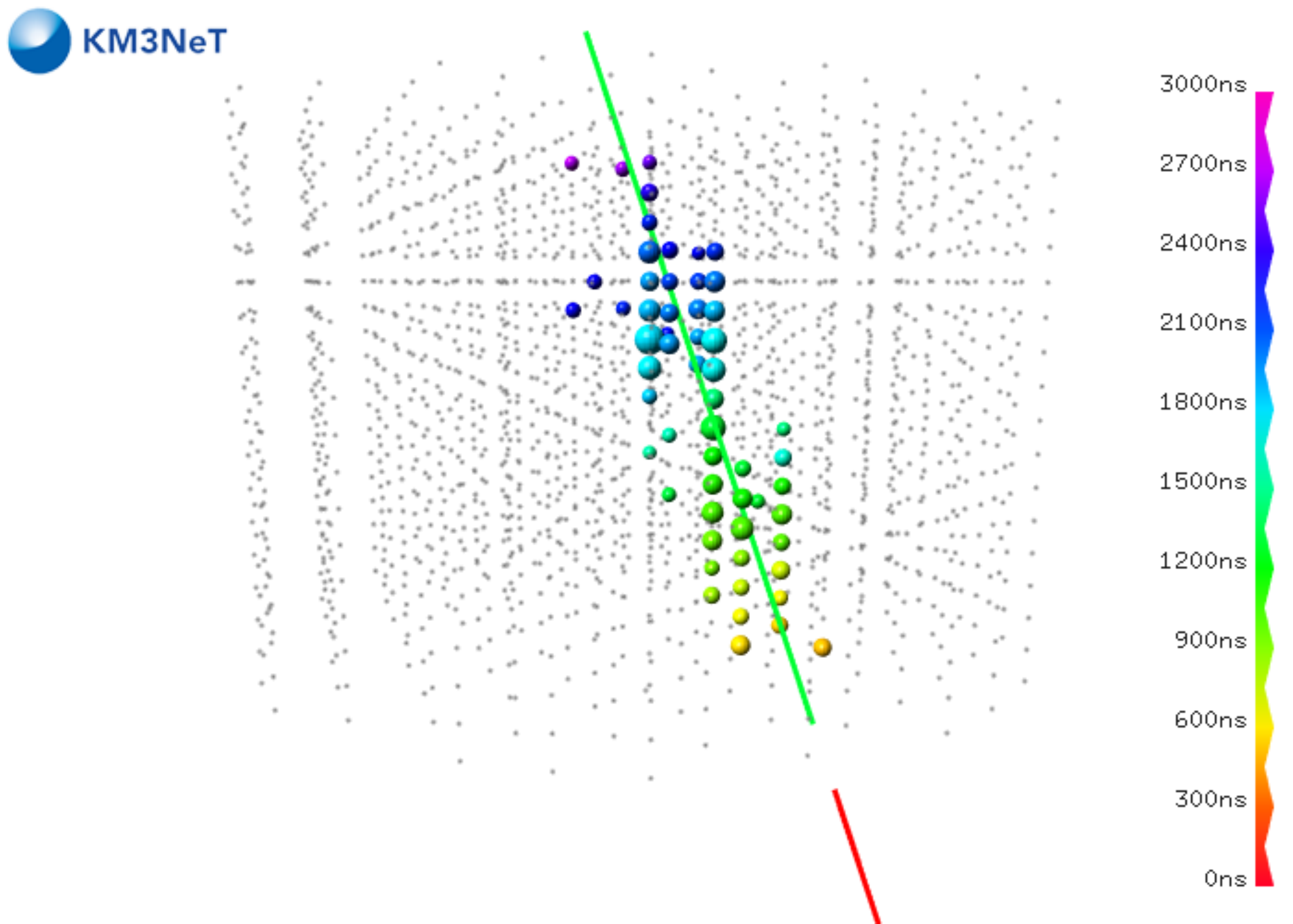}
\includegraphics[width=0.49\textwidth]{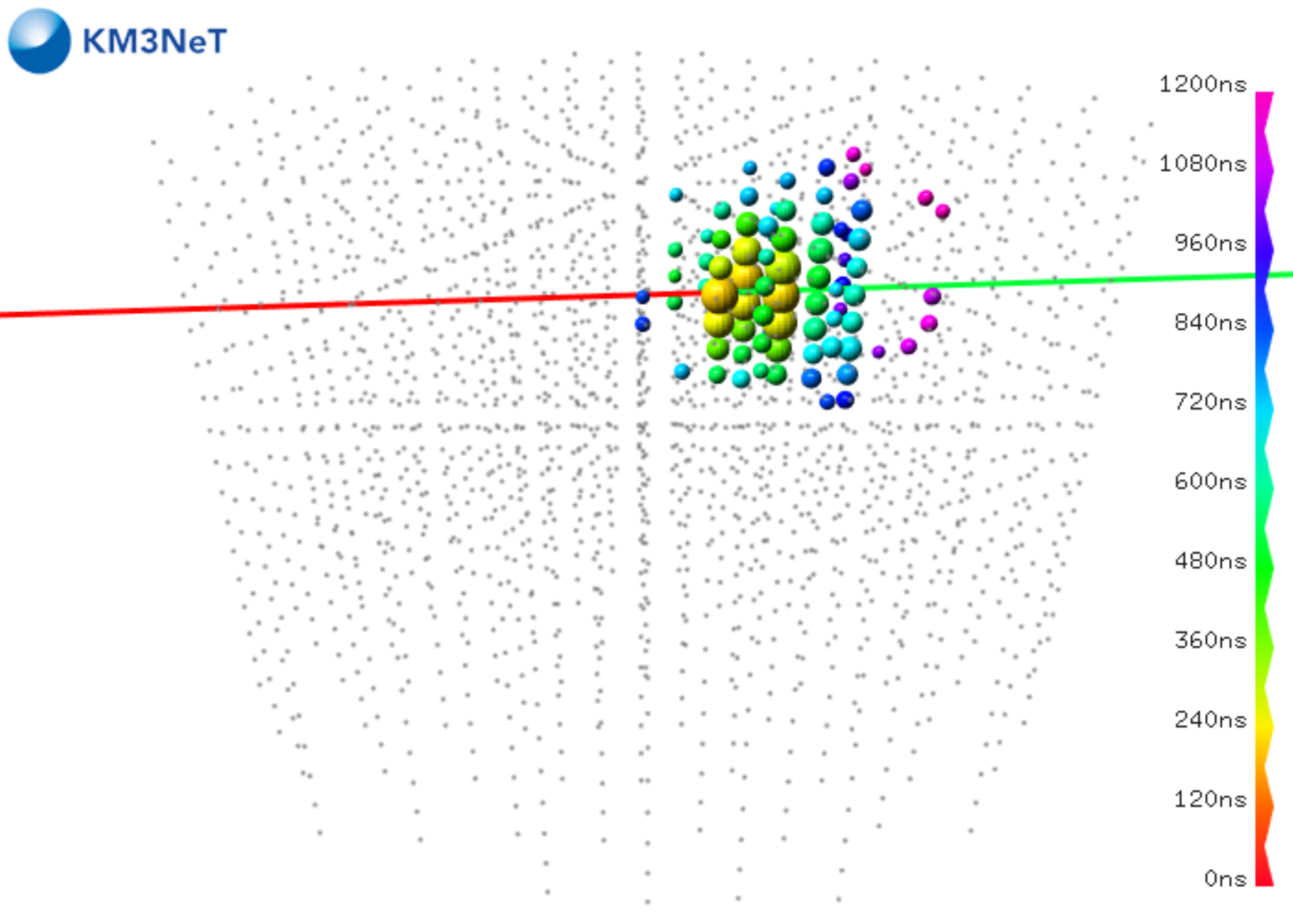}
\end{center}
\caption{
Event displays for a simulated $\nu_\mu$ CC event (left) and a contained
$\nu_\mu$ NC event (right),
showing only DOMs with a total ToT (summed over all
31~PMTs) of more than 30\,ns in a narrow time window. In both cases, the
incoming neutrino is indicated by the red line, and the outgoing lepton (muon or
neutrino) by the green line. The colour scale gives the hit times with respect to the time of the
neutrino interaction, while the size of the circles are proportional to the total
ToT on each DOM. DOMs without hits are shown by grey dots.}
\label{fig.EventDisplay}
\end{figure}

These two events classes produce very different time-space hit patterns in the
detector. The cascade-like events are characterised by a very dense hit pattern
close to the neutrino interaction point. A significant fraction of the neutrino
energy is released in a hadronic shower (and, in the case of
$\nu_e$\,($\overline\nu_e$) CC interactions, the rest in an electromagnetic
cascade), thus allowing for a good estimate of the neutrino energy. A track-like
event is characterised by the Cherenkov light from the emerging muon that can
travel large distances through Earth rock and sea water. The spatial hit pattern
in this case is closely related to the muon direction, thus allowing for a
precise measurement of the latter. Typical hit patterns for track-like and
cascade-like events are shown in \myfref{fig.EventDisplay}

Starting from the ANTARES experience, algorithms that reconstruct direction,
energy and interaction vertex of the neutrinos from the muon tracks or the
showers have been developed. These have been optimised for `pure' track events
($\nu_\mu$ CC events far from the detector, where only a single energetic muon
is observed) and for cascade events ($\nu$ NC and $\nu_e$ CC events, where only
a cascade is observed), respectively. Thus their performance on more complicated
event topologies is not optimal; prospects for improvements are discussed at the
end of this section.

\begin{figure}
\begin{overpic}[width=0.49\textwidth]{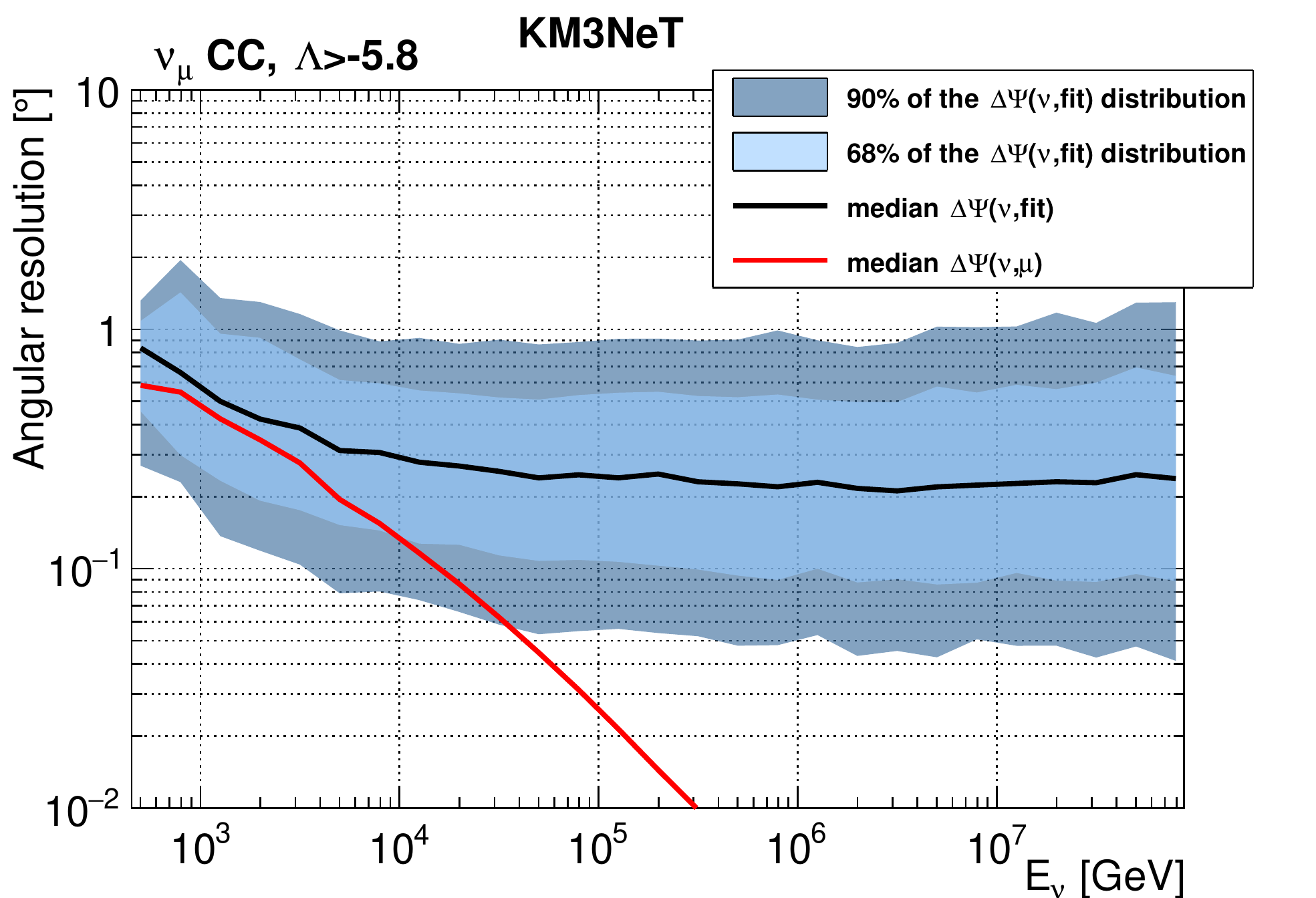}
\end{overpic}
\hfill
\begin{overpic}[width=0.49\textwidth]{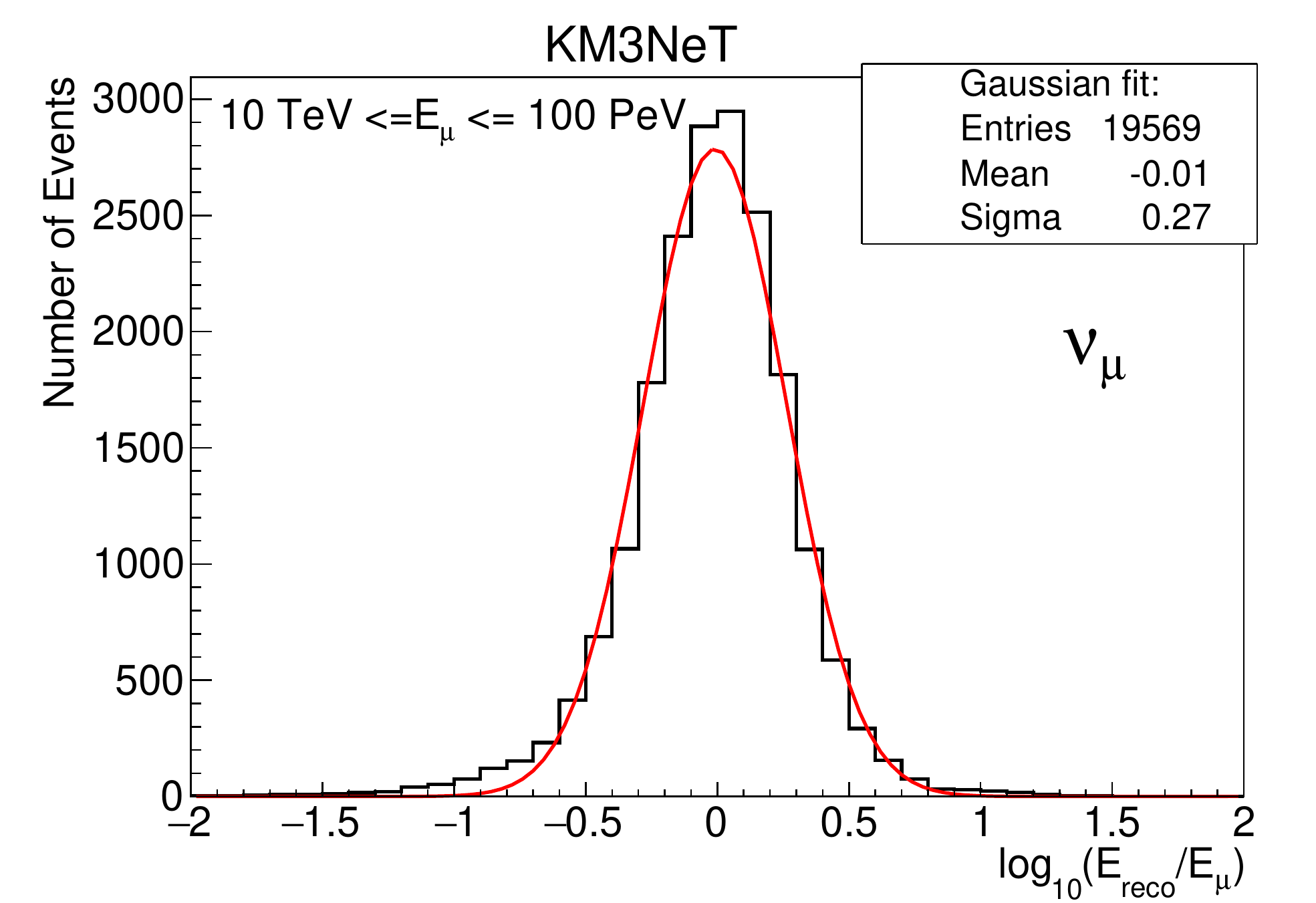}
\end{overpic}
\caption{%
Left panel: median of the angle between the neutrino and the reconstructed muon
direction (black line) and between the neutrino and the true muon direction (red
line), for selected $\nu_\mu$ CC events ($\Lambda>-5.8$, see below). The dark
and light blue bands represent the 90\% and 68\% quantiles of the distributions. 
Right panel: distribution of $\log_{10}(E_\text{reco}/E_\mu)$, where
$E_\text{reco}$ is the reconstructed and $E_\mu$ is the true muon energy for
events with $E_\mu \geq 10\,$TeV that satisfy a containment criterion. The red
line represents the Gaussian fit.}
\label{fig.TrackResolution}
\end{figure}

\paragraph{Track reconstruction}

Muons with energies above 1\,TeV can reach track lengths of the order of
kilometres and have a direction that is nearly collinear with that of the parent
neutrino. To reconstruct the muon direction -- and consequently the neutrino
direction -- an algorithm is used that maximises the likelihood that the
observed space-time PMT hit pattern is consistent with Cherenkov emission from
the fitted muon trajectory. An initial hit selection exploits hit coincidences
between PMTs in the same optical module or between different optical modules to
remove uncorrelated hits from background photons, mostly from $^{40}K$ decays. 
The reconstruction of the muon trajectory starts with a linear fit, followed by
three consecutive fitting steps, each using the results of the previous one as
starting point. A pseudo-vertex position is also estimated, which, however,
usually is related to the entry point of the muon in the detector rather than to
the location of the interaction vertex; this quantity is useful for background
rejection. In addition to the track information (direction and pseudo-vertex) an
estimator of the fit quality, $\Lambda$, and the number of hits associated with
the final track fit, $N_\text{hit}$, are determined. The $\Lambda$ parameter is
used in the analysis to reject badly reconstructed events, in particular
atmospheric muons mis-reconstructed as up-going. The $N_\text{hit}$ parameter is
related to the muon energy and is used to reject low-energy events that are
mainly due to atmospheric neutrino background. A very good angular resolution of
about $0.2^\circ$ is achieved for neutrinos above 10\,TeV, see
\myfref{fig.TrackResolution} (left).

The amount of light collected by the PMTs when a muon travels inside the
detector is correlated with the muon energy. To estimate the muon energy, a
method exploiting this dependence by means of an artificial neural network has
been developed. The first step is the selection of events with a reconstructed
muon track travelling inside the detector for an adequate distance. The second
step is the evaluation of several quantities related with the total event ToT
and with the number of DOMs hit. These quantities are used to feed the neural
network. The energy resolution obtained for well reconstructed (cut on $\Lambda$
applied) and contained events is $\simeq 0.27$ units in $\log_{10}(E_\mu)$
for $10\,\text{TeV} \leq E_\mu \leq 10\,$PeV (see \myfref{fig.TrackResolution}
right); without containment requirement, the resolution slightly worsens to
$\simeq 0.28$ units. Further details on the track reconstruction code can be
found in \cite{icrc_arca_track_recon}.

This energy reconstruction method must be trained on appropriate samples of MC
events and is not yet fully integrated in the reconstruction software for ARCA. 
A simple energy reconstruction using the $N_\text{hit}$ parameter is embedded in
the reconstruction software and gives results of almost equivalent quality. This
method is used for the sensitivity studies presented in the following.

\subsubsection{Cascade reconstruction}
\label{sec-sci-too-rec-casc}

The length of a cascade event depends logarithmically on the cascade energy
and is of the order of 10\,m in the energy range relevant for ARCA analyses. At
the length scale of typical distances between optical modules, cascades thus
produce almost point-like signatures, characterised by vertex position,
direction, and energy. CC interactions of $\nu_\mu$ and $\nu_\tau$, if they
happen in the detector volume, also produce cascades, but the outgoing $\mu$,
$\tau$, or $\tau$ decay products produce a more complex signature. Hence,
cascade reconstruction is optimised for $\nu$ NC and $\nu_e$ CC interactions, and then the
performance is assessed on the latter class. Three independent algorithms have
been applied to reconstruct cascade vertex position, direction and energy. The
first has been specifically developed for ARCA, and exhibits the best
performance. The second and third have been adapted from ANTARES analyses and
have outputs which prove useful in event classification and background
discrimination. All three are described here, although only the performance of
the first is shown.

\begin{figure*}
\begin{center}
\includegraphics[width=0.49\textwidth]{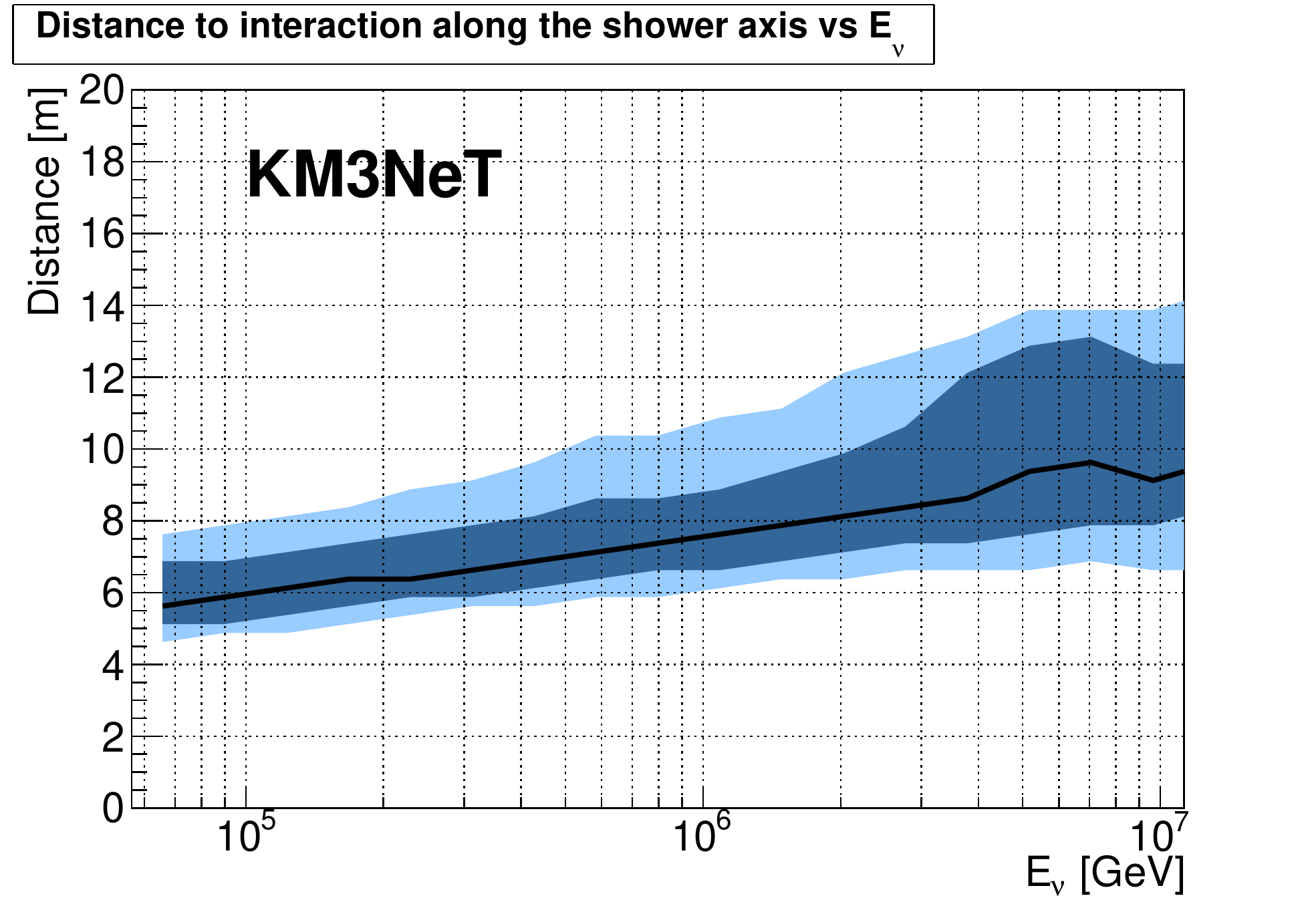}
\hfill
\includegraphics[width=0.49\textwidth]{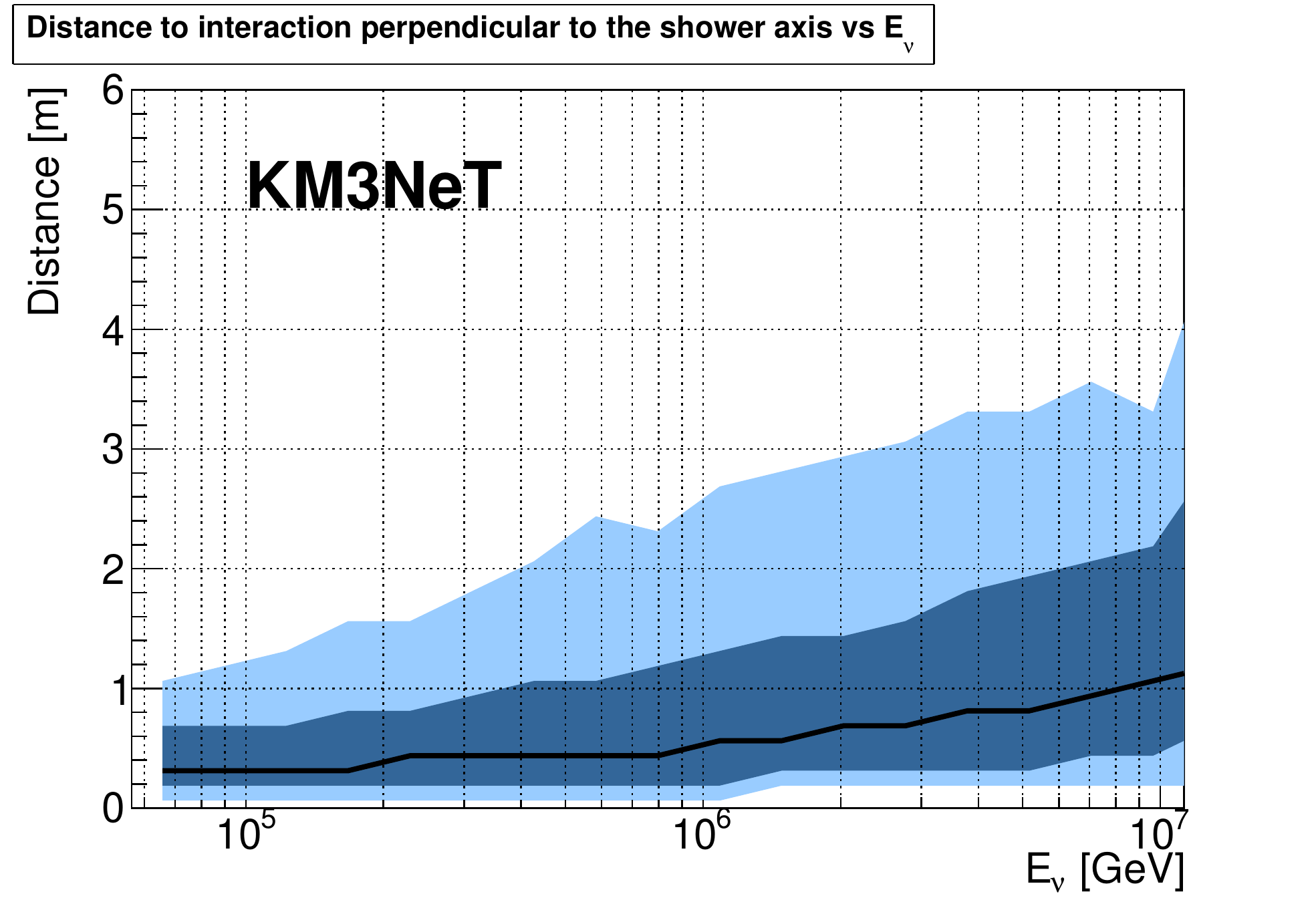}
\end{center}
\caption{%
ARCA vertex resolutions for contained $\nu_e$~CC events using Algorithm 1, after
the event selection of \mysref{sec-sci-too-dif}. Left: resolution in the
longitudinal direction, showing the offset from the MC vertex to the shower
maximum. Right: directional resolution in the lateral direction, which gives the
characteristic accuracy of $\sim 0.5$\,m. For both plots, the black line shows
the median value; dark blue shaded regions give the 68\% range, while light
blue shaded regions give the 90\% range.} 
\label{fig:sim_aa_shwr_vertex}
\end{figure*}

The first algorithm (Algorithm~1) has been specifically developed to exploit the
information provided by the KM3NeT multi-PMT DOM. The hit selection is designed
to be simple and to allow for a fast reconstruction. Hits on the same PMT within
350\,ns are merged using the time of the first hit, and coincidences of two
merged hits within 20\,ns on a single DOM are used for the vertex fit. This fit
minimises time-residuals assuming a spherically expanding shell of light about
an assumed cascade maximum. The offset of this fit from the MC true vertex
position in the longitudinal direction (\myfref{fig:sim_aa_shwr_vertex} left)
mostly measures the shower elongation, while the offset in the lateral direction
(\myfref{fig:sim_aa_shwr_vertex} right) measures the accuracy, reaching a
precision well below 0.5\,m in the high-energy regime.

The direction and energy are reconstructed using maximum-likelihood methods,
applied to the merged hits as described above. All PMT hits within $-100$\,ns to
$+900$\,ns of the expected Cherenkov light-front from the vertex fit are used. 
Thus each PMT only has a 0.2\% chance of receiving a random hit from the optical
background. Rather than fitting the ToT ($\sim$charge) measurement from each
PMT, the algorithm simply fits the probability of a PMT recording one or more
photons within this time-window, making the procedure highly robust. This
probability is estimated from simulations as a function of PMT distance and
pointing direction to the shower, angle from the shower axis to the PMT
position, and electromagnetic-equivalent cascade energy. The strong geometrical
dependence in hit probabilities allows for a very high reconstruction quality:
nearby PMTs facing towards the cascade, close to the Cherenkov angle, will tend
to have a hit probability of unity, while distant PMTs facing away from the
cascade, far from the Cherenkov angle, will tend to have a hit probability of
zero.

\begin{figure*}
\begin{center}
\includegraphics[width=0.49\textwidth]{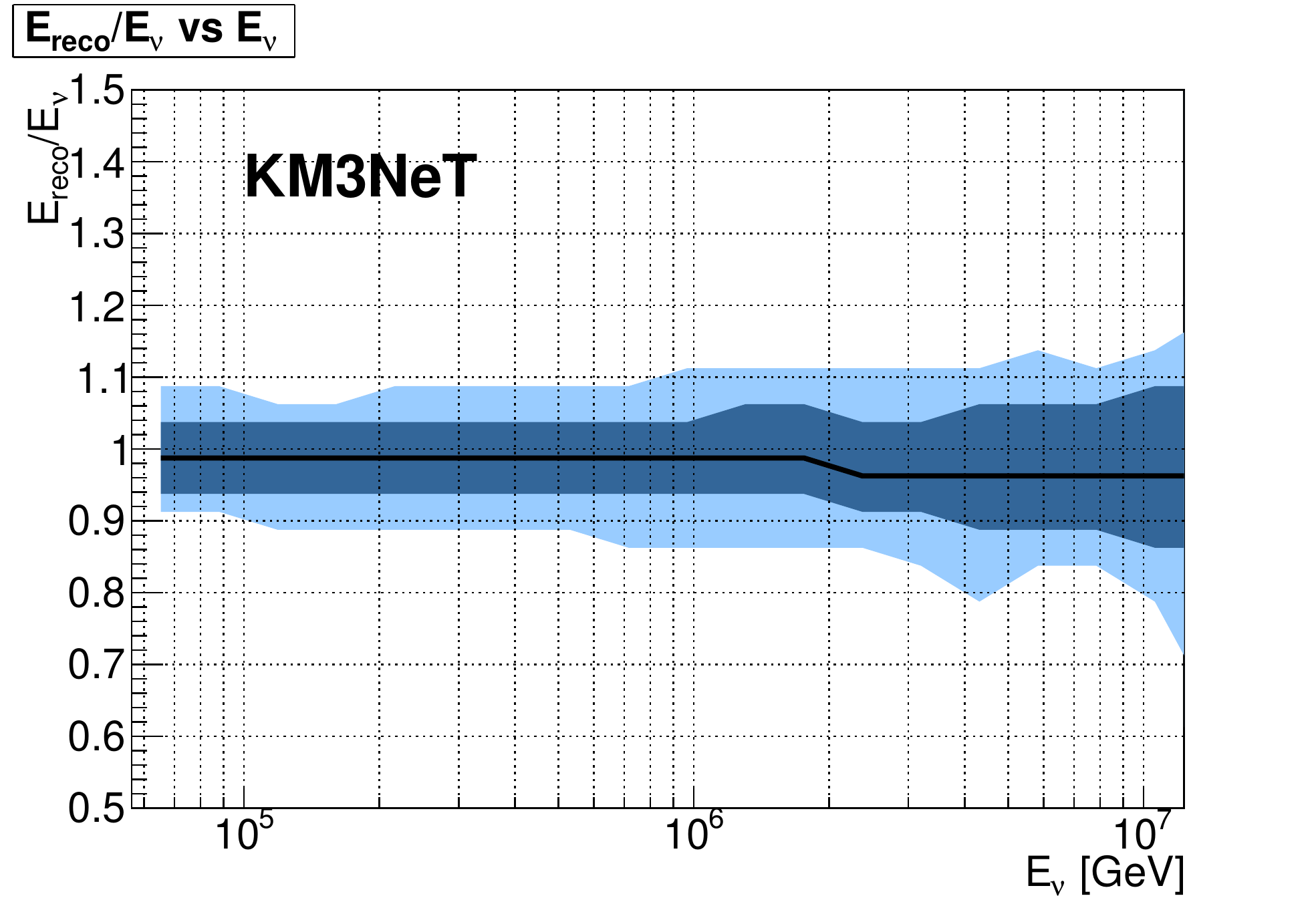}
\hfill
\includegraphics[width=0.49\textwidth]{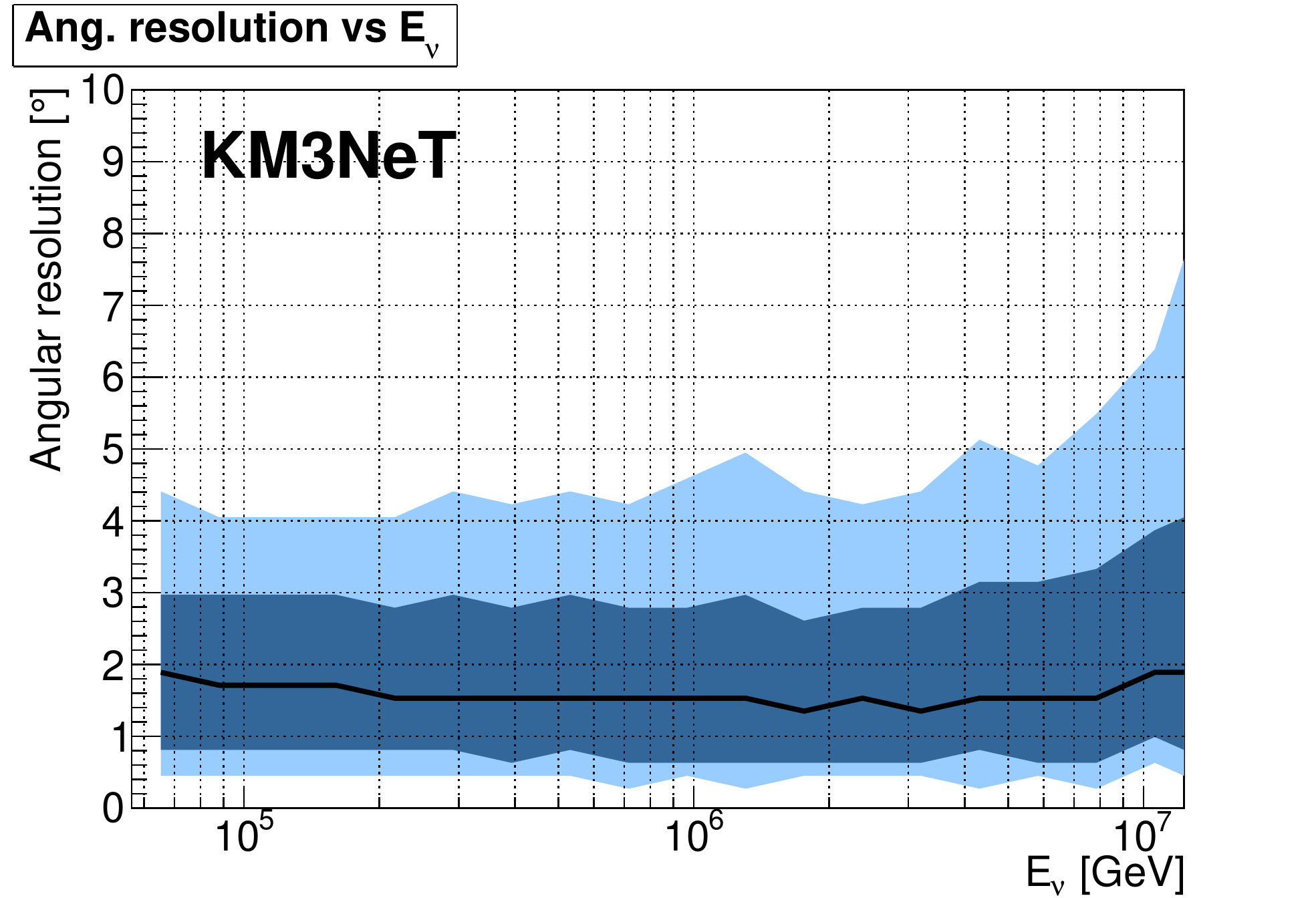}
\end{center}
\caption{%
ARCA resolutions for contained $\nu_e$~CC events using Algorithm 1, after the
event selection of \mysref{sec-sci-too-dif}. Left: energy resolution, right:
directional resolution. For both plots, the black line shows the median value;
dark blue shaded regions give the 68\% range, while light blue shaded regions
give the 90\% range.} 
\label{fig:cascade_reco_resolutions}
\end{figure*}

For contained events above 50\,TeV, the $\pm 1\sigma$ energy and median
direction resolutions achieved with this method are roughly 10\% and $2^\circ$
respectively, with no loss of efficiency. The resolutions after the selection
cuts described in \mysref{sec-sci-too-dif} are shown in
\myfref{fig:cascade_reco_resolutions}. For energy above 60\,TeV, corresponding
to the approximate low-energy threshold of the cut-and-count diffuse-flux
analysis (see \mysref{sec-sci-too-dif}), the $1\sigma$ energy resolution is
characteristically 5\%, while the median directional resolution is $1.5^\circ$.
This energy resolution is close to the limit imposed by variations in
the hadronic cascade component (mostly due to the variable
inelasticity), which yields less Cherenkov light ($\sim$90\% at
100\,TeV) than the electromagnetic component \cite{Kowalski:2004qc}.

The second algorithm fits the vertex position from the positions and the arrival
times of the PMT hits using an M-estimator procedure and applies selection cuts
on the resulting quality parameters. The cascade direction is determined from
the average direction of hits with respect to the vertex position, the energy is
estimated from the observed ToT values, taking into account the expected
relative intensities at given PMT positions. The third algorithm starts from a
simple vertex estimation based on large-amplitude hits, followed by a hit
selection using this vertex and causality relations and finally by two
sequential, independent log-likelihood fits yielding first the vertex position
and then the energy and direction of the event. The algorithms yield similar
accuracy and are fully efficient for events passing the cuts. While they are
less precise than Algorithm~1, they exhibit different responses to non-cascade
events, and their output is useful for background suppression. More details on
the cascade reconstruction codes presented here can be found in
\cite{icrc_arca_casc_recon}.

\subsubsection{Prospects for improved reconstruction}

The main reconstruction goal of ARCA is to precisely determine the parameters of
track-like and cascade-like events, and the methods presented above have been
developed with this in mind. New reconstruction algorithms tuned on $\nu_\mu$
CC and $\nu_e$ CC events are in the testing phase and first results are very
promising. Also reconstructions tuned for different event classes that present
more complex topologies are in the development phase. In particular:

\begin{itemize}
\item
Improved track and cascade reconstructions

The track and cascade reconstructions described above are first-generation
algorithms developed for ARCA, and there are good prospects for improvements in
both. In fact, when the reconstruction algorithms were developed the full PMT response was not yet being implemented in the simulation
chain. Reconstructions based on a more-detailed knowledge of the detector are
currently in development or in the testing phase.

In particular, the best current cascade reconstruction (Algorithm 1) uses very little timing information to fit the cascade energy and direction, and no information from individual PMT signal magnitudes (all time-over-threshold values treated equally). A new cascade reconstruction algorithm that exploits this information in detail is under development. First estimates indicate that a cascade resolutions of $1^\circ$ may be attainable with improved efficiency.

Additionally, a new track reconstruction algorithm has recently been developed. From initial values obtained by a rigorous scan of the full solid angle,
the likelihood is maximised using a multi-dimensional probability distribution function of the arrival time of Cherenkov light from the muon. In \myfref{fig:resolution_JGandalf}, the angular resolution reached for $\nu_\mu$ CC events is reported, showing that an angular resolution better than $0.1^\circ$ is reached for events with energy higher than 100 TeV.
 
 \begin{figure*}
\begin{center}
\includegraphics[width=0.7\textwidth]{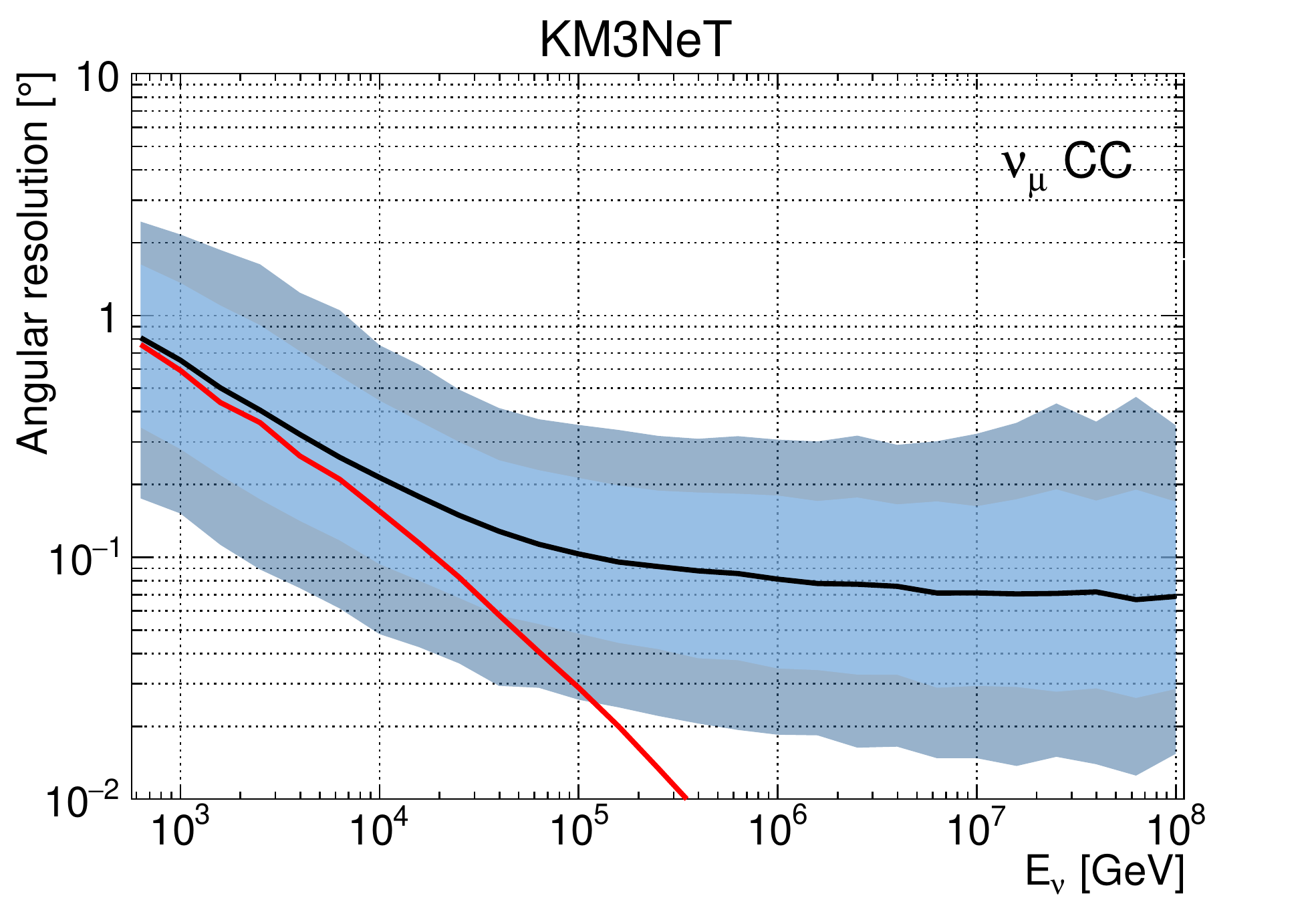}
\end{center}
\caption{%
ARCA resolutions for $\nu_\mu$~CC events using the new track reconstruction algorithm. The black line shows the median value;
dark blue shaded regions give the 90\% range, while light blue shaded regions
give the 68\% range. Quality cuts that remove most of the atmospheric muons are applied.} 
\label{fig:resolution_JGandalf}
\end{figure*}

However, these reconstructions have not yet been
processed through the full Monte Carlo chain described in \mysref{sec-sci-too-sim}, and hence are not used in the analyses presented here. However, since the atmospheric background for point-source studies (\mysref{sec-sci-too-poi}) reduces with the square of the angular resolution, using these reconstructions is expected to significantly improve the sensitivity of such studies in the near future.

\item
$\tau$ `double-bang' events

A $\tau$ produced in a $\nu_\tau$ CC interaction will on average travel
4.89\,cm/TeV before decaying. If the decay is not into a $\mu$ ($\sim 83\%$
probability), the decay products will create a cascade-like signature offset
from the first interaction vertex. At sufficiently high energies ($E \gtrsim
100$\,TeV), this second cascade will be offset from the first by distances
significantly larger than the precision of the cascade reconstruction, creating
a `double bang'. Identifying such double-bang events would be a clear signature
of the flavour of the neutrino primaries.

A preliminary investigation, conducted assuming an initial hadronic cascade
(`bang') energy of $E_1=0.2 E_{\nu_\tau}$, an outgoing tau of energy $E_\tau=0.8
E_{\nu_\tau}$, a tau decay length $\ell_\tau=4.89\,\text{m}\cdot
E_\tau/(100\,\text{TeV})$, and a second `bang' energy of $E_2=0.67 E_\tau$,
showed that cascade reconstruction Algorithm~1 could identify both events when
separated by 10\,m or more, i.e.\ for $\nu_\tau$ at $\sim
250$\,TeV and above. It is expected that an even closer separation will be resolvable.

\item
Starting track events

A $\nu_\mu$ CC interaction in the detector volume will produce a cascade at the
interaction vertex, and an outgoing $\mu$; $\nu_\tau$ CC events with subsequent
$\tau\to\mu\nu\nu$ decays will produce a similar signature. Such interactions
typically do not manifest themselves as either well-reconstructed cascade or
track events, due to the presence of the other component. An optimal
reconstruction method would separate both components and reconstruct them
simultaneously, allowing for improved energy and direction resolution on the
neutrino primary, and a better event selection.

\item
Muon bundles

Groups of muons from the core of an extended air shower (EAS) exhibit a
signature very similar to that of a single high-energy muon in the detector. 
However, their stochastic energy-loss pattern is much more uniform, and their
lateral spread is non-negligible at the characteristic spatial resolution scale
of ARCA. Currently, muon bundles are reconstructed using a single-muon
hypothesis. Identifying such events can be used to reduce the background for
studies searching for an excess of single energetic down-going muons, either
from an astrophysical $\nu_\mu$ flux, or from the prompt decay of charm
particles in EAS.

\item
Coincident EAS

The rate of down-going $\mu$ from EAS above ARCA that produce a detectable
signature in the detector is expected to be about 50\,Hz. With a typical event
duration of $5\,\mu$s, approximately one in 4,000 events will have a
coincidental muon present, corresponding to a double coincidence every
80\,s. Current simulations only model
particles for individual EAS, and current reconstruction methods only return at
most a single track or cascade event. Observe that this effect is much less
important for ARCA than it is for IceCube: the increased detector depth reduces
the rate of coincident down-going muonic background, and
the better time-resolution afforded by the low scattering in sea water allows
photons from different sources to be separated within a much narrower time
window.
\end{itemize}

\subsubsection{Background suppression}
\label{sec-sci-too-bac}

Backgrounds from atmospheric muons, as well as random coincidences of hits from
${}^{40}$K decays, are reduced to acceptable levels by applying selection cuts on
the event reconstruction quality, the reconstructed zenith angle for track-like
events and quantities related to the event energy (such as the number of hits)
or event topologies (e.g.\ using Boosted Decision Tree techniques -- 
see \mysref{sec-sci-sen}). For point-source studies, the main method of reducing
the background event rate of both muon and neutrino events 
is via the excellent angular resolution afforded by seawater,
since the background rate reduces with the square of the resolution.
However, in particular for studies of a diffuse flux, the most problematic
remaining source of background is the atmospheric neutrino flux.

\subparagraph{Self-veto of down-going atmospheric neutrinos}
\label{sec-self-veto}

\begin{figure*}
\begin{center}
\includegraphics[width=0.49\textwidth]{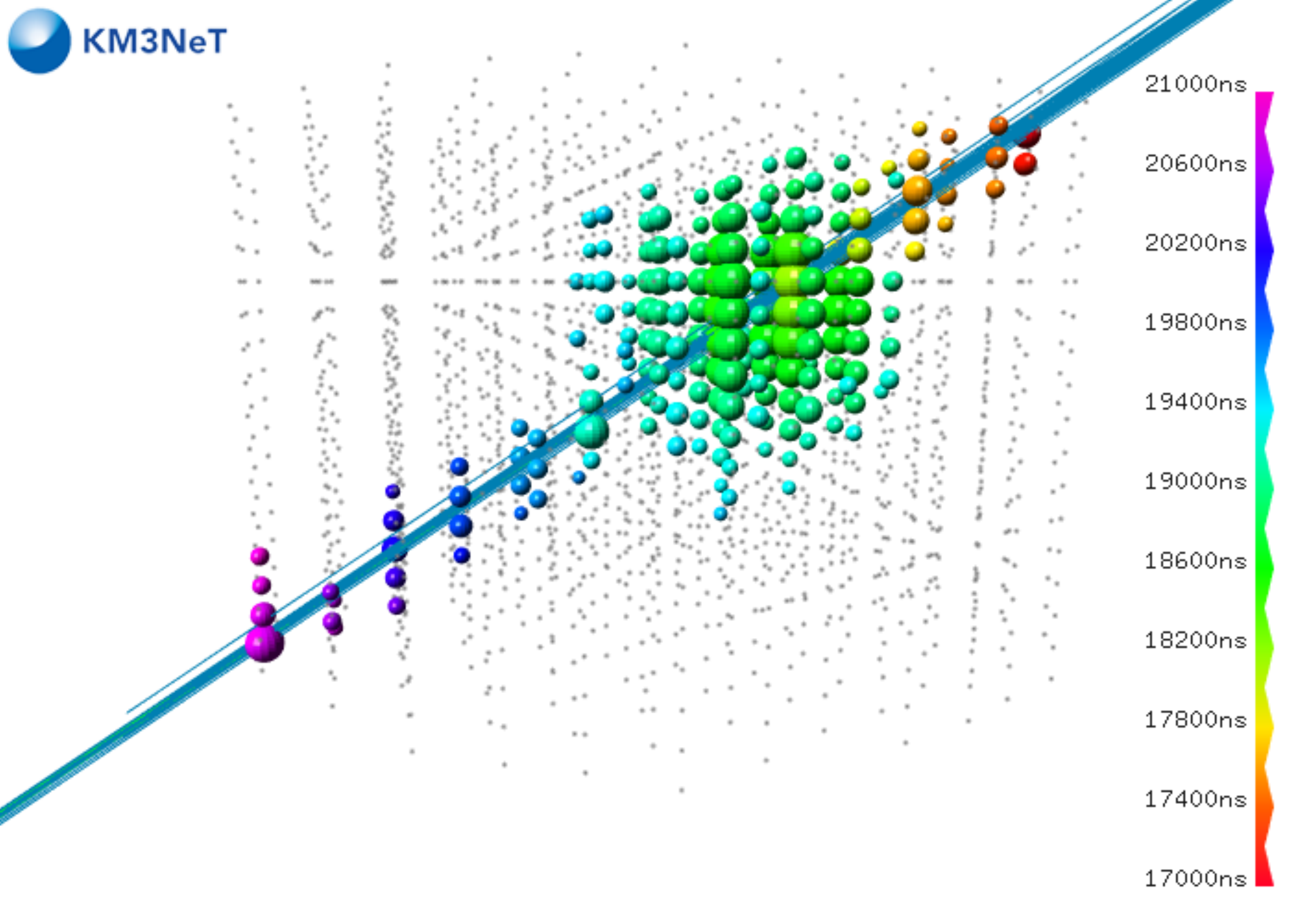}
\end{center}
\caption{
Simulated signature of a self-vetoed event: a $\sim1$\,PeV atmospheric neutrino
(creating the cascade-like signature) accompanied by a muon bundle.
Only DOMs with a total ToT (summed over all
31~PMTs) of more than 30\,ns in a narrow time window are shown. The colour scale
gives the hit times relative to the shower core impacting the sea surface -- only some of the muons
at sea level (shown as blue lines) penetrated to the detector depth. The size of the circles are proportional to the total ToT on each DOM. DOMs without hits are shown by grey dots.
} 
\label{fig:self-veto-event}
\end{figure*}

The interactions of cosmic rays with the atmosphere generates extensive air
showers (EAS) where both neutrinos and muons are produced. Despite the $\sim
3$\,km overburden of water, muons with an energy in the TeV range and above can
reach the detector, either singly, or in multiples (muon `bundles'),
particularly under low zenith angles. These muon bundles can be used to `veto'
any accompanying neutrinos, allowing for a strong reduction of the down-going
atmospheric neutrino background. This technique has been proposed in
\cite{self-veto-2009}, where it is predicted in the context of an IceCube-like
detector that atmospheric neutrinos above 10\,TeV and with zenith angles less
than $60^\circ$ can be vetoed with almost 100\% efficiency. More detailed
calculations in \cite{self-veto-2014} suggest a somewhat lower, but still
significant, veto probability.

Events simulated by CORSIKA (see \mysref{sec-sci-too-sim}) have been used to
estimate the self-veto probability, and some preliminary results for the
high-energy diffuse analysis using the cascade channel (see
\mysref{sec-sci-too-dif}) are shown in \cite{icrc_selfveto}. In the case of
ARCA, accompanying muons make neutrino-induced cascade events less cascade-like,
so that while these events are not explicitly vetoed (as would be the case with
IceCube), their topology is such that they appear more background-like than
signal-like in sensitivity studies targeting down-going cascade events (see
\mysref{sec-sci-sen}). An example of such an event is given in
\myfref{fig:self-veto-event}.

\begin{figure*}
\begin{center}
\includegraphics[width=0.49\textwidth]{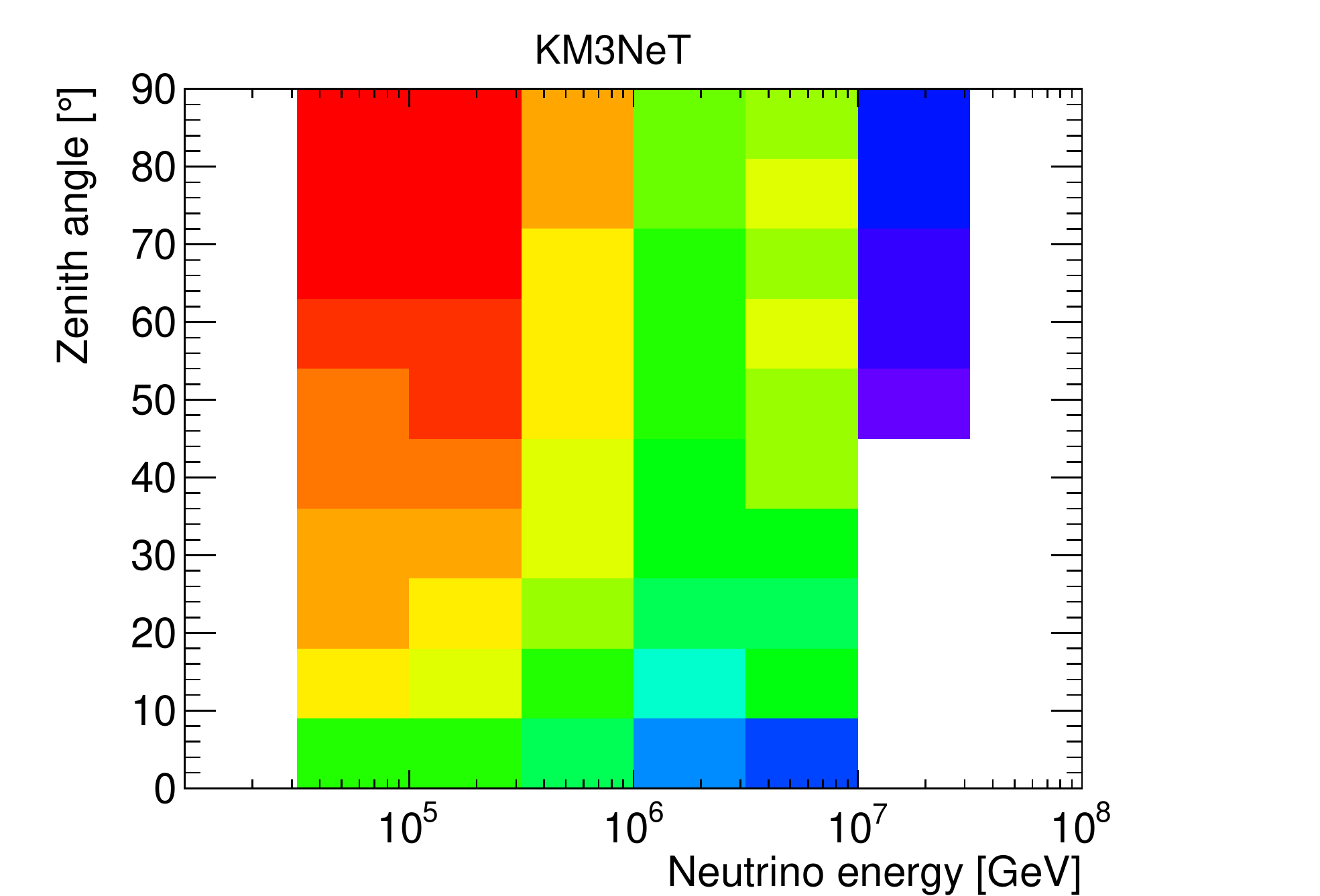} 
\hfill
\includegraphics[width=0.49\textwidth]{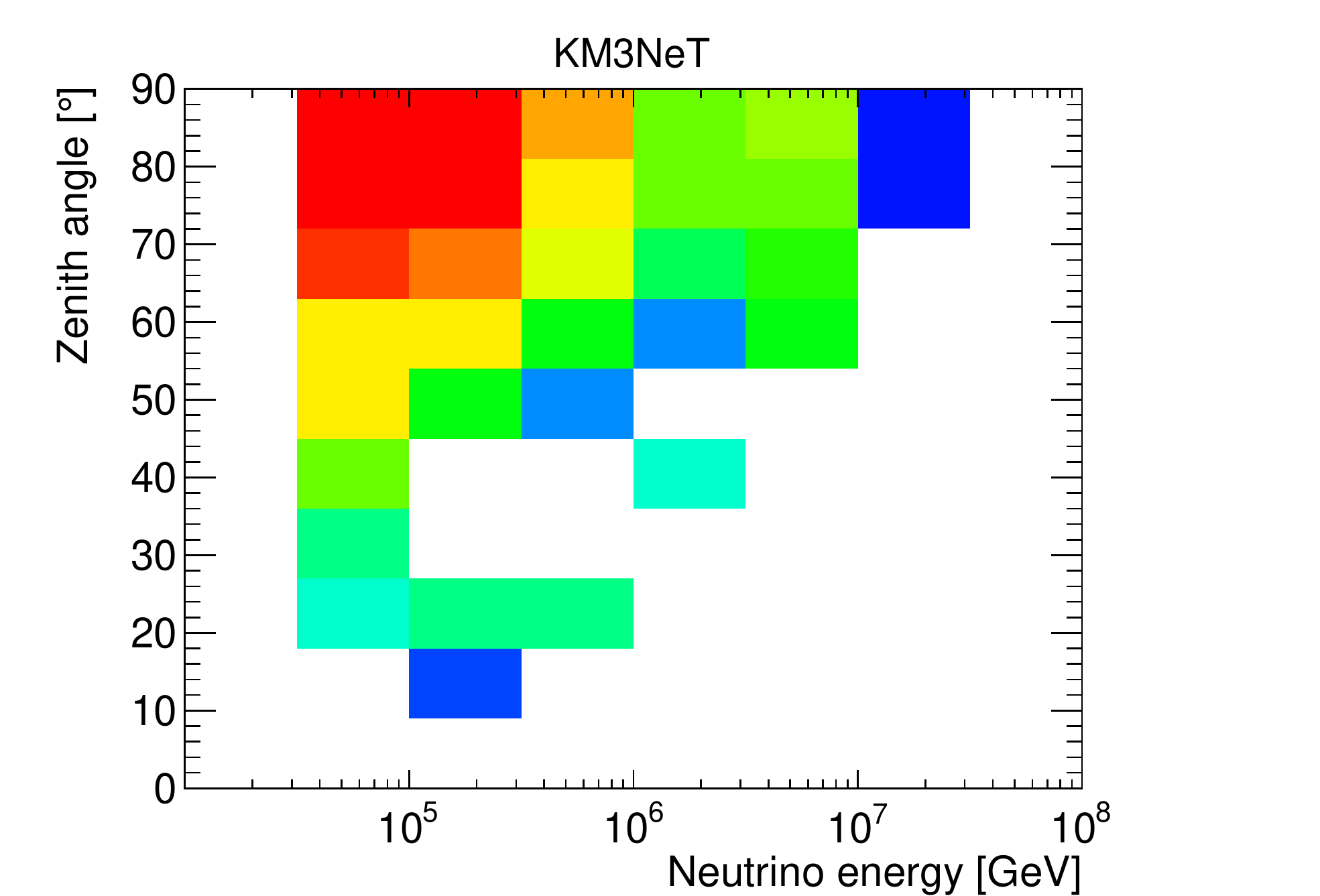}
\end{center}
\caption{%
Effect of the self-veto on down-going $\overline\nu_e$ events as a function of
energy and zenith angle, showing the yearly rate both before (left) and after
(right) the self-veto effect is taken into account. Figure taken from
\cite{icrc_selfveto} (Fig.~4).}
\label{fig:self-veto-effect}
\end{figure*}

An effective `veto' effect can be demonstrated by using the (less sensitive)
`cut-and-count' analysis method, as shown in \myfref{fig:self-veto-effect}. 
Shown are the distributions of atmospheric $\overline\nu_e$ CC events in the event
selection both before (left) and after (right) the self-veto effect has been
taken into account. The total effect is a reduction of the down-going
atmospheric neutrino events in the selection by a factor of about two, or $\sim
25$\% in the all-sky background, with higher-energy events close to the zenith
being more efficiently rejected. It is difficult, however, to compare this
estimate with those of \cite{self-veto-2009} and \cite{self-veto-2014} due to
the different event samples, rejection methods, and detector depths. It should
also be noted that the current estimate suffers from low statistics, and that
the analysis was not optimised with the self-veto effect being taken into
account. Hence, the final self-veto efficiency is expected to be higher, and
improvements of the results for searches of both diffuse and point-like
astrophysical neutrino fluxes in the cascade channel are anticipated.

\subsection{Sensitivity studies}
\label{sec-sci-sen}

In this section, studies of the sensitivity of ARCA to diffuse fluxes and
point-like sources are presented. All the analyses take into account (anti-)neutrinos
of all flavours ($\nu_\mu$, $\nu_e$, and $\nu_\tau$) in equal proportions and
their CC and NC interactions, as simulated according to
\mysref{sec-sci-too-sim}. Each analysis proceeds in the following steps:

\begin{enumerate}
\item 
A preselection of the events to reject most of the atmospheric background,
mostly by cuts on parameters that are provided by the reconstruction algorithms
or that are related to the total ToT or number of hits.
\item 
A multivariate analysis based on the Boosted Decision Tree algorithm (BDT) from
the ROOT TMVA package \cite{root-tmva}, applied to the preselected events for a
more stringent background rejection. This step is not applied in all analyses.
When it is used, the exact input observables vary with each analysis, but always consist of a subset
of the reconstructed event directions, energies, and positions from the track 
and the three cascade reconstruction methods described in \mysref{sec-sci-too-rec}.
Additionally, quality parameters related to the fit procedures, such as
the log-likelihood of each fit, are included, as are measures of the photon
arrival time distribution about the light front -- see \mysref{sec-sci-too-rec}, and
\cite{icrc_arca_casc_recon} and \cite{icrc_arca_diffuse}, for further details.

\item 
A `cut-and-count' analysis method for a fast evaluation of the discovery
potential and a rough estimate of the number of events from background and
signal. This method consists of maximising of the Model Discovery Potential
(MDP) (see e.g.\ the methods of
\cite{icrc_arca_diffuse,km3net-fermibubbles-2013}) by placing cuts on simulated
observables to obtain clean event samples.
\item 
A maximum likelihood method applied to the event sample resulting from step~2 to
calculate the discovery potential at different significance levels. All quoted
significances arise from this method -- however, since only loose cuts are
applied for the likelihood maximisation in order to retain the maximum possible
information, the resulting event sample is very broad. Therefore, the expected
numbers of events passing cuts are quoted using the cut-and-count method above,
reflecting the number of high-quality signal candidate events.
\end{enumerate}

For the last step, the likelihood ratio function:
\begin{equation}
  \mathrm{LR}  = 
  \sum_{k=1}^n \log\frac{\frac{n_\text{sig}}{n}\cdot P_\text{sig}(X_k) + 
  \left(1-\frac{n_\text{sig}}{n}\right) \cdot P_\text{back}(X_k)}{P_\text{back}(X_k)} 
  \label{eq.LR}
\end{equation}
is employed, where $n_\text{sig}$ is the estimated number of signal events, $n$
is the total number of events (and hence, implicitly, $n-n_\text{sig} =
n_\text{back}$ is the number of background events), and $P_\text{sig}$ and
$P_\text{back}$ are the probability distribution functions (PDF) for signal and
background events, respectively. The LR is maximised by altering $n_\text{sig}$
to obtain LR$_\text{max}$. The PDFs are functions of one or more parameters $X$,
such as the BDT output if it is applied, and/or other parameters related to the
specific analysis.

Pseudo-experiments are performed and LR is maximised for each pseudo-experiment. 
The distributions of LR$_\text{max}$ when simulated signals events are present
are compared to distributions in the background-only case to evaluate the
significances of each simulated observation.

Unlike in the high-energy starting event (HESE) analysis of IceCube
\cite{icecube-evidence-2013}, 
no explicit veto to remove atmospheric muon contaminations is used for
KM3NeT/ARCA. Rather, the methods of steps~2 and~4 above assign to each event
likelihoods based on the observed event topology, which is well-preserved in sea
water due to the low light scattering.

\subsubsection{Isotropic diffuse neutrino flux}
\label{sec-sci-too-dif}

The detection and detailed investigation of the astrophysical flux observed by
IceCube is one of the main physics goals of ARCA during KM3NeT Phase-2.0. In the
following, an estimate of the time to detect this flux at the $5\sigma$ level is
presented.

\begin{figure}
\begin{minipage}[tb]{1\textwidth}
\centering
\includegraphics[width=0.6\textwidth]{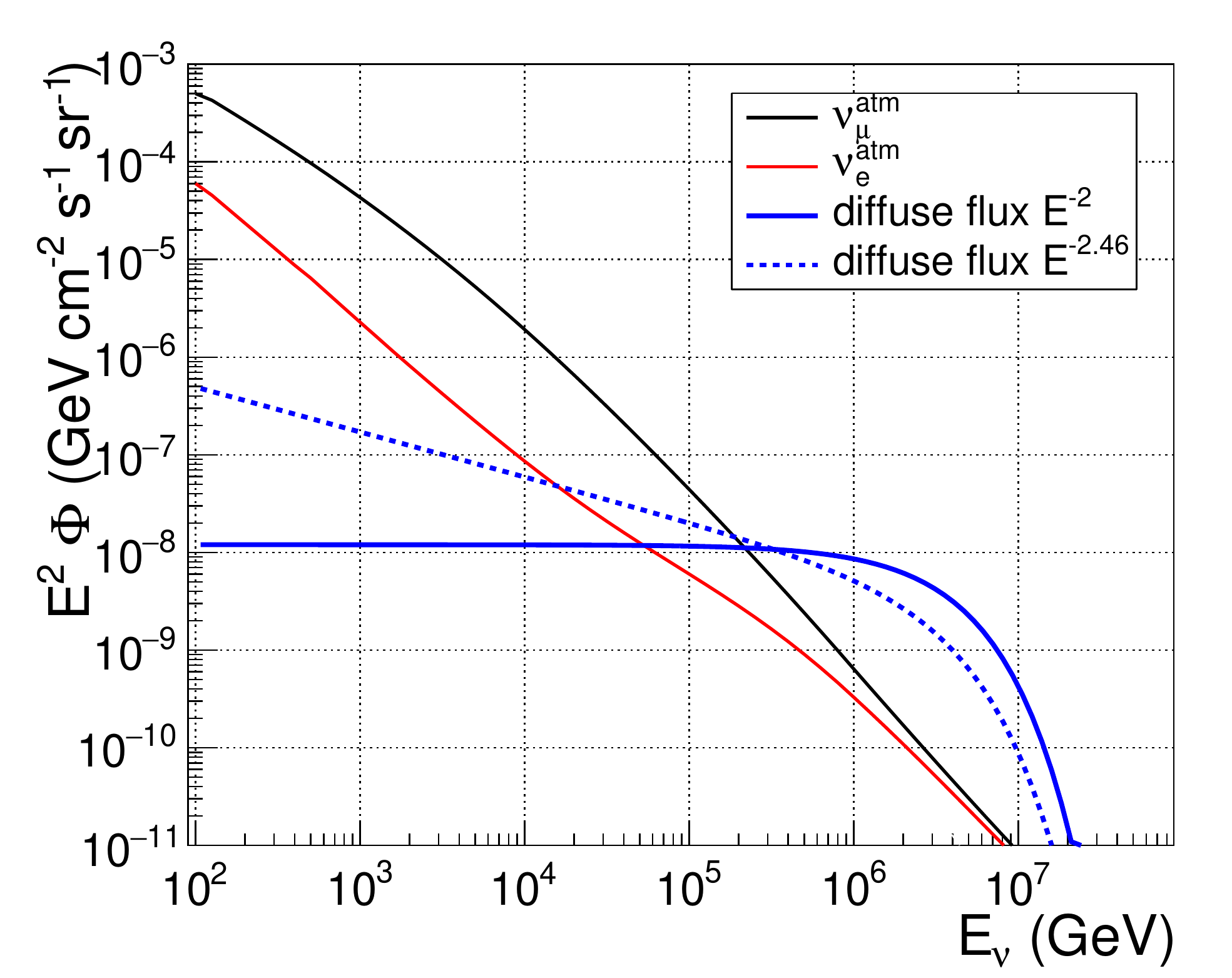}
\end{minipage}
\caption{%
Diffuse signal fluxes according to \myeref{eq.DiffuseFlux1} (full blue line)
and \myeref{eq.DiffuseFlux2} (dashed blue line) as a function of the neutrino
energy. For comparison the atmospheric neutrino fluxes are also reported
(Honda flux \cite{honda-2007} for the conventional component and the Enberg flux \cite{enberg-2008} for the prompt component, see \mysref{sec-sci-ass}).}
\label{fig.DiffuseSpectra}
\end{figure}

This study has been optimised assuming that the IceCube signal originates from
an isotropic, flavour-symmetric neutrino flux following a power law spectrum
with a cut-off at a few PeV. The cutoff -- or a steeper spectrum -- is implied
by the observation of events with a deposited energy exceeding 1 PeV and the
absence of events at about 6.3\,PeV associated with the Glashow resonance (W
production in scattering of $\overline\nu_e$ on electrons). The single-flavour
energy spectrum has been parameterised as:
\begin{equation}
   \Phi(E_\nu) = 1.2\times10^{-8} \cdot 
                 \left(\frac{E_\nu}{\text{GeV}}\right)^{-2} \cdot 
                 \exp\left(-\frac{E_\nu}{3\,\text{PeV}}\right)
                 \;\;\text{GeV}^{-1}\,\text{cm}^{-2}\,\text{s}^{-1}\,\text{sr}^{-1}\;.
  \label{eq.DiffuseFlux1}
\end{equation}
Since the first IceCube discovery \cite{icecube-evidence-2013}, several new analyses
with updated event samples and different event selection strategies have been
published
\cite{icecube-observation-2014,icecube-flavour-2015,icecube-diffusemuon-2014}. 
In these analyses various compatible parameterisations for the cosmic neutrino
flux have been proposed. To check the robustness of our results with respect to
the diffuse neutrino flux assumed we have also calculated the significance of
the KM3NeT/ARCA observation to the following diffuse flux from
\cite{icecube-above-1tev}, which is similar to the results recently reported in
\cite{icecube-max-likelihood}:
\begin{equation}
  \Phi(E_\nu) = 4.11\times10^{-6} \cdot
                \left(\frac{E_\nu}{\text{GeV}}\right)^{-2.46} \cdot 
                \exp\left(-\frac{E_\nu}{3\,\text{PeV}}\right)
                \;\;\text{GeV}^{-1}\,\text{cm}^{-2}\,\text{s}^{-1}\,\text{sr}^{-1}\;.
  \label{eq.DiffuseFlux2}
\end{equation}
Note that for this steeper spectrum, a cut-off to suppress the Glashow resonance
signature is not necessarily required by observations, but is kept here in order
to avoid biasing the analysis by maximising the selection of such events. In
\myfref{fig.DiffuseSpectra} these fluxes are presented together with the
atmospheric neutrino fluxes for comparison.

In the following, the sensitivity studies for diffuse fluxes are presented for
the cascade channel and for the track channel.

\paragraph{Cascade channel}
 
Events simulated as described in \mysref{sec-sci-too-sim} have been
reconstructed with the three available cascade reconstruction codes discussed in
\mysref{sec-sci-too-rec}.

\begin{figure}
\includegraphics[width=0.49\textwidth]{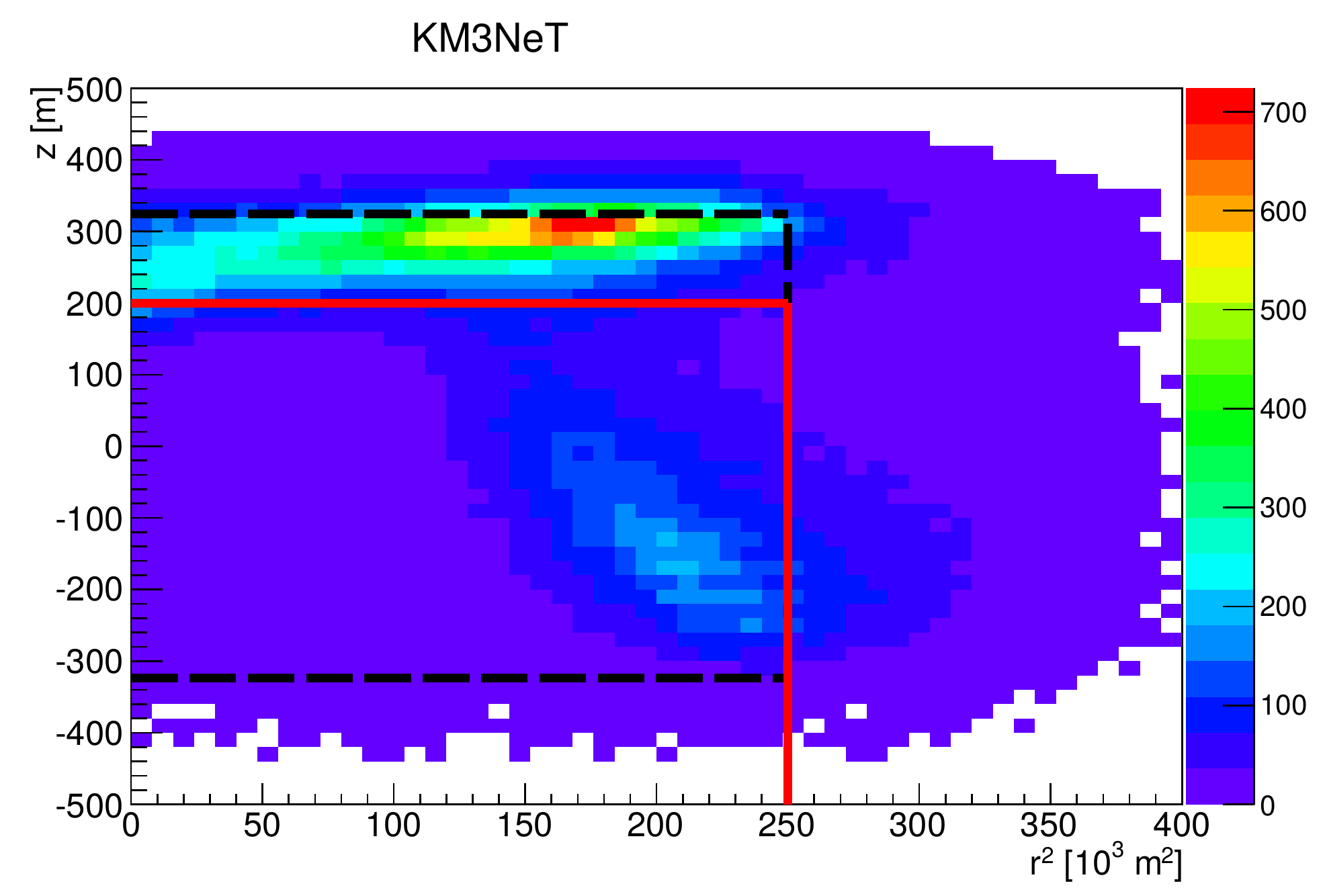}
\hfill
\includegraphics[width=0.49\textwidth]{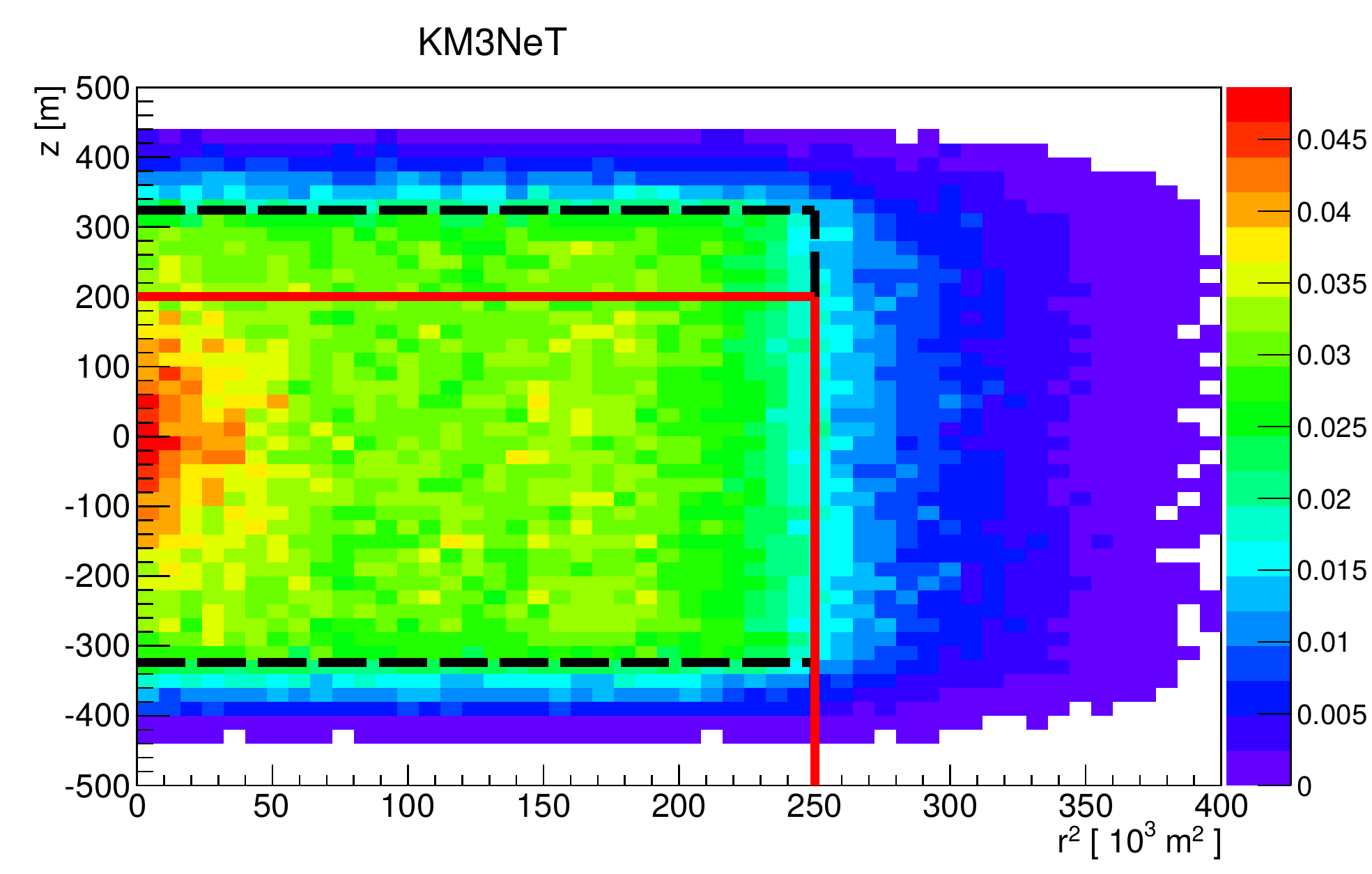}
\includegraphics[width=0.49\textwidth]{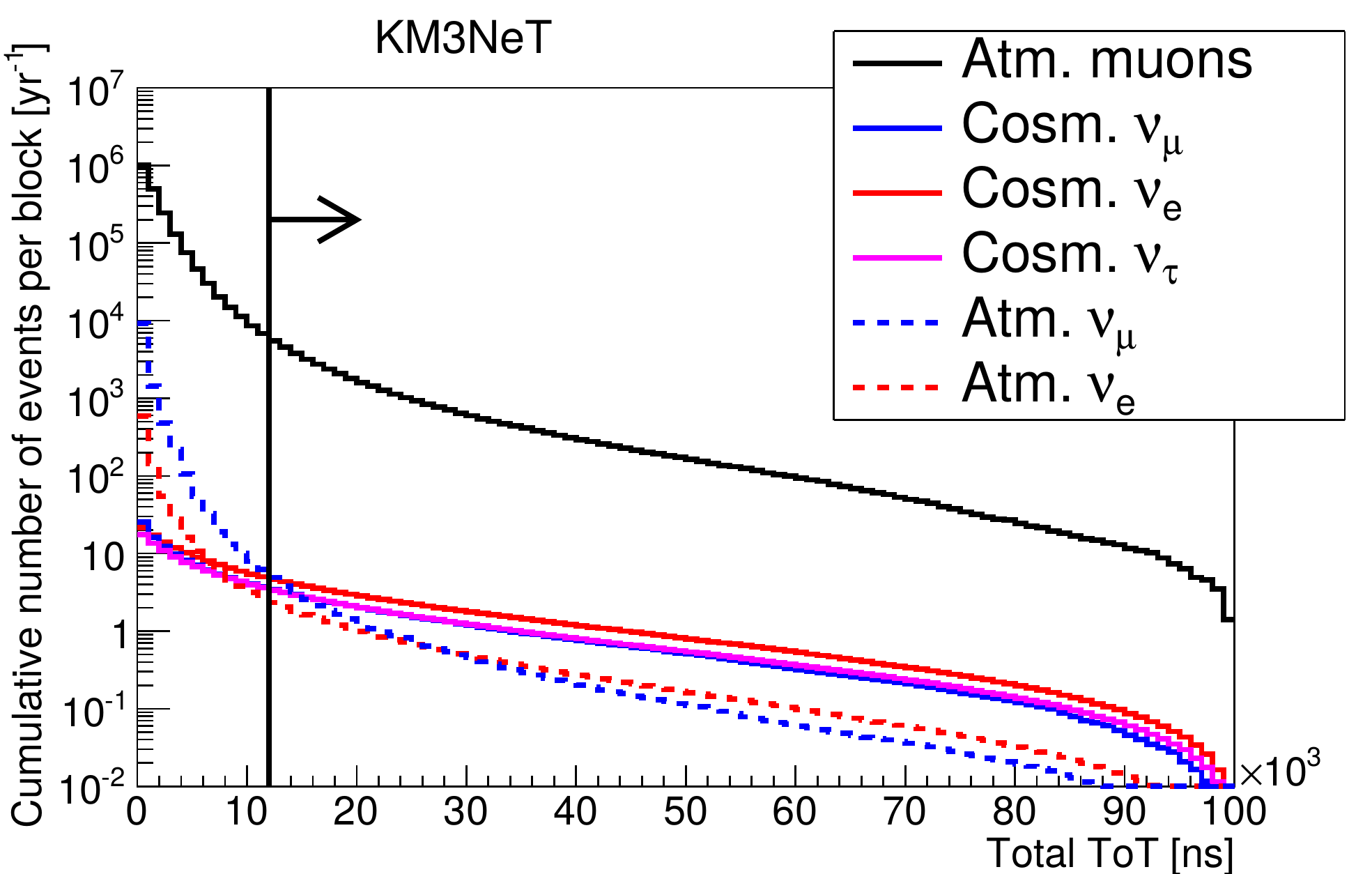}
\hfill
\includegraphics[width=0.49\textwidth]{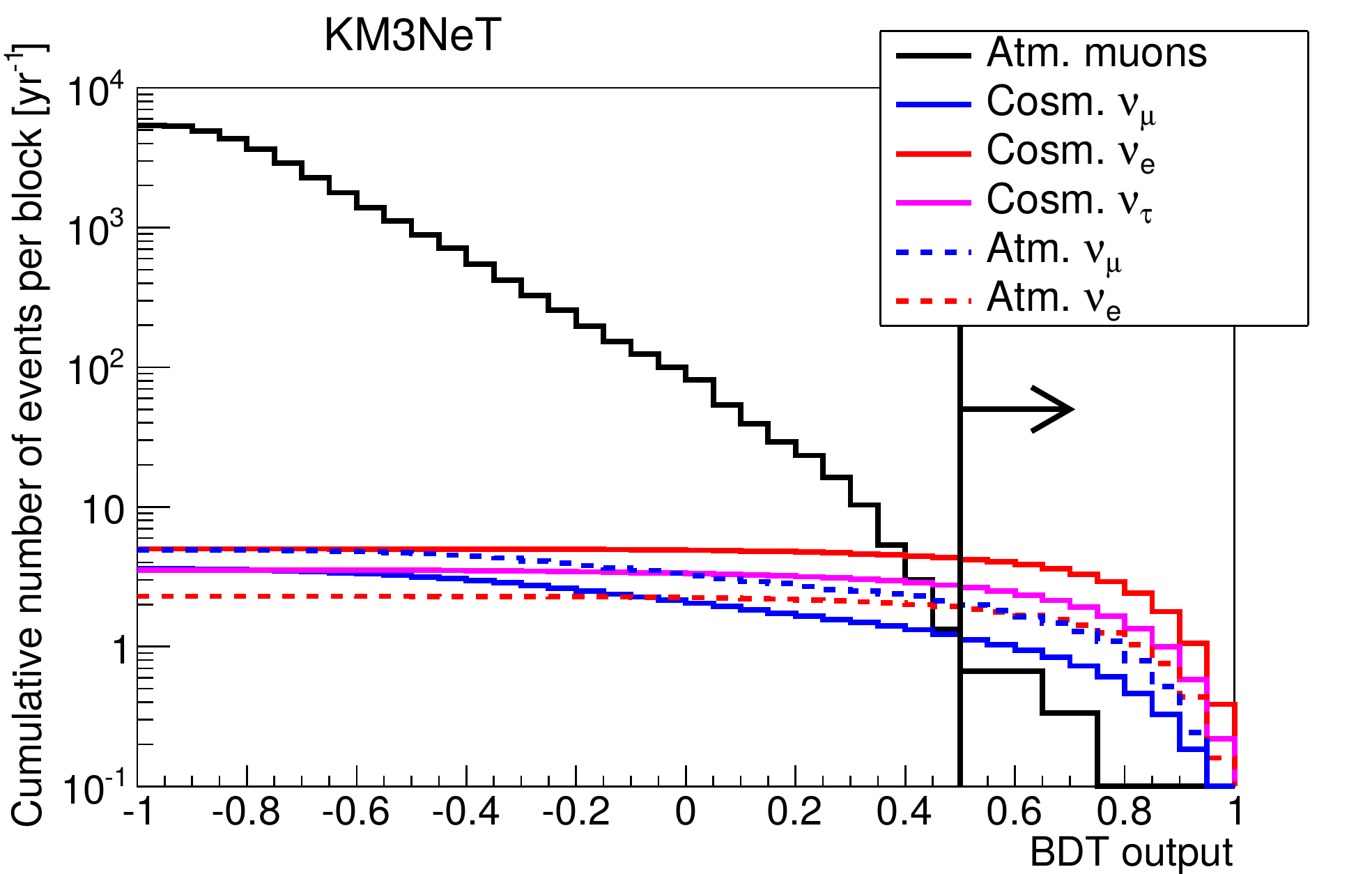}
\caption{%
Reconstructed vertex positions ($r$ and $z$ are the radial and vertical distances
from the detector centre) for atmospheric muon events (left top)
and $\nu_e$ CC events (right top), for one KM3NeT/ARCA building block in one year of operation (events yr$^{-1}$ bin$^{-1}$). The black dashed lines show the
instrumented volume of one building block. The red line shows the selected
fiducial volume defined by $r<500\,$m and $z<200\,$m. Left bottom:
cumulative distribution of $\text{ToT}_\text{evt}$ (see text) for events
contained in the fiducial volume, for the different event
classes. Right bottom: cumulative distribution of the BDT output $\rho$ for the preselected
events (see text) for the different event classes. Vertical lines represent the
selection cuts applied.}
\label{fig.PreselectionCuts}
\end{figure}

The first selection cut requires the containment of the reconstructed vertex in
a cylindrical volume around the detector centre, with radius $r<500\,$m and
height $z<200\,$m. The effect of this cut is illustrated in
\myfref{fig.PreselectionCuts}. It rejects most of the atmospheric muons which,
coming from above, have the reconstructed vertex in the upper part of the
detector. The containment cut reduces the fiducial volume by about 20\% with
respect to the instrumented volume, although this is compensated for by the
included region below the instrumented volume.

\begin{figure}
\includegraphics[width=0.49\textwidth]{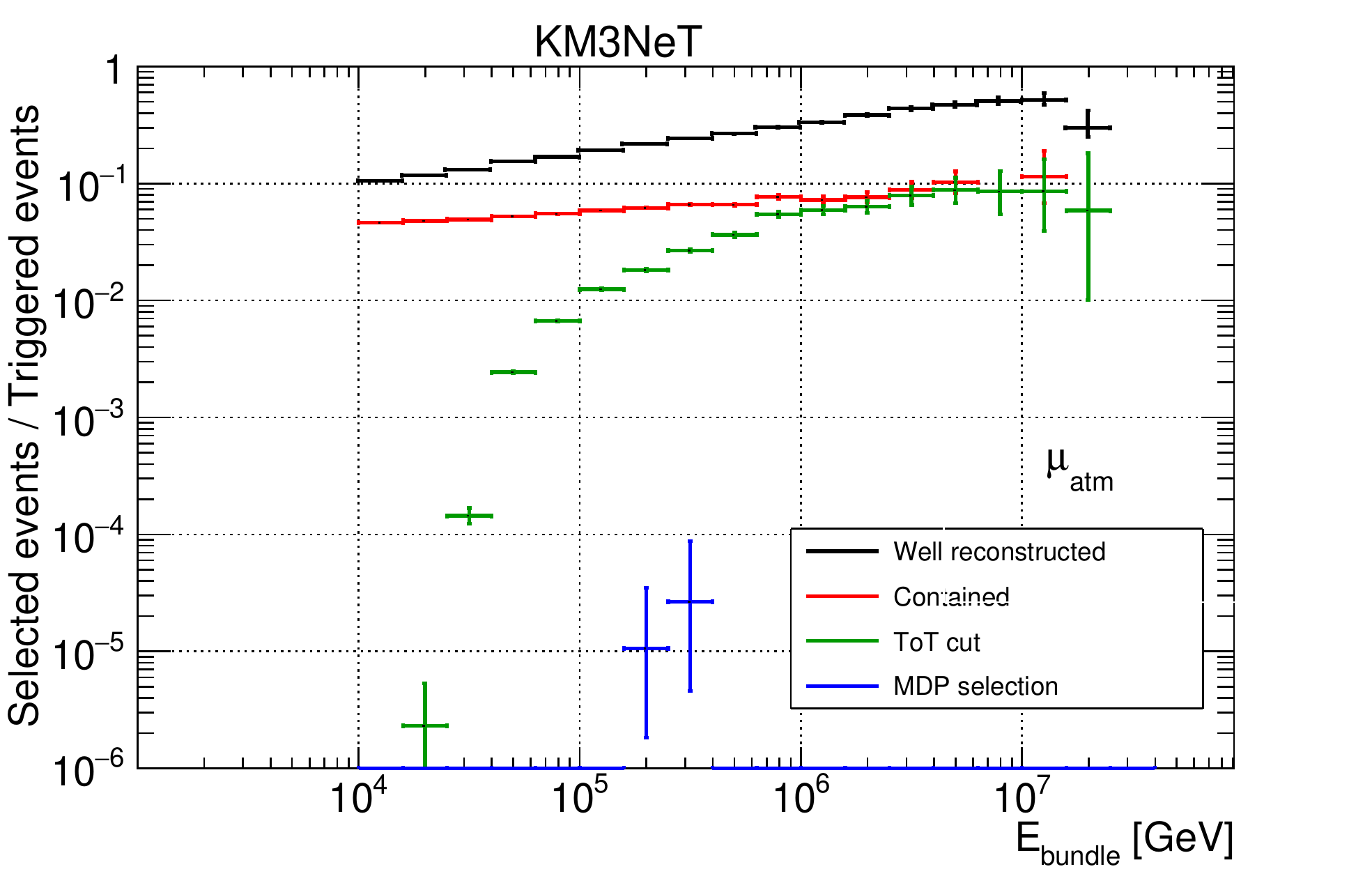}
\hfill
\includegraphics[width=0.49\textwidth]{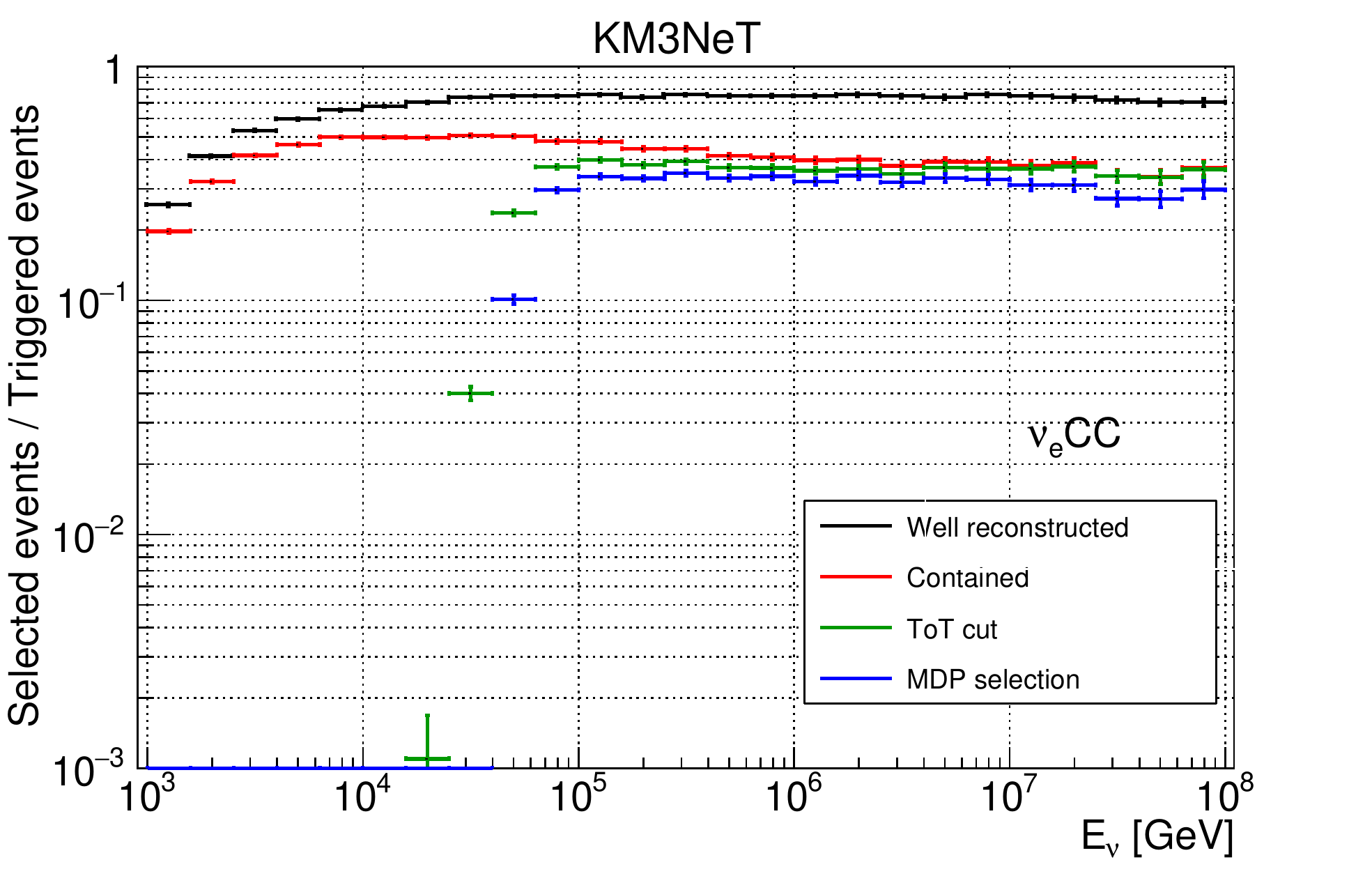}
\caption{%
Ratio of the numbers of selected events and triggered events at each step of the
cascade analysis (see text) for atmospheric muons (left) and $\nu_e$ CC
neutrino reactions (right) as a function of the primary energy.}
\label{fig.CascadeEff}
\end{figure}

The corresponding rejection efficiency is reported in \myfref{fig.CascadeEff}
(red line). A further reduction of the background is obtained by removing
low-energy events. The event ToT, i.e\ $\text{ToT}_\text{evt}=\sum\limits_{k=1}^N
{\rm ToT}_k$ with $N$ being the number of causally connected hits selected by
the cascade reconstruction algorithm, is related to the energy deposited in the
detector. The cumulative $\text{ToT}_\text{evt}$ distribution is shown in
\myfref{fig.PreselectionCuts} (bottom left panel). A cut
$\text{ToT}_\text{evt}>12\,\mu$s is applied and rejects most low-energy
atmospheric muons and a large part of the atmospheric neutrino background, which
is concentrated at lower energies. The corresponding rejection efficiency is
reported in \myfref{fig.CascadeEff} (green points).

As shown in \myfref{fig.CascadeEff} (left panel) the number of reconstructed
atmospheric muons is still too large. To further reduce this background, a BDT
algorithm was applied to the preselected event sample. As input for the BDT
training, several quality parameters from the available shower and track
reconstruction algorithms are used. The BDT is then trained to discriminate
tracks from showers using simulated datasets of atmospheric muons and $\nu_e$ CC
events as training samples. The cumulative distribution of the resulting
discrimination parameter $\rho$ and the cut applied on $\rho$ are shown in
\myfref{fig.PreselectionCuts} (right bottom panel).

\begin{figure}
\begin{minipage}[tb]{1\textwidth}
\centering
\includegraphics[width=0.7\textwidth]{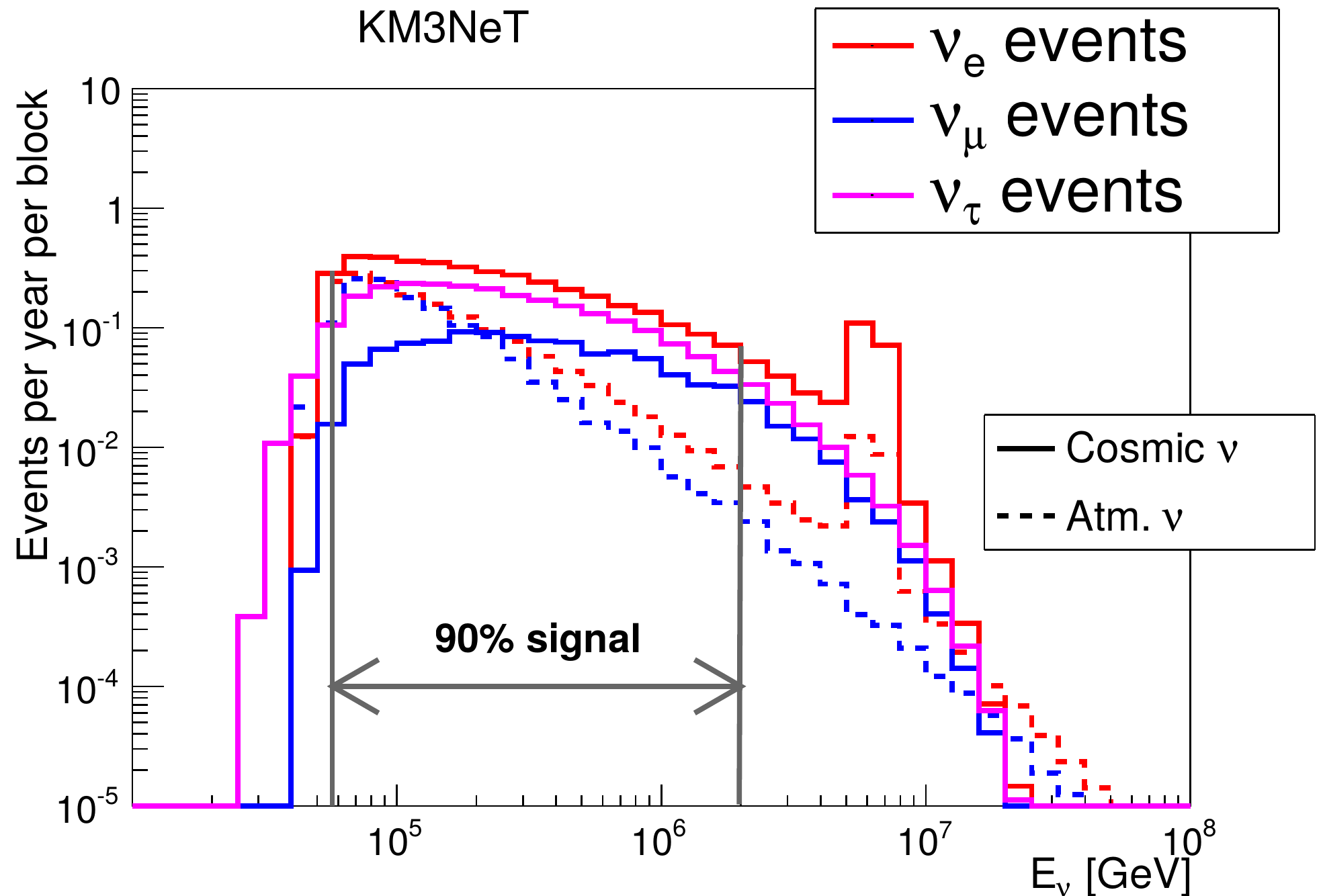}
\end{minipage}
\caption{%
Number of events per year for one building block as a function of the true
neutrino energy for events with the MDP cuts. The black vertical lines show the
energy range where the 90\% of the signal is expected.}
\label{fig.CascadeEnergy}
\end{figure}

A first estimate of the discovery potential, obtained with the cut-and-count
approach, yields final event selection cuts of $\rho>0.5$ and
$E_\text{rec}>10^{4.7}\,\text{GeV}\approx50\,$TeV. The corresponding rejection
efficiency is reported in \myfref{fig.CascadeEff} (blue line). With
these cuts the background due to atmospheric muons is almost completely rejected
for the presently available simulation live-time of three years at high energies. 
\mytref{tab.EventCascade} reports the number of events per 5 years for ARCA at
each step of the analysis for the different event samples. Most of the selected
events are $\nu_e$ and $\nu_\tau$ events, due to the higher cascade energy
deposition of the CC interactions as compared to the NC channel, and because the
BDT identifies $\nu_\mu$ CC events as less ``shower-like'' due to the presence
of the outgoing muon. With these event rates, a $5\sigma$ discovery of the
IceCube flux in the cascade channel can be achieved with 50\% probability after
1.3\,years of ARCA operation. The MC neutrino energy distribution for the final
cut-and-count event sample is shown in \myfref{fig.CascadeEnergy}. The final
cuts preferentially select events in the neutrino energy range from about
50\,TeV to about 2\,PeV ($10^{4.7}\,\text{GeV}<E_\nu<10^{6.3},\text{GeV}$).

To further improve the evaluation of the discovery potential, the maximum
likelihood method (step~4 above) has been applied to the preselected events. The
PDF functions in \myeref{eq.LR} are functions of the reconstructed energy
$E_\text{rec}$ and the BDT output $\rho$. The resulting significance is reported
in \myfref{fig.Significance} as a function of the number of observation years. 
With the KM3NeT/ARCA detector the assumed signal flux will be detectable at
$5\sigma$ in the cascade channel in about one year of observation time.

The estimate of the significance depends on the assumed background and in
particular on the model assumed for the description of the conventional and the
prompt components of the atmospheric neutrino flux (see \mysref{sec-sci-ass}). 
For the cascade channel the maximum variation of the significance, reported as a
red band in \myfref{fig.Significance}, has been obtained assuming the maximum
and minimum flux values of the prompt atmospheric neutrino component reported in
\cite{enberg-2008}. Moreover, the significance has also been estimated taking into
account the new prompt calculation reported in \cite{Gauld2016}
(see \mysref{sec-sci-ass}). In this case, the time to discover the diffuse
flux is reduced by about $30\%$.

\begin{table*}
\small
\begin{center}
\def\arraystretch{1.5}
\begin{tabular}{|c|c|c|c|}
\hline
                       & reconstruction level & after preselection cuts & after final cuts \\
\hline 
$\mu_\text{atm}$       & $2.4 \times 10^{7}$   & $5.5 \times 10^{4}$     & 6 \\
 \hline
 \hline 
$\nu_\text{atm}^\mu$   & $1.0 \times 10^{5}$   & 49                     & 20 \\
\hline
$\nu_\text{atm}^e$     & $7.1 \times 10^{3}$  & 23                    & 19 \\
\hline
\hline
$\nu_\text{cosm}^\mu$  & 352                  & 34                     & 11 \\
\hline
$\nu_\text{cosm}^e$    & 304                  & 49                     & 41 \\
\hline
$\nu_\text{cosm}^\tau$ & 250                  & 34                     & 26 \\
\hline
\end{tabular}
\end{center}
\caption{
Expected number of events for the KM3NeT/ARCA detector (2 building blocks) for the
different event samples in 5 years of observation time. The cosmic events
correspond to the source flux of \myeref{eq.DiffuseFlux1}.}
\label{tab.EventCascade}
\end{table*}

\paragraph{Track channel}

Since energetic muons can have very long tracks (a 10\,TeV muon has a path
length of $\approx5\text{--}6\,$km in water), muon neutrinos with interaction
points far from the instrumented volume can be detected, thus making the
effective volume much larger than the geometrical detector volume. The main
challenge in using the track channel is to distinguish these events from
atmospheric muons. Here we follow the traditional approach to reject atmospheric
muons by using the Earth as a shield, and select track-like events that come
from below the horizon, or a few degrees above it. In this analysis a cut on the
reconstructed zenith angle $\theta_{\text rec}>80^\circ$ is applied.

\begin{figure}
\includegraphics[width=0.49\textwidth]{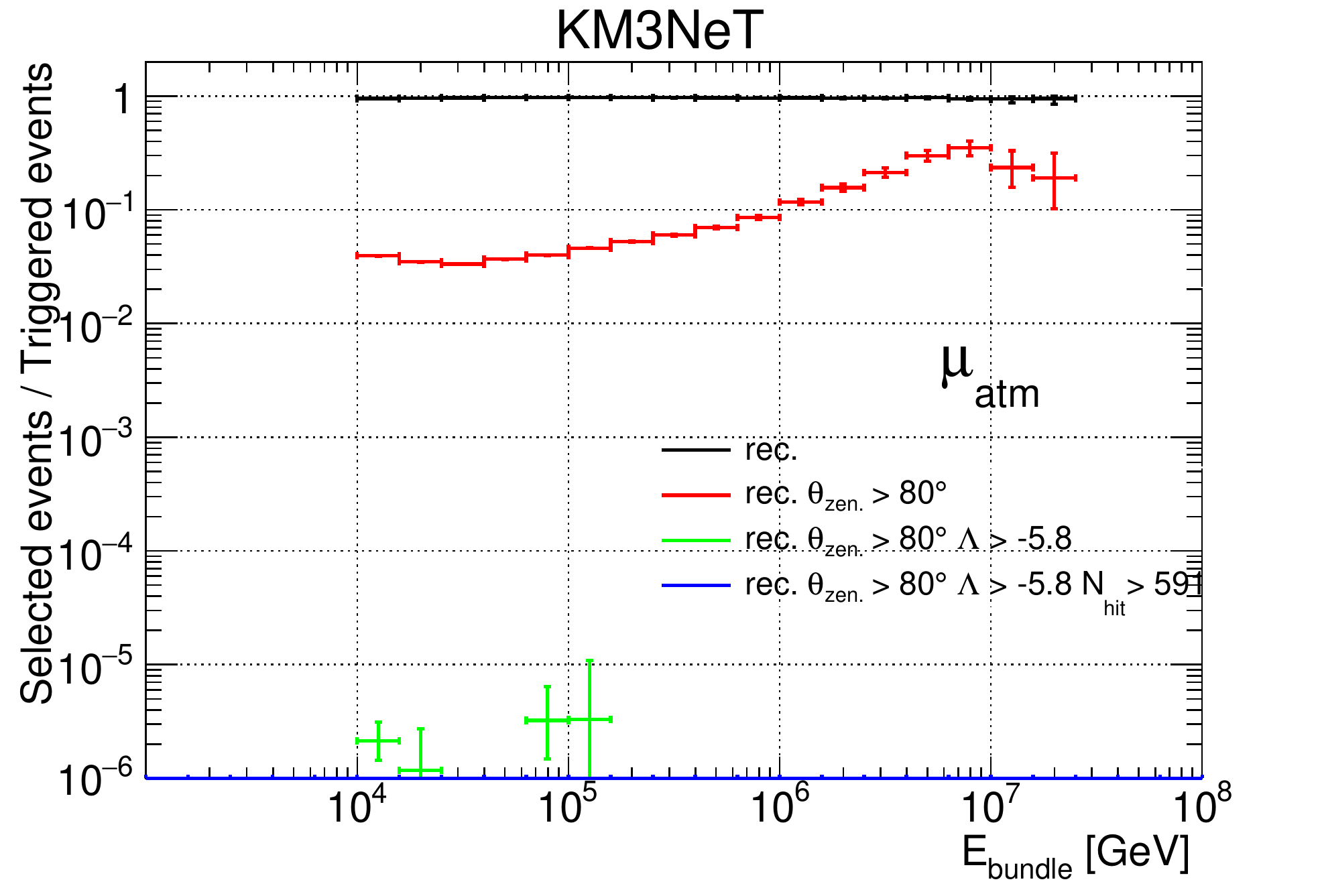}
\hfill
\includegraphics[width=0.49\textwidth]{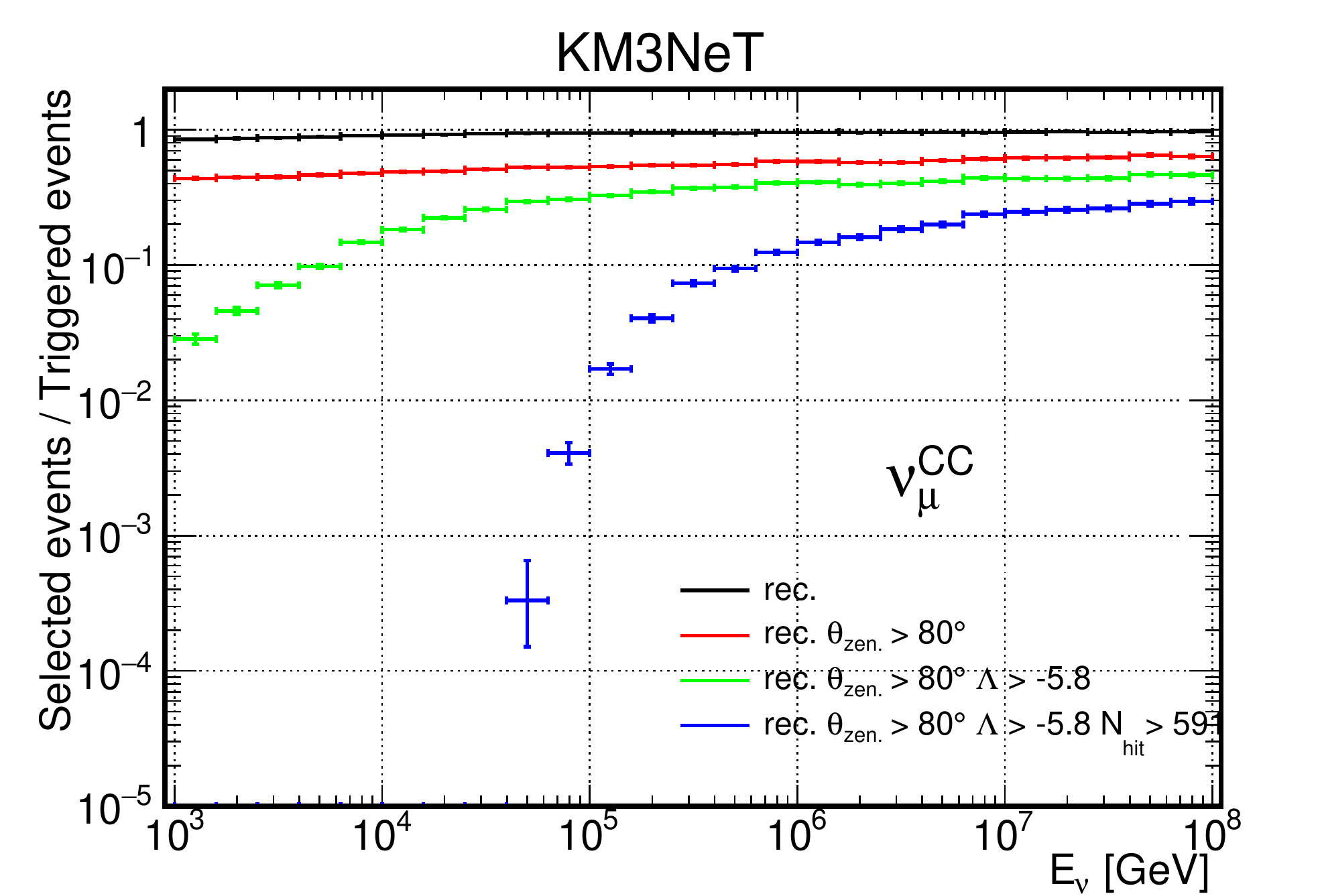}
\caption{
Ratio of the numbers of selected events and triggered events at each step of the
track analysis (see text) for atmospheric muons (left) and $\nu_\mu$ CC
neutrino reactions (right) as a function of the primary energy. }
\label{fig.TrackEff}
\end{figure}

\begin{figure}
\includegraphics[width=0.49\textwidth]{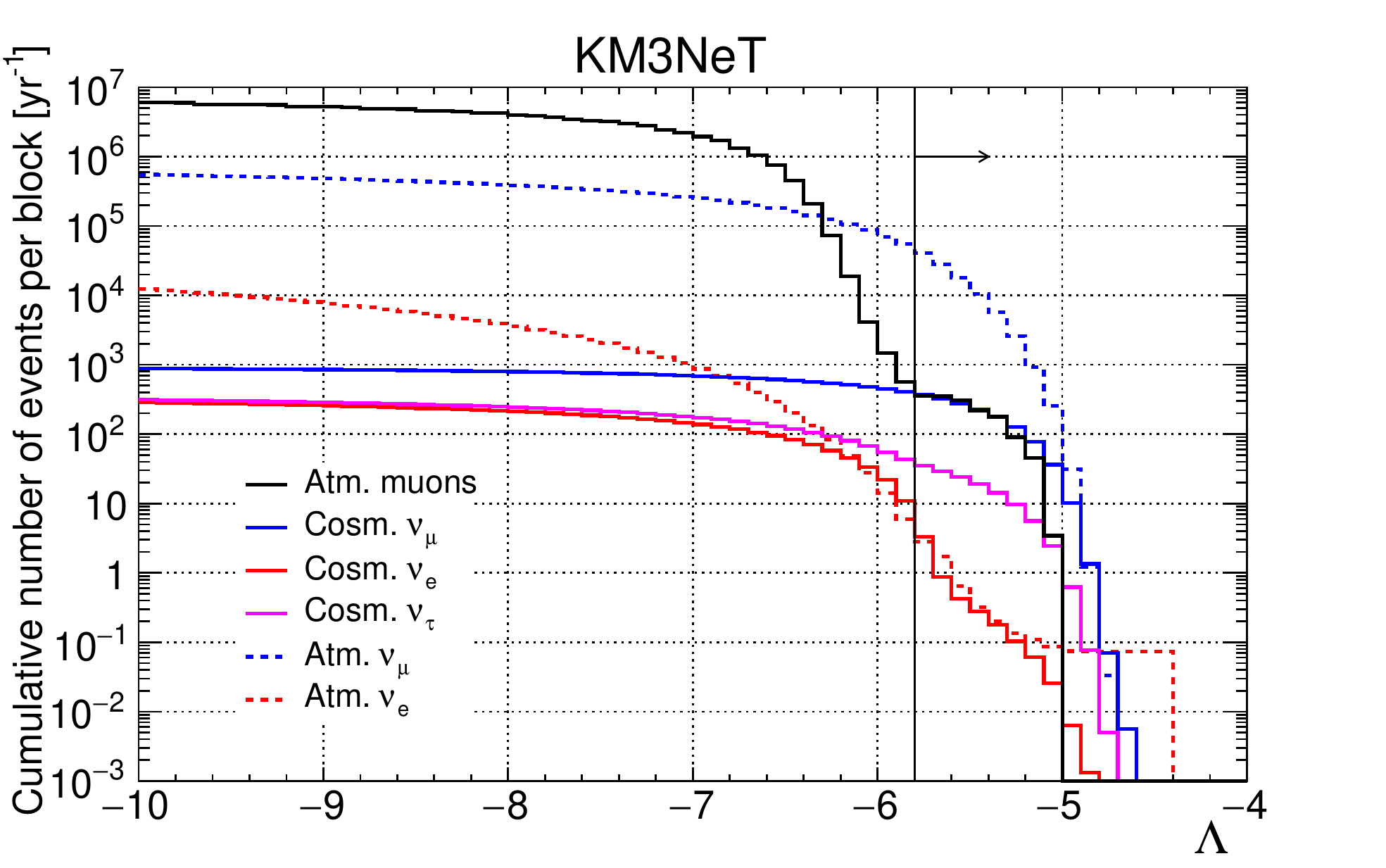}
\hfill
\includegraphics[width=0.49\textwidth]{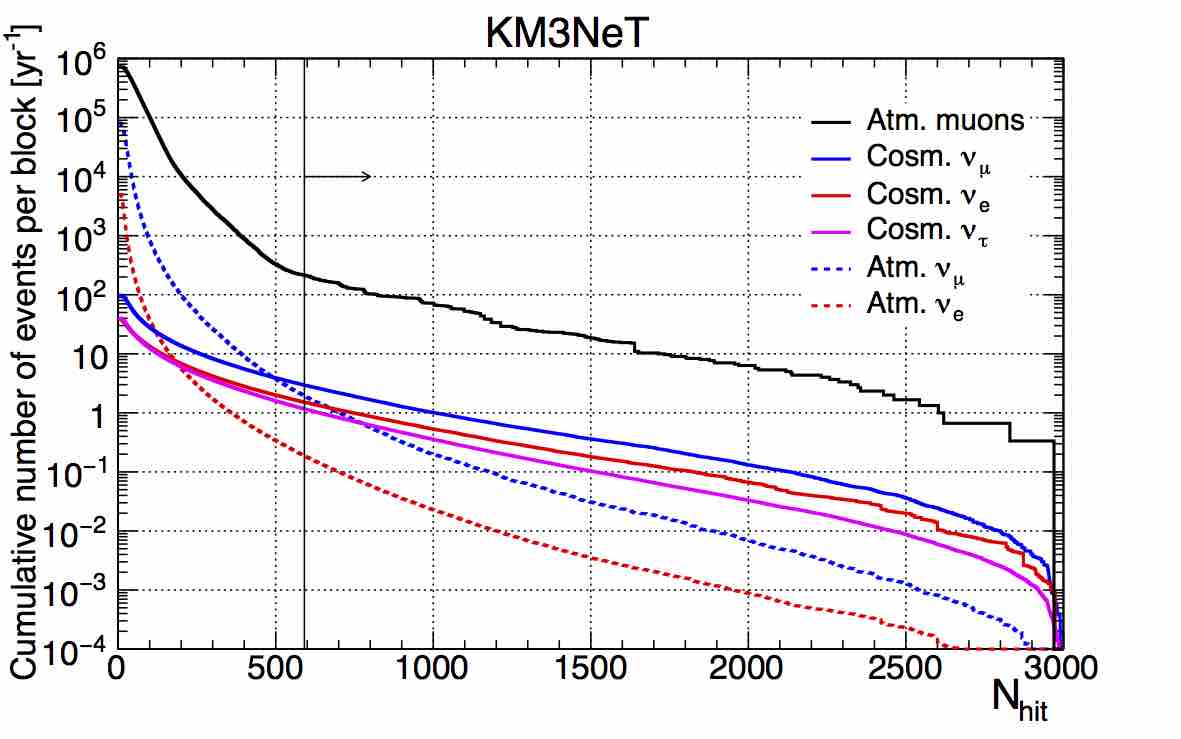}
\caption{%
Cumulative distribution of the $\Lambda$ parameter (left) and of the
number of hits, $N_{\text{hit}}$, (right) for events per KM3NeT/ARCA building block with
$\theta_{\text{rec}}>80^\circ$. The cosmic neutrino signal corresponds to the flux
given in \myeref{eq.DiffuseFlux1}. The vertical lines indicate the final cut
values applied in the analysis (see text).}
\label{fig.Lambda_Nhit}
\end{figure}

In \myfref{fig.TrackEff} the ratio of the numbers of selected events and
triggered events is reported (red lines) for atmospheric muons (left panel) and
$\nu_\mu$ CC events (right panel). The atmospheric muon rate at energies below
$10^6$\,GeV is reduced by more than one order of magnitude. Most of the remaining
atmospheric muons are mis-reconstructed as up-going or are near the horizon. 

To remove the mis-reconstructed events an additional cut on the quality
parameter $\Lambda$ (see \mysref{sec-sci-too-rec}) was applied. In
\myfref{fig.Lambda_Nhit} (left panel) the cumulative $\Lambda$ distribution is
shown for atmospheric muons (black line), for atmospheric neutrinos
(dashed line)
and cosmic neutrinos (solid line), for events with $\theta_\text{rec}>80^\circ$. 
For $\Lambda\gtrsim-6$ the atmospheric muon background is reduced to the level
of the astrophysical neutrino signal.

To reduce the background due to atmospheric neutrinos, a cut on the number of
hits associated with the fitted track, $N_\text{hit}$, is applied (see
\mysref{sec-sci-too-rec}). $N_\text{hit}$ is related to the muon energy loss
in the detector and thus to the primary neutrino energy. The cumulative
$N_\text{hit}$ distribution is presented in \myfref{fig.Lambda_Nhit} (right
panel).
 
\begin{figure}
\centering
\includegraphics[width=0.7\textwidth]{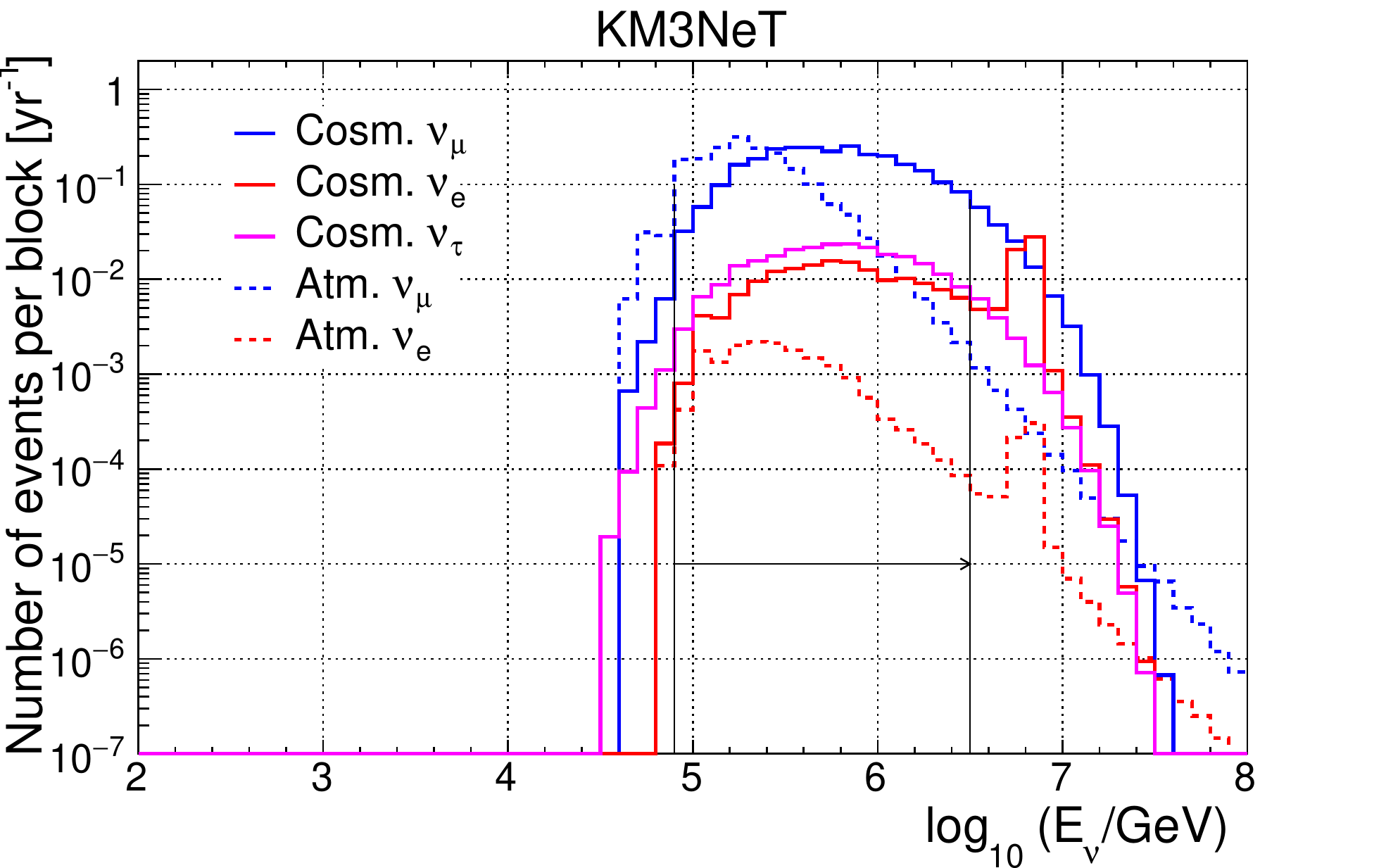}
\caption{%
Number of events passing the MDP cuts (see text), per year for one building
block, as a function of the MC neutrino energy. The black vertical lines show
the energy range where 90\% of the signal is expected.}
\label{fig.TrackEnergy}
\end{figure}

\begin{figure}
\begin{minipage}[tb]{1\textwidth}
\centering
\includegraphics[width=0.7\textwidth]{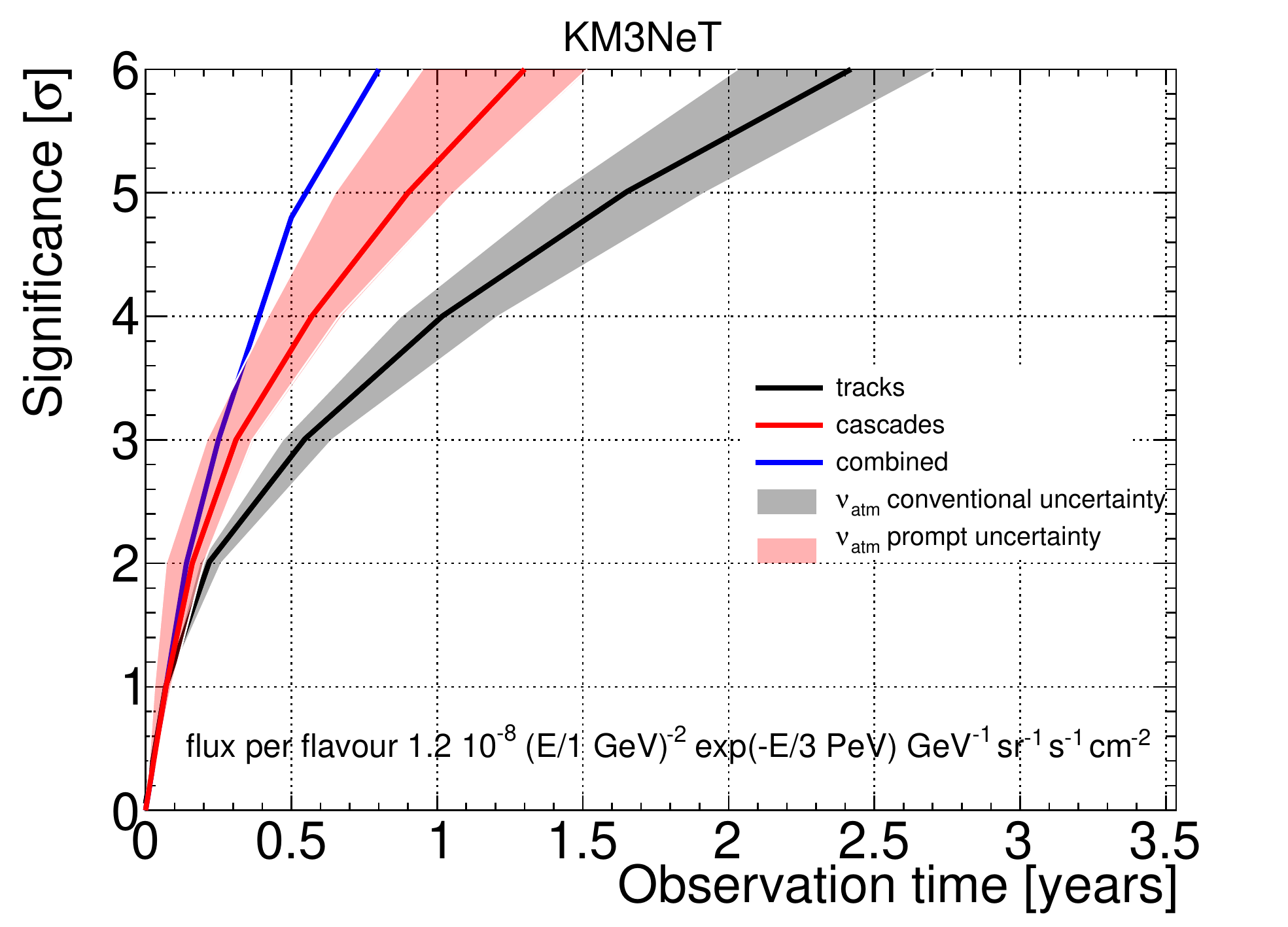}
\end{minipage}
\caption{%
Significance as a function of KM3NeT/ARCA (2 building blocks) observation time for the detection of a
diffuse flux of neutrinos corresponding to the signal reported by IceCube
(\myeref{eq.DiffuseFlux1}) for the cascade channel (red line) and muon
channel (black line). The black and red bands represent the uncertainties due to
the conventional and prompt component of the neutrino atmospheric flux. The blue
line represents the results of the combined analysis (see text).}
\label{fig.Significance}
\end{figure}

The final cut values, obtained by maximising the MDP for 5 years of observation time, are $\Lambda>-5.8$ and
$N_\text{hit}>591$. The resulting numbers of events and the selection
efficiencies are reported in \mytref{tab.EventTrack} and
\myfref{fig.TrackEff}, respectively. 
The number of atmospheric muons surviving the final cuts has been extrapolated from the present statistics (see  \myfref{fig:sim-livetimes}).
The MC neutrino energy distribution for
the event sample after final cuts is shown in \myfref{fig.TrackEnergy}. The
cuts select events in the neutrino energy range from about 80\,TeV to about
3\,PeV. A discovery at $5\sigma$ with 50\% probability is achieved in about
3.2\,years.

As in the cascade analysis, the maximum likelihood method was applied to the
preselected event sample ($\theta_\text{rec}>80^\circ$ and $\Lambda>-5.8$) to
further improve the sensitivities. The likelihood ratio (\myeref{eq.LR}) was
calculated for signal and background using PDFs that were mono-dimensional
functions of $N_\text{hit}$. The resulting significance is reported in
\myfref{fig.Significance} as a function of the observation time. The assumed
signal flux can be detected with KM3NeT/ARCA at $5\sigma$ in the track channel
in about 1.6\,years of observation time with 50\% probability.

For the track channel, the maximum variation of the significance (reported as
a grey band in \myfref{fig.Significance}) has been obtained with the assumed
uncertainties in the intensity of the conventional atmospheric neutrino flux (see \mysref{sec-sci-ass}).

\begin{table*}
\small
\begin{center}
\def\arraystretch{1.5}
\begin{tabular}{|c|c|c|c|}
\hline
                       & reconstruction level  & after preselection cuts  & after final cuts \\
\hline
$\mu_\text{atm}$       & $2.7 \times 10^8$     & $1.1 \times 10^7$        & $\approx 3$    \\
\hline
\hline 
$\nu_\text{atm}^\mu$   & $1.6 \times 10^6$     & $8.0 \times 10^5$        & 18.8 \\
\hline
$\nu_\text{atm}^e$     & $6.0 \times 10^4$     & $4.9 \times 10^4$        & 0.2 \\
\hline
\hline
$\nu_\text{cosm}^\mu$  & $1.9   \times 10^3$    & 977                     & 27.9 \\
\hline
$\nu_\text{cosm}^e$    & 600                   &  381                     & 2.1 \\
\hline
$\nu_\text{cosm}^\tau$ & 655                    &  400                     & 2.6 \\
\hline
\end{tabular}
\end{center}
\caption{
Expected numbers of events for the KM3NeT/ARCA detector (2 building blocks) for the different event samples
in 5 years of observation time. The cosmic events correspond to the source flux
of \myeref{eq.DiffuseFlux1}.}
\label{tab.EventTrack}
\end{table*}

\paragraph{Combined analysis}

To combine the results of the cascade and track analyses, up-going and
down-going events were analysed separately. For down-going events, where the
atmospheric background is very high, only contained events are considered (same
preselection cuts as in the cascade analysis). For up-going events, preselection
cuts on the reconstructed vertices and $\text{ToT}_\text{evt}$ are used to
reject atmospheric muons that are wrongly reconstructed as up-going. BDT
discrimination cuts are then applied to both samples of preselected events. The
BDTs use parameters coming from both the track and shower reconstruction
algorithms and are optimised to reject atmospheric muons. The BDT outputs are
used together with the cascade energy estimate in a maximum likelihood approach
based on \myeref{eq.LR}. The final result is reported in
\myfref{fig.Significance} (blue line) as a function of the observation time. 
Combining the results of the track and cascade analyses, KM3NeT/ARCA is expected
to observe the IceCube flux (\myeref{eq.DiffuseFlux1}) in about 6\,months
with a significance of $5\sigma$ with 50\% probability.

To investigate the sensitivity of these results to the assumed form of the
IceCube diffuse flux, both the cascade and track analyses were repeated for
signal fluxes according to \myeref{eq.DiffuseFlux2} both with and without the
3\,PeV cutoff. In each case, the flux normalisation constant, $\Phi_{5\sigma}$,
required for a $5\sigma$ discovery after 1\,year of observation time, was
calculated. The results are reported in \mytref{tab.IceCubeFlux} in terms of
their ratio to the flux normalisation reported by IceCube, $\Phi_\text{IC}^0$. 
Values larger (less) than unity indicate a $5\sigma$ discovery time of more
(less) than 1\,year. The results show that for flux assumptions with a softer
spectrum and the same cut-off the main results of our analysis do not change,
and in fact a small improvement ($\approx 10\%$) is expected.

\begin{table*}
\small
\begin{center}
\def\arraystretch{1.5}
\begin{tabular}{|c|c|c|}
\hline
$\Phi_\text{IC}^0$  & \multicolumn{2}{c|}{$\Phi_{5\sigma}$/$\Phi_\text{IC}^0$} \\
\cline{2-3}
[GeV$^{-1}$ cm$^{-2}$ s$^{-1}$ sr${^-1}$] & Cascades  & Tracks   \\
\hline 
$1.2  \times 10^{-8}$ (\myeref{eq.DiffuseFlux1})                 & 0.95 & 1.30 \\
\hline 
$4.11 \times 10^{-6}$ (\myeref{eq.DiffuseFlux2})                 & 0.80 & 1.20 \\
\hline
$4.11 \times 10^{-6}$ (\myeref{eq.DiffuseFlux2} without cutoff)  & 0.75 & 0.92 \\
\hline
\end{tabular}
\end{center}
\caption{
Ratios between the flux normalisation needed for a $5\sigma$ discovery in
KM3NeT/ARCA (2 building blocks) within 1\,year with 50\% probability and the different
parameterisations of the IceCube flux (see text).}
\label{tab.IceCubeFlux}
\end{table*}

\subsubsection{Diffuse neutrino flux from the Galactic plane}
\label{sec-sci-too-gal} 

One of the most promising potential source regions of a diffuse astrophysical
neutrino flux is the Galactic Plane (GP). Neutrinos are expected to be produced
in the interactions of the galactic cosmic rays with the interstellar medium and
radiation fields, with a potentially significant excess with respect to the
expected extragalactic background. The observation of diffuse TeV $\gamma$-ray
emission from the GP
\cite{Milagro-2008-galactic-diffuse,HESS-galactic-ridge-2006}, which is expected
to arise from the same hadronic processes that would produce high-energy
neutrinos, strongly supports this hypothesis. Also Fermi-LAT observes, after the
subtraction of known point-like emitting sources, a broad diffuse emission from
the GP, with a spectrum consistent with a significant hadronic component
\cite{fermi-lat-galplane-2012}.

\begin{figure}
\includegraphics[width=0.49\textwidth]{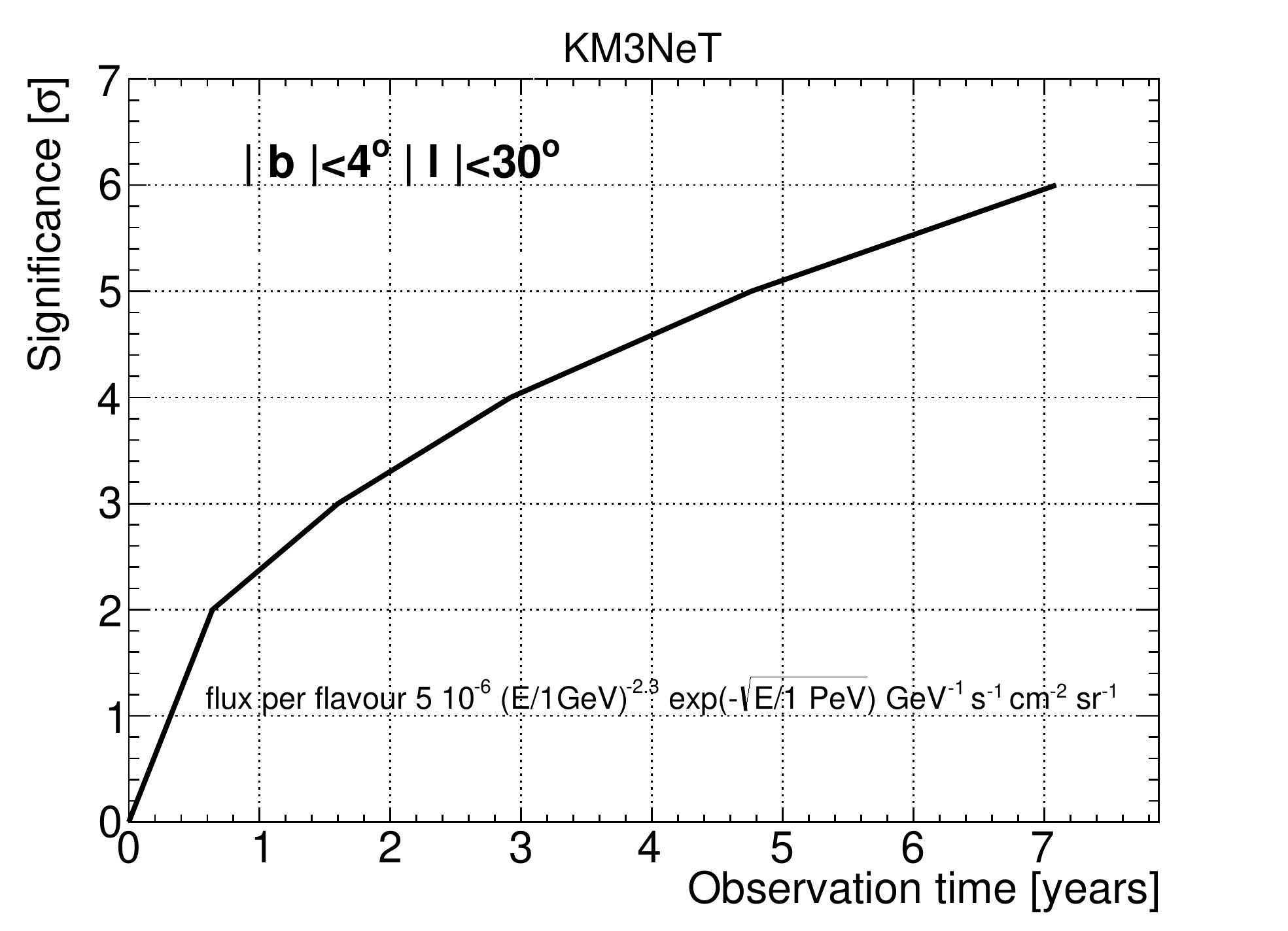}
\hfill
\includegraphics[width=0.49\textwidth]{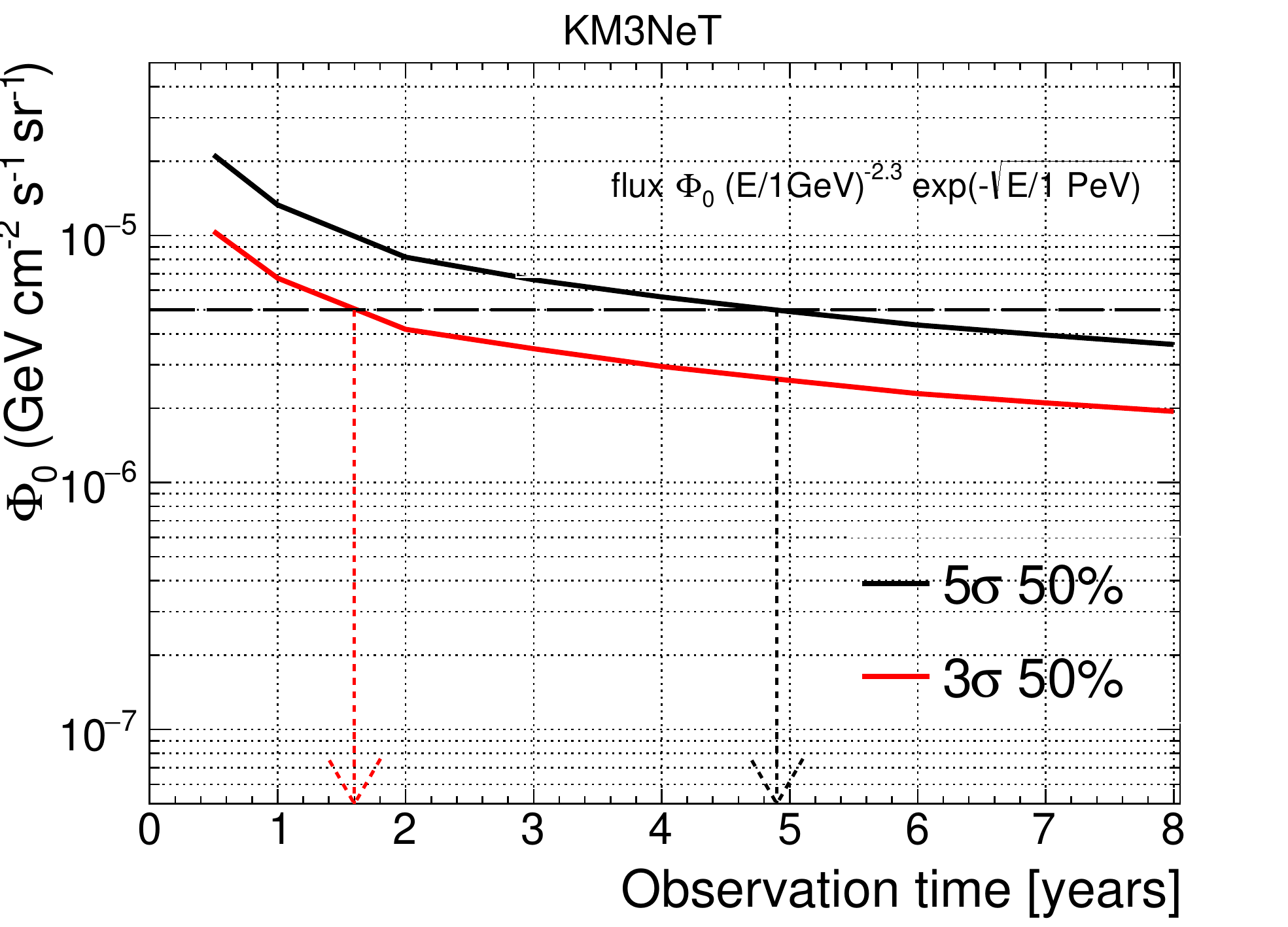}
\caption{%
Significance (left) and $5\sigma$ and $3\sigma$ discovery fluxes (right) for KM3NeT/ARCA (2 building blocks) as a function of the observation time for the detection in the track
channel of a diffuse flux of neutrinos from a selected region of the GP near the
Galactic Center (see text).}
\label{fig.SignificanceGP}
\end{figure}

Recently, also related to the observed IceCube high-energy
neutrino events, new phenomenological models for the diffuse galactic
neutrino emission have been proposed
\cite{AhlersMurase2014,Neronov:2014uma,gaggero-gamma-neutrino,Ahlers-Galactic-Plane-2015,Neronov:2015osa}. In
particular, in \cite{gaggero-galactic-ridge} a non-uniform cosmic-ray (CR)
transport model with a radially dependent diffusion coefficient has been adopted
to explain the high-energy diffuse $\gamma$-ray emission along the whole GP, as
well as the hardening of CR spectra measured by PAMELA and AMS-02 around
250\,GeV and two possible CR cut-offs, at 5\,PeV and 50\,PeV, compatible with
KASKADE and KASKADE-Grande observations. In \cite{gaggero-gamma-neutrino}, these authors
estimate that the astrophysical flux
detected by IceCube in both the HESE \cite{icecube-hese-icrc-fouryears} and diffuse muon \cite{icecube-diffusemuon-2015} analyses is still dominated by an extragalactic diffuse component,
with galactic emission respectively accounting for $15$\% and $10$\% of events.
Using this model, a detailed prediction
of the neutrino emission from the inner galactic plane, i.e.\ for $|l|<30^\circ$
and $|b|<4^\circ$ ($b$ and $l$ being the Galactic latitude and longitude,
respectively) has been obtained (see Fig.~4 of \cite{gaggero-galactic-ridge}). 
This flux is adopted here to estimate the performance of the KM3NeT/ARCA
detector in searching for neutrinos from the GP. The selected region is entirely
located in the Southern hemisphere. The estimated one-flavour neutrino flux has
been parameterised as:
\begin{equation}
 \frac{\text{d}\phi}{\text{d}E_\nu} = 
 5 \times 10^{-6} \left(\frac{E_\nu}{1\,\text{GeV}}\right)^{-2.3}\cdot
 \exp\left(-\sqrt{\frac{E_\nu}{1\,\text{PeV}}}\right)
  \;\;\text{GeV}^{-1}\,\text{cm}^{-2}\,\text{s}^{-1}\,\text{sr}^{-1}\;.
\label{eq.GP}
\end{equation}
An analysis similar to that described for the all-sky diffuse track channel has
been performed to estimate the KM3NeT/ARCA sensitivity to this flux. Events were
preselected requiring the zenith angle to be $\theta_\text{rec}>80^\circ$ and to
point to the sky region with $|l|<30^\circ$ and $|b|<4^\circ$. This sky region
is visible to the KM3NeT/ARCA detector for events up to $10^\circ$ above the
horizon for about 77\% of the time. The final cut values, obtained by minimising
the MDP, were $\Lambda>-5.8$ and $N_\text{hit}>181$. The numbers of events from
the selected GP region are found to be 2.8 background events (muons and
neutrinos) and 3.4 from the source flux of \myeref{eq.GP} in one year of ARCA
operation. The significance as a function of the observation time has been
evaluated by the maximum-likelihood method and is reported in
\myfref{fig.SignificanceGP} (left panel). A discovery at $5\sigma$ with 50\%
probability can be achieved in about 5\,years of observation time with the
KM3NeT/ARCA detector. The discovery flux at $5\sigma$ and $3\sigma$ is reported
as a function of the observation time in \myfref{fig.SignificanceGP} (right
panel). As for the diffuse flux analysis, a reduction on the number of the
observation years is expected if the cascade channel is included in the
analysis. This work is at the moment on-going.

\subsubsection{Point-like neutrino sources}
\label{sec-sci-too-poi}

Due its good angular resolution, KM3NeT/ARCA is a very promising instrument for the
detection of point-like sources. In particular, its location in the Northern
Hemisphere will allow the study of most Galactic sources, as well as
extragalactic sources (which are expected to be approximately uniformly
distributed over the sky) using up-going muon track events. In this section the
sensitivity of the ARCA detector to point-like sources will be
discussed. In particular the two following physics cases will be analysed:

\begin{itemize}

\item 
Neutrino emission by the supernova remnant (SNR) RX\,J1713 and the pulsar wind
nebula (PWN) Vela-X, which are at present the Galactic objects exhibiting the
most intense high-energy emission \cite{Kappes,AharonianReview,MilagroReview}. 
For these sources, the zenith position, angular extension, and neutrino
flux parameterisation are extracted from the measured high-energy $\gamma$-ray
spectra. In both cases, the expected neutrino spectra are evaluated from the
$\gamma$ spectrum under the hypothesis of a transparent source and 100\%
hadronic emission. Although PWN are commonly assumed to be powered by e-/e+ winds, they will entrain ions from the ambient medium, possibly accelerating them to very high energies.

\item 
Sources without significant angular extension, emitting a benchmark $E^{-2}$
neutrino spectrum. These can be viewed as characteristic of extragalactic sites
of hadronic acceleration (e.g.\ AGN) with cut-offs expected at very
high energies.. While the actual spectra of individual
neutrino sources is not expected to follow a simple $E^{-2}$ power-law, and may
exhibit features such as a peak at PeV energies, or a harder spectra extending
to EeV energies \cite{mannheim_2014}, the projected sensitivity to an $E^{-2}$ flux gives a good
indicator of ARCA's ability to study such extragalactic sources with higher-energy fluxes.

\end{itemize}
 
For the detection of neutrinos from point-like sources, the best performance is
expected from a search for track-like events. In fact, as discussed in
\mysref{sec-sci-too-rec}, with long muon tracks an angular resolution of about
$\sim0.2^\circ$ can be achieved. To remove the unavoidable down-going
atmospheric muon background, events are selected that contain tracks
reconstructed as up-going.

\begin{figure}
\begin{minipage}[tb]{1\textwidth}
\centering
\begin{overpic}[width=0.8\textwidth]{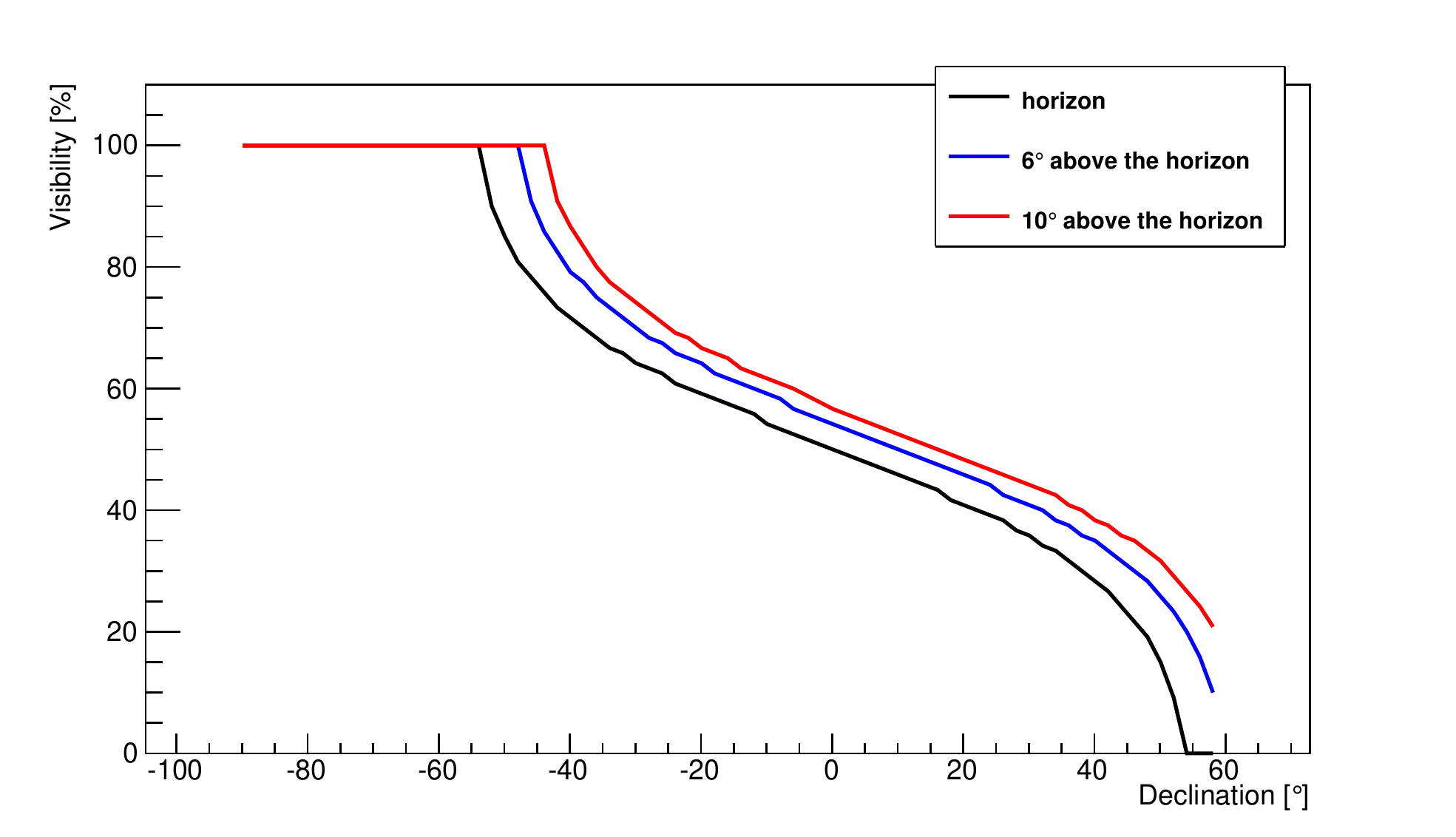}
\put (15,10) {\bf KM3NeT}
\end{overpic}
\end{minipage}
\caption{
KM3NeT/ARCA visibility as a function of source declination for the muon-track
analysis, for $2\pi$ downward coverage, i.e.\ tracks below the horizon (black
line); tracks up to $6^\circ$ above the horizon (blue line); and tracks up to
$10^\circ$ above the horizon (red line).}
\label{fig.Visibility}
\end{figure}

At the latitude of the Mediterranean Sea, selecting tracks that are
reconstructed below or a few degrees above the horizon implies a reduction of
the visibility for source declinations above $-40^\circ$, as shown in
\myfref{fig.Visibility}. On the other hand, it is possible to view
Northern-sky sources below $+ 50^{\circ}$ of declination, giving a total of
$\approx 3.5\pi$\,sr sky coverage.

\paragraph{Galactic sources} 

\begin{figure}
\begin{minipage}[tb]{1\textwidth}
\centering
\includegraphics[width=0.5\textwidth]{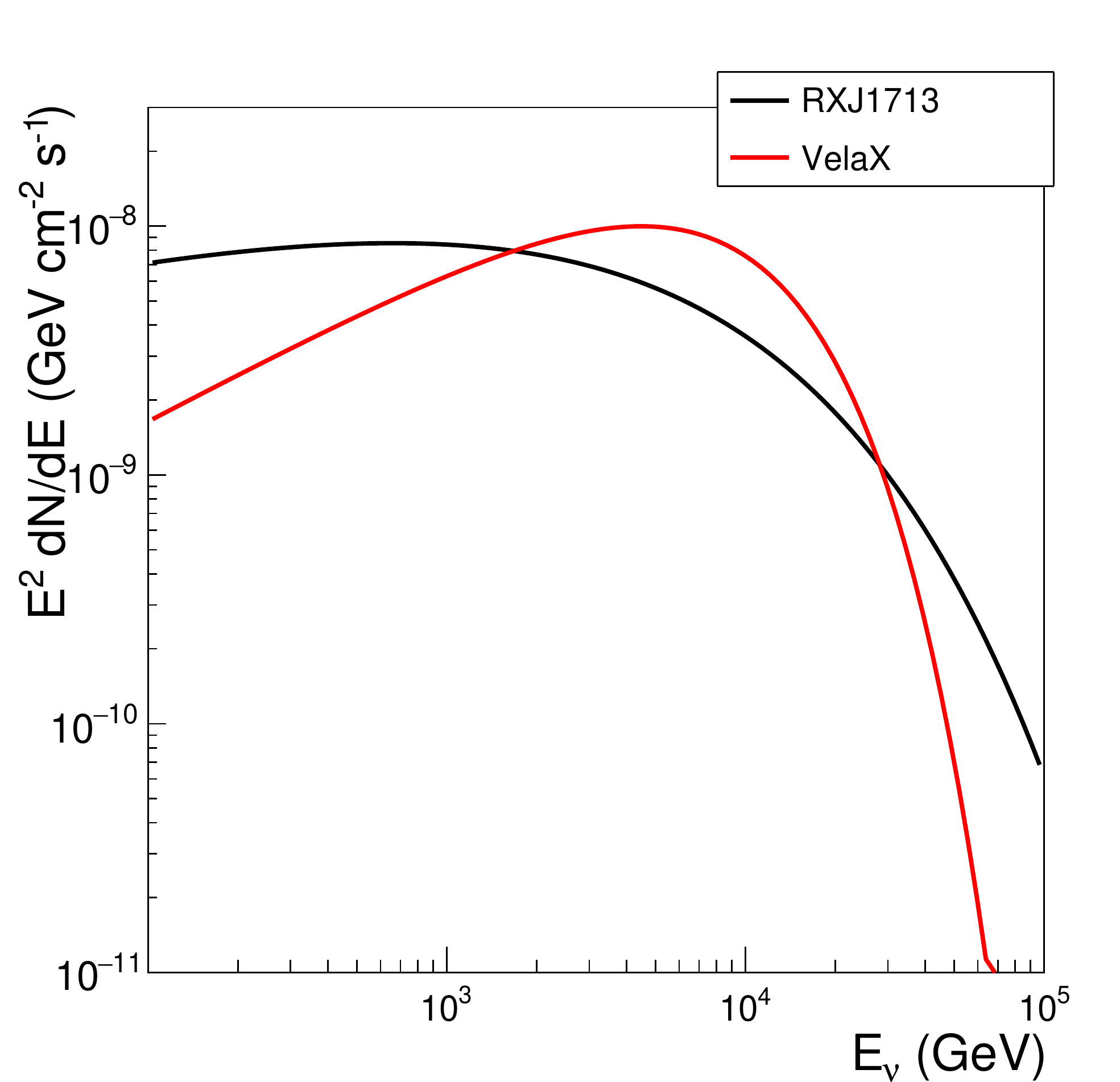}
\end{minipage}
\caption{%
${\nu_\mu}+\bar\nu_\mu$ energy spectra for RX\,J1713 (\myeref{eq.RXJ}, black line) and Vela-X (\myeref{eq.VelaX}, red line).}
\label{fig.GalEnSpectra}
\end{figure}

SNR RX\,J1713.7-3946 (short: RX\,J1713) is a young shell-type supernova remnant
that has been observed by H.E.S.S.\ in several campaigns
\cite{HESS-RXJ1713,HESS-RXJ1713-Corrigendum}. The $\gamma$ rays are emitted from
a relatively large circular region with a radius of about $0.6^\circ$ and a
complex morphology, with an energy spectrum that extends up to 100\,TeV. The
source, at a declination of $-39^\circ\;46'$, is visible for 80\% of the time
when selecting tracks with reconstructed zenith angle
$\theta_\text{rec}>78^\circ$. For the present analysis, homogeneous emission
from a circular region around the measured declination with a radial extension
of $0.6^\circ$ has been assumed. The neutrino flux adopted has been derived from
the measured $\gamma$-ray spectrum and has been parameterised following
\cite{kelner-2006}:
\begin{equation}
  \frac{\text{d}\phi}{\text{d}E_\nu} = 
  16.8 \times 10^{-15} \left[ \frac{E_\nu}{1\,\text{TeV}}\right]^{-1.72}\cdot
  \exp\left(-\sqrt{\frac {E_\nu}{2.1\,\text{TeV}}}\right)
  \;\;\text{GeV}^{-1}\,\text{cm}^{-2}\,\text{s}^{-1}\;.
\label{eq.RXJ}
\end{equation}
This energy spectrum is shown in \myfref{fig.GalEnSpectra} (black line).

In the point source analysis for track-like events, all simulated events ($\nu_\mu$,
$\nu_e$, $\nu_\tau$, $\mu_\text{atm}$) have been reconstructed with the track
reconstruction code described in \mysref{sec-sci-too-rec}.

\begin{figure}
\begin{minipage}[tb]{1\textwidth}
\centering
\includegraphics[width=0.5\textwidth]{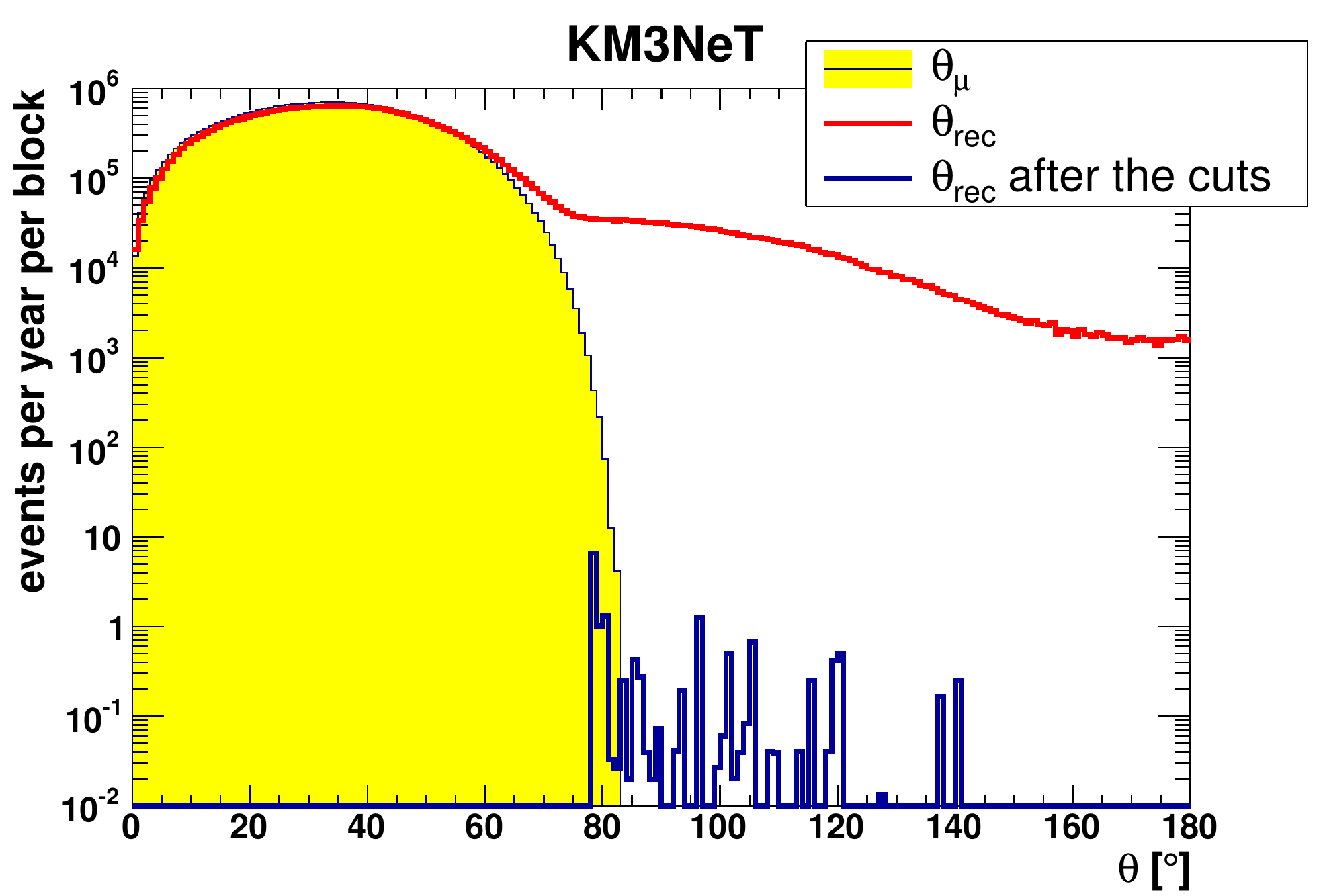}
\end{minipage}
\caption{%
Distributions of the zenith angle $\theta_{rec}$ of atmospheric
muons for one KM3NeT/ARCA building block at generation level (yellow area), reconstruction level (red line), and
after the preselection cut and the cut on $\Lambda$ (blue histogram). 
}
\label{fig.MuAtmRejection}
\end{figure}

Since the maximum elevation for a source at the declination of RX\,J1713 is
$\sim 14^\circ$, and to maximise the signal-to-background ratio, events were
preselected requiring that the reconstructed track has a zenith angle
$\theta_\text{rec}>78^\circ$ and a radial distance from the centre of the source
of $\alpha<10^\circ$. The numbers of events at reconstruction level and after
the preselection cuts are shown in \mytref{tab.EventsRXJ1713}. Even after the
preselection, the numbers of events due to neutrino and muon atmospheric
background largely exceed the number of expected signal events from the source. 
The atmospheric muons can be efficiently removed by imposing a cut on the
$\Lambda$ parameter as shown in \myfref{fig.MuAtmRejection}. Finally, a BDT
trained to discriminate signal events from neutrino background is applied.

\begin{figure}
\includegraphics[width=0.49\textwidth]{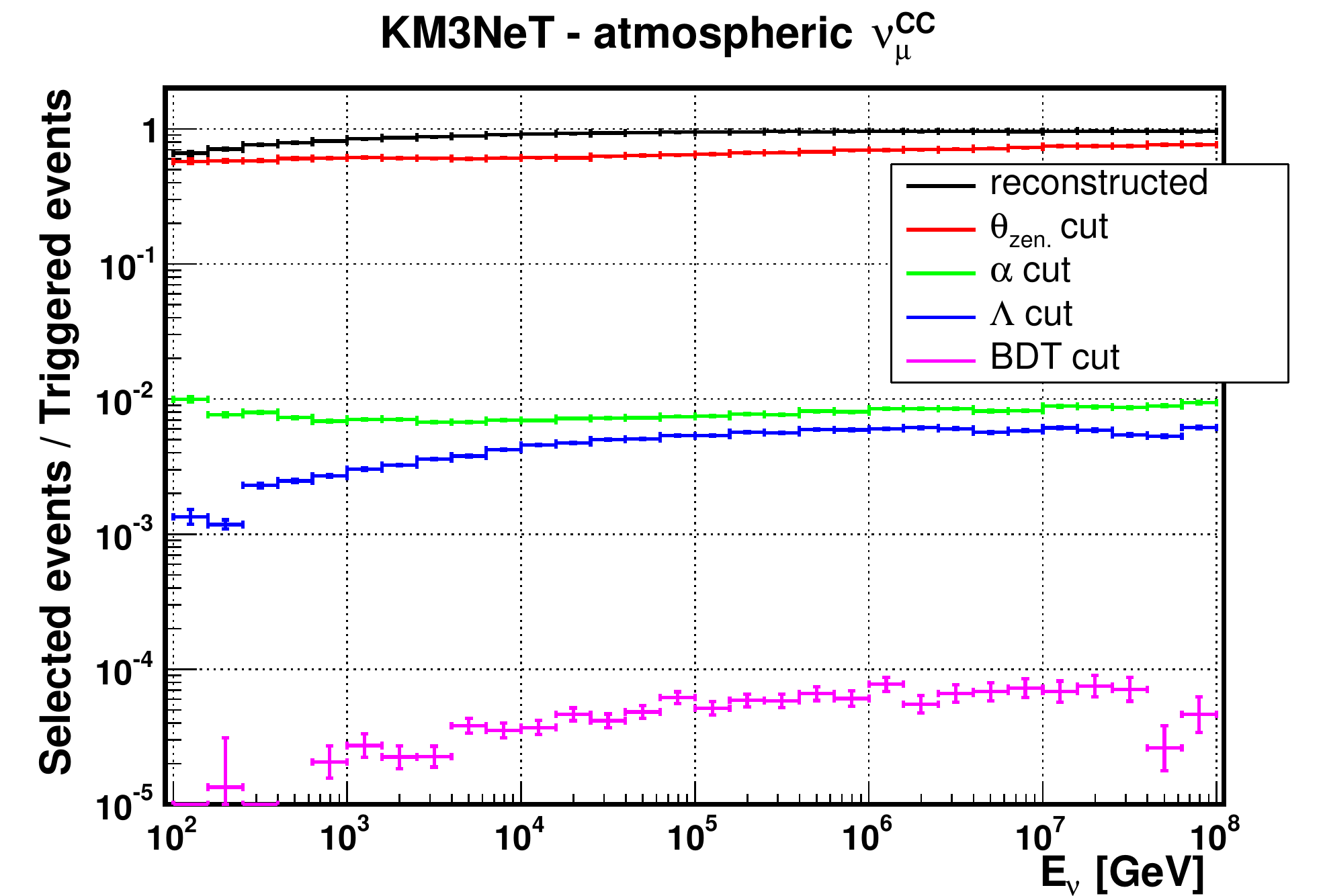}
\hfill
\includegraphics[width=0.49\textwidth]{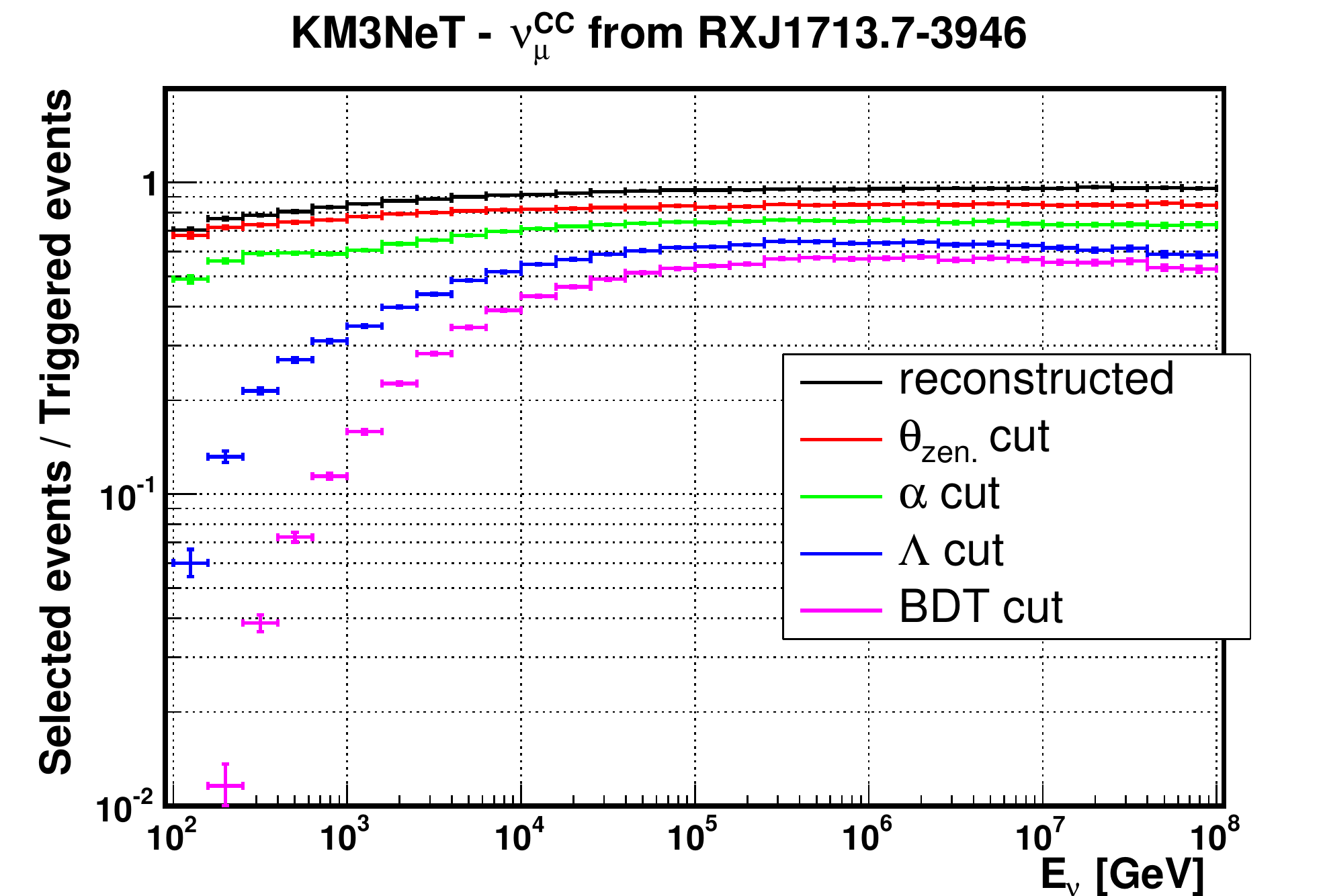}
\caption{%
Ratios between the numbers of selected events and triggered events at
each step of the analysis for atmospheric $\nu_\mu$ CC neutrinos (left panel) and  $\nu_\mu$ CC neutrinos from the source (right panel) as a function of the neutrino energy.}
\label{fig.Eff-Point-like}
\end{figure}

The MDP is then maximised by adjusting the cut on the BDT output value. The
number of events per 5 years of observation time surviving these cuts is indicated in
\mytref{tab.EventsRXJ1713}, together with the number of events expected at
each step of this analysis.  
The ratio between these event numbers and the number
of triggered events is reported as a function of the neutrino energy for
$\nu_\mu$ CC interactions in the right panel of
\myfref{fig.Eff-Point-like}.

\begin{table*}
\small
\begin{center}
\def\arraystretch{1.5}
\begin{tabular}{| c | c  | c | c | }
\hline
& reconstructed tracks &after preselection cuts & after final cuts  \\
\hline 
$\mu_\text{atm}$     & $2.7 \times 10^8$ & $1.5  \times 10^5$ & $\approx 2.0$ \\
\hline 
\hline
$\nu_\text{atm}^\mu$ & $1.6 \times 10^6$ & $1.2\times 10^4$ & 11.6\\
\hline
$\nu_\text{atm}^e$   & $6.0 \times 10^4$ & $545$            & 0\\
\hline
\hline
$\nu_\text{RXJ}^\mu$ & 33.4 &23.5 & 8.1\\
\hline
$\nu_\text{RXJ}^e$   & 12.5 & 0.8 & 0\\
\hline
$\nu_\text{RXJ}^\tau$& 12.3 & 2.55 & 0.57\\
\hline
\end{tabular}
\end{center}
\caption{%
Expected event numbers in 5 years of observation time for KM3NeT/ARCA (2 building blocks) at
different stages of the RX\,J1713 track analysis. The number of surviving atmospheric muons after the final cuts has been extrapolated from the present statistics.}
\label{tab.EventsRXJ1713}
\end{table*}

\begin{figure}
\begin{minipage}[tb]{1\textwidth}
\centering
\includegraphics[width=0.6\textwidth]{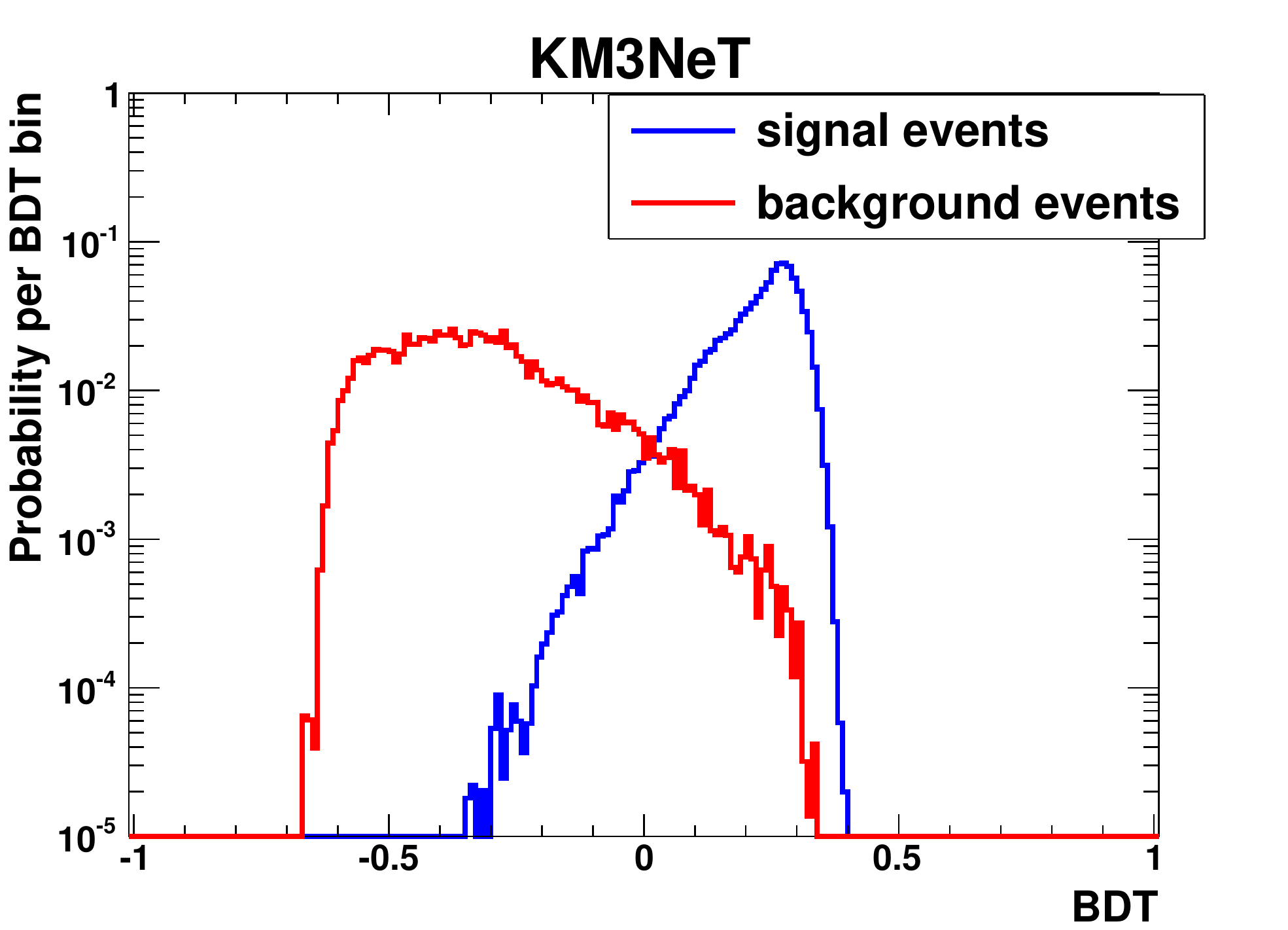}
\end{minipage}
\caption{%
Distributions of the BDT output for background neutrino events and signal
events for the SNR RX\,J1713 analysis.}
\label{fig.PDFRXJ}
\end{figure}

The significance has been evaluated with an unbinned method \cite{Braun2008} by
maximising the likelihood ratio of \myeref{eq.LR}, with PDFs expressed as
functions of the BDT output (\myfref{fig.PDFRXJ}). The result shows that a
3$\sigma$ significance can be reached in about 4\,years of observation time.

\begin{figure}
\begin{minipage}[tb]{1\textwidth}
\centering
\includegraphics[clip=true,trim={0.1cm 0.1cm 0 0},width=0.7\textwidth]{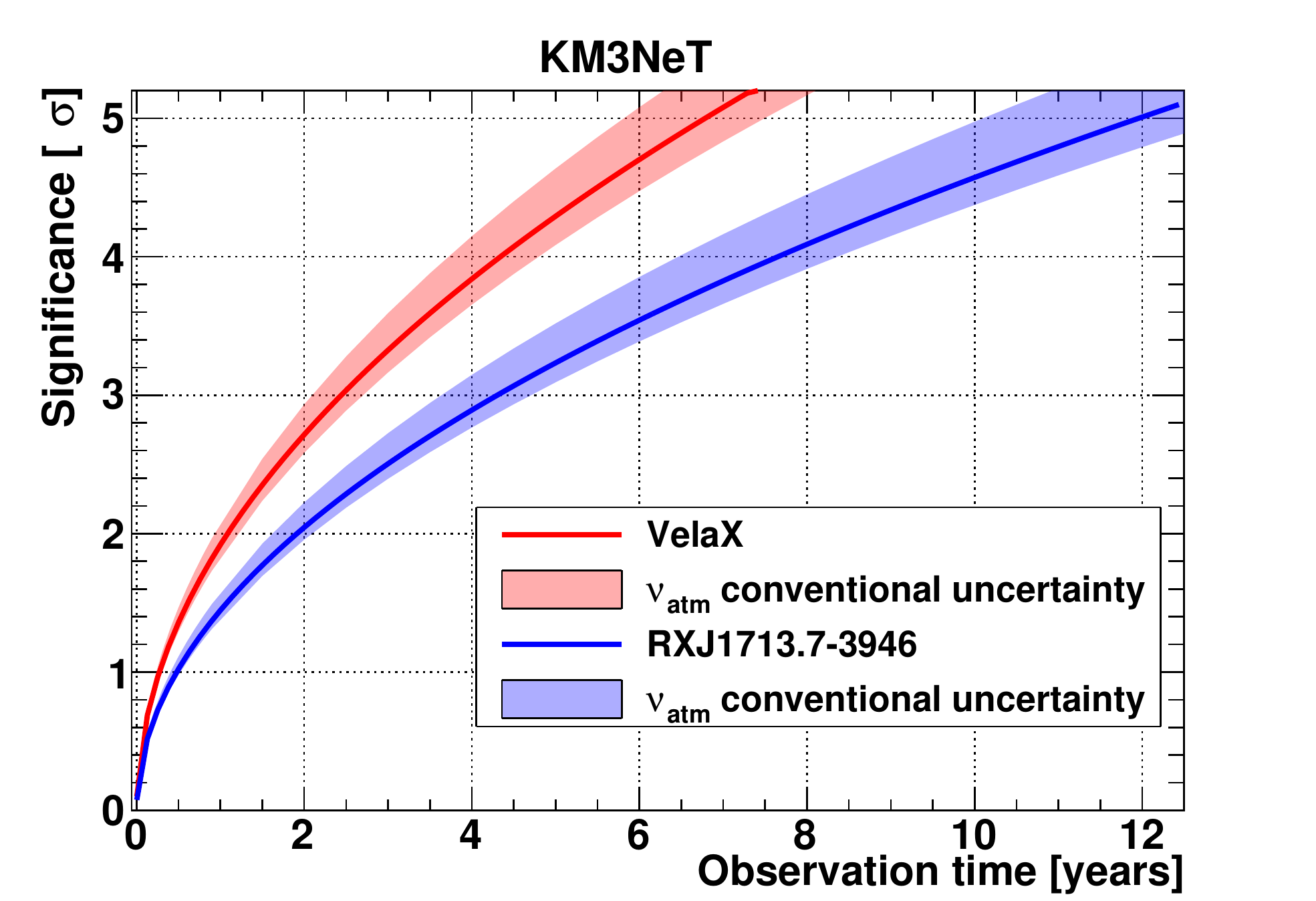}
\end{minipage}
\caption{%
Significance as a function of KM3NeT/ARCA (2 building blocks) observation time for the detection of the Galactic
sources RX\,J1713 and Vela-X. The bands represent the effect of the uncertainties
on the conventional component of the atmospheric neutrino flux.}
\label{fig.SignificanceGalactic}
\end{figure}

The same analysis has been applied to Vela-X, which is one of the nearest and
most intense PWNe (Pulsar Wind Nebulae), and has been extensively studied in TeV $\gamma$ rays by the
H.E.S.S.\ Collaboration \cite{HESS-VelaX2006,HESS-VelaX2012}. Vela-X is located
at a declination of $-45^\circ\;36'$. The neutrino spectrum has been estimated
from the differential energy spectrum using the prescription in
\cite{VissaniPhysRevD2008,VissaniNIM2008,VissaniAstrophJ2006} for an integration
radius of $0.8^\circ$ around the source centre and was
parameterised as:
\begin{equation}
  \frac{\text{d}\phi}{\text{d}E_\nu} = 
  7.2 \times 10^{-15}\cdot\left[\frac{E_\nu}{1\,\text{TeV}}\right]^{-1.36} \cdot
  \exp\left( -\frac{E_\nu}{7\,\text{TeV}}\right)
  \;\;\text{GeV}^{-1}\,\text{cm}^{-2}\,\text{s}^{-1}\;.
\label{eq.VelaX}
\end{equation}
This spectrum is shown in \myfref{fig.GalEnSpectra} (red line). The source has
been simulated as a homogeneously emitting disk of $0.8^\circ$ radius.

The expected sensitivity of ARCA to Vela-X is shown in
\myfref{fig.SignificanceGalactic} as a function of the observation time. 
Owing to the good visibility of the Galactic Plane, a significance of $3\sigma$
can be reached in less than 3\,years of observation time. The bands show the
variation of the significance due to the uncertainty on the normalisation of the
conventional part of the atmospheric neutrino spectrum (see
\mysref{sec-sci-ass}).

\paragraph{Sources with a spectrum $\propto{E^{-2}}$}
\label{sec-sci-sensitivity-2}

The flux required for a $5\sigma$ discovery has also been calculated for a
generic point-like source with a spectrum $\propto E^{-2}$. In the preselection
sample only events with $\theta_\text{rec}>80^{\circ}$ have been selected. In
this analysis, at present, the BDT procedure has not been applied, since the
larger difference in the slopes of the atmospheric and source neutrino energy
spectra eases discrimination between them.

\begin{figure}
\begin{minipage}[tb]{1\textwidth}
\centering
\begin{overpic}[width=0.6\textwidth]{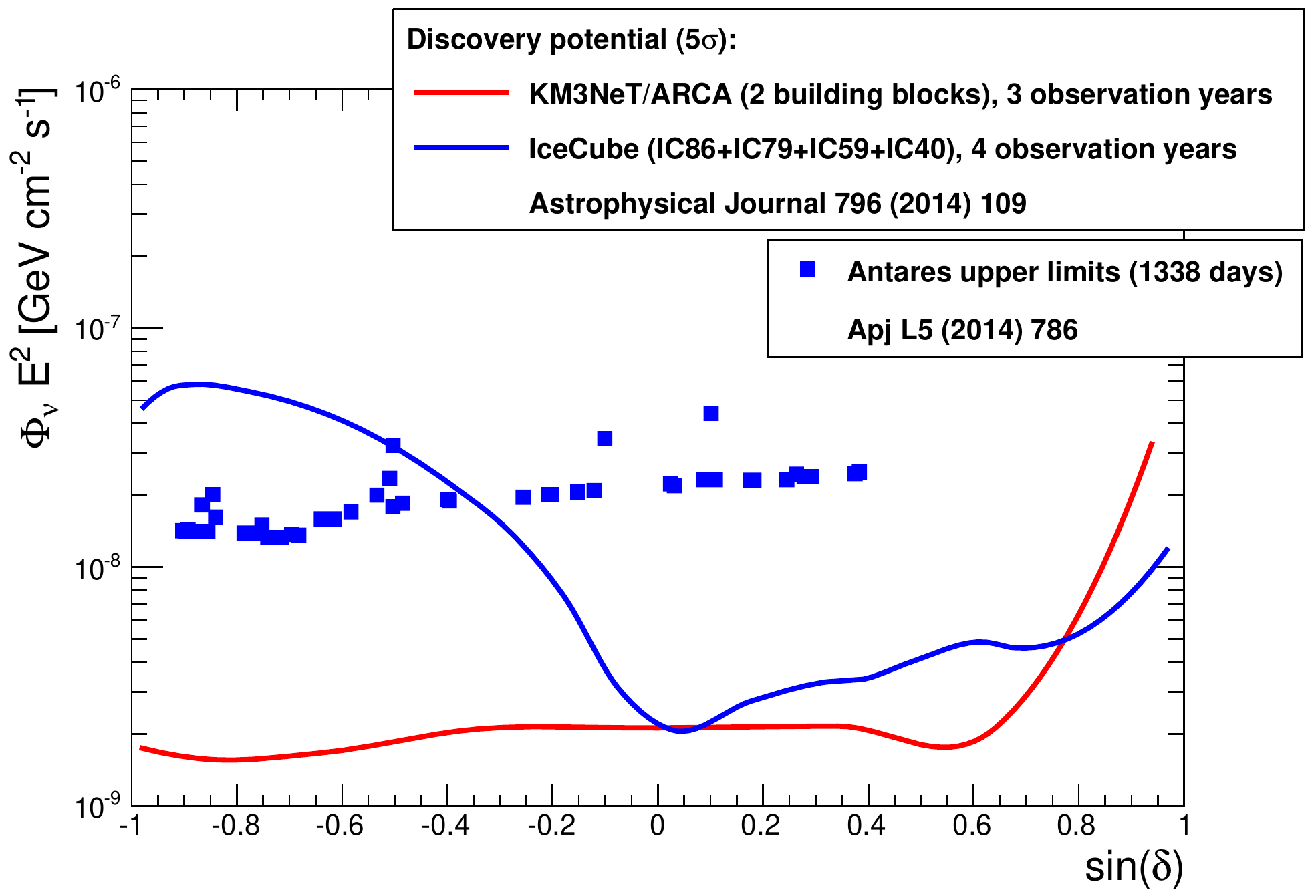}
\put (15,45) {\bf KM3NeT}
\end{overpic}
\end{minipage}
\caption{%
KM3NeT/ARCA (2 building blocks) $5\sigma$ discovery potential as a function of the source
declination (red line) for one neutrino flavour, for point-like sources with a
spectrum $\propto E^{-2}$ and 3\,years of data-taking. For comparison, the
corresponding discovery potential for the IceCube detector
\cite{icecube-muon-fouryear-pointsource} (blue line), and upper limits on
particular sources for the ANTARES detector \cite{ANTARES-PointLike2014} (blue
squares) are also shown.}
\label{fig.DiscoveryE-2}
\end{figure}

After the preselection, an unbinned method has been applied that maximises the
likelihood ratio of \myeref{eq.LR}, with PDFs as functions of the two
parameters $N_\text{hit}$ (related to the energy of the neutrinos) and $\alpha$,
the angular distance from the source centre. The $5\sigma$ discovery flux is
reported in \myfref{fig.DiscoveryE-2} as a function of the declination for
3\,years of observation time, corresponding to the exposure for the current
IceCube result. The upper limit of ANTARES is also reported for comparison.

ARCA's expected resolution on cascades of $\sim1.5^\circ$ (see
\mysref{sec-sci-too-rec-casc}) allows us to also use this channel for a
point-source search, as recently demonstrated by ANTARES
\cite{antares_icrc_cascades}. Since discriminating down-going cascade events from
the muonic background is easier than for tracks, cascade searches have a
$4\pi$\,sr coverage, making this detection channel especially important for
sources with an otherwise limited visibility. First preliminary results for the
cascade channel for generic point-like sources with an $E^{-2}$ spectrum will
also be presented in this section.

The sensitivity of KM3NeT/ARCA to
point-like sources has been evaluated using cascade events. In this analysis all
simulated events have been reconstructed with both the track and cascade
reconstructions. To remove the atmospheric muons, which are the main source of
background, a preselection of events was performed, leading to the two event
samples:

\begin{itemize}
\item 
Sample A: Events reconstructed as down-going with the track reconstruction.
Cuts similar to the cascade diffuse analysis have been applied:
\begin{itemize}
\item
Geometrical containment cuts $z<250$\,m and $r<500$\,m (see
\myfref{fig.PreselectionCuts});
\item
Reconstructed track zenith $\theta_\text{rec}<80^\circ$;
\item
$\text{ToT}_\text{evt}>6\,\mu$s (see \myfref{fig.PreselectionCuts});
\item
$\Lambda<-5.8\,$.
\end{itemize}
\item
Sample B: Events reconstructed as up-going with the track reconstruction.
The following cuts are applied:
\begin{itemize}
\item
Geometrical containment cuts $z<324$\,m and $r<450$\,m (see
\myfref{fig.PreselectionCuts});
\item
Reconstructed track zenith $\theta_\text{rec}>80^\circ$;
\item
$\text{ToT}_\text{evt}>4\,\mu$s (see \myfref{fig.PreselectionCuts}).
\end{itemize}
\end{itemize}

The containment cuts mainly select cascade events that have the interaction
vertex inside the detector volume and remove track-like events. Remaining
track-like events are rejected by the $\Lambda$ cut in Sample~A (removing
well-reconstructed atmospheric muons) and with the ToT cut that removes
lower-energy tracks with the vertex inside the instrumented volume. In both samples,
most of the selected source events are cascade-like events, the track ``contamination''
being of order 10\%.

Since the cut-and-count method has not been applied in this case (i.e.\ no cut
on the distance between the reconstructed direction and the source centre
has been applied, with events reconstructed closer to the source appearing more source-like),
the unweighted number of signal and background events passing the above cuts is
not meaningful, and is not reported.

\begin{figure}
\begin{minipage}[tb]{1\textwidth}
\centering
\begin{overpic}[width=0.49\textwidth]{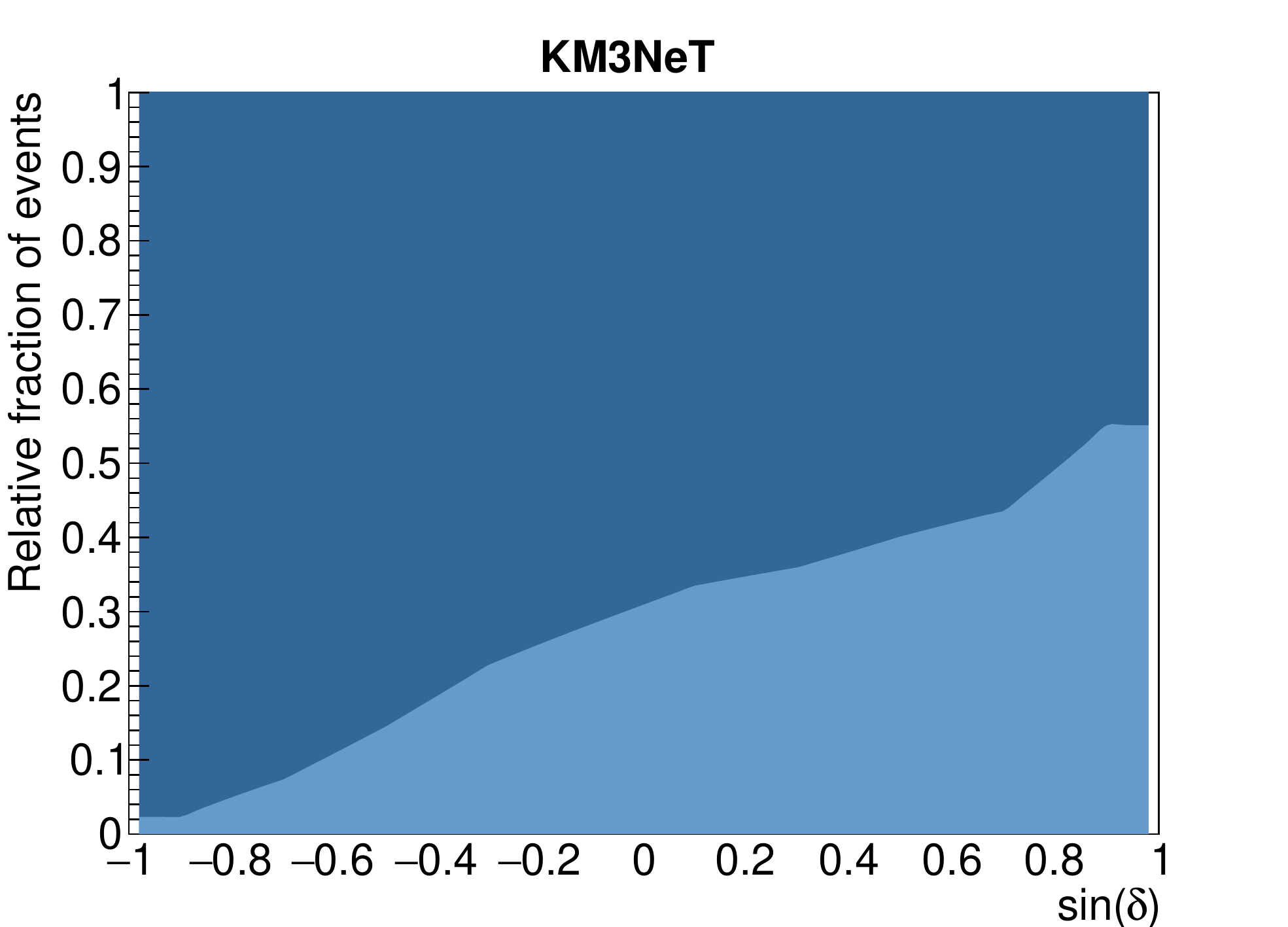}
\end{overpic}
\begin{overpic}[width=0.49\textwidth]{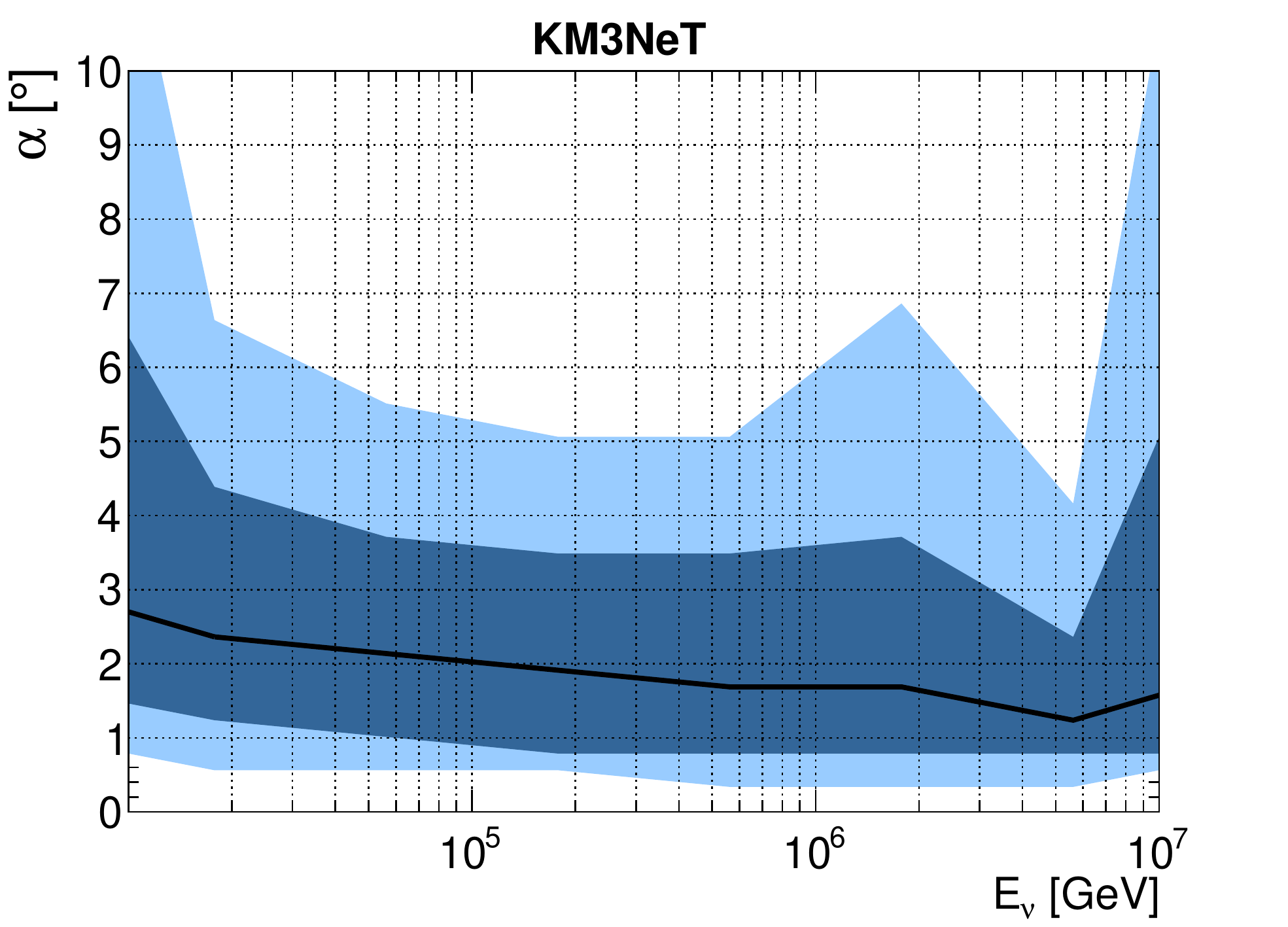}
\end{overpic}
\end{minipage}
\caption{%
Relative fraction of events in Sample~A (dark blue) and in Sample~B
(light blue) as a function of the source declination (left). Angular
resolution of events of the preselected samples for a source at declination
$\delta=45^\circ$. The black line represents the median of the distribution. The
dark and light blue bands show the 90\% and 68\% quantiles of the
distribution (right).}
\label{fig.RelativeFraction}
\end{figure}

The same BDT procedure described in \mysref{sec-sci-too-dif} for the diffuse
cascade analysis, to discriminate tracks from showers, has been applied to the
two samples. An optimal cut on the BDT output variable was found to be $\rho>0.5$.

The discovery potential has been obtained by performing an unbinned
log-likelihood search. The likelihood takes into account the energy and
directional information of each event reconstructed with the cascade
reconstruction. In order to take into account the two different event samples,
the following likelihood ratio, similar to that one of \myeref{eq.LR}, has been
considered:
\begin{equation}
  \text{LR} = 
  \prod_{j=1}^2 \prod_{i=1}^{N^j} 
  \left[ \frac{n_\text{signal}^j}{N^j}\cdot S^j_i  + 
  \left(1- \frac{n_\text{signal}^j}{N^j}\right) \cdot B^j_i \right]\;,
  \label{eq.LR2}
\end{equation}
where $j$ indicates the data sample and $i$ indicates the event in that sample.
$S^j_i$ and $B^j_i$ are the PDFs for the signal and
background of the $j^\text{th}$ sample and are evaluated as functions of the
reconstructed cascade energy and of the distance from the source centre. $N^j$
is the total number of events in the $j^\text{th}$ sample. The estimates number
of signal events $n_\text{signal}^j$ in each sample is related to the total number 
$n_\text{signal}$ by the relative contribution $n_\text{signal}^j= C(\delta)\cdot n_\text{signal}$. In
\myfref{fig.RelativeFraction} (left panel) the relative percentage of events,
$C(\delta)$, of the two selected samples with $\rho>0.5$ is shown as a function
of the declination.

\begin{figure}
\begin{minipage}[tb]{1\textwidth}
\centering
\begin{overpic}[width=0.6\textwidth]{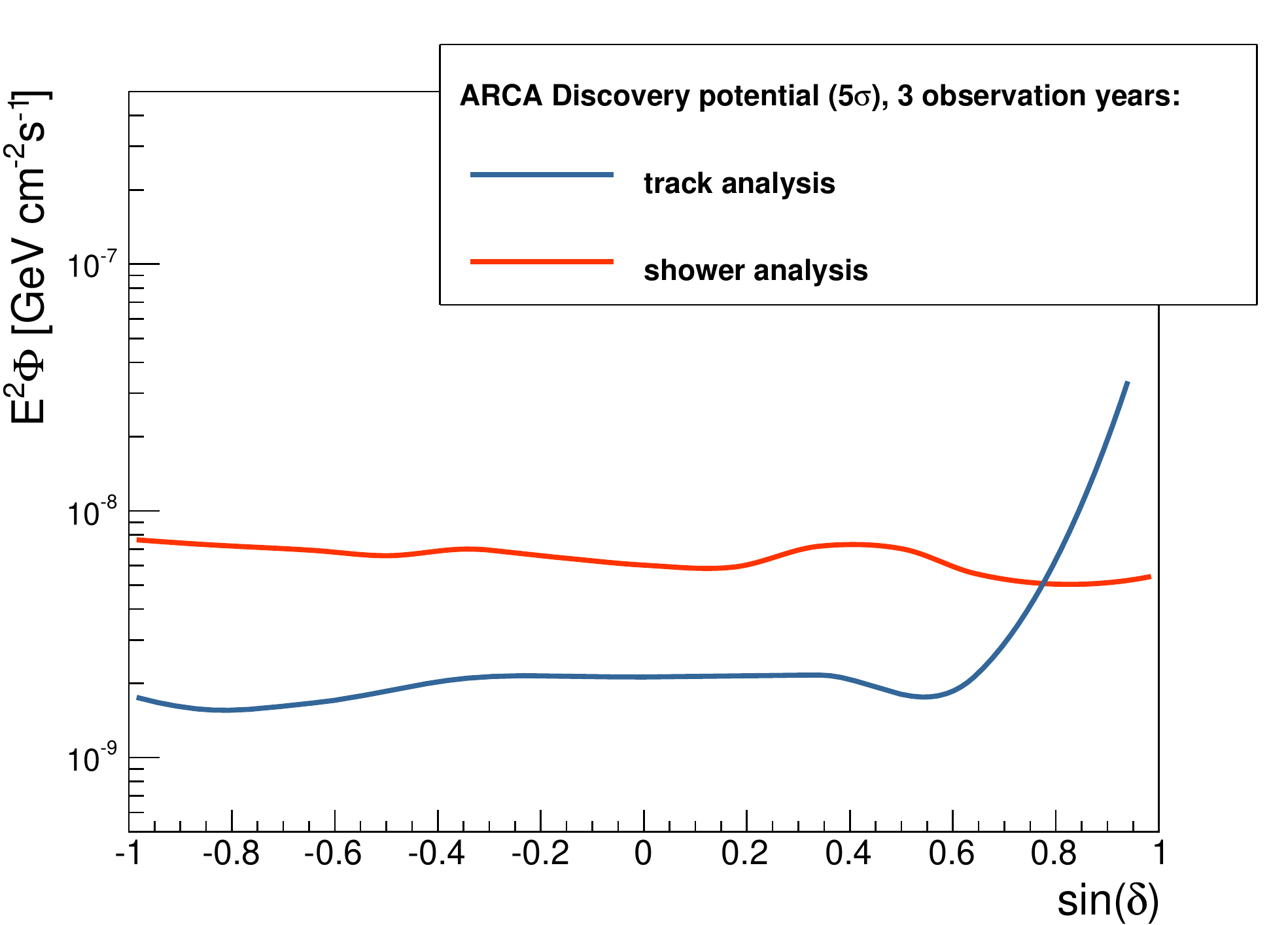}
\put (15,60) {\bf KM3NeT}
\end{overpic}
\end{minipage}
\caption{\
Discovery potential of KM3NeT/ARCA (2 building blocks) at 5$\sigma$ with 50\% probability for point-sources with
an $E^{-2}$ spectrum, for cascade events (red line) for three years of
observation time. For comparison the discovery potential for the track analysis
is also shown (blue curve).}
\label{fig.DiscoveryCascadePointLike}
\end{figure}

The discovery flux at the 5$\sigma$ level is reported in
\myfref{fig.DiscoveryCascadePointLike} as a function of the declination (red
line) for 3\,years of observation time, and is compared with the discovery
flux obtained for the track analysis. For declinations
higher than $50^\circ$, where the visibility for up-going tracks is very poor or
null, a competitive value w.r.t.\ the present IceCube value can be obtained (see
\myfref{fig.DiscoveryE-2}).

The cascade angular resolution of the preselected events for $\delta=45^\circ$
is reported in \myfref{fig.RelativeFraction} right panel, and shows that an
average angular resolution of about $2^\circ$ can be reached.
This includes all events passing the cuts described above (no cuts on reconstructed
angle to the source), showing that a very good angular resolution is
obtained.

Similarly to the diffuse analysis, improvements in point-source sensitivity are
expected when combining the events from the track and cascade channels,
especially for sources located in the Northern sky.

\paragraph{Potential improvements in point-like-source searches}

An improvement in the sensitivity for the search for neutrinos from point-like
sources is expected when the two new reconstruction algorithms (one for track
and one for cascade events, see \mysref{sec-sci-too-rec}), that are
being tested, will be applied to the MC data set. Additionally, the first tests
indicate a higher number of reconstructed events (higher efficiency)
in addition to a better angular resolution.

The search for neutrino sources can be also improved by grouping potential
sources together in a procedure that is known as ``source stacking". This is
usually applied to sources of the same class. In our case, several
potential sources
otherwise too weak to be investigated individually are present in the Galactic and
extragalactic region. This technique has not been yet applied, but an
improvement is expected both for the search of Galactic PeV sources (SNR, PWN,
etc.) and for extragalactic sources (AGN).

\subsubsection{Further physics opportunities}
\label{sec-sci-fur}

In addition to the central science targets of neutrino astronomy, i.e.\
investigating high-energy cosmic neutrinos and identifying their astrophysical
sources, KM3NeT/ARCA will offer a wide spectrum of further physics
opportunities, of which a selection is sketched in the following. Corresponding
physics analyses have been pioneered by the IceCube and ANTARES collaborations.

\begin{itemize}

\item
{\bf Gamma-ray bursts (GRB)}

There is strong evidence that long-duration gamma-ray bursts (GRBs) are
produced from relativistic jets formed in the collapse of a massive star \cite{grb-fireball}.
Shocks generated either within the jet, or when the jet collides with
surrounding material, are potential cosmic ray acceleration sites, with an
associated neutrino flux from subsequent interactions and decays \cite{grb-nu}.
The short duration of GRBs (seconds to minutes) allows a narrow neutrino-search time-window,
effectively reducing the background when compared to a standard point-source search.
This has allowed ANTARES and IceCube to constrain the properties of GRB jets \cite{icecube-grb-2015,antares-grb}.
KM3NeT/ARCA will increase the sensitivity of such searches similarly to that for
$E^{-2}$ point-sources (\mysref{sec-sci-sensitivity-2}).

\item
{\bf Multi-messenger studies:} 

KM3NeT/ARCA will be part of a global alert system able to tag synchronous
observations of different experiments, observing e.g.\ $\gamma$ rays or
gravitational waves, that in themselves are not significant but become so when
combined. Another branch of multi-messenger studies is the creation of alerts for
optical, radio or X-ray telescopes to follow up ``suspicious'' neutrino
observations, such as a doublet of events from the same celestial direction
during a short time period. As per the ANTARES TAToO program \cite{antares-tatoo},
KM3NeT/ARCA will monitor more than half the sky, and will be able to generate alerts
with high angular precision within seconds. As ultra-high-energy cosmic rays
are also expected to retain some directional information, correlation studies
with the arrival directions of events detected by e.g.\ the Pierre Auger Observatory
will also be possible.

\item
{\bf Cosmic ray physics:} 

KM3NeT/ARCA will register a huge number of high-energy atmospheric muons that
reflect the direction of impact of the primary cosmic-ray (CR) particle with
sub-degree precision. This data set will allow us to investigate inhomogeneities
of the CR flux and to complement the corresponding sky maps by IceCube and
dedicated CR experiments.

A further opportunity might be the detailed investigation of muon bundles that
could, via their multiplicity and divergence, be related to the chemical
composition of CRs.

\item
{\bf Particle physics with atmospheric muons and neutrinos:} 

The high-energy end of the atmospheric muon and neutrino spectra are expected
to be dominated by prompt processes, i.e.\ the production of charm or bottom
hadrons in the primary CR reactions in the atmosphere and their subsequent fast
decay to leptons. Little is experimentally known about these reactions, and
theoretical modelling is difficult since it involves QCD processes at the border
line of the non-perturbative regime. Identifying and measuring the muons and
neutrinos from these processes would shed light on the underlying reaction
mechanisms.

\item
{\bf Tau neutrinos:} 

The capability to identify tau neutrino reactions at energies beyond a few
100\,TeV (see \mysref{sec-sci-too-rec}) will not only allow for constraining
the flavour composition of high-energy cosmic neutrino fluxes, but might also provide
an additional handle to investigate prompt neutrino fluxes (see above), which
are the only CR reactions for which a significant production probability for tau
neutrinos is expected.

A further interesting phenomenon of tau neutrinos is their regeneration after CC
reaction in the Earth through the subsequent tau decay (relevant for energies
above a few 10\,TeV). The observation of this phenomenon would be interesting in
itself, but might in addition signal new particle physics, e.g.\ in the context
of supersymmetry.

\item
{\bf Dark matter:} 

Even though the existence of Dark Matter is considered proven and its particle
nature very likely, there is no direct or indirect evidence for the
properties of these particles. Should they have masses in the TeV range or
above, neutrinos from self-annihilation reactions could be the first Dark Matter
signal ever detected. ANTARES and IceCube have already proven the ability of neutrino
telescopes to significantly constrain Dark Matter properties, with searches targeting
accumulations in the Sun \cite{antares-dm-sun,icecube-dm-sun}, the Galactic Centre
\cite{antares-dm-gc,icecube-dm-gc} and halo \cite{icecube-dm-halo}, and nearby galaxies \cite{icecube-dm-gal}.
The corresponding investigations with KM3NeT/ARCA data
will -- as with all indirect searches -- be particularly sensitive to Dark Matter
particles with spin-dependent scattering cross sections on nuclei. The study of neutrino fluxes from the Galactic centre and halo, nearby galaxies and galaxy clusters could also provide constraints on scenarios that invoke the decay of very heavy ($\sim$PeV) dark matter to explain the high-energy neutrino excess observed by IceCube~\cite{Feldstein:2013kka,Esmaili:2013gha,Zavala:2014dla,Murase:2015gea}.

\item
{\bf Exotics:} 

There is a variety of hypothesised stable or quasi-stable particles that would
leave an identifiable, characteristic signature when crossing the detector. 
Amongst these are magnetic monopoles (for which ANTARES and IceCube have already
performed a search \cite{antares-monopoles,icecube-monopole}), strangelets, Q-balls, and nuclearites.

\item
{\bf Violation of Lorentz invariance:} 

Violation of Lorentz invariance (LIV) could lead to oscillation-like interference patterns 
of atmospheric neutrinos in the energy range of TeV and above. Additionally,
LIV would produce a time-delay between neutrinos and photons from distant, time-variable
sources (in particular, GRBs), allowing LIV to be tested by multi-messenger studies.

\end{itemize}

\subsection{Investigation of systematic effects}
\label{sec-sci-sys}

The simulation chain described in \mysref{sec-sci-too-sim} assumes standard values for the detector geometry, water optical properties, bioluminescent rates, and also a perfectly calibrated detector. In the context of KM3NeT/ARCA sensitivity studies, the term `systematic effects' is used broadly to cover all potential deviations from the standard simulated dataset, and this section describes a series of dedicated studies aiming to estimate their potential influence on ARCA event reconstruction and sensitivity to astrophysical neutrino fluxes.

Each systematic was simulated using a data-set of $10$\% that of the standard simulation, with the systematic being inserted at the latest possible point in the chain to ensure the least influence of random variation between the sets. For example, changing the water scattering length required re-simulating the hit-time distribution on the PMTs, while reducing the PMT effective area was performed by keeping $90$\% of the detected photons from the standard simulation. Most systematic effects were simulated as a $10$\% change, and thus did not reflect the expected size of the resulting effect, but rather were used to estimate the change dX/dSys in some relevant quantity $X$ as a function of the systematic Sys.

In each case, the effects of the systematics were first analysed using the `golden channel' approach, i.e.\ by applying the track reconstruction of  \mysref{sec-sci-too-rec} to $\nu_{\mu}$ CC events, and applying the cascade reconstruction Algorithm 1 of  \mysref{sec-sci-too-rec} to $\nu_e$ CC events and with the cut-and-count method. Only when a significant effect was found was the systematic applied to the full analysis chains: the point-like track analysis of \mysref{sec-sci-too-poi}, and the diffuse cascade analysis of \mysref{sec-sci-too-dif}.

Similar effects have been considered for KM3NeT/ORCA, particularly in the case of $\nuan_e$ reconstruction (\mysref{sec:shower_different_water_QE_noise}). However, the effects of systematics on the mass hierarchy sensitivity of KM3NeT/ORCA are treated via the inclusion of nuisance parameters in the calculation described in \mysref{sensitivity}, rather than using fully resimulated data.

\subsubsection{Optics: water properties and DOM response}

The absorption and scattering of light in seawater has been measured at the KM3NeT-It site to within an accuracy of approximately 10\% \cite{NEMO_optical_2007}. The dominant uncertainty is the contribution due to particulates, whereas the scattering and absorption from pure seawater (salt and water) is well-determined. To simulate this effect, the particulate contribution only has been varied so that the scattering/absorption lengths ($\lambda_{\rm scat}$ and $\lambda_{\rm abs}$ respectively) vary by $\pm10$\% at wavelengths near $400$~nm. It is expected that \emph{in-situ} measurements using the KM3NeT calibration system \cite{icrc_calib_units} will be able to significantly improve on this knowledge.

The major uncertainty in the response of a DOM to incident photons is the total effective area, $A_{\rm eff}$, to Cherenkov photons. This is modelled in  GEANT simulations with a high degree of accuracy, as described in \cite{icrc_geant}, and has been measured using ${}^{40}$K coincidences \emph{in-situ} with a precision of $\sim 1$\% \cite{km3net-ppmdu-2015}. In order to model a significant effect, simulations were produced with $A_{\rm eff}$ varied by $\pm 10$\% for all photon wavelengths and incident angles.

\begin{table*}
\small
\begin{center}
\begin{tabular}{| l | c c | c c |}
\hline
\multirow{2}{*}{Effect} & \multicolumn{2}{c|}{Tracks} & \multicolumn{2}{c|}{Cascades} \\
 & $\Delta E/E$& $\Delta \theta$ & $\Delta E/E$ & $\Delta \theta$ \\
 \hline
$\Delta \lambda_{\rm abs} = \pm 10$~\% 	& $\pm 8\%$ 	&  $\pm 0.1^\circ$	& $\pm$30\% 		& $< 0.1^{\circ}$ \\
$\Delta \lambda_{\rm scat} = \pm  10$~\% & $\pm 0.6\%$ 	& $\pm 0.1^\circ$ 	& $< 1\%$ 	& $< 0.1^{\circ}$  \\
$\Delta A_{\rm eff} = \pm  10$~\%       & $\pm 5\% $	& $<0.1^\circ$ 	& $\pm$10\% 		& $< 0.1^{\circ}$ \\
1\% missing DOMs 		& 	 	& $0.01^\circ$	& <0.1\% 	& $+0.02^{\circ}$ \\
1 missing DU 			& 		& $0.01^\circ$ 	& <0.1\% 	&  $+0.025^{\circ}$	\\
\hline
\end{tabular}
\end{center}
\caption{Estimated effects of systematics on event reconstruction accuracy, evaluated on the event samples from $E^{-2}$ point-source searches for tracks ( \mysref{sec-sci-too-poi}) and the diffuse flux search (\mysref{sec-sci-too-dif}). For each sample, the worsening in angular resolution $\Delta \theta$, and percentage change in the mean reconstructed energy $\Delta E$/$E$, are given for the changes listed in the first column. The magnitude of the effects does not reflect the final expected uncertainty.} \label{tab-sci-sys-optic-reco}
\end{table*}

The effects of these systematic uncertainties on reconstruction accuracy are summarised in \mytref{tab-sci-sys-optic-reco}, showing the change in reconstruction variables for each percent systematics uncertainty. The most significant effect for the cascade channel is on the energy reconstruction due to a change in $\lambda_{\rm abs}$, since in the high-energy regime, only after a large distances do PMTs cease to become saturated, so that the energy reconstruction depends on the response after several absorption lengths. In no case was the direction reconstruction affected, since the Cherenkov peak (which contains most of the directional information) remains unobscured by these effects.

In the case of muon reconstruction, changes in absorption and PMT efficiency have similar effects on energy reconstruction, which is smaller than in the case of cascade reconstruction due to the inherent uncertainties. Systematic effects on direction reconstruction depends on the muon ($\sim$neutrino) energy. The difference in the track direction w.r.t the standard value is constant above $\approx$1 TeV (values quoted in the table), and increases with decreasing energy below 1 TeV ( $\approx0.1-0.2^\circ$ at 100 GeV) as expected, since here the reconstruction is photon-limited. Unlike the case of cascade reconstruction,  an increase in water quality, or a larger effective area of the PMTs, improves the directional reconstruction, by increasing the number of Cherenkov photons directly reaching the PMTs.

The effects given in \mytref{tab-sci-sys-optic-reco} describe the best estimates of future systematic effects as a function of future uncertainties in the quantities shown. Another relevant measure is: what range of future performances of ARCA is possible given the current uncertainties in these parameters? For this, the systematic effects above were propagated through the simulation chain, allowing reconstructions, cuts, etc.\ to be re-optimised, i.e.\ assuming the new value of the changed parameter is known. Effects were analysed in the context of the diffuse flux search using cascades, and the RX\,J1713 source search using the track channel. Results are given in  \mytref{tab-sci-sys-optic-analysis}. Note that for the diffuse analyses, the effects are small, since detection efficiency to both signal and background are affected equally.

\begin{table*}
\small
\begin{center}
\begin{tabular}{| l | c c c |}
\hline
Effect & Diffuse (cascades) & Diffuse (tracks)  & RX\,J1731 (tracks)\\
 \hline
$\Delta \lambda_{\rm abs} = -10\%$	& 3.5\%		& 0\%		& 6.5\% \\
$\Delta \lambda_{\rm scat} = -10\%$& <1\%		& 1.5\%		& 1.5\% \\
$\Delta A_{\rm eff} = -10\%$	& 4\%		& 3\%		& 1.5\% \\
$10\%$ less DOMs		& 1.5\%		& 3.0\%		&  \\
1 missing DU			& 0.15\%	& 0.1\%		& \\
\hline
\end{tabular}
\end{center}
\caption{Effects of systematics on the expected sensitivity of the analyses shown, in terms of the change in $5\sigma$ discovery flux after one year. E.g., a reduction in $\lambda_{\rm abs}$ of $10$\% is expected to increase the one-year $5\sigma$ discovery flux of the diffuse cascade analysis by $3.5$\%.} \label{tab-sci-sys-optic-analysis}
\end{table*}

\subsubsection{Detector calibration and alignment}

The suite of calibration and alignment systems described in \cite{icrc_calib_units} have a finite accuracy, and differences from the true DOM positions and orientations might reduce the precision of reconstruction.

Due to the mechanical structure of KM3NeT detection units, the major degrees of freedom for DOM motion are the position in the horizontal plane, and rotation about the vertical axis. The accuracy of acoustic positioning (position in the horizontal plane) is expected to be $20$~cm (corresponding to a hit time uncertainty of about 1 ns in water), while the internal compass for each DOM will measure the rotation angle to within $3^{\circ}$.

To simulate each effect, a false detector was generated with each DOM randomly deviated using Gaussian distributions of width equal to the expected accuracies above, and these were used by reconstruction routines on events generated with the standard simulation chain. 

No detectable effects were observed in the accuracy of either the cascade or track reconstruction. This is partially due to the accuracy of the calibrations, partially the uniform coverage of the DOMs (which make errors in the pointing direction less relevant), and partially the robust nature of the reconstruction algorithms themselves. In the case of orientation angle, the uncertainty was artificially increased until, at $9^{\circ}$ (three times the expected uncertainty), negligible degradation (too small to be measured) in the angular reconstruction accuracy of track-like events was observed. Hence, no further investigation was undertaken.

\subsubsection{Ageing effects}

As KM3NeT ages, some loss of performance due to the degradation or loss of key parts is expected. The effects of PMT ageing are covered by the $A_{\rm eff}$ estimate above. An additional simulation was performed to estimate the effect of both lost DOMs and entire DUs, with the standard simulation re-run once with a random $10$\% sample of DOMs turned off, and once with DUs randomly removed. The effects on both reconstruction and future sensitivity were estimated assuming that the failed units were known, which will be the case due to continual monitoring. The results are shown in \mytref{tab-sci-sys-optic-reco} and \mytref{tab-sci-sys-optic-analysis}. In general, the effects are most important for low-energy muon tracks.

\subsection{Detector geometry studies}
\label{sec-sci-geo}

The chosen geometry of KM3NeT ARCA building blocks, with approximately $90$\,m horizontal spacing between detection units, and $36$\,m vertical spacing between DOMs, was optimised in preliminary studies to target Galactic sources such as RX\,J1713. While some limits on the final layout are imposed through engineering considerations --- in particular, the maximum length of detection units --- the horizontal spacing between detection units can be increased or decreased within a relatively broad range. The discovery by IceCube of a diffuse flux extending above $100$~TeV \cite{icecube-evidence-2013} now motivates revisiting the question of the optimal horizontal spacing. In particular, a larger spacing would be expected to be more optimal when targeting high-energy events.

In order to characterise the effects of a larger spacing, the analyses described above have also been performed by considering a detector block with 120~m spacing between the detection units, giving an approximate $78$\% increase in detector volume. The results are tentative since the analyses have not been fully re-optimised to the alternative geometry. 
Nonetheless, the change in performance gives an indication of the utility of increasing the horizontal spacing.

These tentative results are summarised in \mytref{tab:geometry_optimisation}. 
They have been performed with $10\%$ of the data set, using the fast cut-and-count method. An improvement of about 20--30\% in the discovery flux is observed in the search for a diffuse flux for both channels for the detector layout with increased string spacing. Note that in the cascade channel the sensitivity gain is significantly below the increase of the instrumented volume; one of the reasons is a decrease of the signal detection and reconstruction efficiency with increasing string distances.  For Galactic sources, which present a lower neutrino energy spectrum, the change is of the same order, but as expected, in the reversed direction. 

It is therefore expected that the final optimum, taking into account all channels and their physics priorities, is in or close to the range explored in this first geometry investigation. Clearly, the choice of optimum configuration depends on the targeted science goals --- a larger spacing is better for high-energy diffuse fluxes, and smaller spacings for point-like sources with a low-energy cutoff. Given that ARCA is in a unique position to study Galactic point-like sources of neutrinos, and that the detection of such sources is the most challenging of the sensitivity studies presented here (see \myfref{fig.SignificanceGalactic}), the $90$\,m horizontal spacing has been retained for the two ARCA building blocks in KM3NeT Phase-2.0.

\begin{table*}
\small
\begin{center}
\def\arraystretch{1.3}
\begin{tabular}{| l | c |}
\hline
{\bf Analysis channel} & {\bf Sensitivity change: 90~m $\to$ 120~m} \\
\hline
Cascades: diffuse (\myeref{eq.DiffuseFlux1}) & +27\% \\
\hline
Muons: diffuse (\myeref{eq.DiffuseFlux1}) & +20\% \\
\hline
Muons: point-sources ($E^{-2}$, $\delta = -60^{\circ}$) & +18\%\\
\hline
Muons: RX\,J1713 (TeV cutoff) & -20\% \\
\hline
\end{tabular}
\end{center}
\caption{Sensitivity changes in the different analysis channels when increasing the string distance from 90~m to 120~m. The percentages indicate the variation of the fluxes detectable in one year; positive signs indicate a gain in sensitivity with a $120$~m spacing. The precise calculation is $\Phi^{5 \sigma}_{\rm 90~m} / \Phi^{5 \sigma}_{\rm 120~m} -1$, where $\Phi^{5 \sigma}_{\rm 90~m}$ and $\Phi^{5 \sigma}_{\rm 120~m}$ are the $5~\sigma$ discovery fluxes after one year for 90~m and 120~m spacings respectively. Note that all numbers are approximate, being estimated using analyses which have not been fully optimised (see text).} \label{tab:geometry_optimisation}
\end{table*}

\cleardoublepage
%
%
%
%
%
\section{Oscillation Research with Cosmics in the Abyss (ORCA)}
\subsection{Introduction}
\label{intro}

Important progress has been made in the past two decades on determining the fundamental properties of neutrinos. A variety of experiments using solar, atmospheric, reactor and accelerator neutrinos, spanning energies from a fraction of MeV to tens of GeV, have provided compelling evidence for neutrino oscillations, implying the existence of non-zero neutrino masses (see e.g.~\cite{bib:snowmass} and the review by Nakamura and Petcov in~\cite{bib:PDG} for recent insights  on the subject).  

In the standard $3\nu$ scheme, the mixing of the neutrino flavour eigenstates ($\nu_e$, $\nu_\mu$, $\nu_\tau$) into the mass eigenstates ($\nu_1$, $\nu_2$, $\nu_3$) is described by the Pontecorvo–Maki–Nakagawa–Sakata (PMNS) matrix $U$ which is a product of three rotation matrices related to the mixing angles $\theta_{12}$, $\theta_{13}$ and $\theta_{23}$ and to the complex CP phase\footnote{We have omitted here the two additional Majorana phases $\xi$ and $\zeta$ which are irrelevant in oscillation phenomena.} $\delta$:

\begin{equation}
 \label{eq:matrix}
  U =
   \begin{pmatrix}
  1 & 0 & 0 \\
  0 & c_{23} & s_{23} \\
  0 & -s_{23} & c_{23}
 \end{pmatrix} \times
 \begin{pmatrix}
  c_{13} & 0 & e^{-i\delta}s_{13} \\
  0 & 1 & 0 \\
  -e^{i\delta}s_{13} & 0 & c_{13}
 \end{pmatrix}\times\\
 \begin{pmatrix}
  c_{12} & s_{12} & 0 \\
  -s_{12} & c_{12} & 0 \\
  0 & 0 & 1
 \end{pmatrix},
 \end{equation}
where $c_{ij} \equiv \cos\theta_{ij}$ and $s_{ij} \equiv
\sin\theta_{ij}$.
 
 Oscillation experiments are not sensitive to the absolute value of neutrino masses but they provide measurements of the squared-mass splittings $\Delta m^2_{ij} = m_i^2 - m_j^2$ ($i,j=1,2,3$). In the $3\nu$ scheme, there are two independent squared-mass differences; one is responsible for oscillations observed in solar and long-baseline reactor experiments $(\Delta m^2_{sol} \simeq 7.5\times 10^{-5}\, \mathrm{eV}^2)$, while the other impacts the atmospheric neutrino sector ($\Delta m^2_{atm} \simeq  2.5\times 10^{-3}\, \mathrm{eV}^2)$. 
 
 At present, the values of all mixing angles and squared-mass differences in the $3\nu$ oscillation scheme can be extracted from global fits of available data with a precision better than 15\%, the largest remaining uncertainty being currently on $\sin^2{\theta_{23}}$ and its possible octant (i.e. whether $\theta_{23}$ is smaller or larger than $\pi/4$)~\cite{Capozzi2013,bib:gonzales, bib:Forero}.  \mytref{tab:osc_param} summarises the best fit values of the oscillation parameters and the associated 3$\,\sigma$ uncertainties as published in~\cite{Capozzi2013}\footnote{Updates of global fits presented at conferences, yet unpublished, achieve a precision better than 12\% on $\sin^2\theta_{23}$.}.
 
 The recent observation of  $\overline{\nu}_e$ disappearance in several short-baseline reactor experiments~\cite{Abe:2011fz,Ahn:2012nd,An:2012eh} has  provided the first high-significance measurement of the mixing angle $\theta_{13}$ which drives the $\nu_\mu - \nu_e$ transition amplitude with large mass splitting. The relatively large value of this parameter, $\sin^2(2\theta_{13})\simeq 0.1$, is an asset for the subsequent searches for the remaining major unknowns in the neutrino sector, and in particular for the determination of  the neutrino mass hierarchy (NMH).

\begin{figure}[!hbt] \centerline{
\includegraphics[width=0.5\textwidth]{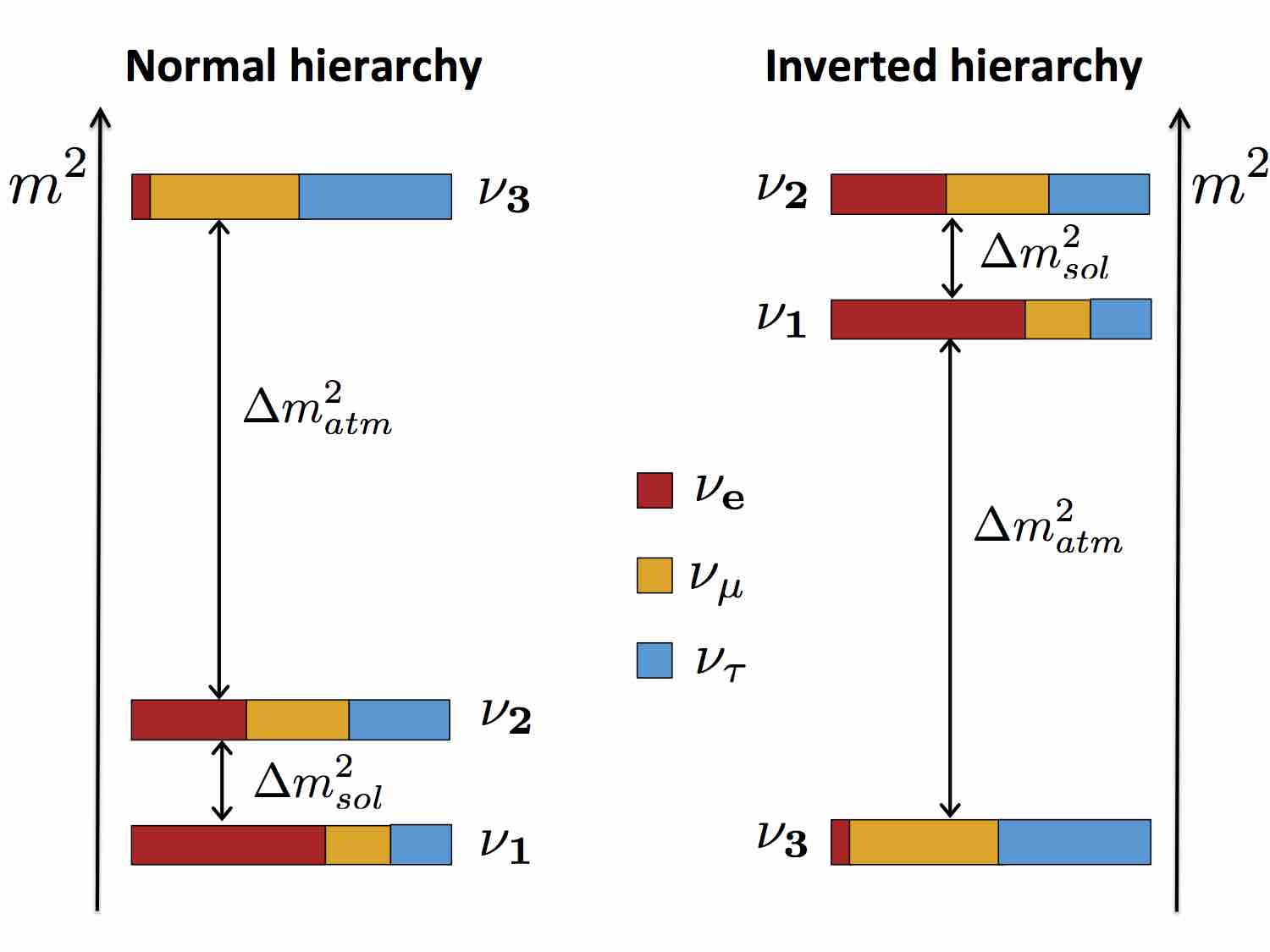} }
\caption{Scheme of the two distinct neutrino mass hierarchies. The colour code indicates the fraction of each flavour ($e, \mu,\tau$) present in each of the mass eigenstates ($\nu_1,\nu_2,\nu_3$).
}\label{fig:hierarchies}
\end{figure}

The ordering of neutrino mass eigenstates has indeed not been determined so far. After fixing $\Delta m^2_{21} = (\Delta m^2)_{sol} >0$, two solutions remain possible depending on the sign of $\Delta m^2_{31}$:  the {\bf normal hierarchy} (NH: $m_1 < m_2 < m_3$) and the {\bf inverted hierarchy} (IH: $m_3 < m_1 < m_2$), as can be seen from \myfref{fig:hierarchies}.

From a theoretical point of view, the determination of the NMH is of fundamental importance to constrain the models that seek to explain the origin of mass in the leptonic sector and the differences in the mass spectrum of charged quarks and leptons~\cite{winter_nmh}. More practically, it has also become a primary experimental goal because the NMH can have a strong impact on the potential performances of next-generation experiments with respect to the determination of other unknown parameters such as the CP phase $\delta$ (related to the presence of CP-violating processes in the leptonic sector), the absolute value of the neutrino masses, and their Dirac or Majorana nature (as probed in neutrinoless double beta decay experiments, or 0$\nu\beta\beta$).  From the astrophysical point of view, the NMH impacts e.g. neutrino flavour conversion in supernovae~\cite{Lunardini:2003eh,Barger:2005it}. Finally, the NMH also affects the precise determination of the PMNS matrix parameters,  as can be seen from \mytref{tab:osc_param}, which summarises the current best fit values and their 3$\sigma$ uncertainties under both hierarchy hypotheses.

 \begin{table}[tb]
\small
\renewcommand{\baselinestretch}{1.50}\normalsize
	\centering
	\begin{tabular}{lc|c|c}
	\multicolumn{2}{c|}{Parameter (hierarchy)} & Best fit & $3\sigma$ range   \\ \hline
	$\sin^2(\theta_{12})/10^{-1}$ &(NH or IH)& $3.08$ & $2.59-3.59$ \\ \hline
	$\sin^2(\theta_{13})/10^{-2}$ &(NH) & $2.34$ & $1.76-2.95$ \\
	$\sin^2(\theta_{13})/10^{-2}$ &(IH)& $2.40$ & $1.78-2.98$ \\ \hline
	$\sin^2(\theta_{23})/10^{-1}$ &(NH)& $4.37$ & $3.74-6.26$ \\
	$\sin^2(\theta_{23})/10^{-1}$ &(IH)& $4.55$ & $3.80-6.41$ \\ \hline
	$\delta_{cp}/\pi$ &(NH)& $1.39$ & - \\
	$\delta_{cp}/\pi$ &(IH)& $1.31$ & - \\ \hline
	$\Delta m^2_{21}/10^{-5}\, \text{eV}^2$ &(NH or IH)& $7.54$ & $6.99-8.18$ \\ \hline
	$\dml /10^{-3}\, \text{eV}^2$ &(NH)& $2.43$ & $2.23-2.61$ \\
	$\dml /10^{-3}\, \text{eV}^2$ &(IH)& $2.38$ & $2.19-2.56$ \\
	\end{tabular}
\renewcommand{\baselinestretch}{1.00}\normalsize
	\caption{The best fit values and $3\sigma$ ranges of the mixing parameters from~\cite{Capozzi2013}, in the normal (NH) or inverted (IH) hierarchy hypothesis. For the large squared-mass difference the following convention is used: $\dml=\Delta m^2_{31}-\Delta m^2_{21}/2$ with $+\dml$ for NH and $-\dml$ for IH.}
	\label{tab:osc_param}
\end{table}

While the combination of 0$\nu\beta\beta$ and direct neutrino mass experiments with cosmological constraints on $\Sigma_\nu m_\nu$ might have an indirect sensitivity to the NMH, most of the efforts currently focus on the determination of NMH via neutrino oscillation experiments (see e.g. Sec.~3.1 of~\cite{bib:snowmass} for an overview of the subject). One option uses medium-baseline ($\sim$50\,km) reactor experiments such as JUNO and RENO-50, which probe the $\nu_e$ oscillation probability at low energies ($\sim$MeV) where matter effects are negligible~\cite{juno}. These experiments may be sensitive to the NMH through the interference effects arising from the combination of the fast oscillations driven by $\Delta m^2_{31}$ and  $\Delta m^2_{32}$. Such a measurement however requires an extreme accuracy both in the energy resolution and in the absolute energy scale calibration. 

Another appealing strategy consists in probing the impact of matter effects in both the $\nu_\mu$ survival probability and in the rate of $\nu_\mu \leftrightarrow \nu_e$ appearance at the atmospheric mass scale. As will be detailed in the next subsection, this option requires long oscillation baselines and matter effects that essentially affect the $\nu_e$-component of the propagation eigenstates, making it possible to determine whether the $\nu_1$ and $\nu_2$ states are lighter or heavier than $\nu_3$. The $\nu_e$ appearance channel is the main focus of current (such as NOvA~\cite{pot} and T2K~\cite{t2k}) and next-generation (such as CHIPS~\cite{chips}, LBNE~\cite{lbne}, LBNO~\cite{bib:Loi_lbno} or more recently DUNE~\cite{bib:dune}) accelerator neutrino experiments. In atmospheric experiments, such as ICAL at INO~\cite{bib:ino_MH}, HyperKamiokande~\cite{bib:HK_LOI}, PINGU~\cite{bib:pingu_sensitivity} and ORCA, both channels are important due to the much longer baselines providing stronger matter effects. This strategy has been extensively discussed both for magnetised detectors~\cite{Bernabeu:2001xn,PalomaresRuiz:2004tk,Indumathi:2004kd,Gandhi:2004bj,Petcov:2005rv,Indumathi:2006gr,Gandhi:2007td,Gandhi:2008zs,Samanta:2009qw,Samanta:2010xm,Blennow:2012gj,Ghosh:2012px,Ghosh:2013mga} and for water-Cherenkov detectors~\cite{bib:parametric,Akhmedov:1998xq,Chizhov:1998ug,Bernabeu:2003yp,Petcov:2004bd,Huber:2005ep,Gandhi:2007td,bib:mena}, including more recently the specific case of the Mton-scale underice/sea detectors PINGU and ORCA~\cite{bib:Akhmedhov,bib:tomo_agarwala,bib:MH_sens,bib:pingu_inelasticity,bib:pingu_winter,Blennow:2013vta,Ge:2013zua,Choubey:2013xqa,bib:stat3,bib:razzaque,Capozzi:2015bxa,bib:JPY_AK}.

\begin{figure}[ht]
\centering
 \includegraphics[width=0.45\textwidth]{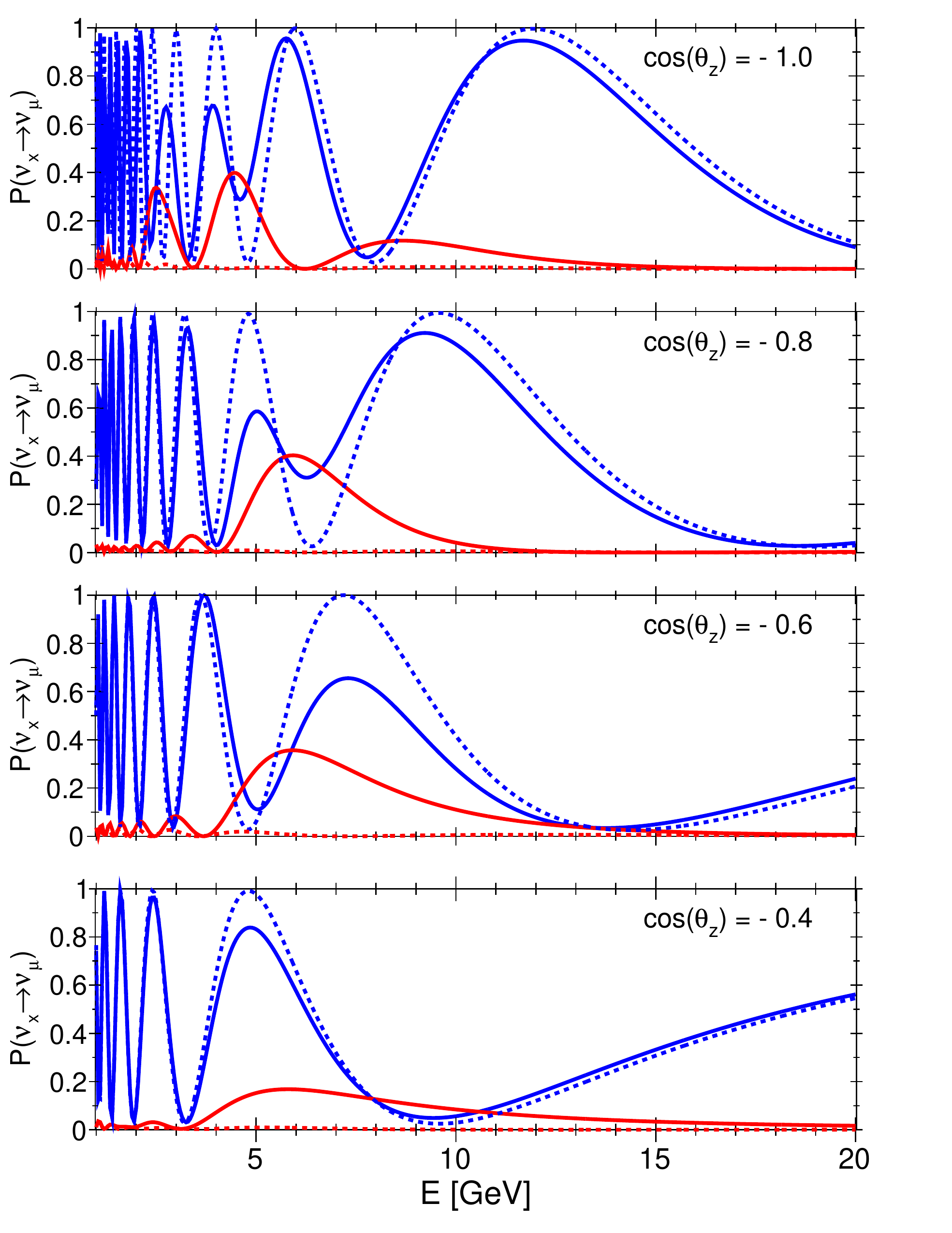}
  \includegraphics[width=0.45\textwidth]{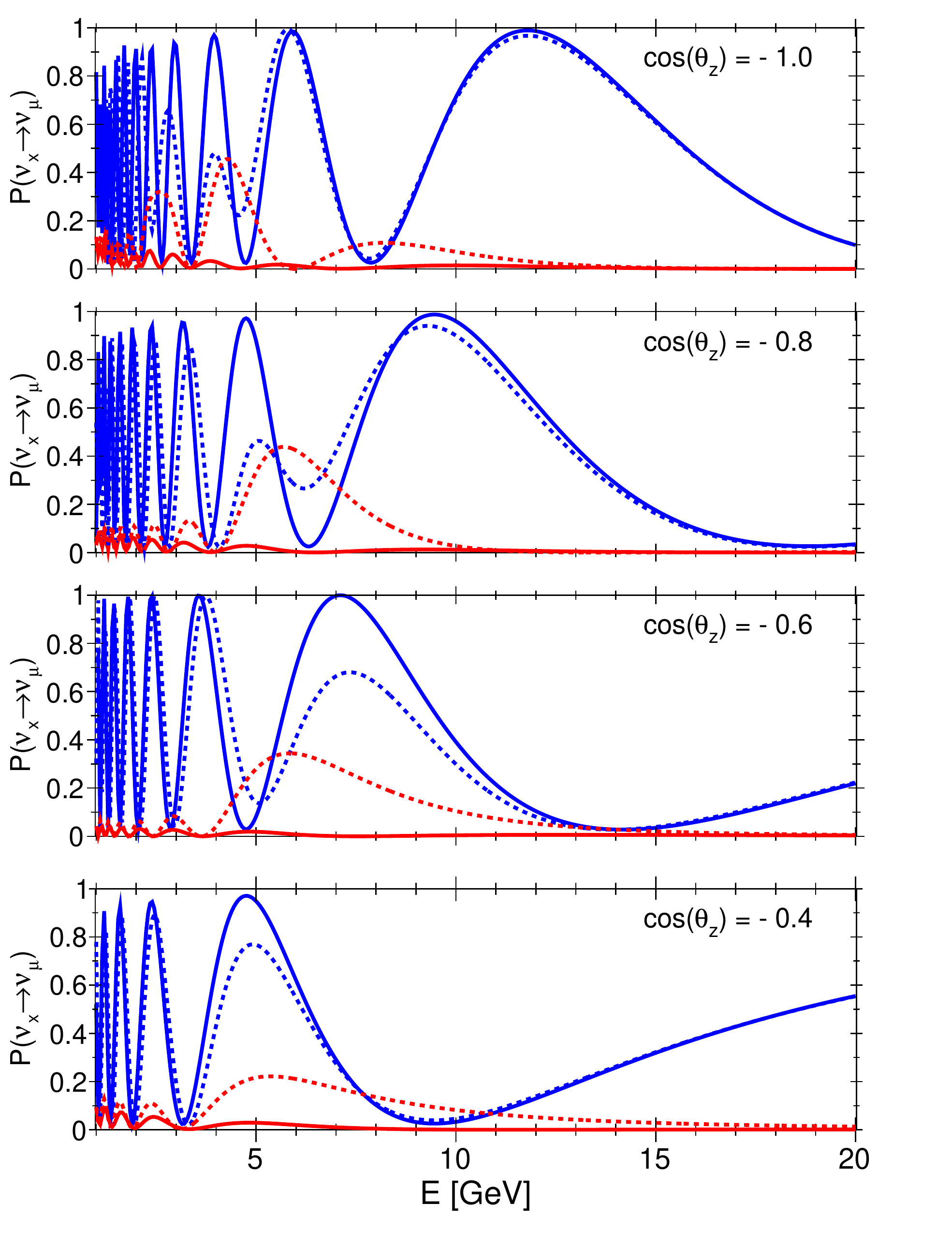}
\caption{
Oscillation probabilities $\nu_\mu \rightarrow \nu_\mu$ (blue lines) and $\nu_e \rightarrow \nu_\mu$ (red lines) as a function of the neutrino
energy for several values of the zenith angle (corresponding to different baselines).  The solid (dashed) lines are for NH (IH). For
neutrinos (left) and for antineutrinos (right).}\label{fig:probas}
\end{figure}

In the $3\nu$ framework, the $\nu_\mu \leftrightarrow \nu_e$  and  $\nu_\mu \leftrightarrow \nu_\mu$ transition probabilities in vacuum can be approximated by the following formulae: 
\begin{equation}
P_{3\nu}(\nu_\mu \rightarrow \nu_e) \approx  \sin^2\theta_{23}\, \sin^22\theta_{13} \, \sin^2 \left(\frac{\Delta m^2_{31} \, L}{4 E_\nu}\right) 
\label{eq:Pnunuevacuum}
\end{equation}
\begin{equation}
P_{3\nu}(\nu_\mu \rightarrow \nu_\mu) \approx 1 - 4\cos^2\theta_{13}\, \sin^2\theta_{23}\, (1-\cos^2\theta_{13}\, \sin^2\theta_{23})\, \sin^2\left(\frac{\Delta m^2_{31} \, L}{4 E_\nu}\right)
\label{eq:Pnumuvacuum}
\end{equation}

where $E_\nu$ is the neutrino energy and $L$ stands for the oscillation baseline. These relations establish the direct link between the transition probabilities and the value of $\theta_{13}$; they also show that the transitions in vacuum are actually insensitive to the sign of $\Delta m^2_{31}$.

This sign can however be revealed once matter effects come into play along the neutrino propagation path~\cite{Wolf_matter,MS_matter}. Contrarily to the other flavours, the $\nu_e$ component can indeed undergo charged-current (CC) elastic scattering interactions with the electrons in matter and consequently acquire an effective  potential $A=\pm \sqrt{2} G_F N_e$, where $N_e$ is the electron number density of the medium, $G_F$ is the Fermi constant and the $+(-)$ sign is for $\nu_e$ ($\overline{\nu}_e$). In the case of neutrinos propagating  in a medium with constant density, the transition probabilities now read (adapted from~\cite{PhysRevD.73.013006}):

\begin{equation}
P^m_{3\nu}(\nu_\mu\rightarrow\nu_e) \approx \sin^2\theta_{23} \sin^22\theta^{m}_{13} \sin^2\left(\frac{\Delta^{m}m^2L}{4E_\nu}\right)
\end{equation}

\begin{eqnarray}
P^m_{3\nu}(\nu_\mu\rightarrow\nu_\mu) &\approx& 1 - \sin^22\theta_{23} \cos^2\theta^{m}_{13} \sin^2\left(\frac{(\Delta{m^2_{31}}+\Delta^{m}m^2)L}{8E_\nu}+\frac{AL}{4}\right)\nonumber\\
&-& \sin^22\theta_{23} \sin^2\theta^{m}_{13} \sin^2\left(\frac{(\Delta{m^2_{31}}-\Delta^{m}m^2)L}{8E_\nu}+\frac{AL}{4}\right)\\
&-& \sin^4\theta_{23} \sin^22\theta^{m}_{13} \sin^2\left(\frac{\Delta^{m}m^2L}{4E_\nu}\right)\nonumber
\end{eqnarray}
as a function of the effective neutrino mixing parameters in matter:  
\begin{eqnarray} 
\sin^22\theta^m_{13} & \equiv &\sin^22\theta_{13} \left(\frac{\Delta m^2_{31}}{\Delta^m m^2}\right)^2 \\ 
\Delta^m m^2 & \equiv & \sqrt{(\Delta m^2_{31}\, \cos2\theta_{13} -2\,E_\nu\,A)^2+ (\Delta m^2_{31} \sin2\theta_{13})^2} \, , \label{eq:deltaeff}   
\end{eqnarray}
where $A$ is positive for neutrinos and negative for antineutrinos. Both the amplitude and the phase of the oscillations can therefore be affected by matter effects. From \myeref{eq:deltaeff}, the resonance condition is met when the effective mixing is maximal, i.e $\Delta^m m^2$ is minimal. This happens for the case of the NH (IH) in the neutrino (antineutrino) channel at the energy:

\begin{equation}
E_{\rm res} \equiv \frac{\Delta m^2_{31} \, \cos2\theta_{13}}{2 \, \sqrt{2} \, G_F \, N_e} \simeq 7 \, \textrm{GeV} \, \left(\frac{4.5 \, \textrm{g/cm}^3}{\rho}\right) \, \left(\frac{\Delta m^2_{31}}  {2.4 \times 10^{-3} \, \textrm{eV}^2}\right) \,\cos2\theta_{13}\, . 
\end{equation}
where $\rho$ is the matter density of the medium. For neutrinos passing through the Earth's mantle (core) the resonance will appear around 7~GeV (3~GeV), which explains why atmospheric neutrinos are an appropriate probe for these effects, in association with the large baselines available.

As an illustration, oscillation curves $P(\nu_x \rightarrow \nu_\mu)$ (x=e,$\mu$) obtained with the ORCA software tools (using the PREM model~\cite{bib:prem} of the Earth density layers) are shown in \myfref{fig:probas} for various zenith angles $\theta$ (i.e various baselines) as a function of the neutrino energy, both for neutrinos and antineutrinos. In each case, both NMH hypotheses are represented. The strongest impact of the NMH to the oscillation probabilities is in the resonance region $E_\nu \sim (4 - 8)$\,GeV.
In the region $\cos\theta \lesssim -0.85$  and $E_\nu< 7$\,GeV, the effect of the resonant enhancement of the oscillations~\cite{Akhmedov:1988kd,Ermilova:1988pw,Krastev:1989ix,Liu:1997yb,bib:parametric,Akhmedov:1998ui,Akhmedov:1998xq,Chizhov:1998ug,Chizhov:1999az,Chizhov:1999he,Bernabeu:2001xn,Akhmedov:2005yj,Akhmedov:2006hb,bib:tomo_agarwala} for the neutrino trajectories crossing the Earth's core can also be seen. Above $\sim$15\,GeV, the $\nu_e \rightarrow \nu_\mu$ transition probability  becomes very small and differences from distinct NMHs tend to disappear as well\footnote{This justifies the approximation of a 2-flavour $\nu_\mu \rightarrow \nu_\tau$ oscillation scheme adopted by high-energy atmospheric neutrino experiments so far~\cite{bib:antares_osc,bib:IC_osc}.}. 

\begin{figure}[ht]
  \centering
  \includegraphics[width=0.49\linewidth]{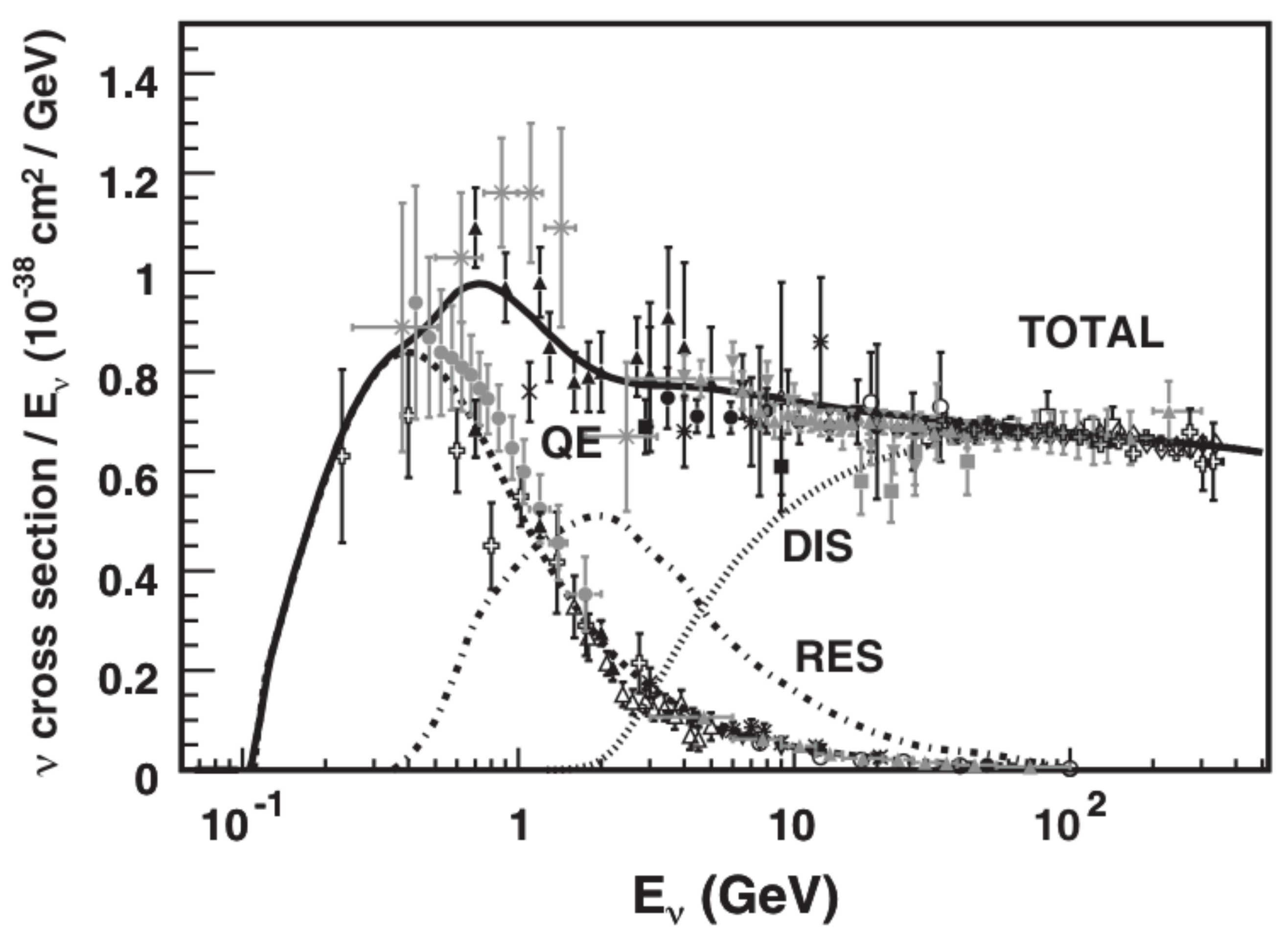}
  \includegraphics[width=0.49\linewidth]{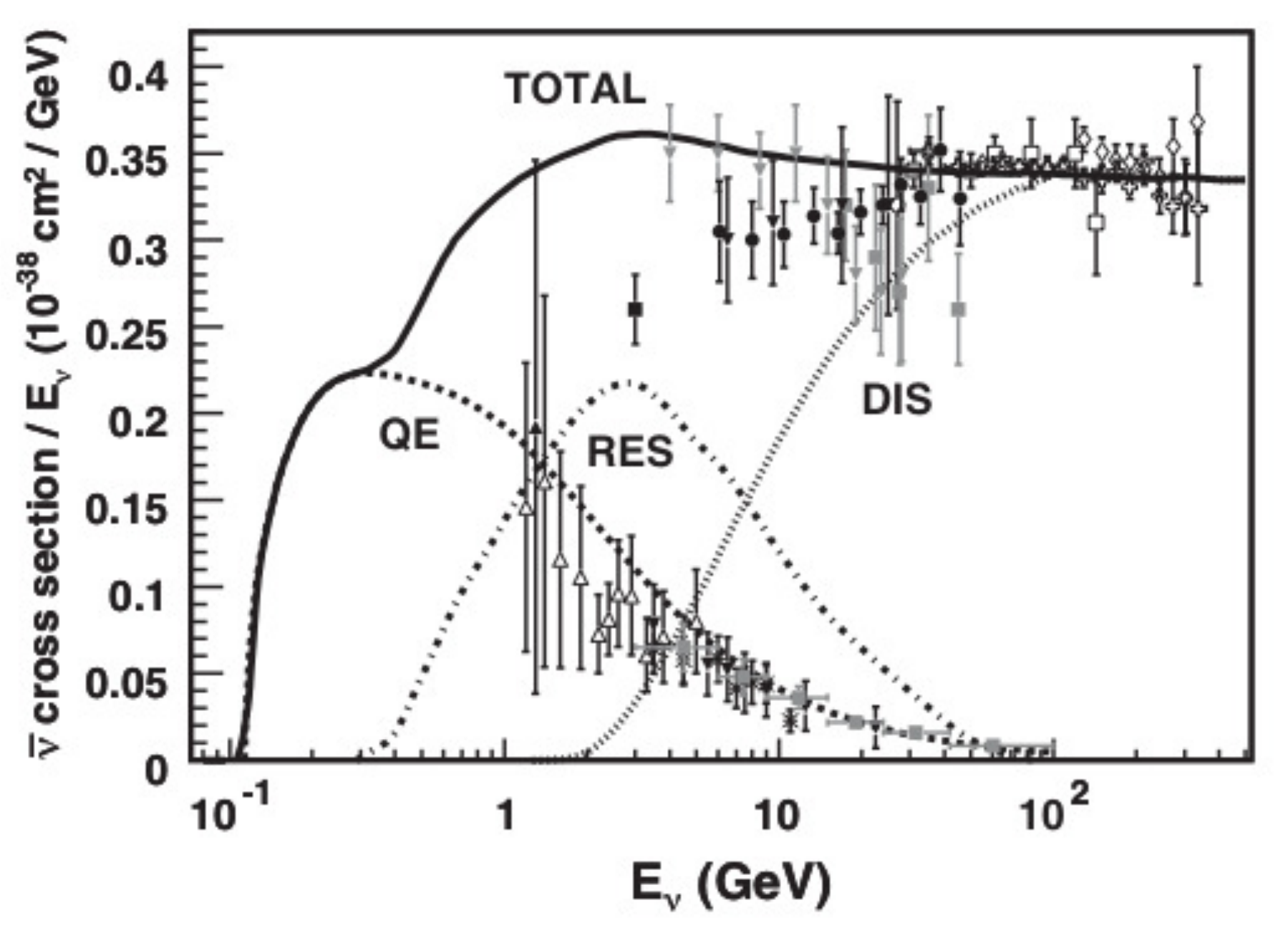}
  \caption{Total neutrino (left) and antineutrino (right)  CC cross sections per nucleon (for an isoscalar target) divided by neutrino energy  and plotted as a function of the energy. Also shown are the various contributing processes: quasi-elastic scattering (dashed), resonance production (dot-dashed), and deep inelastic scattering (dotted). Taken from~\cite{formaggio}.}\label{fig:xsection}
\end{figure}

\begin{figure}[h] \centerline{
\includegraphics[width=0.49\linewidth]{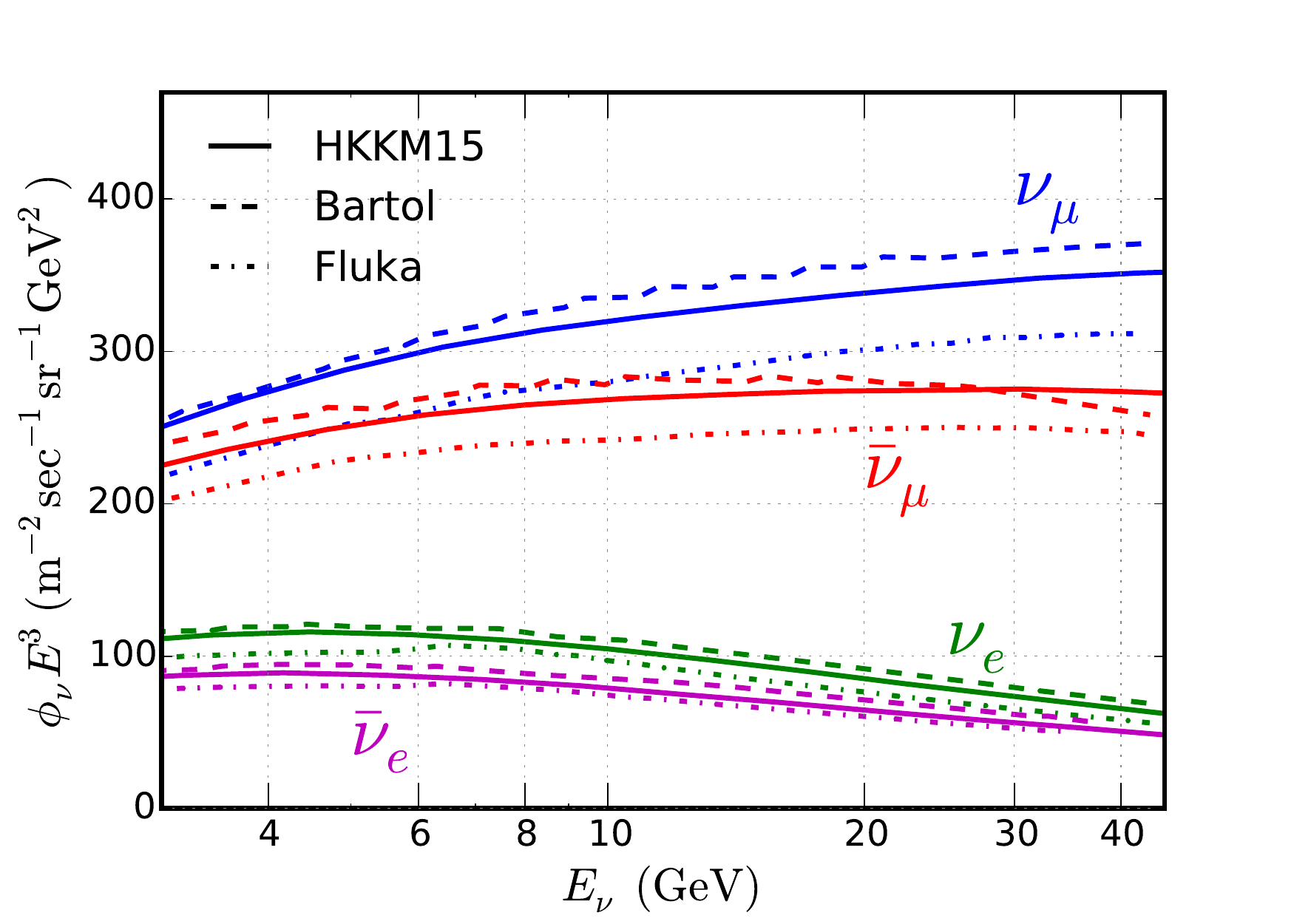}%
\includegraphics[width=0.49\linewidth]{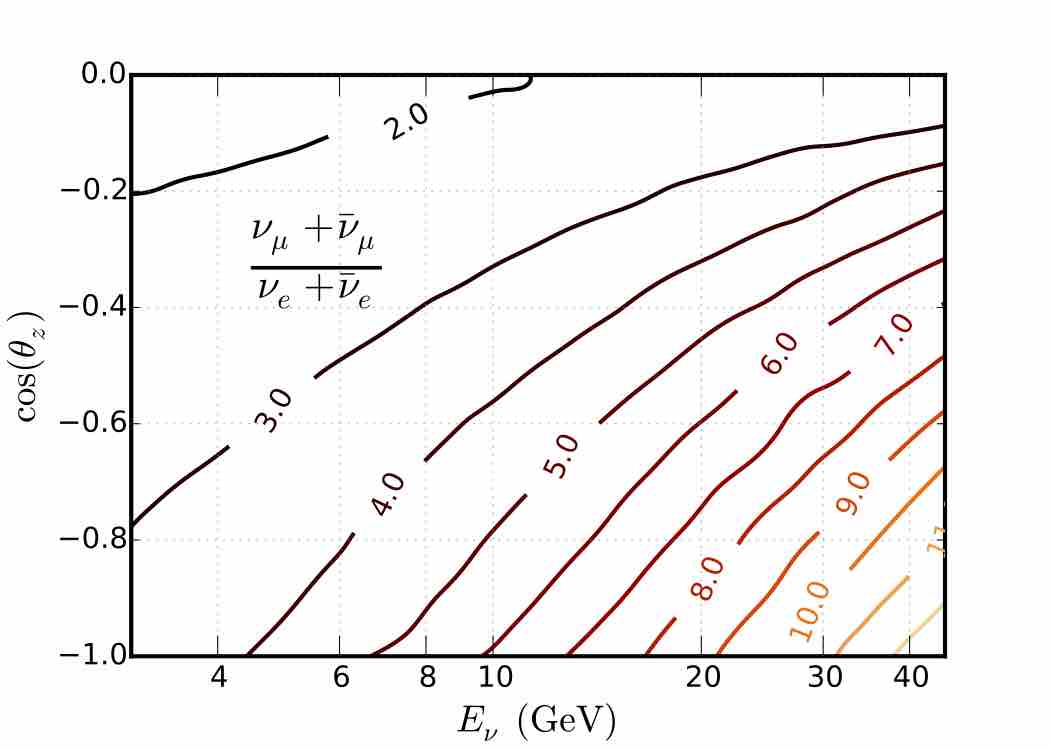}%
}
\caption{
Atmospheric neutrino flux in its different flavour components: absolute values (left) and ratios (right). Taken from~\cite{bib:JPY_AK}. 
}\label{fig:other-nflx}
\end{figure}

\begin{figure}[h] \centerline{
\includegraphics[width=0.49\linewidth]{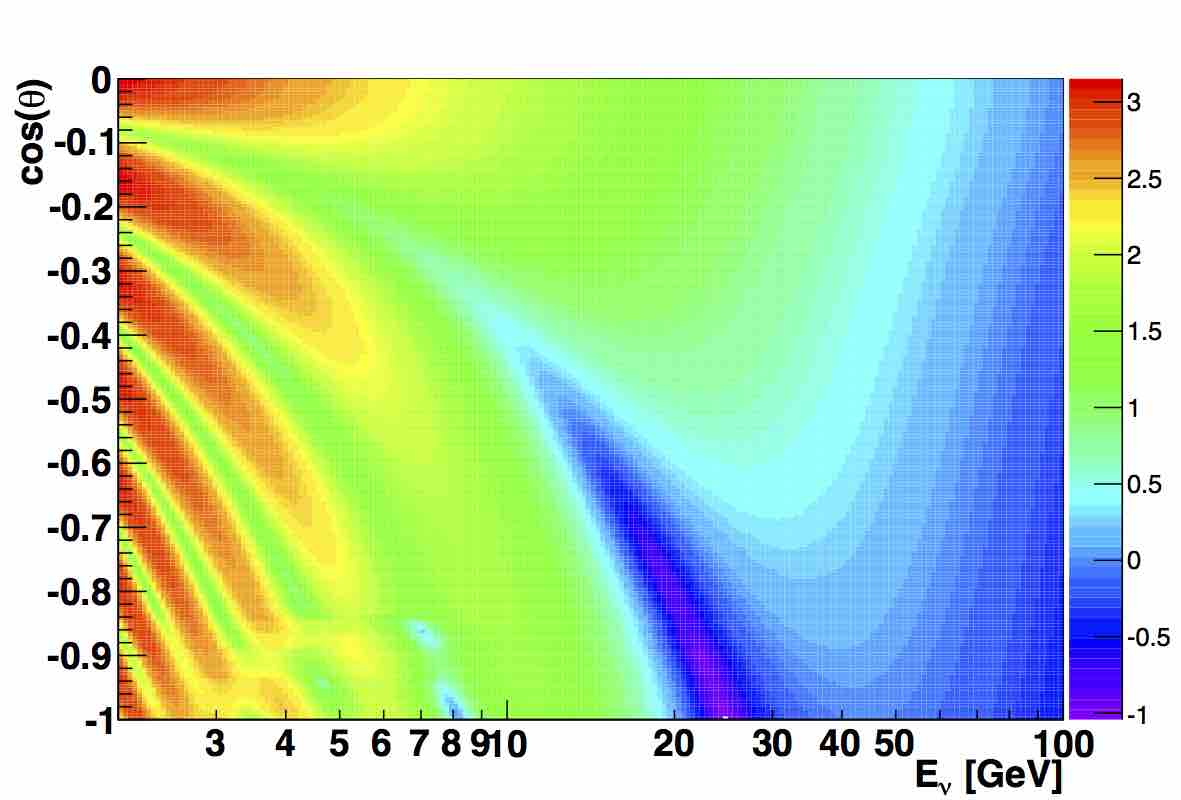}%
\includegraphics[width=0.49\linewidth]{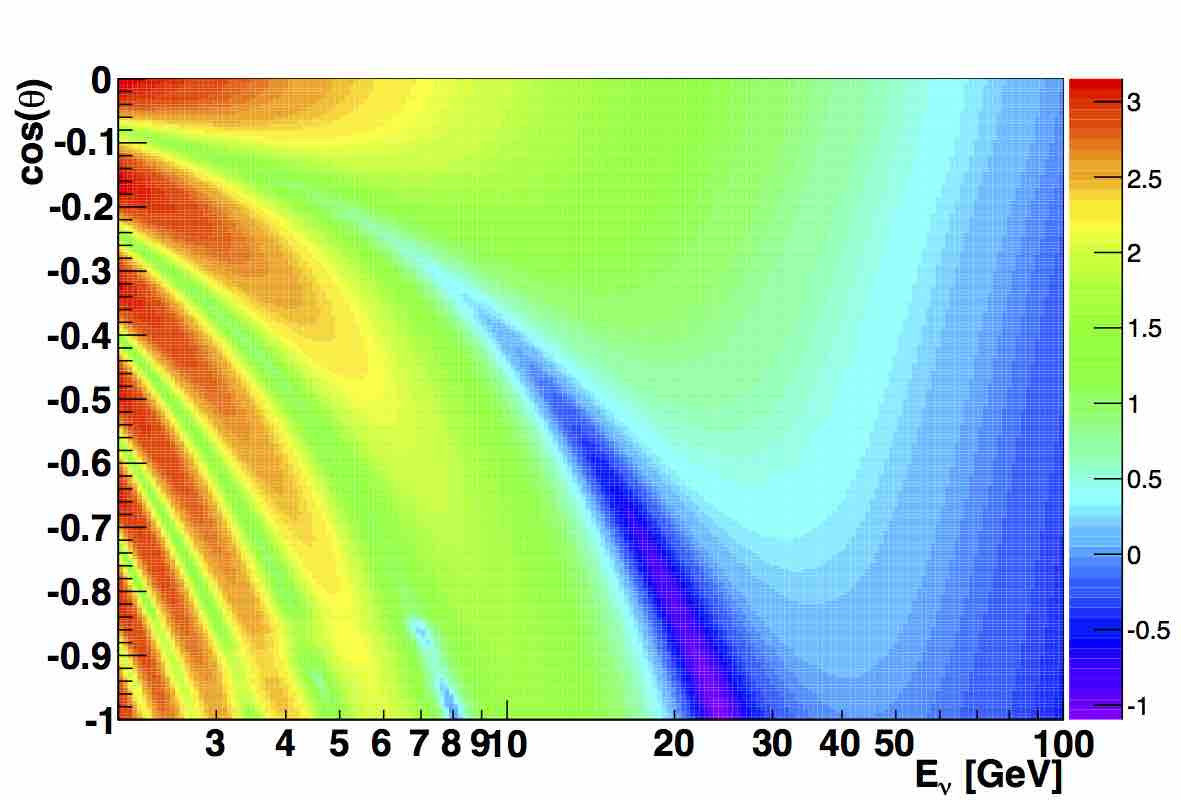}%
}
\caption{
Neutrino ``oscillograms'': $\nu_\mu + \overline{\nu}_\mu$ event rate (in units of $\rm GeV^{-1}\cdot \rm y^{-1}\cdot \rm sr^{-1}$ in log scale) as a function of the neutrino energy and 
cosine of the zenith angle, for a 1~Mton target volume. The left (right) plot shows the distribution for the normal (inverted) mass hierarchy. 
}\label{fig:evtrates}
\end{figure}

\myfref{fig:probas} shows that to first order, the effect for neutrinos in the NH scheme is the same as for antineutrinos in the IH scheme. Nevertheless, and even in the case of non-magnetised detectors (such as ORCA) which do not distinguish $\nu$'s and $\overline{\nu}$'s event-by-event, a net asymmetry in the combined ($\nu + \overline{\nu}$) event rates between NH and IH for a given flavour can be observed. This mainly comes from the fact that in the GeV energy range relevant for atmospheric neutrinos, the CC cross section is different (by about a factor of 2) for neutrinos and antineutrinos, as can be seen from \myfref{fig:xsection}. The relative contribution of $\nu_e$ and $\nu_\mu$ in the steeply falling atmospheric neutrino spectrum, as shown in \myfref{fig:other-nflx}, also affects the number of events of each flavour that can be expected at the detector level. 

Convoluting the oscillation probabilities with the atmospheric neutrino fluxes and the neutrino-nucleon cross section, one can construct bidimensional plots of event rates as a function of the neutrino energy $E_\nu$ and cosine of the zenith angle $\theta$. Such an ``oscillogram'' is represented in \myfref{fig:evtrates} for $\nu_\mu + \overline{\nu}_\mu$, for both NMH hypotheses. Integrating over energies above 4\,GeV, one typically expects of the order of 4650 $\nu_\mu$-induced events and 2850 $\nu_e$-induced events per year in a 1\,Mton detector. The phase space region where the differences between NH and IH are most visible clearly depends on the three ingredients mentioned here above; but other factors also come into play, related both to intrinsic effects (such as the physics of the neutrino interaction) and to the detector performance (such as energy and angular resolutions), that will blur the oscillogram patterns and partly wash out the asymmetry effect.

 \begin{figure}[ht!]
\centering
 \includegraphics[width=0.8\textwidth]{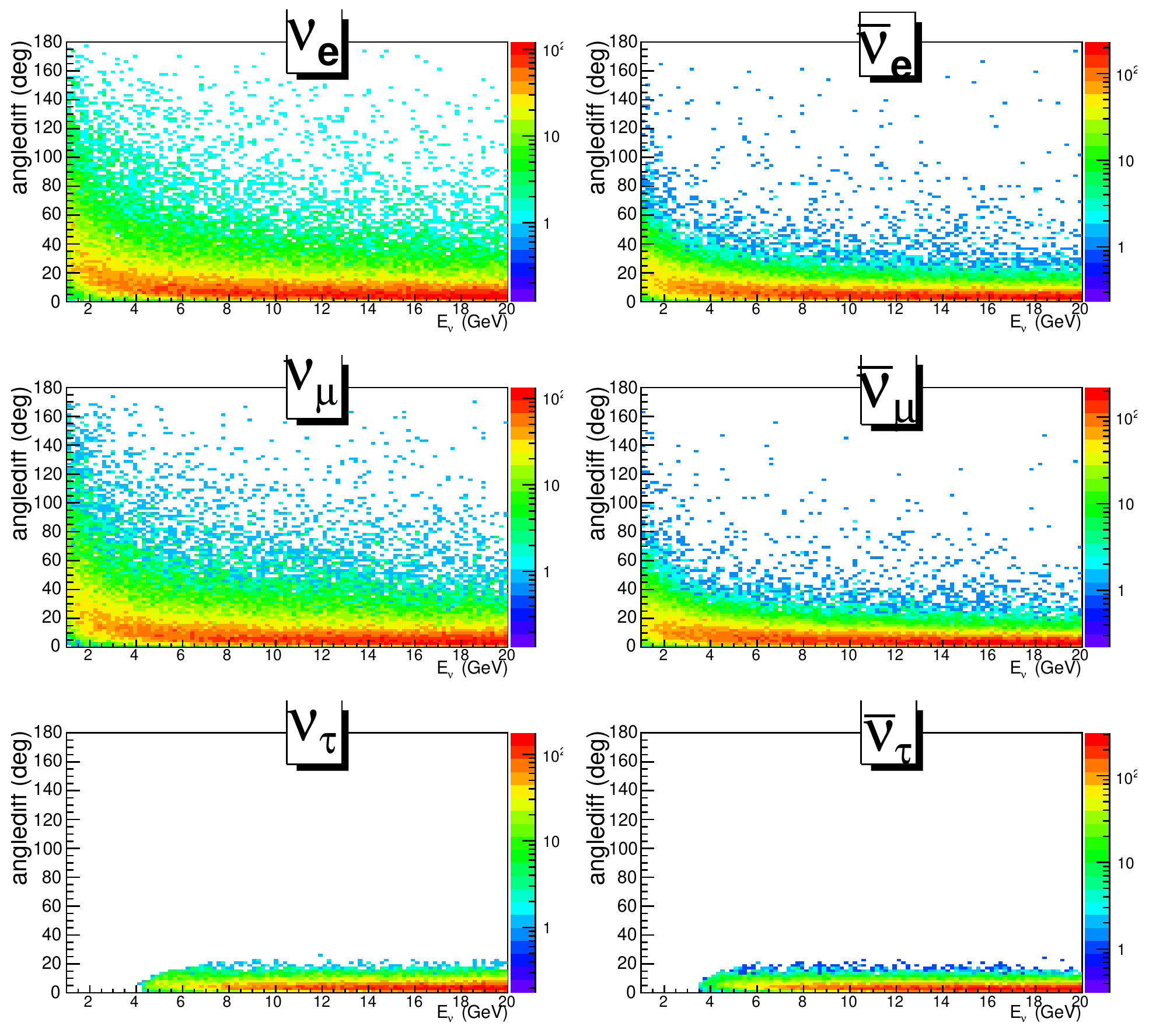}
\caption{Distribution of the angular difference between the out-coming lepton and the parent neutrino as a function of the neutrino energy, for neutrinos (left column) and antineutrinos (right column) of each flavour. [plot obtained with ORCA tools based on GENIE]}
\label{fig:kinspread}
 \centering
  \includegraphics[width=0.45\linewidth]{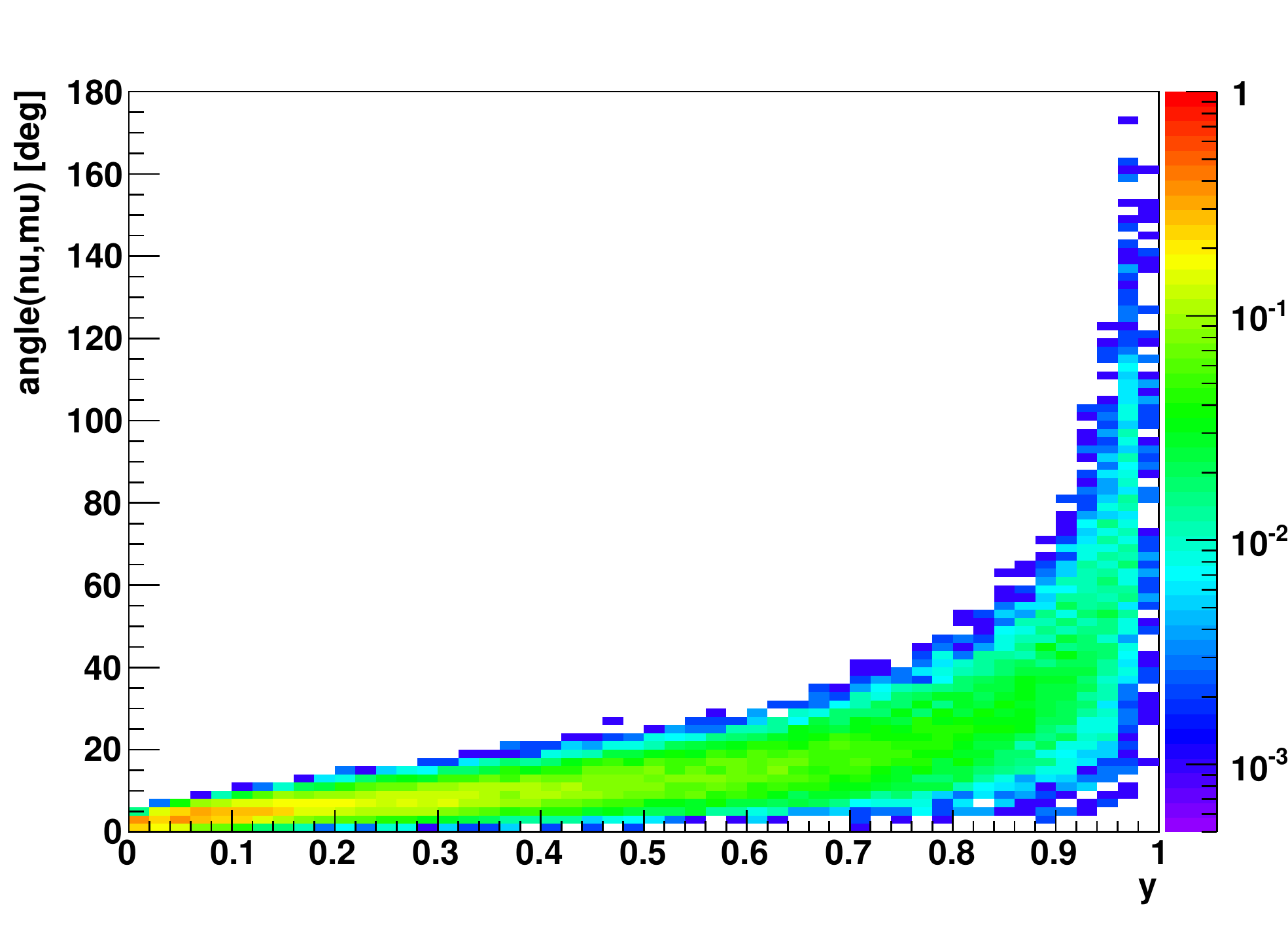}
  \includegraphics[width=0.45\linewidth]{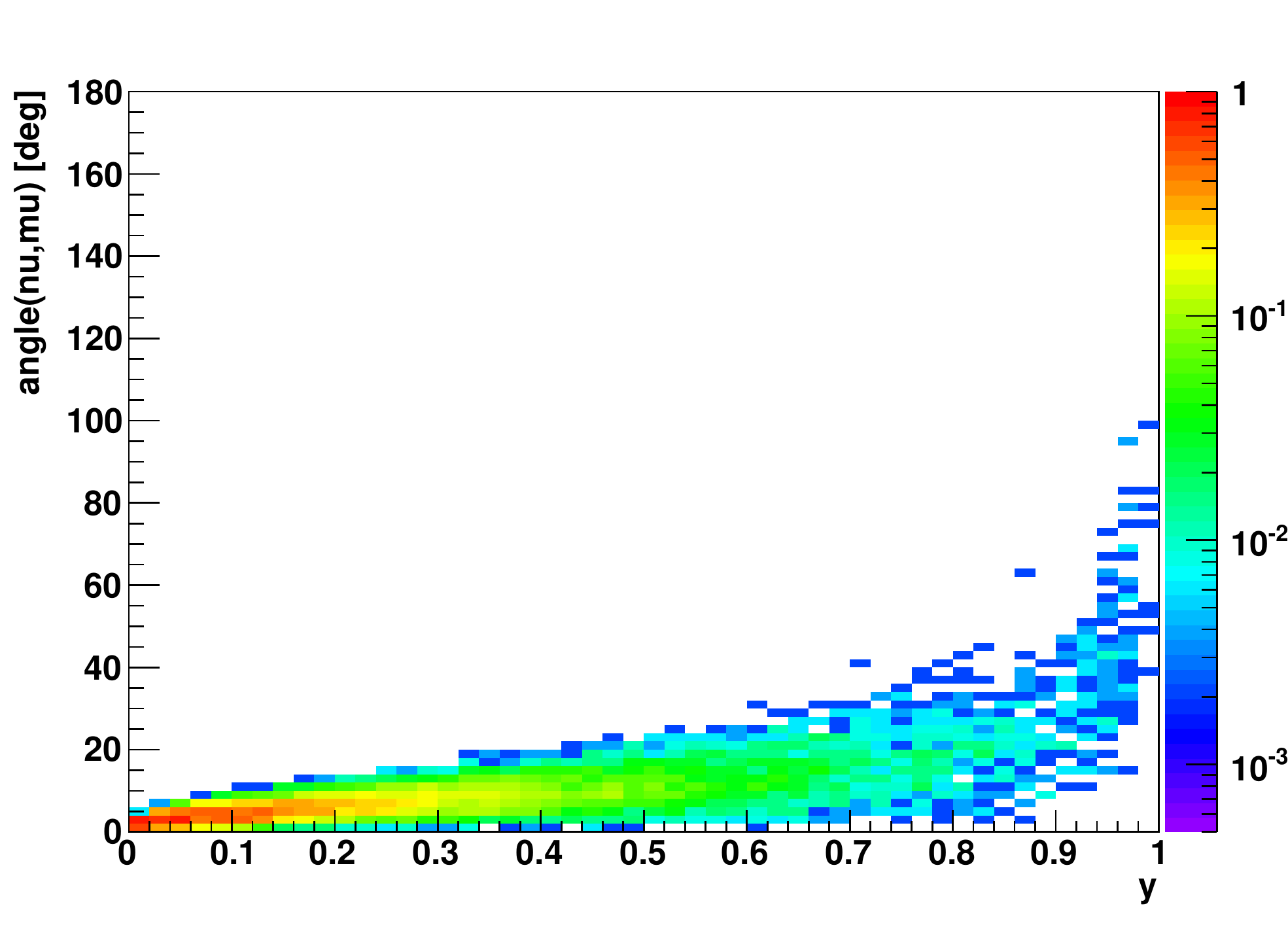}
  \caption{Scattering angle $\phi_{\nu,\mu}$ as a function of Bjorken $y$ for neutrinos (left) and antineutrinos (right) 
           for (anti)neutrino energies of $10\,\mathrm{GeV}$. The distributions are normalised to 1. [plot obtained with ORCA tools based on GENIE]}
  \label{fig:kinBjorken}
\end{figure}

An intrinsic uncertainty in the neutrino energy and direction arises from the kinematics of the neutrino interaction. At the relevant energies, the out-coming lepton can no longer be considered as collinear with its parent neutrino, as can be seen from \myfref{fig:kinspread}\footnote{Note that the angle and energy resolutions can be improved by combining information from the leptonic and hadronic parts of the interaction to better reconstruct the kinematics.}. This smearing can conveniently be expressed in terms of the Bjorken inelasticity parameter 
\begin{eqnarray}
y = \frac{E_\nu-E_l}{E_\nu}
\end{eqnarray}
where $l$ stands for a charged lepton and which represents the fraction of energy transferred to the associated hadronic shower. Since the cross section for neutrino and antineutrino behave differently as a function of $y$, measuring the inelasticity of the neutrino interaction could provide some statistical separation between the $\nu$ and $\overline{\nu}$ channels and therefore enhance the sensitivity to the NMH~\cite{bib:pingu_inelasticity}. This effect could be best exploited in the muon channel, where the lepton track and the hadronic shower can in principle be more easily identified than in the other channels; the difference in the muon angular spread for 10\,GeV $\nu_\mu$ and $\overline{\nu}_\mu$ is illustrated in \myfref{fig:kinBjorken}.
Preliminary studies performed for ORCA using flavour identification tools are presented in \mysref{sec:particle_id} and could be the starting point for a statistical separation between $\nu_\mu$'s and $\overline{\nu}_\mu$'s, providing additional enhancement of the sensitivity to NMH in the track and in the shower channel.

The kinematic smearing described here is only one among other sources of systematics directly related to the physical processes at play; fluctuations in the development of the particle cascades, and in the production and propagation of the associated Cherenkov light, must also be taken into account. These effects are discussed in more detail in \cite{ShowerFluct}.

Uncertainties in the neutrino oscillation parameters can also degrade the sensitivity to the NMH.
These uncertainties are taken into account when evaluating the ORCA sensitivity to the NMH (see \mysref{globalfit}). Other sources of systematics such as the uncertainties on the atmospheric spectra, the uncertainties of the Earth matter density profile, or the unknown $\delta_{CP}$ phase are further discussed in \mysref{syst}. Detector-related effects, and in particular the energy and angular resolution, are presented along with the description of the event selection and reconstruction performances in \mysref{evtselreco}.

\begin{figure}[h!]
\centering
 \includegraphics[width=0.49\textwidth]{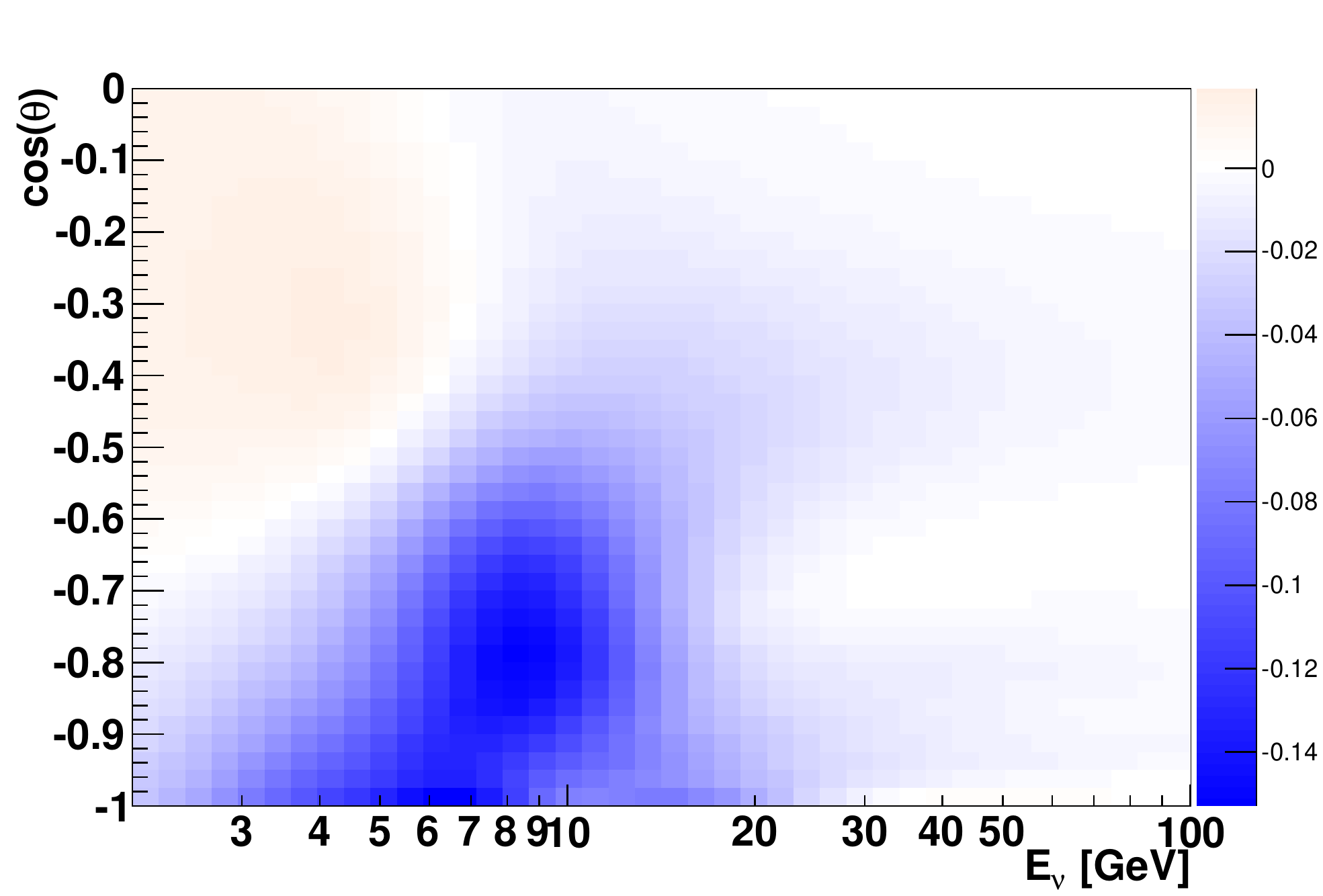}
 \includegraphics[width=0.49\textwidth]{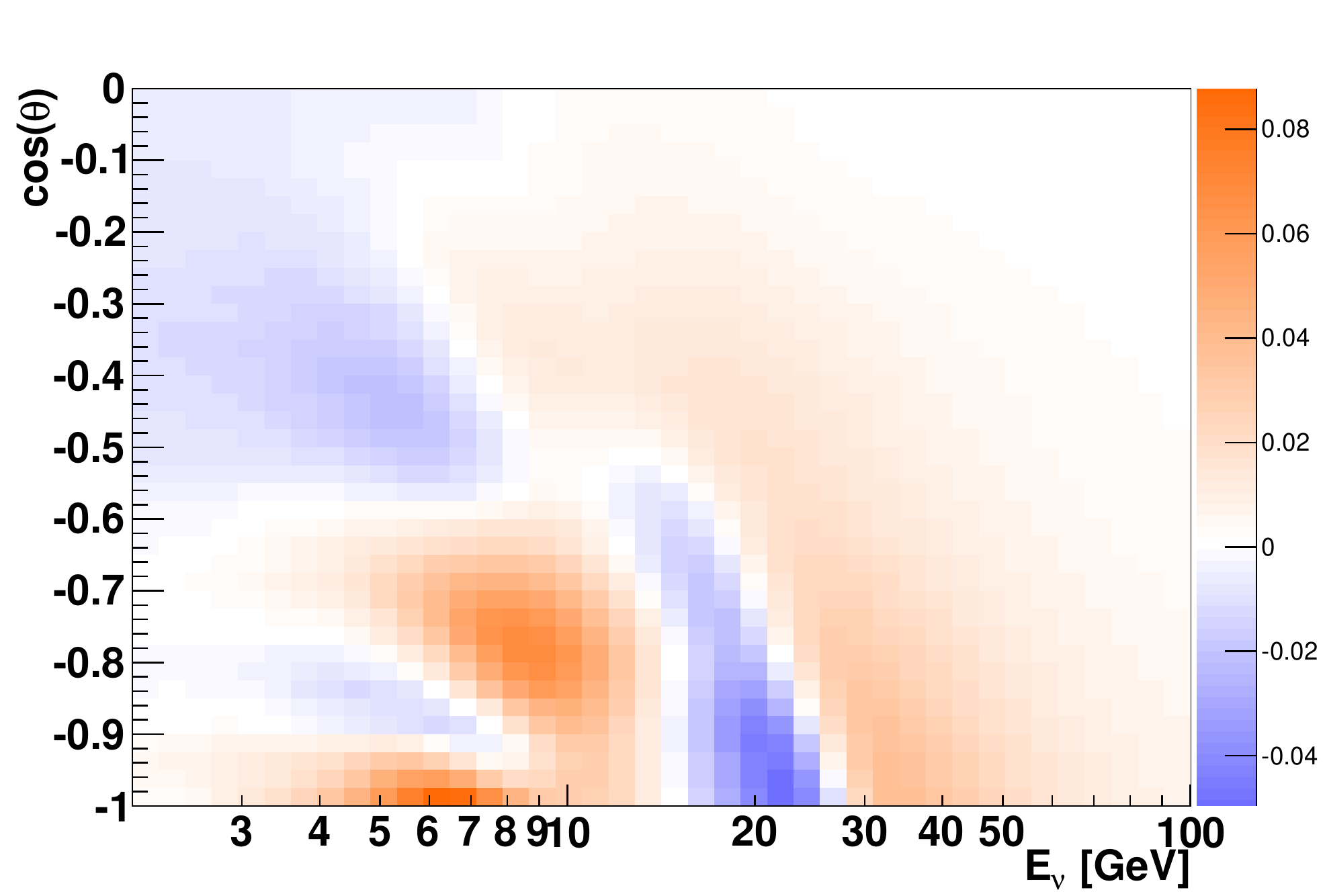}
\caption{
Asymmetry (as defined by \myeref{Aprime}) between the number of $\nu+\bar{\nu}$ CC interactions expected in case of NH and IH, expressed as a function of the energy and the
cosine of the zenith angle. The right (left) plot applies to muon (electron) neutrinos.  A smearing of 25\% is applied on the energy. On the angle, a smearing $\sigma_\theta = \sqrt{\frac{m_p}{E_\nu}}$ is applied, where $m_p$ denotes the nucleon mass and $E_{nu}$ the neutrino energy in GeV.}
\label{fig:asymmetry}
\end{figure}

In order to identify, for each flavour, the phase space region where the effects are larger and therefore the discrimination more powerful, asymmetry variables can be defined such as
\begin{equation}
\label{A}
\mathcal{A} = \frac{N_{IH}-N_{NH}}{\sqrt{N_{NH}}}
\end{equation}
which was used in~\cite{bib:Akhmedhov}, or
\begin{equation}
\label{Aprime}
\mathcal{A'}= \frac{N_{IH}-N_{NH}}{N_{NH}}
\end{equation}
where $N_{NH}$ and $N_{IH}$ are the number of expected events at a given angle and energy for NH and IH respectively. $\mathcal{A'}$ essentially reflects the asymmetry of oscillation probabilities and does not depend on the effective mass of the detector, while $\mathcal{A}$ is useful to provide an estimation of the significance of the hierarchy measurement by summing over all oscillogram entries, as proposed by~\cite{bib:Akhmedhov}. This approach should however be taken with care as it typically overestimates the discrimination power of the experiment. Alternative approaches discussed in~\cite{bib:stat1,bib:stat2,bib:stat3,bib:stat4}, and providing a more rigorous statistical treatment, are followed in \mysref{sensitivity}.

An example of asymmetry plots (following the definition of \myeref{Aprime}) for $\nu_\mu$ and $\nu_e$ obtained with ORCA software tools and a smearing on energy and angle is shown in \myfref{fig:asymmetry}. It is clear that the region where the asymmetry is more evident is above 5\,GeV.
The plots also indicate that comparable levels of asymmetry are reached in both $\nu_\mu$ and $\nu_e$ charged-current interaction channels. Most first-stage studies have concentrated on the $\nu_\mu$ channel (and on the detection of the associated muon) to determine the sensitivity to NMH, anticipating on the larger statistics in the muon channel and the worse angular resolution of deep sea (/ice) Cherenkov detectors for shower-like events (as produced by $\nu_e$)~\cite{bib:mena,bib:Akhmedhov,bib:pingu_sensitivity,bib:pingu_winter,bib:MH_sens}.

In the course of the study it has however been pointed out that this approach may have been too conservative and that the shower channel, and in particular the $\nu_e$-induced events, could also provide a significant contribution to the total sensitivity to NMH\footnote{Experiments like Super-Kamiokande and the proposed Hyper-Kamiokande have indeed mainly focused on the electron neutrino channel, because of the good resolutions they can achieve for this topology in the few GeV energy range~\cite{bib:SK_atm,bib:HK_LOI}. }. To first order, the atmospheric flux of $\nu_{e}$ of energy $E_\nu$ which reach the detector after crossing the Earth along a given trajectory, $\Phi_{\nu_{e}}(E_\nu,\theta)$ is given by~\cite{Akhmedov:1998xq,Chizhov:1998ug}: 
\begin{eqnarray}
 \Phi_{\nu_e}(E_\nu, \theta) &= & \Phi^{0}_{\nu_e} \, P(\nu_e \rightarrow \nu_e) + \Phi^{0}_{\nu_\mu} \, P(\nu_\mu \rightarrow \nu_e)\\
 & \simeq & \Phi^{0}_{\nu_e} \, \left[1 + ( \sin^2\theta_{23} \, r - 1) \, P_{2\nu}\right] 
\label{eq:Phie} 
\end{eqnarray}
where $\Phi^{0}_{\nu_{e(\mu)}} = \Phi^{0}_{\nu_{e(\mu)}}(E_\nu, \theta)$ is the $\nu_{e(\mu)}$ flux at the production point in the atmosphere, \\
$P_{2\nu}=\sin^22\theta^m_{13} \, \sin^2 \left(\frac{\Delta^m m^2_{31} \, L}{4 E_\nu}\right)$ is the $\nu_e$ disappearance probability in a $2\nu$ scheme, and
\begin{equation} 
r \equiv r(E_\nu, \theta) \equiv \frac{\Phi^{0}_{\nu_{\mu}}(E_\nu,
  \theta)}{\Phi^{0}_{\nu_{e}}(E_\nu, \theta)} \, . 
\label{eq:r}
\end{equation}

As can be seen from \myfref{fig:other-nflx}, the ratio $r$ is close to 2 around 2\,GeV and below, which tends to suppress the oscillations. This is referred to as the "screening effect" in~\cite{bib:Akhmedhov}. However, in the energy range of a few GeV the ratio increases, which could on the contrary enhance the asymmetry as stated in~\cite{bib:tomo_agarwala}. The final asymmetry level in the $\nu_e$ channel will also depend on the value of the mixing angle $\theta_{23}$; it could in particular be further enhanced if $\theta_{23}$ is found to be in the second octant (i.e. $\theta_{23}>45^\circ$). The status of electron neutrino studies within ORCA is summarised in \mysref{electron}.

\FloatBarrier

\subsection{Simulations}
\label{simu}
\label{sec:simulations}
\subsubsection{Benchmark detector}
\label{simu:det}
The detector geometry used in the Monte Carlo simulations for KM3NeT/ORCA follows the design as described in \mysref{sec:deepsea_orca}. The simulated ORCA detector corresponds to one building block of 115 string detection units (DUs) with 18 digital optical modules (DOMs) each. The DOMs are made of  glass spheres that are designed to resist the hydrostatic pressure of the deep sea environment, each one containing 31 photomultiplier tubes  (PMTs) of 3 inch diameter and the related electronics. 

\begin{figure}[!hb]
\centering
 \includegraphics[width=10cm]{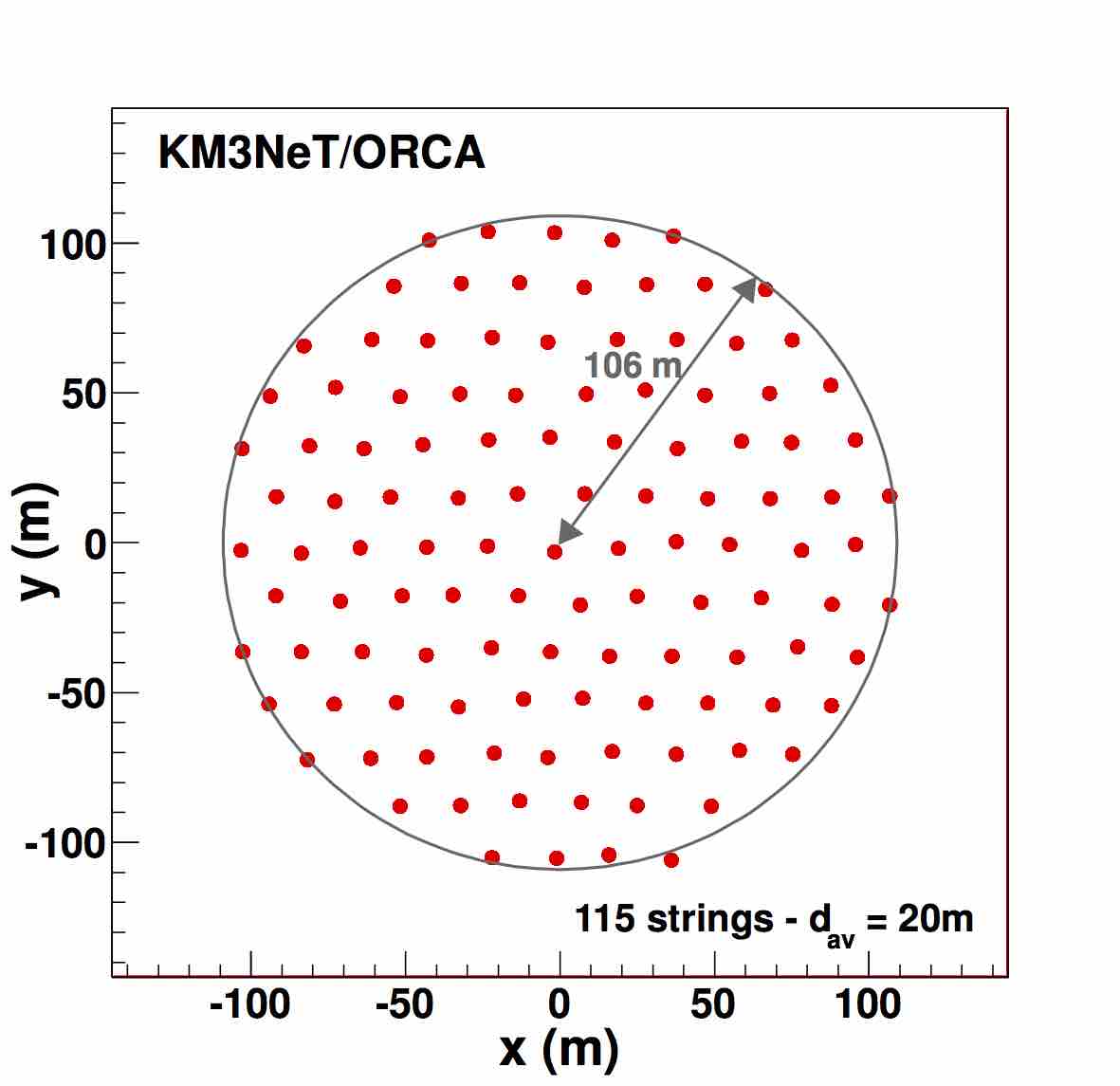}
\caption{Footprint of the ORCA  benchmark detector (top view), with 115 strings (20\,m spacing) with 18 OMs each (6\,m spacing).
The instrumented volume is $3.6\times10^{6}\,\text{m}^3$ (cylinder: R=106\,m, z=102\,m)}
\label{fig:footprint}
\end{figure}

The ORCA simulated detector characteristics rely on reasonable assumptions based on the expertise acquired in the KM3NeT collaboration. To reduce the energy threshold, both the vertical and the horizontal spacing must be reduced with respect to the high-energy KM3NeT design (KM3NeT/ARCA). Vertically this can essentially be done at will whereas horizontally there are limitations due to the deformation of the lines by the sea currents and the unfurling procedure of the strings. For a line
with 6\,m vertical spacing and 18 modules the maximum deviation at the top of the line is about 10\,m
 (corresponding to a sea current of about 30\,cm/s). In addition,
the accuracy with which a string can be placed on the sea bottom is from ANTARES experience a few meters. A 
20\,m distance is therefore assumed to be feasible. 

The collaboration therefore decided to start the simulation study with a detector consisting of 2070 optical modules distributed on 115 DUs placed at a distance of about 20\,m from each other (accounting for the positioning uncertainty at deployment), in a circular pattern of radius 106\,m (\myfref{fig:footprint}). The detector is located at the KM3NeT-France site (2450\,m depth). The DUs host 18 DOMs with 6\,m vertical spacing. In this geometry, the first floor is 50\,m distant from the seabed and the detector has a total instrumented volume of about  $3.6\times10^{6}\,\text{m}^3$
(equivalent to $\sim$ 3.7\,Mt for sea water). Larger vertical inter-DOM spacings have been investigated as well using a masking technique described in~\mysref{sec:masking}. The results obtained in terms of detector performances for the NMH discrimination indicate an optimum inter-DOM distance of about 9\,m (see \mysref{globalfit}).

\subsubsection{Event generation and characterisation}
\label{simu:event}
 This section describes the software packages used for the generation of Monte Carlo events.
Additionally, a selection of event observable distributions 
is used to characterise their typical fundamental and detector physics phenomenology.

The employed software packages generate atmospheric muons and atmospheric neutrinos.
Several codes have been developed for the KM3NeT project and older codes, 
that were developed by the ANTARES collaboration, 
have been modified to take into account the KM3NeT DOM characteristics. 
The codes simulate the particle interactions with the medium surrounding the detector, 
light generation and propagation as well as the detector response.
In the simulation chain a volume surrounding the instrumented volume, called "can", 
is defined. The can volume is a cylinder with height and radius exceeding the instrumented 
volume by about 3 absorption lengths for the atmospheric muon background simulation and 
by 40~m for the neutrino generation. Generated particles are propagated inside the
can and Cherenkov light is generated.

Neutrino and antineutrino induced interactions
in sea water
in the energy range from 1 to 100\,GeV have been
generated with a software package based on the widely used
GENIE~\cite{Distefano:2016bcw, genie1, genie2}
neutrino event generator. 
Electron and muon neutrino events are weighted to reproduce the conventional atmospheric neutrino flux 
following the Bartol model \cite{bib:Bartol}.

All particles emerging from a neutrino interaction vertex are propagated with
the GEANT4 based software package KM3SIM \cite{hours} 
that has been developed by the KM3NeT collaboration.
It generates Cherenkov light from primary and secondary particles in showers
and simulates hits taking into account the light absorption and scattering in water
as well as the DOM and PMT characteristics. 

The background due to down-going atmospheric muons is generated with the 
MUPAGE \cite{mupage_becher,mupage-2008} program. MUPAGE provides a parameterised description
of the underwater flux of atmospheric muons including also multi-muon events. 
The parameterised muon flux was obtained starting from full simulations with 
HEMAS \cite{bib:HEMAS} and cosmic ray data.
These muons are tracked inside the can with the code KM3 which generates and propagates 
the light produced by the muons and their secondary particles, taking into account
the optical properties of the water. For the photon propagation, the code uses tables
containing parameterisations obtained from a full GEANT3 simulation.
The code simulates the PMT hit probabilities and the response of the PMTs.
The PMT photocathode area, quantum efficiency and angular acceptance, as well 
as the transmission of light in the optical module glass sphere and in the 
optical gel are taken into account.
 
In order to reproduce the randomly distributed background PMT hits due to the 
Cherenkov light from $\beta$-decays of $^{40}$K, single photoelectron hits can be added to 
the hits induced by charged particles inside a chosen time window. 
Also the hits in coincidence due to $^{40}$K  between two PMTs inside the 
same DOM are taken into account. 

First measurements of the optical background rate indicate a single PMT noise rate of
8\,kHz and twofold coincidence noise of about 340\,Hz, for details see
\cite{km3net-ppmdom-2014}.
For the simulation results described below a conservative optical background light estimation has been used.
An uncorrelated hit rate of 10\,kHz per 
PMT and time-correlated noise on each DOM 
(500\,Hz twofold, 50\,Hz threefold, 5\,Hz fourfold and 0.5\,Hz fivefold)
was added. The simulated time-correlated noise rates due to $^{40}$K decays have been
verified with a complete simulation based on GEANT4.\\

\begin{figure}[h]
  \centering
 \begin{overpic}[width=0.49\linewidth]{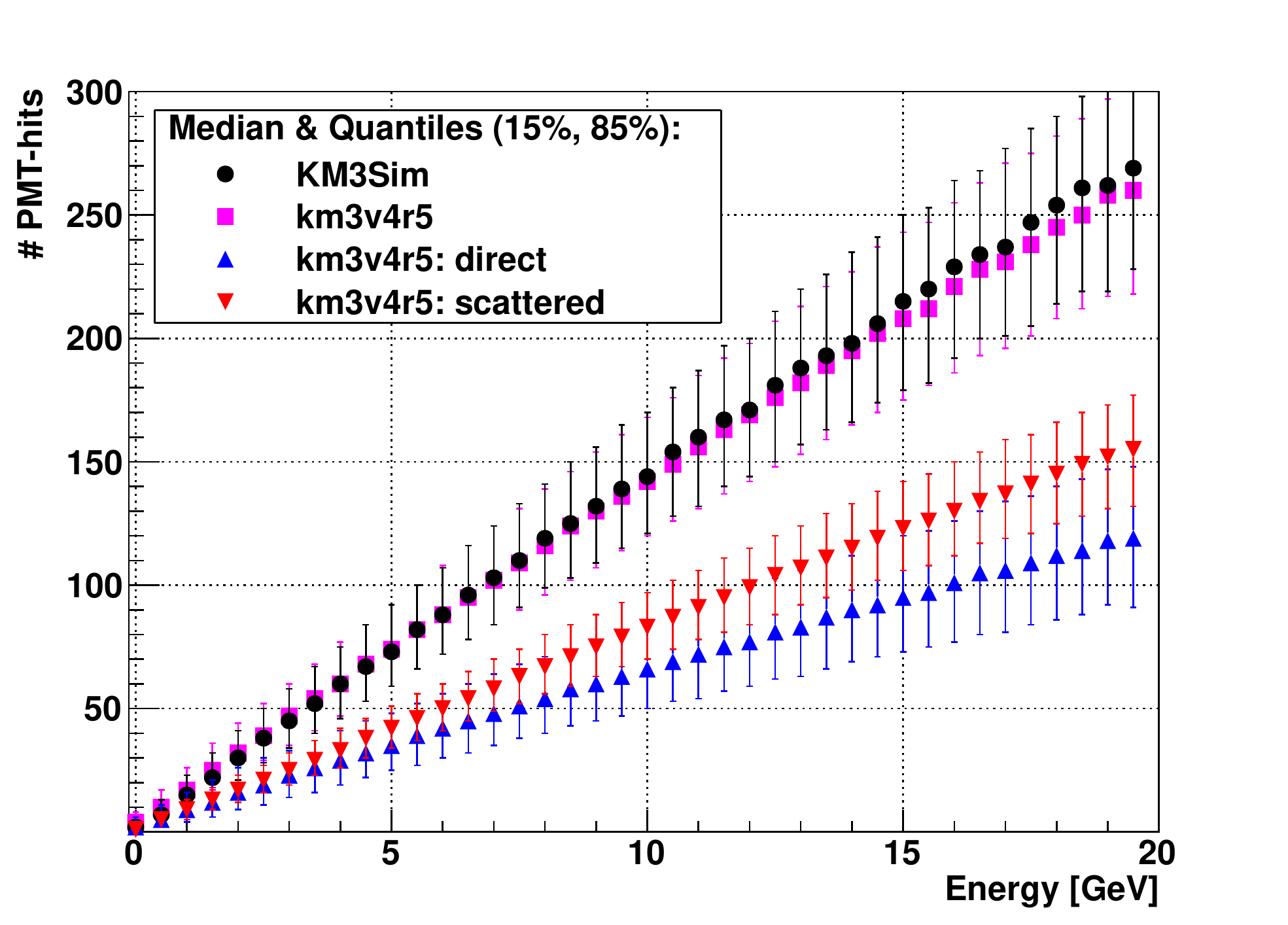}
\put (40,66) {\bf KM3NeT}
\end{overpic}
\begin{overpic}[width=0.49\linewidth]{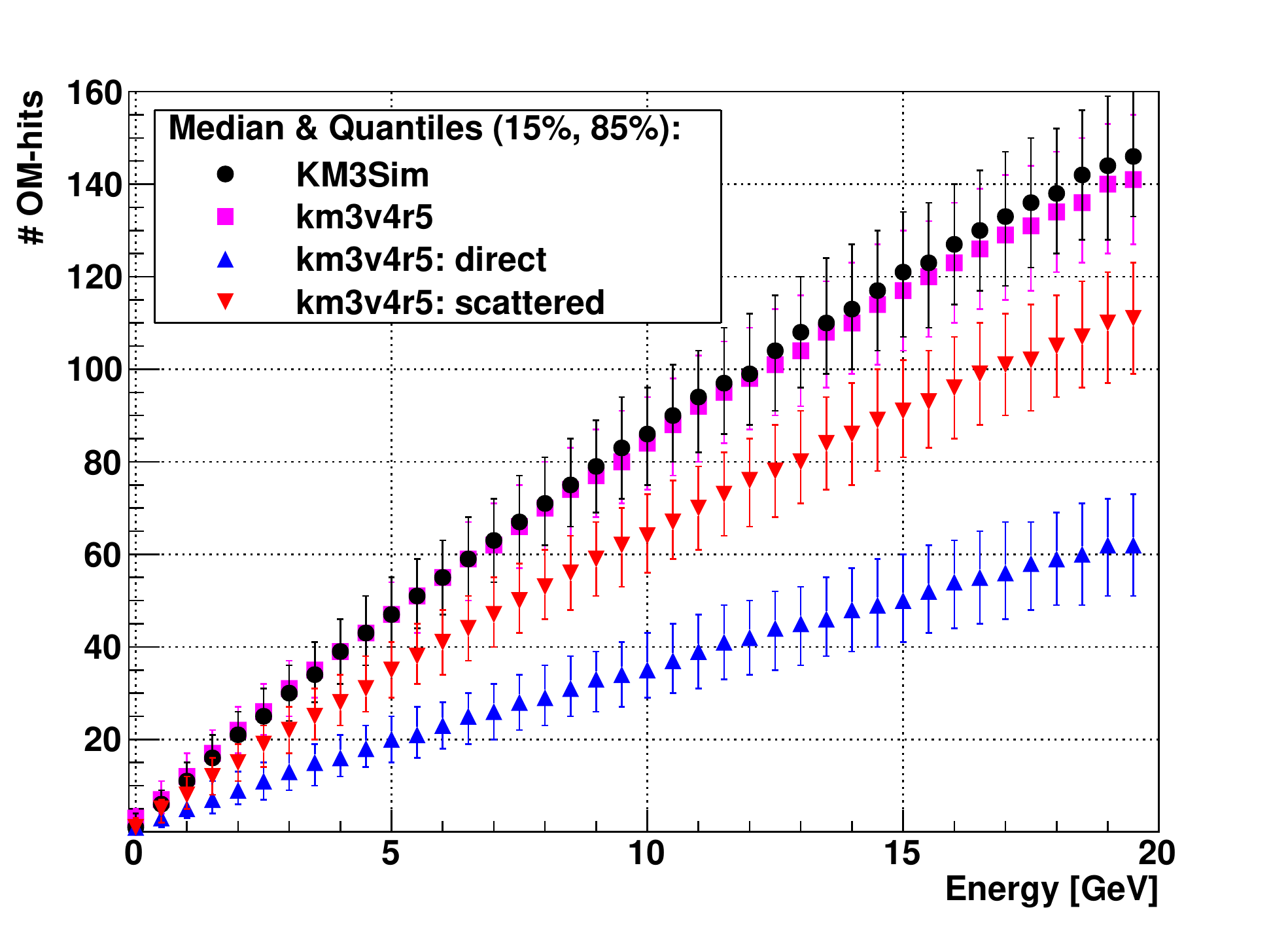}
\put (40,66) {\bf KM3NeT}
\end{overpic}
 \caption{Median and $15\,\%$ / $85\,\%$ quantiles (vertical bars) for the distribution of the number of PMTs with hits (left) and DOMs with at least one hit (right) generated by a muon as a function of its energy. Shown are results for two different simulation packages (KM3 (v4r5) and KM3SIM); 
for the simulation with KM3 the light yield is differentiated into 'direct', 'scattered' and 'all' light.}
  \label{fig:premium_events_muon}
\end{figure}

\myfref{fig:premium_events_muon} shows linearly increasing 
distributions of the total number of hit PMTs and DOMs 
as a function of the muon energy for events for which almost all produced
light is contained within the instrumented volume. 
Also shown is a comparison of the simulated light yield
for the older KM3 and the more recent KM3SIM code. Both simulations agree quantitatively very well.
Roughly 15 detected photons, i.e. PMT hits,
and 8 different hit DOMs can be expected per GeV of a contained muon. 
About 2/3 of the DOMs are hit by scattered light, 1/3 by direct light, while 40\% of all
hit PMTs register unscattered light. 

\begin{figure}[h]
  \centering
\begin{overpic}[width=0.49\linewidth]{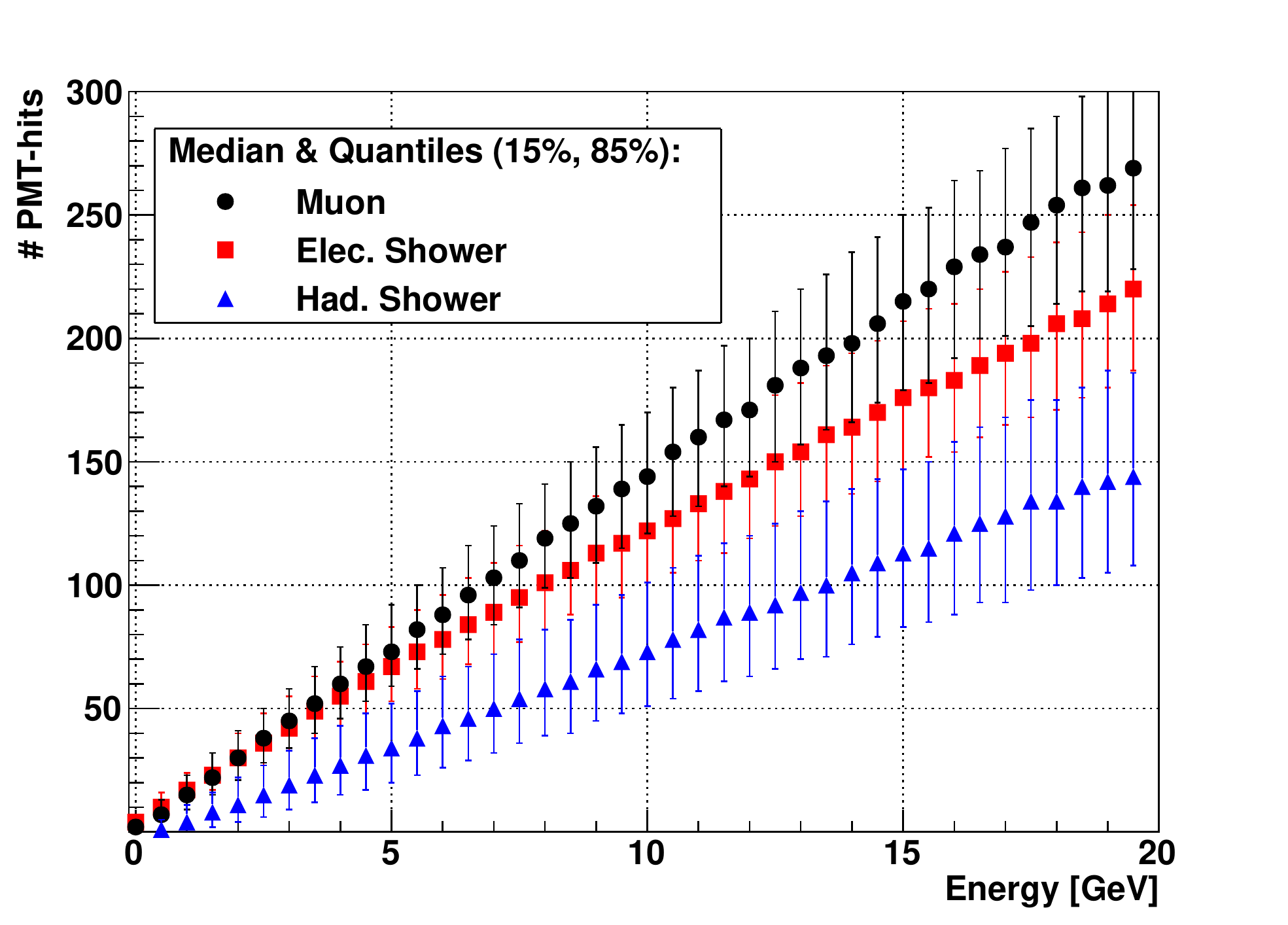}
\put (40,66) {\bf KM3NeT}
\end{overpic}
\begin{overpic}[width=0.49\linewidth]{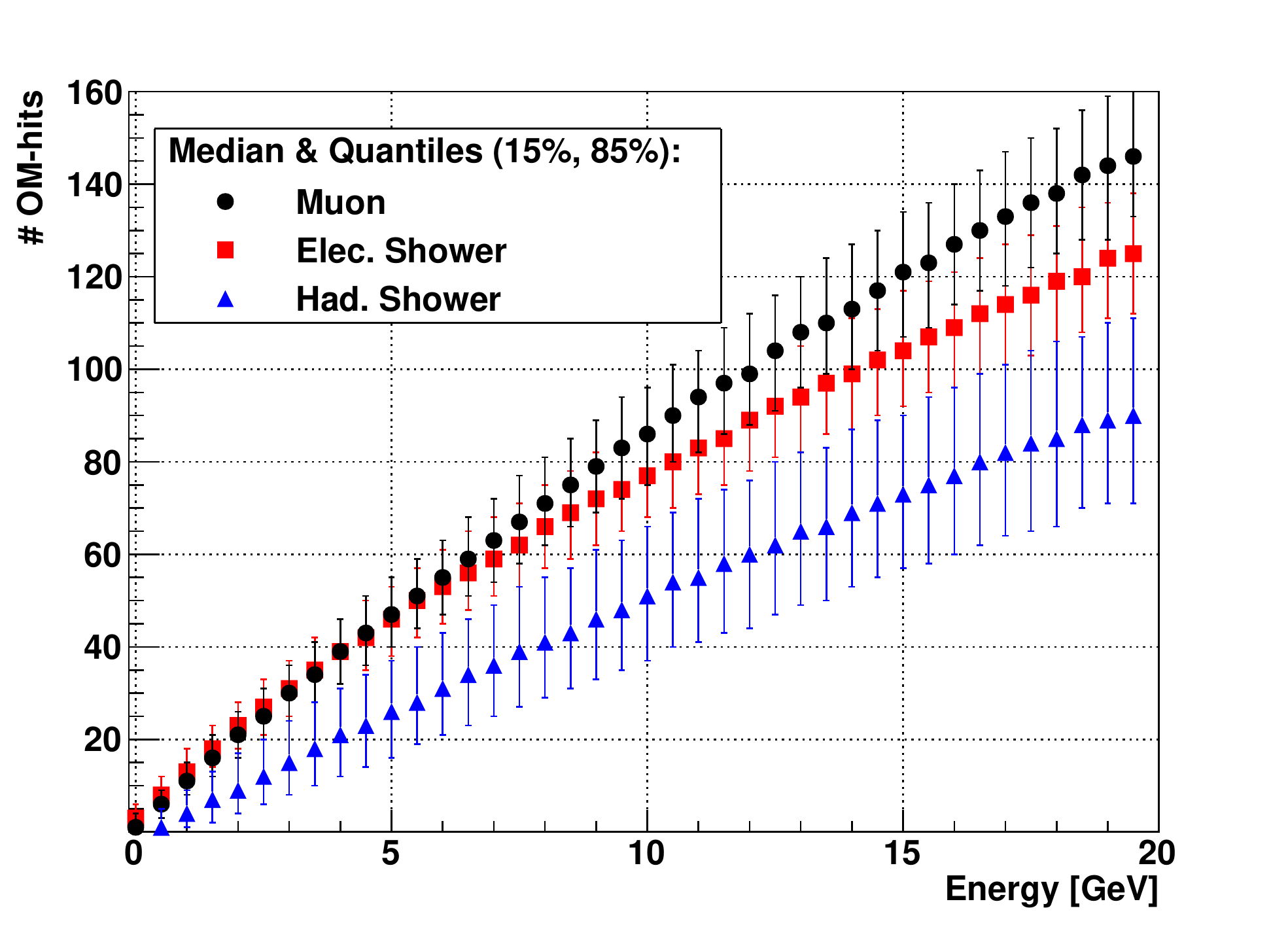}
\put (40,66) {\bf KM3NeT}
\end{overpic}
\caption{Median and $15\,\%$ / $85\,\%$ quantiles (vertical bars) for the distribution of the number of PMT (left) and DOM (right) hits generated by a muon, an electromagnetic and a hadronic shower as a function of their respective energy.  The KM3SIM package has been used.}
  \label{fig:premium_events_showers}
\end{figure}

\myfref{fig:premium_events_showers} compares the number of hit PMTs (DOMs)
due to the Cherenkov light emission from a muon, an electromagnetic and a 
hadronic shower as a function of their respective energy. An electromagnetic 
shower will cause roughly 12 hits per GeV while a hadronic shower is,
as can be expected due to the Cherenkov thresholds of the comparably 
massive hadrons involved, much dimmer with 7 hits per GeV, i.e.
an electromagnetic shower of about 5~GeV energy is almost as bright as
a hadronic shower with 10~GeV. The DOM hit multiplicity scales in a similar manner,
but somewhat more favourably for hadronic showers due to the on average
greater opening angle as compared to electron positron pair cascades.

\begin{figure}[h]
\centering
\begin{overpic}[width=0.49\linewidth]{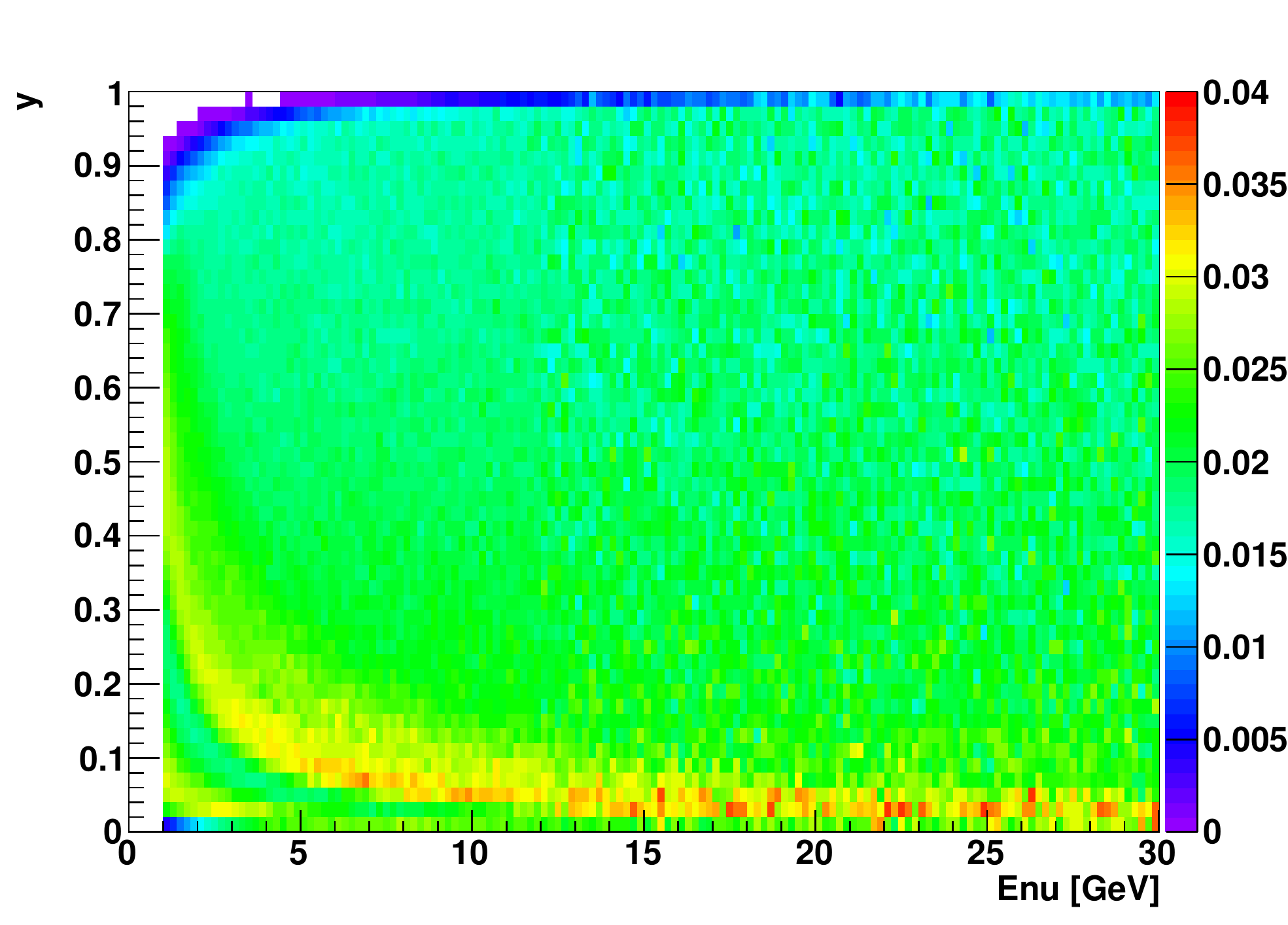}
\put (40,66) {\bf KM3NeT}
\end{overpic}
\begin{overpic}[width=0.49\linewidth]{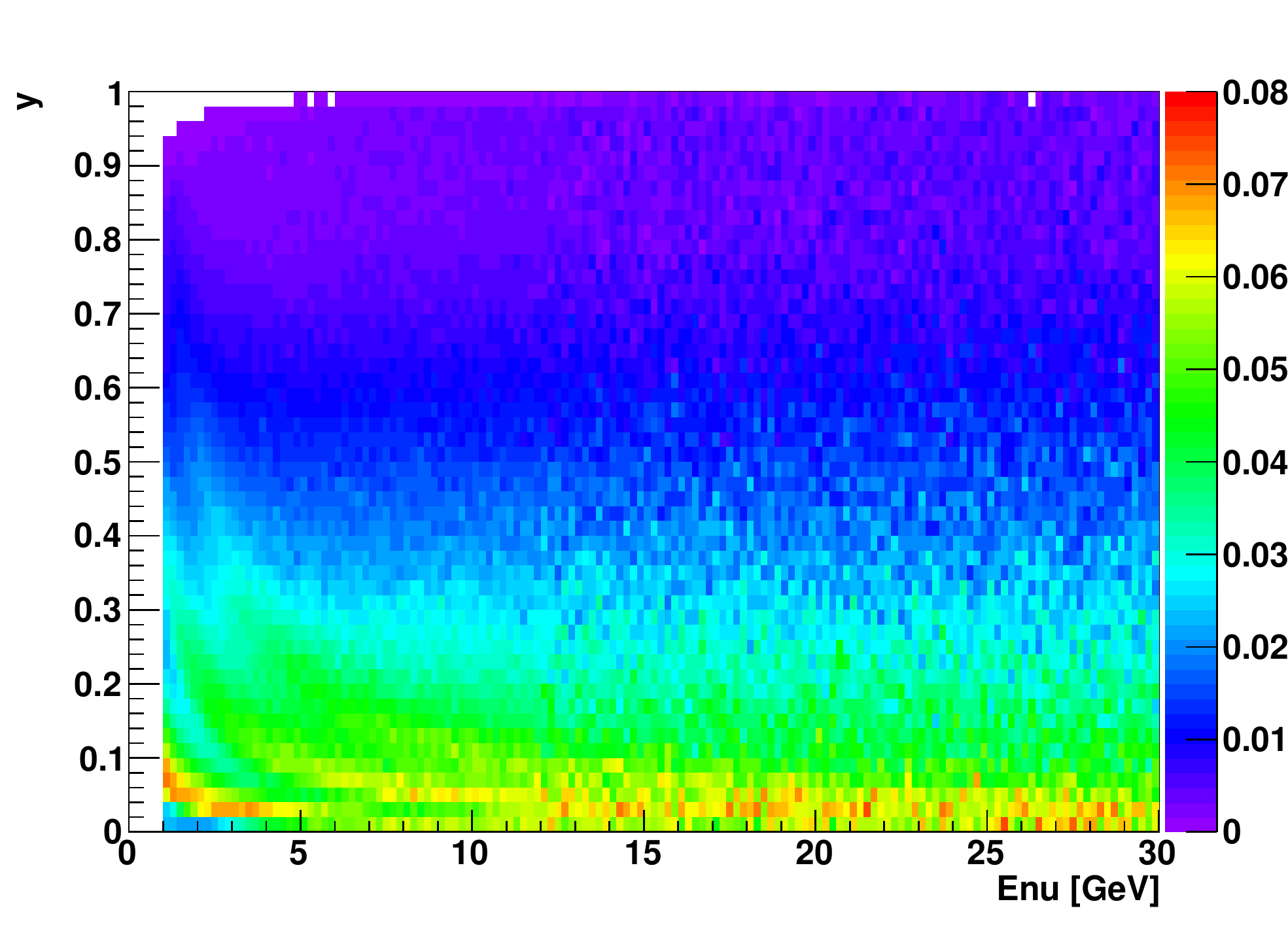}
\put (40,66) {\bf KM3NeT}
\end{overpic}
\caption{Distribution of the interaction inelasticity parameter $y$ as a function of the neutrino energy for neutrinos (left) and antineutrinos (right). Each energy bin is normalised to 1.}
\label{fig:bjorkenY}
\end{figure}

The inelasticity parameter $y$ of a neutrino interaction on the nucleon
critically determines the reaction kinematics as can be seen in \myfref{fig:bjorkenY} and \myfref{fig:kinBjorken}.
At energies below 10 GeV the different strengths of the different interaction channels, 
quasi-elastic, resonant and deep inelastic,
are visible in the $y$-distributions and result in a higher average
inelasticity for neutrinos ($<y> \approx 0.5$) 
compared to antineutrinos ($<y> \approx 0.35$).
The scattering angle $\phi_{\nu,\mu}$ between the incoming neutrino
and the outgoing muon shows a strong dependency and increase with increasing reaction inelasticity. The lower average
inelasticity for antineutrinos leads to on average also lower scattering angles. This indicates the discrimination potential
of this parameter and the importance to get access through event reconstruction.

\paragraph{Muons from hadronic showers\\} 
\label{sec:muon_from_hadShower}
Employing detailed GEANT3 based simulations,
the muon production within the hadronic shower has been studied.
A significant contribution of muons with path lengths in excess of the hadronic shower extension,
i.e. with energies of at least one or several GeV,
could complicate and probably deteriorate the particle flavour identification capabilities 
(see \mysref{sec:particle_id}). 
However, as is shown in  \cite{bib:hofestaedt2016} and summarised in the following,
GeV muons from hadronic showers affect only about 1\,\% of the events.\\

\begin{figure}[htpb]
\centering
\begin{minipage}[c]{0.48\textwidth}
\centering
{
\begin{overpic}[width=\textwidth]{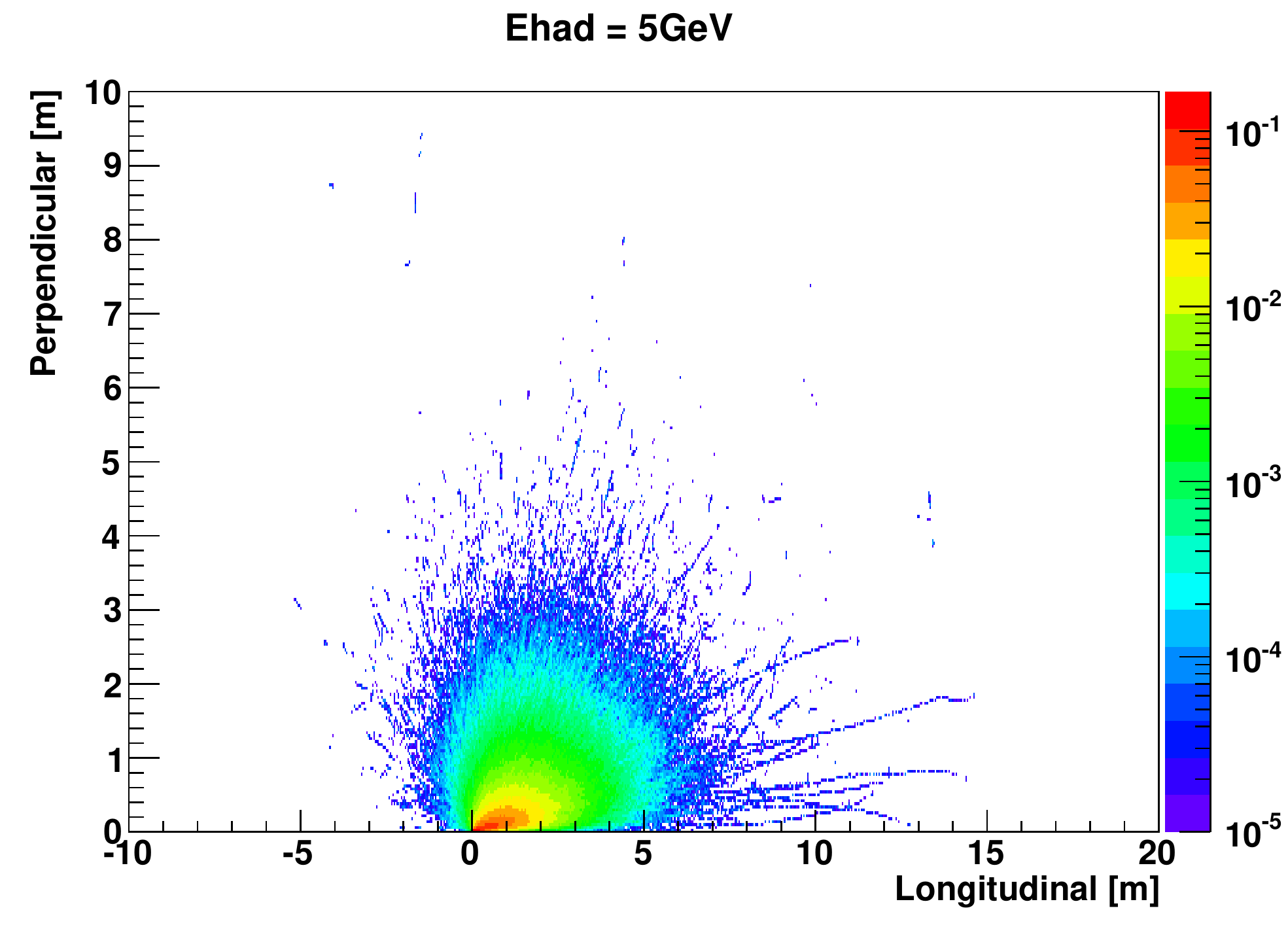}
\put (12,58) {\bf KM3NeT}
\end{overpic}
}
\end{minipage}
\hfill
\begin{minipage}[c]{0.48\textwidth}
\centering
 {
\begin{overpic}[width=\textwidth]{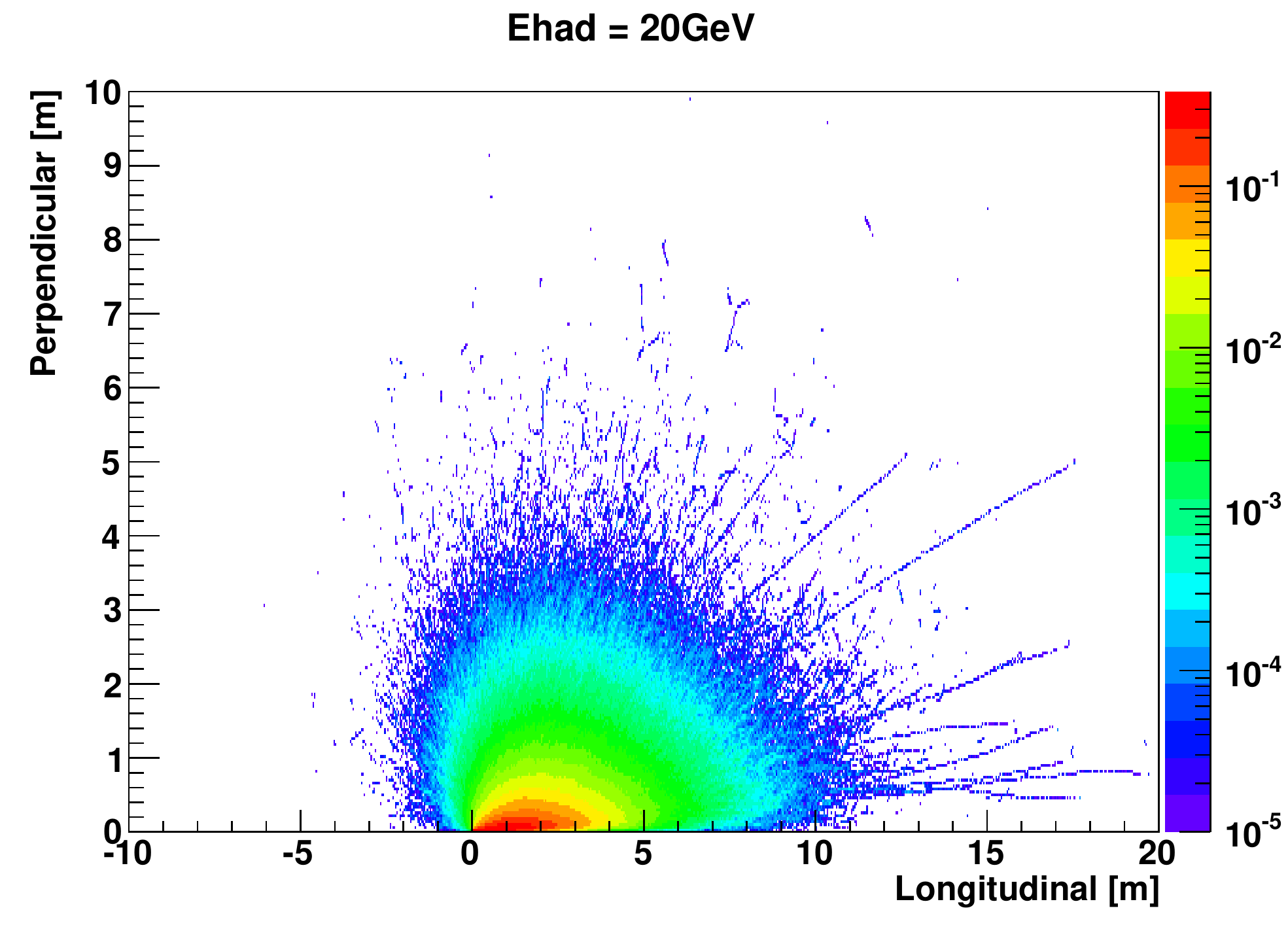}
\put (12,58) {\bf KM3NeT}
\end{overpic}
}
\end{minipage}
\caption{
Simulated Cherenkov photon emission positions along and perpendicular to the direction of hadronic showers from $\nu_e$ CC events. 
3400 superimposed hadronic showers with $E_{\rm had} \approx
5\,\rm{GeV}$ (left).
4000 superimposed hadronic showers with $E_{\rm had} \approx
20\,\rm{GeV}$ (right).
}
\label{fig:muhad_emisPosi2D}
\end{figure}

In \myfref{fig:muhad_emisPosi2D} the Cherenkov photon emission positions along 
and perpendicular to the hadronic shower direction are shown for simulated 
shower energies of $E_{\rm had} \approx 5\,\rm{GeV}$ (left) and 
$E_{\rm had} \approx 20\,\rm{GeV}$ (right). Each Cherenkov photon is weighted 
with its wavelength dependent detection probability taking into account the PMT quantum 
efficiencies and the absorption and scattering in sea water. 
In total 3400 (4000) $\nu_e$ CC events with $8 < E_\nu/\rm{GeV} < 12$ ($30 < E_\nu/\rm{GeV} < 50$)
are used to extract and superimpose their hadronic showers.
For both cases only a few muon tracks can be seen to emit light significantly beyond the
hadronic shower extension.

\begin{figure}[htpb]
\centering
\begin{overpic}[width=0.49\linewidth]{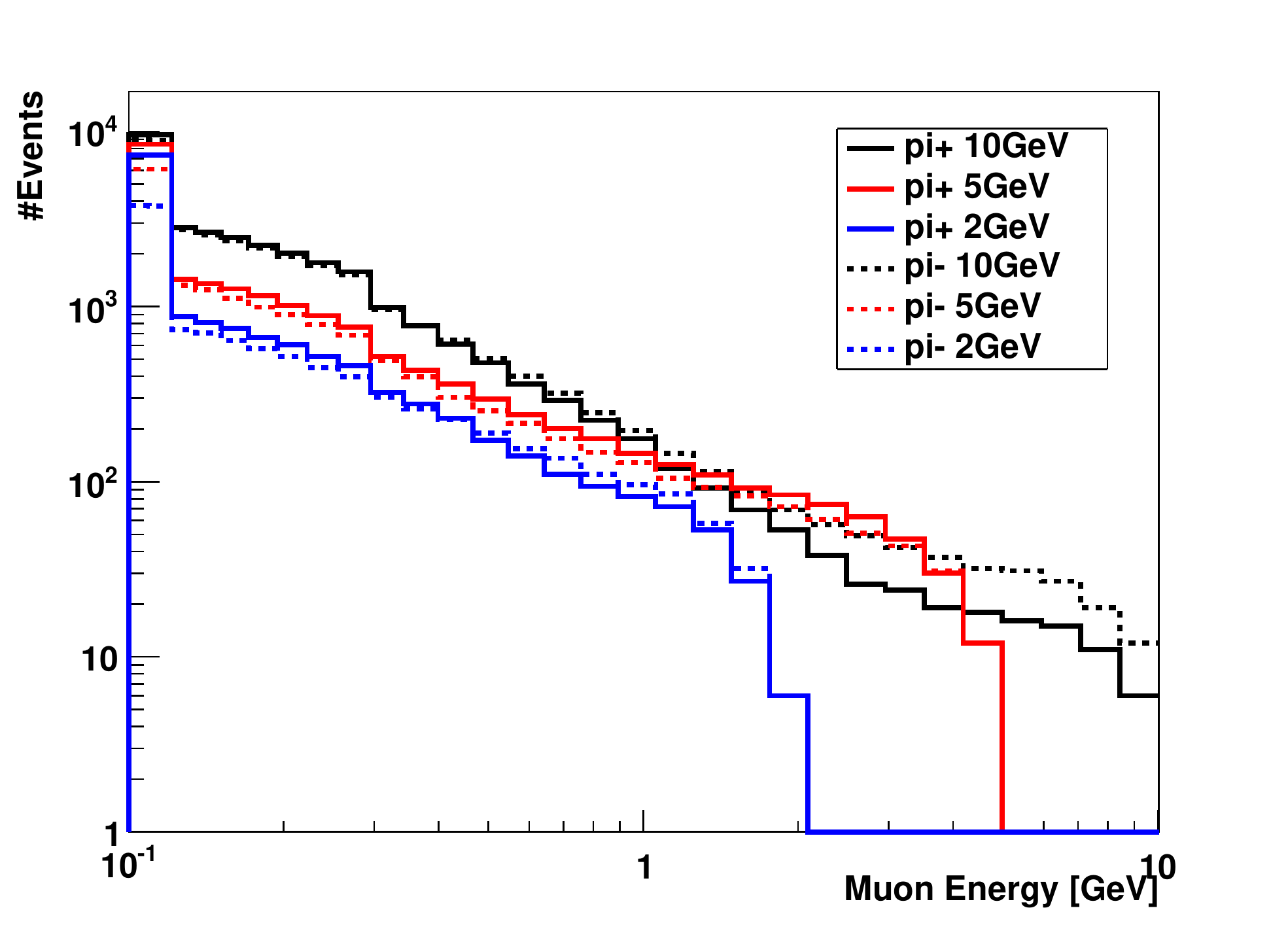}
\put (40,66) {\bf KM3NeT}
\end{overpic}
\begin{overpic}[width=0.49\linewidth]{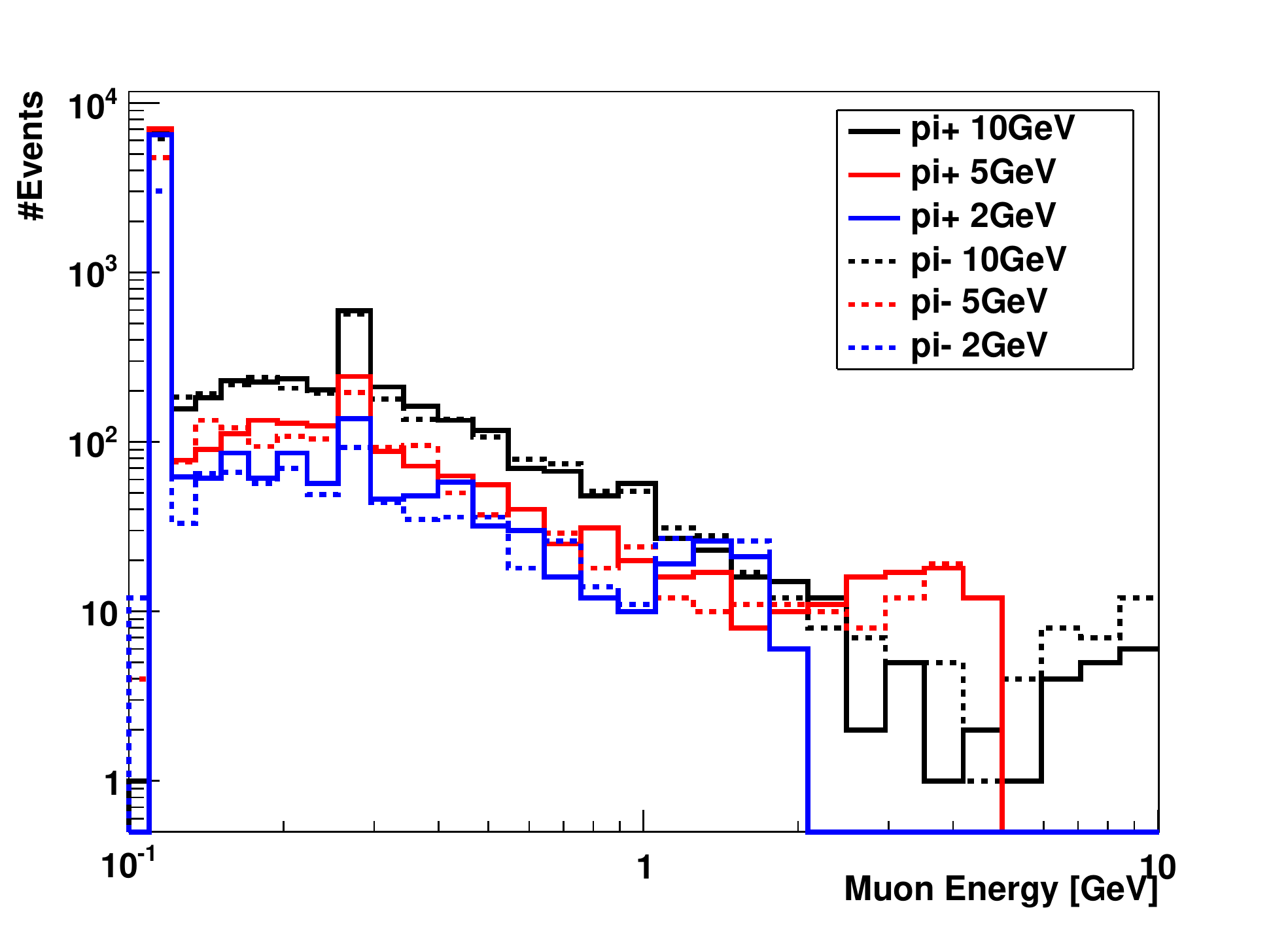}
\put (40,66) {\bf KM3NeT}
\end{overpic}
\caption{
Energy spectrum (left) and cumulative energy distribution (right) of the most energetic muon 
from charged pion $\pi^\pm$ simulations with energies of $E_\pi = 2, 5, 10\,\rm{GeV}$. 
In total, $\rm 10^4$ events for each pion energy are simulated. 
The peak visible at $E_{\mu} \approx 250\,\rm{MeV}$ is due to the decay of kaons produced at rest in pion induced hadronic interactions.}
\label{fig:muhad_leadingEmu_fromPion}
\end{figure}

Most muons in the hadronic shower come from pion decays. 
However, pions with energies in the GeV range will likely interact before they decay,
as the hadronic interaction length for pions in water is approximately 1~m. 
In order to study the muon production from charged pions in greater detail,
$\rm 10^4$ charged pions with energies of $E_\pi = 2, 5, 10\,\rm{GeV}$ have been simulated in sea water. 
The mean number of muons $\langle  N_\mu \rangle$ and the fraction of simulated events 
with at least one muon $N_{\mu \ge 1}$ are summarised in \mytref{tab:muhad_pion_sim}. 
The energy spectrum and cumulative energy distribution of the most energetic muon is shown in \myfref{fig:muhad_leadingEmu_fromPion}. 
For all three pion energies the fraction of events producing a muon with more than 1\,GeV (2\,GeV) is below 2\,\% (1\,\%).

\begin{table}[htpb]
\small
\centering
\begin{tabular}{l c c c c}
\multicolumn{2}{l}{Simulation}  & ~ & $\langle  N_\mu \rangle$    & $N_{\mu \ge 1}$   \vspace{0.1cm} \\
\hline
\multicolumn{1}{ c }{\multirow{3}{*}{$\pi^+$}} & 10\,GeV    & ~ & 2.79  & 0.96    \\
\multicolumn{1}{ c  }{}                        & 5\,GeV     & ~ & 1.44  & 0.84    \\
\multicolumn{1}{ c  }{}                        & 2\,GeV     & ~ & 0.93  & 0.73    \\
\hline
\multicolumn{1}{ c }{\multirow{3}{*}{$\pi^-$}} & 10\,GeV    & ~ & 2.22  & 0.89    \\
\multicolumn{1}{ c  }{}                        & 5\,GeV     & ~ & 0.91  & 0.61    \\
\multicolumn{1}{ c  }{}                        & 2\,GeV     & ~ & 0.42  & 0.38    \\
\end{tabular}
\caption{Mean number of muons $\langle  N_\mu \rangle$ and fraction of simulated pions producing at least one muon $N_{\mu \ge 1}$.}
\label{tab:muhad_pion_sim}
\end{table}

\subsubsection{Implementation of different vertical spacings}
\label{sec:masking}

Different detector configurations have been investigated with the same footprint as the benchmark detector (cf. \myfref{fig:footprint}) but different vertical spacings: 6\,m, 9\,m, 12\,m and 15\,m. All these configurations rely on the same simulation of the neutrino signal, performed on the benchmark detector described in \mysref{simu:det}. The different vertical spacings are achieved by masking parts of the detector:\begin{itemize}
\item for 6\,m vertical spacing, all simulated DOMs in the benchmark detector are used;
\item for 9\,m vertical spacing, every third DOM on each DU is masked, thus alternating vertical spacings of 6\,m and 12\,m. The DUs are masked in three different schemes  (1st scheme: masking DOM 1, 4, \dots; 2nd scheme: masking DOM 2, 5, \dots; and 3rd scheme: masking DOM 3, 6, \dots);
\item for 12\,m vertical spacing, every second DOM on each DU is masked;
\item for 15\,m vertical spacing, five different masking schemes are used that alternate vertical spacings of 12\,m and 18\,m.
\end{itemize}
In the 9\,m and 15\,m configurations, neighbouring DUs use different masking schemes in order to make the masked detector as homogeneous as possible. 
Doing so the instrumented volume stays the same for all detector configurations, but the DOM density changes. 
In order to compare the effective volume of the different detector configurations assuming the same number of DOMs for each vertical spacing, the effective volumes of the masked detectors are scaled accordingly (factor of 1/1.5/2/2.5). 

It should also be noted that the surface to volume ratio for the masked
detectors is larger than it would be for a full detector with 18 DOMs per DU.
Therefore, the presented results overestimate possible surface-related effects.

\subsubsection{Triggering}
\label{sec:trigger}

As described in \mysref{sec-tec-com-tri}, muon and shower events are
extracted from the real-time data stream using causality conditions.
In the case of ORCA, with a simulated 10\,kHz uncorrelated single
noise rate per PMT and about 500\,Hz time-correlated noise from
 $^{40}$K decays on each DOM, the estimated L1 rate (coincidences
 on the same DOM in a short time-window) per optical module is about 1.5\,kHz.

The trigger algorithms described in \mysref{sec-tec-com-tri} were
optimised for ORCA by considering the effective volume and the event
purity. The effective volume is the volume in which a neutrino
interaction would trigger the event to be written to disk and the
event purity is the fraction of triggered events that contain a
neutrino interaction or at least one atmospheric muon. 
The trigger rate from neutrino interactions is $\mathcal{O}(\rm{mHz})$
and is negligible compared to the rate from atmospheric muons ($\mathcal{O}(40\,\rm{Hz})$).

The trigger settings correspond to a L1 time window of $\Delta T = 10\,\rm{ns}$, 
a maximum angle between the PMT axes of 90 degrees (L2), 
and a minimum number of L1 hits of 3 for the shower trigger and 4 for the muon trigger
\footnote{
Muon and shower triggers with larger minimum numbers of L1 hits in conjunction with larger distance parameters $R$ and $D$ have also been studied. However, these triggers show smaller effective volumes than those used in this document.
}.
Both triggers run in parallel and one of them or both must fire to flag an event (logical {\it OR}).
For the different considered vertical spacings the distance parameters ($R$ and $D$) 
of the muon and shower triggers have been adjusted such that each of the triggers has a rate of 
$\sim 10\,\rm{Hz}$ from pure noise. The rate of atmospheric muon events is evaluated at a depth of 2450\,m using the 
simulations described in \mysref{sec:simulations} and amounts to about 36\,Hz (6\,m) - 55\,Hz (15\,m) 
depending on the vertical spacing of the ORCA detector. 

In order to estimate the trigger rates, dedicated simulations for each vertical spacing have been performed, 
i.e. the detector masking described in \mysref{sec:masking} has not been applied.
Trigger rates from pure noise and atmospheric muons are summarised in \mytref{tab:trigger_rates} 
for the various vertical spacings. The trigger event purity is 65\% - 73\%.\\
It should be noted that during periods of high
bioluminescence~\cite{ANT-Biolum-2013},
the trigger conditions 
(minimum number of L1 hits and distance parameters $R$ and $D$) can be tightened in order 
to reduce the output data rate and match the available data transfer bandwidth.

\begin{table}[htpb]
\small
\centering
\begin{tabular}{c | c c | c c | c }
detector configuration & \multicolumn{2}{c}{trigger configuration} & \multicolumn{2}{|c|}{trigger rates [Hz]} & ~  \\
vertical spacing [m] & $R$ [m] &  $D$ [m] & pure noise  & atm. muons & event purity  \\
\hline
6 & 35 &  40  & 19  & 36 & 0.65   \\
9 & 39 &  43  & 18  & 41 & 0.69  \\
12 & 42 &  46  & 19  & 47 & 0.71   \\
15 & 44 &  50  & 20  & 55 & 0.73   \\
\end{tabular}
\caption{Expected trigger rates from pure noise and atmospheric muons for the trigger configurations used for different detector configurations with 6\,m, 9\,m, 12\,m and 15\,m vertical spacing.}
\label{tab:trigger_rates}
\end{table}

\begin{figure}[h!]
\centering
\includegraphics[width=0.49\linewidth]{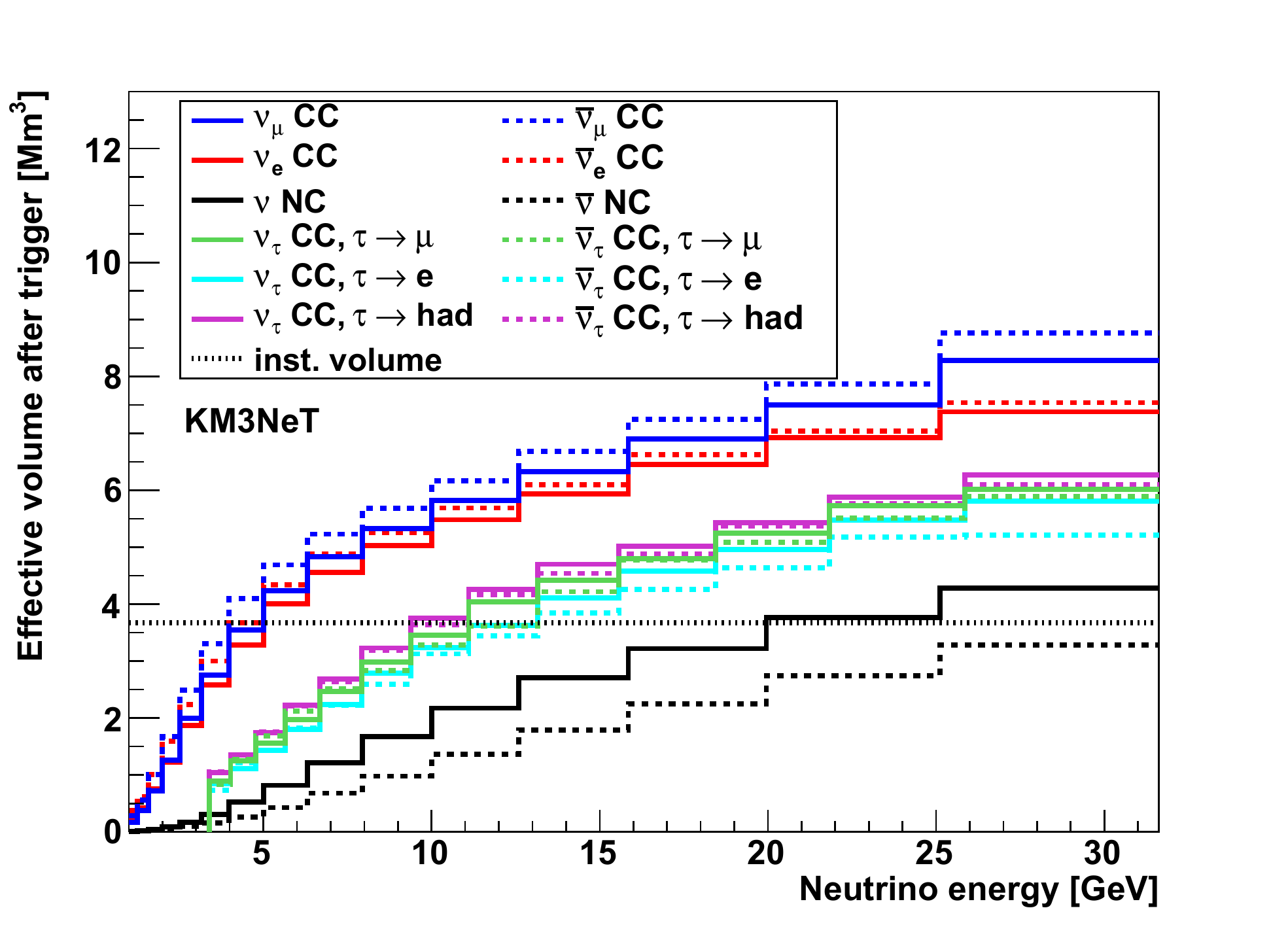}
\includegraphics[width=0.49\linewidth]{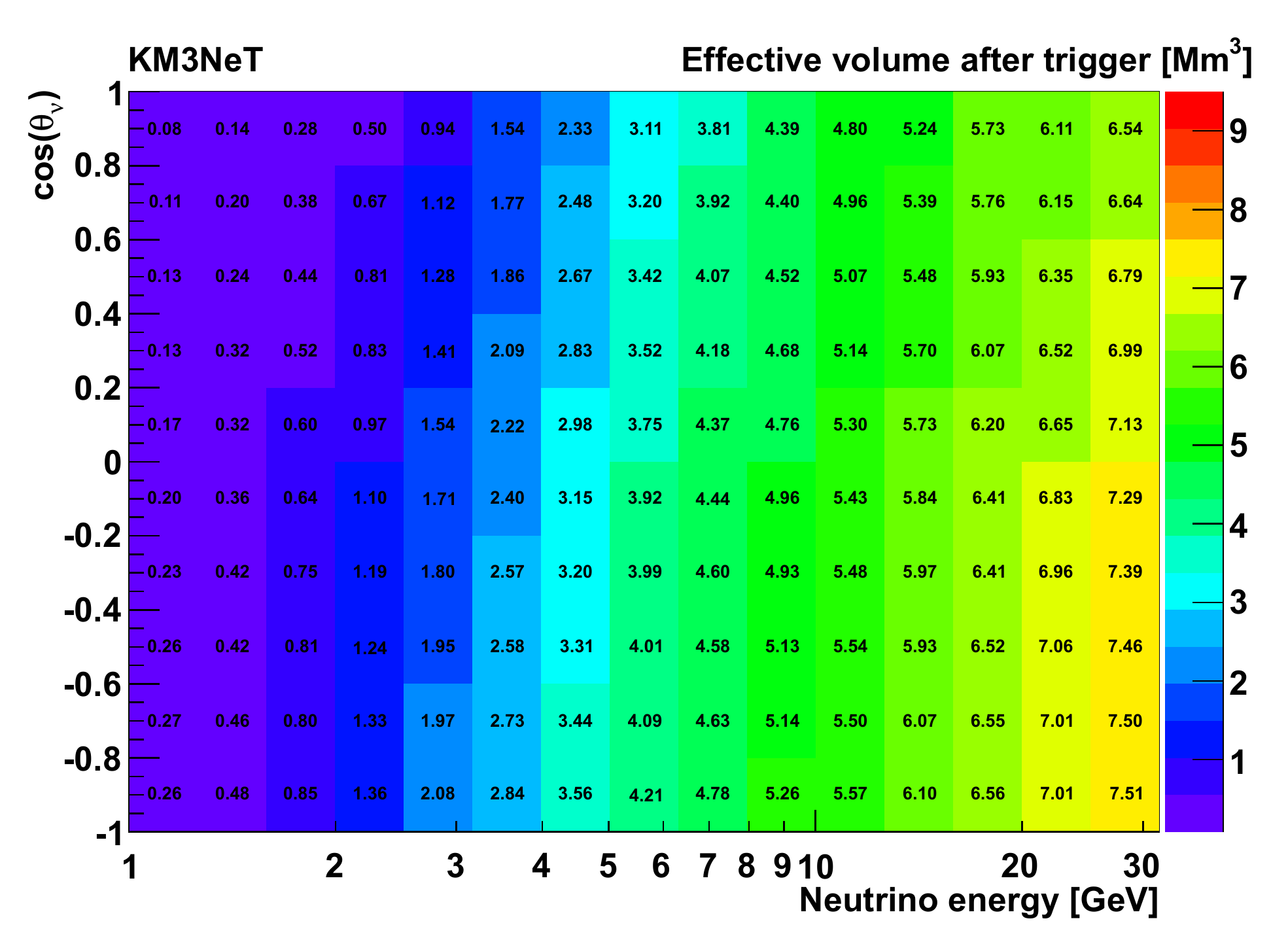}
\caption{
Effective volume at trigger level (left) for 6\,m vertical spacing as
a function of neutrino energy for different neutrino flavours (up-going
events only) and effective volume at trigger level for $\nu_e$ 
CC events as a function of neutrino energy and cosine of
the neutrino zenith angle $\theta_\nu$ (right).
}
\label{fig:trigger_effVol_differentTypes}
\end{figure}

The effective volume at trigger level for 6\,m vertical spacing is shown for different neutrino flavours in \myfref{fig:trigger_effVol_differentTypes} (left) as a function of neutrino energy. Events are weighted to reproduce the conventional atmospheric neutrino flux following the Bartol model \cite{bib:Bartol} and only up-going neutrinos are considered.
The effective volume is smaller for $\nuan$ NC and $\nuan_\tau$ CC than for $\nuan_{e,\mu}$ CC events as the outgoing neutrinos are invisible to the detector. For $\bar \nu_{e,\mu}$ CC events the effective volume is larger than for $\nu_{e,\mu}$ CC due to the lower average inelasticity and the resulting higher average light yield (at the considered energies hadronic showers have a smaller average light yield than electromagnetic showers).\\
The effective volume depends also on the neutrino direction as
\myfref{fig:trigger_effVol_differentTypes} (right)
shows for $\nu_e$ CC events.
Other neutrino flavours exhibit a similar zenith angle dependency. For
vertical up-going events ($\cos \theta_\nu \approx -1$) the effective
volume rises more steeply with energy than for horizontal events ($\cos \theta_\nu
\approx 0$) as more PMTs are oriented downward than upward in an DOM
and the density of DOMs is higher in vertical than in horizontal direction.

\begin{figure}[h!]
\centering
\includegraphics[width=0.49\linewidth]{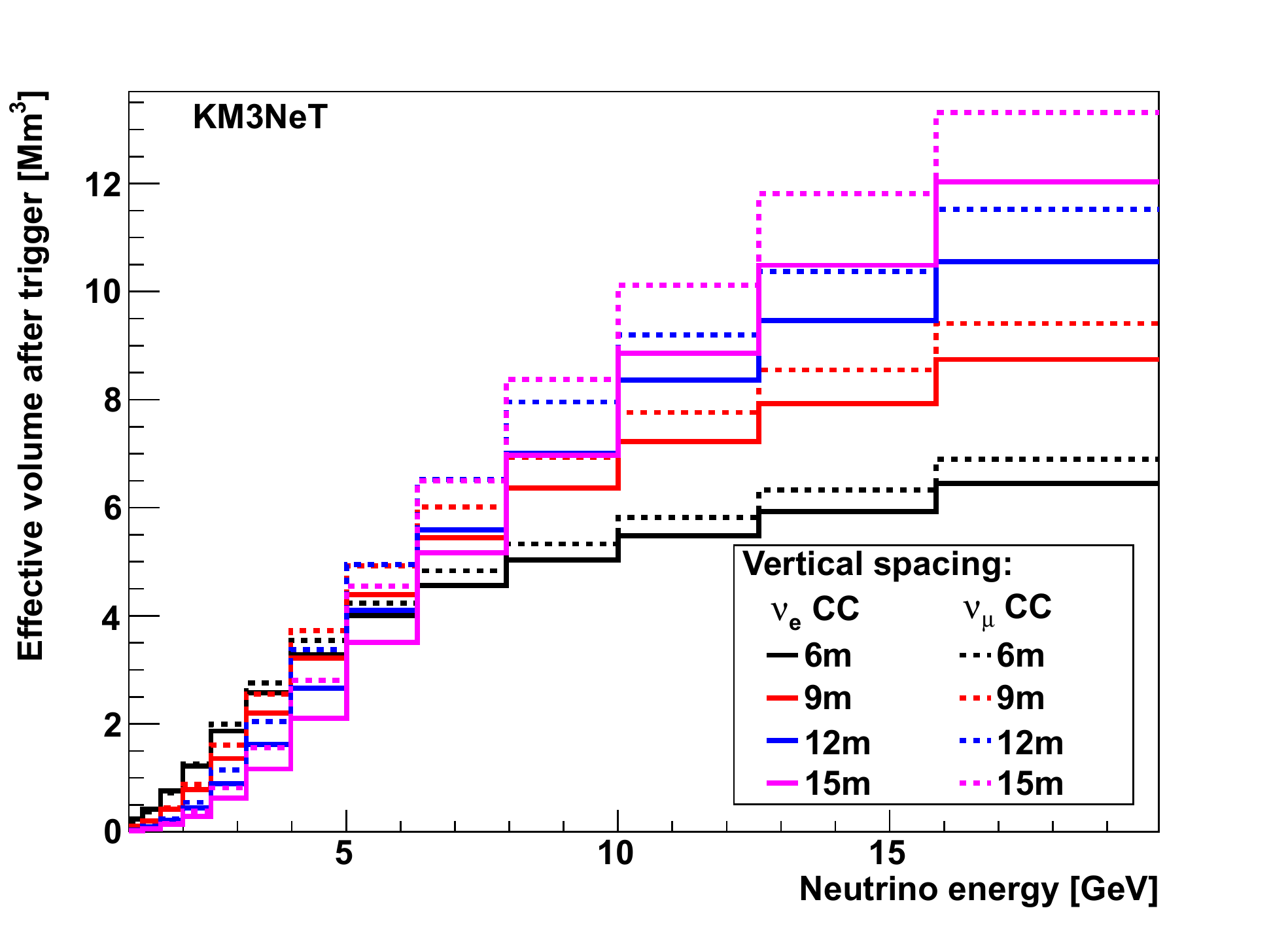}
\caption{
Effective volume at trigger level for different vertical spacings
(6\,m/9\,m/12\,m/15\,m) as a function of neutrino energy for up-going
$\nu_e$ and $\bar \nu_e$ CC (solid lines) and $\nu_\mu$ and $\bar
\nu_\mu$ CC events (dashed lines). 
}
\label{fig:trigger_effVol_differentSpacing}
\end{figure}

The effective volumes at trigger level for $\nuan_e$ CC and $\nuan_\mu$
CC events for different vertical spacings are shown in
\myfref{fig:trigger_effVol_differentSpacing} as a function of neutrino
energy. For 9\,m, 12\,m and 15\,m vertical spacing the simulation of
the benchmark detector with a 6\,m spacing is masked and the resulting
effective volumes are scaled to the same number of DOMs per DU as
described in \mysref{sec:masking}. Further details on the
triggering studies can be found in \cite{bib:hofestaedt2016}.

\subsection{Muon neutrino studies}
\label{evtselreco}
\label{muon}

This section presents the strategy adopted to reconstruct muon neutrino charged current events with ORCA, and its current performance.
All results shown in this section are based on the Monte Carlo simulations presented in the previous sections.

\subsubsection{Muon direction reconstruction} \label{sec:LNSreco}

The track reconstruction algorithm presented here permits to estimate muon (and consequently neutrino) directions using the combined information of the PMT spatial positions and the Cherenkov photon arrival times.
The reconstruction code used is based on the strategy developed for the ANTARES telescope and described in reference~\cite{bib:aartstrategy}. This algorithm has been modified to exploit the multi-PMT peculiarities taking into account the directional sensitivity of the KM3NeT optical module. 

\begin{figure}[!hbpt]
\centering
\includegraphics[width=\linewidth,trim=0 3cm 0 0]{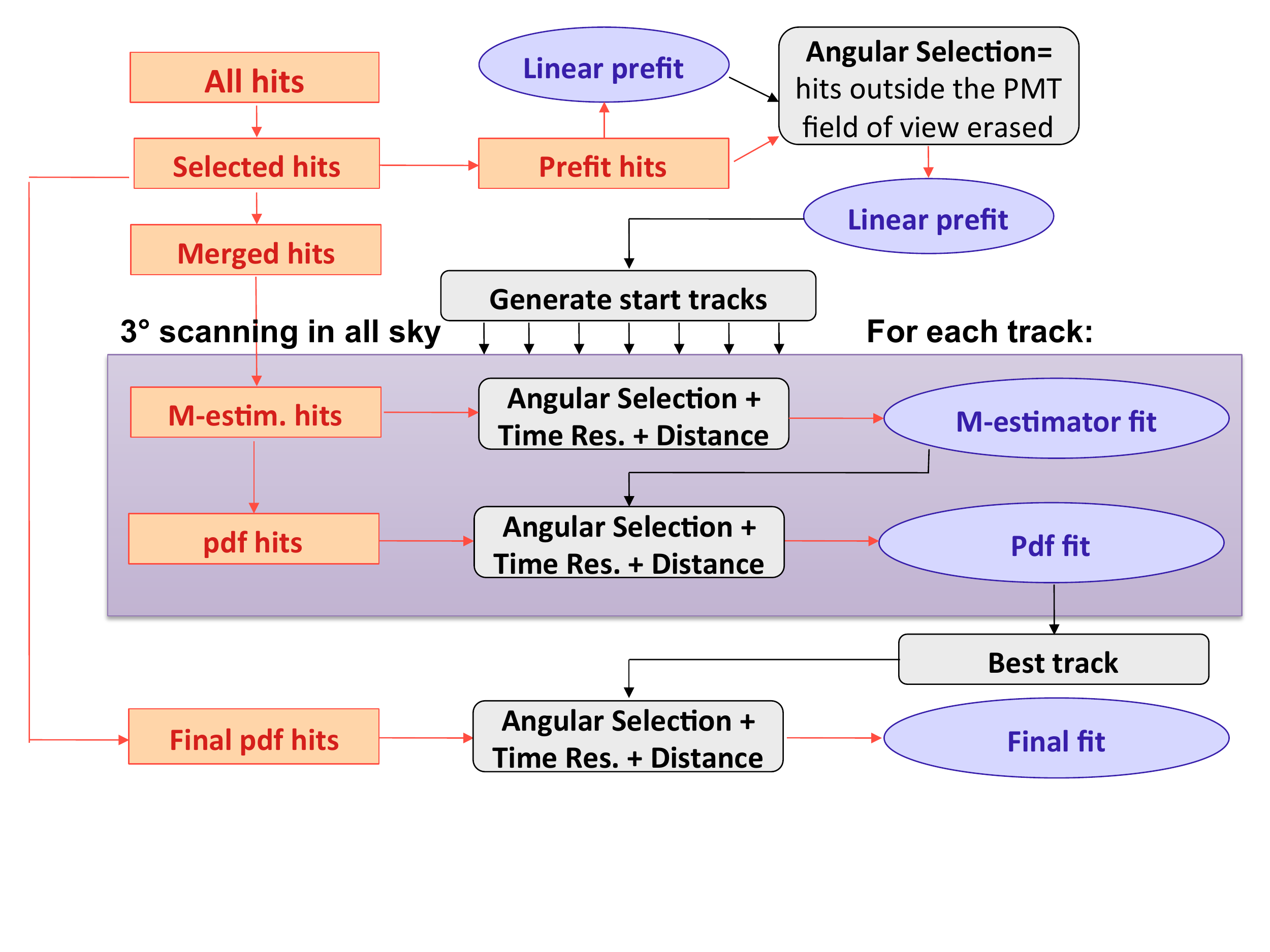}
\caption{Schematic depiction of the reconstruction algorithm.}
\label{fig:reco_schema}
\end{figure}

After an initial hit selection, requiring space-time coincidences between hits, the reconstruction proceeds through four consecutive fitting procedures, each using the result of the previous one as starting point. 
Each fitting stage improves the result, but the last fit produced, that provides the most accurate result, works well only if the input parameters  of the muon track are not too far from the true track parameters.
Moreover, the efficiency of the algorithm is improved with a scanning of the entire sky in steps of 3$^{\circ}$ starting from the prefit track, thus generating 7200 tracks. 
A scheme of the overall procedure is shown in \myfref{fig:reco_schema}.

As described in \mysref{sec:simulations}, the optical background 
induced by $^{40}$K decays has been simulated adding an uncorrelated hit 
rate of 10~kHz per PMT and a time-correlated hit rate of 500~Hz per DOM 
(two coincident hits in different PMTs inside the same DOM).
To remove the hits from $^{40}$K decays, the requirement of space-time 
coincidences between hits is used, since hits due to optical background 
are mostly uncorrelated.

\begin{figure}
\centering%
\begin{overpic}[width=.6\textwidth]{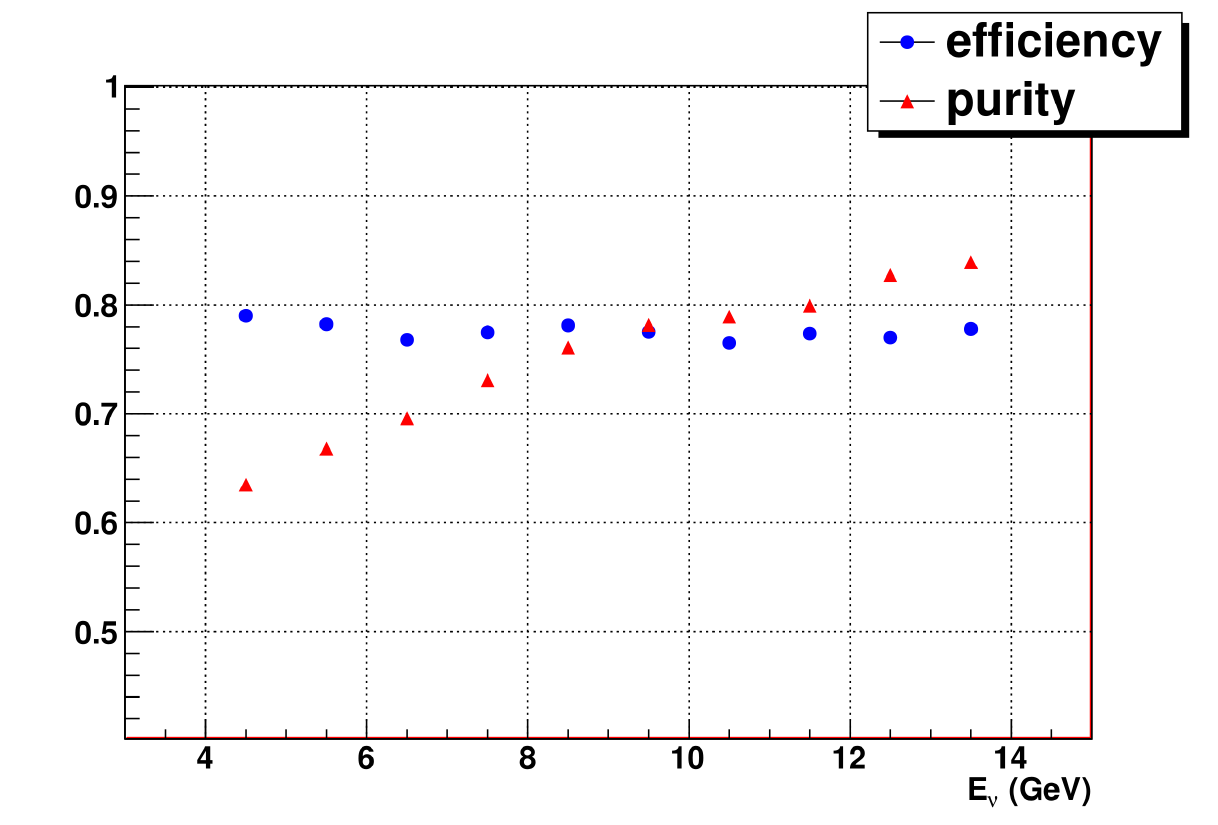}
\put (12,55) {\bf KM3NeT}
\end{overpic}
\caption{Efficiency and purity of the hit selection adopted by the 
track reconstruction algorithm as a function of the interacting neutrino energy.}
\label{fig:hitselperformance}
\end{figure}

In particular, the hit selection proceeds by first selecting all the local coincidences,
i.e. coincidences of hits within the same DOM, 
in a time window of 10~ns and for which the PMTs involved are less
than 
$90^\circ$ apart. Among them, a cluster is selected 
such that any hit in the cluster is causally related to all the 
remaining ones, according to the following causality relation:
\begin{equation}
  |\Delta t| < d/c_{water} + 20 \mathrm{ns} 
\end{equation}
where $\Delta t$ is the time difference between the two hits, $d$ is the 
distance between the two PMTs and $c_{water}$ is the group velocity of light in 
water. The cluster of hits obtained is further extended  by including
the yet unselected hits which fulfil all the following conditions:
\begin{itemize}
\item are causally connected to at least 75\% of all the hits in the cluster,
\item are closer than 50\,m to at least 40\% of all the hits in the cluster,
\item are all causally connected among them.
\end{itemize}
The latter extension procedure is iterated twice. The resulting 
performance of the hit selection for $\nu_\mu$-CC events in terms of 
efficiency and purity is shown in 
\myfref{fig:hitselperformance}, where the efficiency is
the fraction of signal hits selected among all the signal hits, whereas 
the purity is the fraction of signal hits among all the selected ones. 
The resulting set of hits is referred to as \textit{Selected hits} in the following.

The \textit{Selected hits} serve as input of the first step of the track reconstruction 
procedure, referred to as ``linear prefit'', which is a linear fit 
through the positions of the hits. Once a first estimate of the track is
obtained, the evaluation for each hit of the expected angle of incidence
$\theta_{i}$ of the photon on the PMT is possible. 
An ``angular selection'' is then applied discarding all the hits with 
$\cos \theta_{i}>-0.5$. The prefit is then repeated with the new hit set. \\
Additional starting tracks are obtained by rotating the prefit track by 
step of 3$^\circ$ over the whole sky. For each starting track, the two 
fits called \emph{M-estimator fit} and \emph{Pdf fit} are performed. 
These fits are based on the maximum likelihood method and use probability 
distribution functions (PDF) that depend on the time residuals, i.e.
the difference between the time of the hits and the expected times 
according to the track hypothesis and Cherenkov light emission. \\
The \emph{M-estimator fit}, is a maximum likelihood fit based on a 
function that describes the data for small time residuals. The behaviour 
of this function for large residuals is a trade-off between a reproduction 
of the data and the ease of finding the global maximum. The input set of hits, 
called  \emph{M-estim hits} is chosen among the \emph{Selected hits} 
with conditions on the time residual and the orthogonal distance from 
the starting track and discarding all hits with $\cos \theta_{i}>-0.5$.
The PDF used for the \emph{Pdf fit}  has been parameterised by fitting 
a set of spectra obtained from Monte Carlo simulations of muons traversing
the detector without including background hits. The input track is the 
track resulting from the \emph{M-estimator fit} and the input hits are 
chosen among the \emph{M-estim hits} with conditions on the time residual
and the orthogonal distance from the M-estimator track and discarding all
hits with $\cos \theta_{i}>-0.5$. Once the fitting procedures are 
performed for each starting direction, the solution with the highest 
likelihood per degree of freedom is chosen as the best one.
A further adjustment of the track direction is then achieved with the 
\emph{Final fit}, using the best track as starting point. This fit relies 
on the maximum likelihood method and the PDF is obtained taking into 
account the contributions from both the background hits and the signal hits.
The quality of the final reconstructed track is estimated by the quantity:
\begin{equation}
\Lambda = \frac{\mathcal{L}}{N_{\rm hits} -5} 
\label{eq:lambda}
\end{equation}
where $N_{\rm hits}$ is the number of hits used in the final fit and  $\mathcal{L}$  
is the maximum value of the likelihood.

\subsubsection{Neutrino energy estimate}

The neutrino energy estimation is performed in two steps: first the muon
energy is estimated by reconstructing the muon track length and the 
interaction vertex, then the neutrino energy is estimated depending on 
the reconstructed muon length and the number of hits used by the track 
reconstruction algorithm. These two procedures are described in detail 
in the following sections.

\paragraph{Reconstruction of the muon track length and the interaction vertex\\}

A dedicated algorithm for the muon energy estimate has been developed
relying on the length of the reconstructed muon track. In case of events
interacting sufficiently close to the instrumented volume, this algorithm
also reconstructs the neutrino interaction vertex as the starting point
of the reconstructed muon track.

\begin{figure}[t]
\centering
\includegraphics[width=12cm,trim=0 3cm 0 0]{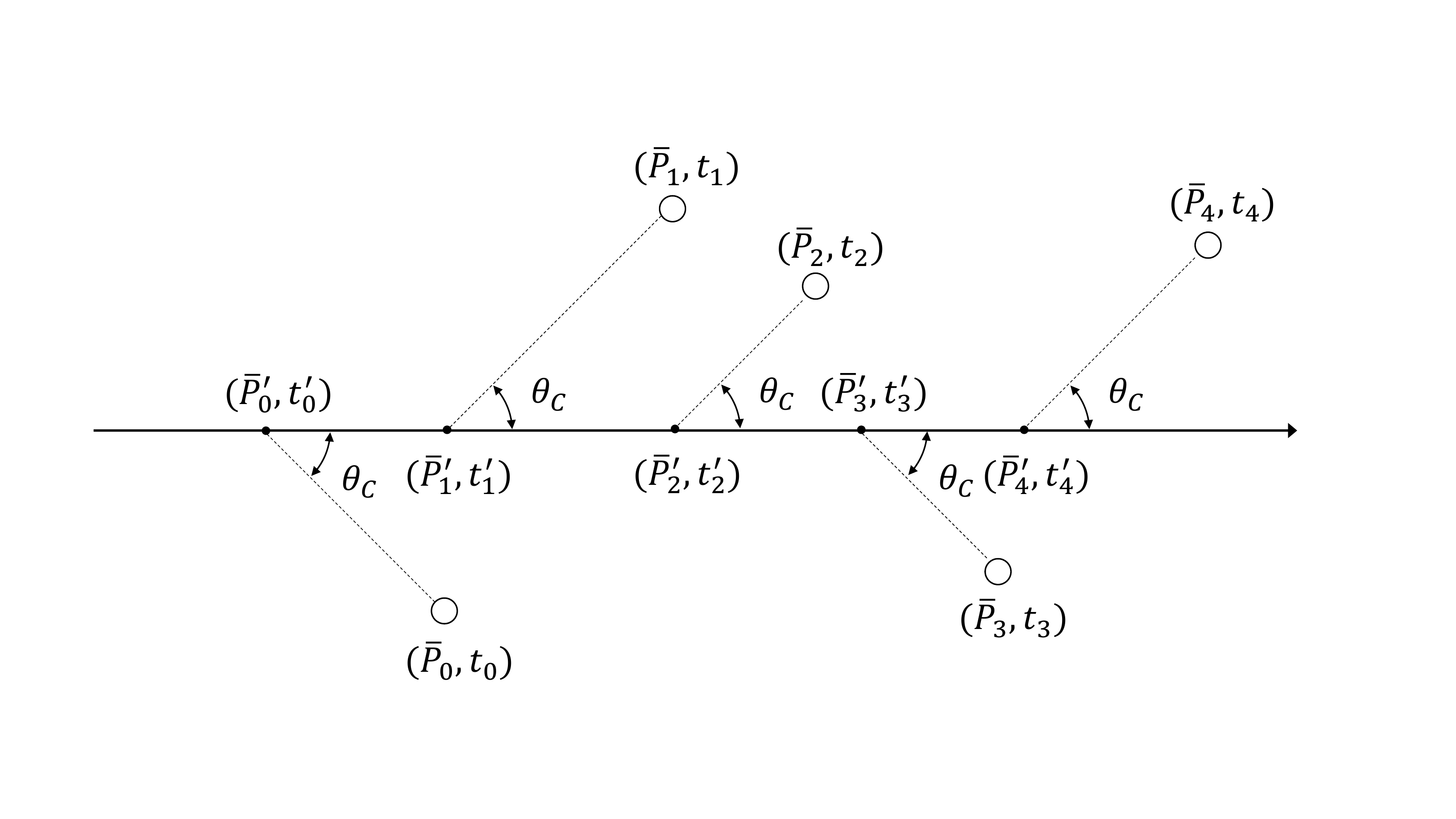}
\caption{Schematic of the track length estimation procedure.
The horizontal arrow indicate the reconstructed muon trajectory,
the open circles the PMTs which recorded a hit and 
the solid dots their projections on the muon trajectory according the Cherenkov hypothesis.}
\label{fig:emission_points}
\end{figure}

The estimate of the track length and of the vertex position proceeds 
through different phases:
\begin{enumerate}
\item The detected photons are projected back to the track according to 
        the Cherenkov angle. The first track length estimate, $l_{\mu}^{\prime}$
        is then defined as the distance between the position of the first
        and last projected photon on the track. The first projected photon
        is the first vertex estimate $\mathbf{V}^{\prime}$.
        If the muon is generated inside or near the instrumented volume, 
        $\mathbf{V}^{\prime}$ is an estimate of the interaction vertex, 
        otherwise it indicates the first photon seen by the detector. 
        For these reasons in the following the vertex estimate will be 
        referred to as the ``pseudo-vertex" estimate.
\item Some specific features of the hits from the hadronic shower are 
        identified and used to select a set of hits around the first 
        pseudo-vertex estimate.
\item The selected hits are fitted with the hypothesis that they originate
        isotropically from a single point. This fit gives a second pseudo-vertex 
        estimate $\mathbf{V}^{\prime\prime}$ and a second track 
        length estimate  $l_{\mu}^{\prime\prime}$.
\item The final pseudo-vertex estimate $\mathbf{V}$ is chosen between
        the first and the second according to the likelihood value of 
        the fit. The corresponding $l_\mu$ is kept.
\end{enumerate}
In the following each stage is described in details. The reader only 
interested in the obtained performances can jump to \mysref{sec:muon:performance}. \\
The procedure to estimate $\mathbf{V}^{\prime}$ and $l_{\mu}^{\prime}$
is sketched in \myfref{fig:emission_points}.
It is assumed that the track direction has already been reconstructed and 
a subset of hits correlated to the track, called \textit{track-hits}, have been selected.
From the position $P_i$ and the time $t_{i}$ of each hit, the corresponding
photon emission point $P^{\prime}_i$ and the emission time $t^{\prime}_i$ can be easily calculated.
The emission points $P^{\prime}_i$ are ordered on the basis of their occurrence
time $t^{\prime}_i$, and the first point $P^{\prime}_0$ is the first pseudo-vertex estimate
$\mathbf{V}^{\prime}$. If $P^{\prime}_n$ is the last emission point identified,
$|P^{\prime}_n - P^{\prime}_0|$ corresponds to the first track length estimate $l_{\mu}^{\prime}$.
Due to the contamination of the optical background and hadronic shower 
photons, a strict selection is needed to identify the \textit{track-hits}.
The following conditions are applied to perform this hit selection:
\begin{itemize}
\item A maximum orthogonal distance from the reconstructed track of 50 m;
\item A time residual with respect to the reconstructed track in the interval [-10;10]~ns;
\item $\cos \theta_{i} < 0$, where $\theta_{i}$ is the expected angle of incidence of the photon on the PMT 
        ($\cos \theta_{i} = -1$ corresponds to a photon hitting head-on the PMT);
\item a minimum density of one point $P^{\prime}_i$  for each 2 meters along the track segment  $\overrightarrow{P^{\prime}_0 P^{\prime}_n}$. 
\end{itemize}
The percentage of background hits contained in the set of \textit{track-hits} is below 2\%.
The set of \textit{track-hits} contains 60$-$70\% of the total 
amount of hits coming from the track. On the other hand, the 
contamination due to the hits produced by the hadronic shower increases with the
inelasticity $y$. For low values of $y$, the largest part of the neutrino
energy is transferred to the muon and almost all the selected hits are 
hits produced by the true muon track. In this case the purity of the 
\textit{track-hits} reach about 98\%. When $y \approx 1$ the hadronic
shower takes almost all the neutrino energy and most of the detected hits
are due to the shower. Consequently, the purity of the selection 
decreases to about 20\%, the track length is overestimated and the 
estimated vertex position is some meters away from the real interaction vertex.
In such a case, the particles produced at the vertex may even travel 
backwards with respect to the muon direction.
To overcome this problem,  a study of 
the distribution in time and space of hits produced at the interaction 
vertex has been performed, with the goal of identifying specific features
in the reconstruction phase which could be used to distinguish hadronic
shower hits among hits due to the optical background and to the muon.

The parameters analysed are the distance $d$ from the estimated 
pseudo-vertex to the hit position, the transverse and longitudinal
projection of $d$ with respect to the reconstructed muon track direction,
called $k$ and $l$ respectively. Moreover, the time evolution of the 
shower hits can also be studied. Under the simplistic assumption that all 
the hits are emitted from the vertex at a time $t_V$, a hit with 
distance $d$ from the vertex should occur at a time $t_V + d/v$, if $v$
is the speed of light in the medium. A ``time residual" can be thus defined
as $\Delta t = t_ i - (t_V + d/v)$, where $t_i$ is the time of the hit.
Finally, the conditions applied to select hits from the shower are:
$l <$ 120 m, $k<$ 100 m, $|\Delta t| <$ 50 ns, and $(k-l)/k>-$2. The 
first two conditions are intended to reject the optical background hits
and identify a region where the shower is likely to be. The other two 
are used to distinguish the shower hits from the hits due to the muon track.

The used cuts are chosen in order to distinguish as much as possible 
shower hits from muon and background hits but trying to keep the few 
hits that are produced by the shower at low energy. Hits selected in 
this way are called \emph{shower-hits}. In this hit set the contamination
due to the background hits is around 2-3\%. The purity of the 
\emph{shower-hits} increase with the inelasticity reaching about 75\%
when $y \sim 1$. The set of \emph{shower-hits} contains about 50\% of
the total number of hits coming from the shower. To find the vertex 
position a maximum likelihood fit applied to the selected \emph{shower-hits}.
A function obtained from the $\Delta t$ distribution for the simulated
shower hits is used as PDF and the final estimate of vertex position is
chosen among the first emission point and the result of the fit. Once 
the vertex has been identified, the track length is scaled according to 
the distance from the estimated vertex and the last back projected photon
on the track. The muon energy is estimated as $E_R = 0.24 \, l_\mu^{rec}$~GeV,
if the estimated track length $l_\mu^{rec}$ is expressed in meters.

A selection of the events based on containment
conditions is needed for this analysis. Such conditions are based 
on the results of the reconstruction. In particular, all the events for 
which the muon reconstructed vertex lies within a volume defined by 
$\vert z\vert < 52$m and $r<107$m, which roughly
corresponds to the instrumented volume, are selected.

\paragraph{Bjorken $y$ estimation\\}\label{sec:recoLNS_By}

\begin{figure}
\centering%
\begin{overpic}[width=0.8\textwidth]{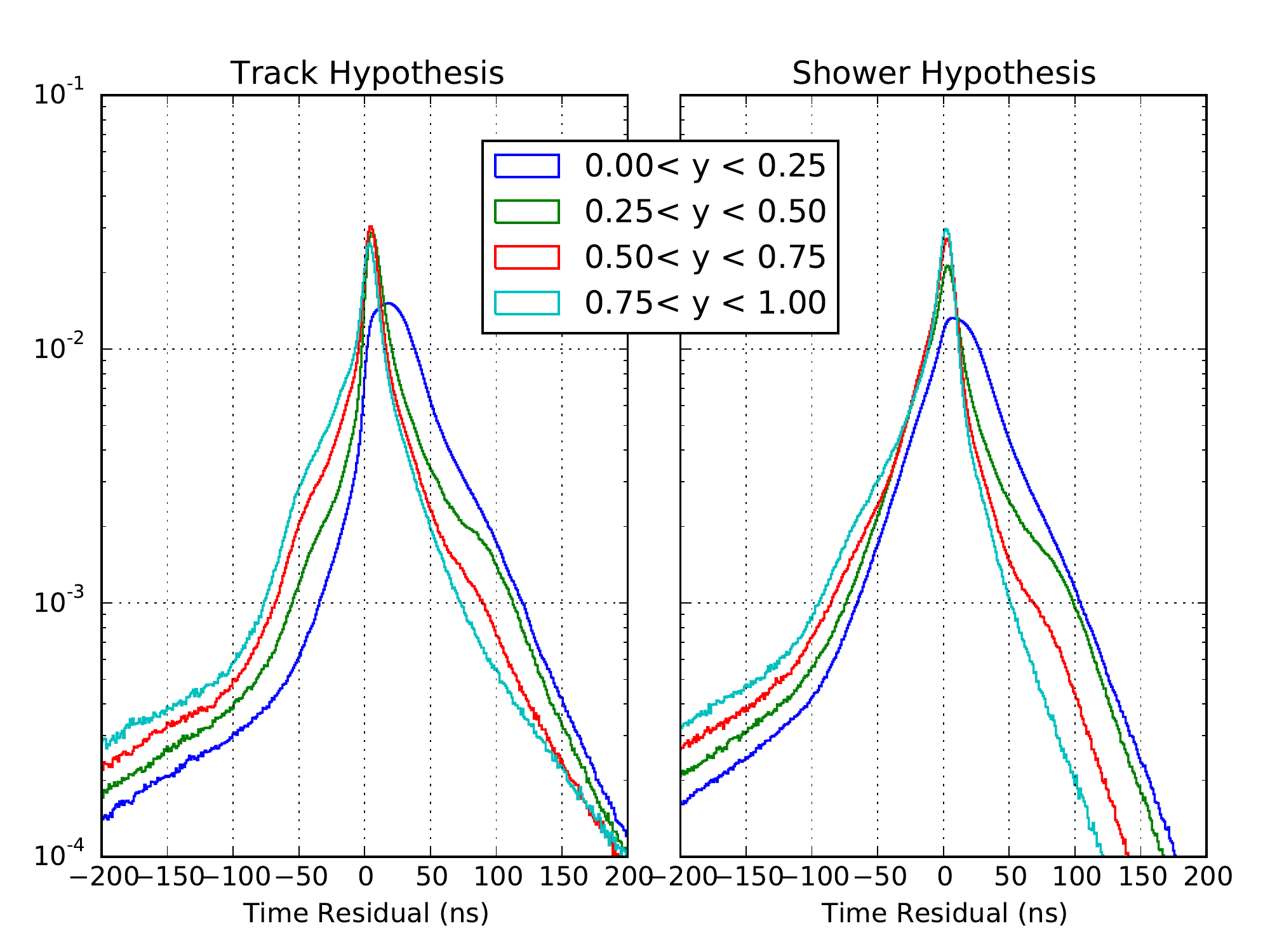}
\put (9,64) {\bf KM3NeT}
\put (81,64) {\bf KM3NeT}
\end{overpic}
\caption{Normalised distributions of the time residuals for muon neutrino 
charged current events with energy higher than 5~GeV and whose vertex is 
reconstructed within the instrumented volume for 4 Bjorken $y$ intervals, 
with respect to the reconstructed muon track on the left (track hypothesis) 
and with respect to the reconstructed vertex on the right (shower hypothesis).
For the blue curve, the peak at $\rm Time Residual = 0$ 
is less sharp due to the lower resolution of the vertex reconstruction at
low Bjorken $y$ interactions, caused by the lower amount of light emitted at the
interaction vertex.}
\label{fig:numu_tres_vs_y}
\end{figure}

\begin{figure}[htpb]
\centering
\begin{minipage}[c]{0.48\textwidth}
\centering
{\begin{overpic}[width=\textwidth]{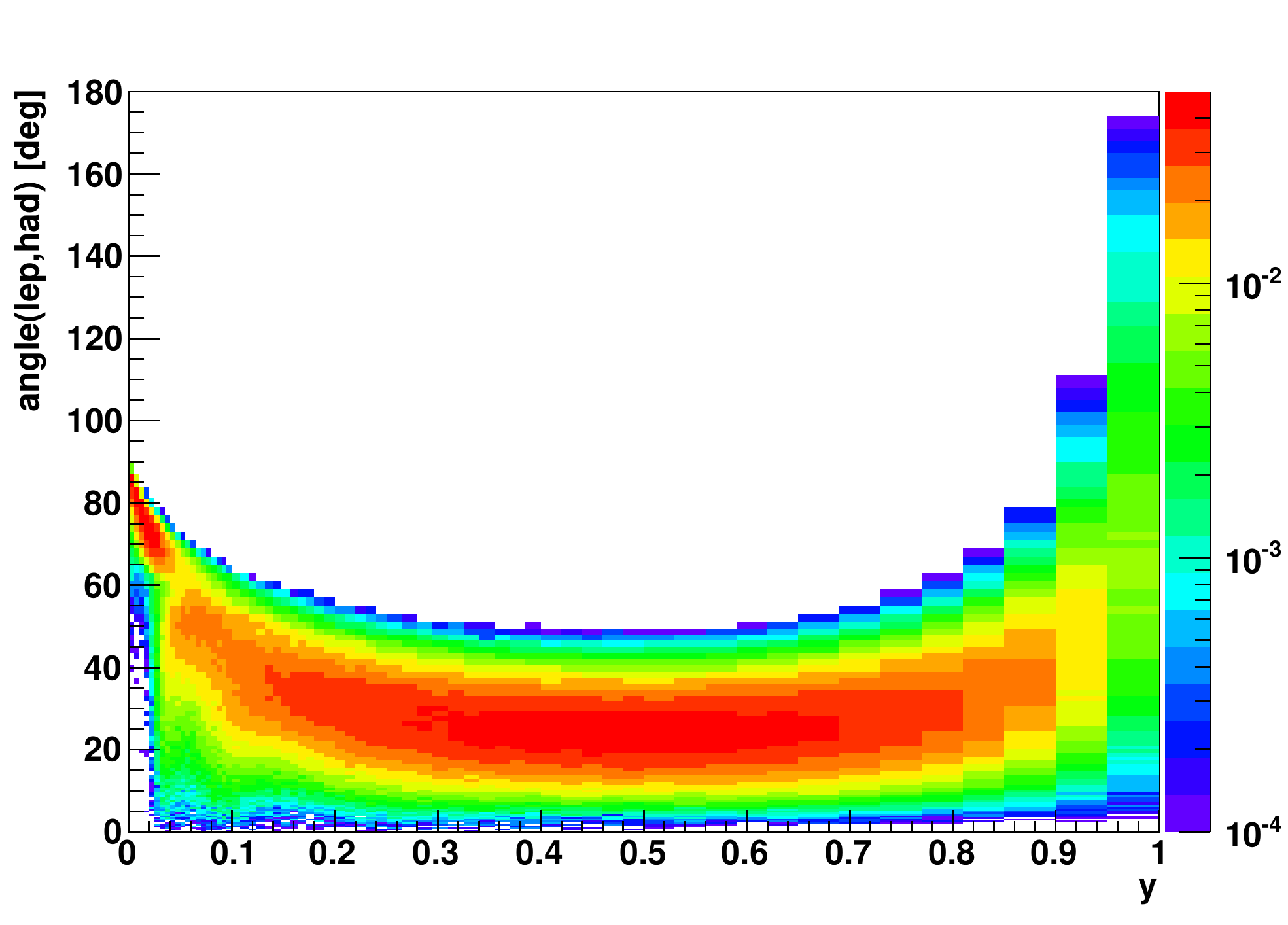}
\put (40,66) {\bf KM3NeT}
\end{overpic}
}
\end{minipage}
\hfill
\begin{minipage}[c]{0.48\textwidth}
\centering
{\begin{overpic}[width=\textwidth]{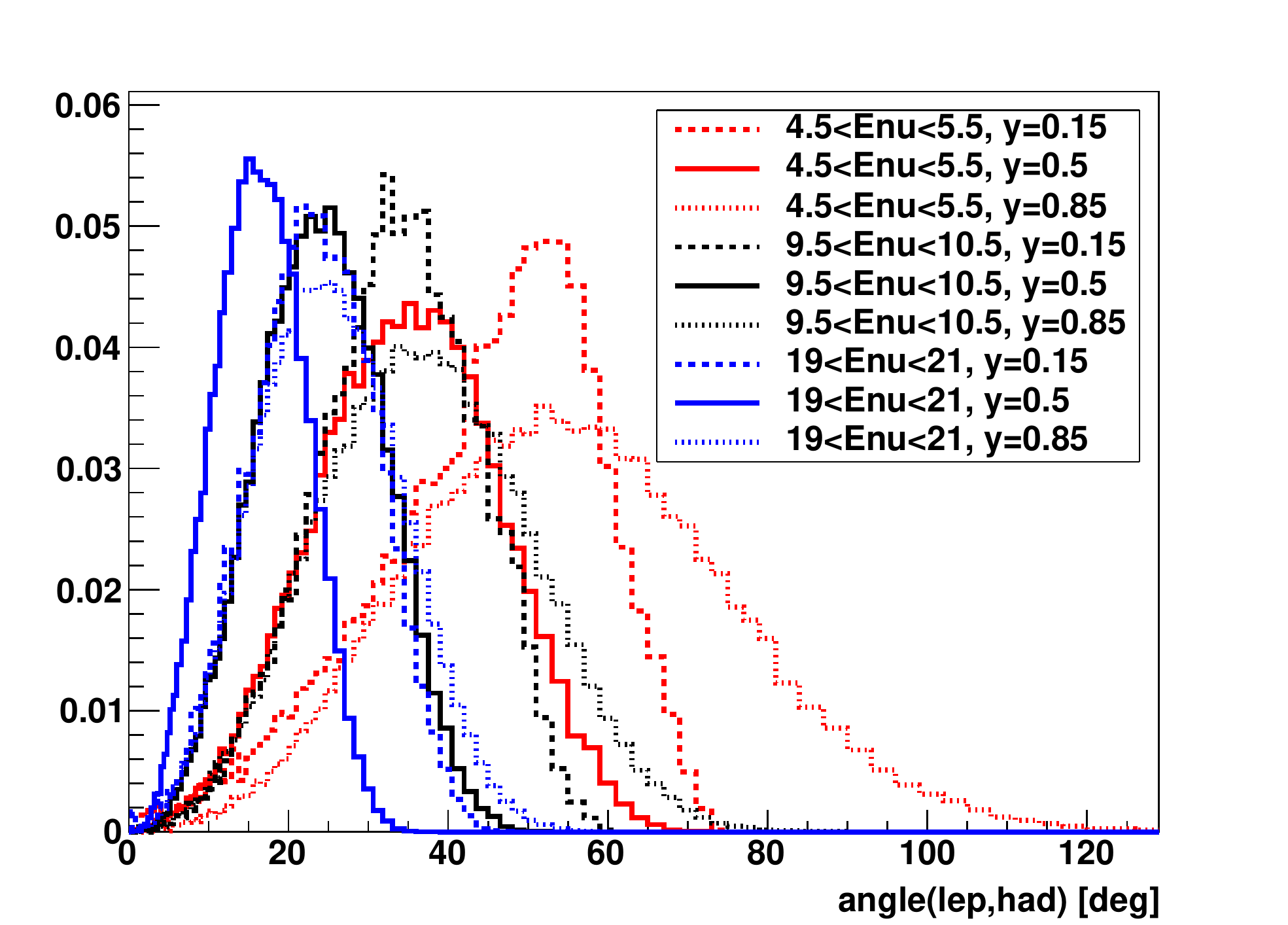}
\put (40,66) {\bf KM3NeT}
\end{overpic}
}
\end{minipage}
\caption{
Angle between the outgoing muon and the outgoing hadronic shower
$\phi_{\mu,\mathrm{lep}}$ from $\nu_\mu \mathrm{CC}$ interactions with
$9.5< E_\nu/\mathrm{GeV} <10.5$ as a function of inelasticity $y$. The
kinematics looks very similar for $\nu_e \mathrm{CC}$ interactions. The
features at $0<y<0.2$ are due to the different neutrino interaction
channels (cf. \myfref{fig:bjorkenY}). Each column of bins in $y$ is
normalised to 1 (left).
Distribution of the opening angle $\phi_{\mu,\mathrm{lep}}$ in
$\nu_\mu \mathrm{CC}$ interactions for three different inelasticities
$y$ and neutrino energies (right). 
}
\label{fig:shower_angleLepHad}
\end{figure}

\begin{figure}
\centering%
{\begin{overpic}[width=0.6\textwidth]{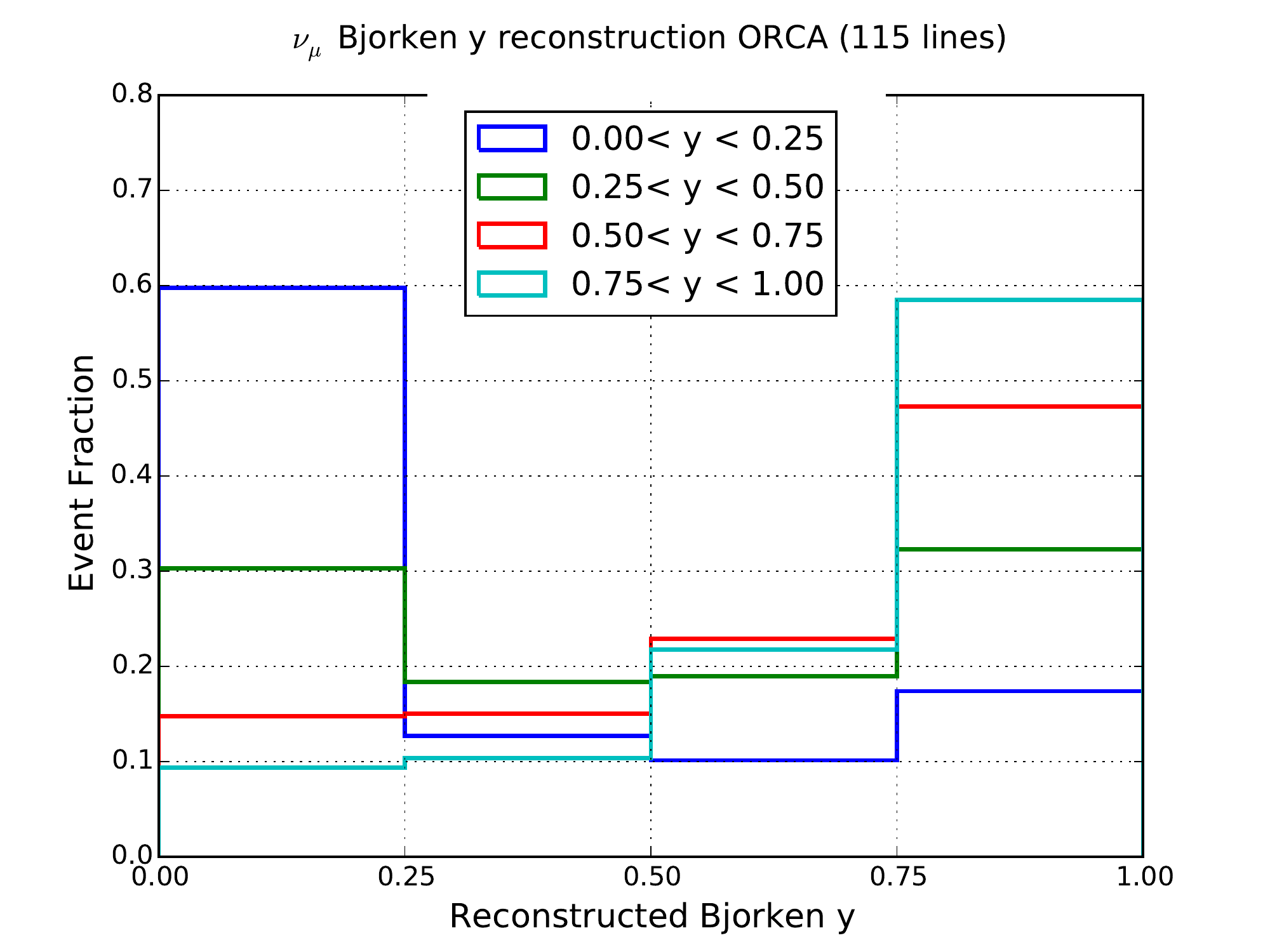} 
\put (43,46) {\bf KM3NeT}
\end{overpic}
}
\caption{Distribution of the reconstructed Bjorken $y$ (horizontal axis)
for four bins of true Bjorken $y$ (colour code). Each histogram is 
normalised to 1. The plot refers only to muon neutrinos interacting by
charged current (no antineutrinos).}
\label{fig:yresolution}
\end{figure}

The Bjorken $y$ is estimated on the basis of the distribution of the
time residuals of the \emph{selected hits} (cf. \mysref{sec:LNSreco})
with respect to the reconstructed track and with respect to the 
reconstructed vertex, according to a track and a shower hypothesis 
respectively (\myfref{fig:numu_tres_vs_y}).
The simulated angle between the outgoing muon and the outgoing hadronic shower $\phi_{lep,\mathrm{had}}$ 
is shown in \myfref{fig:shower_angleLepHad} 
for neutrinos with $E_\nu \approx 10\,\mathrm{GeV}$. 
 The rationale 
for this approach relies on the fact that a different repartition of the
total neutrino energy among the hadronic shower and the muon influences 
the distribution of the time residuals. The estimation is performed
among four Bjorken $y$ intervals: 0-0.25, 0.25-0.50, 0.50-0.75 and 0.75-1.
For each interval, the log-likelihood of the time 
residuals is calculated for a track hypothesis, if the tested Bjorken $y$ 
is $<0.5$, or a shower hypothesis, if the tested Bjorken $y$ 
is $>0.5$. The Bjorken $y$ interval corresponding to the 
highest likelihood is chosen. The performance of the algorithm is shown
in \myfref{fig:yresolution}.

\paragraph{Neutrino energy estimation\\}

The estimation of the neutrino energy is obtained by combining the 
estimated track length and estimated Bjorken $y$ with the number of hits
used by the muon track reconstruction.
Depending on the reconstructed track length, the true 
neutrino energy can be related to the number of hits used by the track 
reconstruction ($n_{\rm hits}^{\rm fit}$), or equivalently the number of 
degrees of freedom of the fit 
$\mathrm{NDoF}_{\rm fit} = n_{\rm hits}^{\rm fit}-5$.
The relation between $\mathrm{NDoF}_{\rm fit}$ 
and the energy of the interacting neutrino, for a certain interval of 
reconstructed muon track length, is obtained 
by fitting the median distribution of
 $E_{\nu}$ as a function of  $\mathrm{NDoF}_{\rm fit}$. 
In order to further improve the accuracy,
two different estimations are used, taking into account the
reconstructed Bjorken $y$ being higher or lower than 0.5.

\subsubsection{Performance}
\label{sec:muon:performance}
The simulations have been performed with the 3.7\,Mt benchmark detector presented in \mysref{simu:det}. The performances of the reconstruction algorithm have also been studied with configurations that mimic vertical spacings of $9\,\mathrm{m}$,
$12\,\mathrm{m}$ and $15\,\mathrm{m}$ according to the masking
procedure described in \mysref{sec:masking}.
The instrumented volume is the same for all the
mentioned configurations.

\begin{figure}[h!]
\centering%
{\begin{overpic}[width=0.48\linewidth]{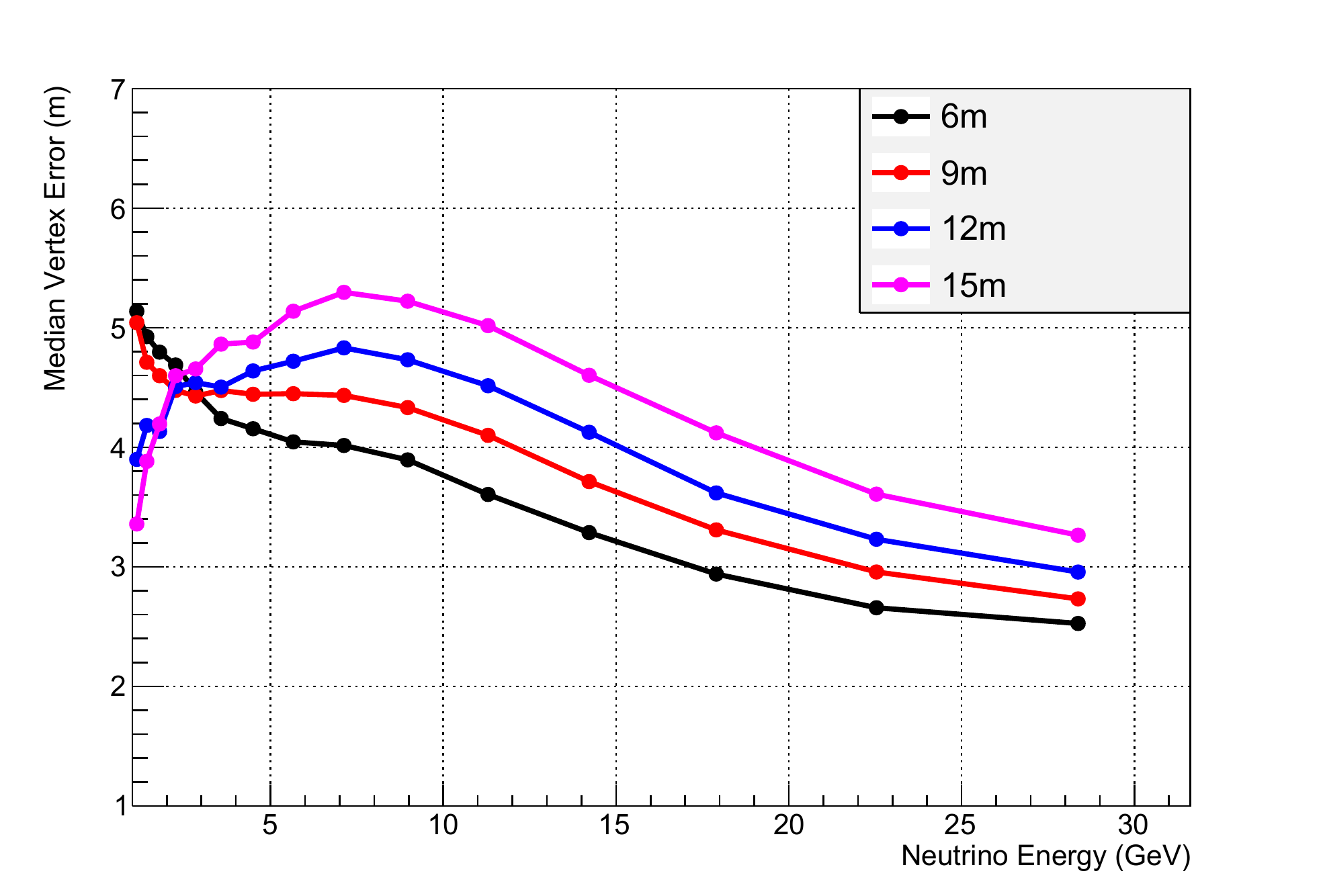}
\put (12,56) {\bf KM3NeT}
\end{overpic}
}
{\begin{overpic}[width=0.48\linewidth]{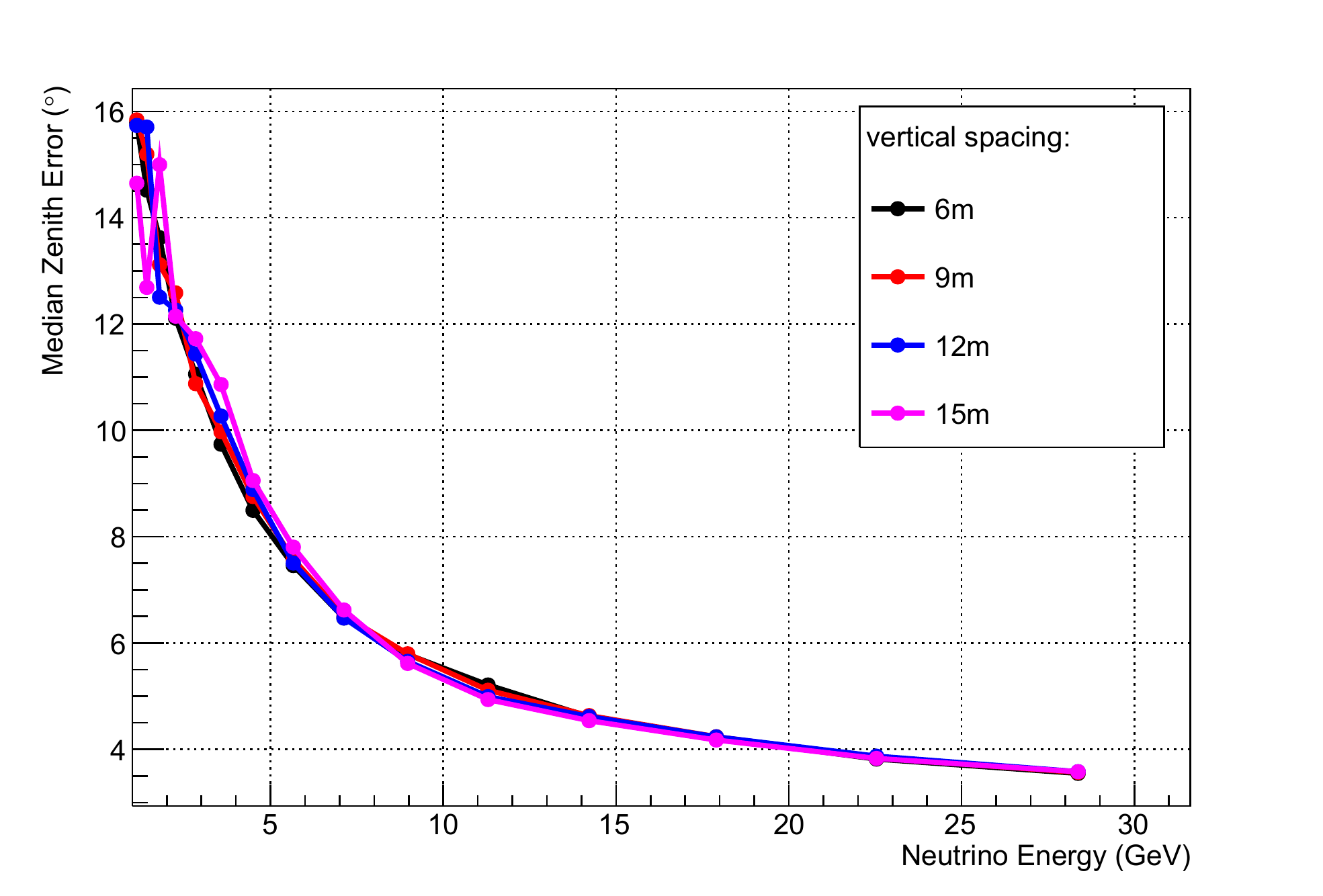} 
\put (12,56) {\bf KM3NeT}
\end{overpic}
}
{\begin{overpic}[width=0.48\linewidth]{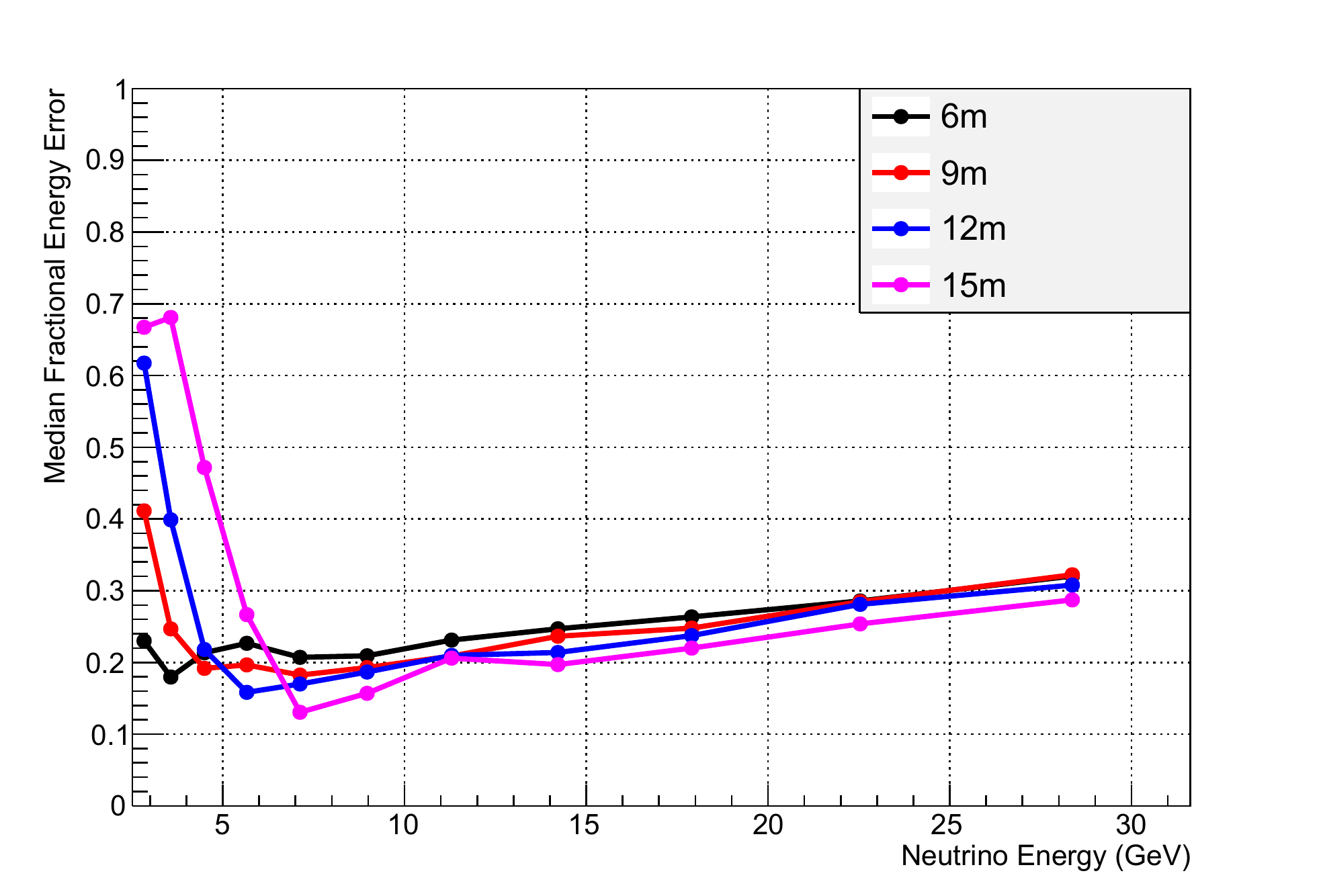}
\put (12,56) {\bf KM3NeT}
\end{overpic}
}
\caption{Median resolution as a function of the true neutrino energy, for 
various vertical spacings, of: the distance between the true interaction vertex and reconstructed one
(top left), the absolute value of the difference between the reconstructed
zenith angle and the true neutrino zenith (top right), and the fractional
energy error (bottom). For muon neutrino and antineutrino events 
weighted according to the atmospheric spectrum, reconstructed as up-going,
with vertex reconstructed within the instrumented volume and $\Lambda>-5.0$.}
\label{fig:reso_numu}
\end{figure}

\myfref{fig:reso_numu} shows the performances for events reconstructed as up-going, whose 
vertex is reconstructed within the instrumented volume, with quality cut of the
reconstruction algorithm of $\Lambda > -5.0$ (see \myeref{eq:lambda}). The top left plot 
shows the median distance between the true and estimated vertex position,
$distance(P_{vertex}^{true},P_{vertex}^{reco})$, as a function of the 
neutrino energy. The value of the $distance(P_{vertex}^{true},P_{vertex}^{reco})$ 
is of the order of a few meters for all reconstructed events. 
The top right plot in \myfref{fig:reso_numu} shows the resolution 
on the reconstructed neutrino zenith angle and the bottom plot shows the
fractional energy resolution, which is defined as $|E_\nu - E_{\rm rec}|/E_\nu$.

\begin{figure}[t]
\centering
{\begin{overpic}[width=0.48\linewidth]{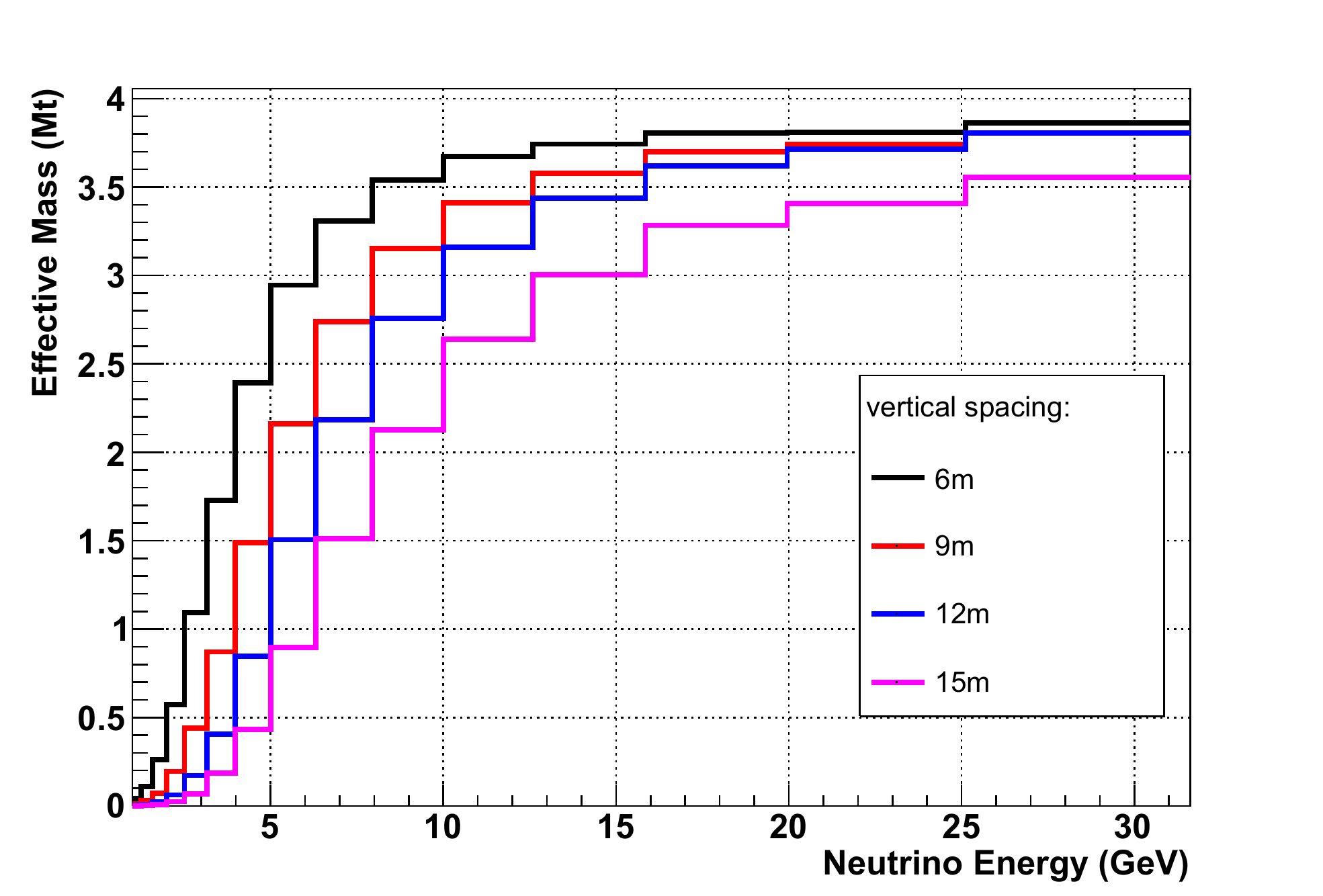}
\put (12,56) {\bf KM3NeT}
\end{overpic}
}
{\begin{overpic}[width=0.48\linewidth]{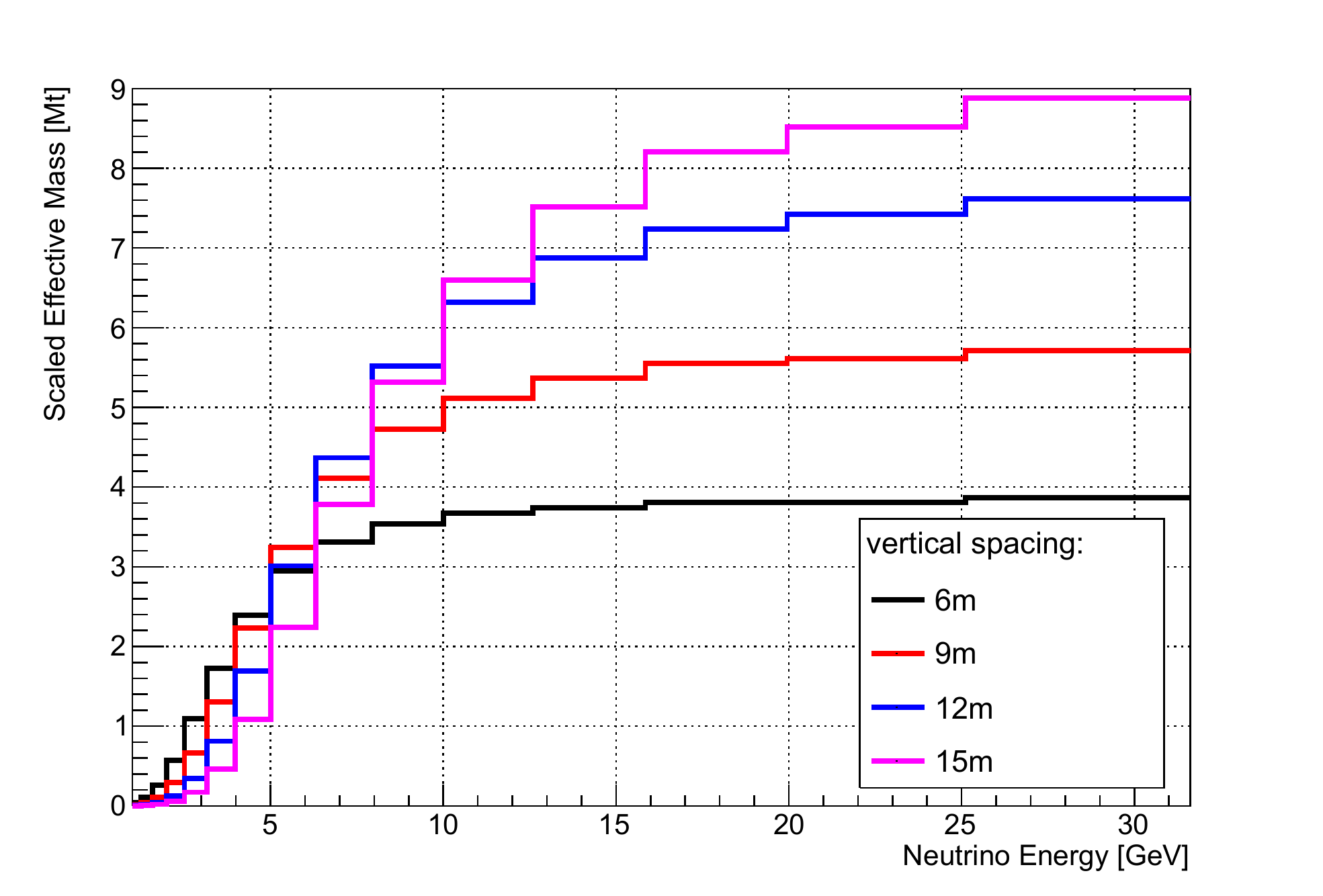}
\put (12,56) {\bf KM3NeT}
\end{overpic}
}
{\begin{overpic}[width=0.48\linewidth]{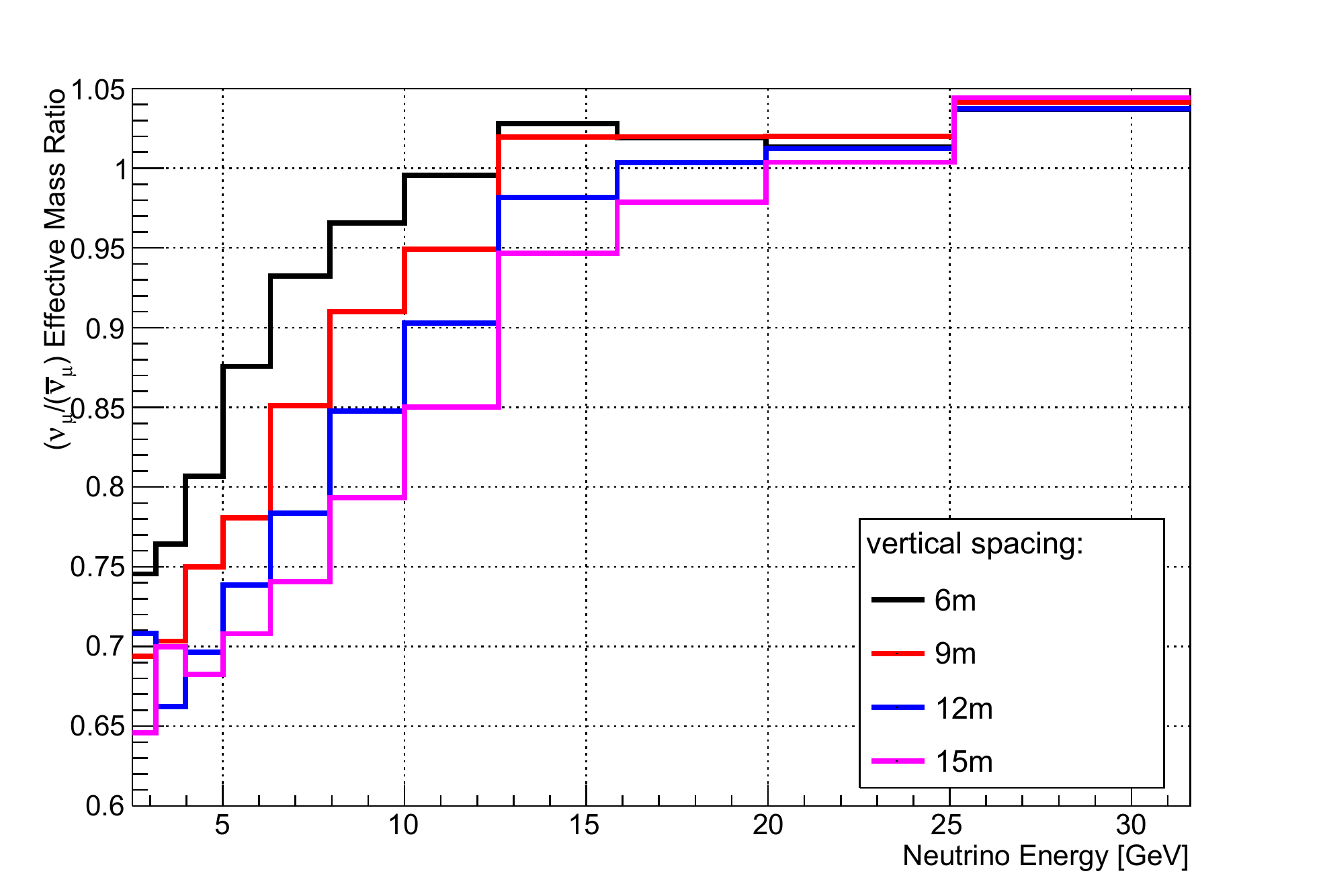}
\put (12,56) {\bf KM3NeT}
\end{overpic}
}
\caption{Top left: effective mass for $\nu_\mu$ and $\bar{\nu}_\mu$
 as a function of the neutrino energy for events
reconstructed as up-going and whose vertex is reconstructed inside the 
instrumented volume with a reconstruction quality $\Lambda>-5.0$, for
4 different vertical spacings. Top right: effective masses
for 6\,m/9\,m/12\,m/15\,m scaled by a factor
1/1.5/2/2.5.
Lower plot: ratio between neutrino
and antineutrino effective masses as a function of the neutrino energy.}
\label{fig:recoLNS_EffMass}
\end{figure}

Another parameter needed to evaluate the reconstruction performance as 
well as to calculate the sensitivity for the measurement of the neutrino
mass hierarchy is the detector effective volume. The effective volume
$V_{\rm eff}$ can be defined as the volume of a 100\% efficient detector
for observing neutrinos that interact within that volume, for a set of
specified quality cuts. In the simulation adopted, described in
\mysref{sec:simulations}, all the neutrinos interacting within a
volume larger than the instrumented volume and surrounding the detector,
that can be referred to as generation volume $V_{\rm gen}$, are kept
for the subsequent steps of the simulation and, eventually, the
reconstruction. The effective volume is then obtained by scaling
$V_{\rm gen}$ with the ratio of the reconstructed events $N_{\rm rec}$
(or selected according to a given criterion) and the generated events
$N_{\rm gen}$:
\begin{equation}
 V_{\rm eff}(E_{\nu}, \theta_{\nu}) = \dfrac{N_{\rm rec}(E_{\nu}, \theta_{\nu})}{N_{\rm gen}(E_{\nu}, \theta_{\nu})} V_{\rm gen}.
 \label{eq:veff}
\end{equation}
Assuming a seawater density of 1.025 g/cm$^{3}$, the effective volume is
converted into an effective mass $M_{\rm eff}$. The $M_{\rm eff}$
calculated for events with a quality parameter $\Lambda>-5.0$ and whose
vertex is reconstructed within the instrumented volume is plotted in
\myfref{fig:recoLNS_EffMass} as a function of the neutrino energy and
for various intervals of the direction of the incoming neutrino.

\subsection{Electron neutrino studies}
\label{electron}
This section describes the methodology and performance of a 
reconstruction strategy that has been developed for neutral-current and charged-current shower-like
events in ORCA \cite{bib:hofestaedt2016}. Electron neutrino events will play a crucial role for the
envisaged mass hierarchy measurement, good angular and energy
resolutions are therefore mandatory.

 \subsubsection{Phenomenology of shower events} 
\label{sec:shower_phenomenology}
Charged-current (CC) interactions of electron (anti)neutrinos $\nuan_e$ with nucleons constitute a very important 
signal class for the neutrino mass hierarchy measurement. They result in a particle shower:
\begin{equation}
\nuan_e + N \rightarrow e^\pm + h ,
\label{eq:nueCC_interaction}
\end{equation}
where $N$ refers to the target nucleon and $h$ to the hadronic system in the final state. 
The outgoing electron initiates an \textit{electromagnetic shower} while the hadronic system
develops into a \textit{hadronic shower} with a possibly complex structure of 
hadronic or electromagnetic sub-showers,
depending on the decay modes of individual particles in the shower.  

In the following, the energy $E_{\rm had}$ and momentum $\vec p_{\rm had}$ of the hadronic shower
are defined by the difference of the respective energy and momentum of the neutrino and the 
electron:
\begin{equation}
(E_{\rm had}, \vec p_{\rm had}) = (E_\nu, \vec p_\nu) - (E_e, \vec p_e) .
\label{eq:def_hadShower}
\end{equation}
The inelasticity $y$ (``Bjorken $y$'') of the reaction
is defined as:
\begin{equation}
y = \frac{ E_{\rm had} } { E_\nu } = \frac{ E_\nu - E_e } { E_\nu } .
\label{eq:def_y}
\end{equation}
In events induced by the neutral-current (NC) weak interaction of a neutrino on a nucleon
only a hadronic shower is visible.

\paragraph{Kinematics\\}

The kinematics for $\nu_e \mathrm{CC}$ interactions is similar to
that presented in \myfref{fig:shower_angleLepHad}. 
The angle is minimal for $y=0.5$ with a mean value of roughly $25\,^{\circ}$. 
For $y \rightarrow 0$ ($y \rightarrow 1$) the angle between the incoming neutrino 
and the outgoing hadronic shower (lepton) becomes larger, 
leading to larger $\phi_{lep,\mathrm{had}}$. 
For increasing neutrino energies the angle $\phi_{lep,\mathrm{had}}$ becomes smaller.

\paragraph{Light production in showers\\}

Some information about the Cherenkov light production of showers can be found 
in the literature,  e.g. in \cite{bib:kopper} and references therein. 
Mostly, however, previous studies have focused on energies well above those relevant for ORCA. 
Therefore, the most important characteristics of showers, as obtained from Monte-Carlo simulation studies, 
in the relevant energy range for ORCA are briefly summarised in the following.

In general, an electromagnetic shower consists of a cascade of $e^\pm$ emitting 
photons via brems\-strahlung, which interact with matter and again produce 
$e^\pm$-pairs via pair production. 
The evolution of a hadronic shower is similar but the initial particles 
are hadrons and the developing cascade will show significantly larger fluctuations 
as it is dominated by particle decays.
In water the electromagnetic and nuclear 
interaction lengths are roughly $36\,\mathrm{cm}$ and $83\,\mathrm{cm}$ \cite{bib:PDG}, respectively. 
Therefore, compared to muon tracks, showers appear in first approximation as a point-like 
burst of light in the detector. The light is emitted by charged particles with energies
above their Cherenkov threshold.

\begin{figure}[h!]
\centering
\begin{minipage}[c]{0.48\textwidth}
\centering
    {\begin{overpic}[width=\textwidth]{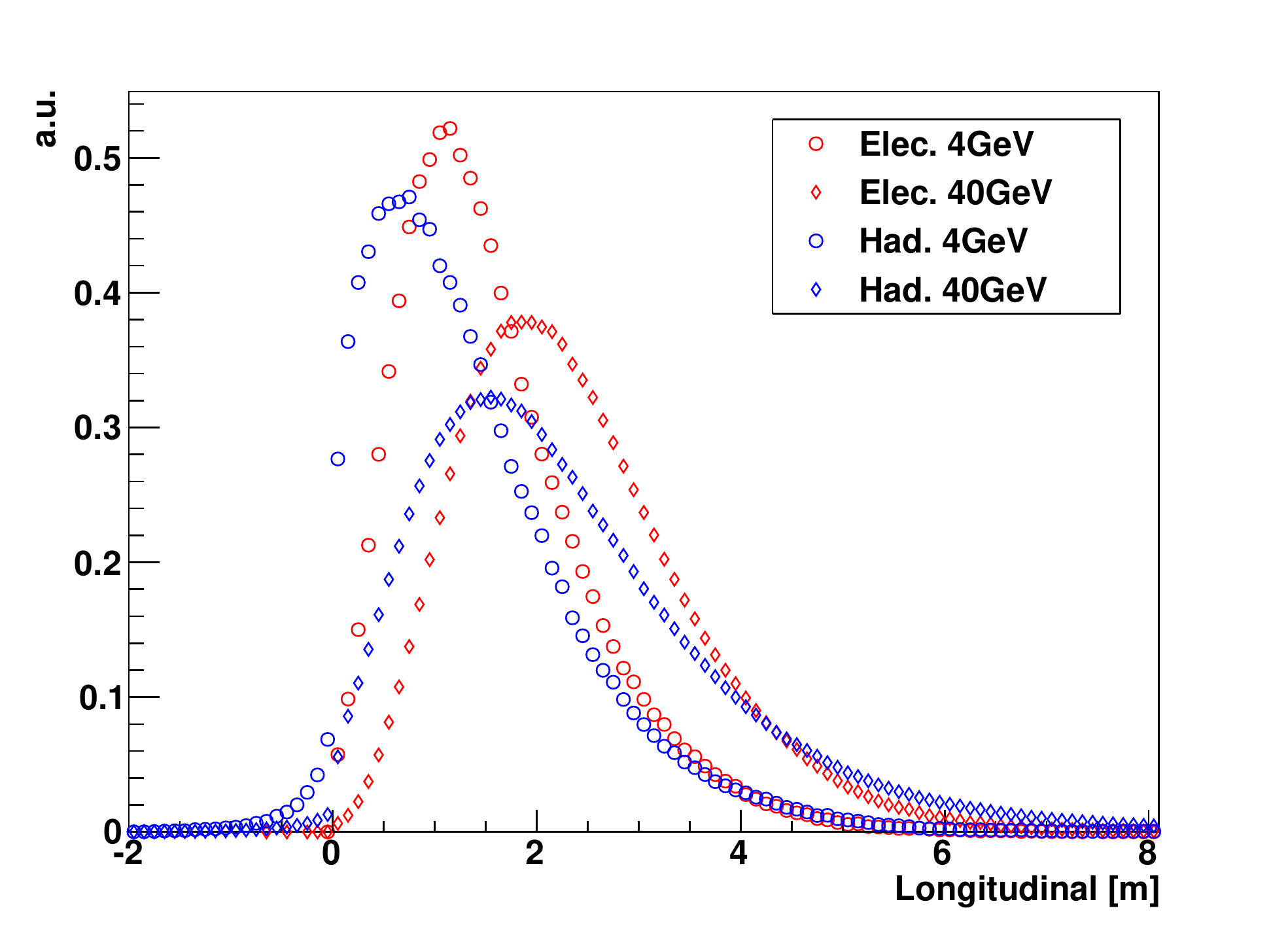}
\put (40,66) {\bf KM3NeT}
\end{overpic}
}
\end{minipage}
\hfill
\begin{minipage}[c]{0.48\textwidth}
\centering
{\begin{overpic}[width=\textwidth]{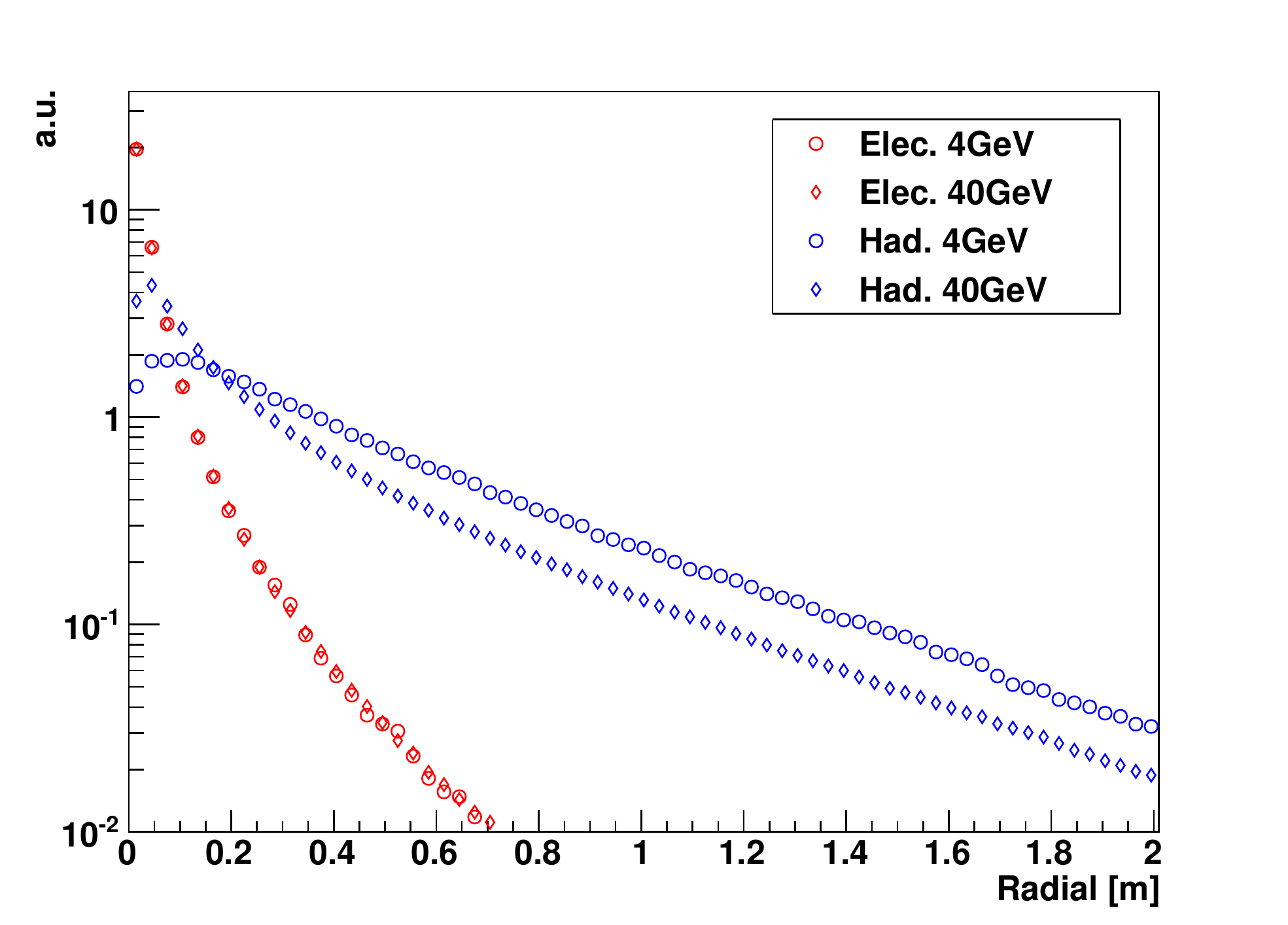}
\put (40,66) {\bf KM3NeT}
\end{overpic}
}
\end{minipage}
\caption{Light emission profiles of electromagnetic (red) and hadronic showers (blue) with 4\,GeV and 40\,GeV energy, depicted in shower direction (left, longitudinal) and perpendicular to the shower direction (right, transverse).}
\label{fig:shower_emisPosi_comparison}
\end{figure}

The longitudinal and transverse light emission profile of electromagnetic and hadronic 
showers can be seen in \myfref{fig:shower_emisPosi_comparison}. 
For the energies of interest the brightest point of a shower is offset roughly 1-2\,m 
in the shower direction. The longitudinal extension of the showers increases with $\log(E)$. 
In spite of the larger interaction length the longitudinal offset for hadronic showers 
is smaller than for electromagnetic showers with the same shower energy $E_e = E_{\rm had}$, 
since they are initiated by several hadrons, each with an energy below $E_{\rm had}$, 
and the initial hadrons have different directions reducing the longitudinal extension 
when projecting onto the shower axis. 
The transverse extension of the showers is negligible compared to the longitudinal.

\begin{figure}[h!]
\centering
\begin{minipage}[c]{0.325\textwidth}
\centering
    {\includegraphics[trim={0cm 0 0 0}, clip=true, width =\textwidth]{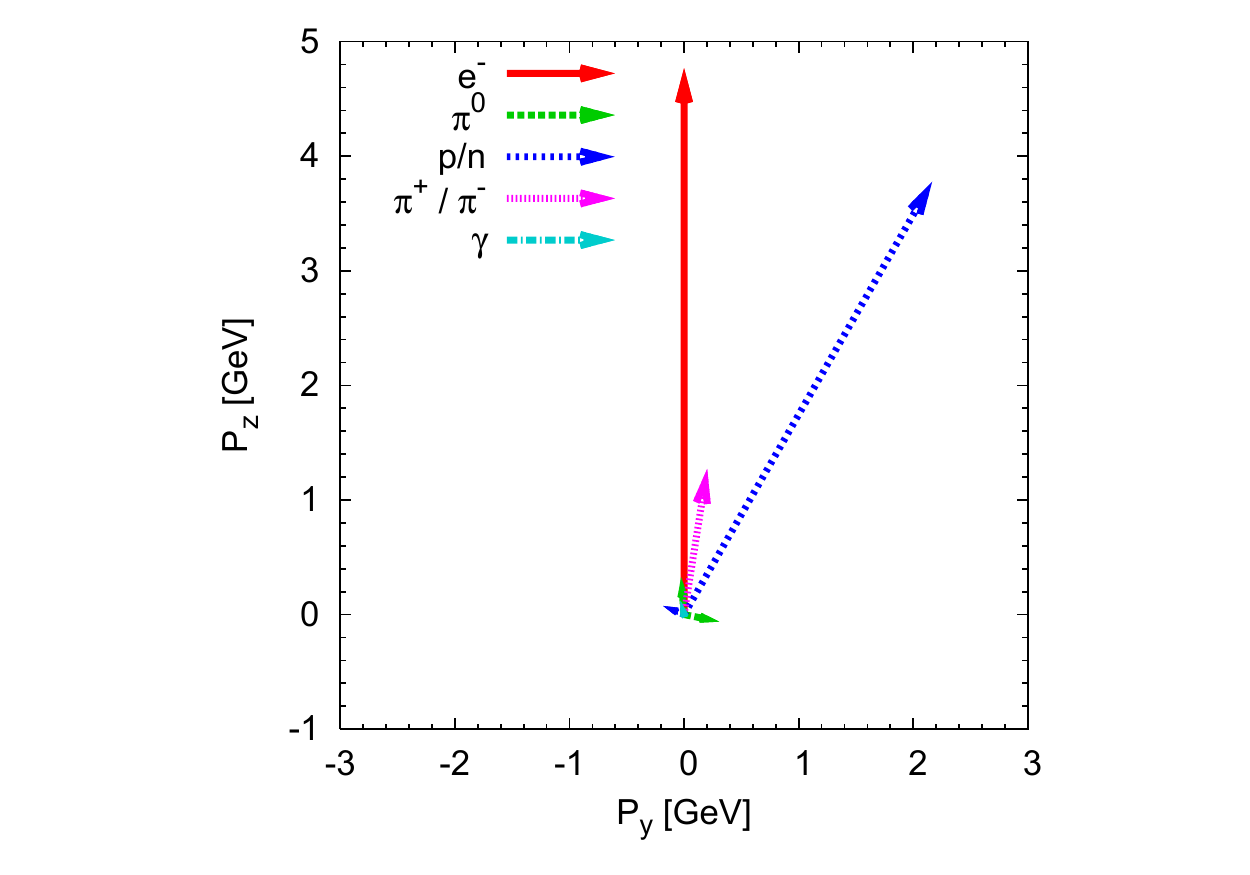}}
\end{minipage}
\hfill
\begin{minipage}[c]{0.325\textwidth}
\centering
    {\includegraphics[trim={0cm 0 0 0}, clip=true, width=\textwidth]{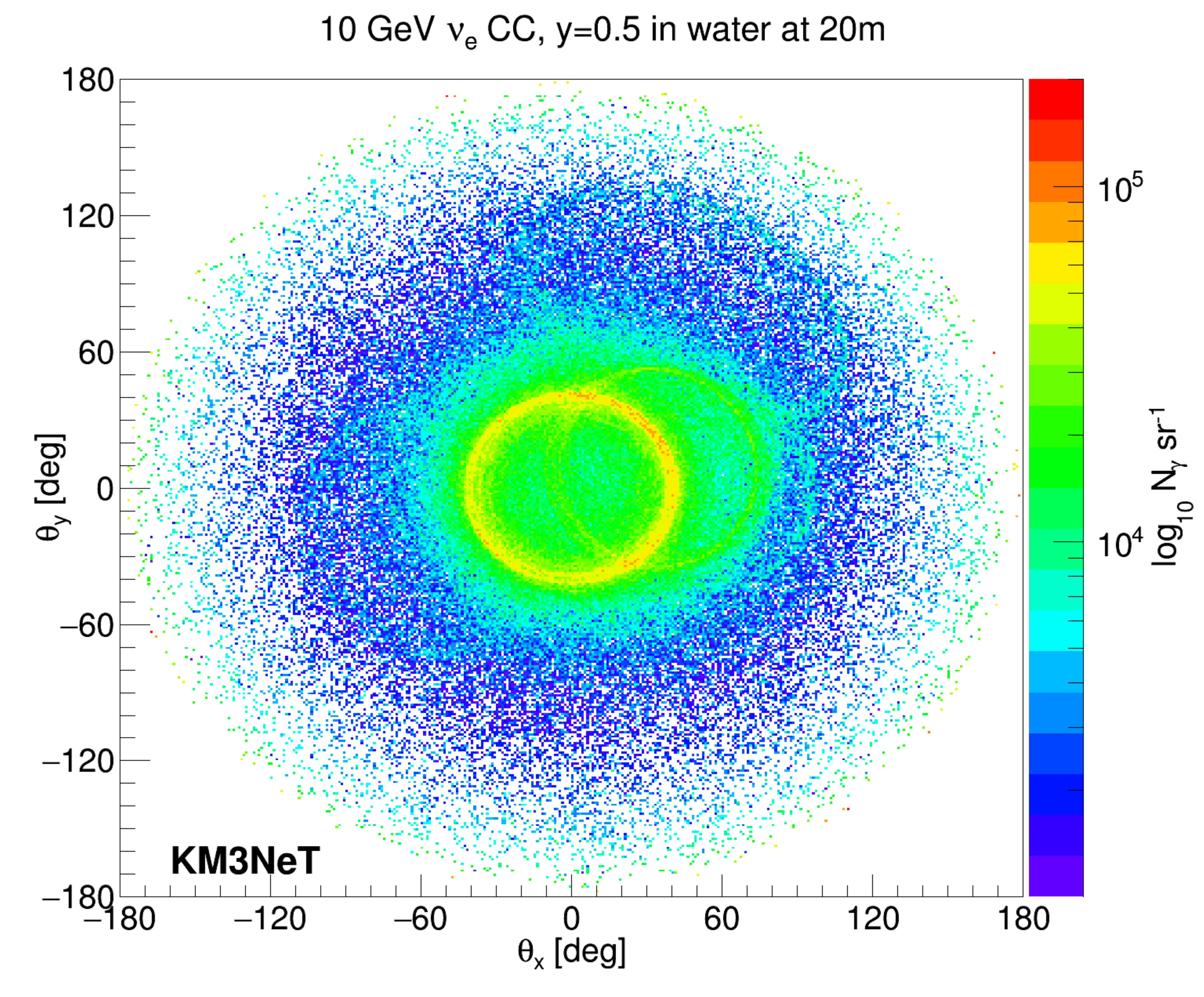}}
\end{minipage}
\hfill
\begin{minipage}[c]{0.325\textwidth}
\centering
    {\includegraphics[trim={0cm 0 0 0}, clip=true, width=\textwidth]{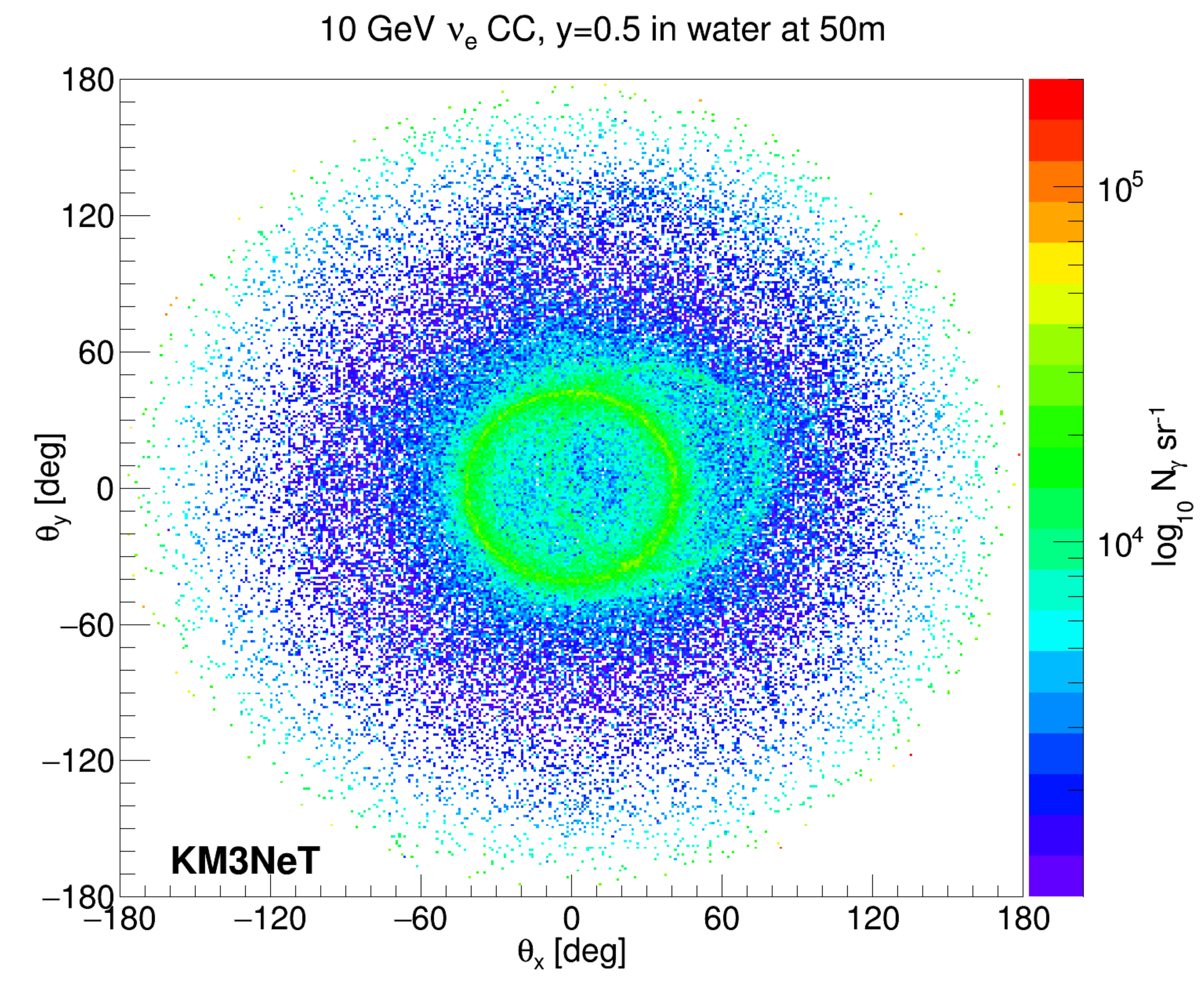}}
\end{minipage}
\vfill
\vspace{0.25cm}
\begin{minipage}[c]{0.325\textwidth}
\centering
    {\includegraphics[trim={0cm 0 0 0}, clip=true, width =\textwidth]{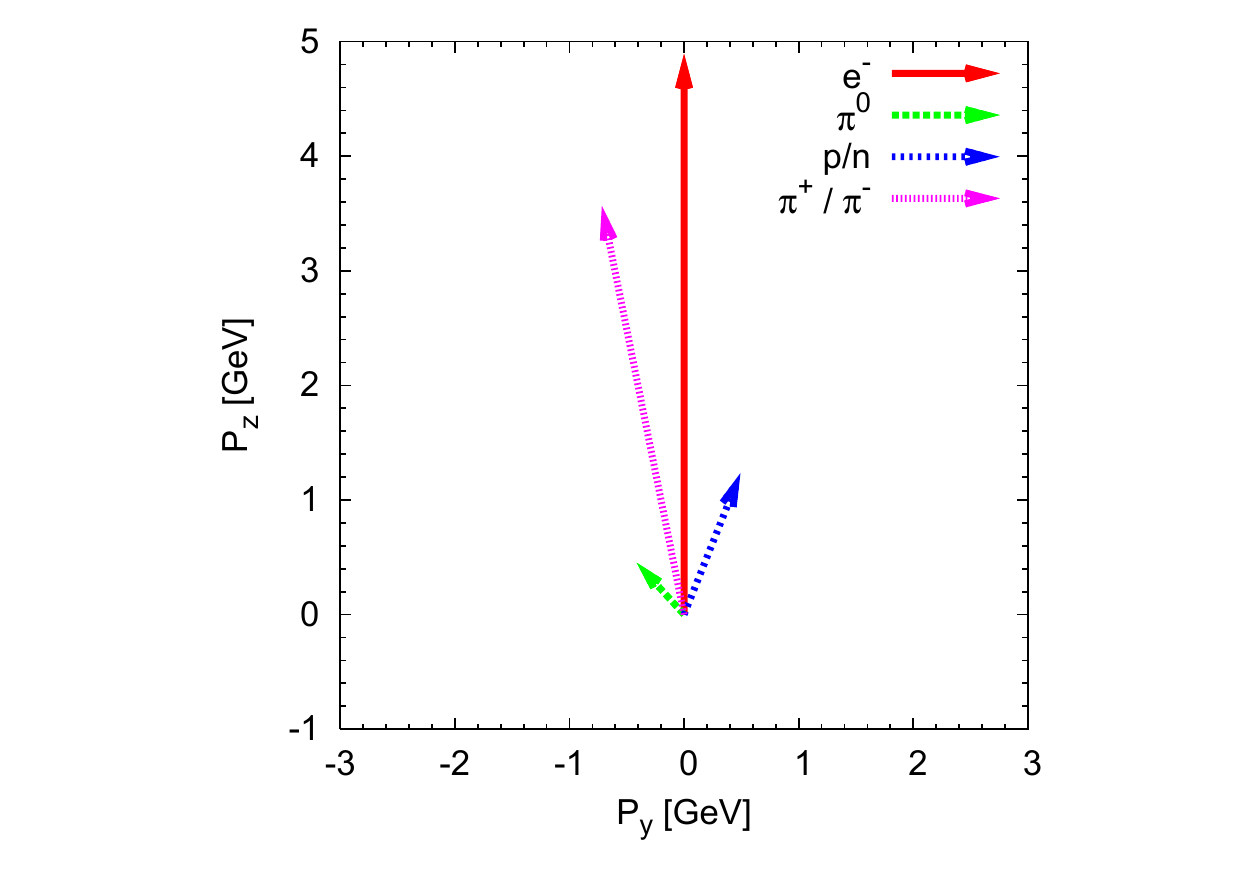}}
\end{minipage}
\hfill
\begin{minipage}[c]{0.325\textwidth}
\centering
    {\includegraphics[trim={0cm 0 0 0}, clip=true, width=\textwidth]{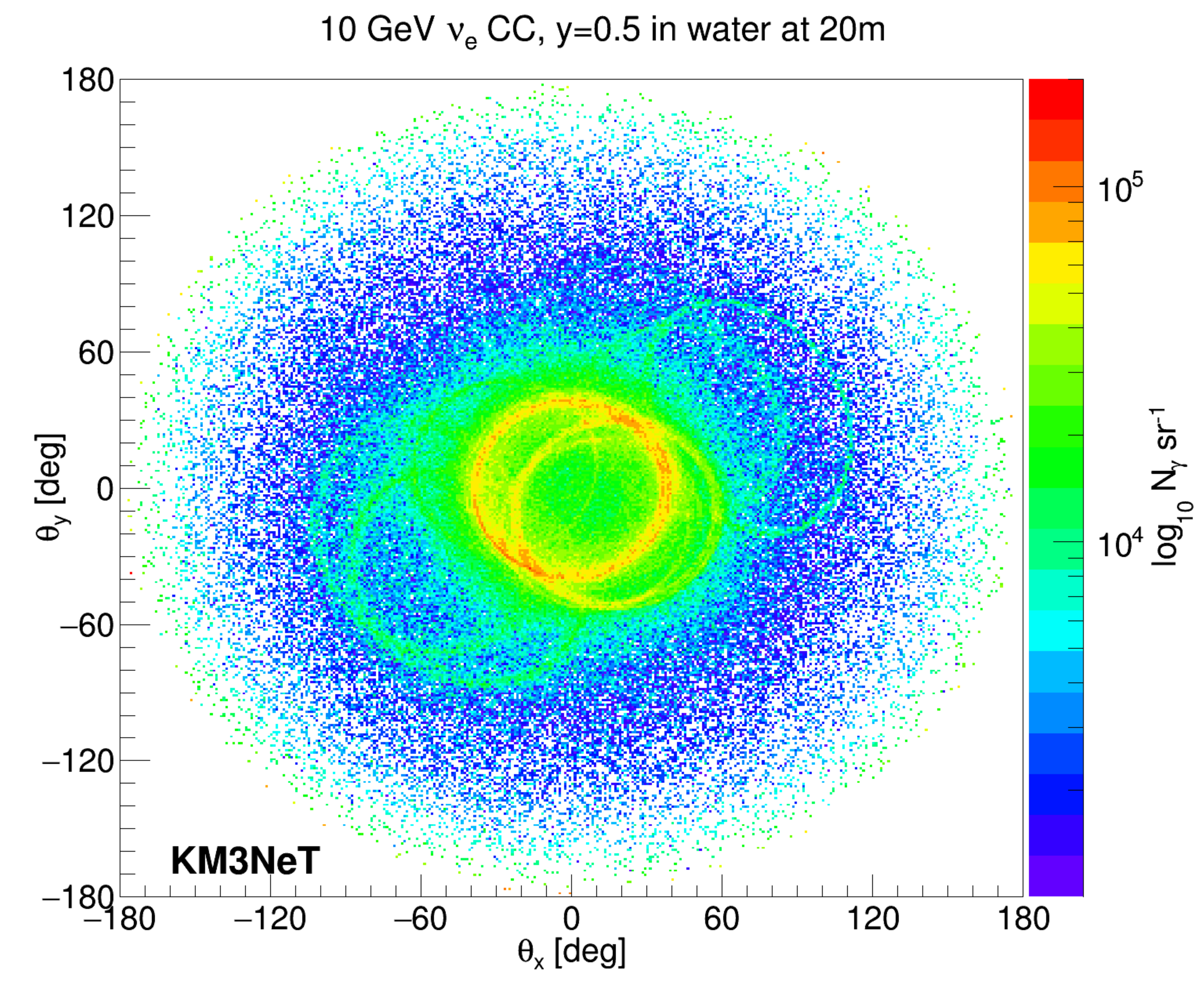}}
\end{minipage}
\hfill
\begin{minipage}[c]{0.325\textwidth}
\centering
    {\includegraphics[trim={0cm 0 0 0}, clip=true, width=\textwidth]{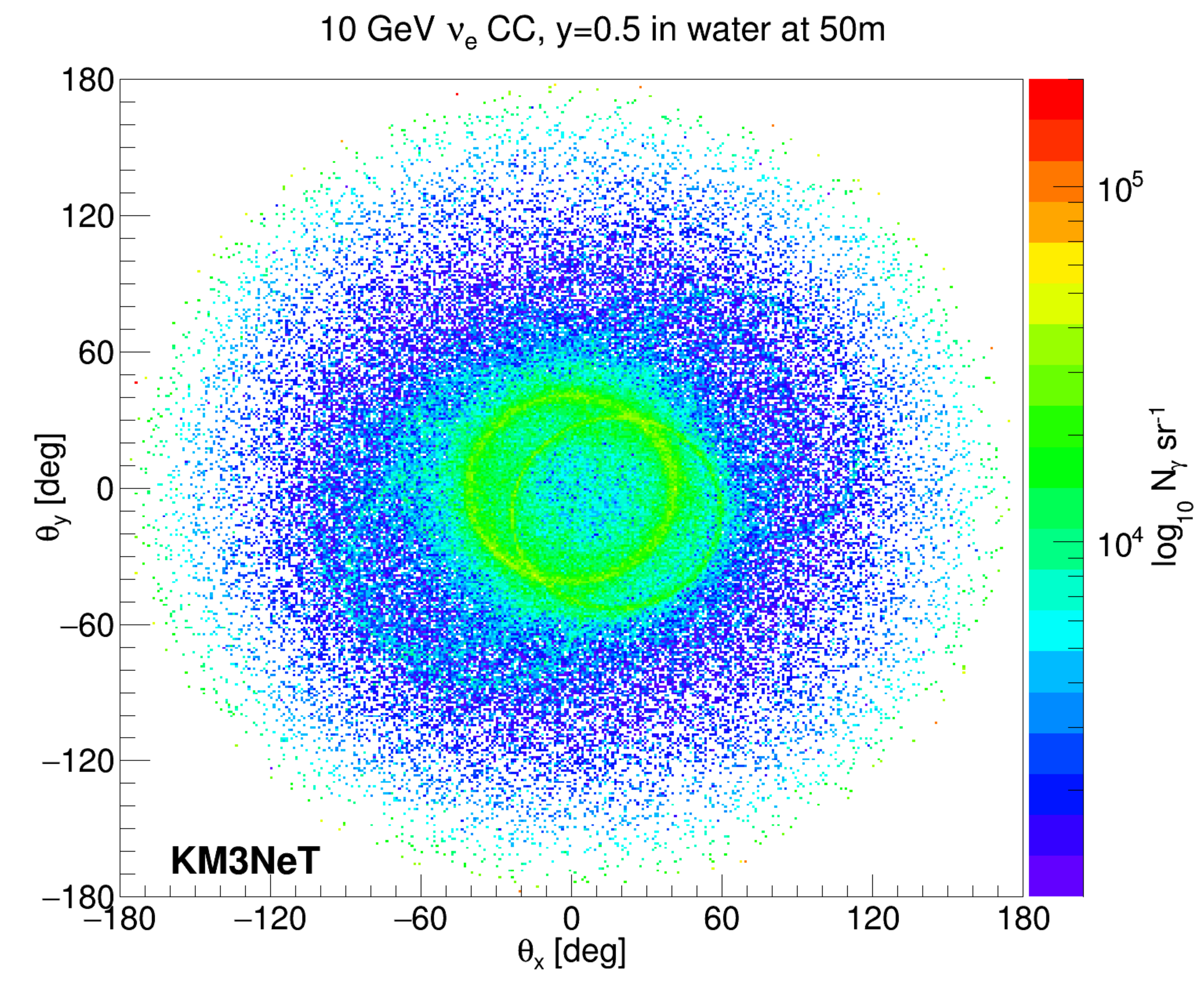}}
\end{minipage}
\caption{Two different simulated $\nu_e \mathrm{CC}$ events with
  $E_\nu \approx 10\,\mathrm{GeV}$ and $y \approx 0.5$ in the upper
  and lower row. Each event is rotated in such a way that the electron
  is in the $z$-direction.
Left: Illustration of the particles produced in the two events. Each arrow represents one particle. The arrow direction and length correspond to the particle momentum in the $p_y$-$p_z$-plane, and the arrow colour indicates the particle type.
Middle and right: Photon distributions in sea water recorded on shells
at 20\,m and 50\,m around the neutrino interaction vertex. Each photon
is weighted with the solid angle averaged effective area of a PMT for
the photon wavelength. The Cherenkov ring from the electron is centred
around $(0,0)$ with an opening angle of $42^{\circ}$, as the electron
moves in the $z$-direction.
}
\label{fig:shower_example_event}
\end{figure}

Although an electromagnetic shower consists of many $e^\pm$-pairs with rather short path lengths 
and overlapping Cherenkov cones, the small pair opening angle preserves the Cherenkov angle peak
of the effective angular light distribution which 
results in a single Cherenkov ring in a projection onto a plane perpendicular to the shower axis. 
Similarly, each hadronic shower particle with energy above the Cherenkov threshold
will produce a Cherenkov ring. Therefore, hadronic showers show a huge variety of 
different signatures due to the various possible combinations of initial hadron types, their
momenta and the diversity of their  hadronic interactions in the shower evolution.\\
Two simulated electron neutrino event examples with $E_\nu \approx 10\,\mathrm{GeV}$ 
and $y\approx 0.5$ each 
are shown in \myfref{fig:shower_example_event}. 
The Cherenkov photon ring from the electron is clearly visible together with fainter rings 
from hadronic shower particles. 
Due to the large scattering length in water, the angular profile of the emitted light
is well conserved over large distances, which leads to the different visible, distinct Cherenkov rings.


While electromagnetic showers show only negligible fluctuations in the number of emitted 
Cherenkov photons and in the angular light distribution, 
hadronic showers show significant intrinsic fluctuations in the relevant energy range. 
These intrinsic fluctuations of hadronic showers and the resulting limitations for the 
energy and angular resolutions have been studied in detail, see \cite{ShowerFluct}.\\
In hadronic showers also muons can be produced via charged pions, 
which can lead to a wrong flavour classification of the event (see \mysref{sec:particle_id}).
The relevance of muons leaking out of hadronic showers and their energy distribution has 
been studied in detail, see \mysref{sec:muon_from_hadShower}.

\begin{figure}[h!]
\centering
\begin{minipage}[c]{0.48\textwidth}
\centering
{\begin{overpic}[width=\textwidth]{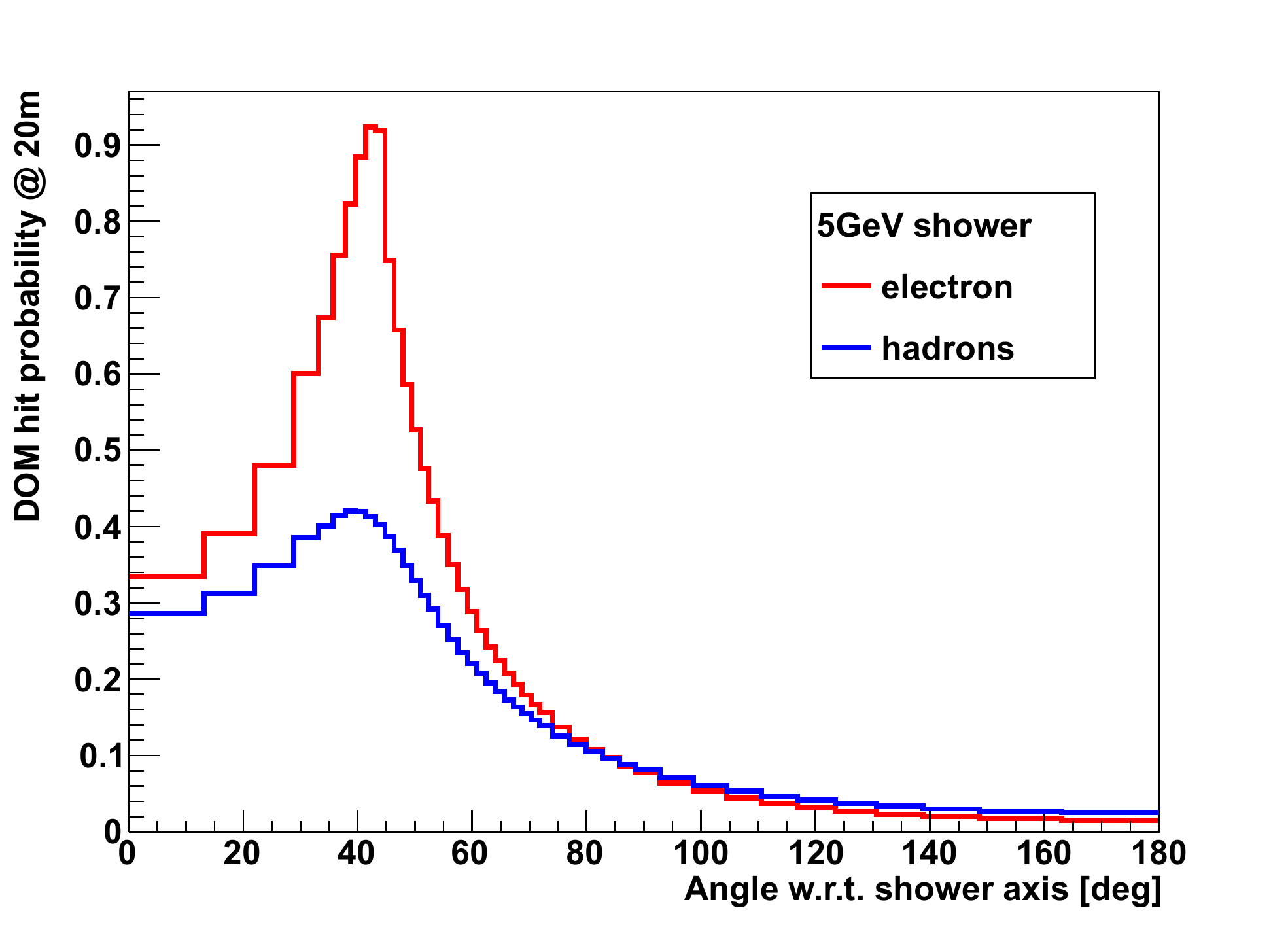}
\put (40,66) {\bf KM3NeT}
\end{overpic}
}
\end{minipage}
\hfill
\begin{minipage}[c]{0.48\textwidth}
\centering
{\begin{overpic}[width =\textwidth]{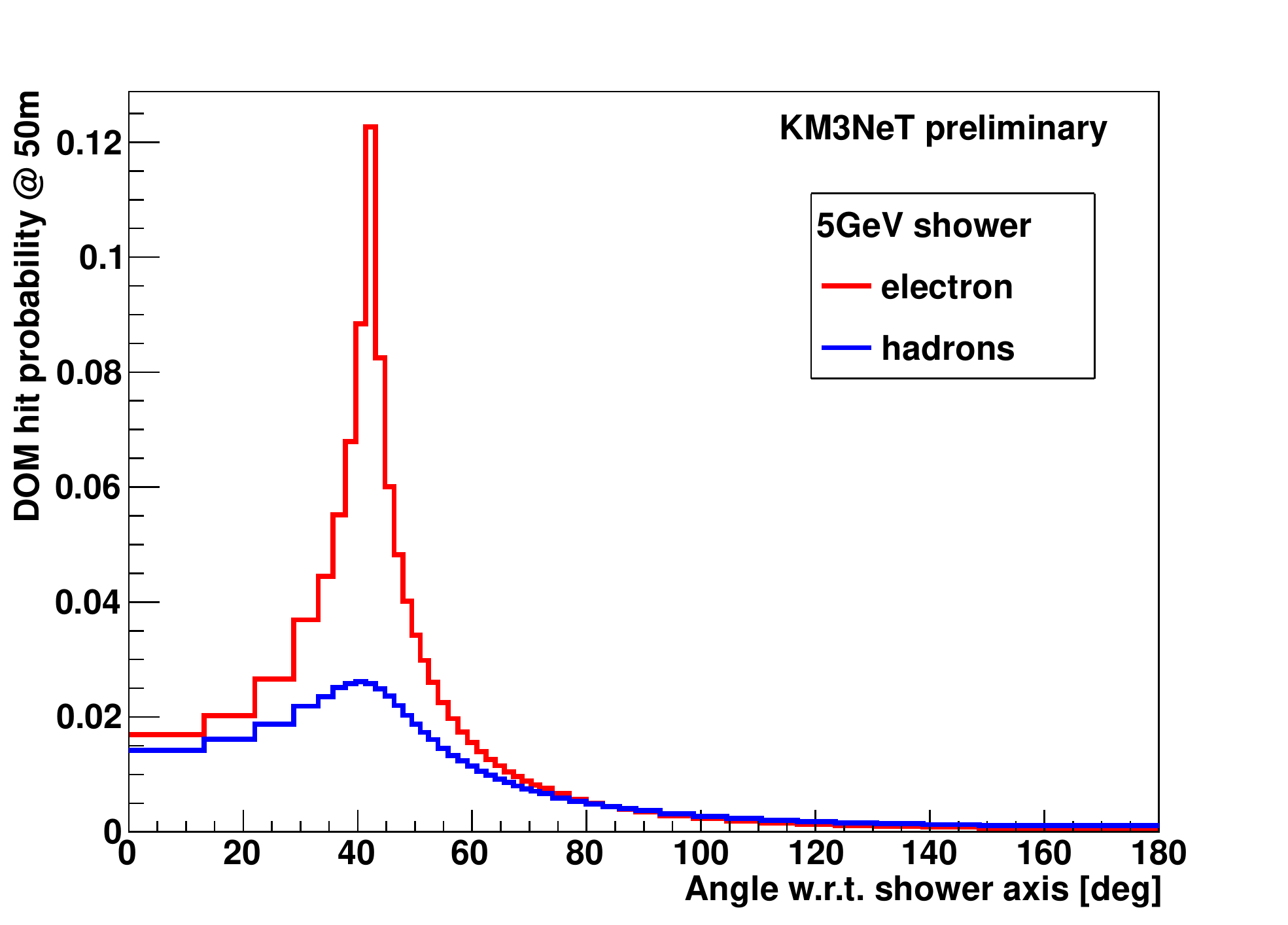}
\put (40,66) {\bf KM3NeT}
\end{overpic}
}
\end{minipage}
\caption{
DOM-hit probability (probability to detect at least one photon in an entire multi-PMT optical module) at a distance of 20\,m (left) and 50\,m (right) away from the brightest point for showers with $E_e = E_{\rm had} = 5\,\mathrm{GeV}$ as a function of the angle $\theta$ between shower direction and the vector from the brightest point to the DOM centre. 
}
\label{fig:shower_OMhitProb_elecVShad}
\end{figure}

The averaged angular light distribution for electromagnetic and hadronic showers is shown in 
\myfref{fig:shower_OMhitProb_elecVShad} for $E_e = E_{\rm had} = 5\,\mathrm{GeV}$. 
For both shower types the 
probability to detect at least one photon within one DOM (DOM-hit probability) 
is maximal at the Cherenkov angle of $42^{\circ}$, but it is more peaked for electromagnetic 
than for hadronic showers. At smaller distances the Cherenkov peak becomes washed out 
due to the extension of the shower in conjunction with the small lever arm for the definition of the angle with respect to the shower direction. 
Note that the light distribution for a single hadronic shower 
event will not be as smooth as shown in these plots due to the distinct Cherenkov rings from each hadron.\\

\paragraph{Sensitivity to the reaction inelasticity $y$\\}

Electromagnetic and hadronic showers induced by neutrino interactions in the energy range 
relevant for the neutrino mass hierarchy measurement show slightly different light emission characteristics
in the detector. Due to the large scattering length in water these differences are conserved 
over sufficiently large distances, so that information from a large detector volume 
can contribute to the discrimination between the two shower types. 
In electron neutrino charged-current events, in which both an electromagnetic and a hadronic shower 
are present at the same time and partly overlapping, the angular separation $\phi_{e,\mathrm{had}}$ 
of both showers can help to distinguish between them. 
This can make an estimation of the reaction inelasticity $y$ in $\nuan_e \mathrm{CC}$ events feasible. 
Additionally, it might allow for a partial separation of $\nuan_e \mathrm{CC}$ and NC events on a 
statistical basis.\\

However, with an ORCA-like 
detector\,\footnote{Detector with a spacing between optical sensors of several metres up to few tens of metres.} 
it seems impossible to distinguish a shower induced by a single electron 
from a shower induced by a single hadron, 
since both resulting Cherenkov light cones will be of the same
intensity for the same particle energy. 
\myfref{fig:shower_example_event} (bottom) shows a simulated example event, 
in which the electron ($E_e=4.77\,\mathrm{GeV}$) and the pion ($E_\pi=3.71\,\mathrm{GeV}$) induce 
Cherenkov rings of similar intensity.

The most intense Cherenkov ring in $\nuan_e \mathrm{CC}$ events is seen in most cases from the electron, as can be 
inferred from the distribution of the inelasticity parameter $y$ in \mysref{sec:simulations} and 
keeping in mind that the hadronic shower energy
$E_{\rm had}$ is often shared between many hadrons. 
A measure for the intensity of a Cherenkov ring $E_x^{\rm cher}$ induced by a particle $x$ with 
energy $E_x$ can be defined by:
\begin{equation}
E_x^{\rm cher} = \begin{cases} E_x - m_p, & \mathrm{if~particle}~x~ \mathrm{is~a~baryon} \\ E_x, & \mathrm{else} \end{cases}
\label{eq:def_Echer}
\end{equation}
where $m_p$ is the proton mass.\\

\begin{figure}[h!]
\centering
\begin{minipage}[c]{0.48\textwidth}
\centering
{\begin{overpic}[width=\textwidth]{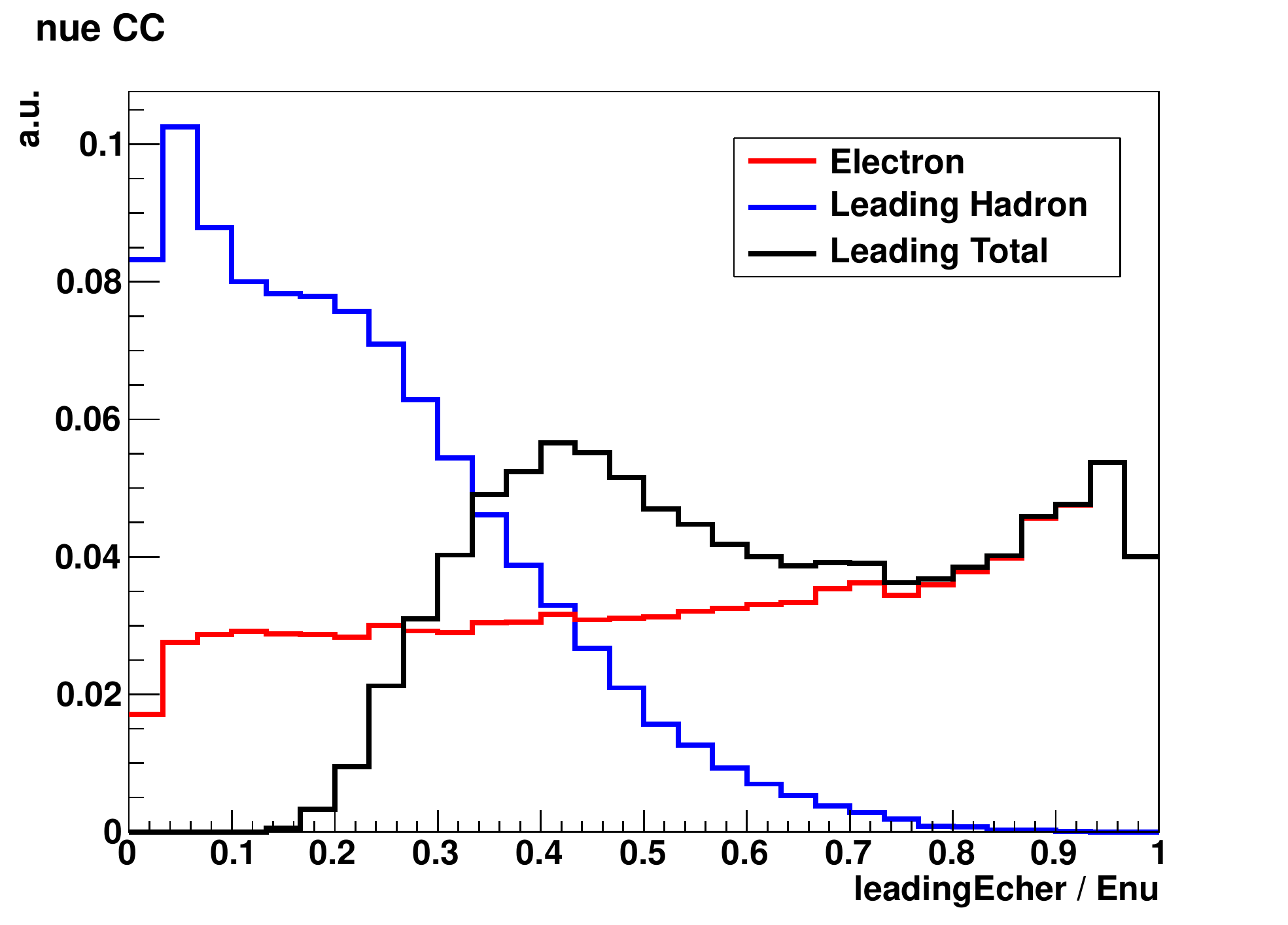}
\put (40,66) {\bf KM3NeT}
\end{overpic}
}
\end{minipage}
\hfill
\begin{minipage}[c]{0.48\textwidth}
\centering
{\begin{overpic}[width=\textwidth]{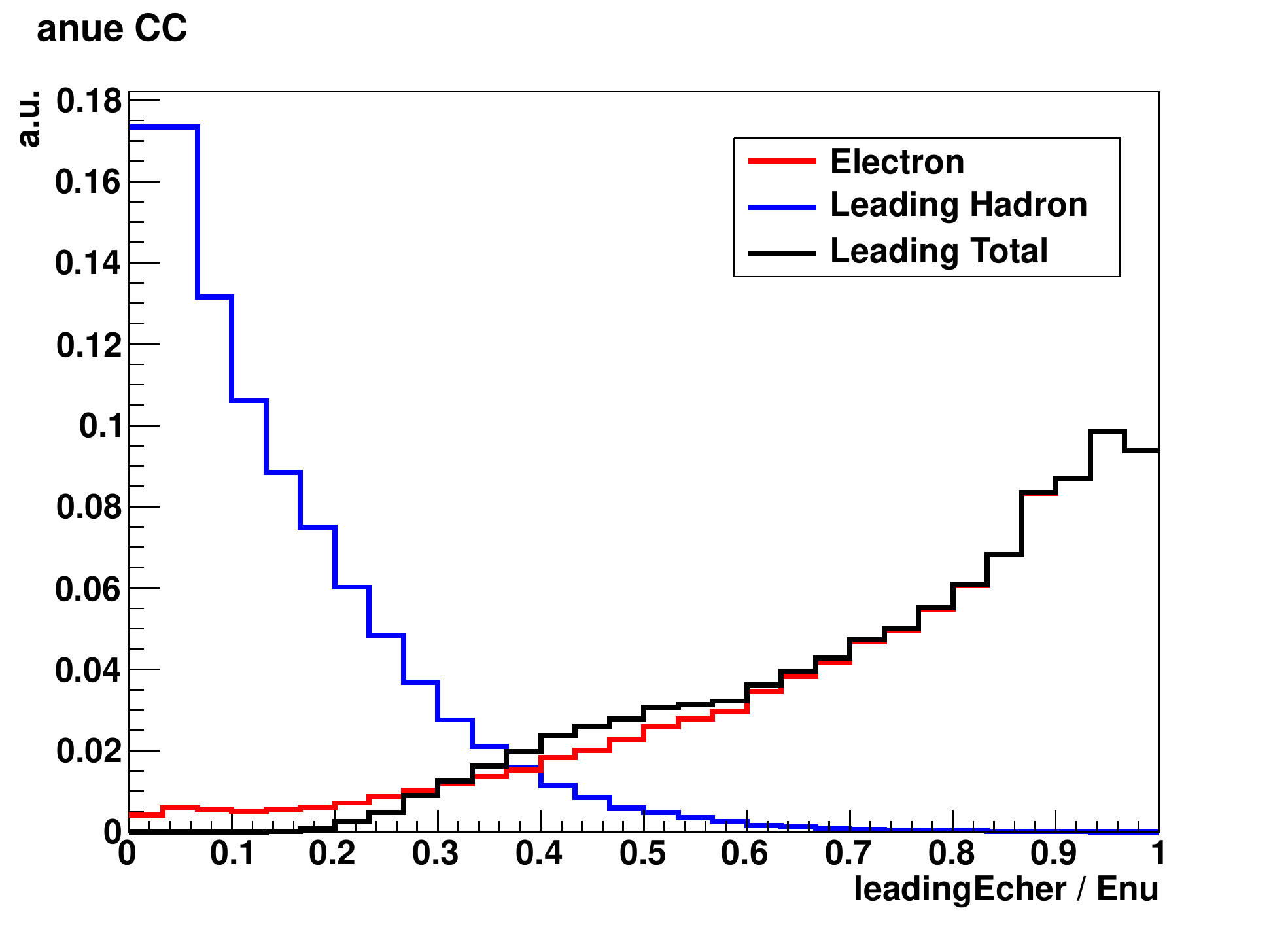}
\put (40,66) {\bf KM3NeT}
\end{overpic}
}
\end{minipage}
\vfill
\vspace{0.5cm}
\begin{minipage}[c]{0.48\textwidth}
\centering
{\begin{overpic}[width=\textwidth]{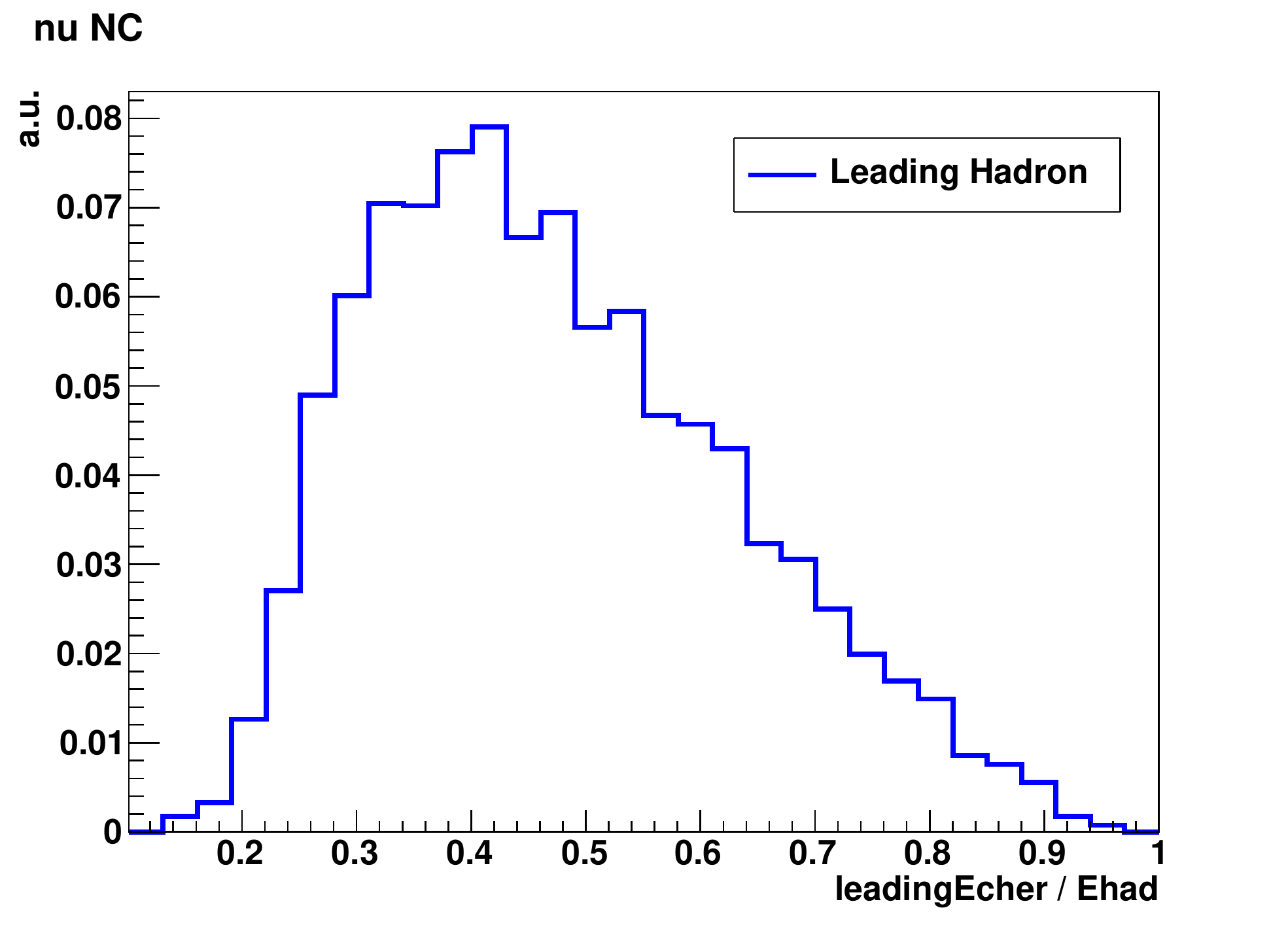}
\put (40,66) {\bf KM3NeT}
\end{overpic}
}
\end{minipage}
\caption{Distribution of $E_e/E_\nu$ (red), the leading $E_{\rm had}^{\rm cher} / E_\nu$ in the hadronic shower (blue) and the leading $E_{\rm tot}^{\rm cher} / E_\nu$ of the total event (back). Top left: $\nu_e \mathrm{CC}$ events with $9\,\mathrm{GeV}<E_\nu < 11\,\mathrm{GeV}$. Top right: $\bar \nu_e \mathrm{CC}$ events with $9\,\mathrm{GeV}<E_\nu < 11\,\mathrm{GeV}$. Bottom:  $\nu_e$ and $\bar \nu_e \mathrm{NC}$ events with $9\,\mathrm{GeV}<E_{\rm had} < 11\,\mathrm{GeV}$.}
\label{fig:shower_measureable_BjorkenY}
\end{figure}

In the example event in \myfref{fig:shower_example_event} (bottom), the most intense Cherenkov ring is 
from the electron ($E_e^{\rm cher} = E_e$) and the relative intensity is 
$E_e/E_\nu = 4.77\,\mathrm{GeV} / 9.82\,\mathrm{GeV} = 0.49$, 
while the leading Cherenkov ring in the hadronic shower $E_{\rm had}^{\rm cher}$ 
is from the pion ($E_\pi^{\rm cher} = E_\pi$) with a relative intensity of 
$E_{\pi} / E_\nu = 3.71\,\mathrm{GeV} / 9.82\,\mathrm{GeV} = 0.38$. 
The distribution of $E_e/E_\nu$, the leading $E_{\rm had}^{\rm cher} / E_\nu$ 
in the hadronic shower and the leading $E_{\rm tot}^{\rm cher} / E_\nu$ of 
the total event is shown in \myfref{fig:shower_measureable_BjorkenY} 
for $\nuan_e \mathrm{CC}$ events with $9\,\mathrm{GeV}<E_\nu < 11\,\mathrm{GeV}$. 
Additionally, the distribution of leading $E_{\rm had}^{\rm cher} / E_\nu = E_{\rm tot}^{\rm cher} / E_\nu$ 
in $\nuan_e \mathrm{NC}$ events with $9\,\mathrm{GeV}<E_{\rm had} < 11\,\mathrm{GeV}$ is shown.\\
Generically, the measurable inelasticity $y$ is given by $1 - E_{\rm tot}^{\rm cher} / E_\nu$. 
Therefore, it is expected that all measured events will show an inelasticity of $y \lesssim 0.8$, 
as $E_{\rm tot}^{\rm cher} / E_\nu \gtrsim 0.2$ for all neutrino interaction types 
(cf. \myfref{fig:shower_measureable_BjorkenY}). This is even the case for NC events, 
which are in principle very 
similar\,\footnote{Small differences are due to different characteristics of hadronic showers induced by $W$ or $Z$ bosons.} 
to $\nuan_e \mathrm{CC}$ events with the same neutrino energy as the hadronic energy 
in the NC events and an inelasticity of $y=1$.

\subsubsection{Shower reconstruction algorithm} 
\label{sec:shower_reco_algo}
A neutrino-induced shower-like event is characterised by 8 free parameters: 
vertex position $\vec x_{\rm vtx}$ and time $t_{\rm vtx}$, energy $E$, 
direction $\hat e_s$ and inelasticity $y$. The shower direction is characterised by 2 angles.\\
The shower reconstruction is performed in two steps. In the first step the vertex is reconstructed based on the recorded time of the PMT signals, commonly called hits, and in the second step the direction, energy and inelasticity are reconstructed based on the number of hits and their distribution in the detector. In both steps a maximum likelihood fit is performed for many different starting shower hypotheses and the solution with the best likelihood is chosen.\\
This factorisation of the fitting procedure works well due to the 
homogeneity of water and its large scattering length which allows 
for a precise vertex reconstruction
independent of the shower direction.

\paragraph{Vertex reconstruction\\} 
\label{sec:shower_vtx_reco}
The majority of the Cherenkov light from electromagnetic and hadronic showers is emitted 
within a few metres around the neutrino interaction vertex, cf. \mysref{sec:shower_phenomenology}. 
Therefore, direct hits from a shower are characterised by a small time residual $t_{\rm res}$:
\begin{equation}
t_{\rm res} = t_{\rm hit} - t_{\rm vtx} - d / c_{\rm water} ,
\label{eq:shower_tres}
\end{equation}
where $d$ is the distance between the vertex position $\vec x_{\rm vtx}$ and the PMT position,
$t_{\rm vtx}$ is the vertex time and $c_{\rm water}$ is the speed of light in water.
The vertex position and time are here defined as the brightest point and its corresponding time 
in the shower evolution and not by the neutrino interaction itself, 
because the brightest point is what is actually seen by the detector.\\ 
The vertex reconstruction is performed in two successive maximum likelihood fits. 
For both fits, the likelihood for the vertex hypothesis $(t_{\rm vtx}, \vec x_{\rm vtx})$ is given by:
\begin{equation}
L = \sum_{\rm hits} g(t_{\rm res} | (t_{\rm vtx}, \vec x_{\rm vtx})) , 
\label{eq:shower_vtx_lh}
\end{equation}
where $g(t_{\rm res} | (t_{\rm vtx}, \vec x_{\rm vtx}))$ is a function
of the hit time residuals for a given shower hypothesis.

The first vertex fit (prefit) is designed to be very robust against 
noise hits and an imprecise initial vertex hypotheses.
The initial hit selection is optimised for low energetic shower-like events and is described below 
together with the choice of the initial vertex hypothesis. In the prefit, the following function $g$ is used:
\begin{equation}
g(t_{\rm res}) =  1/\sqrt{ 4 + (t_{\rm res} / \mathrm{ns} )^2 }
\label{eq:shower_prefit}
\end{equation}
Based on the initial hit selection in total 15 starting vertex hypotheses for the prefit are generated.
The fitted vertex with the best likelihood is chosen as result of the prefit.

\begin{figure}[h!]
\centering
\begin{overpic}[width=0.7\linewidth]{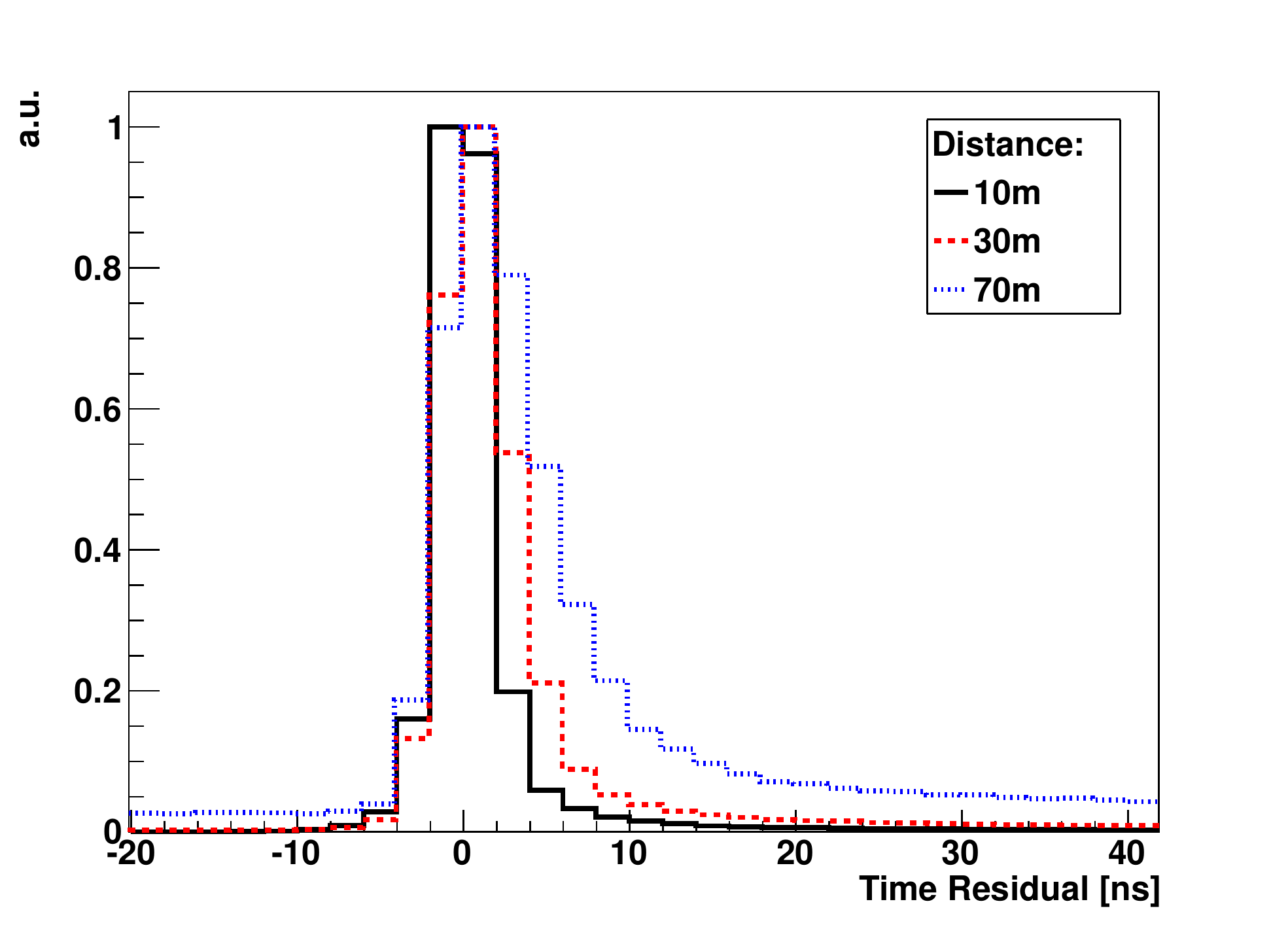}
\put (40,66) {\bf KM3NeT}
\end{overpic}
\caption{Time residual distribution for three different distances $d$ between the vertex and the PMT, obtained from simulations of fully contained  $\nu_e \mathrm{CC}$ and $\bar \nu_e \mathrm{CC}$ events with 5\,kHz noise rate. The time residual is defined with respect to the brightest point of the shower. The distributions are normalised so that the maximum is 1.}
\label{fig:shower_vtx_pdf}
\end{figure}

The second vertex fit is more precise but needs a hit selection with higher signal purity 
and a good starting vertex hypothesis. The result of the prefit is used to generate in 
total 10 starting vertex hypotheses (result of the prefit and 9 vertex hypotheses around it
with time shifts of $\pm 25\,\mathrm{ns}$ and position shifts of 5\,m in a random direction). 
A rather pure signal hit selection is achieved by selecting hits according to the following criteria:
\begin{itemize}
    \item $10\,\mathrm{m} < d < 80\,\mathrm{m}$
    \item $-50\,\mathrm{ns} < t_{\rm res} < 50\,\mathrm{ns}$
    \item $-1< \cos(\psi) < 0.1$
\end{itemize}
where $\psi$ is the angle between the PMT direction (vector normal to the photocathode plane) 
and the vector from the vertex to the PMT, i.e. only PMTs which are orientated towards the 
vertex and can be hit by unscattered photons are taken into account. 
The fit uses a function $g(t_{\rm res})$ obtained from simulated
$\nuan_e \mathrm{CC}$ events,
 and which is dependent on the distance $d$. Such distributions are shown for three different distances $d$ in
\myfref{fig:shower_vtx_pdf}. With increasing distances the peak of direct hits 
becomes broader due to scattering and dispersion, and the hit probability decreases 
due to absorption leading to a relative increase of the noise level.\\
The fitted vertex with the best likelihood and within 10\,m and 50\,ns around 
the result of the prefit is chosen as final vertex.

\subparagraph{Initial hit selection for first vertex fit\\} 
For the initial selection of shower-like hits the following hit patterns are defined:
\begin{description}
    \item[L1] coincidence between hit times of 2 PMTs on the same DOM in a time window $\Delta t \le 10\,\mathrm{ns}$
    \item[L2] L1 with an angle between the hit PMTs smaller than
      $90\,^\circ$, note that these are the same definition as used in the triggers, cf. \mysref{sec:trigger}
    \item[L3] coincidence between hits on 3 PMTs on the same DOM in a time window $\Delta t \le 10\,\mathrm{ns}$
    \item[V2L2] coincidence between two L2 hits on different DOMs which are closer than 35\,m and within a time window $\Delta t \le 10\,\mathrm{ns} + t_{D}$, where $t_{D}$ is the time required by the light to travel the distance $D$ between the two DOMs
    \item[T0L0] coincidence between two hits on adjacent or next-to-adjacent DOMs on the same string in a time window $\Delta t \le 10,\mathrm{ns} + t_{D}$
\end{description}

The general strategy is to find first a \textit{reference hit} that is very likely a signal hit and close to the neutrino interaction vertex. The position/time of this reference hit is then used as an initial vertex hypothesis to select additional hits based on their time residual and further requirements to suppress noise hits.\\
Firstly, the largest cluster of causally connected L2 hits is selected by requiring $\Delta t \le D / c_{\rm water} + 10\,\mathrm{ns}$ for all L1 hits within the cluster.
From these causally connected L2 hits the subset of hits that additionally satisfy the L3 or V2L2 criteria is selected. 
These L3 or V2L2 hits are ranked according to their hit multiplicity (number of coincidences on the same DOM) as well as the number and multiplicity of causally connected hits in the vicinity of $25\,\mathrm{m}$. The most signal-like hit is chosen as 'reference hit'.\\ 
Secondly, all hits around the reference hit are selected that are
closer than $100\,\mathrm{m}$, within a time window of $-250\,\mathrm{ns} <
t_{\rm res} < 10\,\mathrm{ns}$ and causally connected with most L3
or V2L2 hits.
The loose lower time cut allows for distances up to about $50\,\mathrm{m}$
between the true neutrino interaction vertex and the reference hit,
e.g. because the neutrino interaction is outside the detector volume. 
The drawback of this relatively large time window is a contamination with noise hits. 
Therefore, hits are discarded that do not satisfy the L1 criterion,
or are either causally connected with an adjacent L3 or V2L2 hit on
the same string or fulfil the T0L0 criterion in addition to being
causally connected with a L3 or V2L2 hit in the vicinity of $25\,\mathrm{m}$. 
The hits selected by this procedure are used in the first vertex fit and the position/time of the 15 most signal-like hits are used as initial vertex hypotheses.

\paragraph{Reconstruction of energy, direction and inelasticity\\} 
\label{sec:shower_EdirY_reco}
Once the shower vertex is fixed, the remaining parameters which can be fitted are the 
shower energy $E$, direction $\hat e_s$ and the reaction inelasticity $y$. 
In principle all of these parameters can be inferred from the angular light 
distribution (cf. \myfref{fig:shower_OMhitProb_elecVShad}): the shape is 
sensitive to the inelasticity $y$, the integral is in first order proportional 
to the energy (as the light yield is in first order proportional to the shower energy) 
and the direction in which this angular light emission profile is 
present gives the shower direction.\\
In the following, the shower energy $E$, direction $\hat e_s$ and inelasticity $y$ are 
reconstructed using a maximum likelihood fit 
based on the probability that the hit pattern is created by a trial shower hypothesis 
$\vec \alpha = (t_{\rm vtx}, \vec x_{\rm vtx}, E, y, \hat e_s)$.\\
As discussed in \mysref{sec:shower_phenomenology}, the electron mostly 
is the dominant particle in $\nu_e \mathrm{CC}$ events and produces the brightest Cherenkov ring. 
Therefore, the reconstruction is designed to find the electron direction 
$\hat e_e$ and not the neutrino direction.\\ 
The final hit selection, the definition of the likelihood function and the fitting 
procedure are described in the following.

\subparagraph{Final hit selection\\} 
Based on the result of the vertex fit, hits are selected according to the following criteria:
\begin{itemize}
    \item $10\,\mathrm{m} < d < 80\,\mathrm{m}$
    \item $-25\,\mathrm{ns} < t_{\rm res} < 25\,\mathrm{ns}$
    \item $-1< \psi < 0.1$
\end{itemize}
For simplification\,\footnote{Besides reducing the computation time for the fit, this simplification is justified by the fact that each DOM in principle measures the intensity of the shower event at a given position. As the PMTs on the same DOM are nearly at the same position and direct light from a shower arrives at the DOM nearly at the same time, the information of the individual PMTs --- which PMT is hit at which time --- is not needed. Of course, the multi-PMT structure is needed to estimate the shower intensity from the number of hit PMTs $N_{\rm hits}^{\rm DOM}$.}, all PMT-hits on the same DOM are merged and the times of the individual hits 
are not taken into account, so that the event is quantified by $N_{\rm hits}^{\rm DOM}$ for each DOM. 
For the fit all DOMs with $10\,\mathrm{m} < d < 80\,\mathrm{m}$ are taken into account, 
that includes also the DOMs without any selected hit.

\begin{figure}[h!]
\centering
\includegraphics[width=0.5\linewidth,trim={5cm 2.5cm 9cm 3cm},clip]{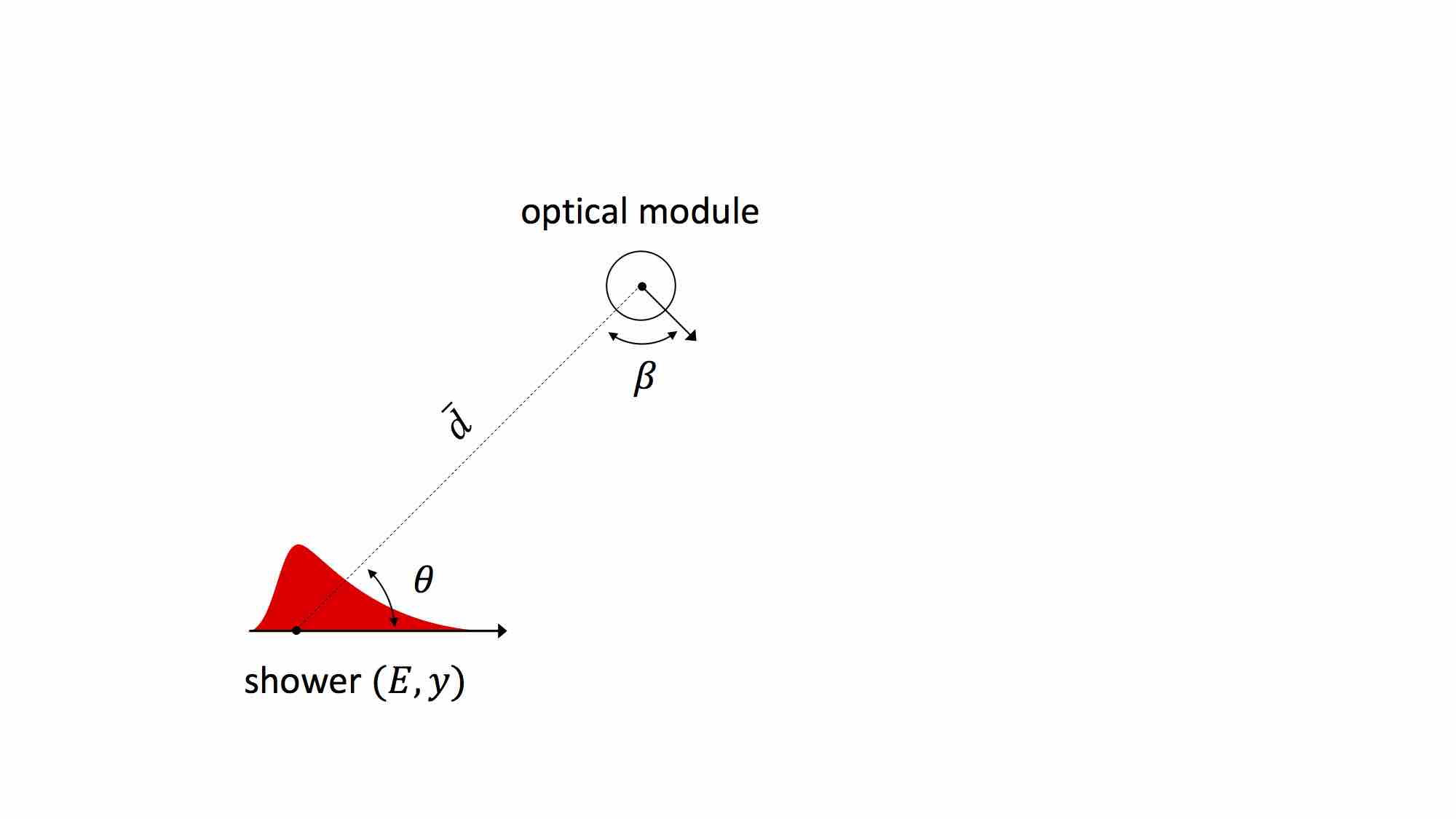}
\caption{Illustration of the quantities relevant for the probability $P(N_{\rm hits}^{\rm DOM})$.}
\label{fig:shower_variable_definition}
\end{figure}

\subparagraph{Likelihood\\} 
Ignoring shower-to-shower fluctuations, the probability $P(N_{\rm hits}^{\rm DOM})$
to detect $N_{\rm hits}^{\rm DOM}$ on a given DOM depends on: 
$E$, $y$, the distance $d$ between the vertex and the DOM, the angle $\theta$ 
between shower direction $\hat e_s$ and the vector $\vec d$ from the vertex to the DOM, 
and the DOM orientation. The DOM orientation can be described by a single angle $\beta$ 
between $\vec d$ and the DOM direction, because the angular acceptance of the entire DOM 
(sum of all PMT angular acceptances) shows in first order a rotational symmetry due to 
the multi-PMT structure, cf.\ \mysref{sec-tec}. 
All of these quantities are illustrated in \myfref{fig:shower_variable_definition}.

The likelihood is computed as follows:
\begin{equation}
L = \prod_{\rm selected ~ DOMs} P \left ( N_{\rm hits}^{\rm DOM}(E, y, d, \theta, \beta) | \vec \alpha ) \right ).
\label{eq:shower_EdirY_lh}
\end{equation}
To define the probability $P(N_{\rm hits}^{\rm DOM})$ two auxiliary quantities are introduced: 
the number of expected photons $\langle N_\gamma \rangle$ 
and the variance $var \left (\langle N_\gamma \rangle \right )$ of the $\langle N_\gamma \rangle$ distribution. 
To take fluctuations in the hadronic shower into account, 
the variance of the expected number of photons has been introduced.\\
The dependency on the DOM orientation and the distance are parameterised. 
For the number of expected photons the dependency on the DOM orientation is assumed to follow 
the angular acceptance of the entire DOM. For the final hit selection, 
the attenuation of $\langle N_\gamma \rangle$ with distance is well described 
by $\exp (-d / \lambda_{\rm att}(d)) \cdot d^{-2}$, where the first term 
describes the effective attenuation due to absorption and scattering and the latter 
term describes the geometrical reduction of solid angle coverage of the DOM 
on a sphere with radius $d$. 
The effective attenuation length has been derived from a fit to the MC simulations and can be parameterised as 
$\lambda_{\rm att}(d) = a + b \cdot d$ with $a=32.2\,\mathrm{m}$ and $b=0.034$.\\ 
Taking these parameterisations into account, $\langle N_\gamma \rangle$ and $var \left( \langle N_\gamma \rangle \right )$ 
depend on $(E, y, \theta, d)$. Although the $d$ dependency is already taken into account via 
the parameterisation above, the shape of the $\theta$ distribution changes with distance 
(see \myfref{fig:shower_OMhitProb_elecVShad}) so that a coarse binning in $d$ is needed.\\
The probability density function $P(N_{\rm hits}^{\rm DOM})$ depends on 
$(N_{\rm hits}^{\rm DOM}, \langle N_\gamma \rangle, var \left( \langle N_\gamma \rangle \right ), \beta)$. 
The quantities $ \langle N_\gamma \rangle$, $var \left( \langle N_\gamma \rangle \right)$ and the probability 
 $P(N_{\rm hits}^{\rm DOM})$ are obtained from MC simulations of $\nuan_e \mathrm{CC}$ events.

\begin{figure}[h!]
\centering
\includegraphics[width=0.7\linewidth]{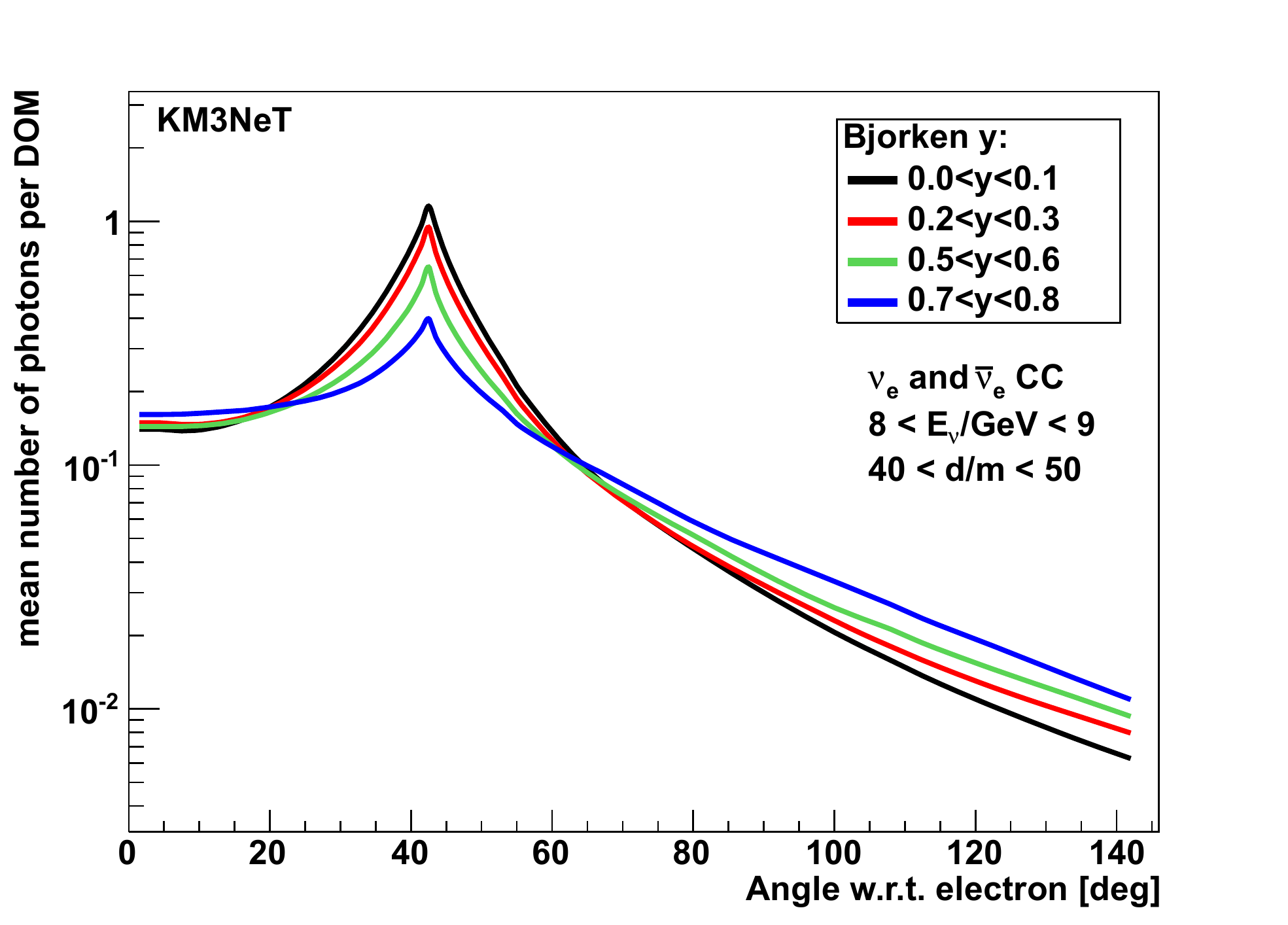}
\caption{Number of expected photons $\langle N_\gamma \rangle$ as a function of the angle $\theta$ between the shower direction (which is the electron direction) and the vector from the vertex to the DOM for different inelasticity $y$ intervals, and for shower energies of $8\,\mathrm{GeV}<E<9\,\mathrm{GeV}$ and at distances of $40\,\mathrm{m}<d<50\,\mathrm{m}$.}
\label{fig:shower_EdirY_pdf_refDirElec}
\end{figure}

An example distribution of the expected number of photons $\langle N_\gamma \rangle$ as a function of the 
angle $\theta$ for different inelasticity $y$ intervals is shown in \myfref{fig:shower_EdirY_pdf_refDirElec}. 
As the angle $\theta$ is defined with respect to the electron direction, 
a clear Cherenkov peak of the electron at $42\,^{\circ}$ is visible. 
With higher inelasticity $y$ this peak becomes fainter due to less energetic electrons, 
while the number of expected photons in the 'off-peak region' ($\theta \gtrsim 60\,^{\circ}$) 
increases due to the more energetic hadronic showers. Therefore, these PDF tables gain 
sensitivity to the reaction inelasticity $y$ from the ratio of the peak to the off-peak region.

\subparagraph{Fitting procedure\\} 
For technical reasons, each event is fitted with 9 different fixed inelasticity $y$ assumptions\footnote{The last inelasticity $y$ bin (0.8-1) is chosen larger than the other bins, as the MC statistics decreases for $y \rightarrow 1$.} 
($y=[0-0.1, 0.1-0.2, \dots , 0.7-0.8, 0.8-1]$). For each fixed $y$ the likelihood 
maximisation is performed for 5 different starting shower hypotheses. 
The initial shower hypothesis is calculated from the selected hits. 
The direction is estimated by the sum of all vectors from the vertex to the DOM 
weighted by  $N_{\rm hits}^{\rm DOM}$ and the energy is estimated empirically by $(\sum N_{\rm hits}^{\rm DOM} - 20)/4$. 
The other four seeds are randomly chosen perpendicular to the first starting shower hypotheses with the same energy.\\
Finally, the result with the best likelihood of all 45 fits is selected. 
Thus, the final result has a discrete value for the reconstructed inelasticity $y$.

\subsubsection{Event selection} 
\label{sec:shower_reco_evt_sel}
The final event selection criteria are:
\begin{itemize}
    \item $E_{\rm reco} > 1\,\mathrm{GeV}$
   \item result of both vertex fits is similar in space and time: distance $<4\,\mathrm{m}$ and time difference $<20\,\mathrm{ns}$
    \item a minimum of 7 (3) out of the 15 (10) reconstructed vertices
      from different seeds in the first (second) vertex fit are
      similar to the selected vertex of this fit: distance $<2\,\mathrm{m}$ and time difference $<10\,\mathrm{ns}$   
\item $\mathrm{cov}_{20\,^{\circ}} \ge 0.4$, \item $\mathrm{cov}_{45\,^{\circ}} \ge 0.4$, $\mathrm{cov}_{60\,^{\circ}} \ge 0.4$ and $\mathrm{cov}_{75\,^{\circ}} \ge 0.4$
\end{itemize}
The final hit selection (including also hits with $d<10\,\mathrm{m}$)
 must again also fulfil the shower trigger, cf. \mysref{sec:trigger}.
 The \textit{coverage} $\mathrm{cov}_{x}$ is defined as the fraction of directions on a cone with the opening angle $x$ 
around the reconstructed direction at the reconstructed vertex position 
that satisfy the following containment condition: $L_{\rm inVol} > 20\,\mathrm{m}$, 
where $L_{\rm inVol}$ is the path length inside the instrumented volume for 
distances away from the vertex between 10\,m and 70\,m.\\
This coverage cut is introduced to ensure that a reasonable fraction 
of the expected hit pattern from the reconstructed shower is contained 
in the instrumented volume. 
Therefore the coverage cut is in principle a containment cut for the 
reconstructed vertex depending on the reconstructed shower direction.

\subsubsection{Reconstruction performance} 
\label{sec:shower_reco_performance}
\begin{figure}[h!]
\centering
\begin{minipage}[c]{0.48\textwidth}
\centering
    {\includegraphics[width=\textwidth]{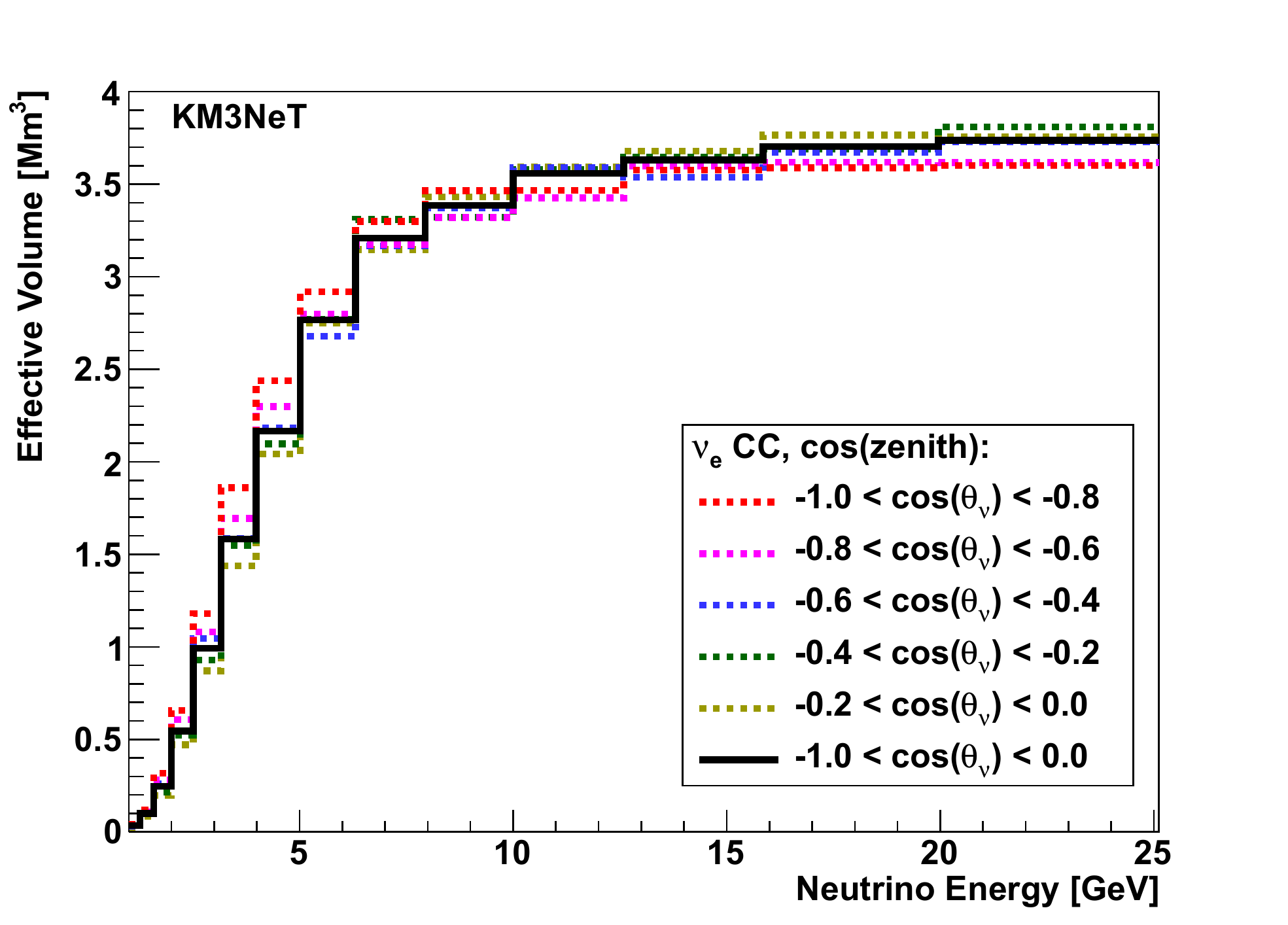}}
\end{minipage}
\hfill
\begin{minipage}[c]{0.48\textwidth}
\centering
    {\includegraphics[width =\textwidth]{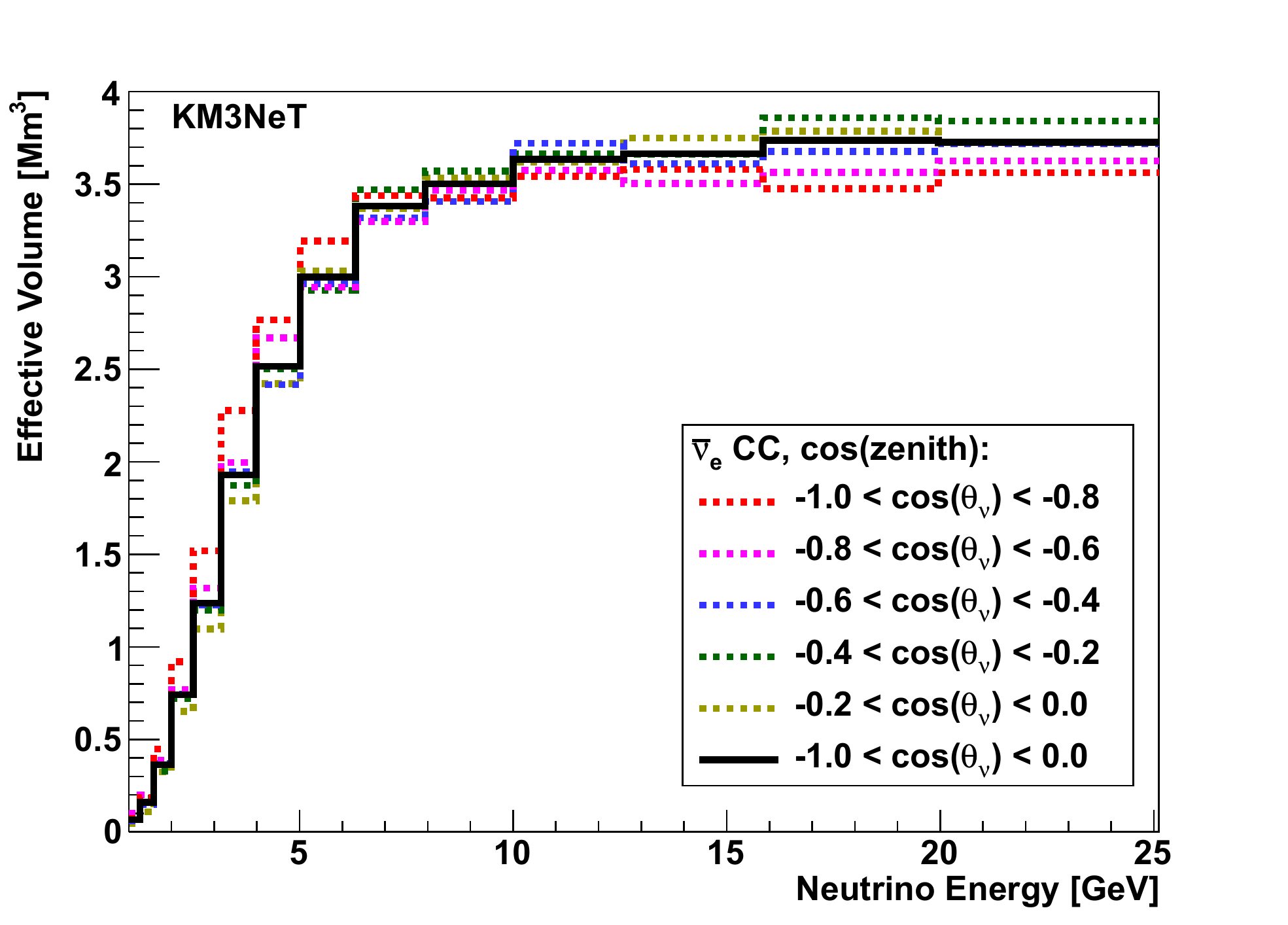}}
\end{minipage}
\caption{Effective volumes in $\mathrm{Mm}^3$ ($10^6\,\mathrm{m}^3$) as a function of neutrino energy for different true neutrino cos(zenith) ranges, where cos(zenith)=-1 means vertical up-going and cos(zenith)=0 means horizontal. The solid black line corresponds to up-going neutrinos weighted according to the Bartol atmospheric neutrino flux model. Left: $\nu_e \mathrm{CC}$. Right: $\bar \nu_e \mathrm{CC}$.}
\label{fig:shower_reso_effVol}
\end{figure}

The performance of the shower reconstruction is studied on MC simulations described in \mysref{sec:simulations} 
and events are selected according to the criteria described in \mysref{sec:shower_reco_evt_sel}. 
The vertical spacing between the DOMs is 6\,m if not stated otherwise.
For all following results the events are weighted to reproduce the conventional atmospheric neutrino flux 
following the Bartol model \cite{bib:Bartol}.
\paragraph{Performance for charged-current electron neutrino events\\} 
\label{sec:shower_reso_eCC}
\subparagraph{Effective volume:}
The effective volume for up-going $\nu_e \mathrm{CC}$ and $\bar \nu_e \mathrm{CC}$ events is shown 
in \myfref{fig:shower_reso_effVol} as a function of neutrino 
energy for different neutrino zenith angle ranges. 
Depending on the zenith angle the plateau reaches $3.8\,\mathrm{Mm}^3$
(horizontal), $3.6\,\mathrm{Mm}^3$ (vertical up-going) and around $3.7\,\mathrm{Mm}^3$ for all up-going 
$\nu_e$  and $\bar \nu_e$. 
The turn-on is slightly steeper for vertical up-going than for horizontal events as more PMTs are oriented downward than upward in a DOM.
90\% of the plateau is reached around $E_\nu = 8\,\mathrm{GeV}$ $(7\,\mathrm{GeV})$ for $\nu_e$ ($\bar \nu_e$). 
The turn-on is slightly steeper for $\bar \nu_e$ than for $\nu_e$  due to the lower average inelasticity.

\begin{figure}[h!]
\centering
\begin{minipage}[c]{0.325\textwidth}
\centering
{\begin{overpic}[width=\textwidth]{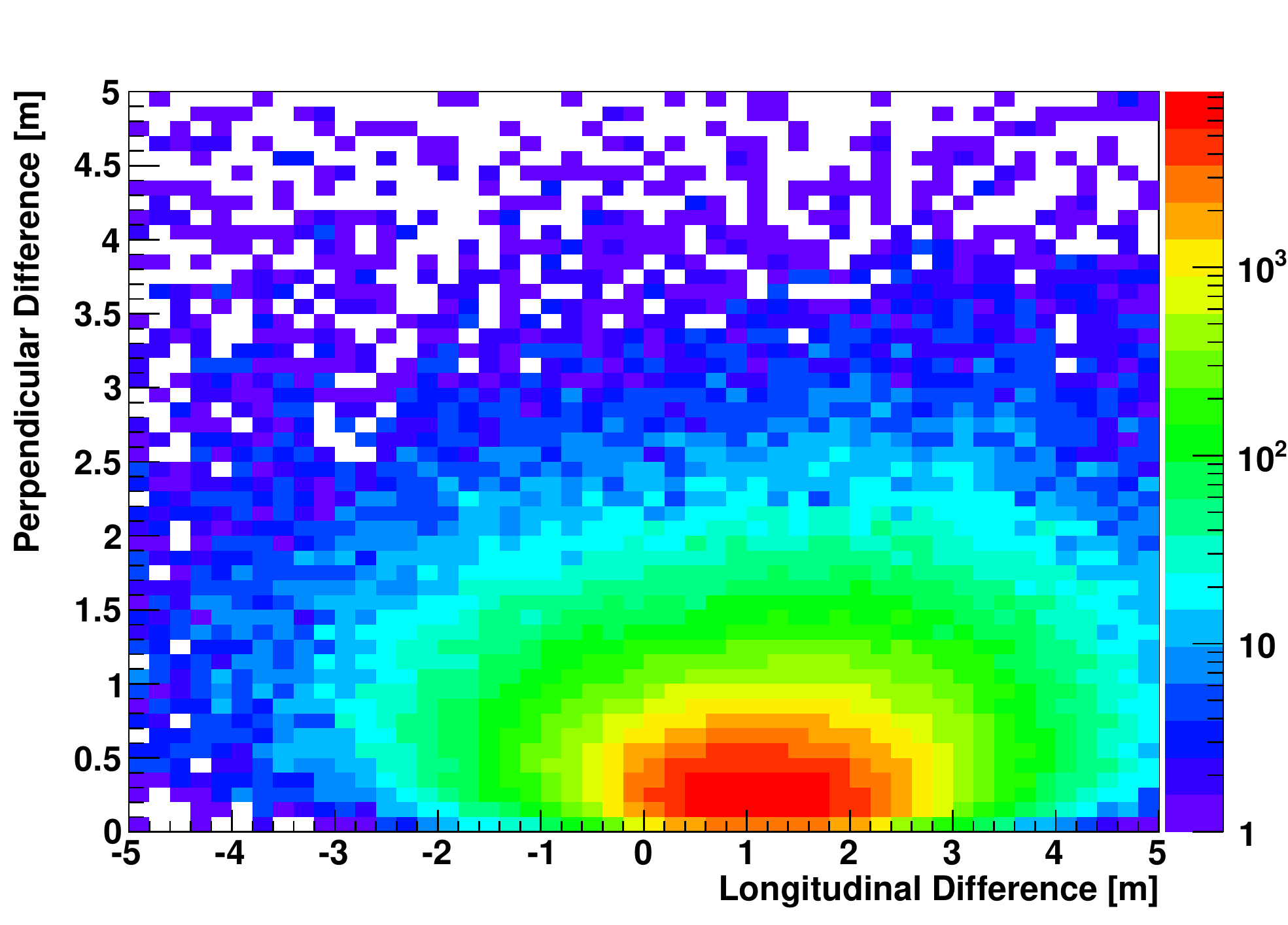}
\put (10,66) {\tiny KM3NeT}
\end{overpic}
}
\end{minipage}
\hfill
\begin{minipage}[c]{0.325\textwidth}
\centering
 {\begin{overpic}[width=\textwidth]{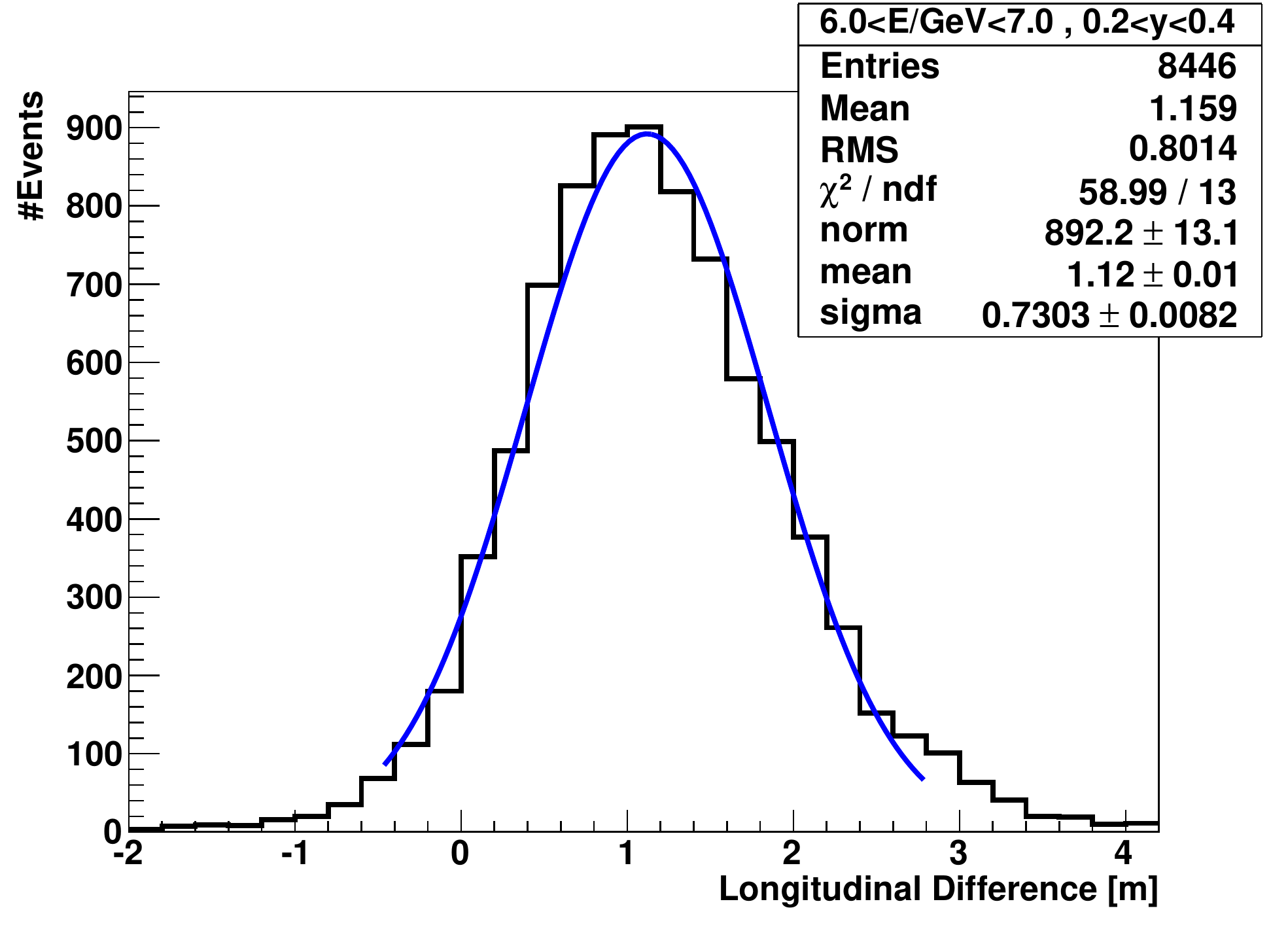}
\put (10,66) {\tiny KM3NeT}
\end{overpic}
}
\end{minipage}
\hfill
\begin{minipage}[c]{0.325\textwidth}
\centering
 {\begin{overpic}[width=\textwidth]{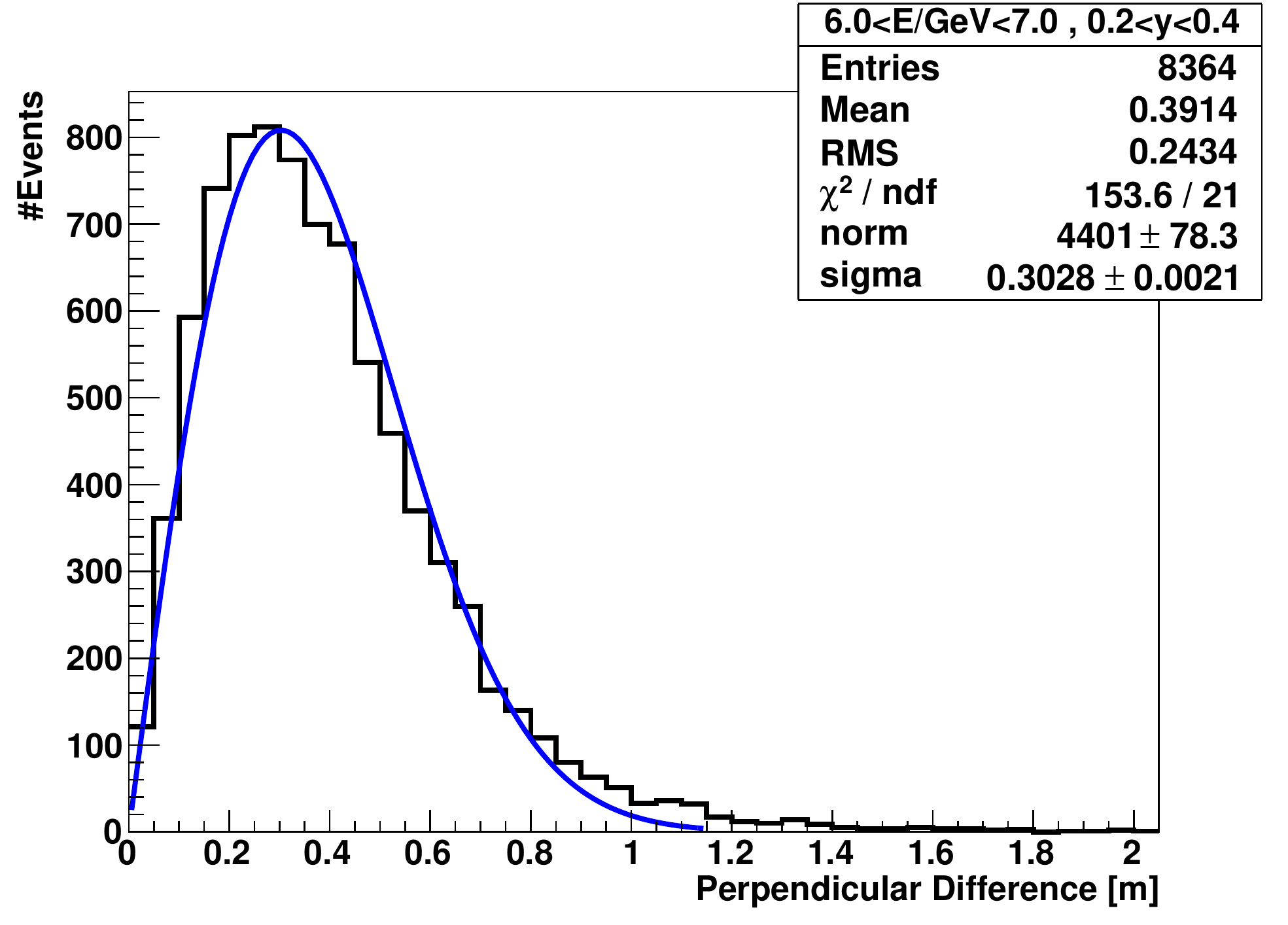}
\put (10,66) {\tiny KM3NeT}
\end{overpic}
}
\end{minipage}
\caption{
Longitudinal and perpendicular distance between the neutrino
interaction position and the reconstructed vertex position for all
selected events with $E_\nu = 2-30\,\mathrm{GeV}$ (left).
Longitudinal distance for $6\,\mathrm{GeV} < E_\nu < 7\,\mathrm{GeV}$
and $0.2 < y < 0.4$ fitted with a Gaussian (middle).
Perpendicular distance for the same $E_\nu$ and $y$ range fitted with
'distance $\times$ a Gaussian' (right).
}
\label{fig:shower_reso_vtx_fitting}
\end{figure}

\subparagraph{Vertex resolution:}
The distance between the neutrino interaction position and 
the reconstructed vertex position is shown 
\myfref{fig:shower_reso_vtx_fitting} (left) for all selected events 
in the energy range of $E_\nu = 2-30\,\mathrm{GeV}$. 

The distance is split in a longitudinal and a perpendicular component with respect to the neutrino direction.
An offset in neutrino direction is clearly visible and expected, 
since the brightest point of the shower is reconstructed and not the interaction vertex position. 
As discussed in \mysref{sec:shower_phenomenology}, the brightest point  of the shower is 
offset by $0.5\,\mathrm{m}-2\,\mathrm{m}$ in the relevant energy range 
(cf. \myfref{fig:shower_emisPosi_comparison}). 
The longitudinal and perpendicular distances are fitted with 
Gaussian functions for different neutrino energy and inelasticity bins. 
As an example, the distributions and the Gaussian fits are shown for 
$6\,\mathrm{GeV} < E_\nu < 7\,\mathrm{GeV}$ and $0.2 < y < 0.4$ in \myfref{fig:shower_reso_vtx_fitting} (middle and right).\\
The mean of the Gaussian fit to the distribution of the longitudinal
distances corresponds to the shift between the brightest point and the
neutrino interaction. The vertex resolution corresponds to the
resolution on the brightest point and is given by the fitted
widths. The longitudinal and perpendicular width can be combined
into a 3-dimensional resolution on the vertex by $\sigma_{\rm 3D} =
\sqrt { \sigma_{\rm long.}^2 + \sigma_{\rm perp.}^2 } $. 
The combined vertex resolution is about
$0.5\,\mathrm{m}-1\,\mathrm{m}$
and is dominated by the longitudinal vertex resolution.
This precise vertex reconstruction justifies the factorisation
of the shower reconstruction into a vertex reconstruction and
a shower energy, direction and inelasticity reconstruction. 

\begin{figure}[h!]
\centering
\begin{minipage}[c]{0.48\textwidth}
\centering
{\begin{overpic}[width=\textwidth]{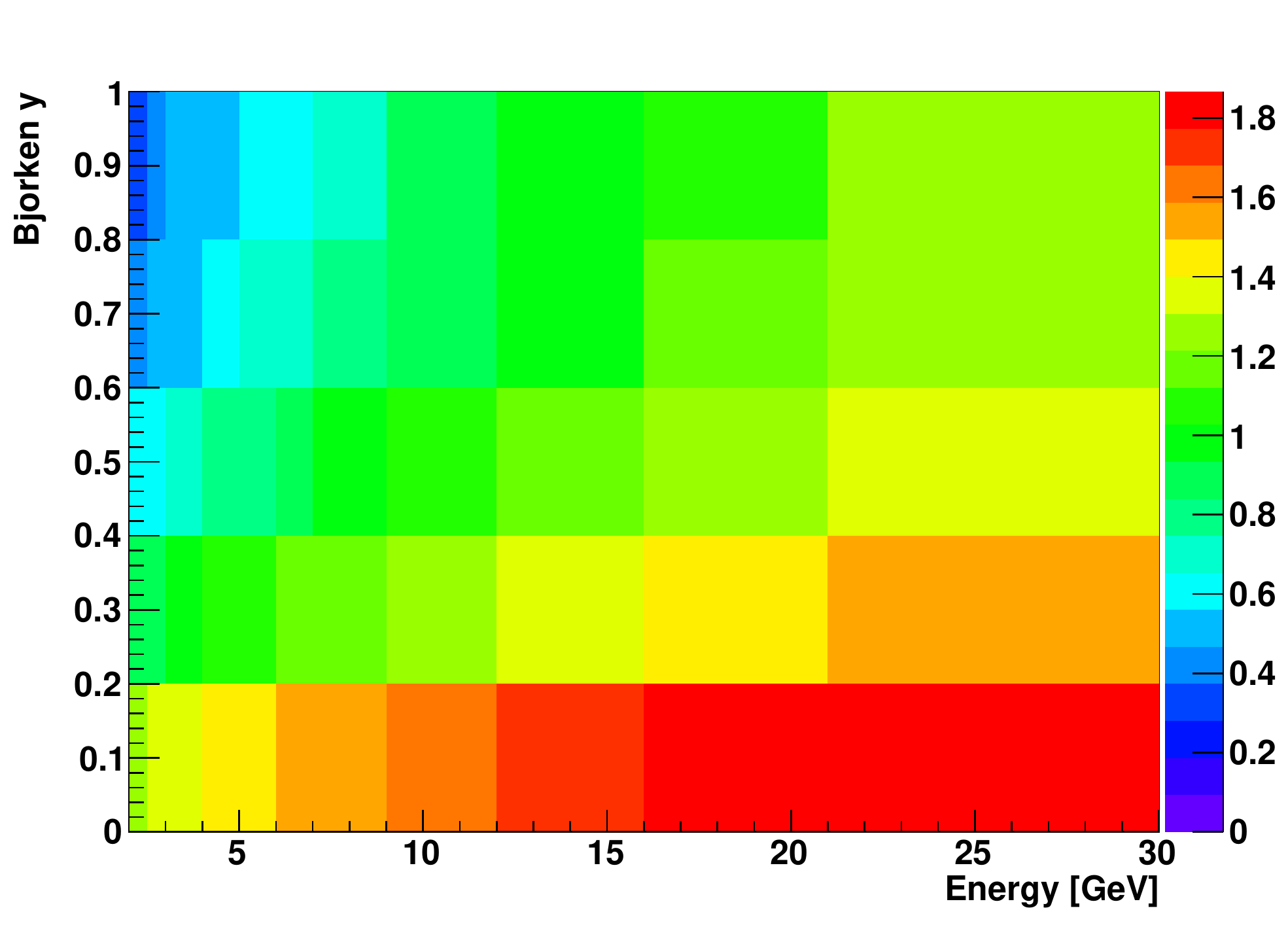}
\put (40,66) {\bf KM3NeT}
\end{overpic}
}
\end{minipage}
\caption{
Mean value in meter (from Gaussian fits) of the longitudinal distance
between the neutrino interaction vertex and the reconstructed
brightest point as a function of $E_\nu$ and $y$.
}
\label{fig:shower_reso}
\end{figure}

The fitted mean longitudinal vertex shift (in meter) 
is shown in \myfref{fig:shower_reso} as
a function of $E_\nu$ and Bjorken $y$. The increasing distance of the
reconstructed shower bright point from the interaction vertex
with increasing neutrino energy is clearly visible.

\begin{figure}[h!]
\centering
\begin{minipage}[c]{0.48\textwidth}
\centering
    {\includegraphics[width=\textwidth]{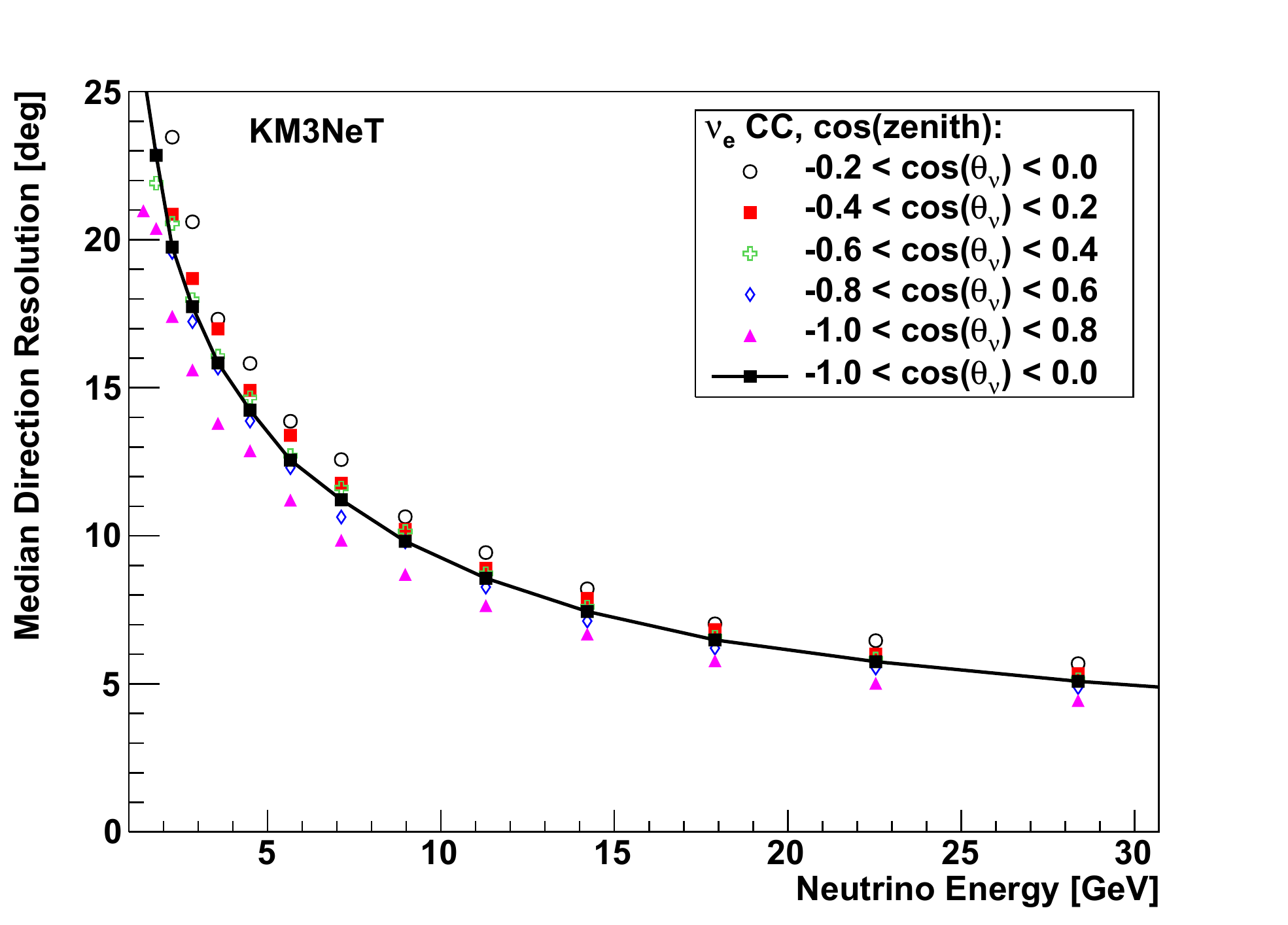}}
\end{minipage}
\hfill
\begin{minipage}[c]{0.48\textwidth}
\centering
    {\includegraphics[width =\textwidth]{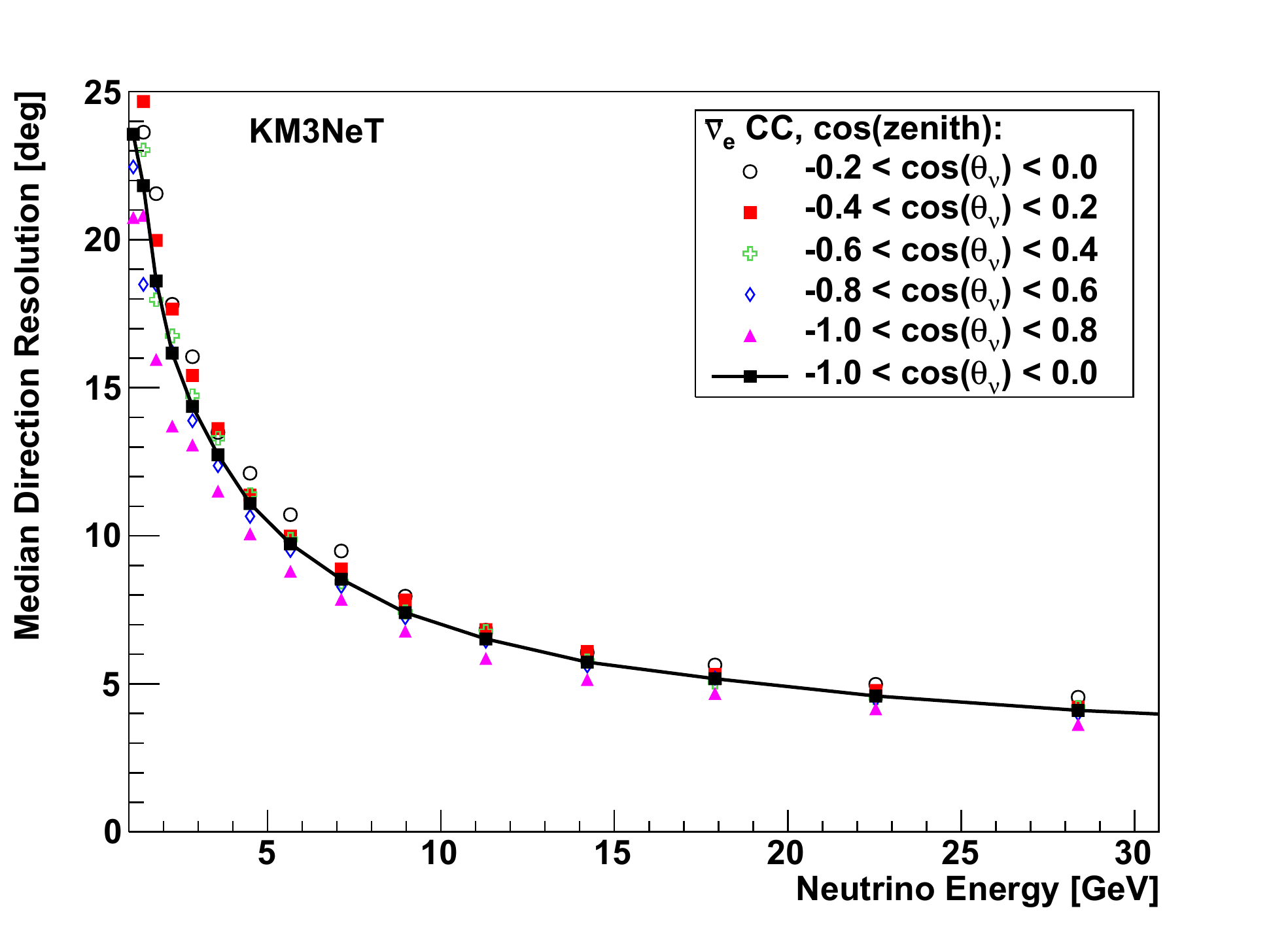}}
\end{minipage}
\caption{Median neutrino direction resolution (angle between
  reconstructed direction and neutrino direction) as a function of
  neutrino energy for different true neutrino cos(zenith) ranges,
  where cos(zenith)=-1 means vertical up-going and cos(zenith)=0 means
  horizontal. The black line corresponds to up-going neutrinos weighted
  according to the Bartol flux model. $\nu_e
  \mathrm{CC}$ (left) and $\bar \nu_e \mathrm{CC}$ (right).}
\label{fig:shower_reso_dirnu_forZenith}
\end{figure}

\subparagraph{Direction resolution:}
The median neutrino direction resolution (the angle between
reconstructed direction and neutrino direction) as a function of
neutrino energy is shown in \myfref{fig:shower_reso_dirnu_forZenith}
for different neutrino zenith angle ranges and for $\nu_e$ and $\bar
\nu_e$  separately.
For events weighted with the Bartol flux model the median directional
resolution is better than $10\,^{\circ}$ for energies above
$8.5\,\mathrm{GeV}$
for $\nu_e \mathrm{CC}$ and above $5.5\,\mathrm{GeV}$
for $\bar \nu_e \mathrm{CC}$ events. The resolution is slightly better
for vertical up-going than for horizontal neutrinos as more PMTs are oriented downward than upward in a DOM.

\begin{figure}[h!]
\centering
\begin{minipage}[c]{0.6\textwidth}
\centering
{\begin{overpic}[width=\textwidth]{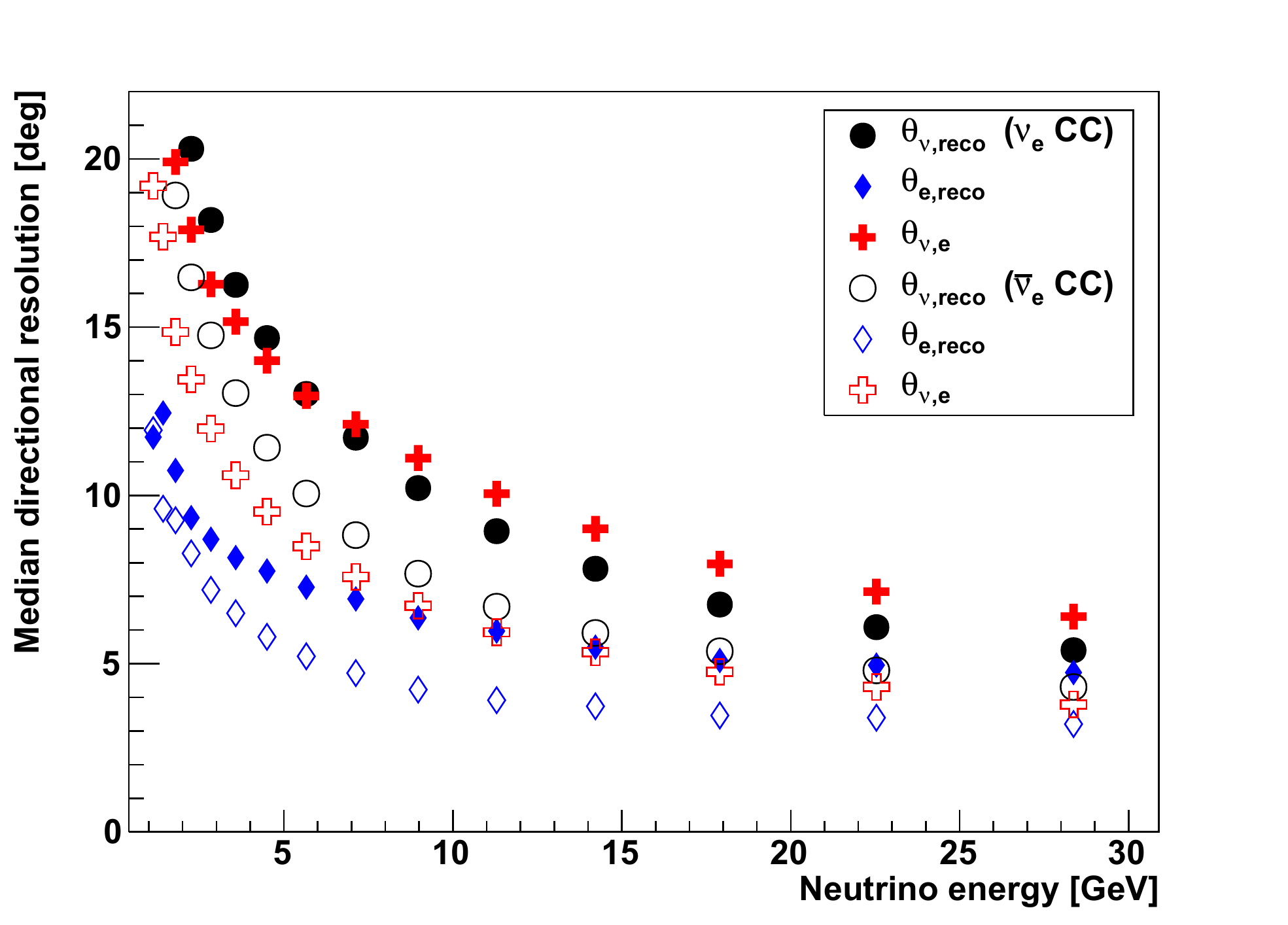}
\put (40,56) {\bf KM3NeT}
\end{overpic}
}
\end{minipage}
\caption{Median intrinsic scattering angle (red crosses), median electron direction resolution (blue diamonds) and the median neutrino direction resolution (black filled circles) as a function of neutrino energy for up-going $\nu_e \mathrm{CC}$ (solid marker) and $\bar \nu_e \mathrm{CC}$ (hollow marker) events weighted according to the Bartol flux model.}
\label{fig:shower_reso_dir_diffobj}
\end{figure}

As the reconstruction is designed to find the electron direction, the resolution is better for $\bar \nu_e$ than for $\nu_e$ due to the smaller average inelasticity for $\bar \nu_e$ leading on average to a smaller intrinsic scattering angle between the neutrino and the electron. The median intrinsic scattering angle, the median resolution with respect to the electron direction and the neutrino direction as a function of neutrino energy are shown in \myfref{fig:shower_reso_dir_diffobj}. For the relevant energy range the median electron direction resolution is smaller than the intrinsic scattering angle and the median neutrino direction resolution, verifying that the reconstruction actually has the ability to find the electron in $\nuan_e \mathrm{CC}$ events.

\begin{figure}[h!]
\centering
\begin{minipage}[c]{0.6\textwidth}
\centering
{\includegraphics[width=\textwidth]{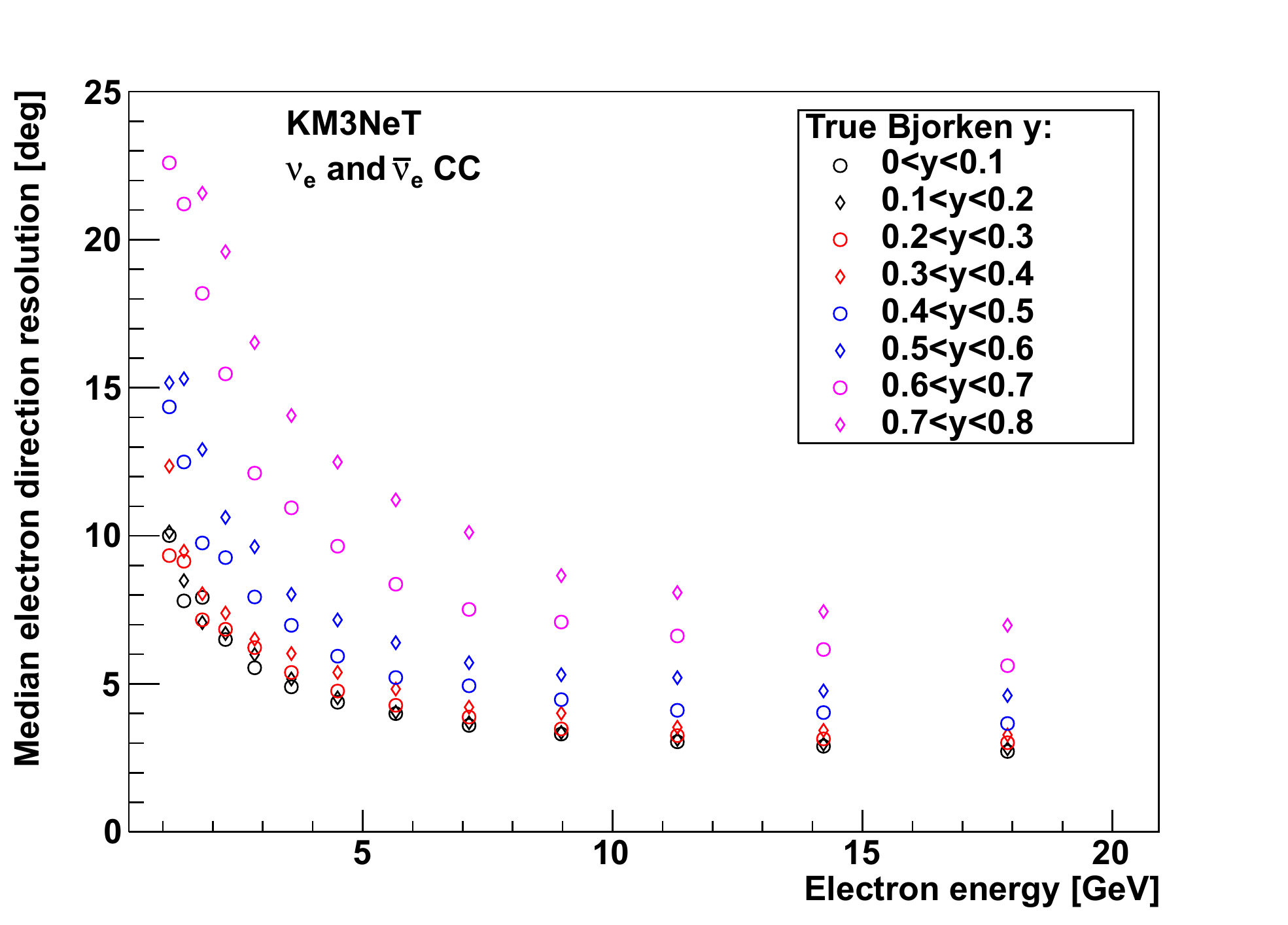}}
\end{minipage}
\caption{
Median electron direction resolution (angle between reconstructed direction and electron direction) as a function of electron energy for up-going $\nu_e$ and $\bar \nu_e \mathrm{CC}$ events weighted according to the Bartol flux model. Different marker colours and styles represent different true inelasticity $y$ ranges.
}
\label{fig:shower_reso_elecDir_diffBy}
\end{figure}

\myfref{fig:shower_reso_elecDir_diffBy} shows the median electron direction resolution as a function of electron energy for different true inelasticity $y$ ranges. The reconstruction of the electron direction is only slightly affected by the additional light from the hadronic shower up to $y \approx 0.5$. For $y \gtrsim 0.6$ the reconstruction can additionally be confused by high energetic particles in the hadronic shower producing a brighter Cherenkov ring than that from the electron. Due to momentum conservation the most energetic particles produced in neutrino interactions tend to have smaller scattering angles with respect to the neutrino direction. Therefore, by sometimes reconstructing the dominant particle from the hadronic shower the median neutrino direction for $\nu_e \mathrm{CC}$ events is slightly better than the intrinsic scattering angle between neutrino and electron for neutrino energies above $\sim 5\,\mathrm{GeV}$, as can be seen from \myfref{fig:shower_reso_dir_diffobj}.

\begin{figure}[h!]
\centering
\begin{minipage}[c]{0.48\textwidth}
\centering
    {\includegraphics[width=\textwidth]{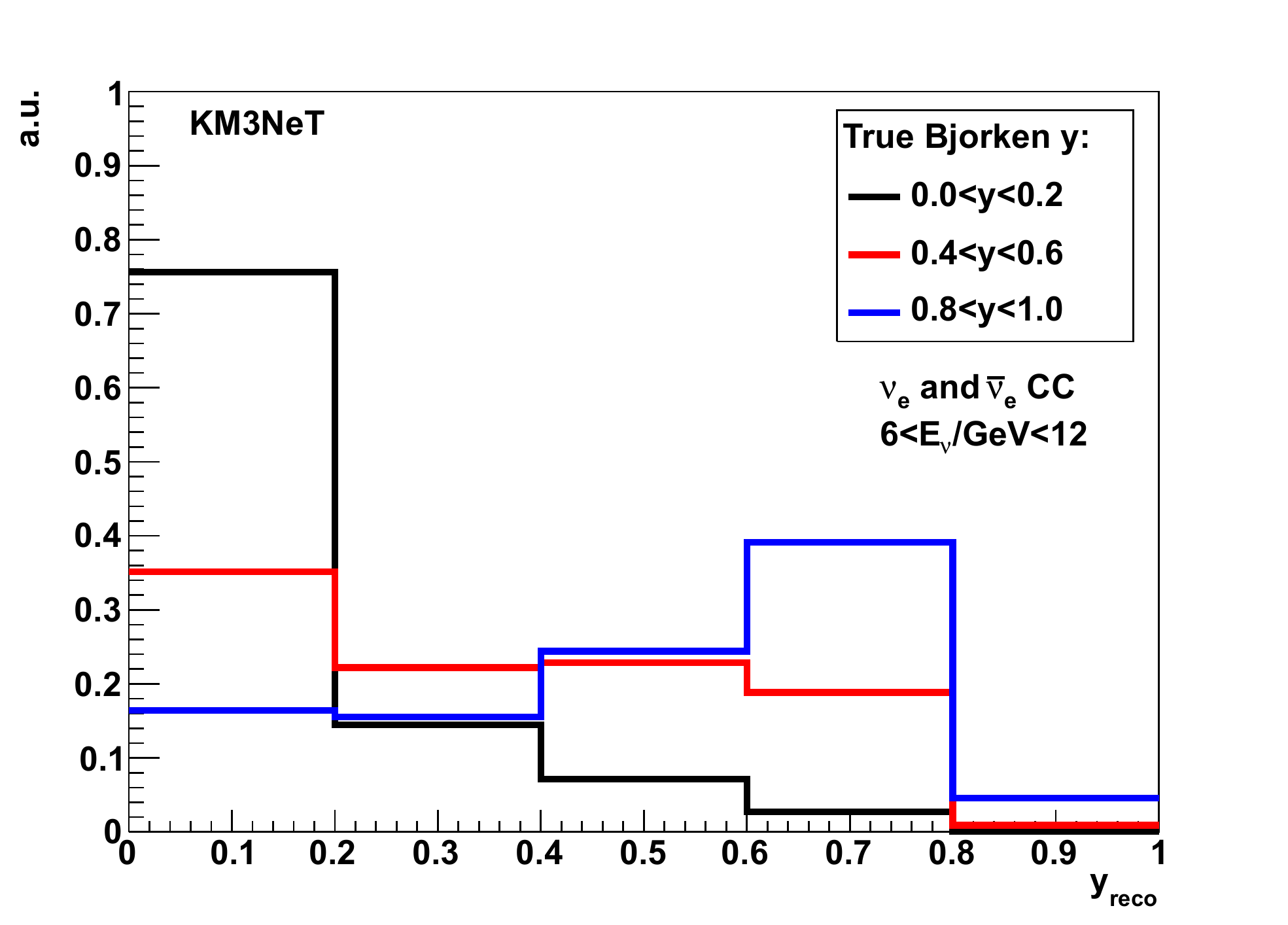}}
\end{minipage}
\hfill
\begin{minipage}[c]{0.48\textwidth}
\centering
    {\includegraphics[width=\textwidth]{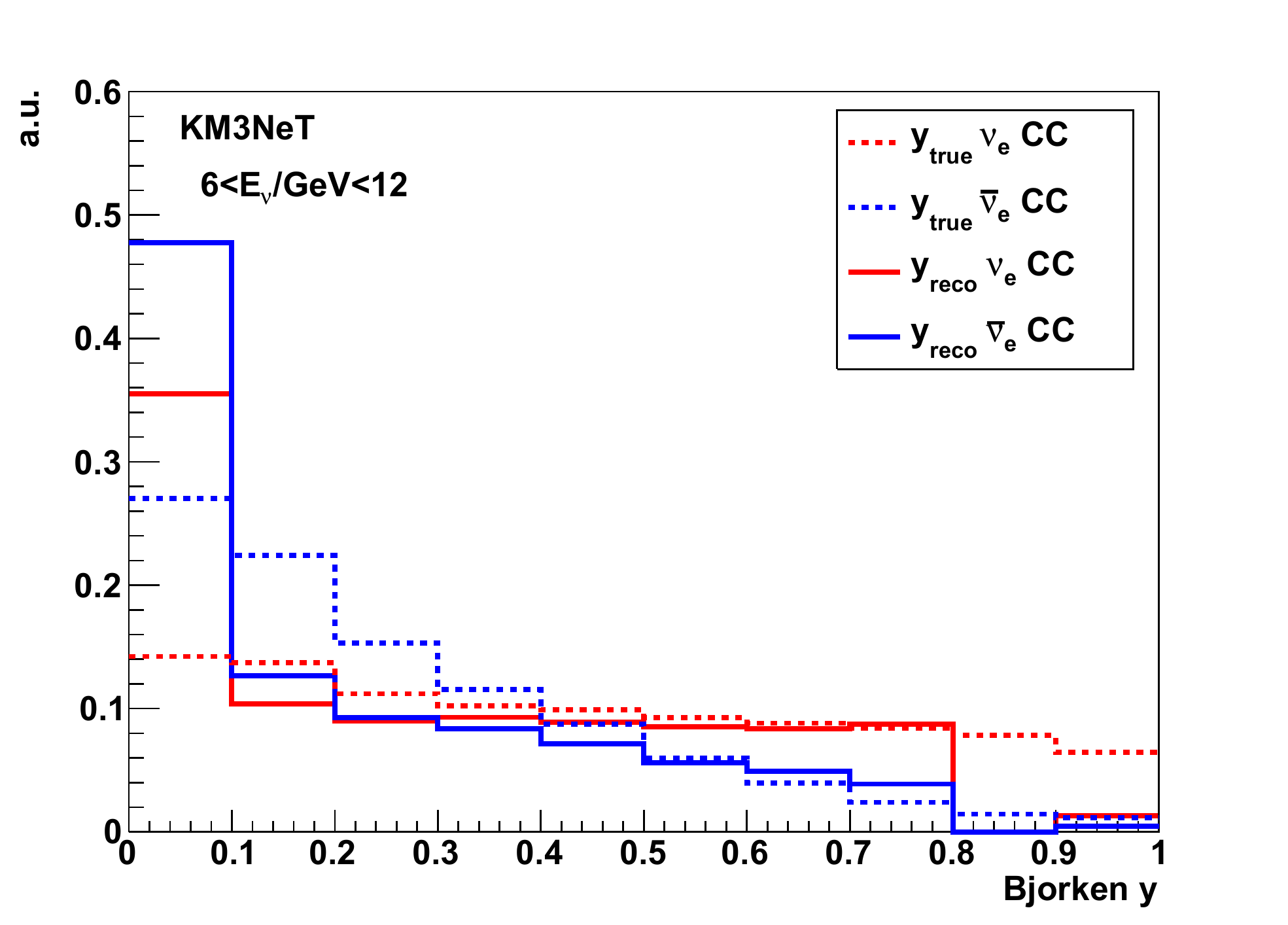}}
\end{minipage}
\caption{
Distribution of reconstructed inelasticity $y_{\rm reco}$ for three
different true $y$ ranges ($0<y<0.2$, $0.4<y<0.6$ and $0.8<y<1$) for
$\nu_e \mathrm{CC}$ and $\bar \nu_e \mathrm{CC}$ events with
$6\,\mathrm{GeV} < E_\nu < 12\,\mathrm{GeV}$ weighted according to the
Bartol flux model (left).
Distributions of the reconstructed inelasticity $y_{\rm reco}$ (solid lines) and
true inelasticity $y_{\rm true}$ (dashed lines) for $\nu_e \mathrm{CC}$ and $\bar
\nu_e \mathrm{CC}$ events (right).
}
 \label{fig:shower_reso_y}
\end{figure}

\subparagraph{Inelasticity resolution:}
The resolution on the inelasticity $y$ for a low, medium and high $y$ range is shown in \myfref{fig:shower_reso_y} (left) for $6\,\mathrm{GeV} < E_{\nu} < 12\,\mathrm{GeV}$.
The distributions of the reconstructed inelasticity $y_{\rm reco}$ and true inelasticity $y_{\rm true}$ for $\nu_e \mathrm{CC}$ and $\bar \nu_e \mathrm{CC}$ events are shown in \myfref{fig:shower_reso_y} (right) for $6\,\mathrm{GeV} < E_\nu < 12\,\mathrm{GeV}$.\\
The absence of $y_{\rm reco} > 0.8$ can be explained by dominant particles in the hadronic shower mimicking a lower inelasticity as discussed in \mysref{sec:shower_phenomenology}. 
The accumulation of events at low $y_{\rm reco}$ is larger than expected from the MC inelasticity distribution and visible for $\nu_e \mathrm{CC}$ and $\bar \nu_e \mathrm{CC}$ events. This is a feature of the reconstruction algorithm.\\
Due to the sensitivity to $y$ the $y_{\rm reco}$ distribution is different for $\nu_e$ and $\bar \nu_e \mathrm{CC}$ events leading to a separation power between both channels. This sensitivity to $y$ can also be used to separate $\nuan_e \mathrm{CC}$ events from $\nuan \mathrm{NC}$ events.

\begin{figure}[h!]
\centering
\begin{minipage}[c]{0.48\textwidth}
\centering
    {\includegraphics[width=\textwidth]{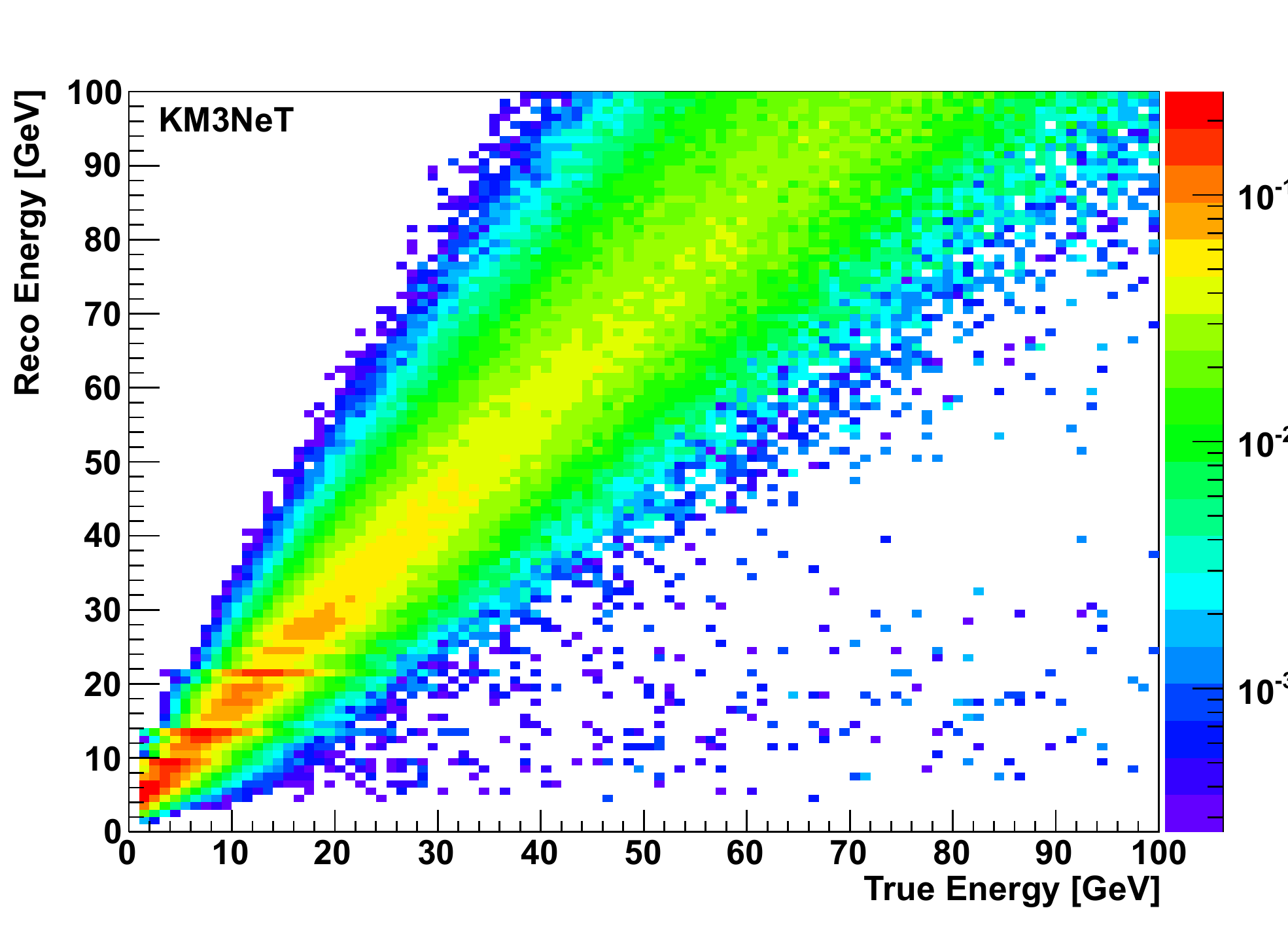}}
\end{minipage}
\hfill
\begin{minipage}[c]{0.48\textwidth}
\centering
    {\includegraphics[width=\textwidth]{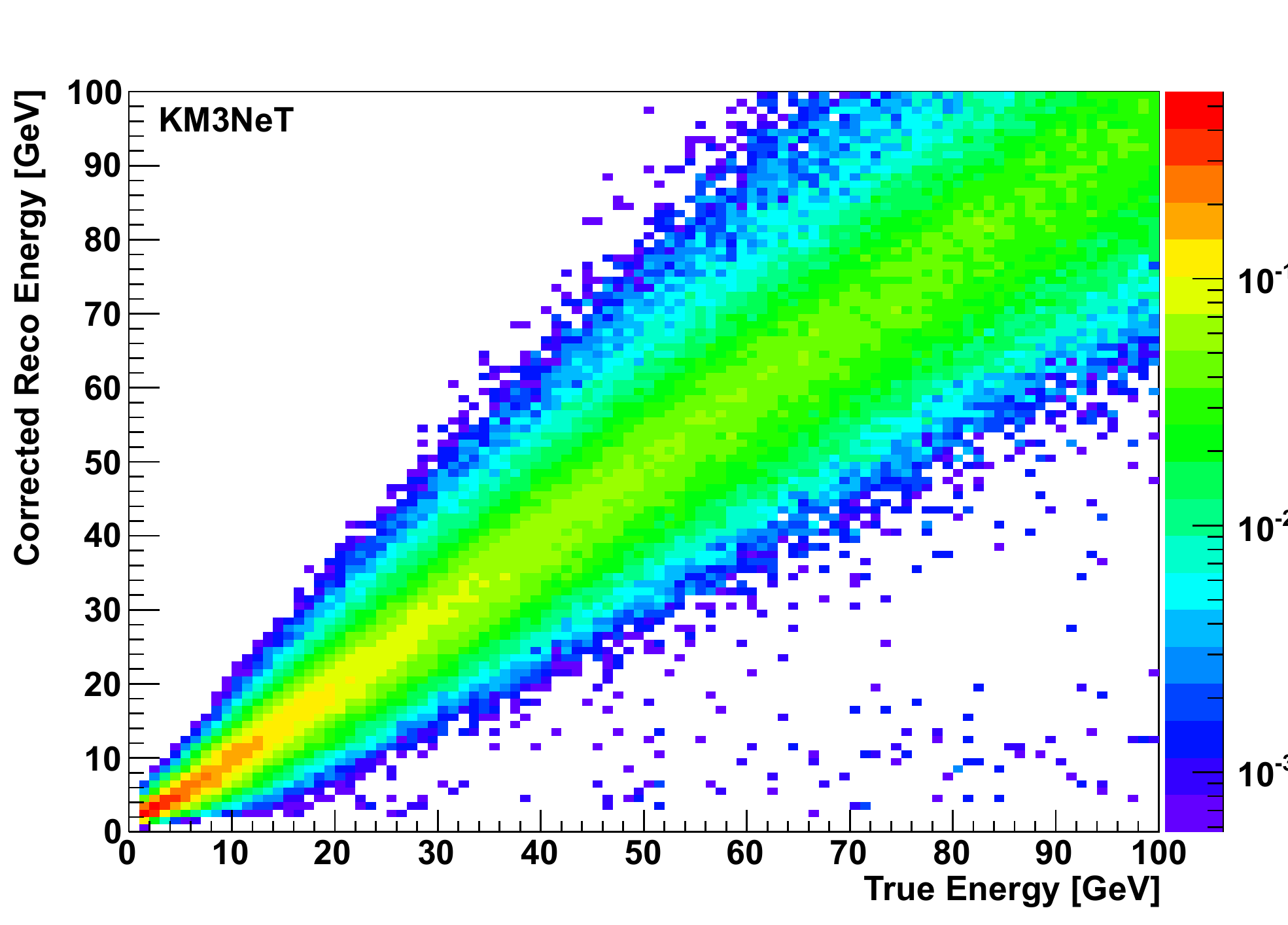}}
\end{minipage}
\caption{
Reconstructed energy as a function of true neutrino energy for $\nu_e
\mathrm{CC}$ and $\bar \nu_e \mathrm{CC}$ events weighted according to
the Bartol flux model (left). 
Corrected reconstructed energy as a function of true neutrino energy
for the same events (right).
}
\label{fig:shower_reso_Ereco_Etrue}
\end{figure}

\subparagraph{Energy resolution:}
In \myfref{fig:shower_reso_Ereco_Etrue} (left) the reconstructed energy is shown as a function of the neutrino energy for $\nuan_e \mathrm{CC}$ events weighted according to the Bartol flux model. The reconstructed energy is systematically higher than the neutrino energy. Therefore, an energy correction depending on the reconstructed zenith angle $\theta_{\rm reco}$, inelasticity $y_{\rm reco}$ and reconstructed energy $E_{\rm reco}$ is applied. The corrected reconstructed energy $E_{\rm reco}^{\rm corr}$ is given by
\begin{equation}
E_{\rm reco}^{\rm corr} = f(y_{\rm reco}, \theta_{\rm reco}, E_{\rm reco}) \cdot E_{\rm reco}, 
\label{eq:shower_energy_correction}
\end{equation}
where the 3-dimensional correction function  $f(y_{\rm reco}, \theta_{\rm reco}, E_{\rm reco})$ has been calculated from MC such that the median reconstructed energy is equal to the neutrino energy assuming a Bartol flux model. The corrected reconstructed energy as a function of the neutrino energy is shown in \myfref{fig:shower_reso_Ereco_Etrue} (right).

\begin{figure}[h!]
\centering
\includegraphics[width=\textwidth]{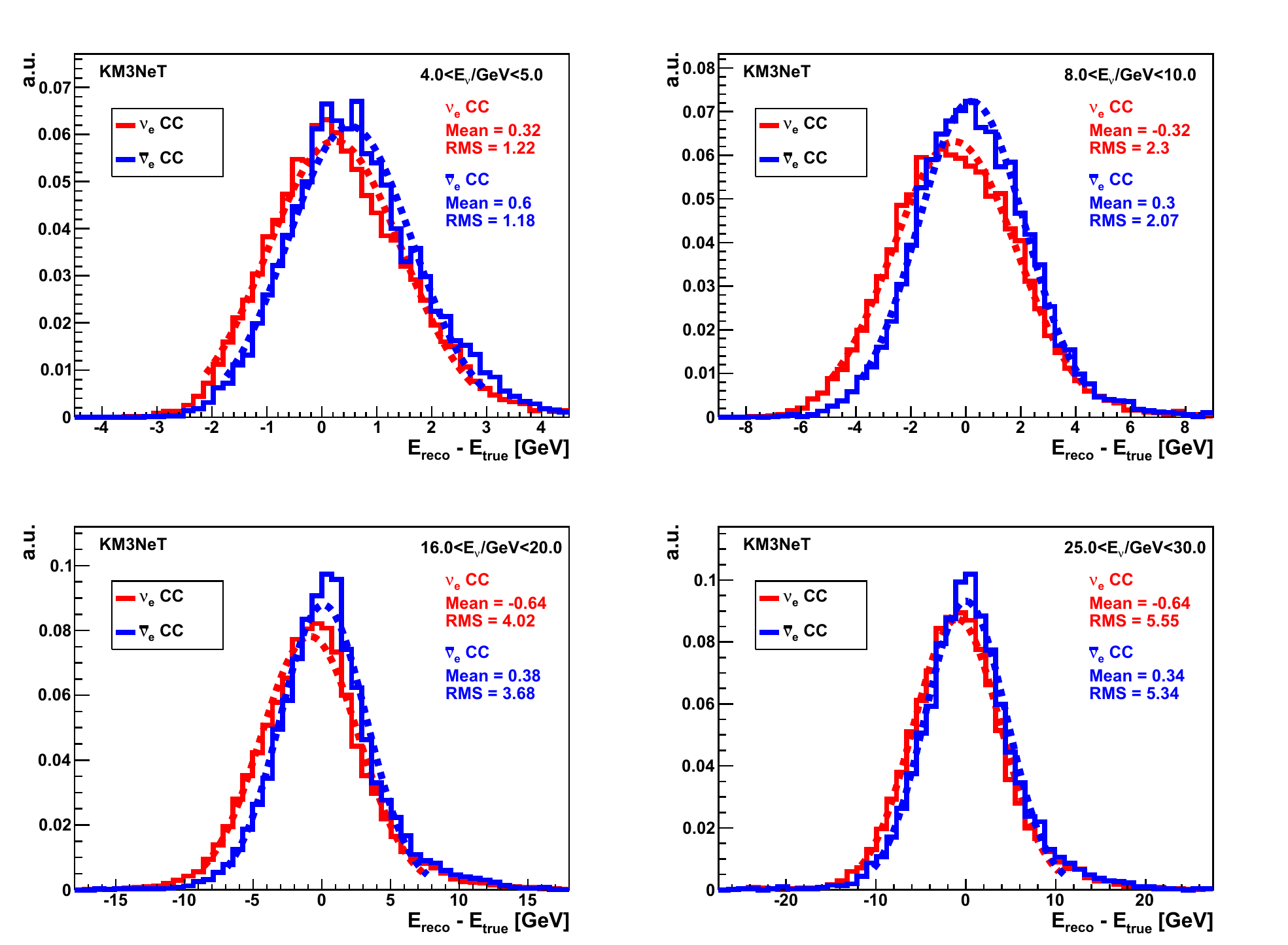}
\caption{Difference between corrected reconstructed energy and neutrino energy in different neutrino energy bins for $\nu_e \mathrm{CC}$ (red) and $\bar \nu_e \mathrm{CC}$ (blue) events weighted according to the Bartol flux model. Dashed lines show Gaussian fits.}
\label{fig:shower_reso_dE_nueCC_anueCC}
\end{figure}

The difference between reconstructed and neutrino energy in different neutrino energy bins is shown in 
\myfref{fig:shower_reso_dE_nueCC_anueCC} for $\nu_e \mathrm{CC}$ and $\bar \nu_e \mathrm{CC}$ events separately. 
These distributions are very well described by Gaussians.\\

\begin{figure}[h!]
\centering
{\begin{overpic}[width=0.5\textwidth]{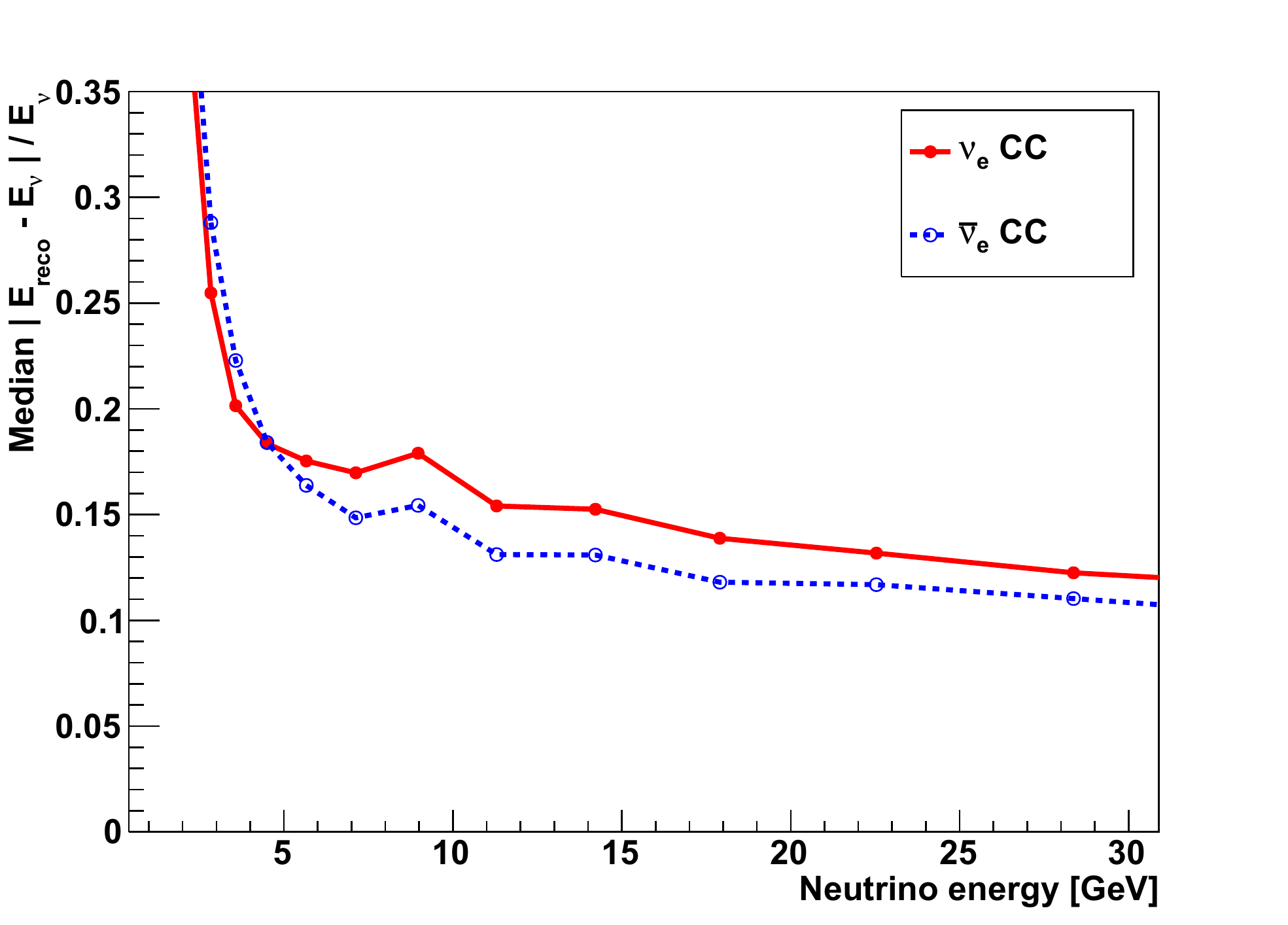}
\put (40,56) {\bf KM3NeT}
\end{overpic}
}
\caption{
Median fractional energy resolution ($|E_{\rm reco} - E_\nu|/E_\nu$) as a function of neutrino energy for $\nu_e \mathrm{CC}$ (red) and $\bar \nu_e \mathrm{CC}$ (blue) events weighted according to the Bartol flux model.}
\label{fig:shower_fracEreso_nueCC_anueCC}
\end{figure}

\begin{figure}[h!]
\centering
\begin{minipage}[c]{0.48\textwidth}
\centering
    {\includegraphics[width=\textwidth]{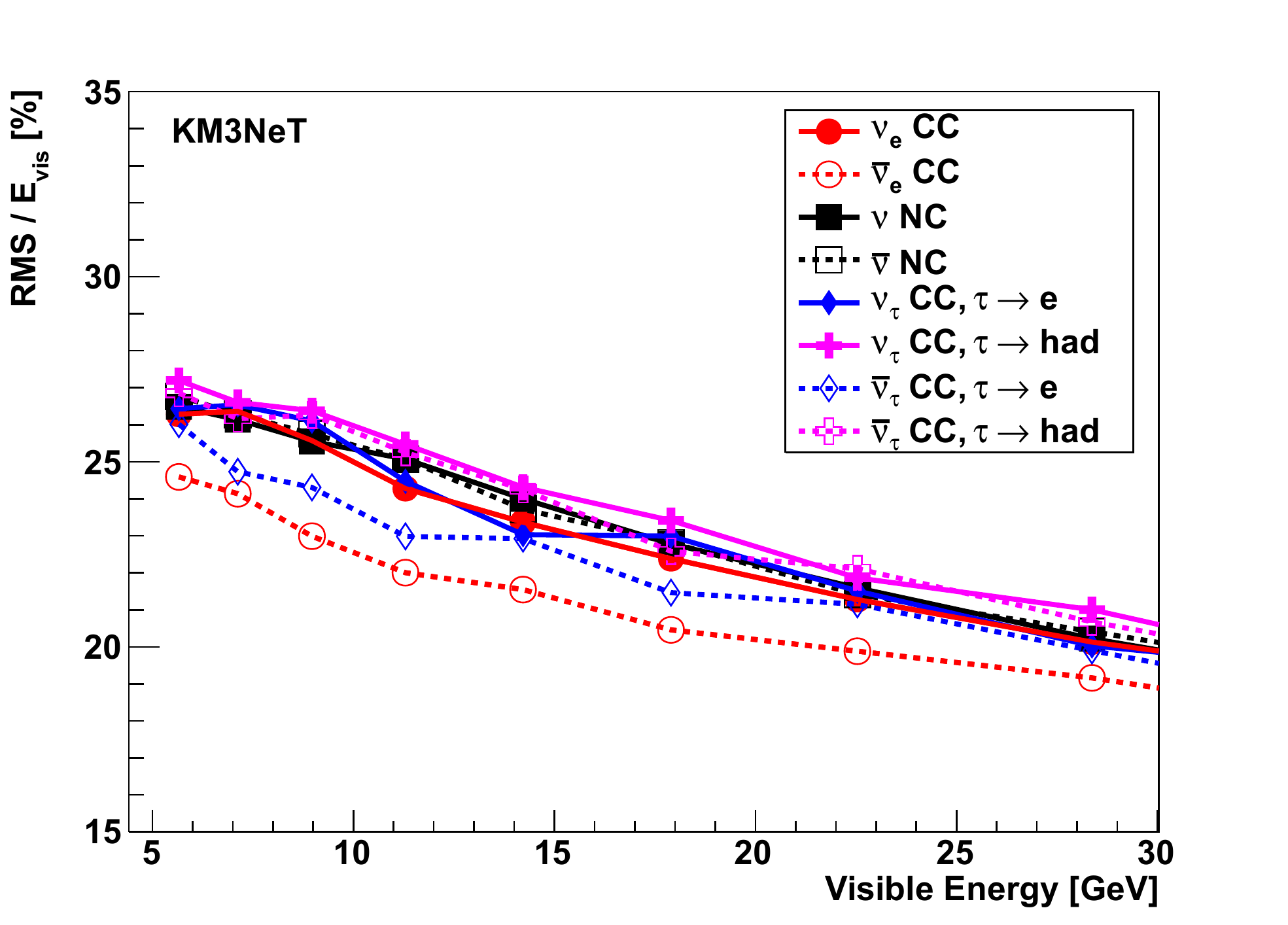}}
\end{minipage}
\hfill
\begin{minipage}[c]{0.48\textwidth}
\centering
    {\includegraphics[width=\textwidth]{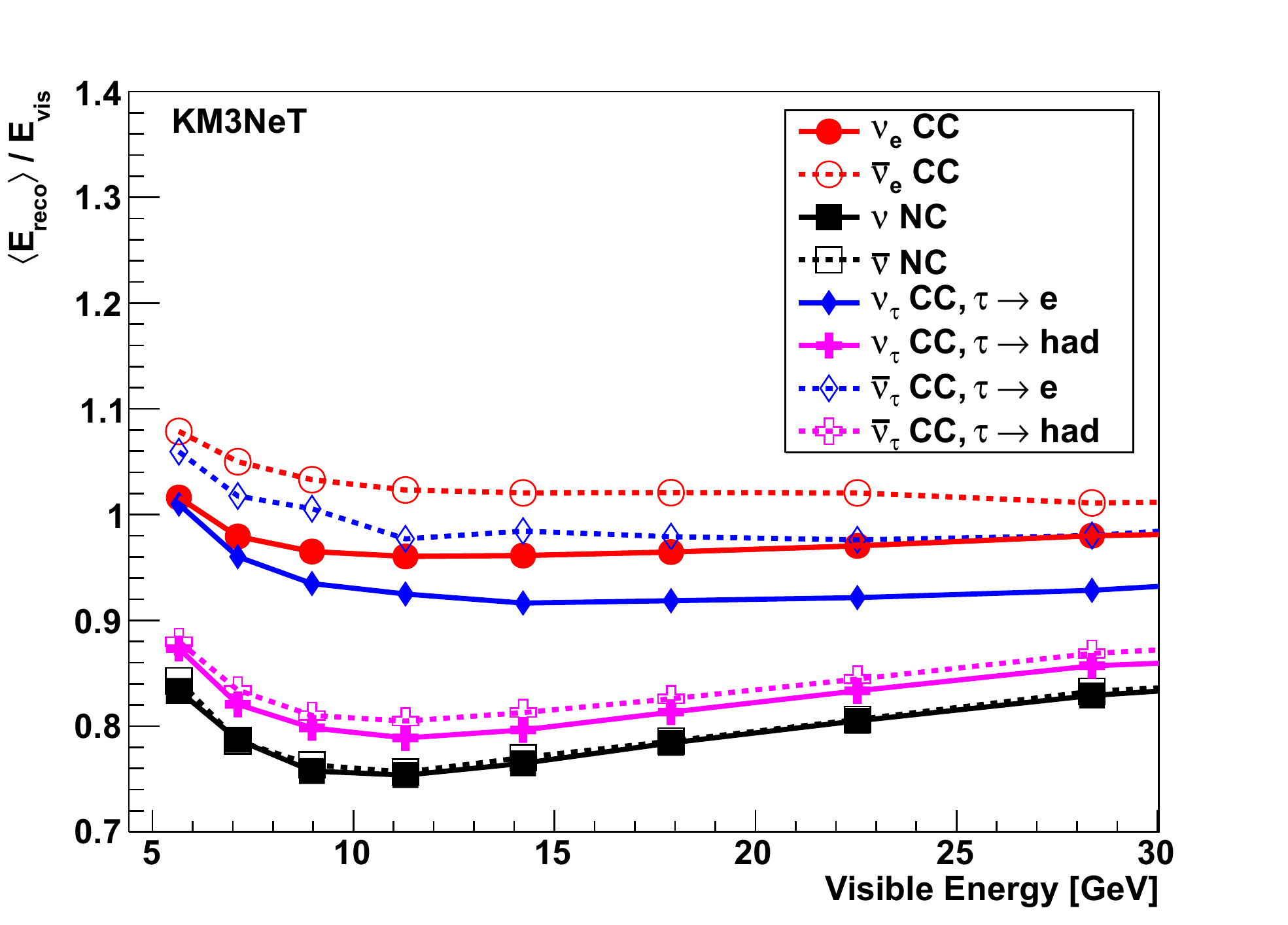}}
\end{minipage}
\caption{
Relative energy resolution RMS/$E_{\rm vis}$ as a function of the
visible energy $E_{\rm vis}$ for shower-like neutrino interaction
channels (left) and mean relative offset in reconstructed energy --
given as $\langle E_\mathrm{reco} \rangle / E_{\rm vis}$ -- as a
function of $E_{\rm vis}$ (right). 
}
\label{fig:shower_allflavours_energyReso}
\end{figure}

The median fractional energy resolution -- given as $| E_{\rm reco} - E_\nu | / E_\nu$ -- is better than 18\% for neutrino energies above $5\,\mathrm{GeV}$ for up-going $\nu_e \mathrm{CC}$ and $\bar \nu_e \mathrm{CC}$ events and is shown as a function of neutrino energy in \myfref{fig:shower_fracEreso_nueCC_anueCC}. 
The relative energy resolution -- given as the RMS of $(E_{\rm reco} - E_\nu)$ distributions (cf. \myfref{fig:shower_reso_dE_nueCC_anueCC}) over neutrino energy -- is better than 26\% (24\%)
for neutrino energies above $7\,\mathrm{GeV}$ for up-going $\nu_e$ ($\bar \nu_e$) CC events and is shown as a function of visible energy $E_{\rm vis}$ in \myfref{fig:shower_allflavours_energyReso} (left) together with the resolution for the other shower-like
 neutrino interaction channels. For $\nuan_e \mathrm{CC}$ events the visible energy is equal to the neutrino energy.
The resolution is better for $\bar \nu_e \mathrm{CC}$ events than for $\nu_e \mathrm{CC}$ events due to the lower 
average contribution from the hadronic shower which shows larger fluctuations
than electromagnetic showers \cite{ShowerFluct}.
\myfref{fig:shower_allflavours_energyReso} (right) shows the mean relative offset between the mean reconstructed energy and the visible energy.
At energies corresponding to the effective volume turn-on region the
reconstructed energy is overestimated for $\bar \nu_e \mathrm{CC}$ and
$\nu_e \mathrm{CC}$ events as only events  pass the event selection
criteria that appear more energetic than
they actually are.
Above $\sim 9\,\mathrm{GeV}$ the reconstructed energies are slightly overestimated (underestimated) for $\bar \nu_e$ ($\nu_e$) CC due to the smaller light yield of hadronic showers compared to electromagnetic showers.

\paragraph{Performance for shower-like neutrino events\\} 
\label{sec:shower_reso_allflavours}
The performance of the shower reconstruction is evaluated separately on different shower-like neutrino interaction event samples:
$\nu_{e}$ and $\bar \nu_{e}$~CC events,
$\nu_{e/\mu}$ and $\bar \nu_{e/\mu}$~NC events,
$\nu_{\tau}$ and $\bar \nu_{\tau}$~CC events where the $\tau$ lepton
decays in an electron or hadrons.

\begin{figure}[h!]
\centering
\begin{minipage}[c]{0.48\textwidth}
\centering
{\begin{overpic}[width=\textwidth]{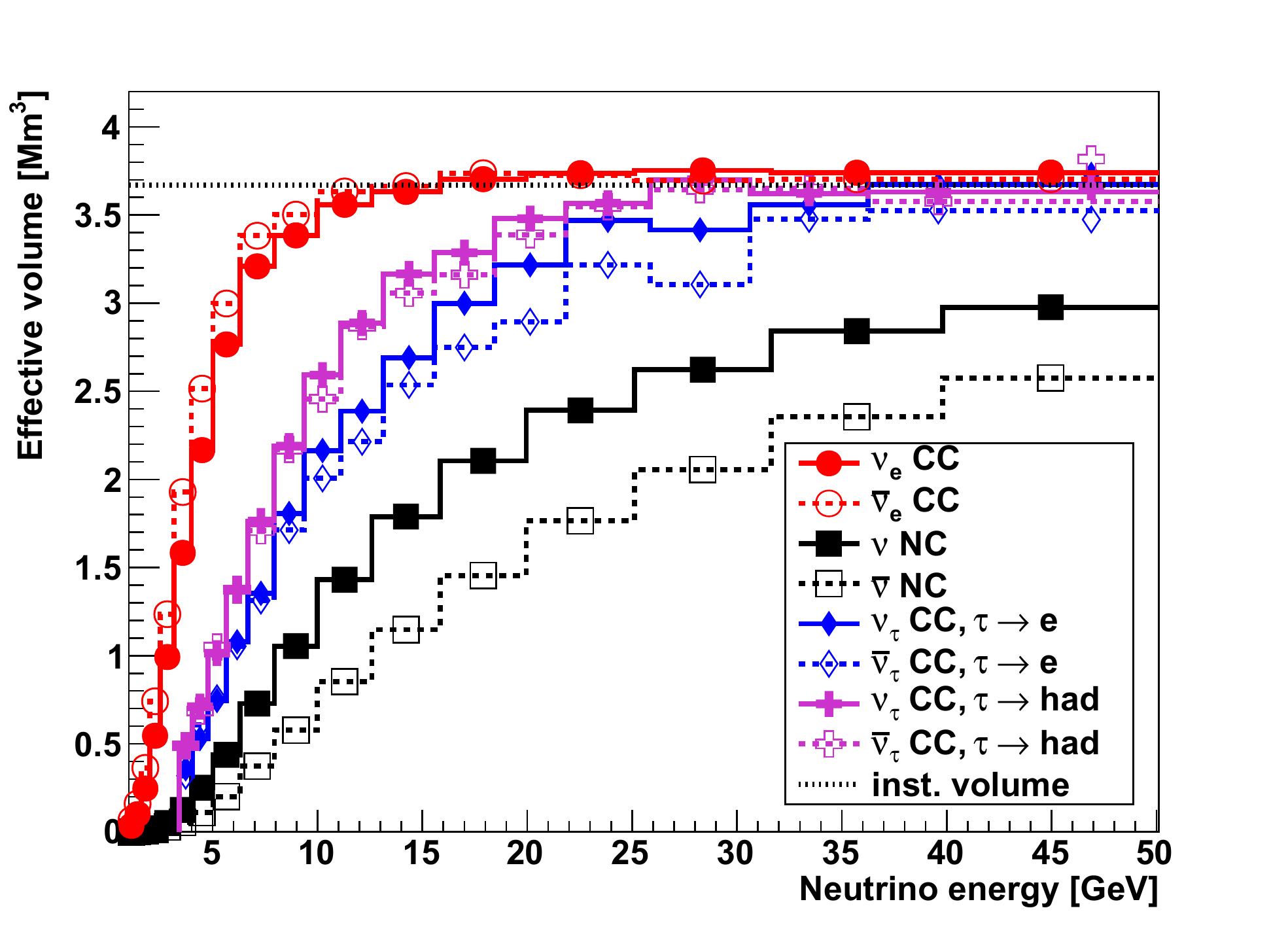}
\put (14,59) {\tiny \bf KM3NeT}
\end{overpic}
}
\end{minipage}
\hfill
\begin{minipage}[c]{0.48\textwidth}
\centering
    {\includegraphics[width=\textwidth]{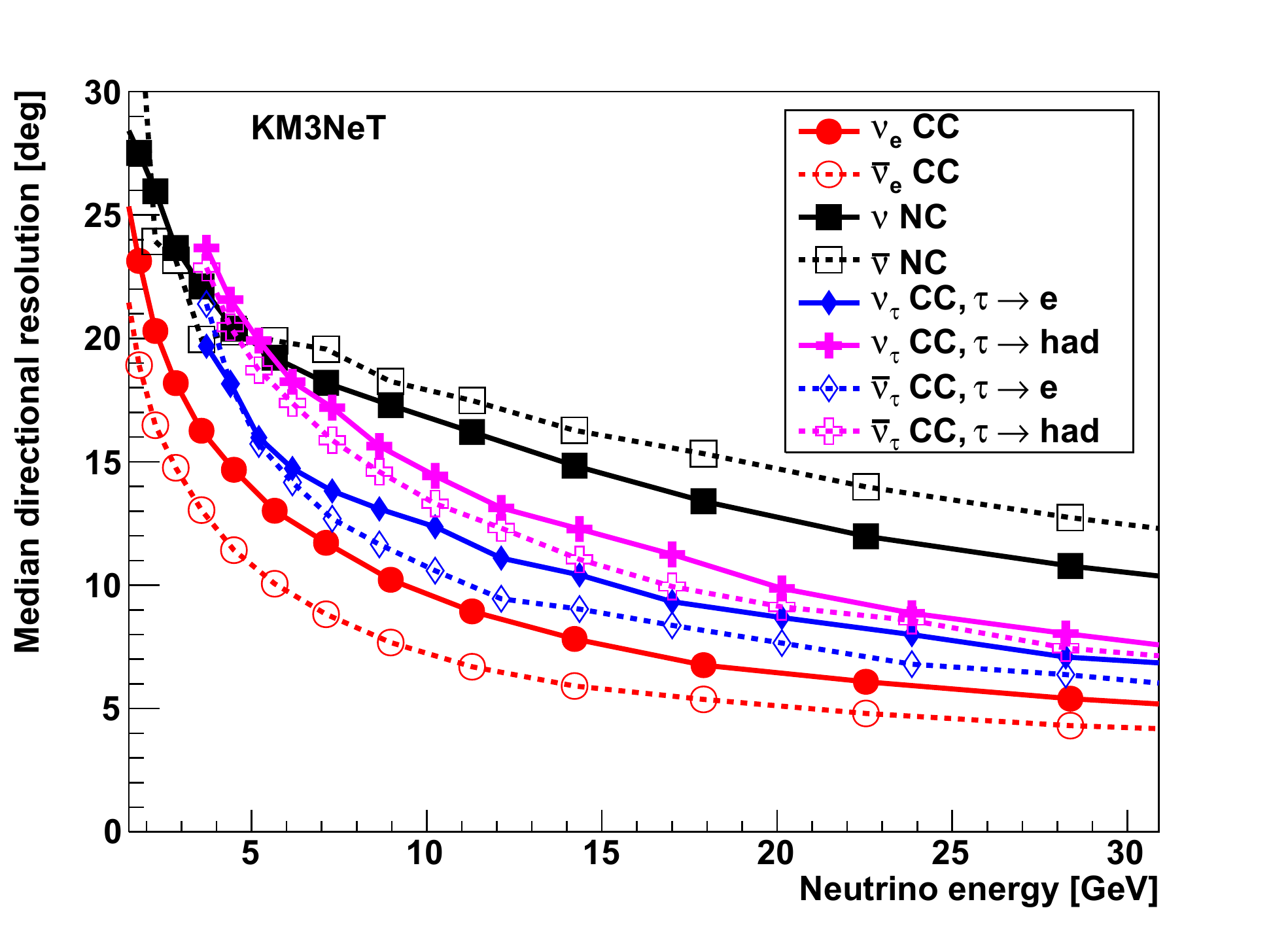}}
\end{minipage}
\caption{
Effective volumes (left) and median neutrino direction resolution (right) as a function of neutrino energy for different up-going shower-like neutrino event types. 
}
\label{fig:shower_allflavours_effVol_dirReso}
\end{figure}

The effective volume for up-going shower-like neutrino events is shown in \myfref{fig:shower_allflavours_effVol_dirReso} (left) as a function of neutrino energy.
The turn-on is much less steep for $\nuan \mathrm{NC}$ and $\nuan_\tau \mathrm{CC}$ events than for $\nuan_e \mathrm{CC}$ events, as the outgoing neutrinos are invisible to the detector. 
For $\nuan_\tau \mathrm{CC}$ events the turn-on is steeper than for $\nuan \mathrm{NC}$ events as on average the visible energy in $\nuan_\tau \mathrm{CC}$ events is larger than in $\nuan \mathrm{NC}$ events.
In $\nu \mathrm{NC}$ events the average inelasticity is higher than in $\bar \nu \mathrm{NC}$ events leading to more energetic hadronic showers and a steeper turn-on.

The median directional resolution is shown in \myfref{fig:shower_allflavours_effVol_dirReso} (right) as a function of neutrino energy. 
The directional resolution for $\nuan \mathrm{NC}$ and $\nuan_\tau \mathrm{CC}$ events is clearly worse than for $\nuan_e \mathrm{CC}$ events as the information of the outgoing neutrinos is unavailable. 
As the angle between the hadronic shower and the neutrino is smaller for $\nu \mathrm{NC}$ than for $\bar \nu \mathrm{NC}$ events due to a higher average inelasticity, the directional resolution is better.

The relative energy resolution -- given as RMS over visible energy $E_\mathrm{vis}$ -- for up-going shower-like neutrino events is shown as a function of $E_\mathrm{vis}$
in \myfref{fig:shower_allflavours_energyReso} (left). $E_\mathrm{vis}$ is defined as the difference between the energy of the incoming neutrino and the outgoing neutrino(s) from the primary neutrino interaction (NC events) or $\tau$-decay ($\nuan_\tau \mathrm{CC}$ events).
The resolution is worse for events with higher average contribution
from hadronic showers which show larger fluctuations \cite{ShowerFluct}.

Due to the smaller light yield of hadronic showers compared to electromagnetic showers, the ration $\langle E_\mathrm{reco} \rangle / E_\mathrm{vis}$ is different for each neutrino interaction channel and energy dependent. 
This can be seen in \myfref{fig:shower_allflavours_energyReso} (right).
The higher the fraction of electromagnetic shower component in the event the higher is the mean reconstructed energy. 
This leads also to different turn-on behaviours in the effective volume for both shower types, and consequently to different compositions (in terms of electromagnetic and hadronic shower components) of well reconstructed neutrino events.
The latter explains 
the behaviour
below $E_\mathrm{vis} \lesssim 10$\,GeV. 

\begin{figure}[h!]
\centering
\includegraphics[width=0.5\linewidth]{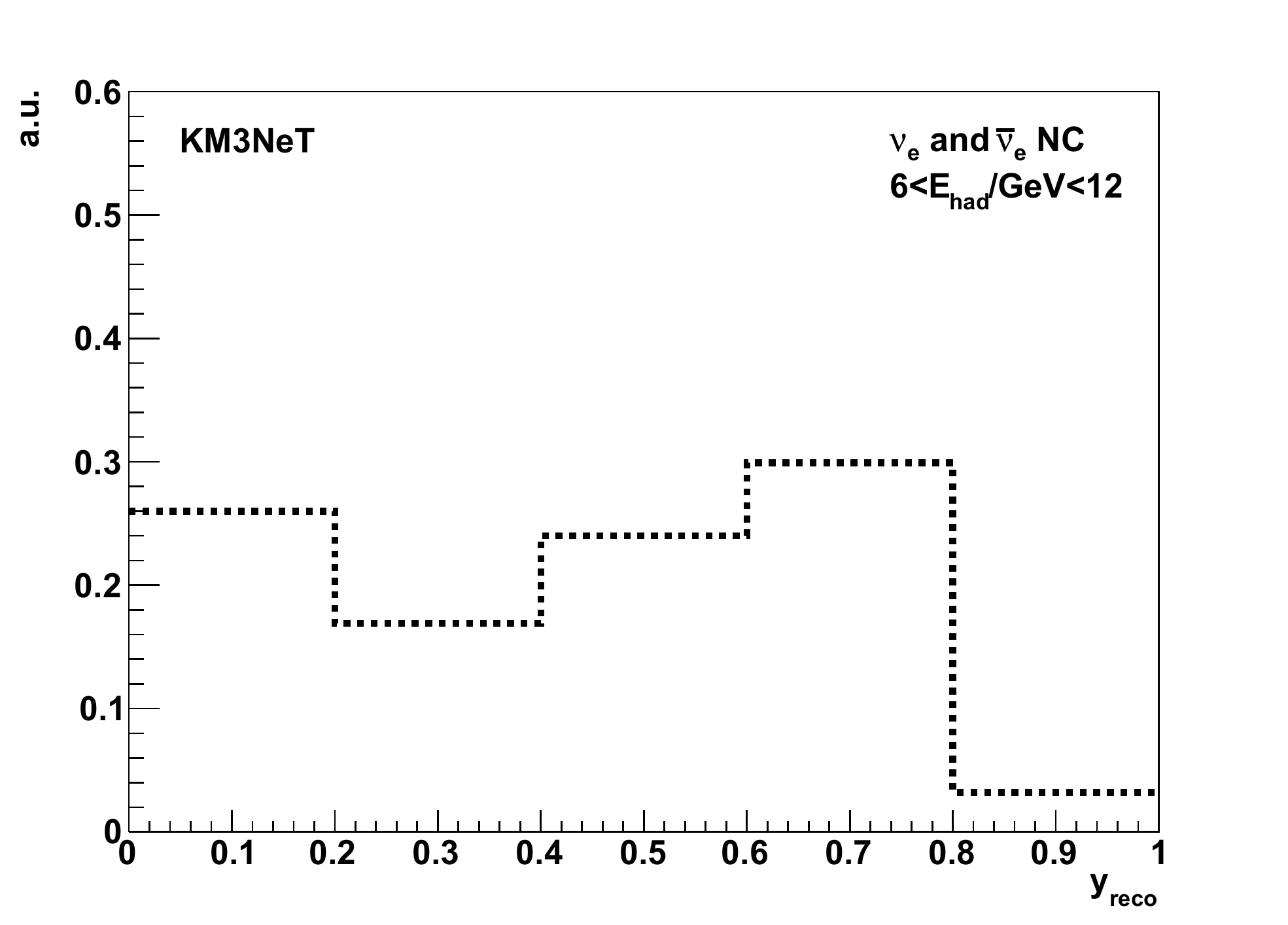}
\caption{Distribution of reconstructed inelasticity $y_{\rm reco}$ for NC events with 
hadronic shower energies of $6\,\mathrm{GeV} < E_{\rm had} < 12\,\mathrm{GeV}$ and an arbitrary true inelasticity.}
\label{fig:shower_resoNC_y}
\end{figure}

The distribution of the reconstructed inelasticity $y_{\rm reco}$ for $\nuan \mathrm{NC}$ events with hadronic shower energies of $6\,\mathrm{GeV} < E_{\rm had} < 12\,\mathrm{GeV}$ is shown in \myfref{fig:shower_resoNC_y}. As expected, the $y_{\rm reco}$ distribution for NC events looks similar to the distribution for $\nuan_e \mathrm{CC}$ events with $0.8<y<1$ (cf. \myfref{fig:shower_reso_y}), but different to the other $y$ ranges, leading to a separation power between shower-like events from $\nuan_e \mathrm{CC}$ and $\nuan \mathrm{NC}$ events.


\subsubsection{Performance for different vertical spacings}
\label{sec:shower_different_spacings}
The different detector configurations studied in this section are
described in \mysref{sec:masking}.
The performance for different vertical spacings is studied on up-going $\nuan_e \mathrm{CC}$ events weighted according to the Bartol atmospheric neutrino flux model.
Events are selected according to the same criteria as described in \mysref{sec:shower_reco_performance}. 
For each detector configuration the respective energy correction is applied (cf. \mysref{sec:shower_reco_performance}).

\begin{figure}[h!]
\centering
\includegraphics[width=0.49\linewidth]{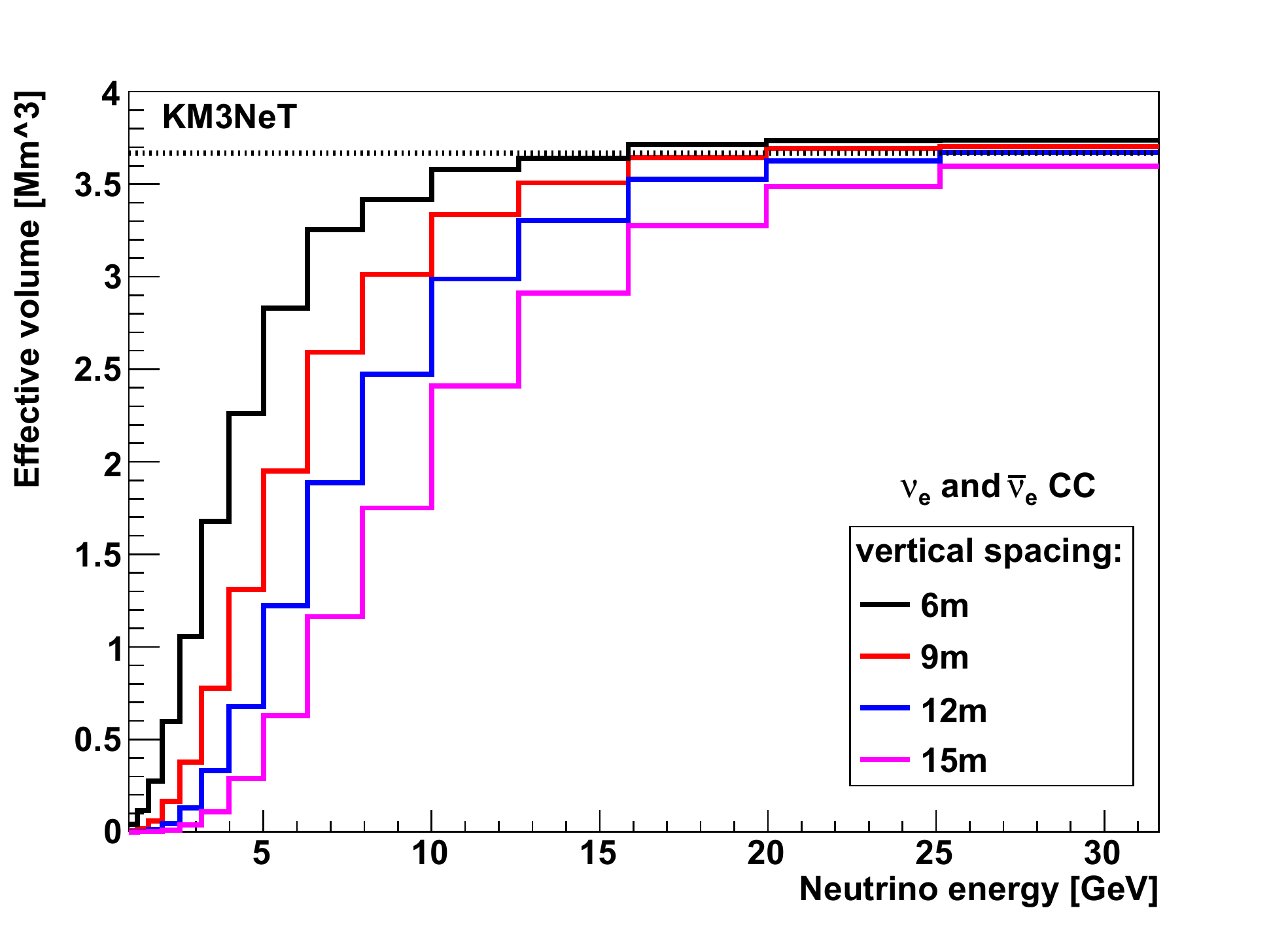}
{\begin{overpic}[width=0.49\linewidth]{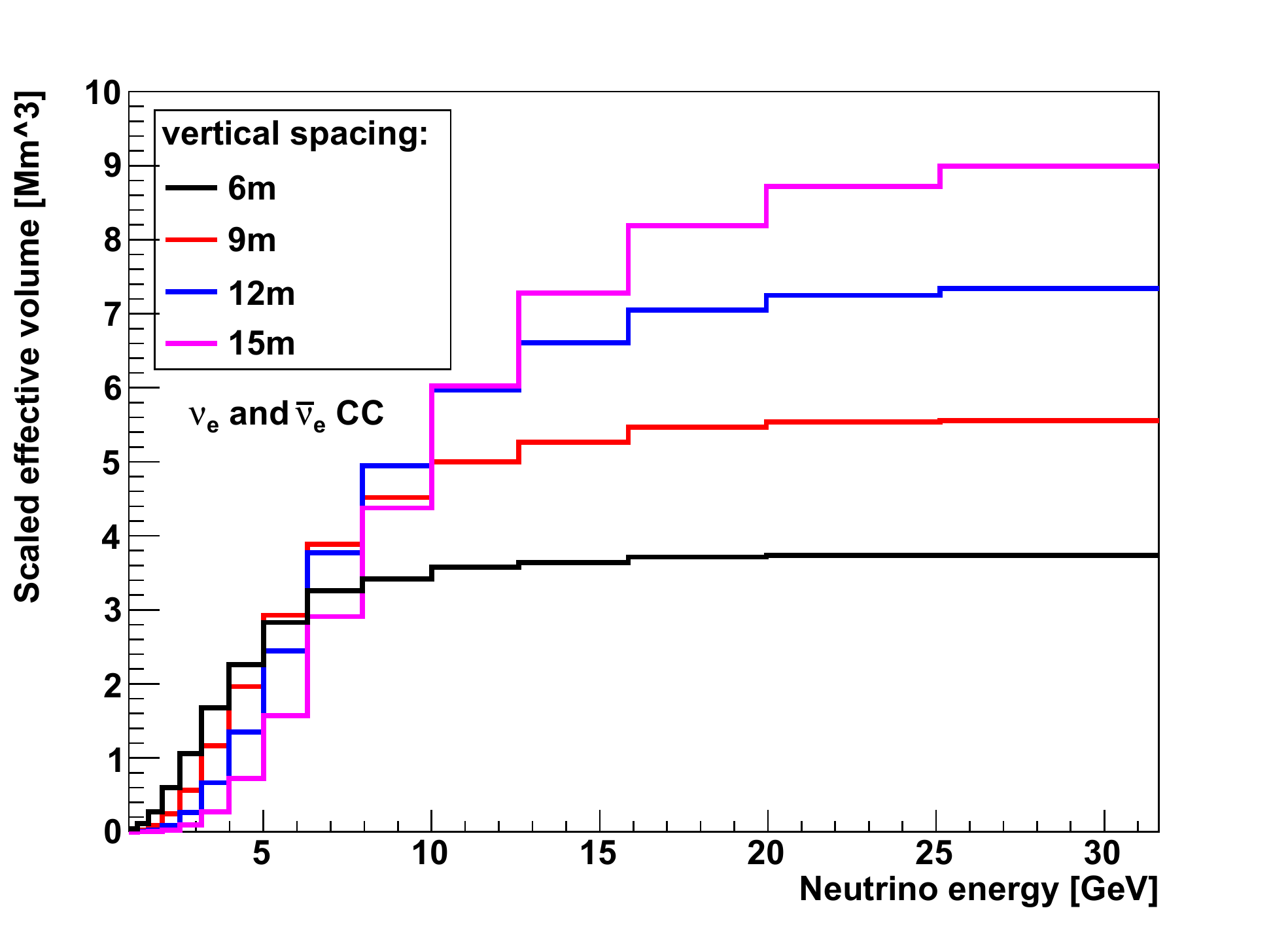} 
\put (37,60) {\tiny \bf KM3NeT}
\end{overpic}
}
{\begin{overpic}[width=0.49\linewidth]{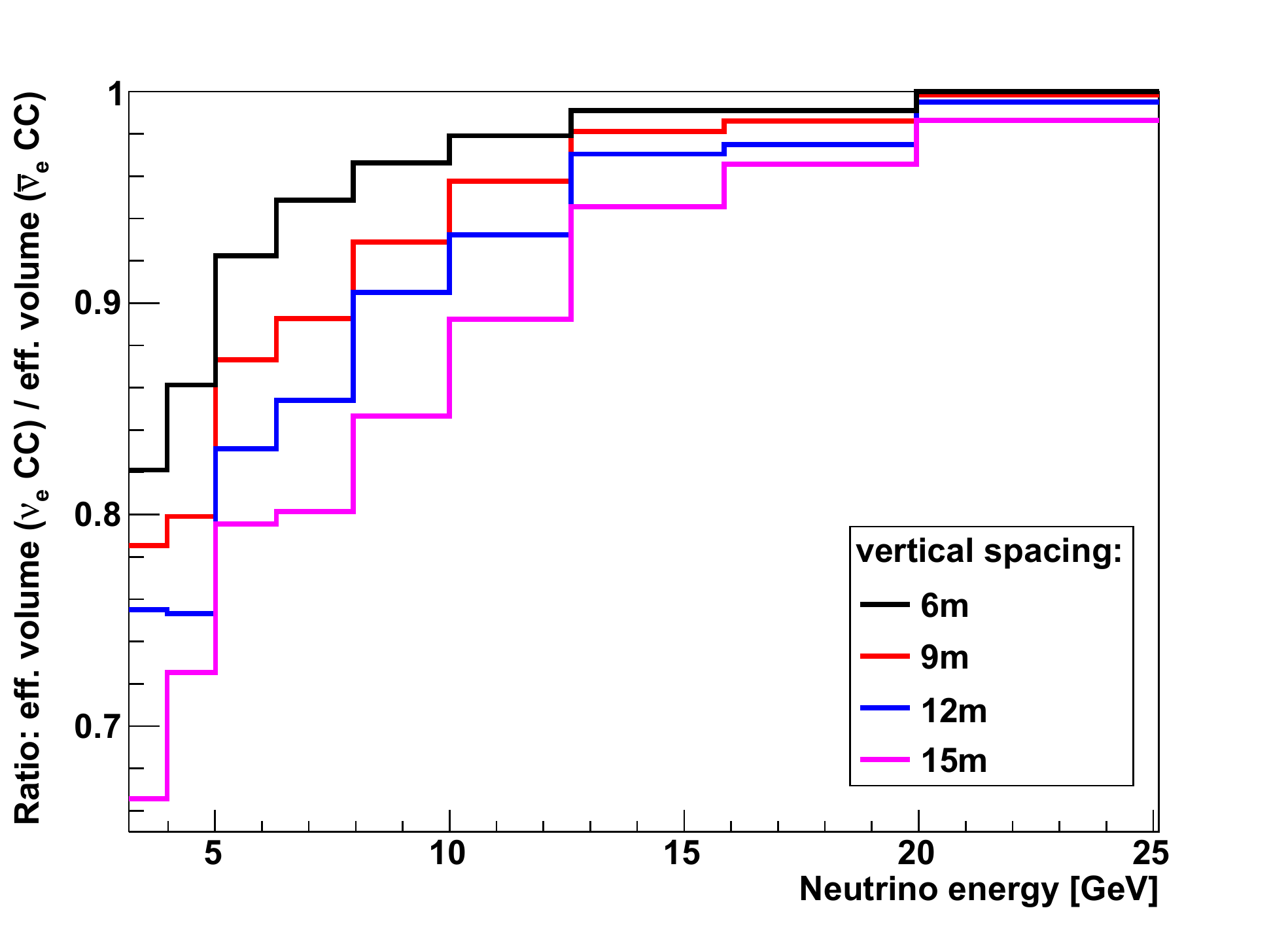}
\put (72,33) {\tiny \bf KM3NeT}
\end{overpic}
}
\caption{
Effective volumes for the different vertical spacings (top left).
Effective volumes for 6\,m/9\,m/12\,m/15\,m scaled by a
factor 1/1.5/2/2.5 (top right).
Ratio of neutrino and antineutrino effective volumes for
charged-current electron neutrino events for the different vertical
spacings (bottom).
}
\label{fig:shower_diffSpacings_effVol}
\end{figure}

The effective volumes
for the masked detectors with
different vertical spacings are shown in 
\myfref{fig:shower_diffSpacings_effVol} (top left). 
For all detector configurations a similar plateau value is reached, 
but the turn-on is less steep for smaller DOM density (larger vertical spacing). 
Assuming the same number of DOMs for each vertical spacing, 
these effective volumes can be scaled accordingly 
as shown in \myfref{fig:shower_diffSpacings_effVol} (top right). 
The ratio of effective volumes for $\nu_e \mathrm{CC}$ and $\bar \nu_e \mathrm{CC}$ 
is shown in \myfref{fig:shower_diffSpacings_effVol} (bottom).\\
\begin{figure}[h!]
\centering
\includegraphics[width=0.49\linewidth]{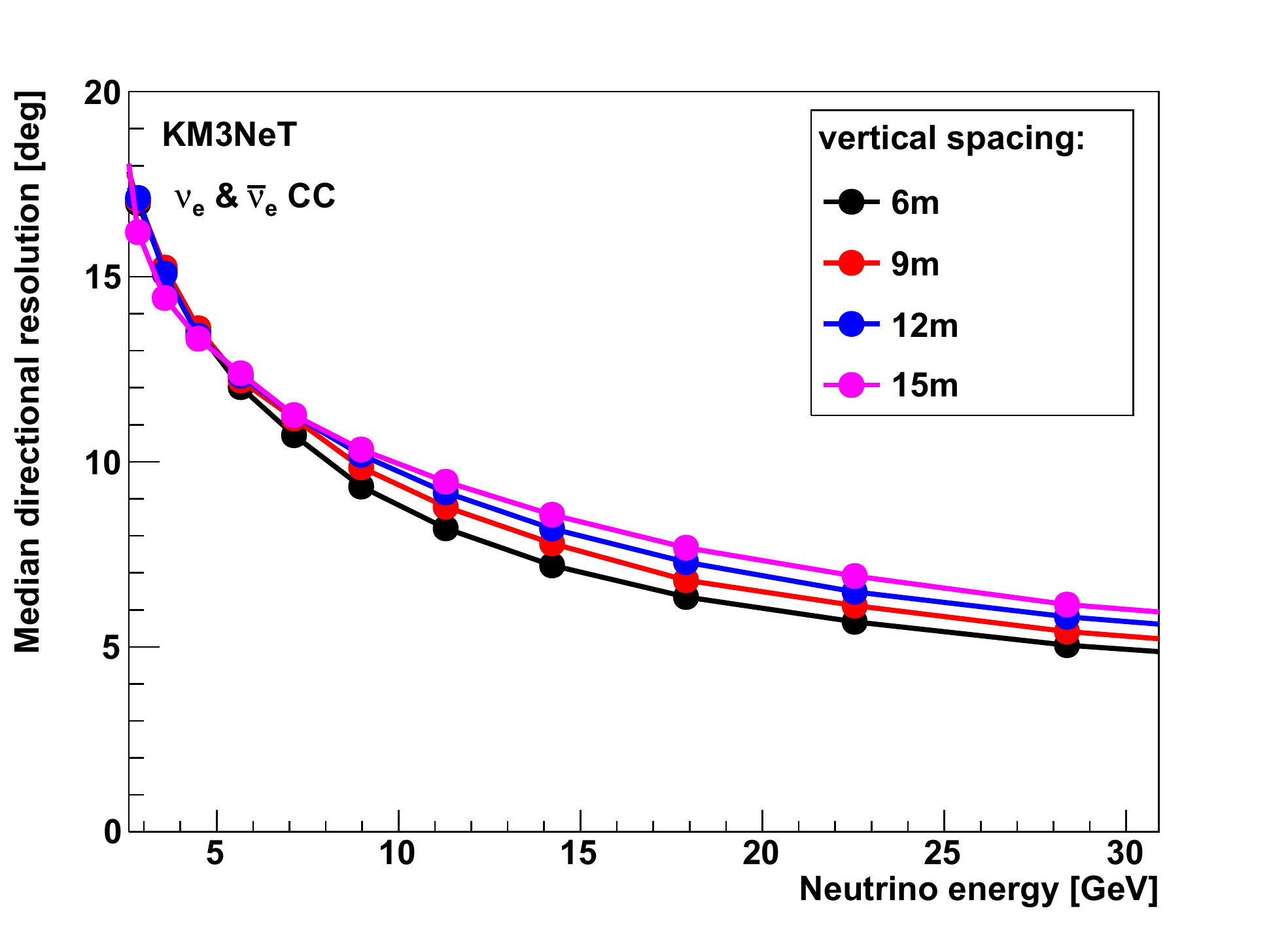}
\includegraphics[width=0.49\linewidth]{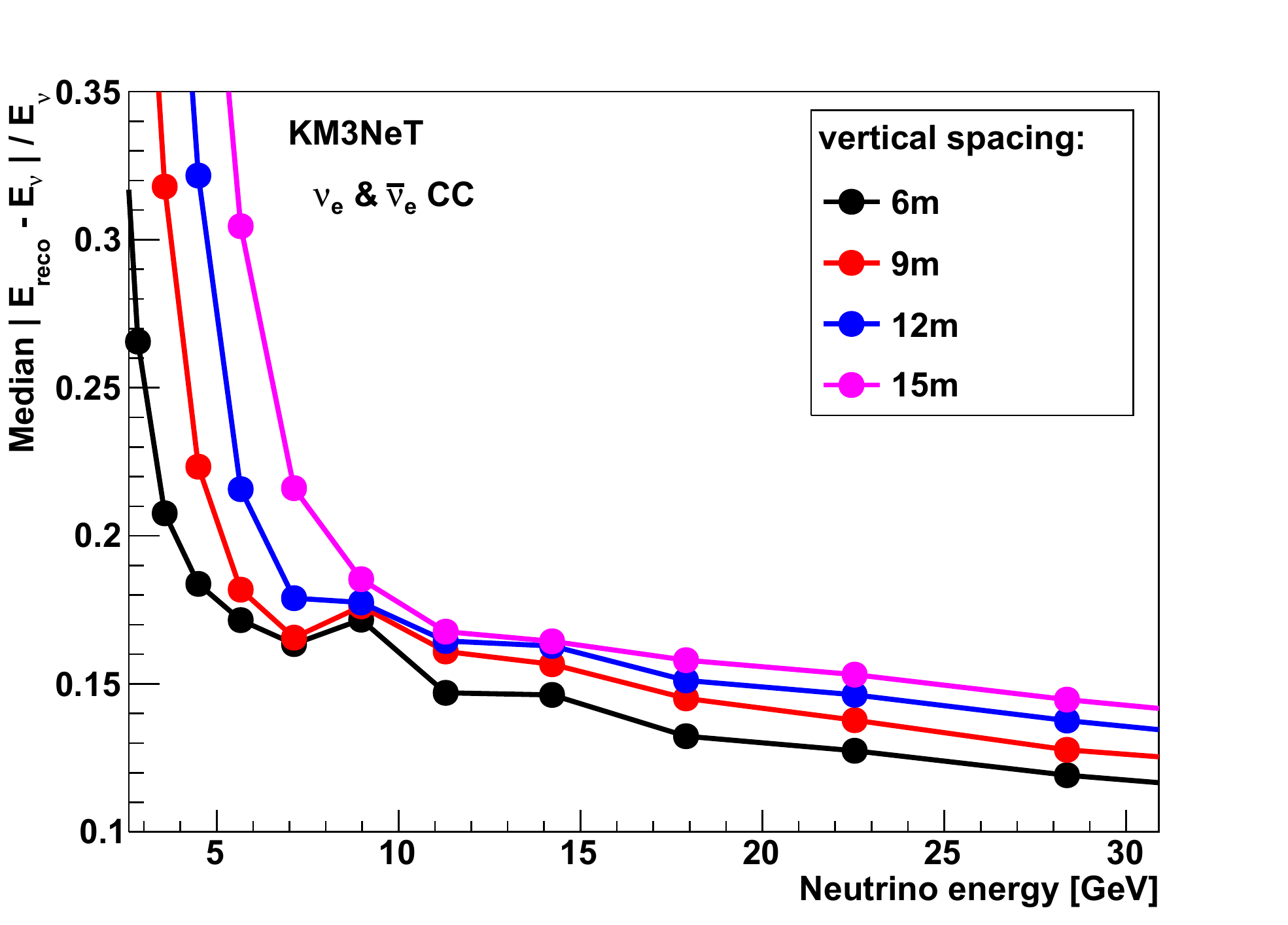}
\includegraphics[width=0.49\linewidth]{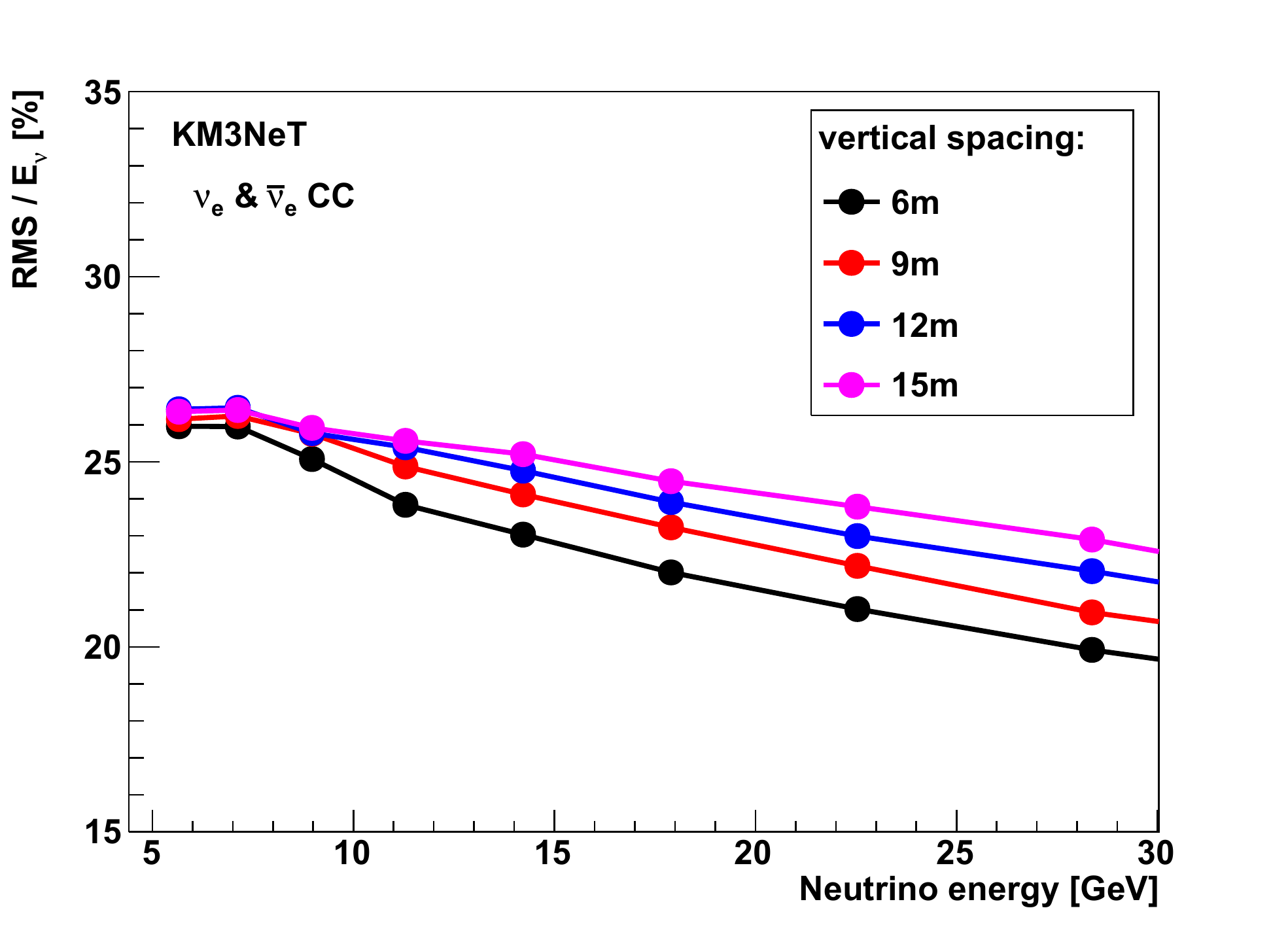}
\includegraphics[width=0.49\linewidth]{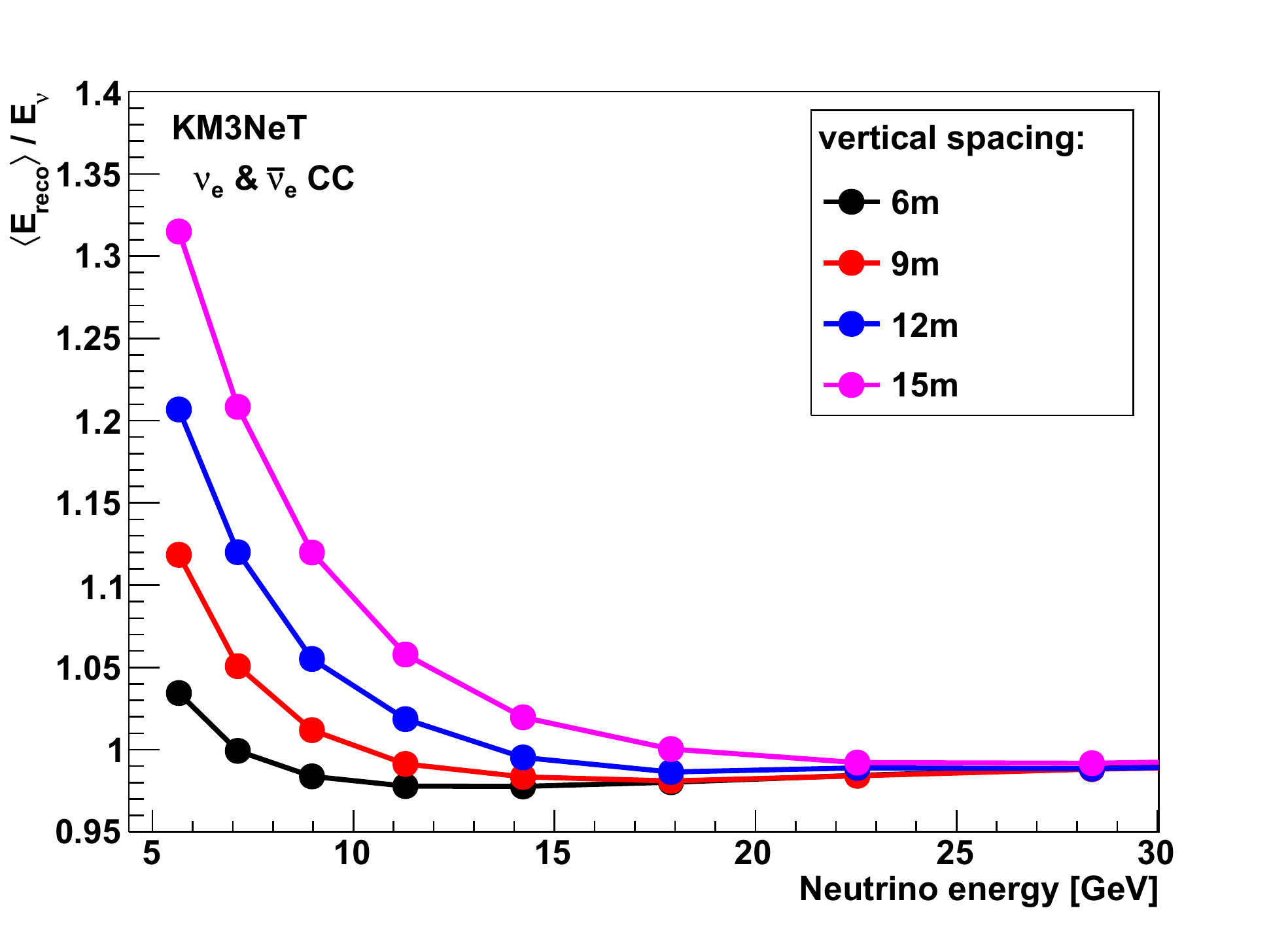}
\caption{
Resolution of the shower reconstruction for different vertical spacings for 
up-going $\nu_e \mathrm{CC}$ and $\bar \nu_e \mathrm{CC}$ events
as a function of neutrino energy.
Top left: median neutrino direction.
Top right: median fractional energy resolution ($|E_{\rm reco} - E_\nu|/E_\nu$).
Bottom left: relative energy resolution RMS/$E_\nu$.
Bottom right: mean relative offset in reconstructed energy $\langle E_\mathrm{reco} \rangle / E_\nu$.
}
\label{fig:shower_diffSpacings_reso}
\end{figure}

In \myfref{fig:shower_diffSpacings_reso} the resolutions for the different 
vertical spacings are summarised.
The resolution on both the neutrino direction and energy deteriorates slightly
for larger vertical DOM spacings.
The performance for other shower-like neutrino events for different 
vertical spacings is similar as described previously.

\subsubsection{Effect of variation in water/PMT properties and noise level on reconstruction performance}
\label{sec:shower_different_water_QE_noise}
The reconstruction performances have been studied for a variation in water properties,
PMT quantum efficiencies (QE) and optical background noise.
For this purpose, the absorption and scattering lengths $\lambda_{\rm abs}$ and
$\lambda_{\rm scat}$ have been changed by $\pm 10\%$, while the QE has
been changed by $-10\%$ -- a fuller discussion of these parameters is given in
\mysref{sec-sci-sys}. To test the influence of the optical background, the single noise
rate is increased from an already conservative 10\,kHz to 20\,kHz in the whole detector.
Bioluminescence does not produce correlated noise apart from random
coincidences and can be simulated by increasing single noise rates.

\paragraph{Effect of known parameter variations}
It is assumed that the true water, PMT and noise properties are known
so that they can be accounted for in the reconstruction.
The trigger conditions are unchanged compared to the nominal values\footnote{
For 20\,kHz single noise rate the trigger rate from pure noise would be too high, 
so that the trigger conditions would have been tightened. However, the
purpose of this study is to demonstrate the robustness of the
reconstruction with respect to an increased noise rate.
} and
events are selected according to the same criteria as for the nominal values.
This study has been performed for the detector
with 6\,m vertical spacing --- similar effects are expected for larger spacings.

\begin{figure}[h!]
\centering
\includegraphics[width=0.49\linewidth]{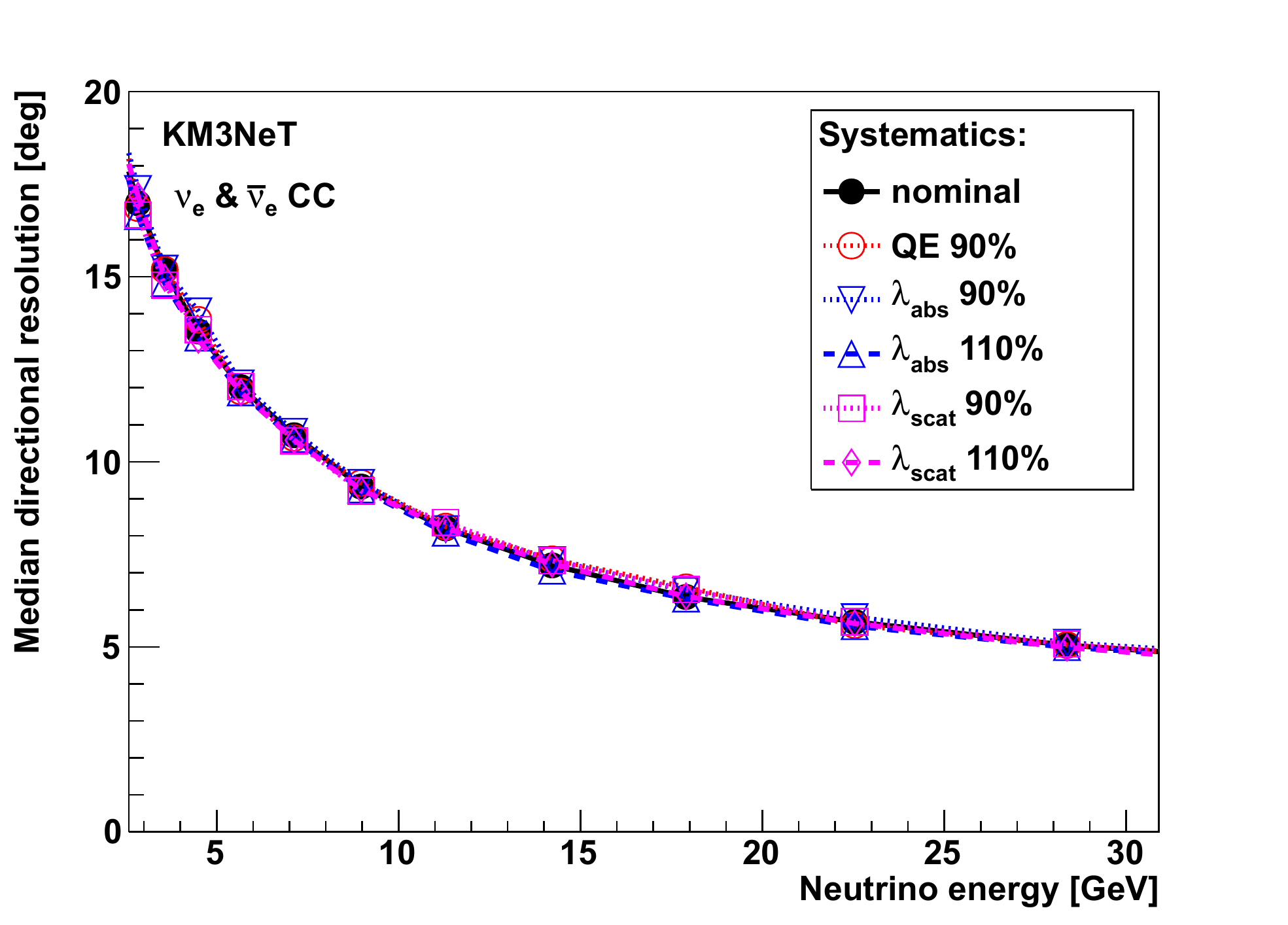}
\includegraphics[width=0.49\linewidth]{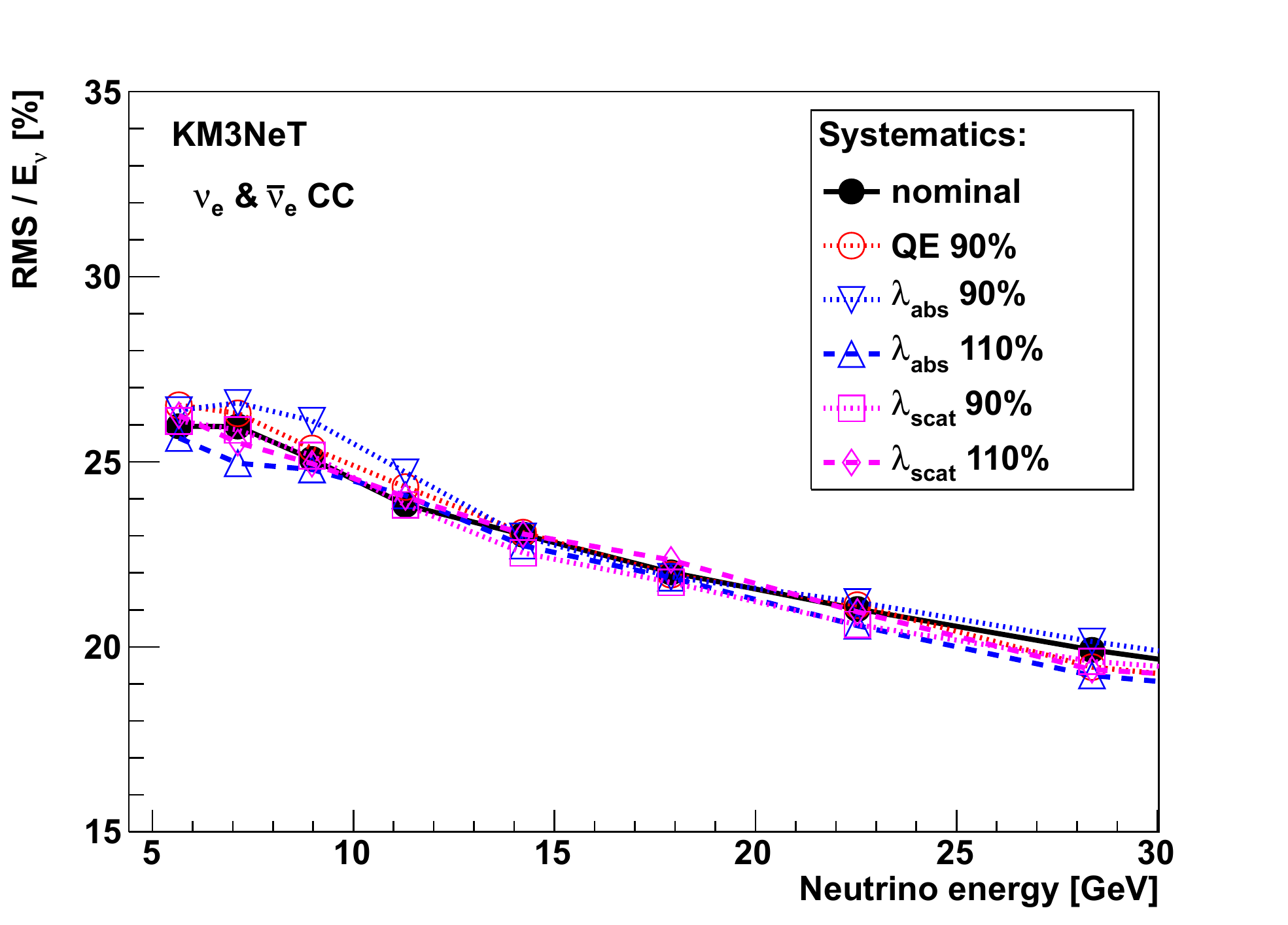}
\caption{
Resolution of the shower reconstruction for different water properties ($\lambda_{\rm abs}$ and $\lambda_{\rm scat}$) and quantum efficiencies (QE) for 
up-going $\nu_e \mathrm{CC}$ and $\bar \nu_e \mathrm{CC}$ events.
Median neutrino direction (left) and
relative energy resolution RMS/$E_\nu$ (right).
}
\label{fig:shower_diffProp_reso}
\end{figure}

\begin{figure}[h!]
\centering
\includegraphics[width=0.49\linewidth]{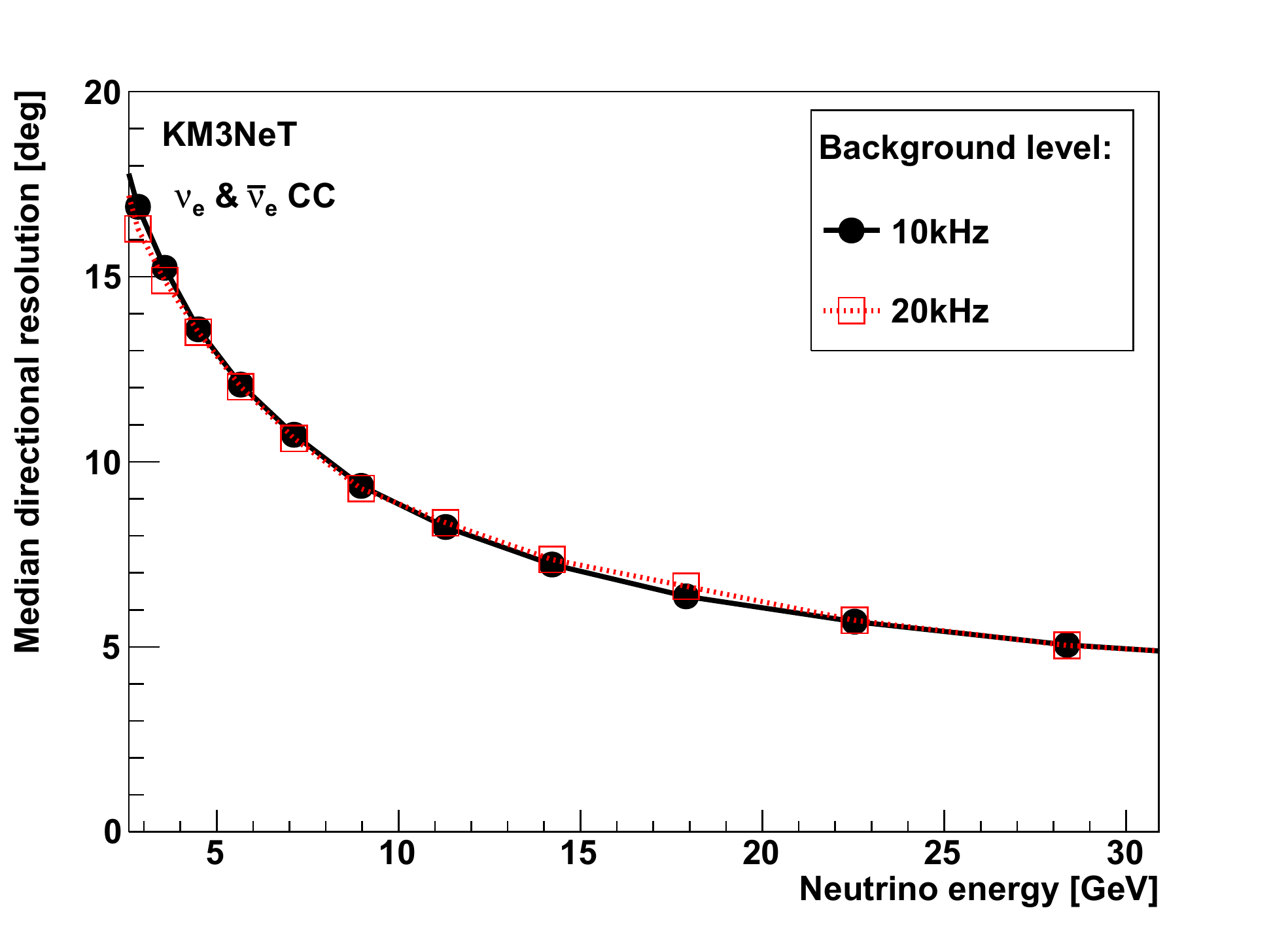}
\includegraphics[width=0.49\linewidth]{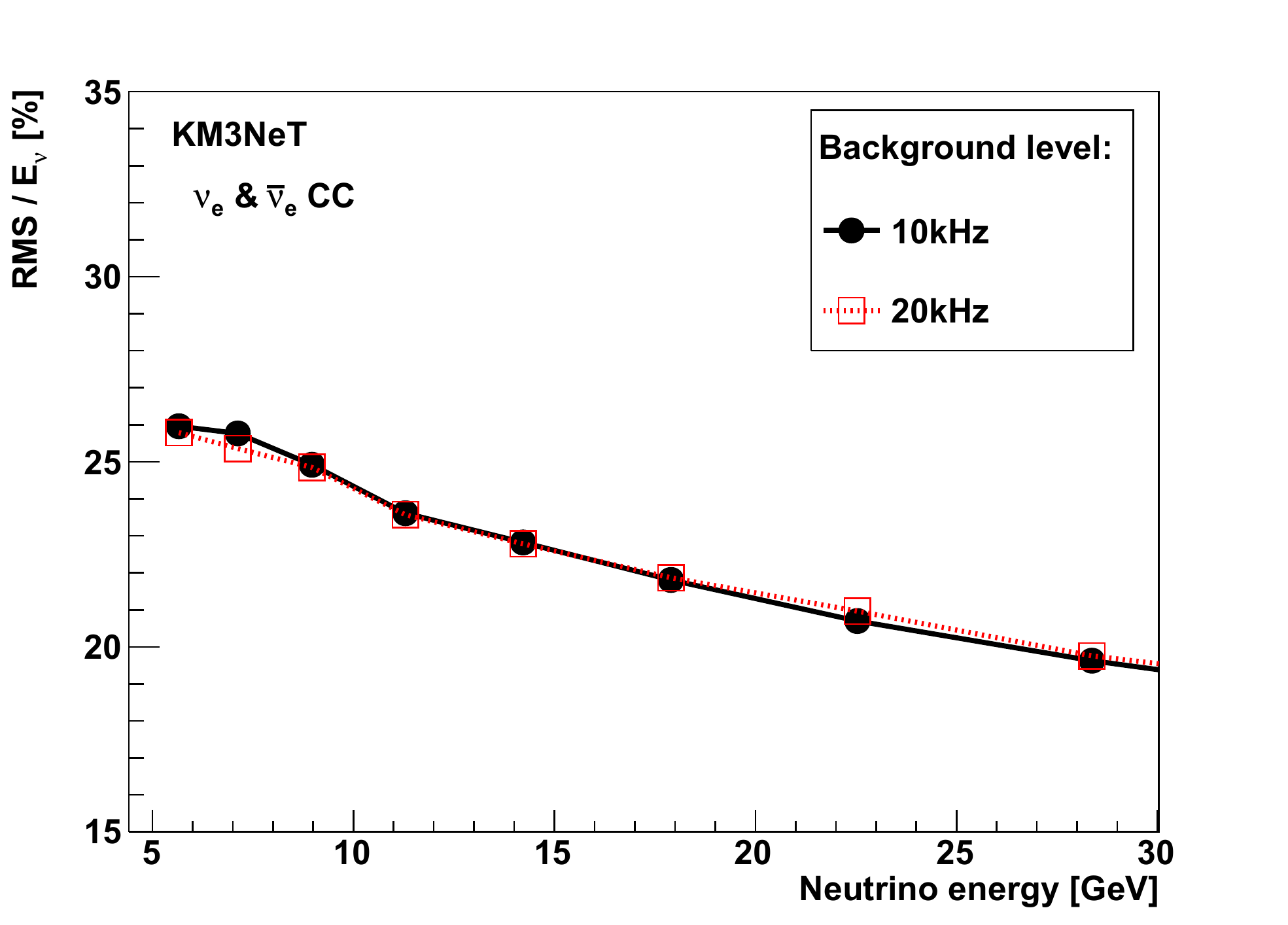}
\caption{
Resolution of the shower reconstruction for 10\,kHz and 20\,kHz single optical noise rates for 
up-going $\nu_e \mathrm{CC}$ and $\bar \nu_e \mathrm{CC}$ events.
Median neutrino direction (left) and
relative energy resolution RMS/$E_\nu$ (right).
}
\label{fig:shower_diffNoise_reso}
\end{figure}

The energy and direction resolution for a known variation in water,
PMT and noise properties is shown in \myfref{fig:shower_diffProp_reso}
and \myfref{fig:shower_diffNoise_reso} together with the performance
for the nominal values.
For all studied variations the direction resolution is unaffected, as
the direction resolution is dominated by the intrinsic scattering
angle and not by detector effects. The energy resolution deteriorates
slightly for a lower number of detected photons, i.e. reduced
$\lambda_{\rm abs}$ or QE.

For 20\,kHz single noise rates the resolutions are as good as for
10\,kHz, confirming the good signal-to-noise ratio due to small
time windows in the hit selections (cf. \mysref{sec:shower_reco_algo})
allowed by the large scattering length in water.


\begin{figure}[h!]
\centering
\includegraphics[width=0.49\linewidth]{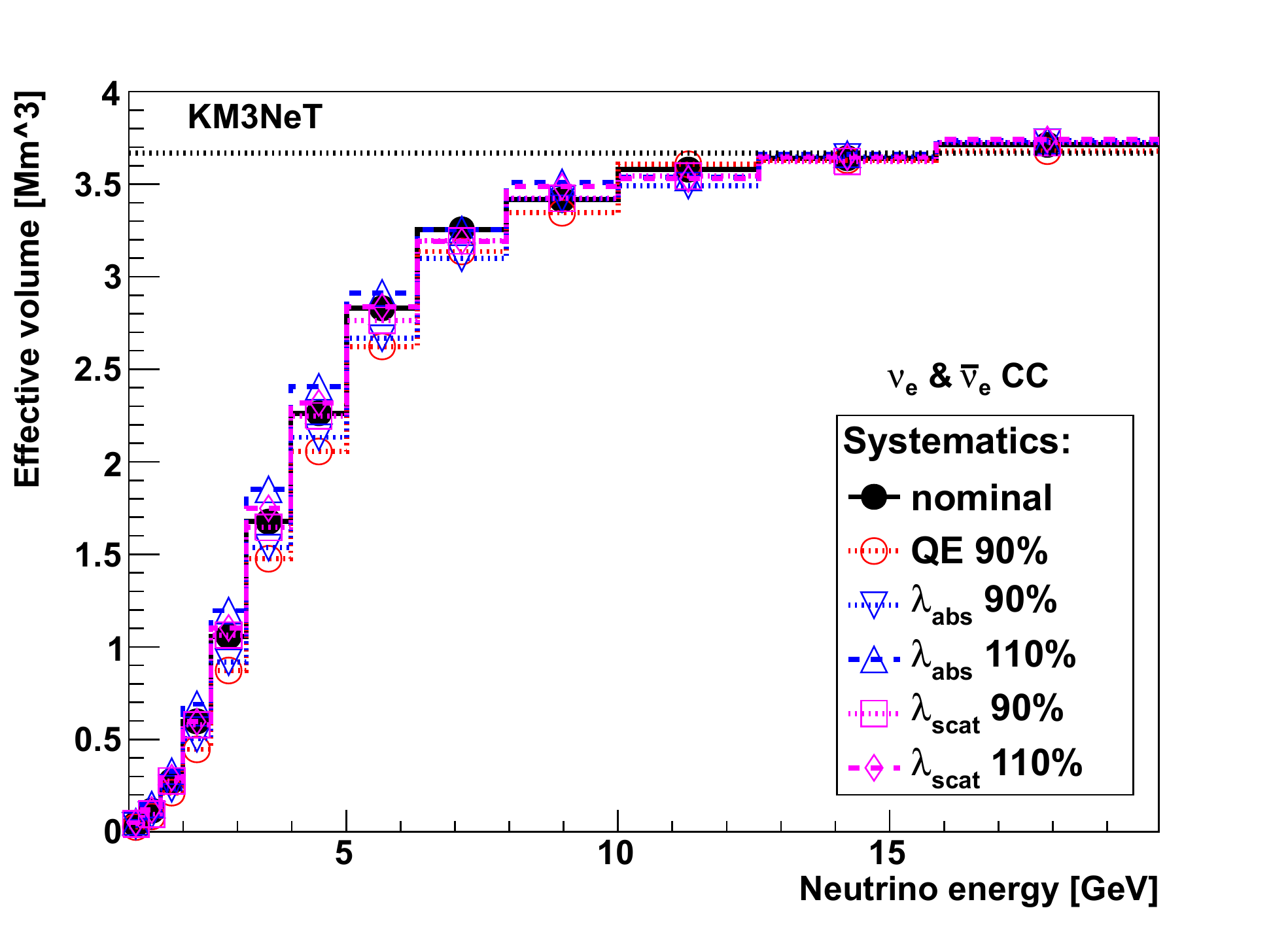}
\includegraphics[width=0.49\linewidth]{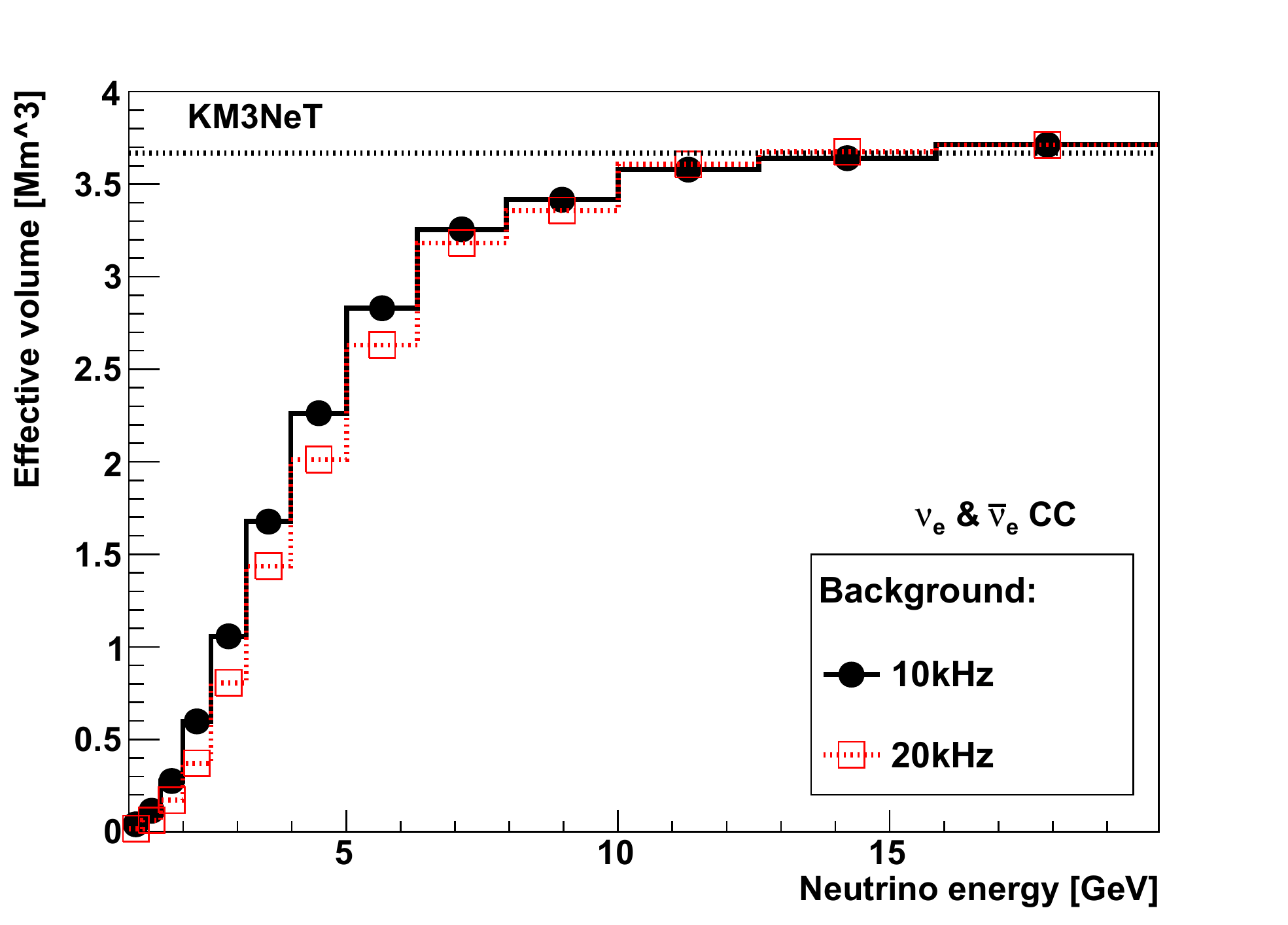}
\caption{
Effective volumes for different water properties, quantum efficiencies
(QE) and optical background noise rates for 
up-going $\nu_e \mathrm{CC}$ and $\bar \nu_e \mathrm{CC}$ events.
Different absorption $\lambda_{\rm abs}$ and scattering length
$\lambda_{\rm scat}$, and quantum efficiencies (left) and
10\,kHz and 20\,kHz single optical noise rates (right).
}
\label{fig:shower_diffProp_effVol}
\end{figure}

The effective volumes are shown in \myfref{fig:shower_diffProp_effVol}.
For all studied variations in water, PMT and noise properties a similar plateau value is reached, 
but the turn-on is less steep for less detected photons, i.e. reduced $\lambda_{\rm abs}$ or QE.
For 20\,kHz single noise rates the effective volume is only slightly
lower compared to a 10\,kHz noise rate.

The negligible deterioration in direction and energy resolution in
conjunction with the relatively modest loss in effective volume for an
increase in single noise rates by a factor of
 two\footnote{This is even a factor of 2.5 compared to the
   measured 8\,kHz, cf. \mysref{sec:simulations}.}
 compared to the nominal assumed rate of 10\,kHz demonstrates the
 robustness of the reconstruction against higher noise rates.
 Consequently, it is expected that the assumed performance can
 be achieved for most of the data taking time.

\begin{figure}[htpb]
\centering
{\begin{overpic}[width=0.5\linewidth]{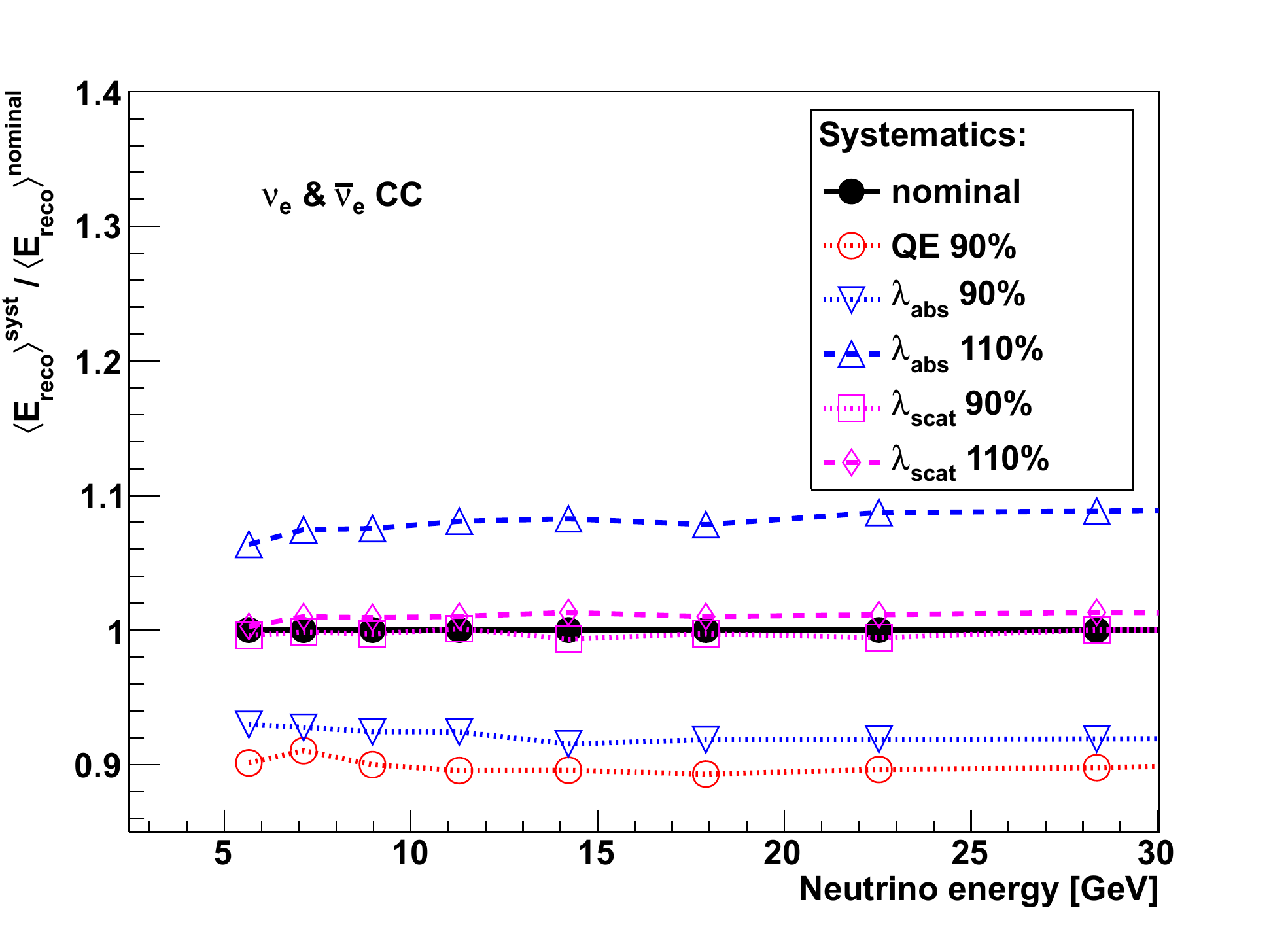} 
\put (20,60) {\tiny \bf KM3NeT}
\end{overpic}
}
\caption{Ratio of the mean reconstructed energy for nominal and varied
  water properties ($\lambda_{\rm abs}$ and $\lambda_{\rm scat}$) and quantum efficiencies (QE) for 
up-going $\nu_e \mathrm{CC}$ and $\bar \nu_e \mathrm{CC}$ events.}
\label{fig:shower_diffProp_unknown}
\end{figure}

\paragraph{Effect of undetected parameter variations}
While the direction and energy resolutions are unaffected, 
\myfref{fig:shower_diffProp_unknown} depicts the ratio of the mean
reconstructed energy for nominal and varied water and PMT
properties. Variations of the same properties and magnitude 
as above have been used for this study, but the
underlying assumption is now that the variation 
relative to the nominal values is not known and not accounted for
in the reconstruction. An exemplary $\pm$10\% variation in scattering length has a 
negligible effect on the mean reconstructed energy, while the same
variation in the absorption length induces a corresponding shift in 
reconstructed energy of $\pm$8\%. A decrease in quantum efficiency
of 10\% results in a corresponding downward shift in energy of 10\%.

\subsection{Flavour identification and muon rejection}
\label{sec:particle_id}
 The determination of the NMH requires a precise estimate of the neutrino energy and zenith angle and a high-purity event sample. In addition, since neutrino events of all flavours are reconstructed, the discrimination between neutrino flavours is necessary.
In this section an event type discrimination algorithm is developed and its performance is outlined. The algorithm is conceived with the distinction between three classes of events in mind. These classes are ``atmospheric muons'', ``shower-like'' and ``track-like'' neutrino events. In particular the atmospheric muon event class is induced by the passage of downward-going muon bundles coming from a cosmic ray air shower which is misreconstructed as upward-going, i.e. neutrino induced, event. Track-like events are those that are induced by charged current muon neutrino interactions, having the signature of a straight track passing through or nearby the instrumented volume. Finally, shower-like events are those coming from all other neutrino interaction channels and flavours: all neutral current interactions and the charged current interactions of electron and tau neutrinos\footnote{except for those roughly 18\% of $\tau$ decays producing a muon.}.

\subsubsection{Methodology}
\label{sec:classification_method}

In order to optimally exploit the information imprinted in the light emission of the events, several machine learning algorithms, so called classifiers, have been evaluated. Finally, a classification algorithm known as {\it Random Decision Forest} (RDF) \cite{breiman2001} has been used in this study.

A RDF consists of many decision trees that individually categorise an event into different classes. Each decision tree consists of several nodes. During classification a number of features, i.e. observables contributing discrimination power, are calculated for an event. At each node a decision in favour of a class is taken and the event is pushed to a child node according to the result of the decision. The individual node decisions in a tree are found by a cut on one of the calculated features. The cuts are chosen so that they maximise performance key figures such as the signal class purity. In this way the event is classified as more likely to be a track, a shower or an atmospheric muon. The decision process is repeated until the event reaches a leaf, a node without children, and the classification into one of the classes is finished.

A decision tree is trained on Monte-Carlo event data. A major disadvantage of single trees, however, is the low ability to generalise the trained tree, i.e. the ability to not only reproduce the features in the training sample. Several methods are proposed in the literature to improve the performance of single decision tree methods and we use the RDF approach. For an RDF many of the above described decision trees are trained simultaneously, a total of 101 in our study. Instead of using all features at once, for each tree a predefined fraction of features and events is selected. Finally, the classification is done by a majority decision of the trees as described above. The purity and efficiency of the classification can be set by defining cuts different from a simple 50\% majority decision.

\subsubsection{Event preselection}\label{Preselection}

Even if the detector will be located under more than 2000\,m of sea water, the number of atmospheric muons arriving at the detector and being triggered (cf. \mysref{sec:trigger}) is larger than that of atmospheric neutrinos by several orders of magnitude.
However, since the atmospheric muon flux is fully shielded by the Earth, looking at upward going events will allow to search for neutrinos. Nonetheless, Cherenkov photons from atmospheric muons can produce a hit pattern in the detector such that reconstruction algorithms still reconstruct the event as upward-going.
 A pre-selection of events is necessary before training the RDF, since in any case it would not be able to handle such a large contamination of atmospheric muons. Both the reconstruction strategies described in the previous section can produce a proper rejection of the atmospheric background without significantly reducing the amount of good neutrino events.

At first each event is requested to be reconstructed as upward going. This holds both for the track and the shower reconstruction algorithm. Then two different sets of quality criteria are applied, one for the muon and one for the shower reconstruction method. The logical ``OR'' of the two chosen criteria is used to define the input sample for the RDF.

Concerning the shower reconstruction algorithm, a preliminary event selection is implicitly done in the reconstruction itself, cf. \mysref{sec:shower_reco_evt_sel}.
Through-going atmospheric muons release a large amount of light in the detector and can easily be separated from low energy neutrino showers.
This is done by requiring a proper hit selection in the shower reconstruction itself. This is not the case for the muon track reconstruction algorithm, for which both the signal and the background events show the same hit topology.
For bright reconstructed shower events it is required that the hit pattern is compatible with a point-like emission. Here, bright events are defined as events with more than 15 causally connected L2 hits (defined as in \mysref{sec:shower_reco_algo}) and the compatibility with a point-like emission is evaluated based on the time residuals of these L2 hits with respect to the reconstructed vertex. If the difference between the 80\% and 20\% quantiles of the time residuals is smaller than 15\,ns, the event is considered as a shower event candidate.
This requirement results in a preselection of shower event candidates and efficiently focuses the time-consuming part of the shower reconstruction to neutrino-like events.
Additionally, the shower reconstruction algorithm provides many different event-by-event quality parameters, which provide further rejection power for atmospheric muons and are used as features in the RDF.

\begin{figure} [!hbt]
\centering
{\begin{overpic}[width=0.48\textwidth]{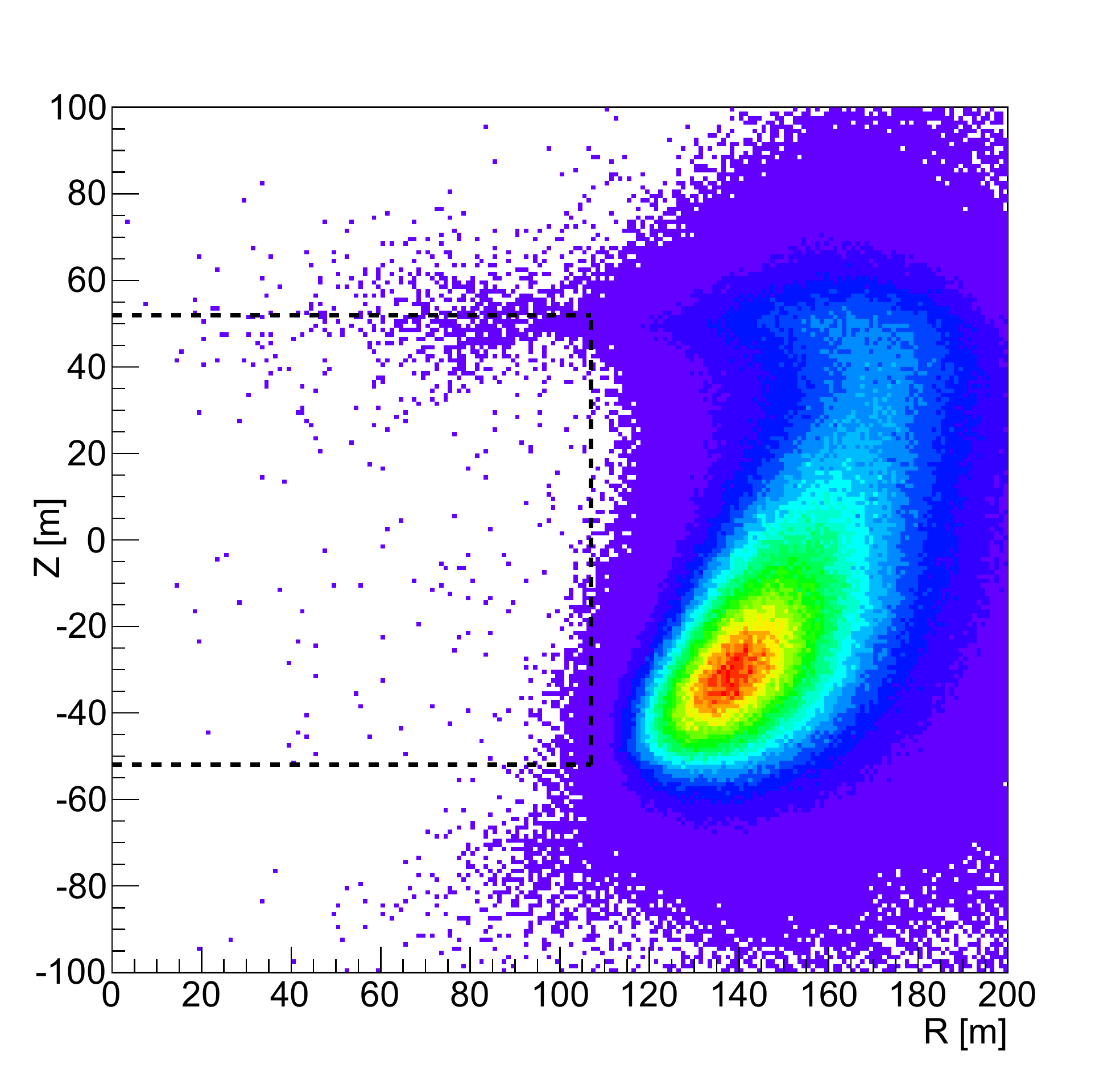}
\put (40,88) {\bf \small KM3NeT}
\end{overpic}
}
{\begin{overpic}[width=0.48\textwidth]{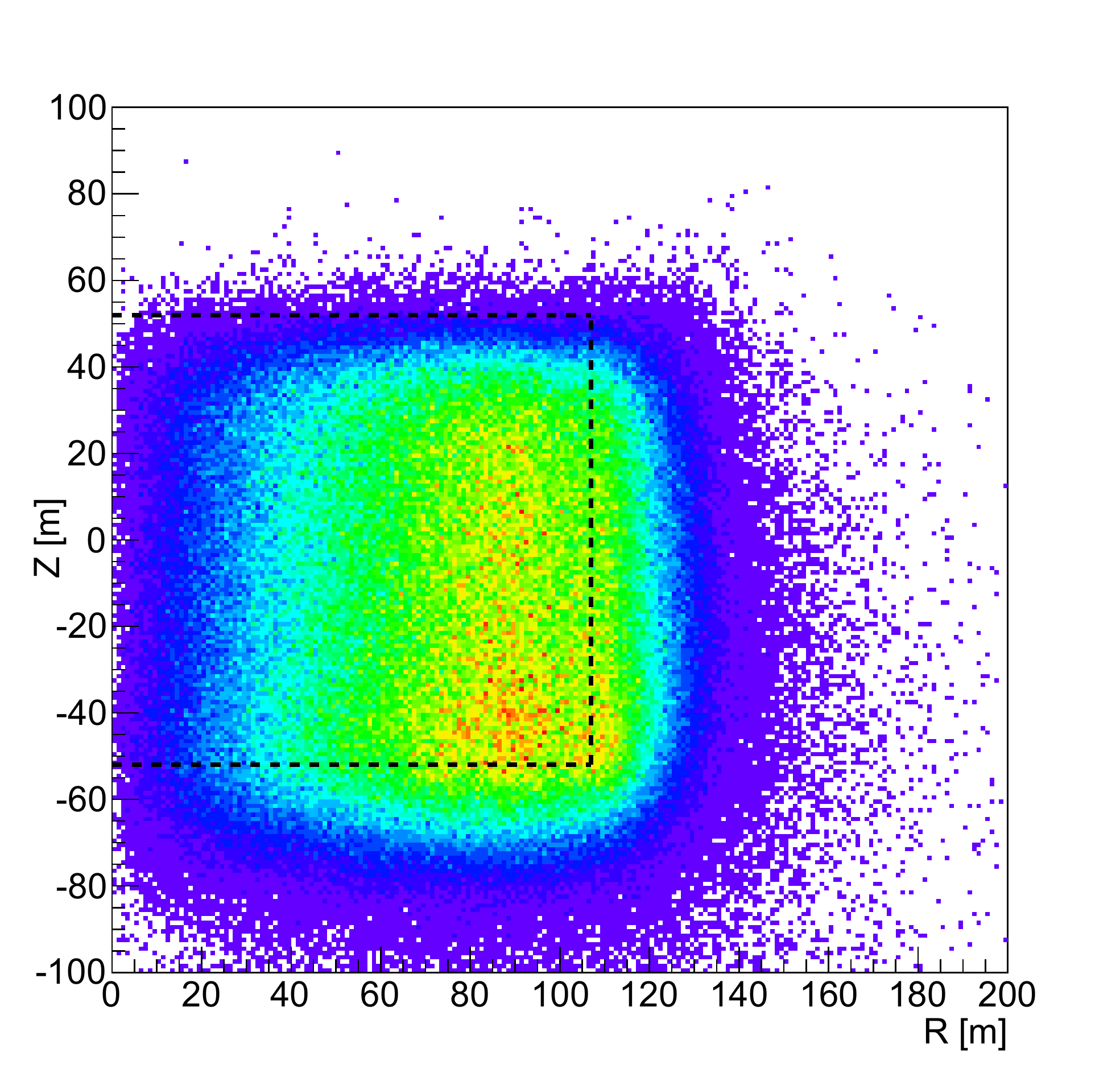}
\put (40,88) {\bf \small KM3NeT}
\end{overpic}
}
\caption{r-z distribution of the reconstructed track starting point for upward going reconstructed tracks: Left: atmospheric muons. Right: atmospheric neutrinos below 20 GeV. Up-going events with $\Lambda$ > -6 are shown. The black lines represent the contour of the instrumented volume.}
\label{fig:vertex_RZ}
\end{figure}

\begin{figure} [!hbt]
\centering
{\begin{overpic}[width=0.48\textwidth]{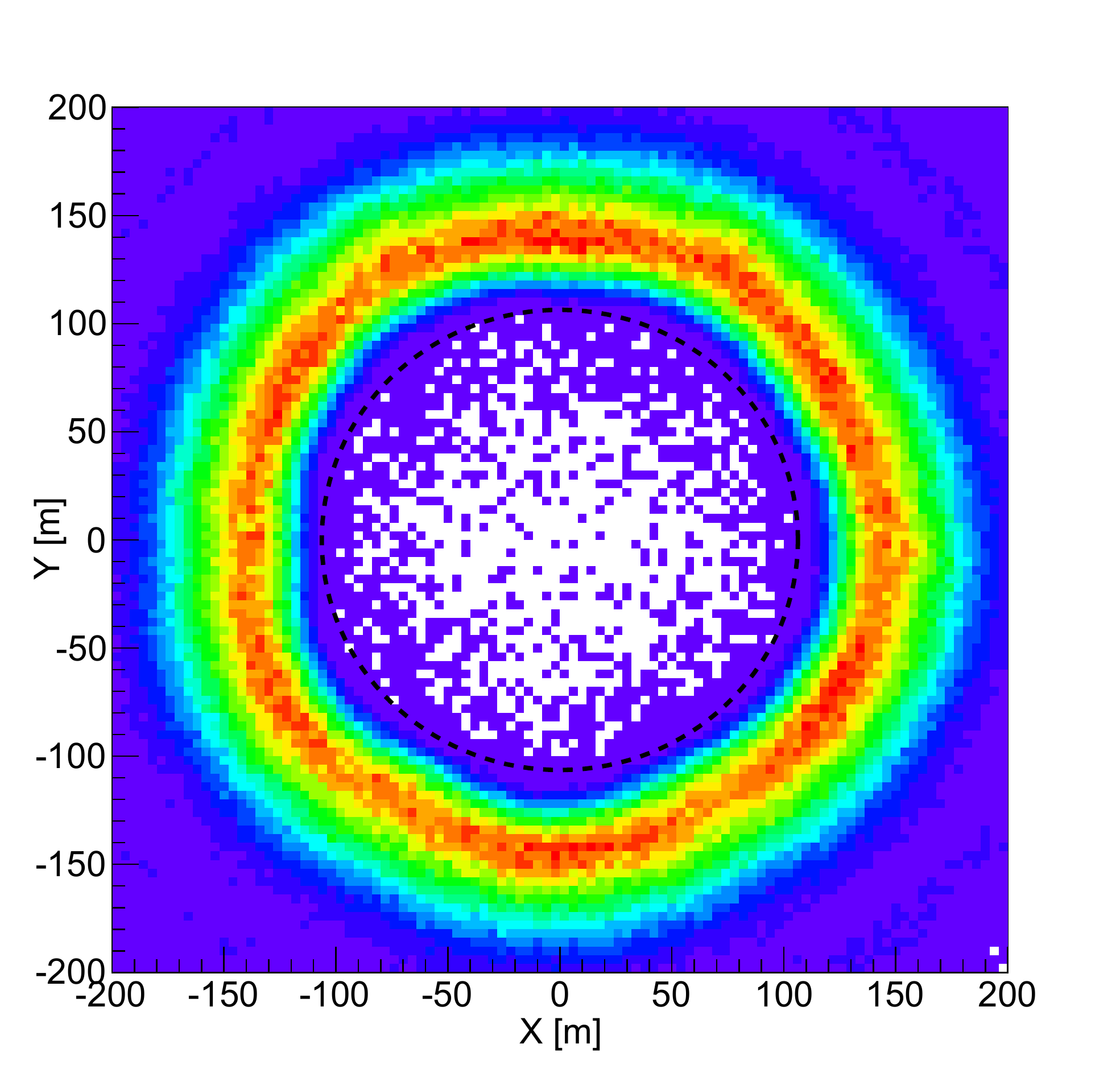}
\put (40,88) {\bf \small KM3NeT}
\end{overpic}
}
{\begin{overpic}[width=0.48\textwidth]{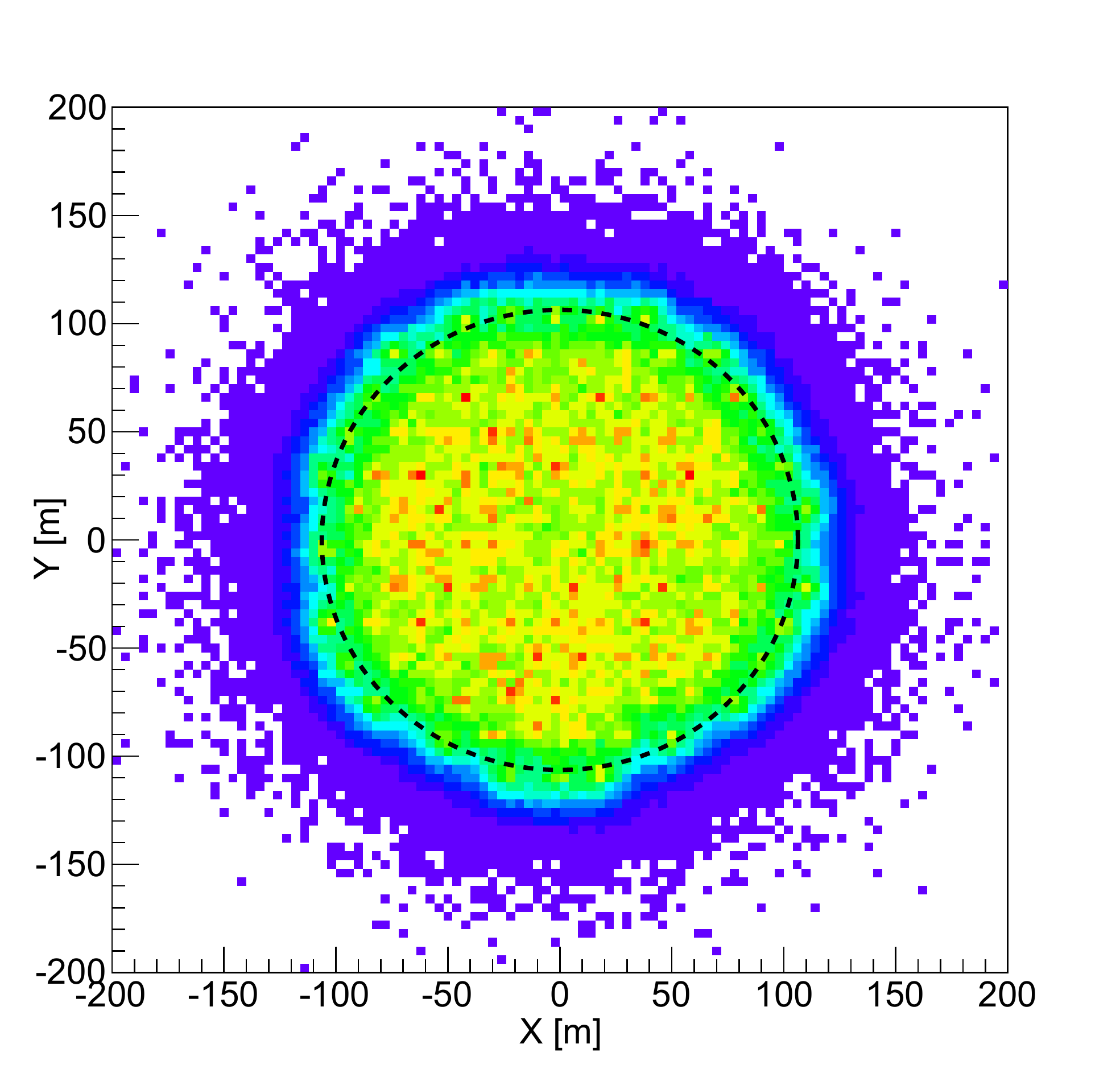}
\put (40,88) {\bf \small KM3NeT}
\end{overpic}
}
\caption{x-y distribution of the reconstructed track starting point for upward going reconstructed tracks: Left: atmospheric muons. Right: atmospheric neutrinos below 20 GeV. Up-going events with $\Lambda$ > -6 are shown. The black lines represent the contour of the instrumented volume.}
\label{fig:vertex_XY}
\end{figure}

As far as the track reconstruction is concerned, the $\Lambda$ parameter described previously can provide a first rejection of atmospheric muons; however, acting on this parameter alone would also suppress a large part of the neutrino sample at lower energy if high purity is requested.  
Adding also the reconstructed track starting point information allows an improved rejection of wrongly reconstructed atmospheric muon tracks. \myfref{fig:vertex_RZ} and \myfref{fig:vertex_XY} show the distribution of the reconstructed track starting point for atmospheric muons and low energy (E$_\nu$ $<$ 20 GeV) atmospheric muon neutrinos. A variable R$_\nu$, the radius of a "fiducial cylinder", has been defined and, in combination with the $\Lambda$ quantities, has been tested in order to achieve a preliminary selection cut. The chosen value for R$_\nu$ is equal to the radius of the instrumented volume,
i.e. 106 m. 
The number of wrongly reconstructed atmospheric muons can be reduced by more than 3 orders of magnitude
when applying a preliminary selection cut on R$_\nu$ and the track quality parameter $\Lambda$.

\subsubsection{Classification input}

\label{ObservablesSorting}\label{Ranking}
 
As described above, a decision tree relies on cuts on observables, so called features, that are chosen to discriminate well between the different classes and that are calculated for each event. In the following, the best performing features are ranked according to their discrimination power and very briefly explained.

The ranking is done using the overall classification rate under a majority decision of 50\%. For the ranking, the trees of the RDF are trained with each individual feature and the overall RDF performance is evaluated. In the next step the algorithm adds one more feature to the best one and does the training once more. This is done for every possible configuration. The best configurations are chosen to do more iterations in the same way and this process iterates as long as the performance increases. In \mytref{Table:Ranking} the ranking of the best features used in the classification is listed. 

\begin{table}[!hbt]
\small
\def\arraystretch{1.3}
\begin{tabularx}{\textwidth}{|c|X|}
   \hline
   Rank& Feature Description \\ \hline
   1& normalised eigenvalue of tensor of inertia of the hit distribution   \\ \hline
   2& RMS of time residual distribution with respect to a shower hypothesis \\ \hline
   3& $\chi^2$ of linear fit to the cumulative time residual distribution \\ \hline
   4& reconstructed neutrino energy from shower reconstruction \\ \hline
   5& coverage $\rm{cov}_{20\,^{\circ}}$ as defined in \mysref{sec:shower_reco_evt_sel} \\ \hline
   6& number of hit DOMs within $<10^\circ$ around the reconstructed shower direction and vertex \\ \hline
   7& median of time residual distribution of hits selected under a shower hypothesis\\ \hline
   8& ratio between number of selected hits for a track hypothesis and shower hypothesis \\ \hline
   9& ordinate intercept of a linear fit to cumulative time residual distribution with respect to a shower hypothesis \\ \hline
   10& Bjorken $y$ as reconstructed by the shower reconstruction \\ \hline
\end{tabularx}
\caption{Ten of the best performing features used in the Random Decision Forest classification algorithm.}
\label{Table:Ranking}
\end{table}

\subsubsection{Classification performance}
\label{Quality}

In the following the performance of the classification algorithm is evaluated using all events passing the selection criteria shown above.

\paragraph{Definitions}
\label{subsec:particleid:def}

It is desirable to maximise the number of correctly classified events for all channels. The following definition is used to evaluate the 
performance of the classification algorithm. The \textit{fraction of correctly classified} events $R^{A}_{corr}$ of class $A$ 
is defined as the ratio of correctly classified events with succeeded reconstruction $N^{A}_{corr,rec}$ 
with respect to the total number of events in this class $N^{A}_{all}$: 
\begin{equation}
R^{A}_{corr} = \frac{N^{A}_{corr,rec}}{N^{A}_{all}}
\end{equation}

\paragraph{Classification results}

\begin{figure} [!hbt]
\centering
{\includegraphics[width=0.48\textwidth]{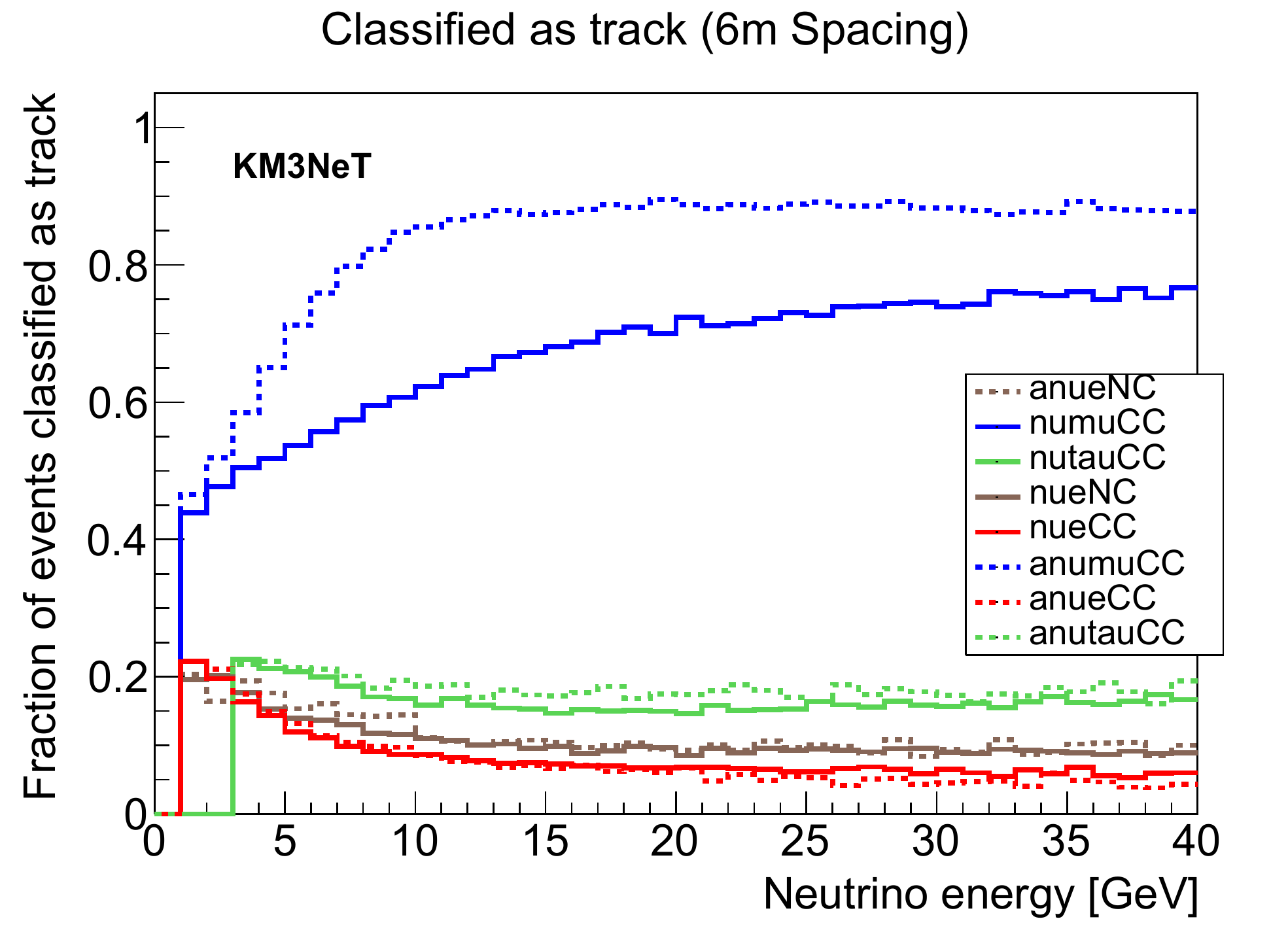}}\quad
{\includegraphics[width=0.48\textwidth]{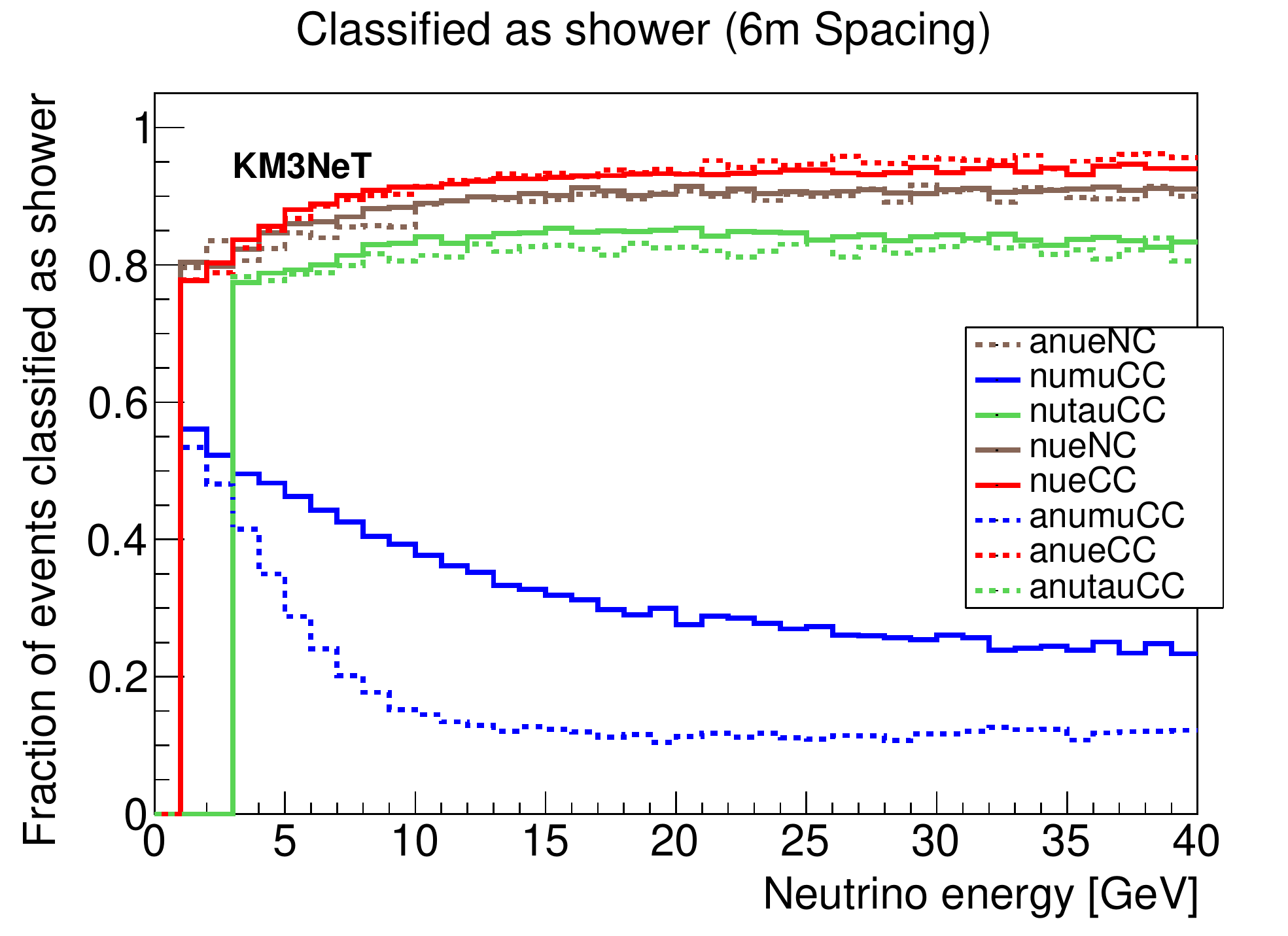}}
\caption{Fraction of events classified as tracks (left) or showers (right) for a detector with 6\,m vertical DOM spacing.}
\label{fig:fid_performance6m}
\end{figure}

\myfref{fig:fid_performance6m} shows the result of the RDF classification. The fraction of correctly classified events per interaction channel
is plotted versus the MC neutrino energy. Here the majority vote of the random decision forest was set to 50\% as this was the best compromise between all classes. Each colour depicts the result for neutrinos and antineutrinos of one flavour in one interaction channel.

The left plot in \myfref{fig:fid_performance6m} shows the fraction of events classified as track-like for the different flavours and interaction channels versus the MC neutrino energy. The shown results are obtained for all events used in the last classification step. Classification results for charged current tau neutrino interactions are shown without distinction between track-like and shower-like decay topologies of the resulting tau lepton\footnote{Note that tau neutrino interactions have been excluded from the event set used for the RDF training.}.
The high energy range shows an expected increase in identification power for long-track muons from muon neutrinos undergoing a charged current interaction.
As can be seen, antineutrinos can be identified more easily than neutrinos. This is expected due to the different reaction inelasticities for neutrinos and antineutrinos. The fraction of interactions with a resulting shower signature wrongly identified as track-like falls below 20\% above 10 GeV.
Electron neutrinos undergoing a charged current interaction are identified more easily as shower-like than neutral current reactions as they yield more light.

In the right plot the fraction of events recognised as showers is depicted.
Most efficiently recognised are electron neutrino charged current interactions. Above a neutrino energy of 15 GeV
 the fraction of correctly classified events reaches more than 90\%. At 6 GeV the fraction reaches 85\%.
Charged current muon (anti-)neutrino events are falsely classified at a rate of 35\% (15\%) at 10 GeV.

\begin{figure} [!hbt]
\centering
{\includegraphics[width=0.48\textwidth]{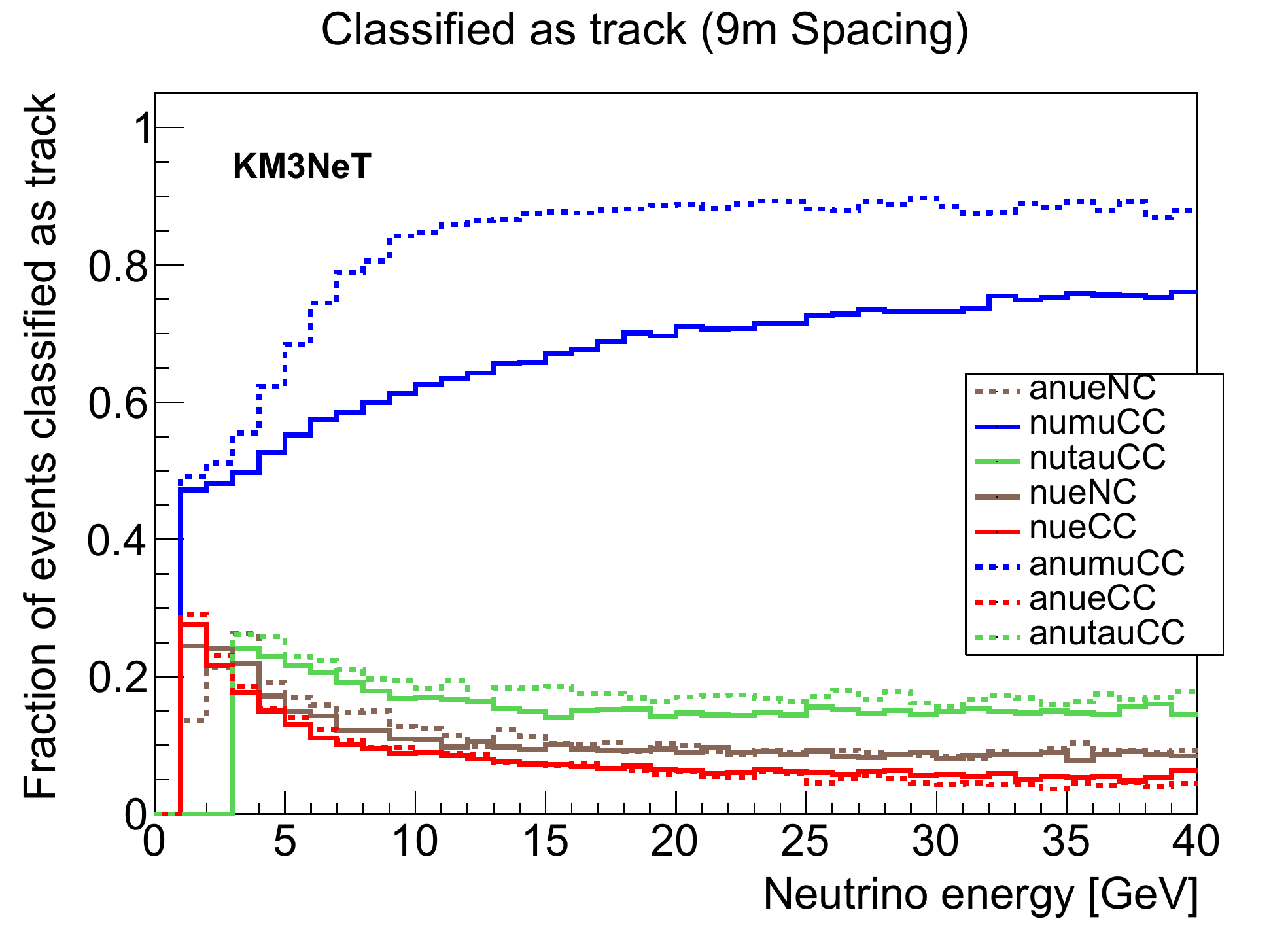}}\quad
{\includegraphics[width=0.48\textwidth]{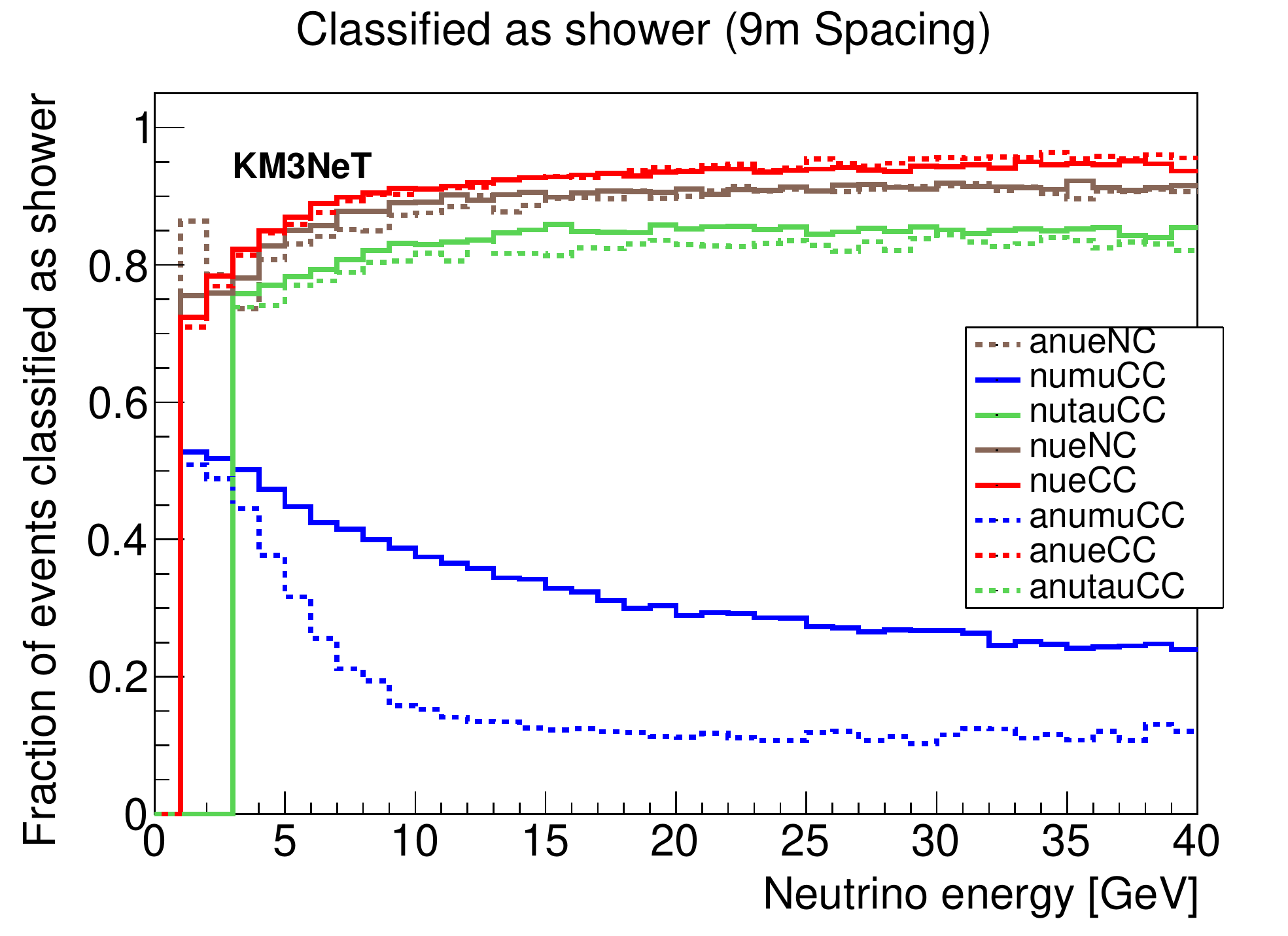}}
\caption{Fraction of events classified as tracks (left) or showers (right) for a detector with 9\,m vertical DOM spacing.}
\label{fig:fid_performance9m}
\end{figure}

The results for the configuration with a spacing of 9\,m (\myfref{fig:fid_performance9m}) show a drop of around 5\% in the identification power for track-like events. The general shape of the distribution remains.
The fraction of misclassified shower events stays nearly the same.
However, shower events need more energy now to result in a clear signature and successful classification as shower-like events.
Therefore, the response curve is shifted by 5\,GeV to higher energies.

\begin{figure} [!hbt]
\centering
{\includegraphics[width=0.48\textwidth]{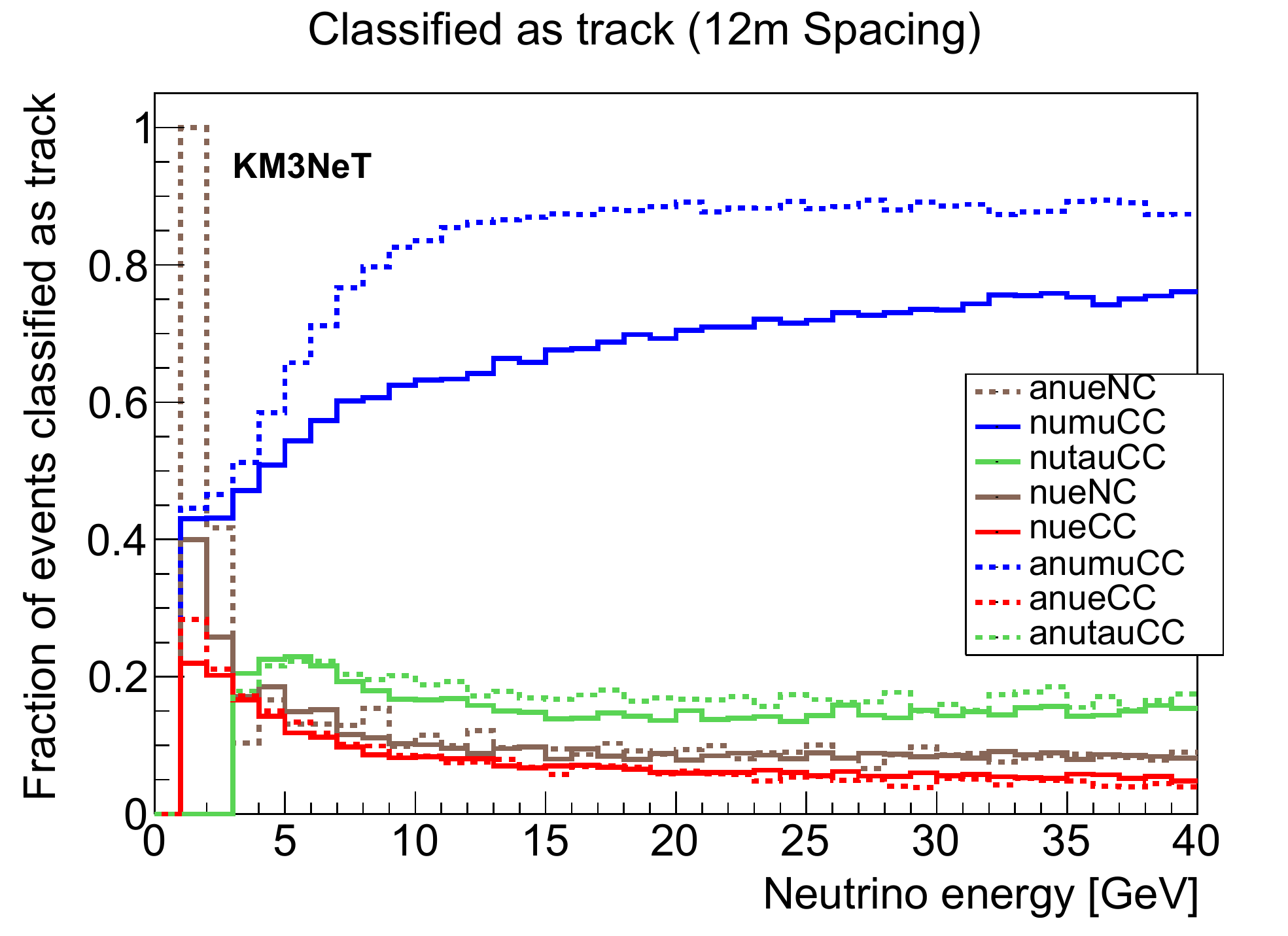}}\quad
{\includegraphics[width=0.48\textwidth]{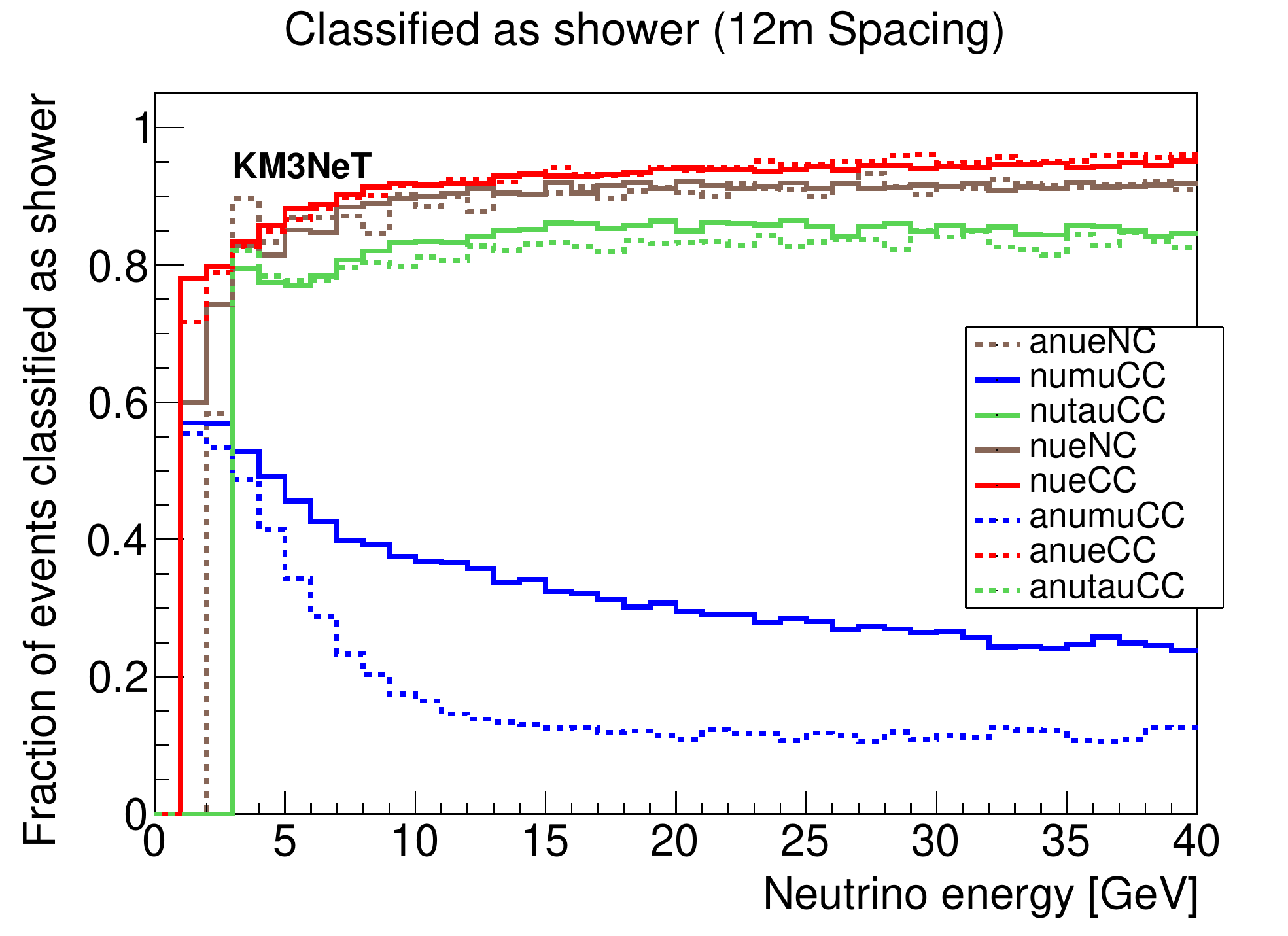}}
\caption{Fraction of events classified as tracks (left) or showers (right) for a detector with 12\,m vertical DOM spacing.}
\label{fig:fid_performance12m}
\end{figure}

\myfref{fig:fid_performance12m} shows the particle identification performance for a detector configuration with 12\,m vertical spacing.
The identification power for the charged current muon neutrinos drops significantly. 
The fraction of misclassified shower events stays below 20\%. 
Again a shift to higher energies of the response curve for shower-like events is observed.

The contamination of atmospheric muons in the neutrino sample, i.e. downward-going atmospheric muons which are reconstructed as up-going and classified either as neutrino induced tracks or as showers is of the order of a few percent.
 These wrongly identified muons have equal probability of ending up either in the ``showers''  or ``tracks'' sample. This surviving background is taken into account in the subsequent calculation of the ORCA sensitivity.

\subsection{Sensitivity studies for the neutrino mass hierarchy}
\label{sensitivity}
 \subsubsection{Global fit}
\label{globalfit}
 This section describes the main mass hierarchy sensitivity calculation based on pseudo-experiments and log likelihood ratios. It is divided into three parts. First, the modelling of the physics and detector is detailed. This model is used to calculate the expected event rates for given values of the oscillation parameters and systematics. Then, the statistical method for the mass hierarchy sensitivity calculation is described. Finally, an overview is given of the current results using this method.
An independent study based on Asimov-sets is described at the end of this section.

\paragraph{Rate calculation}

\begin{figure}[h]
\scriptsize
\begin{center}

\tikzstyle{block} = [rectangle, draw, fill=blue!20, 
    text width=11em, text centered, rounded corners, minimum height=4em]
\tikzstyle{line} = [draw, line width=1pt, -latex']
\tikzstyle{cloud} = [rectangle, draw, fill=red!20, text width=11em, text centered, rounded corners, minimum height=4em]
\tikzstyle{input} = [rectangle, draw, fill=green!20, text width=11em, text centered, rounded corners, minimum height=4em]

\begin{tikzpicture}[node distance = 1.25cm, auto]
    \node [block] (atmflux) {(a) \textbf{Atmospheric Neutrino Fluxes}\\$E_\text{true}$, $\theta_\text{true}$};
    \node [block, below of=atmflux] (detflux) {(b) \textbf{Neutrino Fluxes at Detector}\\ $E_\text{true}$, $\theta_\text{true}$};
    \node [block, below of=detflux] (inter) {(c) \textbf{Interaction Rates at Detector}\\ $E_\text{true}$, $\theta_\text{true}$};
    \node [block, below of=inter] (detected) {(d) \textbf{Number of Detected Events}\\$E_\text{true}$, $\theta_\text{true}$};
    \node [block, below of=detected] (class) {(e) \textbf{Classified Events}\\$E_\text{true}$, $\theta_\text{true}$};
    \node [block, below of=class] (reco) {(f) \textbf{Reconstructed Events}\\$E_\text{reco}$, $\theta_\text{reco}$};
    \node [input, left of=atmflux, node distance=4cm] (bartol) {\textbf{Atmospheric Flux Histograms}};
    \node [cloud, right of=atmflux, node distance=4cm, anchor=north] (osc) {(1) \textbf{Oscillation and Earth Model}};
    \node [cloud, right of=detflux, node distance=4cm, anchor=north] (cs) {(2) \textbf{Interaction Cross Section}};
    \node [input, left of=detected, node distance=4cm] (bg) {\textbf{Backgrounds}\\Misreconstructed atm. muons};
    \node [cloud, right of=inter, node distance=4cm, anchor=north] (effm) {(3) \textbf{Effective Mass}};
    \node [cloud, right of=detected, node distance=4cm, anchor=north] (pid) {(4) \textbf{Classifier (=Particle ID)}};
    \node [cloud, right of=class, node distance=4cm, anchor=north] (res) {(5) \textbf{Detector Resolutions}};
    \path [line] (bartol)  -- (atmflux);

    \path [line] (atmflux)  -- (osc);
    \path [line] (osc)  -- (detflux);

    \path [line] (detflux)  -- (cs);
    \path [line] (cs)  -- (inter);

    \path [line] (inter)  -- (effm);
    \path [line] (effm)  -- (detected);

    \path [line] (bg)  -- (detected);

    \path [line] (detected)  -- (pid);
    \path [line] (pid)  -- (class);

    \path [line] (class)  -- (res);
    \path [line] (res)  -- (reco);
\end{tikzpicture}
\caption{A flowchart showing the different steps in the computation chain. The blue blocks show the intermediate results consisting of sets of histograms as a function of the two variables written below the title. The red blocks describe the steps to go from one intermediate result to the next. The green blocks describe additional inputs that do not use the result from the previous step.}\label{fig:ORCAsim_flowchart}
\end{center}
\end{figure}
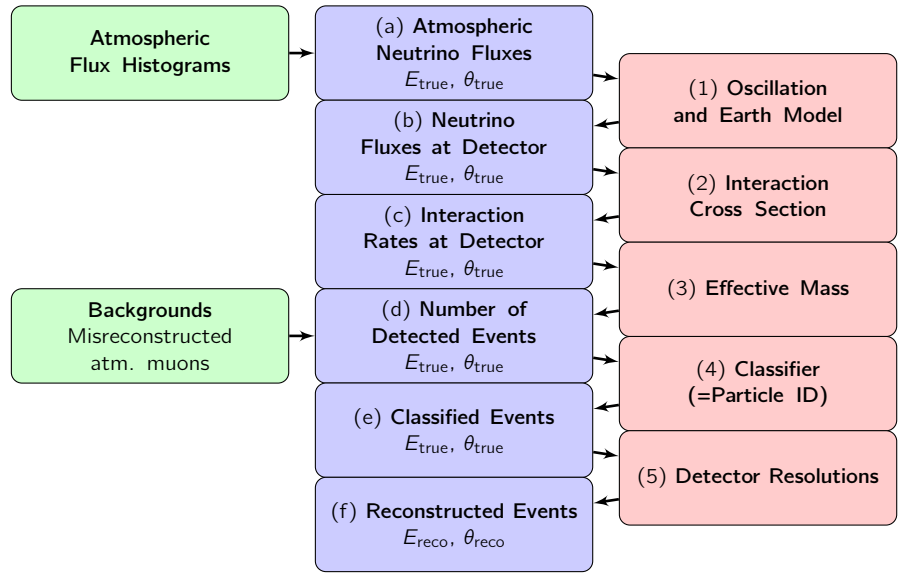

KM3NeT/ORCA's data will consist of observed event rates as a function of the reconstructed neutrino energy and zenith angle. By comparing these to the expected rates it will be possible to distinguish between the two mass hierarchy cases. The rate computation is separated into two parts. First the expected neutrino interaction rate at the detector site is calculated as a function of the true neutrino energy and zenith angle. Secondly, the response of the detector itself is modelled, leading to the rates of reconstructed events as a function of the \emph{reconstructed} energy and zenith angle. 

As shown in \mysref{muon} and \mysref{electron}, KM3NeT/ORCA is sensitive to the inelasticity (Bjorken $y$), potentially adding a third dimension to the rate histograms. At the moment, the inelasticity is not yet included in the sensitivity study. Doing so will likely improve the mass hierarchy significance, due to its power to discriminate neutrinos and antineutrinos on a statistical basis.

The whole computation chain is summarised in \myfref{fig:ORCAsim_flowchart}. Each step is described in detail in the following paragraphs.

{\parskip=0pt \textbf{Detector-independent part}}

The first half of the simulation chain (leading to intermediate result (c) in \myfref{fig:ORCAsim_flowchart}) can be summarised as:
\begin{equation}
  R_a(E,\theta) = \frac{\rho_\text{water}}{m_\text{nucleon}} \times \sum_{b} \sigma_a(E) \times P^\text{osc}_{a,b}(E,\theta) \times \Phi^\text{atm}_b(E,\theta),
\end{equation}
where
\begin{itemize}
 \item $R_a$ is the interaction rate per unit volume at the detector site of (anti)neutrinos of flavour $a$ as a function of the neutrino energy and direction.
 \item The initial flavour $b$ is summed over $\nu_e$, $\nu_\mu$, $\bar\nu_e$ and $\bar\nu_\mu$.
 \item $\Phi^\text{atm}_b$ is the atmospheric neutrino flux for neutrinos of flavour $b$.
 \item $P^\text{osc}_{a,b}$ is the oscillation probability for a neutrino passing through Earth.
 \item $\sigma_a$ is the charged-current neutrino-nucleon cross section for a neutrino of flavour $a$.
\end{itemize}

A consistent binning is used throughout the calculation. The energy axis is binned linearly in $\log_{10}(E)$ from 2 to 100 GeV in 40 bins. The zenith angle axis is binned linearly in $\cos(\theta)$ from -1 to 0 in 40 bins, where we use the convention that $\cos(\theta)=-1$ corresponds to vertically up-going neutrinos.

The \emph{atmospheric neutrino fluxes} are modelled by the HKKM2014 simulations \cite{honda1}. The given flux values are tabulated as a function of energy and zenith angle, averaged over the azimuth angle. In order to deal with the rather coarse binning the values are interpolated. To be more precise, a two-dimensional spline interpolation is made of the \emph{cumulative} tables. The spline's derivatives then yield a \emph{bin-integral conserving interpolation} of the flux tables.
The chosen tables are for the Fr\'ejus site (without mountain) at solar minimum, since the Fr\'ejus site is expected to be most similar to the KM3NeT/ORCA detector site.

The \emph{oscillation probabilities} depend on the mixing parameters (including the hierarchy) and the Earth density profile. They are calculated by evaluating the neutrino propagation time evolution operator in a constant density medium (see \cite{ohlsson}) at small steps along the trajectory.
The Earth's \emph{density profile} is given by the Preliminary Reference Earth Model (PREM)~\cite{bib:prem}. To speed up calculations the model is approximated by 42 constant-density shells. The electron density (an ingredient for the oscillation probability calculation) is approximated to be half of the nucleon density.

We use the charged current and neutral current neutrino-nucleon cross sections from the GENIE Monte Carlo generator \cite{genie1, genie2} for an oxygen nucleus and two protons.

The results of the first half of the simulation chain (i.e. at intermediate result (c)) are eight histograms of neutrino interaction rates per unit volume at the detector as a function of the true energy and zenith angle: six for the charged current (CC) interactions (three flavours, neutrinos and antineutrinos) and two for neutral current (NC) interactions of neutrinos and antineutrinos. Throughout the simulation, NC events are approximated as equal for all three flavours. 

{\parskip=0pt \textbf{Detector-dependent part}}

The second part of the simulation chain models the detector response to neutrino interactions. Each step is based on the results presented in the previous sections of this document.

The energy- and zenith angle-dependent \emph{effective mass} determines how many of the interacting events can be reconstructed. This is step (3) in the flowchart. It is derived from MC simulations as
\begin{equation}
  M_\text{eff} := V_\text{gen.} \times \rho_\text{water} \times N_\text{sel.} / N_\text{gen.},
\end{equation}
where $N_\text{gen.}$ is the total number of generated events in a large generation volume $V_\text{gen.}$. Events that are successfully reconstructed by \emph{either one} of the two reconstruction algorithms are counted in $N_\text{sel.}$. The density of sea water $\rho_\text{water}$ is assumed to be 1025 kg$/$m$^3$. The effective mass is binned as a function of the true neutrino energy and zenith angle, and is evaluated for each of the eight event classes separately.\\
At this step an additional histogram is created, representing the expected background from \emph{misreconstructed atmospheric muons}. As shown in \mysref{sec:particle_id}, the contamination of such events can be effectively reduced to a few percent by applying
cuts. Due to the high suppression efficiency it is increasingly difficult to generate high statistics samples for this type of background. Therefore, the distribution from a looser cut is used and rescaled to the total number of events found for a stricter cut. This is a conservative estimate as for looser cuts the event distribution turns out to be mostly centered around our signal area (up-going, around 10 GeV) while the distribution becomes more uniform as we apply stricter cuts.\\
At the end of this step there are a total of nine histograms.

The next step ((4) in \myfref{fig:ORCAsim_flowchart}) is \emph{particle identification}. Each input histogram is the basis for two new ones, representing events identified as `tracks' and `showers', respectively. The identification probabilities are based on the RDF study described in  \mysref{sec:particle_id}, and depend on the true neutrino energy only. Neutrino events identified as atmospheric muons are discarded. The probability for atmospheric muon background events to be identified as a track/shower has not been determined due to lack of statistics; a simple 50/50 separation is applied. After this step we have eighteen histograms.

In the final step ((5) in \myfref{fig:ORCAsim_flowchart}) the \emph{energy resolutions} and \emph{angle resolutions} are applied. They are implemented as response matrices filled from simulated data.
First the zenith angle is `smeared out' using a three-dimensional response matrix that provides binned $\cos(\theta_\text{reco})$-distributions as a function of $\cos(\theta_\text{true})$ and $E_\text{true}$. Then a two-dimensional energy response matrix providing $E_\text{reco}$-distributions as a function of $E_\text{true}$ is used to smear the energy.

The resolutions are evaluated separately for each of the sixteen neutrino event classes. We have, for example:
\begin{itemize}
 \item $\nu_\mu$ CC interactions identified as tracks
 \item $\bar\nu_e$ CC events identified as showers
 \item NC $\nu$ events identified as showers
 \item $\nu_e$ CC events misidentified as tracks
 \item \ldots
\end{itemize}
Each is smeared using dedicated response matrices. In particular, neutrinos and antineutrinos are smeared differently. So are correctly and wrongly identified events.

The response matrices use a coarser binning than the rate histograms. Depending on the available MC statistics, the number of bins is reduced from 40 to 20 or 10 to avoid artefacts.

After reconstruction all histograms are combined in  two final event histograms representing the track channel and the shower channel, respectively.

\paragraph{Sensitivity calculation}

The sensitivity to the mass hierarchy is calculated using likelihood ratio distributions from \emph{pseudo-experiments} (PEs). The procedure works as follows:
\begin{enumerate}
 \item Pick a set of true values for the oscillation parameters and other systematics\label{enum:S:true}.
 \item Calculate the expected number of events for a given period of data taking, using the simulation chain described above.
 \item Generate pseudo-data by randomly drawing a detected number of events for each bin based on Poisson statistics. The two histograms thus attained constitute the PE.
 \item Find the best-fit likelihoods $\mathcal{L}_\text{NH}$ and $\mathcal{L}_\text{IH}$ for the NH and IH assumption, by maximising the likelihood with respect to the other free parameters in both cases.  \label{enum:S:L}
 \item Calculate the log likelihood ratio $\text{LLR}:=\log(\mathcal{L}_\text{NH}/\mathcal{L}_\text{IH})$. This discriminating variable indicates which hierarchy is favoured by the PE.
\end{enumerate}

\begin{table}
 \begin{center}
  \small
  \begin{tabular}{lllll}
    \emph{parameter} & \emph{true value distr.} & \emph{initial value distr.} & \emph{treatment} & \emph{prior} \\ \hline
    $\theta_{23}$ $[^\circ]$ & $\{40, 42, \ldots, 50\}$  & uniform over $[35,55]$ $\dagger$ & fitted & no \\
    $\theta_{13}$ $[^\circ]$ &  $8.42$ & $\mu=8.42$, $\sigma=0.26$ & fitted & yes\\
    $\theta_{12}$ $[^\circ]$ &  $34$ & $\mu=34$, $\sigma=1$ & nuisance & N/A\\
    $\Delta M^2$ $[10^{-3}$ eV$^2]$ & $\mu=2.4$, $\sigma=0.05$ & $\mu=2.4$, $\sigma=0.05$ & fitted & no \\
    $\Delta m^2$ $[10^{-5}$ eV$^2]$ & 7.6 & $\mu=7.6$, $\sigma=0.2$ & nuisance & N/A\\
    $\delta_\text{CP}$ $[^\circ]$ &  0 & uniform over $[0,360]$ & fitted  & no\\
    overall flux factor &  1 & $\mu=1$, $\sigma=0.1$ & fitted  & yes\\
    NC scaling &  1 & $\mu=1$, $\sigma=0.05$ & fitted  & yes\\
    $\nu/\bar\nu$ skew &  0 & $\mu=0$, $\sigma=0.03$ & fitted  & yes\\
    $\mu/e$ skew &  0 & $\mu=0$, $\sigma=0.05$ & fitted  & yes\\
    energy slope &  0 & $\mu=0$, $\sigma=0.05$ & fitted  & yes\\
  \end{tabular}
 \end{center}
 \caption{Default parameter settings used for the LLR analysis. Where $\mu$ and $\sigma$ are given, they refer to a Gaussian distribution. The $\dagger$ indicates that the initial values for $\theta_{23}$ are generated in a special way: a total of seven initial values is tried. They are $x+i\times 5^\circ$, where $x$ is the randomly drawn value and $i\in[-3,-2,\ldots,3]$.}
  \label{tab:parsettings}
 \end{table}

The \emph{likelihood} $\mathcal{L}$ of the PE given the model is defined as
\begin{equation}
  \mathcal{L} = \prod_{i\in\text{bins}} \mathcal{P}\big(N_i | \mu_i\big),
\end{equation}
where $N_i$ and $\mu_i$ are, respectively, the observed and expected number of events in bin $i$ and 
\begin{equation}
  \mathcal{P}(n|\lambda)  = \frac{\lambda^n e^{-\lambda}}{n!}
\end{equation}
is the Poisson probability to observe $n$ events when the expectation value is $\lambda$. The expected event numbers $\mu_i$ depend on the parameter values (oscillation parameters and systematics) so that maximising the likelihood corresponds to finding the parameter values that best fit the PE.

The default parameter settings are summarised in \mytref{tab:parsettings}. It shows the \emph{true parameter values} used to generate PEs. Most of these are fixed at some nominal value. The \emph{initial values} are those used as starting values by the minimiser in the fitting procedure that finds the likelihood maximum. These values are chosen randomly for each PE to avoid systematic biases. Most parameters are fitted, meaning they are left free in the minimiser. The likelihood is multiplied by Gaussian priors for some parameters (see \mytref{tab:parsettings}). The mean and width of the Gaussian priors correspond to those of the matching initial value distributions. Two parameters ($\theta_{12}$ and $\delta m^2$) are treated as \emph{nuisance parameters}. This means that, rather than leaving them free in the fit, a random `best fit' value from the initial value distribution is assigned to each PE. It emulates the fact that these parameters will be constrained almost exclusively by external measurements.  

The first six parameters listed in \mytref{tab:parsettings} are the oscillation parameters, where the large mass-squared difference $\Delta M^2$ is defined as 
\begin{equation}
  \Delta M^2 := \frac{\Delta m_{32}^2 + \Delta m_{31}^2}{2}.
\end{equation}
The last five entries in the table are \emph{systematics}. The \emph{overall flux factor} and \emph{NC scaling} simply scale the total number of (NC) events by an energy- and zenith-independent factor. The \emph{skew parameters} introduce an additional asymmetry in the ratio of one event type to the other, while conserving the total number of events. They relate to the ratio of neutrinos to antineutrinos and the ratio of $\mu$-flavour events to $e$-flavour events. Finally, the \emph{energy slope} $\alpha$ introduces an energy-dependent scaling of the number of events of the form  $E^\alpha$. 

Because $\theta_{23}$ generally has two likelihood maxima, special steps are taken to avoid ending up in the wrong one. The likelihood maximisation is repeated several times, starting from a different $\theta_{23}$ value each time. Only the best-fit result is considered for either hierarchy.

The distributions for each parameter are uncorrelated and based on the current world uncertainties \cite{Capozzi2013,bib:Forero,DayaBay}.

\begin{figure}
 \begin{center}
\includegraphics[width=0.6\textwidth]{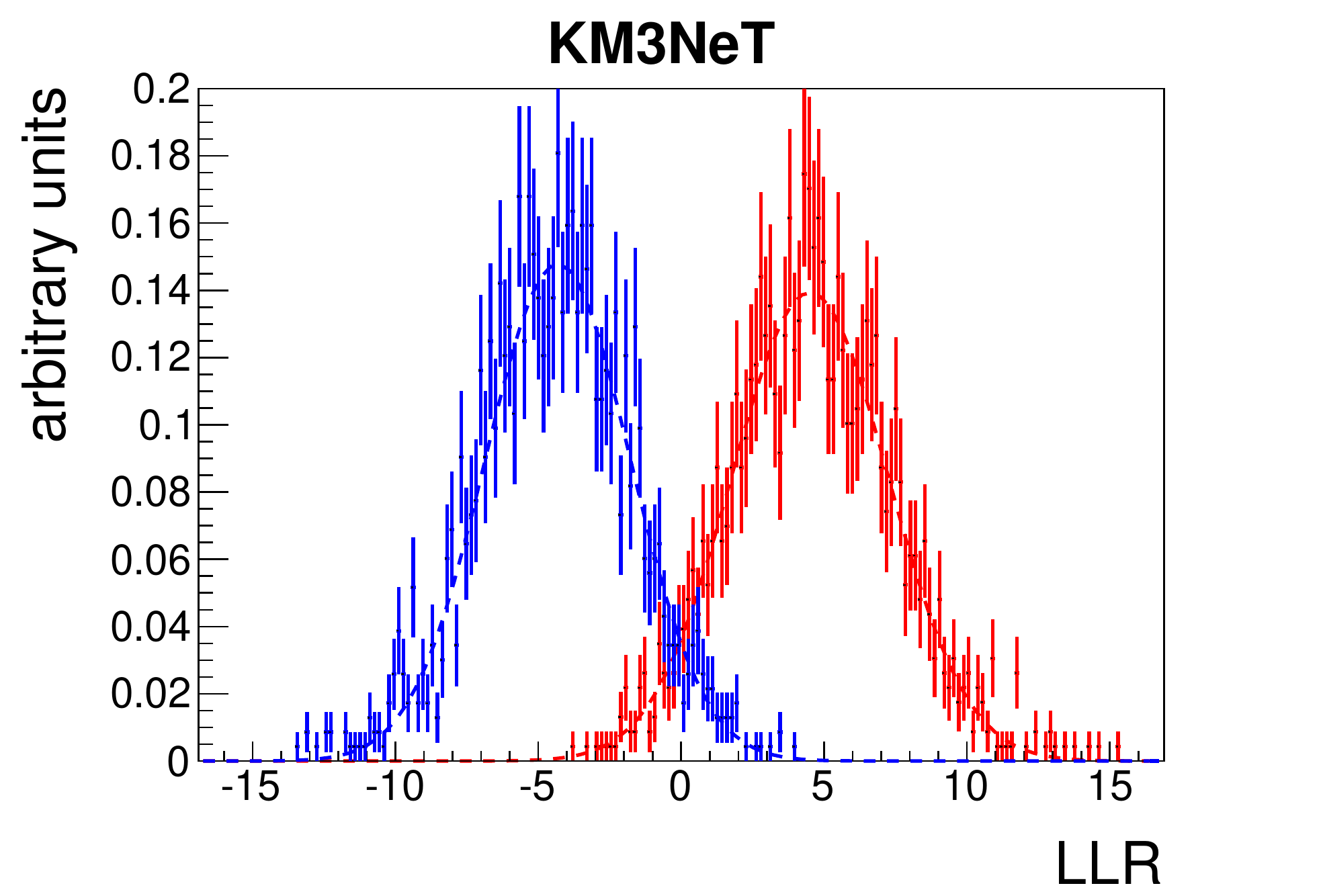}  
 \end{center}
 \caption{Example of log likelihood ratio (LLR) distributions for true NH (red) and true IH (blue) pseudo-experiments. The dashed curves represent Gaussian fits and the dashed vertical lines mark the median of the distributions.}\label{fig:LLR_example}
\end{figure}

The final figure of merit is the \emph{median significance}, computed by comparing LLR distributions for true NH and true IH PEs. An example is shown in \myfref{fig:LLR_example}.  The further these two distributions are apart, the higher the significance. We quote the significance with which the `wrong hierarchy' can be excluded at the median of the `true hierarchy' LLR. In all our simulations the LLR distributions can be excellently approximated by Gaussians. This allows the median significance to be expressed in the following simple form:
\begin{equation}
  S_\text{NH} := \frac{\mu_\text{NH}-\mu_\text{IH}}{\sigma_\text{IH}}.
\end{equation}
This gives the median significance in $\sigma$'s to exclude the IH, given true NH, where the $\mu$'s and $\sigma$'s here refer to the means and widths of the LLR distributions. This number can be interpreted as the minimum significance in $\sigma$'s with which the wrong hierarchy can be excluded in at least half of the pseudo-experiments.

\subparagraph{Alternative Hypothesis}
Initially, the mass hierarchy sensitivity was calculated by comparing LLR distributions generated with identical true oscillation parameter values (other than the hierarchy). However, this approach does not take into account the strong correlation between the measurement of $\theta_{23}$ and the hierarchy. From simulations it follows that the best-fit value of $\theta_{23}$ depends strongly on the assumed hierarchy. In many cases, the best-fit values for the two hierarchy assumptions are not in the same octant. In the actual measurement we therefore have to distinguish between two cases: NH with some best-fit value $\theta_{23}^\text{NH}$ and IH with a \emph{different} best-fit value $\theta_{23}^\text{IH}$. Since the two values can be very far apart, the mean and width of the corresponding LLR distributions can be noticeably different, leading to a different mass hierarchy sensitivity. Note that this effect does not occur for the other parameters, which typically have very similar best-fit values for the two hierarchy assumptions.

\begin{figure}
 \begin{center}
  
\includegraphics[width=0.49\textwidth]{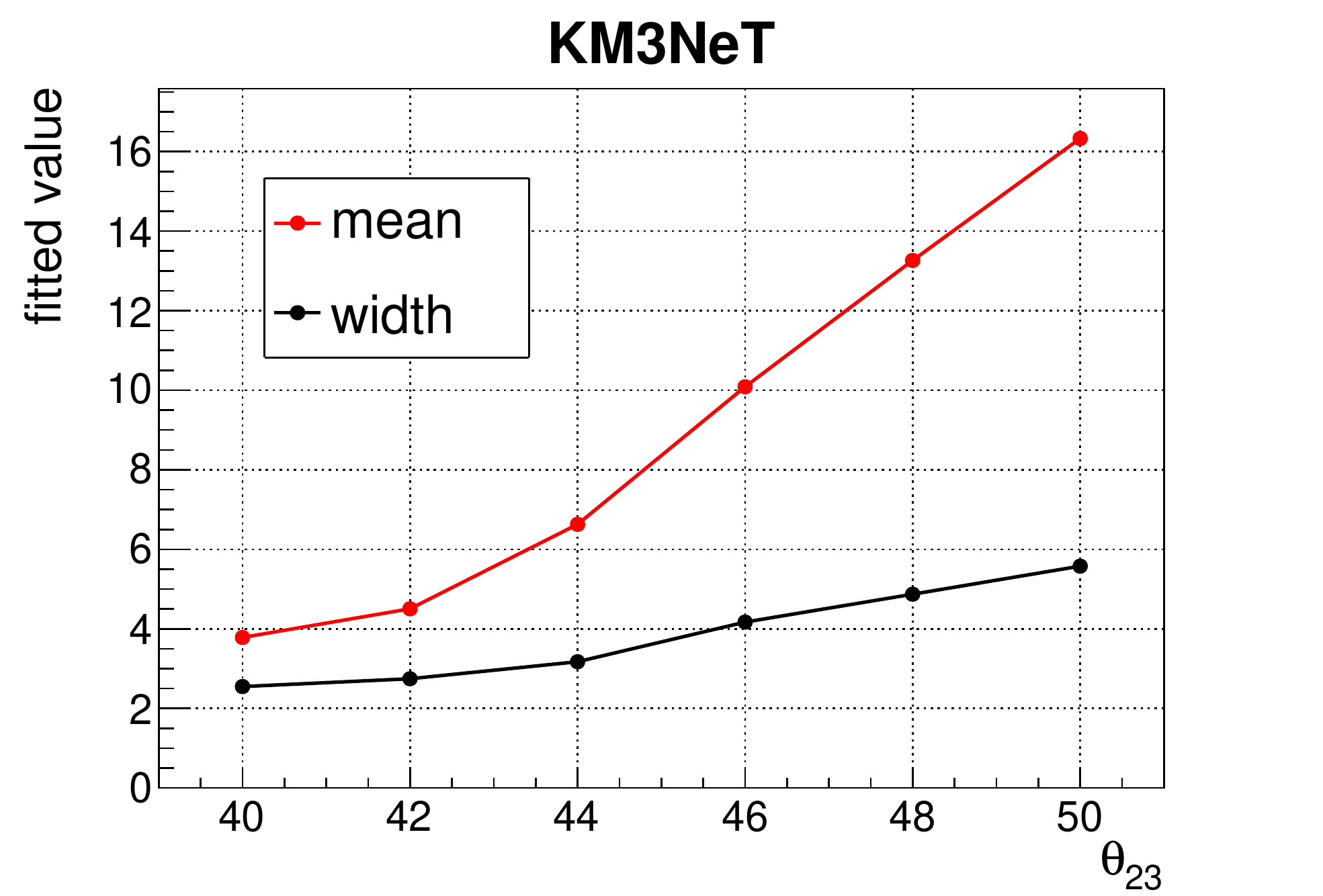}  
\includegraphics[width=0.49\textwidth]{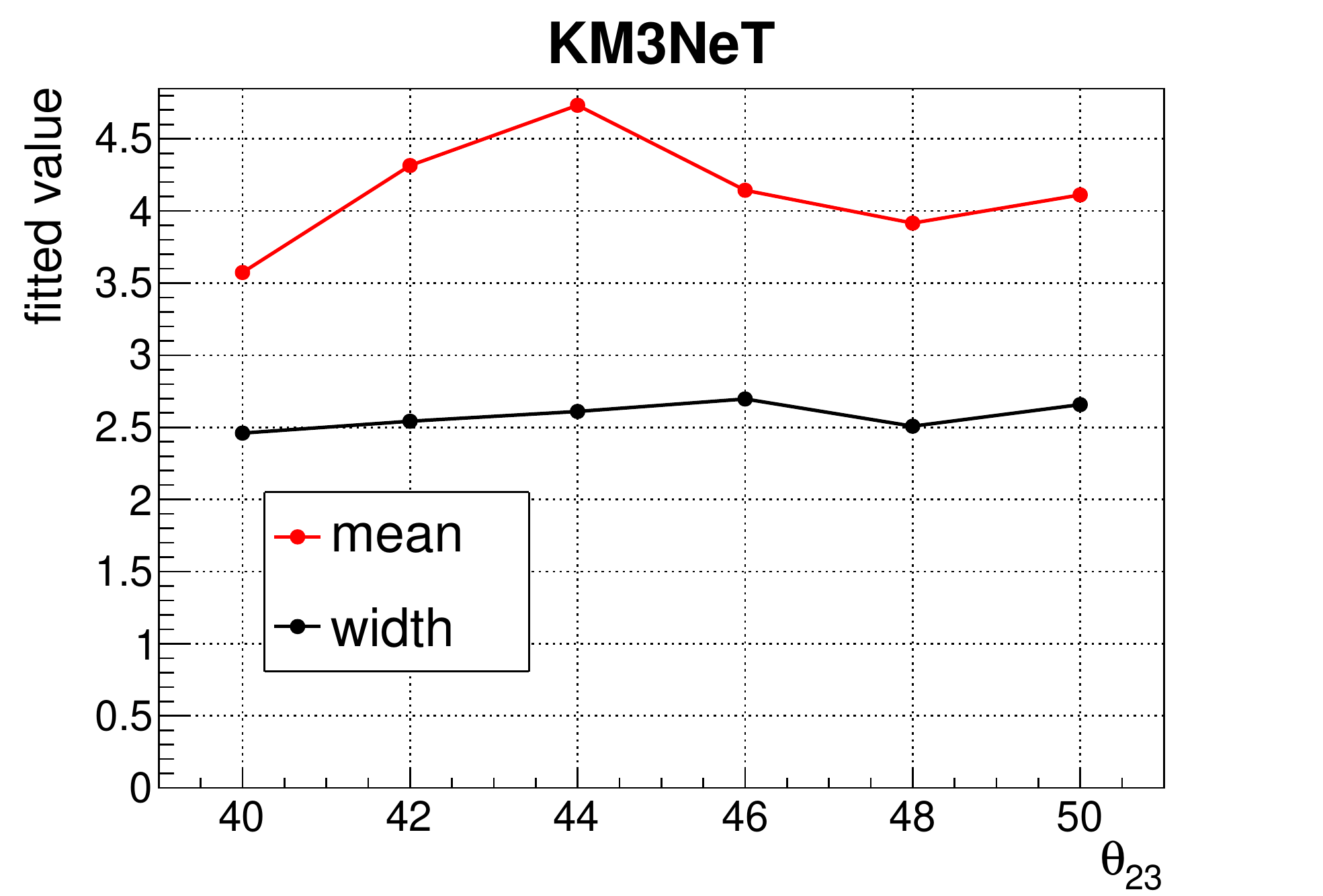}  
 \end{center}
 \caption{Mean and width of the LLR distributions for NH (left) and IH (right) pseudo-experiments as a function of $\theta_{23}$. These values were obtained from a Gaussian fit of the distributions.}\label{fig:LLRdistfits}
\end{figure}

\begin{figure}
 \begin{center}
   \includegraphics[width=0.49\textwidth]{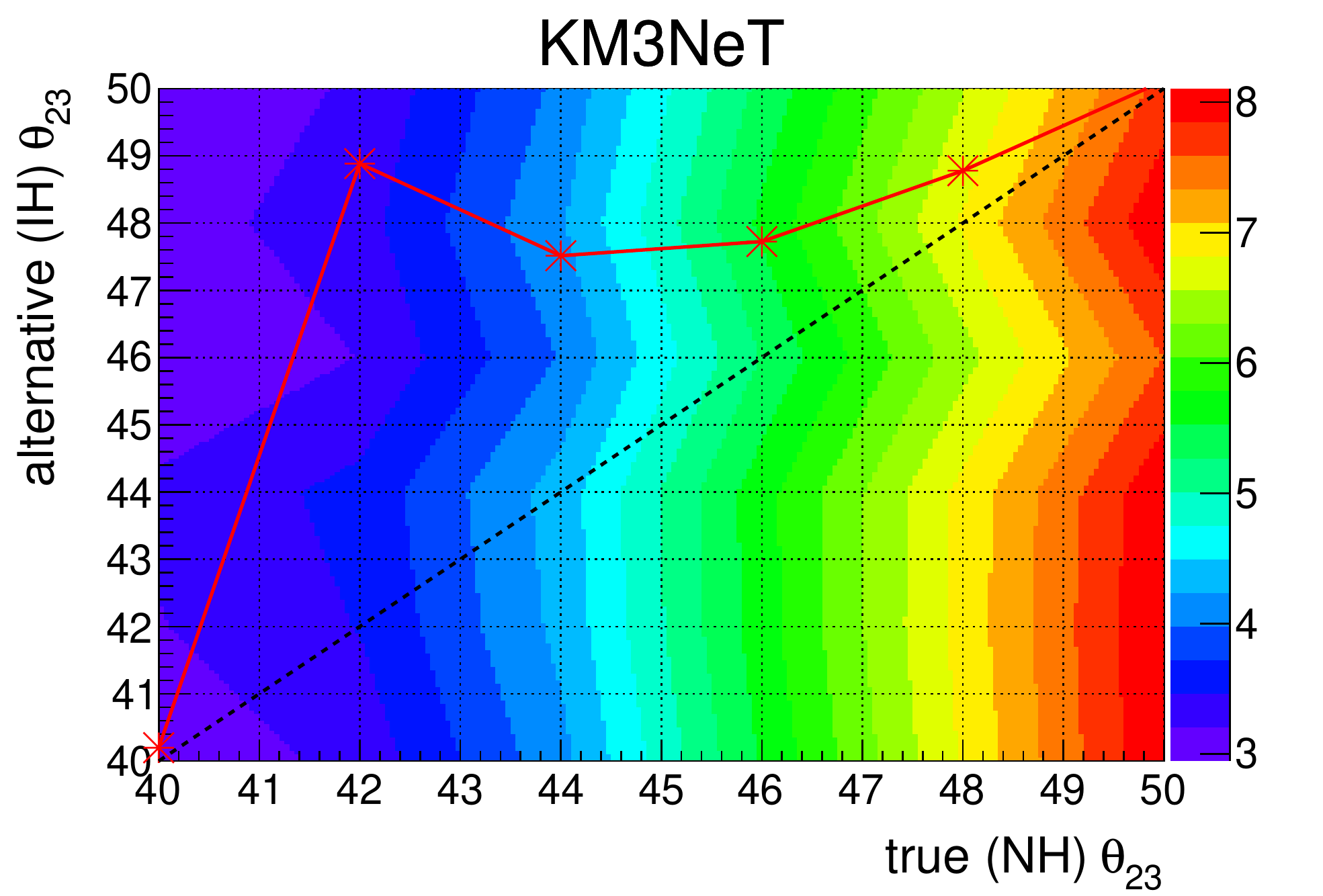}  
   \includegraphics[width=0.49\textwidth]{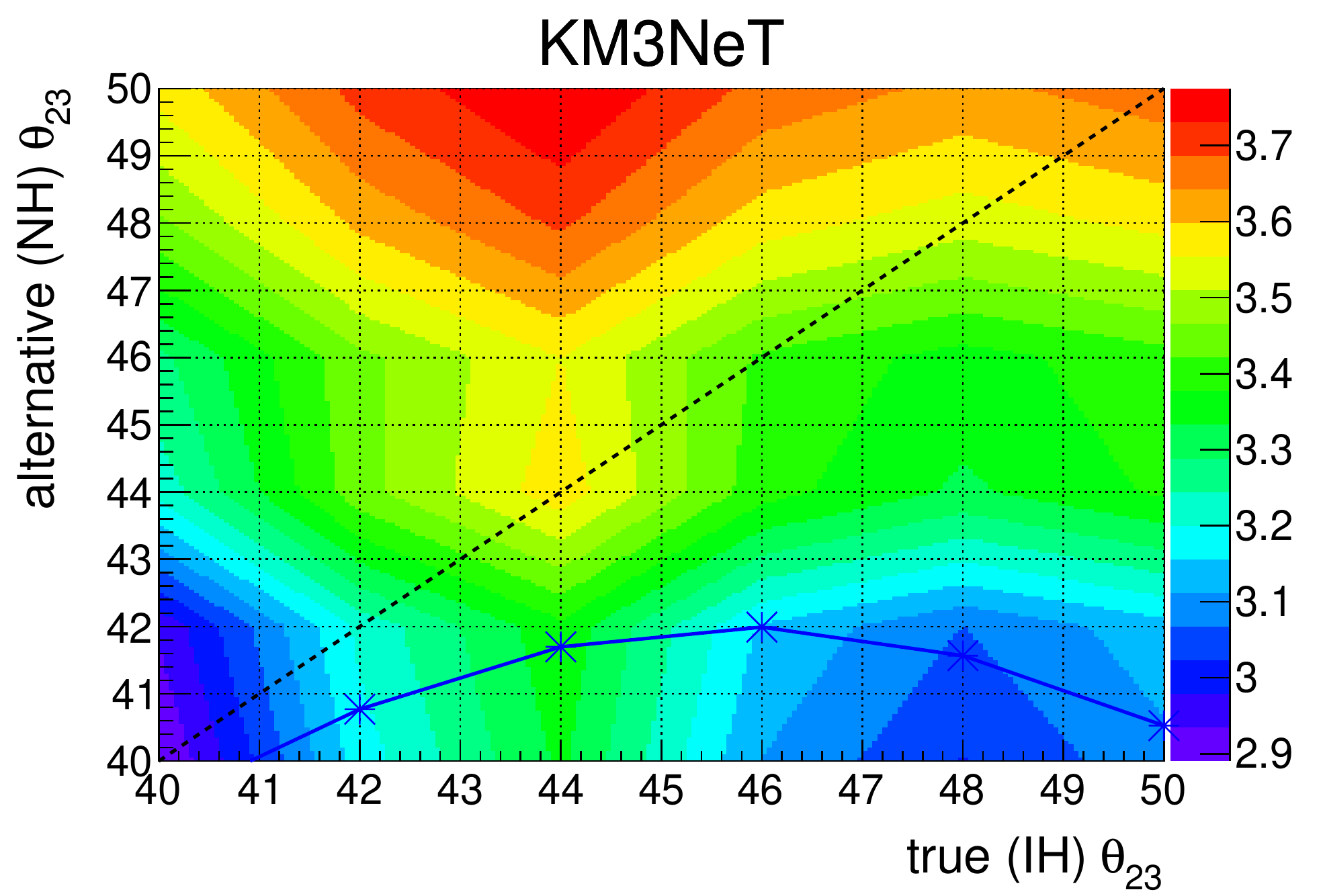}  
 \end{center}
 \caption{The mass hierarchy sensitivity for the true normal (left) and inverted (right) hierarchy. The horizontal axis indicates the true value of $\theta_{23}$. The vertical axis indicates the `alternative' value of $\theta_{23}$: the value belonging to the hypothesis that is being rejected. The diagonal dashed lines indicate the position where the alternative $\theta_{23}$ is the same as the true one. Along these lines the mass hierarchy sensitivity according to the original method can be read off. The solid red and blue lines show the most likely alternative value for each true $\theta_{23}$. They are the most likely value when fitting $\theta_{23}$ under the wrong hierarchy assumption. Along these lines the mass hierarchy sensitivity according to the new method can be read off.}\label{fig:alternativehypoplot}
\end{figure}

To take this effect into account the following procedure was adopted. For each true hypothesis (true hierarchy TH with $\theta_{23\text{,true}}$) the most likely \emph{alternative hypothesis} (other hierarchy OH with $\theta_{23\text{,alt}}$) is determined from the $\theta_{23}$ best-fit distribution of PEs generated with the true hypothesis and fitted assuming the OH. The median significance to reject the alternative hypothesis is then calculated:
\begin{equation}
  S_\text{true} := \frac{|\mu_\text{true}-\mu_\text{alt}|}{\sigma_\text{alt}}
\end{equation}
A technical issue arises because the LLR distributions were only simulated for certain given values of $\theta_{23}$, while the alternative hypothesis $\theta_{23}$'s can take any value. To overcome this we notice that the LLR distributions' fitted widths and means as a function of $\theta_{23}$ look rather smooth, so that they can be reasonably approximated by interpolating between the already calculated values. This is shown in \myfref{fig:LLRdistfits}. This method enables us to calculate the mass hierarchy sensitivity for any value of the true and alternative $\theta_{23}$. \myfref{fig:alternativehypoplot} illustrates the values of the alternative $\theta_{23}$ and the effect on the mass hierarchy sensitivity. All the LLR-method mass hierarchy sensitivity results in this document are produced using this method, unless explicitly stated otherwise.

\begin{figure}
 \begin{center}
   \includegraphics[width=0.8\textwidth]{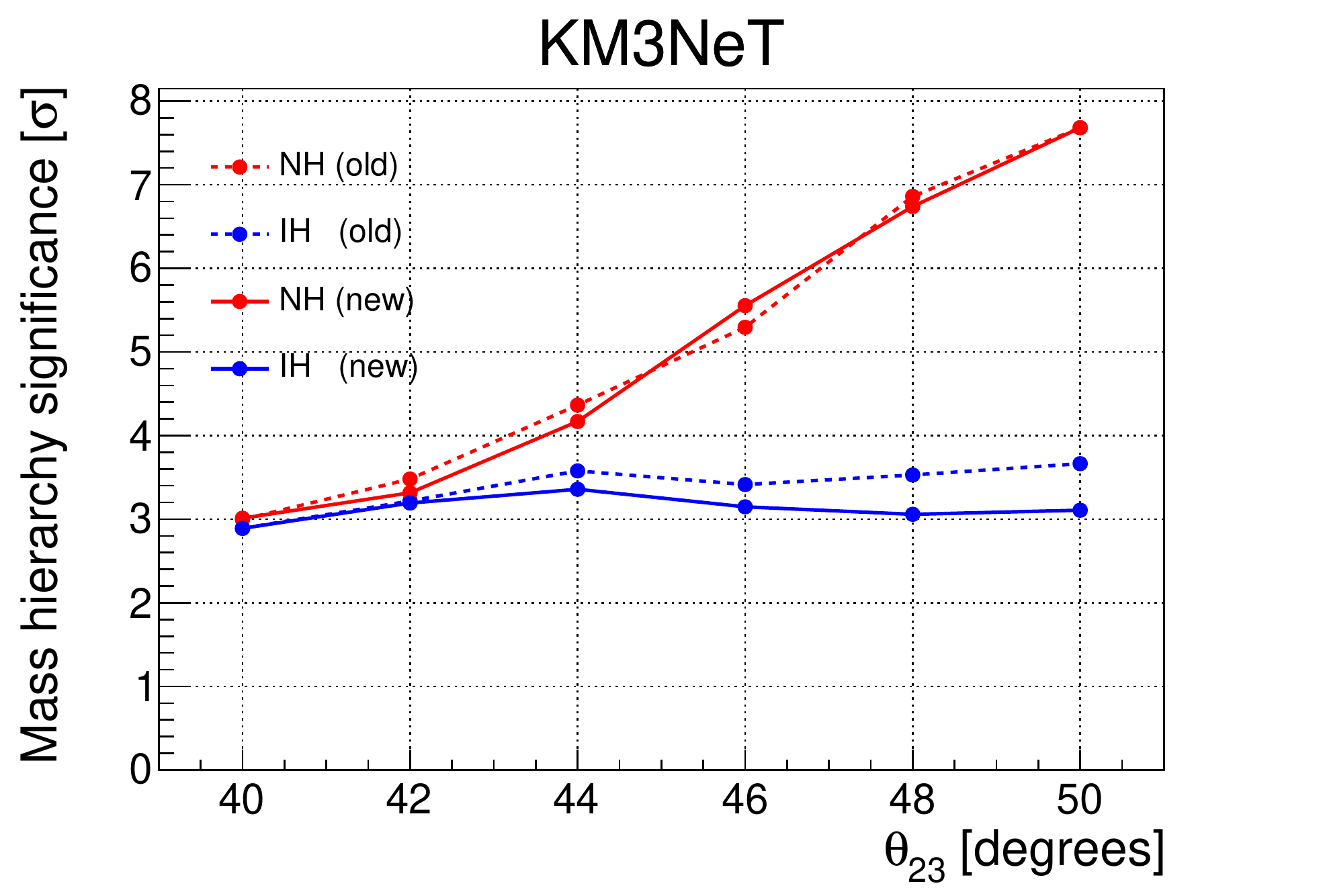}  
 \end{center}
 \caption{Comparison of the mass hierarchy sensitivity calculated using the old method (dashed lines) and the new method (solid lines). In the former, the significance is calculated to reject the other hierarchy at the same $\theta_{23}$, whereas in the latter the alternative hypothesis has a different $\theta_{23}$. The differences are rather small, but there is a noticeable decrease in the second octant IH mass hierarchy sensitivity. This is for the 9\,m spacing and three years of operation time, using the default settings (particularly, $\delta_\text{CP,true}=0^\circ$).}\label{fig:methodComparison}
\end{figure}

\myfref{fig:methodComparison} shows the effect of the new method on the mass hierarchy sensitivity. 

\paragraph{Results}

\begin{figure}[htp]
  \begin{center}
  \begin{overpic}[width=0.49\textwidth]{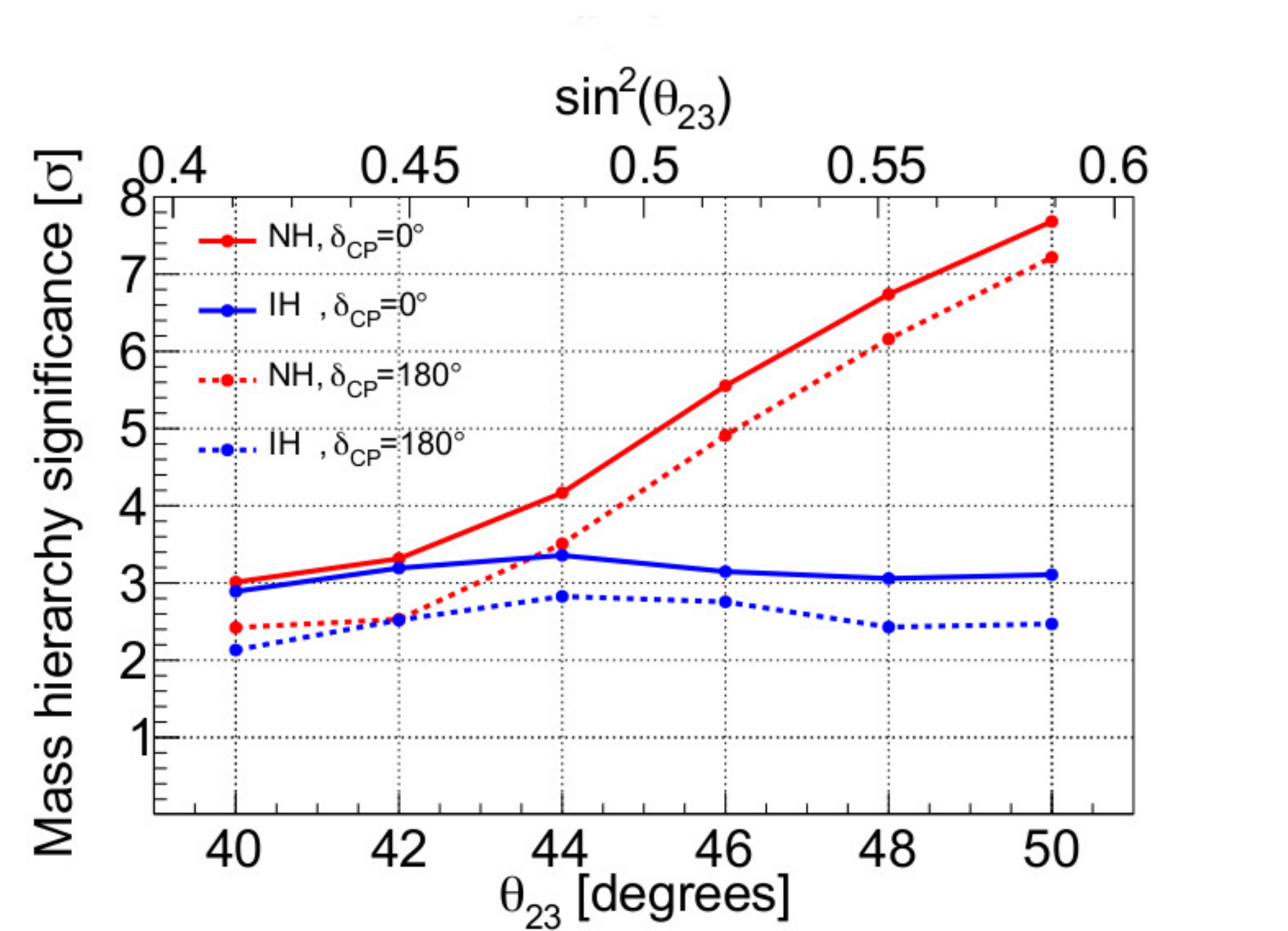}
  \put (20,12) {\bf KM3NeT}
\end{overpic}
  \begin{overpic}[width=0.49\textwidth]{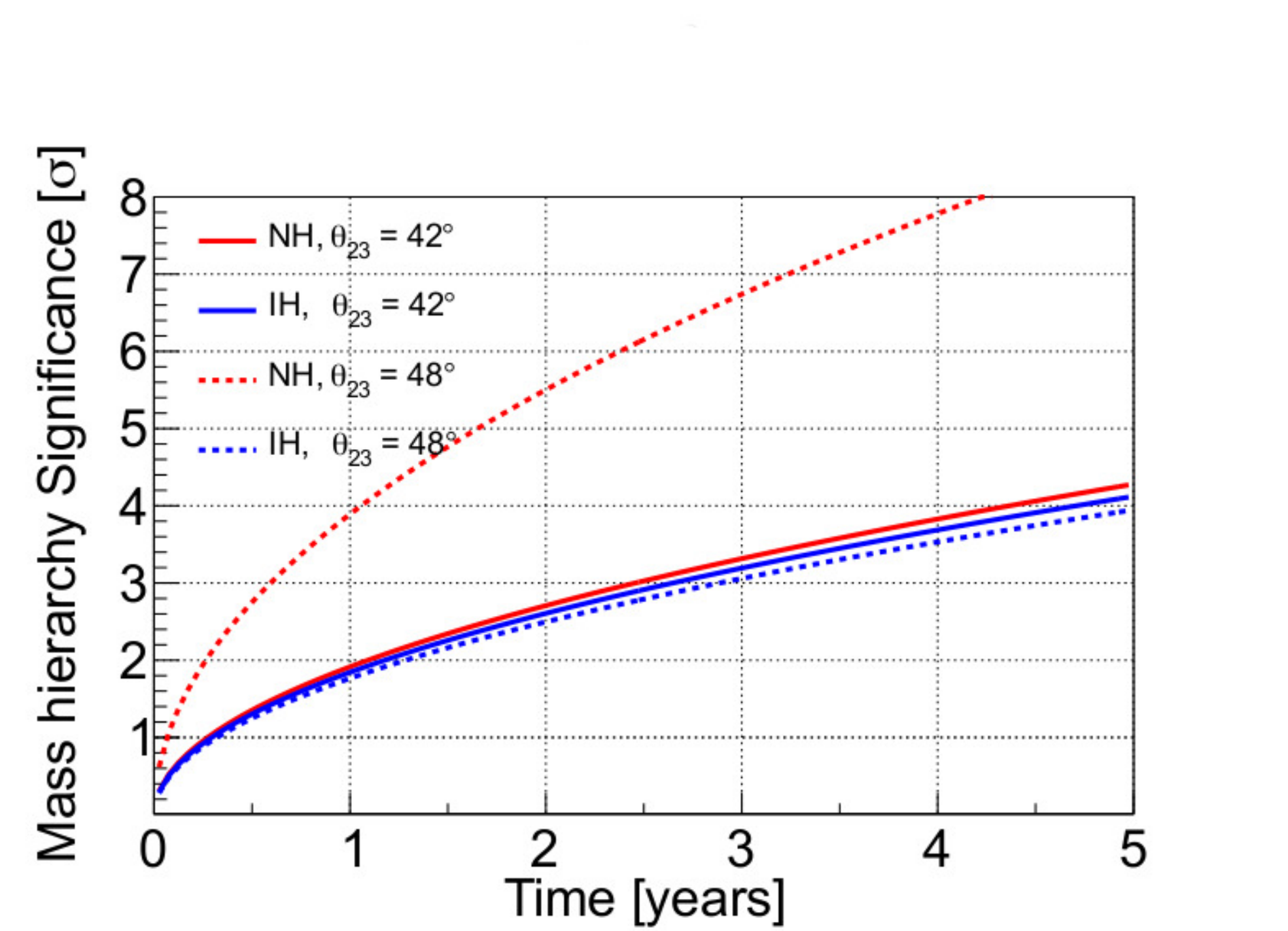}
   \put (30,12) {\bf KM3NeT}
  \end{overpic}
  \caption{The mass hierarchy sensitivity for 9\,m spacing, 
  using the default settings. This includes a fit of $\theta_{23}$, $\Delta M^2$, $\delta_\text{CP}$ and the five systematics. The left plot shows its dependency on $\theta_{23}$ for two values $\delta_\text{CP}$ for three years of operation time whereas the right plot illustrates its improvement over time for two selected values of $\theta_{23}$ and $\delta_\text{CP}=0^\circ$.}\label{fig:MHS}  
  \end{center}
\end{figure}

\myfref{fig:MHS} shows the latest mass hierarchy significance plot. The expected significance depends strongly on the true value of $\theta_{23}$ and $\delta_\text{CP}$. Without CP-violation, the neutrino mass hierarchy can be measured with more than 3$\sigma$ in three years at the current world best fit values of $\theta_{23}$

\subsubsection{Spacing studies}

Whereas the LLR method (described above) provides the most accurate description  
of the planned experiment, its application to certain problems might sometimes be
prohibitive due to the large number of pseudo-experiments to be generated. 
Therefore a simplified approach is used to answer dedicated questions. 

\begin{figure}
 \centering
    \includegraphics[width =0.8\textwidth]{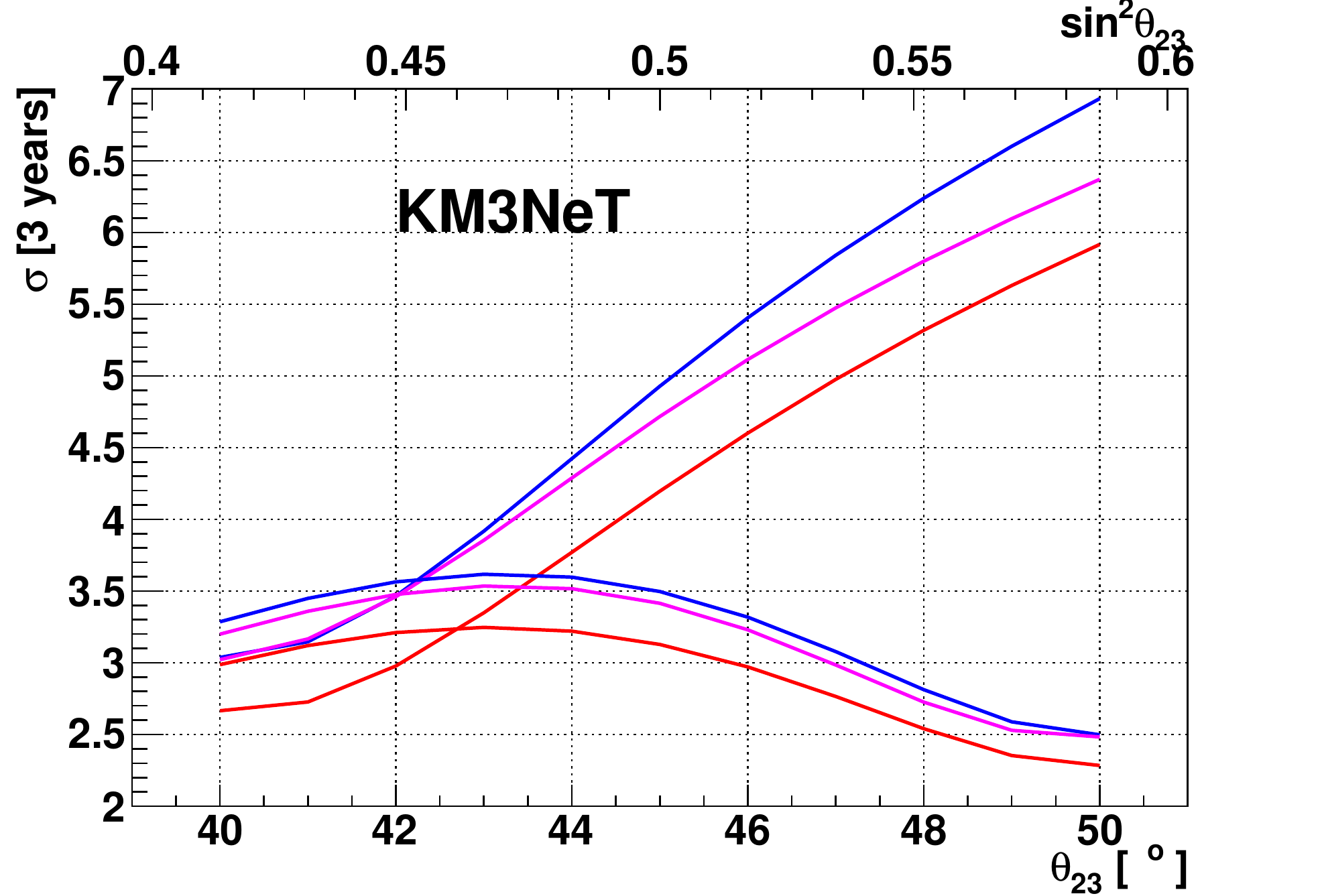}
\caption{NMH Sensitivity for 3 years of data taking as function of the true mixing angle $\theta_{23}$ for both
hierarchy hypotheses. Different vertical distances between adjacent DOMs are simulated: red 6\,m, blue 9\,m, magenta 12\,m.}
\label{spacing}
\end{figure}

The starting point is again the set of two histograms (for
tracks and showers) in the reconstructed quantities $E_{reco},\theta_{reco}$.
In each bin $i$, the expected number of events $(\mu_i^{TH})$ for a given 
true hierarchy (TH) hypothesis is calculated. A $\chi^2$ minimisation is performed
assuming the wrong hierarchy (WH) marginalising over the parameters given 
in \mytref{tab:parsettings}. Contrary to the description in \mytref{tab:parsettings},
$\theta_{12}, \theta_{13}$
and $\Delta m^2$ are fixed to their true values, whereas all other 8 parameters are fitted
unconstrained, {\it i.e.} without adding any priors. The true value of the CP-phase is fixed to $0$. 
As a result of the minimisation, a $\chi^2_{min}$ is obtained
\begin{equation}
\chi^2_{min} = \sum_i \frac{(\mu_i^{TH}-\mu_i^{WHfit})^2}{\mu_i^{TH}}
\end{equation}
which determines the probability to refute the WH hypothesis. It can be simply expressed as
\begin{equation}
\sigma = \sqrt{\chi^2_{min}}
\end{equation}
It has been verified, that the results obtained with this method are generally rather close to those from the full
LLR treatment. The simplified method is used to optimise the vertical distance of the DOMs on the
DUs. Whereas the horizontal spacing between DUs is determined by deployment constraints 
(20m distance between DUs is considered a minimum), the vertical distance is a free parameter with little
constraints from a technical point of view. Simulations have been performed with DOM distances of 6\,m, 9\,m and 12\,m.
The detector performance for these different setups have been shown before. \myfref{spacing} shows the expected
NMH sensitivity after three years of data taking for both hierarchy hypotheses as function of the true mixing angle 
$\theta_{23}$. An optimal distance is found close to 9\,m, as both for 6\,m and 12\,m the NMH 
sensitivity degrades, at least in some regions of the parameter space. 

\subsubsection{Measurement of $\Delta M^2$ and $\theta_{23}$}  

\begin{figure}
 \centering
    \includegraphics[width =0.49\textwidth]{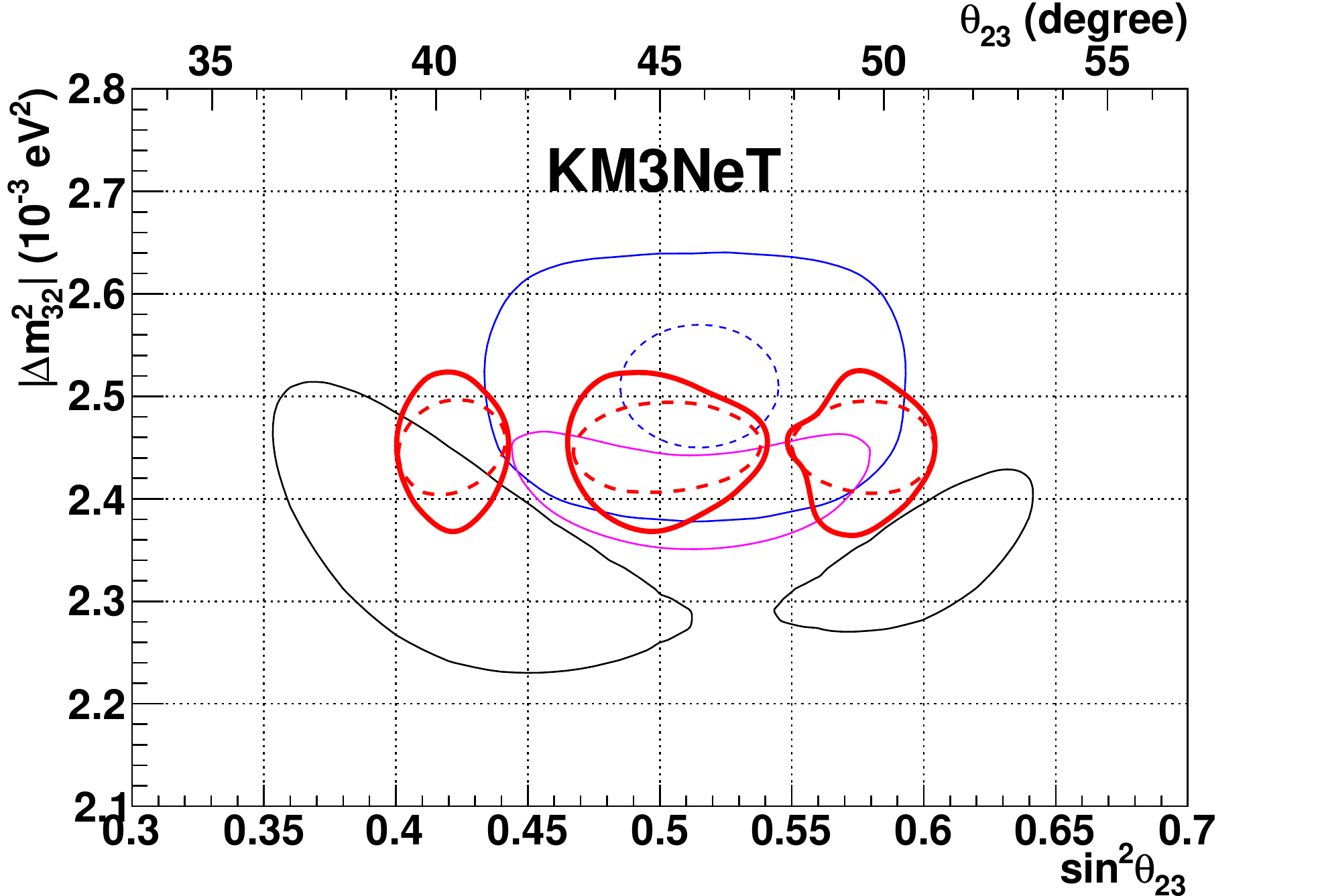}
    \includegraphics[width =0.49\textwidth]{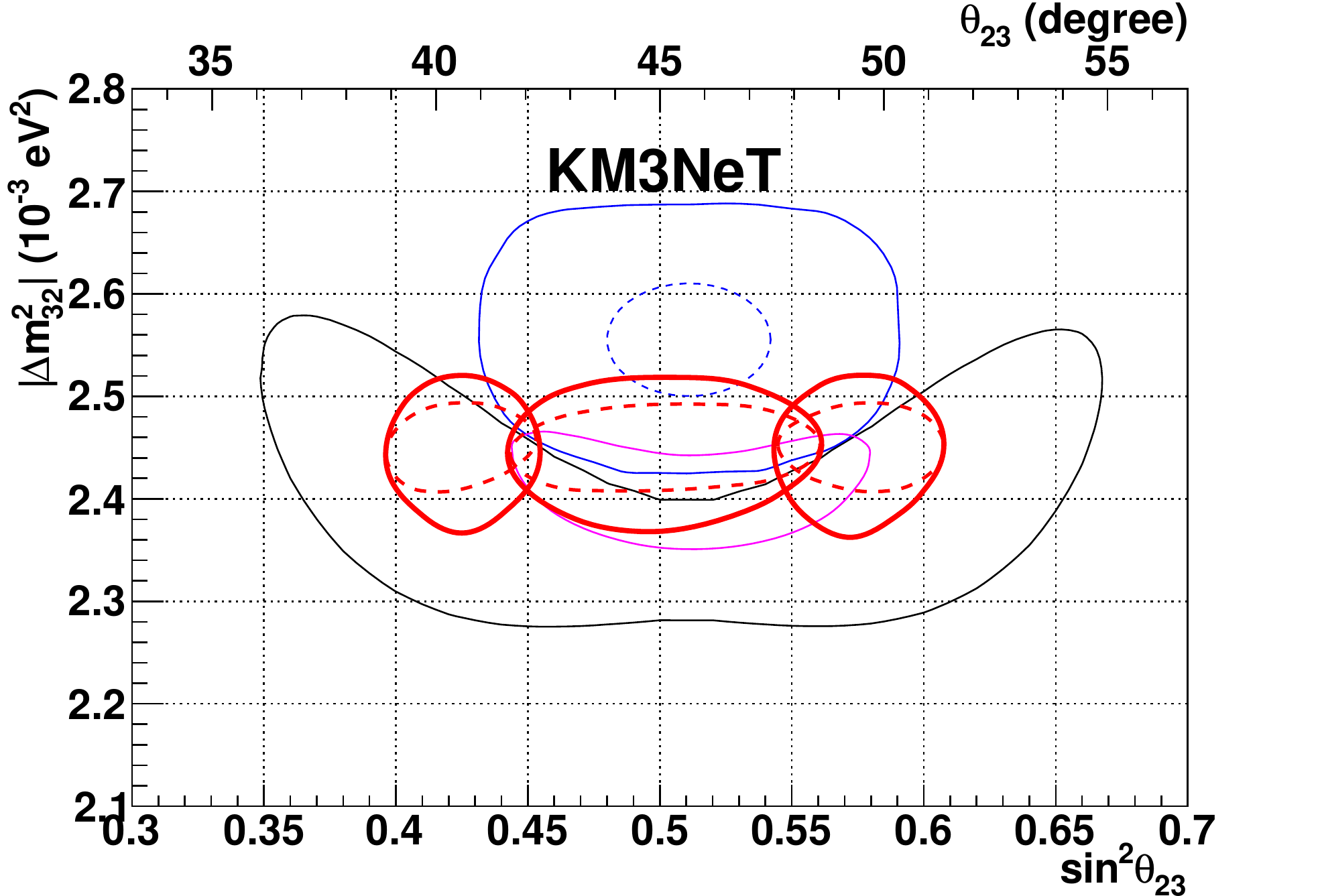}
\caption{Measurement precision in $\Delta M^2$ and $\sin^2\theta_{23}$ after three years of data taking with ORCA with
(solid red) and without (dashed red) energy scale uncertainty
for three test points compared to present results from MINOS~(black)~\cite{minos} and T2K~(blue)~\cite{t2k} and predicted performance
of NOvA~(magenta)~\cite{nova} and T2K~(blue, dashed)~\cite{t2k} in 2020. All contours are at $1\sigma$, left for NH, right IH.}
\label{dm-sin-measure}
\end{figure}

The derivation of measurement contours for the oscillation parameters is done
as well with the simplified procedure, which had been used already
for the spacing study. The same set of nuisance parameters is applied. Optionally an energy scale shift is added
as additional systematic uncertainty. It is implemented as a free scaling of the neutrino energy 
in all detector related distributions such as effective mass, particle identification, angular and energy resolution.
All nuisance parameters are fitted unconstrained, {\it i.e.} without priors.
Both $\Delta M^2$ and $\theta_{23}$ are determined under the assumption that the correct NMH has been already identified.
The $1\sigma$ measurement contours obtained after three years of data taking for three test points 
($\Delta M^2= 2.45\ 10^{-3} \mbox{eV}^2, \sin^2\theta_{23} = 0.42, 0.50, 0.58$) 
are shown on \myfref{dm-sin-measure}.
They are compared to current world best measurements~\cite{minos,t2k} as well as to extrapolations 
of final results from NOvA and T2K~\cite{nova,t2k}, to be expected around 2020. For T2K, the extrapolation is performed
by exploiting the published likelihood shape of the present measurement~\cite{t2k} assuming the planned complete 
beam exposure of $7.8\ 10^{21}$ protons on target.
A precision of $3\%$ in $\Delta M^2$ is reached after three years which can be reduced to $2\%$ when suppressing the energy scale uncertainty.
The precision in $\theta_{23}$ varies between 4\% and 10\%, depending on its true value and the NMH.

\subsubsection{Systematic uncertainties}
\label{syst}
A substantial list of possible uncertainties is already taken into account while fitting the NMH by marginalising
over the related nuisance parameters, as indicated in \mytref{tab:parsettings}. 
Some of these parameters -- such as $\theta_{23}$ and $\Delta M^2$ -- can be determined together with the NMH 
with high accuracy, as shown above.

It is crucial to determine reliable priors for the chosen nuisance parameters. The currently used priors are
listed as well in \mytref{tab:parsettings}. However, it has been verified that loosening the prior conditions or even 
totally suppressing them has only a small impact on the final NMH sensitivity. Therefore, in future studies 
some of them might by treated as unconstrained fit parameters, {\it i.e.} without priors.

Contributions to the uncertainties come from the neutrino flux~\cite{honda1}, cross section~\cite{formaggio}
and from the detector performance. For the latter, a main contribution is expected from the
uncertainty in the photon detection efficiency by the PMTs and the related readout
electronics. However, as demonstrated in ANTARES and also with the KM3NeT prototype 
module~\cite{km3net-ppmdom-2014}, the measurement of $^{40}$K coincidences between adjacent PMTs of the
same DOM allows the photon detection efficiency to be monitored in real time with high precision.
The variable nature of optical noise due to bioluminescence is controlled by sampling it for
each individual PMT with a frequency of 10\,Hz. The results of these measurements are
directly injected into the simulation, as is done in ANTARES. This excludes
bioluminescence as a source of systematic uncertainty of any measurement. 
Apart from the optical noise due to bioluminescence, sea water is a very stable and
homogeneous medium, as monitored over many years by ANTARES. Current knowledge of its light propagation 
properties are discussed in \mysref{sec-sci-sys}. The residual uncertainties of quantities such as absorption
and scattering length have less effect (due to the closer spacing) for ORCA than for ARCA, and are
well-covered by the nuisance parameters discussed above, so no separate investigation
has been performed.

Additional systematic uncertainties, not yet included in the present study, comprise systematic shifts
in the reconstructed energy and zenith angle. These will be considered in the near future.
However, it is believed that the energy scale is well-constrained through the knowledge of the absolute PMT
efficiency and the water parameters.   
The angular resolution of neutrino telescopes in sea water is excellent, and it remains
better than 10 degrees down to the energies relevant for the NMH determination.
Systematic angular offsets are at most in the sub-degree region, as shown by the study of the moon shadow 
in the cosmic ray signal in ANTARES. A deterioration of the angular resolution due to the
movement of the detector elements in the sea current is excluded by permanently re-calibrating
them via an acoustic positioning system. Such a system provides a precision of better than
about 10\,cm for all detector elements, which makes its influence on the angular resolution of
reconstructed neutrino events negligible.
 
The energy and angular resolutions are a crucial input to the sensitivity calculation.
Both are estimated from simulations which are subject to uncertainties on their own.
These can be parametrised by applying a scaling 
({\it i.e.} broadening or narrowing) to these resolution functions, as is planned for the near future.  

Finally, an independent study has been performed to study the variations of the 
Earth model~\cite{bib:prem} on the NMH sensitivity. Both the thickness and density
of each individual layer have been varied within the tolerance of the model,
as well as the sharpness of the layer boundaries. The impact of these variations 
is found to be negligible for the present study and is therefore ignored as a relevant
systematic effect.

\subsection{Outlook}
\label{outlook}
\label{sec:outlook}

The previous sections provide details on the performances of the KM3NeT/ORCA detector in establishing the neutrino mass ordering and improving the precision on the oscillation parameters in the atmospheric sector. While the results obtained so far rely on full Monte Carlo studies and incorporate the leading systematic effects, possible refinements have already been identified and will be scrutinised in the near future. This includes notably the usage of the achieved sensitivity to the interaction inelasticity as a statistical tool to discriminate between neutrino and antineutrinos and to further reject the background from neutral current interactions. As a detailed and ongoing investigation shows, see \cite{ShowerFluct} for first results, event-by-event fluctuations intrinsic to the development of the hadronic system resulting from a neutrino interaction on a nucleon or nucleus limit the achievable resolutions on direction, energy and inelasticity for a given detector geometry. However, some improvements, in particular for the reconstruction of the reaction inelasticity, can still be expected to arise from the development of new reconstruction strategies. Envisaged lines of work comprise e.g. an attempt to identify the leading particle in the hadronic system to constrain its overall momentum, a combined fit of the hadronic shower and the charged lepton in CC interactions, improved energy estimation techniques, etc. 

In addition to the oscillation measurements discussed in the previous section, the size and energy range covered by the KM3NeT/ORCA detector allow for the search of CC interactions of tau neutrinos produced in the oscillation of atmospheric electron and muon neutrinos. While these events can hardly be distinguished on an event-by-event basis, their presence could be revealed by a statistical excess of cascade-like events over the baseline from atmospheric NC interactions and electron neutrino CC interactions. This effect is expected to be seen with high confidence level and statistical power within the first years of operation, but a precise study remains to be carried out.
The energy and flavour distributions of observed events in the ORCA detector could in principle also reveal sizeable discrepancies from expectations due to non standard physics interactions (NSI)~\cite{ohlsson_NSI,gonzalez_NSI}. While strong deviations from expectations (e.g enhanced CP violation effects) might deteriorate the sensitivity to the NMH, a more likely scenario is that KM3NeT/ORCA will be able to invalidate many NSI processes.

The studies presented here indicate that the current unknown value of the Dirac CP violating phase in the neutrino sector mildly impacts the sensitivity to the neutrino mass ordering. However, the knowledge of the mass ordering could reversely bring sensitivity to the CP phase in the (0.2 -- 1) GeV regime~\cite{bib:razzaque}. This would imply a denser instrumentation than what is currently envisaged for ORCA, but considering the importance of measuring the CP phase, sensitivity studies could be performed for a further step of the ORCA project. In the same spirit, sensitivity studies for both the NMH and the CP phase have been proposed relying on a putative upgraded neutrino beam to be sent to ORCA from Protvino~\cite{brunner_beam,brunner_beam2}. Such a strategy would in particular allow for a confirmation of ORCA-only results on the NMH with high statistical power on a short (< 1 yr) timescale~\cite{bourret_beam}. It would require a new beam-line to be setup but would offer the advantage to rely on an already built detector.
 
	With its low energy threshold the KM3NeT/ORCA detector offers the possibility to extend searches started with ANTARES (e.g~\cite{antares-dm-sun,antares-dm-gc}) (and likely to be pursued with KM3NeT/ARCA as well) for extra-terrestrial neutrinos as a signature of the presence of Dark Matter in the centre of the Earth, the Sun and the central region of the Galaxy for which the detector is particularly well located. The low energy threshold of KM3NeT/ORCA is particularly well suited to constrain low-mass WIMP (Weakly Interacting Massive Particle) Dark Matter models. All neutrino flavours could be used for such studies, considering the encouraging first performances in the shower reconstruction channel. 
	
	GeV neutrinos are also likely to be emitted by several classes of astrophysical objects like low-energy Gamma-Ray Bursts~\cite{asano_grb} or Colliding Wind Binaries~\cite{becker_cwb}. Another promising topic is the ability of KM3NeT/ORCA to detect neutrinos from supernovae (SN) explosions.  The use of segmented optical modules closely placed to one another indeed offers new detection capabilities: asking for coincidences of many phototubes on individual storeys is expected to strongly reduce the optical background potentially providing high sensitivity to SN up to few tens of kpc.
These results will possibly be incorporated in an update of the present document, together with the prospects for several other physics studies that can be undertaken with KM3NeT/ORCA. These  span a wide range of scientific fields, including the Earth and Sea Sciences which are not addressed here but are part of the scientific scope of deep-sea neutrino observatories. As an example, a detailed study of the neutrino energy and angular distributions could provide tomographic information on the electron density~\cite{winter_tomo1,rott_tomo,winter_tomo2}, and thus on the composition, of the different Earth layers traversed. Such an approach is  complementary to the standard methods used in geophysics, which do not univocally constrain the chemical composition of the Earth, in particular of its innermost layers (mantle and core). 

\cleardoublepage
\section{Organisation}
\label{sec-org}

KM3NeT federates and unifies the various smaller European efforts in the field of Neutrino Astronomy. The process of convergence was supported by an EU funded Design Study (2008--2009) and Preparatory Phase (2008--2012). The KM3NeT consortium has now formed a collaboration with an elected management. The funding agencies (or funding authorities) involved have installed the Resources Review Board (RRB) which oversees the project. The RRB is advised by an international Scientific and Technical Advisory Committee (STAC). A project organisation is setup with the objective to implement the first phase (Phase-1) of the KM3NeT Research Infrastructure. To this end, a Memorandum of Understanding (MoU), covering the total available budget of about 31~M\euro, has been signed by the members of the RRB. The purpose of this MoU is to define the programme of work to be carried out for this phase and the distribution of charges and responsibilities among the Parties and Institutes for the execution of this work. The MoU sets out i) the organisational, managerial and financial guidelines to be followed by the collaboration, ii) the external scientific and technical review processes and iii) the user access policy. At present, the collaboration consists of more than 240 persons from 52 institutes. The first phase has already started and comprises the final prototyping and preproduction, engineering, construction, calibration, transportation, assembly, installation and commissioning of the elements which form the basis of the KM3NeT neutrino detector and the seafloor and shore station infrastructures as well as the operation of the installed neutrino detectors. The installation is proceeding in two places, off-shore Toulon, France and off-shore Capo Passero, Italy. A third suitable site is available off-shore Pylos, Greece. 
The construction of the Phase-1 detector has already started with the successful deployment of the first string off-shore Capo Passero and will be completed by 2017.

The Collaboration will offer open access for external users to the KM3NeT Research Infrastructure (Article 15 of the MoU). The KM3NeT Research Infrastructure will also provide user ports for continuous Earth and Sea science measurements in the deep-sea environment. The needs for the Earth and Sea sciences are partly incorporated in the present KM3NeT MoU and other needs will be detailed in designated MoUs between KM3NeT and individual Earth and Sea science groups or more generally with EMSO.

The Phase 1 MoU is a first step towards the intended establishment of a European Research Infrastructure Consortium (ERIC). The collaboration has agreed to host the KM3NeT ERIC in the Netherlands. The neutrino signal recently reported by IceCube has led the KM3NeT Collaboration to propose an intermediate phase (i.e. Phase-2.0). The required actions for the next phase(s) are being taken which include the preparation of requests for additional ERDF funds in France, Italy and Greece as well as requests for national funds. Other support options, e.g.\ within the framework of Horizon 2020, will also be explored.

Recently, the KM3NeT and ANTARES Collaborations have agreed to organise each general assembly jointly (typically 3--4 times per year). This agreement fosters the scientific progress and the exchange of know-how and limits travel times and expenses. Following a sequence of joint meetings between ANTARES (Mediterranean Sea), IceCube (South Pole), Lake Baikal (Russia) and KM3NeT (Mediterranean Sea), a Memorandum of Understanding for a Global Neutrino Network (GNN) has been signed on 15 October 2013 by the representatives of each project. This step formalises the active collaboration between these projects. Once infrastructures of similar scale are operational on the three continents, the stated aim of the GNN is a worldwide Global Neutrino Observatory.

\cleardoublepage
\section{Data Policy}
\label{sec-data-policy}

The KM3NeT Collaboration has developed a data policy based on the research, educational and outreach goals of the facility. 
The first exploitation of the data is granted to the collaboration members as they build,
maintain and operate the facility and to priority users.
Accordingly, each collaboration member has full access rights to all data, software and know-how. 
Access for non-members is restricted, as long as methods and results have not yet been published. 
The prompt dissemination of scientific results, new methods and implementations is a central goal of the project, as is education. 
High-level data (event information enriched with quality information) will be published after 
an embargo time of two years under an open access policy on a web-based service.
Exceptional access rights that correspond to these goals can be granted.

The Collaboration has developed measures to ensure the reproducibility and usability of all scientific results 
over the full lifetime of the project and in addition 10 years after shutdown. 
Low-level data (as recorded by the experiment) and high-level data will be stored in parallel at central places. 
A central software repository, central software builds and operation system images are provided and
continuously maintained until the end of the experiment.

The storage and computing needs of the KM3NeT project are highly advanced. 
The Collaboration has developed a data management plan and a corresponding computing model to answer those needs. 
The latter is based on the LHC computing models utilising a hierarchical data processing system with different layers (tiers). 
Data are stored on two main storage centres (CCIN2P3-Lyon, CNRS and CNAF, INFN); 
those large data centres are fully interfaced with the major European e-Infrastructures, 
including GRID-facilities (ReCaS, HellasGRID provide resources to KM3NeT). 
The main node for processing of the neutrino telescope data is the computer centre in Lyon (CCIN2P3-Lyon). 
A corresponding long-term and sustainable commitment has already been made by CNRS, 
which is consistent with the needs for long-term preservation of the data.
A specialised service group within the Collaboration will process the data from low-level to high-level and 
will provide data-related services (including documentation and support on data handling) to the Collaboration and partners. 
WAN (GRID) access tools (e.g. xrootd, iRODS, and gridFTP) provide the access to high-level data for the Collaboration. 
The analysis of these data will be pursued at the local e-Infrastructures of the involved institutes (both local and national). 
The chosen data formats allow for the use of common data analysis tools (e.g. the ROOT data analysis framework) and 
for integration into e-Infrastructure common services.

The central services are mainly funded through CNRS and INFN that have pledged resources of their main computing centres to the project. 
Additional storage space and its management are provided by the partner institutes 
(e.g. INFN has provided 500 TB of disk space for KM3NeT at the ReCaS GRID infrastructure, 
the Hellenic Open University has pledged 100 TB of disc space and 300 cores to the project). 

In addition to the major storage, networking and computing resources provided by the partner institutions and 
their computing centres, grid resources have been pledged and will be used by KM3NeT (ReCaS, HellasGRID). 
These will provide significant resources to be used for specialised tasks (as e.g. for special simulation needs). 
The major resources, however, will be provided by the partners.
External services are employed to integrate the KM3NeT e-Infrastructure into the European context of the GRID 
-- in the fields of data management, security and access; services will be implemented in collaboration with EGI. 

One of the aims of the KM3NeT data management plan is to play an active role in the development and 
utilisation of e-Infrastructure commons. 
KM3NeT will therefore contribute to the development of standards and services in the e-Infrastructures both 
in the specific research field and in general. 
In the framework of the Global Neutrino Network (GNN), KM3NeT will cooperate with the ANTARES,
IceCube and GVD collaborations to contribute to the open science concept by providing access to high-level
data and data analysis tools, not only in common data analyses but also for use by citizen scientists.

In the framework of the ASTERICS project, KM3NeT will develop an interface to the Virtual Observatory
including training tools and training programmes to enhance the scientific impact of the neutrino telescope
and encourage the use of its data by a wide scientific community including interested citizen scientists.
Data derived from the operation of the experiment (acoustics, environmental monitoring) will be of
interest also outside of the field. Designated documentation and courses for external users will therefore
be put in place to facilitate the use of the repositories and tools developed and used by the KM3NeT Collaboration.

\cleardoublepage
\section{Cost and Time Schedule}
\label{sec-cos}

The investment budget for the construction of the first phase (Phase-1) of the KM3NeT research infrastructure, 
which is fully funded, amounts to about 31\,M\euro. 
During 2015-2017, 31 strings equipped with 558 optical modules will be assembled and deployed at the French and Italian sites. 
The overall size of the initial Phase-1 arrays corresponds to about 0.2 building blocks. 

The next phase (i.e.\ KM3NeT 2.0) comprises a complete ARCA and ORCA detector, 
consisting of 2 and 1 building blocks, respectively.
The additional budget for KM3NeT 2.0 is estimated at 95\,M\euro. 
The cost estimates of KM3NeT 2.0 are based on the actual prices of Phase-1 and 
thus can be considered accurate. 

\begin{figure}[!hbt]
\centering
\includegraphics*[height=5cm,trim=5cm 4cm 5cm 4cm]{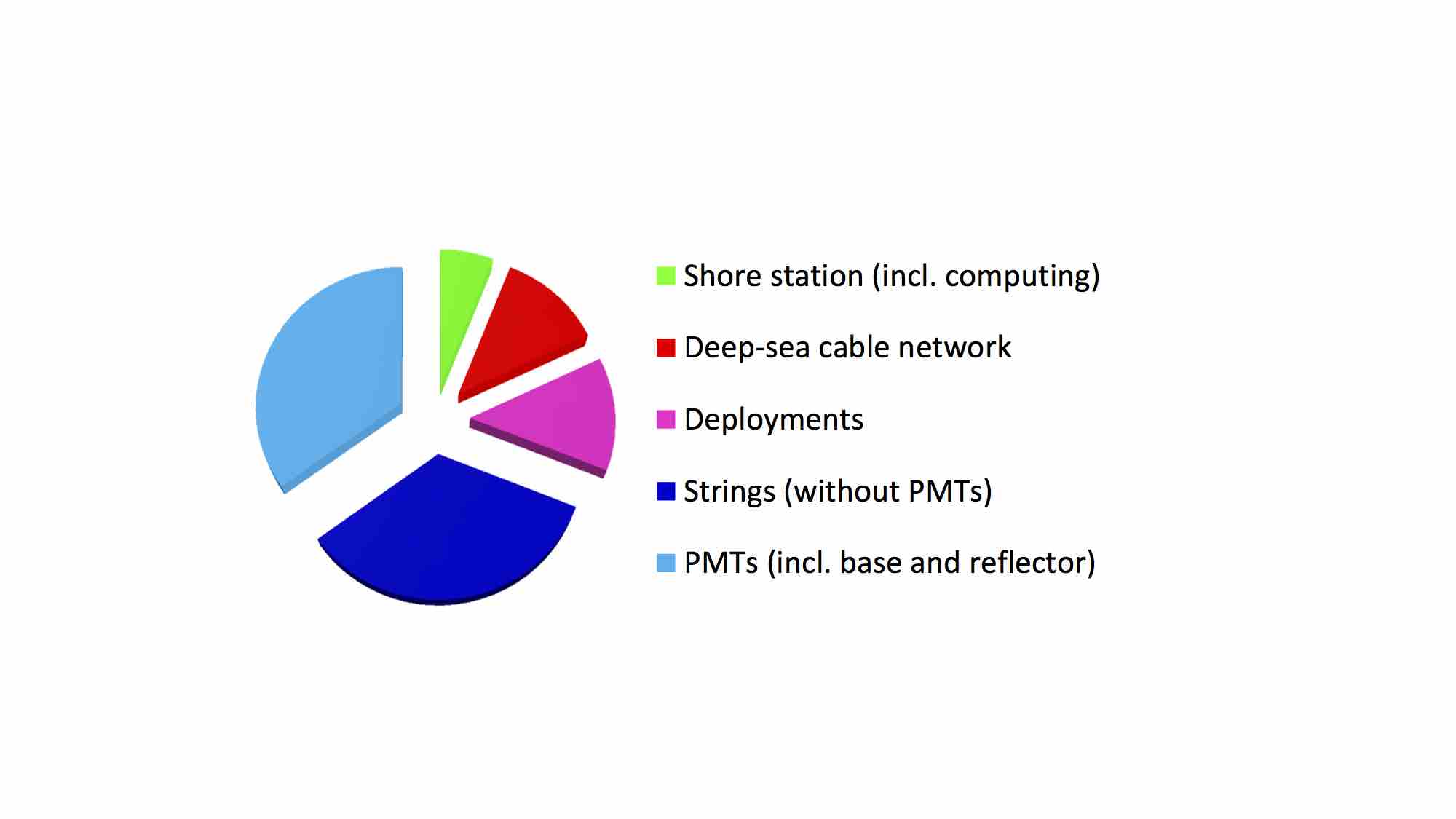}
\caption{Breakdown of costs amongst the major items. } 
\label{fig:costs}
\end{figure}

The breakdown of the cost amongst the major items is illustrated in \myfref{fig:costs}. 
They are consistent with the estimations stated in the KM3NeT Technical Design Report published in 2011 
and represent a factor of four cost reduction compared to that previously achieved for the ANTARES detector. 
The cost of a single KM3NeT string is about 230\,k\euro, 
an additional 90\,k\euro~ is needed for the interlink cable, the string deployment and the ROV connection. 

Once the funds for Phase-2.0 are available, the array could be constructed within three years. 
Thus, if funds were forthcoming in 2017, the full array could be completed in 2020. 
Note that physics studies would already be possible as the array is being constructed, 
thus reducing the overall time needed to obtain a specified precision.
The cost for operation and decommissioning of the infrastructure have been evaluated 
and amount to about 2\,M\euro~ per year and 5\,M\euro, respectively. 
Hence, the total cost for 10 years of operation and decommissioning of the KM3NeT 2.0 infrastructure 
adds about 25\% to the total budget.

\clearpage

\section*{Acknowledgements}

The authors acknowledge the financial support of the funding agencies:
Centre National de la Recherche Scientifique (CNRS), 
Commission Europ\'eenne (FEDER fund and Marie Curie Program),
Institut Universitaire de France (IUF),
IdEx program and UnivEarthS Labex program at Sorbonne Paris Cit\'e (ANR-10-LABX-0023 and ANR-11-IDEX-0005-02),
France;
The General Secretariat of Research and Technology (GSRT),
Greece;
Istituto Nazionale di Fisica Nucleare (INFN),
Ministero dell'Istruzione, dell'Universit\`a e della Ricerca (MIUR),
Italy;
Agence de  l'Oriental and CNRST,
Morocco;
Stichting voor Fundamenteel Onderzoek der Materie (FOM), Nederlandse
organisatie voor Wetenschappelijk Onderzoek (NWO),
the Netherlands;
National Authority for Scientific Research (ANCS),
Romania; 
Plan Estatal de Investigaci\'on (refs. FPA2015-65150-C3-1-P, -2-P and -3-P, (MINECO/FEDER)), Severo Ochoa Centre of Excellence and MultiDark Consolider (MINECO), and Prometeo and Grisol\'ia programs (Generalitat Valenciana), Spain.

The KM3NeT collaboration has received funding from the European Community Sixth Framework Programme
under Contract 011937 and the Seventh Framework Programme under Grant Agreement 212525.
\vspace*{0.5cm}


\cleardoublepage
%
%
\addcontentsline{toc}{section}{References}
{
\chardef\usc=95
\chardef\til=126
\bibliographystyle{common/LoI}
{\raggedright\fontsize{10.pt}{11.pt}\selectfont
\bibliography{common/LoI}}
}
\label{lastpage}
\end{document}